%% file: phys_150_all_8.tex
\documentclass[11pt]{book}
\usepackage{amsmath,graphicx}
\usepackage{dcolumn,bbm,mathtools,xcolor}
\usepackage{enumitem}
\usepackage[cc]{titlepic}
\usepackage{appendix,imakeidx}
\usepackage[nottoc]{tocbibind}
\usepackage{hyperref}
\newcommand{\smfrac}[2]{\mbox{\small $#1 \over #2$}}
\newcommand{\Tr}{\mathrm{Tr\,}}
\newlist{problems}{enumerate}{1}
\setlist[problems]{label={\thechapter.\arabic*.}, ref={\thechapter.\arabic*}}

\makeindex[title=Index, intoc]
\begin{document}
\frontmatter

%
%
\title{\textcolor{red}{\LARGE{\bf An Undergraduate Course on Quantum Computing}}\\ 
\medskip
\small{Fifth Edition}\\
}
\author{\textcolor{blue}{\Large Peter Young}, \\
e-mail: \textcolor{blue}{\texttt{\underline{petery@ucsc.edu}}}\\
\medskip
\small{University of California Santa Cruz, CA 95064}}


\date{\parbox{\linewidth}{\centering%
\endgraf\bigskip\bigskip
\textcolor{purple}{\Large\textit{``Anyone who is not shocked by quantum mechanics hasn't understood
it."}}\endgraf
\small{(Attributed to Niels Bohr)}\endgraf\medskip
\textcolor{purple}{\Large\textit{``I think I can say that nobody understands quantum mechanics."}}\endgraf
\small{(Richard Feynman)}
}}



\titlepic{\includegraphics[width=7.5cm]{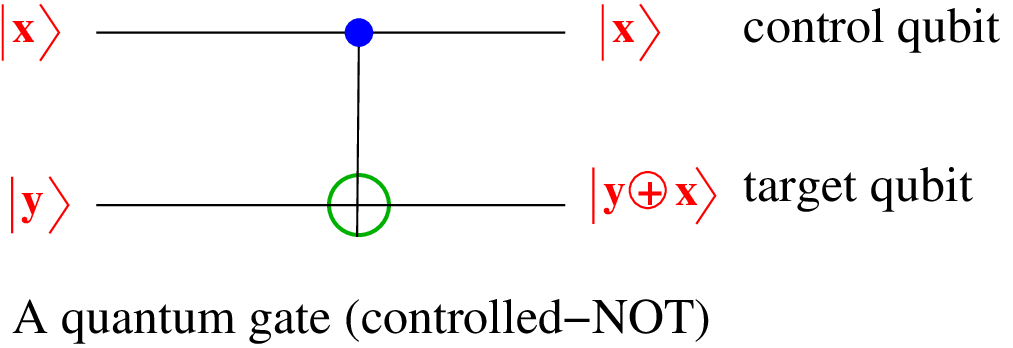}\ \ \includegraphics[width=7.5cm]{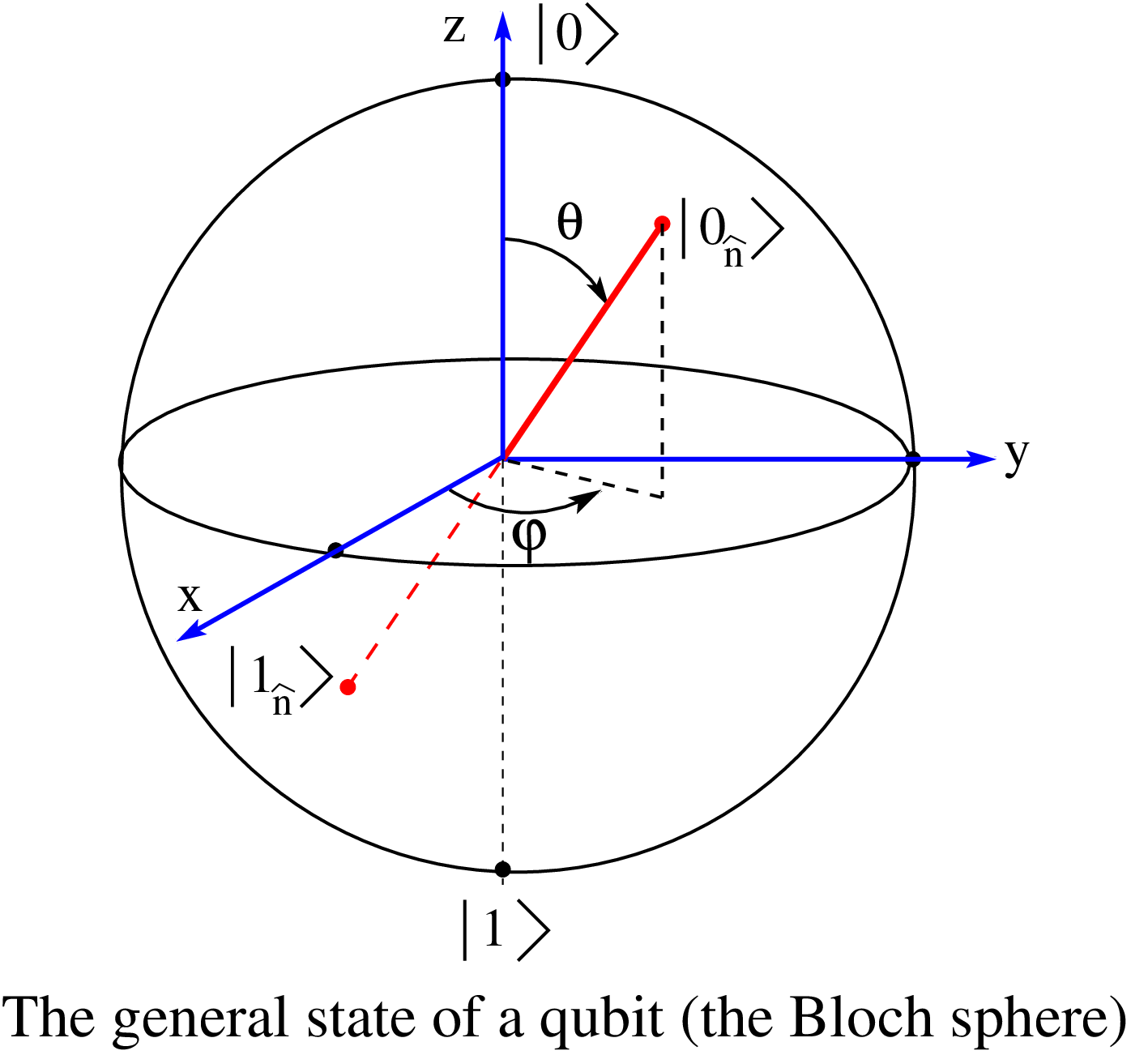}}


\maketitle

\tableofcontents

\chapter{Preface}
\input{preface8.tex}
\mainmatter

\chapter{The Strange World of Quantum Mechanics}
\label{ch:strange}
\input{strange8.tex}

\chapter{Review of Linear Algebra}
\label{ch:lin_alg}
\input{lin_alg8.tex}

\chapter{Introduction to Quantum Mechanics}
\label{ch:qu_intro}
\input{qu_intro8.tex}

\chapter{General state of a qubit, no-cloning theorem, entanglement and Bell
states}
\label{ch:qubits}
\input{qubits8.tex}

\chapter{The Density Matrix}
\label{ch:den_mat}
\input{density_matrix8.tex}

\chapter{Einstein-Podolsky-Rosen (EPR), Bell's inequalities, and \textit{Local
Realism}}
\label{ch:EPR}
\input{EPR8.tex}

\chapter{Classical and Quantum Gates}
\label{ch:gates}
\input{gates8.tex}

\chapter{Generating and measuring Bell States}
\label{ch:bell}
\input{bell8.tex}

\chapter{Quantum Functions}
\label{ch:functions}
\input{functions8.tex}

\chapter{Deutsch's Algorithm}
\label{ch:deutsch}
\input{deutsch8.tex}

\chapter{The Bernstein-Vazirani Algorithm}
\label{ch:bv}
\input{bv8.tex}

\chapter{Simon's Algorithm}
\label{ch:simon}
\input{simon8.tex}

\chapter{Factoring and RSA (Rivest-Shamir-Adleman) Encryption}
\label{ch:rsa}
\input{rsa8.tex}

\chapter{Using Period Finding to Factor an Integer}
\label{ch:period}
\input{period8.tex}

\chapter{The Fourier Transform and the Fast Fourier Transform (FFT)}
\label{ch:fft}
\input{FFT8.tex}

\chapter{The Quantum Fourier Transform (QFT)}
\label{ch:qft}
\input{QFT-FFT-all8.tex}

\chapter{Shor's Algorithm}
\label{ch:shor}
\input{shor8.tex}

\chapter{Coherent Superposition Versus Incoherent Addition of Probabilities}
\input{coherent8.tex}

\chapter{Quantum Error Correction}
\label{ch:err_corr}
\input{error_corr8.tex}

\chapter{Grover's Search Algorithm}
\input{grover8.tex}

\chapter{Quantum Protocols Using Photons}
\label{ch:qkd}
\input{qkd8.tex}

\chapter{Epilogue: Quantum Simulators}
\label{ch:qu_s}
\input{qu_sim8.tex}

\bibliography{refs}

\cleardoublepage
\phantomsection


\printindex

\end{document}

%% file: preface8.tex
This material has been given as a one-quarter course on quantum computing for undergraduates in
the physical sciences at the
University of California Santa Cruz. 

In order that the course be accessible
to majors other than physics, the rules of quantum mechanics were taught
from scratch in the first part. While some of my physics colleagues were surprised
that this could be done, it is perfectly feasible because much of what is
included in a traditional physics course on quantum mechanics concerns
continuous degrees of freedom so one has to cover complicated topics such as
partial differential
equations, boundary conditions, angular momentum, and a plethora of special functions. All this can be omitted in a
quantum computing course which is focused on 2-state systems. A solid
background in linear algebra \textit{is} required. A brief review of this is given at
the start, but the treatment is fast and it is assumed that the students will have
seen the material before.

The aim of the course is to get students to the level where they can understand
the two most important topics covered: Shor's algorithm in Chapter
\ref{ch:shor}
and quantum error correction in Chapter \ref{ch:err_corr}.
Unlike quantum algorithms proposed previously,
Shor's algorithm for factoring integers gives a spectacular speedup on a
problem of \textit{practical importance} (encryption of data sent down a public
channel). Considerable experimental challenges remain to implement Shor's
algorithm for a large number of qubits but quantum error
correction will be essential in order to achieve this, because qubits are
highly susceptible to noise. Incorporating quantum error correction still
leaves huge experimental challenges before achieving the goal of factoring
integers larger than what is possible classically, but without quantum error
correction it would clearly be impossible.

The goal, then, is to present a course at the undergraduate level, but which
still goes into enough depth to give a good understanding of Shor's algorithm and
the basics of quantum error correction. The appendices on the Fast Fourier
Transform (FFT) and the connection of the FFT to the Quantum Fourier Transform
are very detailed and not needed for the course. They are included for the
benefit of interested students, and because they are not found in other
books on quantum computing. No details will be given on the many
experimental approaches to building a quantum computer, which is a huge topic
that would merit a separate course in its own right. A knowledge of the
density matrix, discussed in Ch.~\ref{ch:den_mat} is important for understanding more
advanced material, but is not strictly necessary for the rest of this course,
so instructors may choose to omit it. Quantum computing gets a lot of
attention in the press. One measure of this is that two recent Nobel prizes
in Physics, those in 2022 and 2025, have been for work related to quantum
computing. I give a brief discussion of those prizes in Sections \ref{sec:nobel}
and \ref{sec:nobel_2025}.

There are, of course, excellent more
advanced texts, such as the monumental classic by Nielsen and
Chuang~\cite{nielsen:00}, the books by Mermin~\cite{mermin:07} and 
Rieffel and Pollack~\cite{rieffel:14}, and the online lecture material by
Preskill~\cite{preskill:15}. The
book closest in level and spirit to the present text is the one by
Vathsam~\cite{vathsan:16}, which I found very useful when preparing this
material. Whereas these books, and mine, focus mainly or entirely on theory,
the book by LaPierre~\cite{lapierre:21} also devotes a substantial amount of
material to experimental implementations of quantum computers. The book by
Majidy et al.~\cite{majidy:25} focuses \textit{primarily} on experimental
implementations, which it discusses in detail.

My hope is that this text will take students to a level
where they can follow the rapidly-moving advanced literature in the field.


\vspace{0.2cm}
\noindent Peter Young 

\noindent University of California, Santa Cruz

\noindent \date{\today}

%% file: strange8.tex
\section{Introduction}
The quantum world is strange, and different from the classical world that we see
around us. Our intuition obtained from everyday experience is for objects that
we can see.  It does not apply to the quantum world where we are dealing with
very small objects, objects that (in most cases)
are too small to see. 
I give two quotations, from eminent physicists, which illustrate the
strangeness of the quantum world:
\begin{quotation}
\noindent ``Anyone who is not shocked by quantum mechanics hasn't understood it."\\
(Attributed to Niels Bohr).\index{Bohr, Niels}
\end{quotation}
\begin{quotation}
\noindent ``I think I can say that nobody understands quantum mechanics."\\
(Richard Feynman).
\end{quotation}

The big question which we will address in this course is whether we can use the
difference between the quantum and classical worlds to find more efficient
algorithms to solve certain problems by treating the data in a quantum
computer
in which it is processed according to quantum
rules rather than classical rules. We shall see that for some problems the
answer is ``yes". I should mention now that 
there is a \textit{practical} question of whether we can actually
build a useful quantum computer. The difficulties of
building such a device have not yet been overcome, though much progress has
been made.  In this course, which focuses on theory, we will not describe
the many experimental approaches that are being implemented to try to
achieve this goal. However, we will discuss
in Chapter~\ref{ch:err_corr}
how
one can reduce errors caused by an imperfect device, a topic called
``Quantum Error Correction".

A quantum computer, then, is one in which data is processed by quantum, rather
than classical rules. What do we mean by this? In a classical computer the
data is stored in bits, which take two values 0 and 1. A quantum computer also
uses 2-state systems called qubits\index{qubit}. We indicate these two states by 
$|0\rangle$ and $|1\rangle$, a notation introduced by the physicist Paul
Dirac\index{Dirac notation}. The difference from classical bits is that the general state of a qubit,
which we will write as $|\psi\rangle$, is a
\index{superposition}
\textit{superposition} of states $|0\rangle$ and $|1\rangle$:
\begin{equation}
|\psi\rangle = \alpha|0\rangle + \beta |1\rangle ,
\label{super}
\end{equation}
where $\alpha$ and $\beta$ are numbers (complex in general). For reasons that
will be explained later, we need the condition $|\alpha|^2 +|\beta|^2 = 1$. One
sometimes says loosely that a qubit in the state described by Eq.~\eqref{super}
is simultaneously in states $|0\rangle$ \textit{and} $|1\rangle$. This is to be
contrasted with a classical bit which takes value 0 \textit{or} 1.

Our main goal in this course will be to see if one can gain
computationally from superposition of states.

In the next two sections of this chapter I describe experiments which
illustrate the
strangeness of the quantum world.
More information on this topic can be found in
Refs.~\cite{nielsen:00,mermin:07,vathsan:16} and in Ch.~1, Vol.~3 of the Feynman
Lectures on Physics~\cite{feynman:64}\index{Feynman, Richard}.

\section{The Two-Slit Experiment}
You are probably familiar with experiments involving light going through slits which
demonstrate that light, being a wave, shows interference. 
\index{interference!of light}

\begin{figure}[htb]
\begin{center}
\includegraphics[width=8cm]{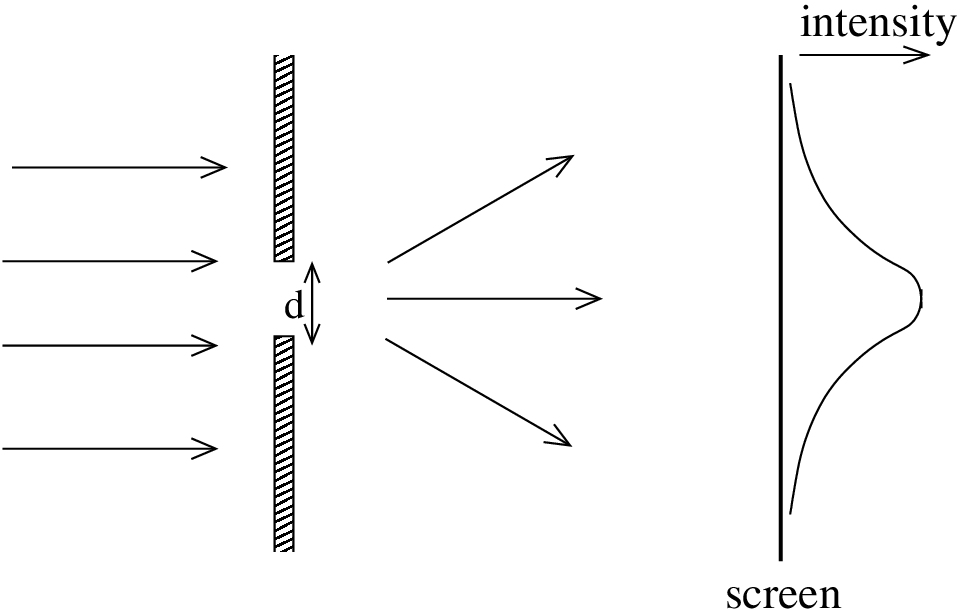}
\caption{A beam of light spreads out (diffracts) when passing through a slit of
width $d$ which is comparable to, or smaller than, the wavelength of the light
$\lambda$. The figure shows a sketch of the intensity of the beam on a screen
after it has passed through the slit.
\label{one_slit}
}
\end{center}
\end{figure}

First consider just one slit. If the slit width $d$ is very large compared
with the wavelength of light $\lambda$ (the geometrical optics limit) then, to
a good approximation, the light continues in a straight line.  However, if the
slit width is comparable to, or less than, $\lambda$, the light spreads out
after passing through the slit, which is called diffraction.
Figure \ref{one_slit} sketches the intensity of light observed on a screen behind the
slit.

\begin{figure}[htb]
\begin{center}
\includegraphics[width=8cm]{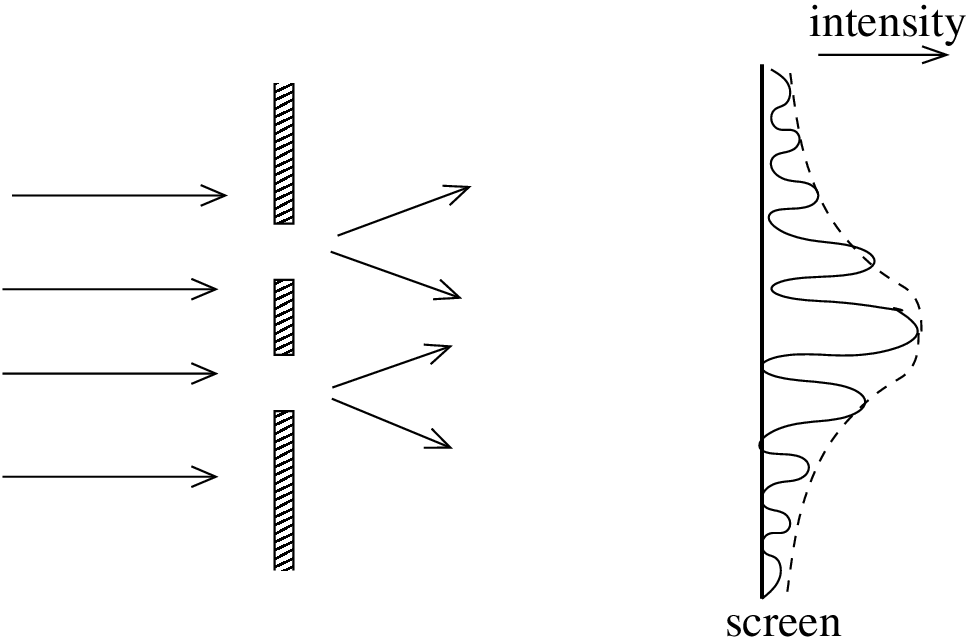}
\caption{A two slit experiment. Interference fringes, oscillations of strong
and weak intensity, are seen due destructive
and constructive interference. The overall envelope of the intensity has a
similar form to that from a single slit shown in Fig.~\ref{one_slit}. 
\label{two_slits}
}
\end{center}
\end{figure}

\begin{figure}[htb]
\begin{center}
\includegraphics[width=5cm]{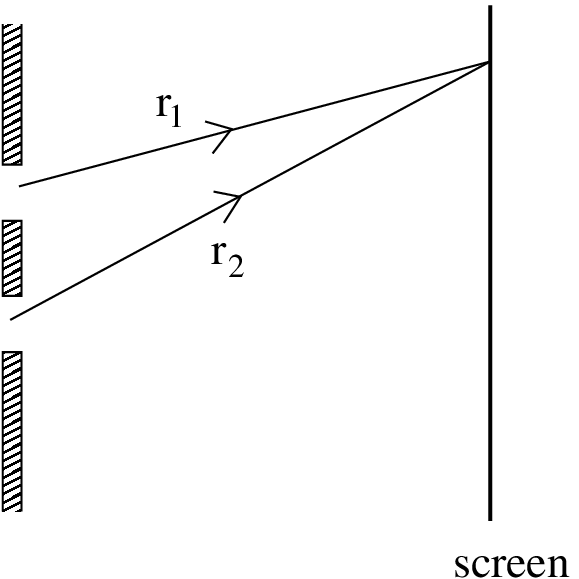}
\caption{The difference in the length of the paths taken by the beams going
through the two slits is $ |r_1 - r_2|$. This varies as a function of the
location on the screen, so the interference changes from constructive, where
$|r_1 - r_2| = n \lambda$, to destructive, where $ |r_1 - r_2| = (n
+\smfrac{1}{2})\lambda$ with $n$ an integer. 
\label{delta_r} }
\end{center}
\end{figure}

If the light beam passes through two slits, as shown in Fig.~\ref{two_slits} one
observes interference fringes, oscillations of strong and weak intensity,
due to interference between between the beams
going through the two slits.  If the difference in path length $|r_1 - r_2|$
(see Fig.~\ref{delta_r})
satisfies $|r_1 - r_2| = n \lambda$ (for integer $n$)
one has constructive interference and a maximum
intensity, whereas if $|r_1 - r_2| = (n +\smfrac{1}{2})\lambda$ one has a minimum
intensity. Hence, as one moves along the screen one alternately gets regions
of low
intensity and high intensity. These are called interference fringes.

This is the classical picture. That is, we shine a beam at the two slits, some of it
goes through one slit, some goes through the other slit, and when these two
beams recombine they interfere.
 
\begin{figure}[htb]
\begin{center}
\includegraphics[width=8cm]{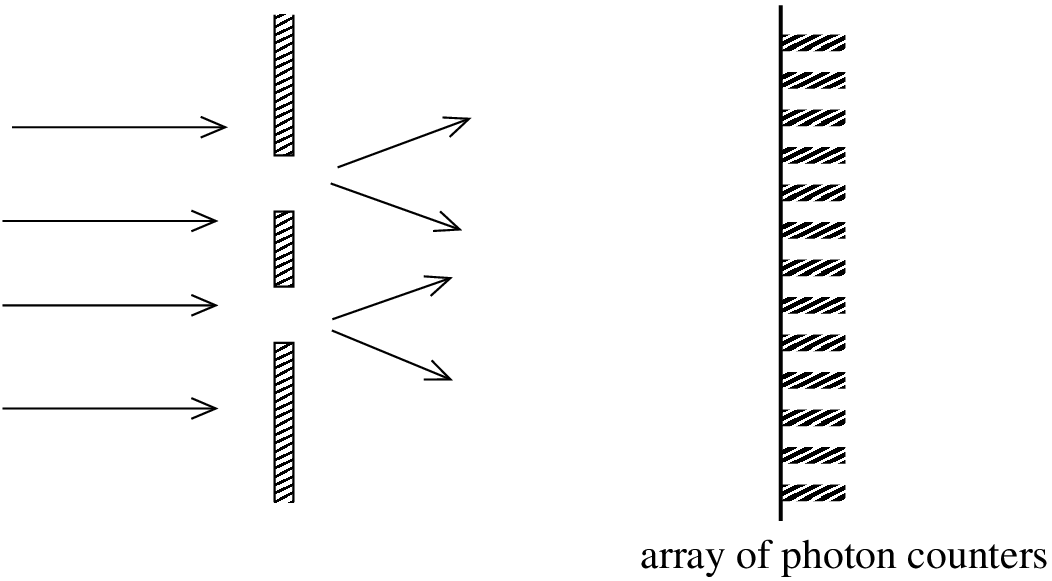}
\caption{The two slit experiment where individual photons are detected by an
array of counters which count the number of photons at their location.
\label{detectors}
}
\end{center}
\end{figure}

Now we reduce the intensity of the light. At some point we notice that light
is not a continuous wave but consists of discrete bunches of energy called
\textit{photons}. To detect individual photons, we place an array of photon
counters
on the screen and count the number of discrete clicks in each counter, see
Fig.~\ref{detectors}.
We record the number of clicks for counters placed at
different points on the screen.

Suppose we reduce the intensity so much that the time between emitting photons
is greater than the time it takes a photon to pass through the experimental
setup, i.e.~the photons go through \textit{one at a time}.
Do we see an interference pattern?  Using our classical intuition we would say
``no" because surely each photon ``must" either go through the upper slit or the
lower one and can therefore not interfere with itself.  In other words we
would expect the intensity of clicks in the counters to vary smoothly along
the screen, as in the classical single slit experiment shown in Fig.~\ref{one_slit}.

Amazingly, this is not so and \textit{we do see an interference pattern}. In
other words
the number of clicks in the counters varies rapidly and in a oscillatory manner as we move
along the screen, just as in the classical two-slit experiment shown in
Fig.~\ref{two_slits}. It looks as though a single photon \textit{does} go through both
slits. You may already be feeling (correctly)
that this looks suspiciously like a superposition state such as the one we
we wrote down in Eq.~\eqref{super}, where now $|0\rangle$ refers to photon
through the
upper slit and $|1\rangle$ to photon through the lower slit. 

You might ask ``why don't we just \textit{look} and see which slit the photon
went through''.  Well, photons being electrically neutral are hard to observe
unless we absorb them (which we want to do only when they reach the screen).
The rate of scattering of one photon by another is immeasurably small. So, with
photons we can't observe which slit they went though. However, we can
do the same experiment with electrons rather than photons. Like photons, electrons
have both particle and wave-like properties, but, being charged, they readily
scatter light so we can see observe them by shining light on them. 
The discussion which follows is
based on Ch.~1, Vol.~3 of Feynman~\cite{feynman:64}.

In this new version of the experiment we send electrons through the slits
one at a time. To see which slit they went through we shine light of
wavelength $\lambda$ at the slits and observe a flash of light every time an
electron goes through.

Suppose that we choose a light source that has a wavelength $\lambda$ which is
bigger than the slit spacing $d$. We do see a flash every time an electron passes
through, and observe that there is \textit{still} an interference
pattern but, the flash of light is of size $\lambda$ which is greater
than the separation of the slits, so we can't tell
which slit the electron went through. Clearly we need to
use a light source with wavelength less than $d$. When we do this, indeed
we see a flash at either the upper slit or the lower slit every time an
electrons passes, so we've achieved our goal of observing which slit each
electron goes through. But alas, when we look at the counts registered on the
detectors we see that the interference fringes have been washed out, and we
have just a smooth variation in the number of clicks along the screen.
Observations such as these show that it is 
not possible to determine which slit each electron goes through \textit{and} observe
interference fringes.

This observation guides us to a second piece of intuition regarding quantum
mechanics (the first, mentioned above, is that a quantum system can be in
a superposition state), namely that a measurement can unavoidably change a
quantum state, and in particular can destroy a superposition.

Classically, measurements are passive, and can be done in a delicate way
so they simply reveal a reality which is already present whether we observe it or
not. Quantum mechanically, measurements play a much more active role and can
change the state of the system. In particular, we shall see that if we observe
a system in a particular state, we can't necessarily say that it was in that state
before the measurement.
 
\section{Stern-Gerlach Experiment}
\index{Stern-Gerlach experiment}
We will now discuss a second experiment which gives additional insight into
superposition states.

Consider the hydrogen atom, which consists of one proton (the nucleus), which has a positive
electric charge,
and one electron which has a negative charge. In its ground
state the electron has a symmetric distribution of velocities and so there is
no net circulating electric current around the proton. Hence the orbital
motion of the electron
does not give rise to a magnetic moment which could interact
with an external magnetic field. However, the electron has an internal state,
called \textit{spin}, which does give rise to a magnetic moment\footnote{The proton
also has a spin and hence a magnetic moment but, because of its much larger
\index{magnetic moment}
mass, its magnetic moment is much smaller than that of the electron and so does not
play a role in our discussion.} $\vec{\mu}$, proportional to the spin angular
momentum.

\begin{figure}[htb]
\begin{center}
\includegraphics[width=4.0cm]{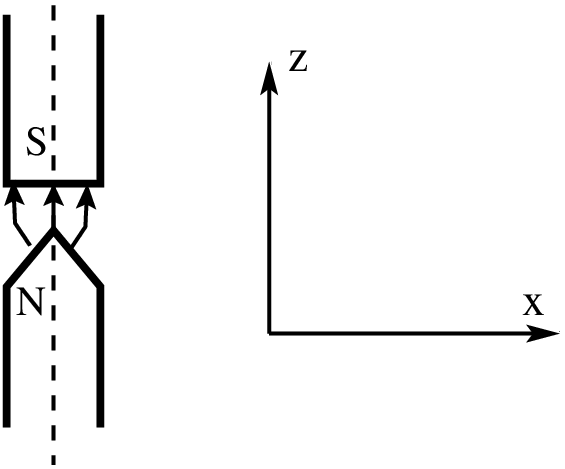}
\caption{A cross section of the magnet in the Stern-Gerlach experiment. The beam goes between the poles of
the magnet, into the plane i.e.~in
the $y$-direction, and intersects the symmetry axis (which is in the
$z$-direction and shown by the dashed line).
\label{sg_mag}
}
\end{center}
\end{figure}

There is a
force on a magnetic moment in a field if the field is non-uniform.  To see this, 
recall that the energy of a
magnetic moment in a magnetic field $\vec{B}$ is $-\vec{\mu} \cdot \vec{B}$ 
and therefore the force, which is minus the spatial gradient of the energy, is given by 
\begin{equation}
\vec{F} = \vec{\nabla}\left(\vec{\mu} \cdot \vec{B}\right) 
\end{equation}
so
\begin{equation}
F_z
= \vec{\mu} \cdot {d \vec{B} \over d z} ,
\end{equation}
where we have assumed, without loss of generality, that the field changes as
function of 
$z$. Hence a beam of hydrogen atoms in a non-uniform field varying in the
$z$-direction will be deflected in the $z$-direction.
For simplicity we assume that the field itself is also (predominantly)
along the  $z$-direction, see Fig.~\ref{sg_mag}, so
\begin{equation}
F_z
= \mu_z \, {d B_z \over d z} ,
\end{equation}
and hence the deflection will be proportional to $\mu_z$.

\begin{figure}[htb]
\begin{center}
\includegraphics[width=9cm]{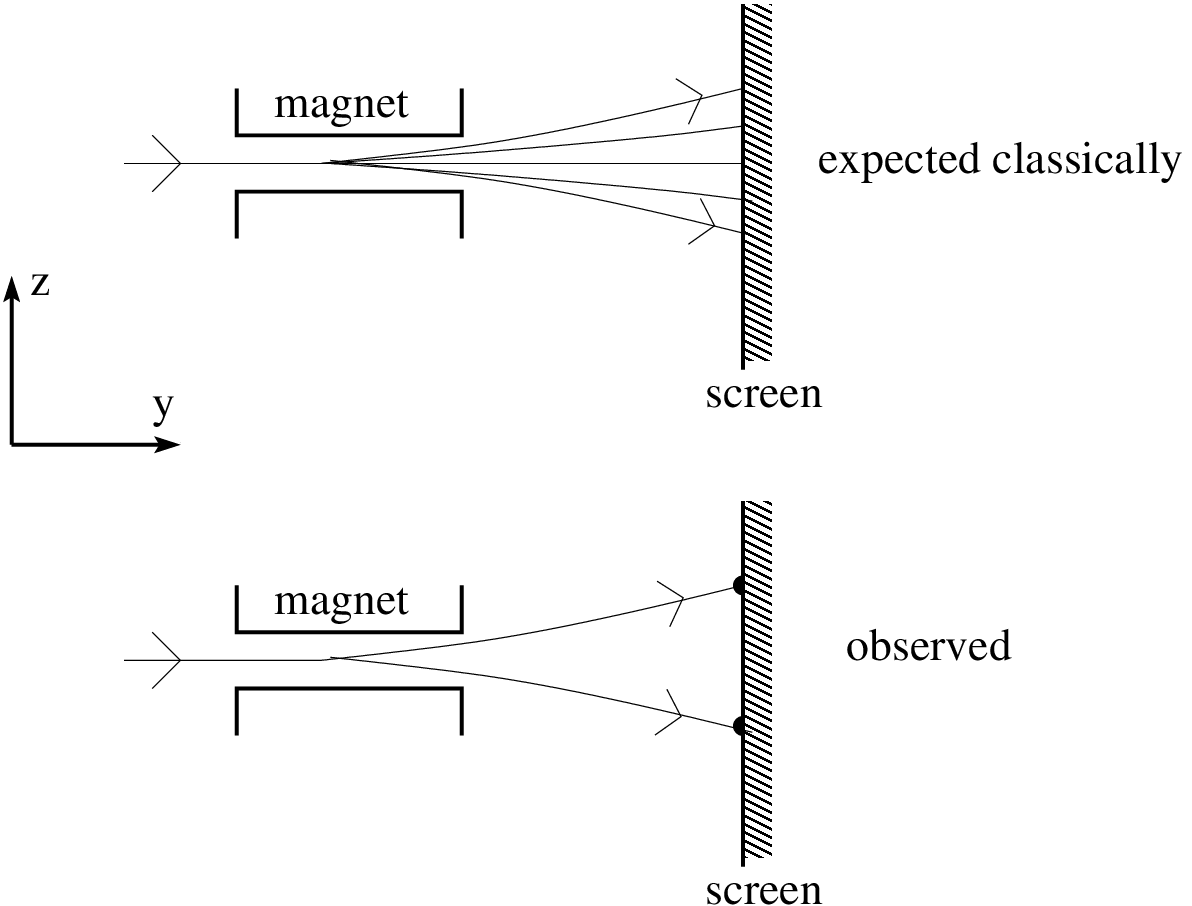}
\caption{The Stern-Gerlach apparatus.
\label{SG_beam}
}
\end{center}
\end{figure}

We send in a beam of unpolarized hydrogen atoms into a non-uniform field. This is the
famous Stern-Gerlach (SG) experiment. Since the direction of $\vec{\mu}$ is
random, classically
$\mu_z$ takes a range
of values, so we would expect a continuous range of deflections. However, it is
found that only two beams
emerge, which are deflected in opposite directions, see Fig.~\ref{SG_beam}.
Since $\vec{\mu}$ is proportional to the spin it seems that the spin
component along $z$ has only two components, corresponding to states which we
might label as\footnote{The electron is the simplest two-state system.} 
$|\uparrow_z\rangle$ and $|\downarrow_z\rangle$, or alternatively as 
$|0\rangle$ and $|1\rangle$ respectively.

Now suppose that we orientate the magnet so the field and its gradient 
are in the $x$-direction.
Again we will see two beams emerging, indicating that $\mu_x$ has only two
possible values $|\uparrow_x\rangle$ and $|\downarrow_x\rangle$.

\begin{figure}[htb]
\begin{center}
\includegraphics[width=10.5cm]{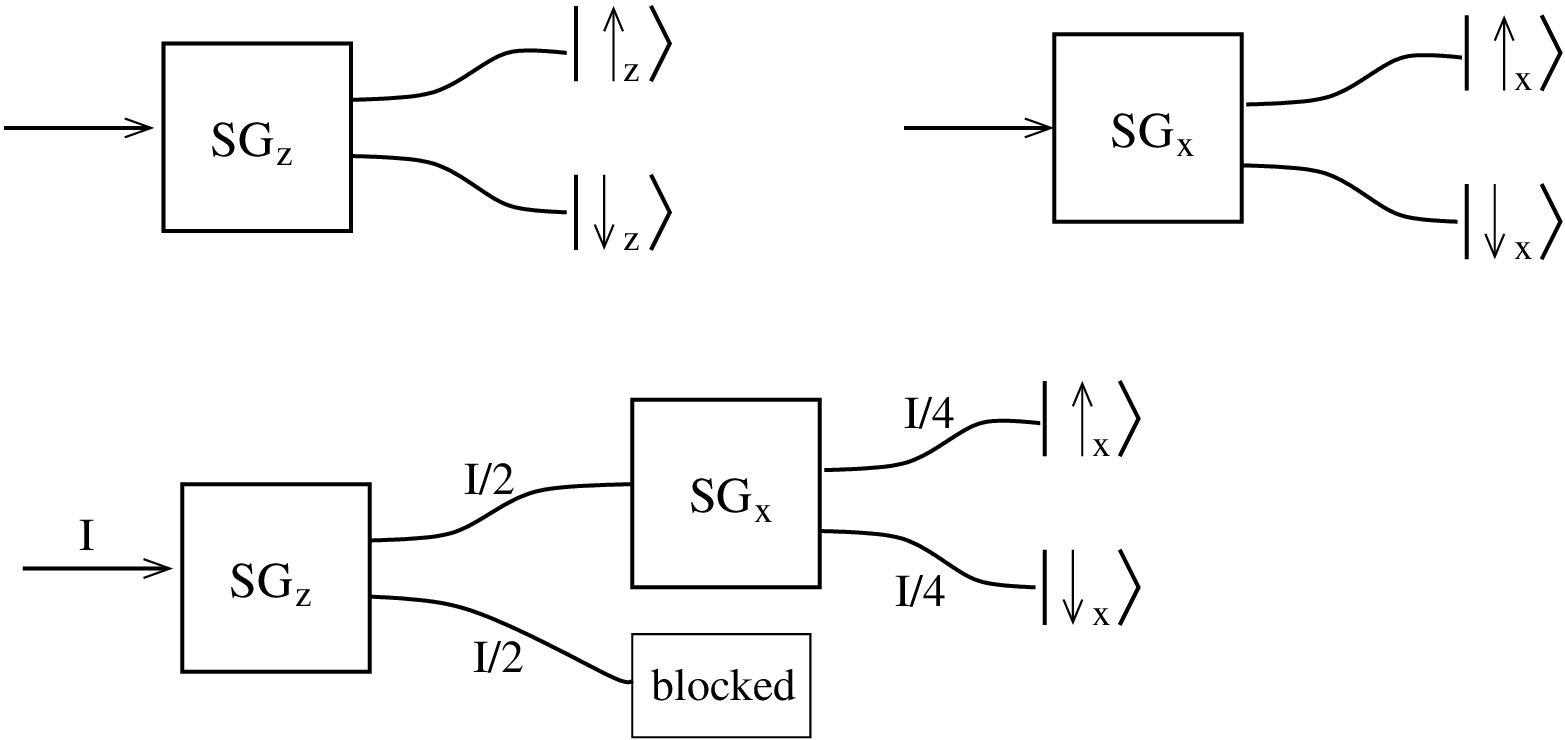}
\caption{The upper figure shows schematically separate Stern-Gerlach experiments with
the field in the $z$-direction (SG$_z$) and in the $x$ direction (SG$_x$).
The lower figure shows a double Stern-Gerlach experiment in which the beam is
passed first through an SG apparatus with a field in the $z$-direction and
then
one of the beams is passed
through an SG apparatus with the field in the $x$-direction.
\label{SG}
}
\end{center}
\end{figure}

How are $|\uparrow_x\rangle$ and $|\downarrow_x\rangle$ related to
$|\uparrow_z\rangle$ and $|\downarrow_z\rangle$? We can get an idea of this if
we run our beam first through a SG setup with the field in the $z$-direction
and then pass one of the resulting beams through an SG setup in the
$x$-direction as shown in Fig.~\ref{SG}. The final result is found to be two beams of
equal intensity.

It looks as though $|\uparrow_z\rangle$ can be thought of as 
$|\uparrow_x\rangle$ with probability $1/2$ and $|\downarrow_x\rangle$ with
probability $1/2$.  We will see in a future lecture that $|\uparrow_z\rangle$ is actually a
superposition of $|\uparrow_x\rangle$ and $|\downarrow_x\rangle$ as follows:
\begin{equation}
|\uparrow_z\rangle = {1 \over \sqrt{2}} \left(|\uparrow_x\rangle +
|\downarrow_x\rangle\right) ,
\end{equation}
where we say that there is an \textit{amplitude}\footnote{Sometimes called a
probability amplitude} $1/\sqrt{2}$ for
\index{amplitude}
$|\uparrow_z\rangle$ to be $|\uparrow_x\rangle$ and amplitude $1/\sqrt{2}$ for
it to be
$|\downarrow_x\rangle$. As we shall also see later, the probability that a
measurement gives a certain result is the square of the modulus of corresponding
\index{amplitude}
amplitude\footnote{The fact that probabilities add to 1, is why 
$|\alpha|^2 + |\beta|^2 = 1$ in Eq.~\eqref{super}.} so the probability of
measuring $|\uparrow_x\rangle$ after the $SG_x$ apparatus is 1/2 (as observed)
and the same for $|\downarrow_x\rangle$.

\begin{figure}[htb]
\begin{center}
\includegraphics[width=12cm]{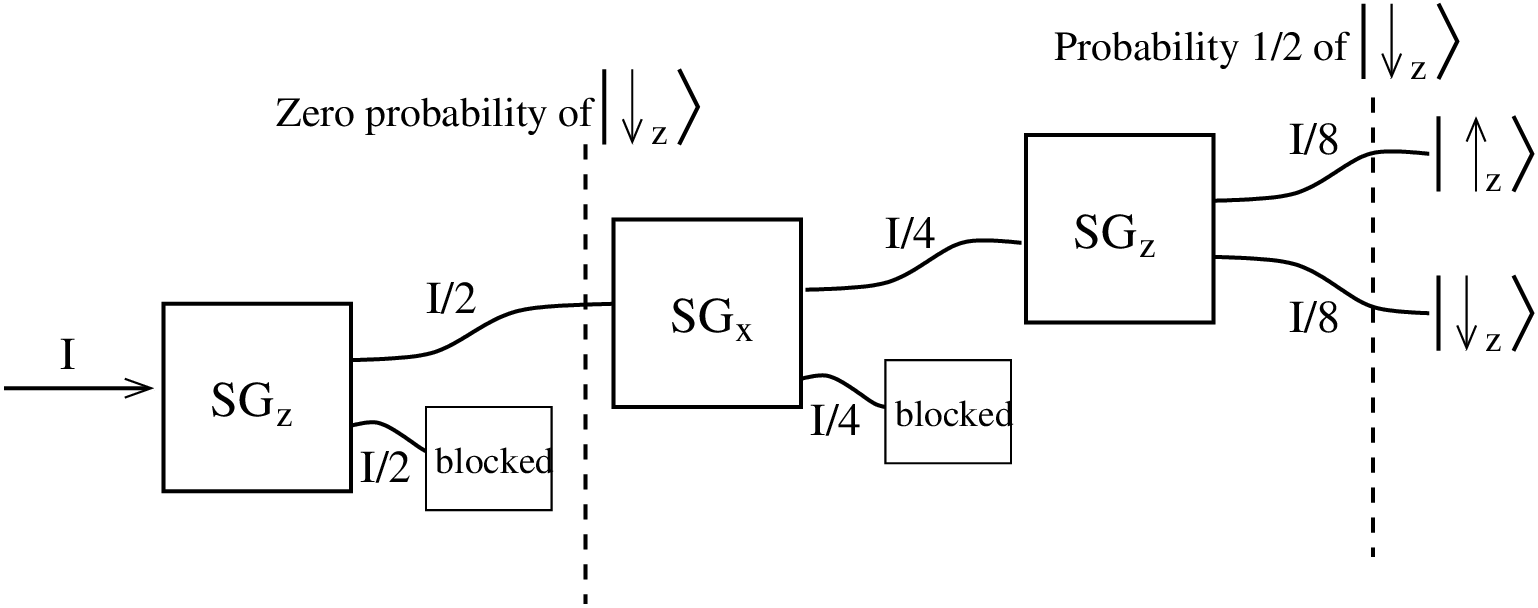}
\caption{We now add another SG$_z$ apparatus after the SG$_x$ apparatus in
Fig.~\ref{SG}. The result is equal intensity in the beams for
$|\uparrow_z\rangle$ and $|\downarrow_z\rangle$. After each SG apparatus the
upper line is for the ``up'' spin and the lower line for the ``down'' spin.
\label{SG2}
}
\end{center}
\end{figure}

It is also true that
\begin{equation}
|\uparrow_x\rangle = {1 \over \sqrt{2}} \left(|\uparrow_z\rangle +
|\downarrow_z\rangle\right) ,
\end{equation}
so if we run one of the beams from the SG$_x$ apparatus in Fig.~\ref{SG}
through another SG$_z$ apparatus we will get beams with equal intensity for
$|\uparrow_z\rangle$ and $|\downarrow_z\rangle$, see Fig.~\ref{SG2}.
Note a surprising aspect of this result. After the first SG$_z$ apparatus,
there is zero probability for getting $|\downarrow_z\rangle$ (because we 
blocked it off), but after the
SG$_x$ apparatus there is a $50\%$ probability for finding
$|\downarrow_z\rangle$. In other words, a non-zero probability for getting
$|\downarrow_z\rangle$ has been \textit{generated} by the measurement.
This is a clear example of a measurement (in this case
that done by the SG$_x$ apparatus) affecting the state of the system.

\section{Photons}
\label{sec:photons}
In the previous section we noted that the spin of the electron is a two-state
quantum system. Here we discuss another two-state quantum system,
the photon, the quantum of light. 
\index{photon}

Light is an oscillating transverse electromagnetic
field, in which the electric field $\vec{E}$
and magnetic field $\vec{B}$ are perpendicular both to each other and to
the
direction of propagation specified by the wavevector $\vec{k}$. For example,
if $\vec{E}$ is in the $x$ direction, $\vec{B}$ in the $y$ direction, and
$\vec{k}$ in the $z$ direction we have\footnote{We understand that the
physical fields are the real parts of these expressions.}
\begin{equation}
\begin{split}
\vec{E} &= E_0\, \hat{x}\, e^{i(kz - \omega t)}, \\
\vec{B} &= B_0\, \hat{y}\, e^{i(kz - \omega t)}.
\end{split}
\label{EBk}
\end{equation}

\index{photon!polarization}
The direction of $\vec{E}$ is called the polarization direction. There are two
distinct polarizations which we can call
``horizontal'' (along $\hat{x}$)
\begin{equation}
\label{horiz}
|\leftrightarrow\rangle, \qquad \mathrm{equivalent \ to}\quad
|\uparrow_z\rangle \equiv |0\rangle ,
\end{equation}
and ``vertical", (along $\hat{y}$)
\begin{equation}
\label{vert}
|\updownarrow\rangle,
\qquad \mathrm{equivalent \ to}\quad
|\downarrow_z\rangle \equiv
|1\rangle .
\end{equation}

What are the analogs of $|\uparrow_x\rangle $ and $|\downarrow_x\rangle$?
The answer is diagonal polarizations:
\begin{equation}
\begin{split}
|\rotatebox[origin=c]{45}{\Large$\leftrightarrow$} \rangle &\equiv {1\over
\sqrt{2}}\left( |\updownarrow\rangle +  |\leftrightarrow\rangle \right),
\qquad \mathrm{equivalent \ to}\quad
|\uparrow_x\rangle \equiv {1\over
\sqrt{2}}\left( |0\rangle + |1\rangle  \right) , \\
|\rotatebox[origin=c]{45}{\Large$\updownarrow$} \rangle &\equiv {1\over
\sqrt{2}}\left( |\updownarrow\rangle -  |\leftrightarrow\rangle \right),
\qquad \mathrm{equivalent \ to}\quad
|\downarrow_x\rangle \equiv {1\over
\sqrt{2}}\left( |0\rangle - |1\rangle  \right) .
\end{split}
\label{ph:diag}
\end{equation}
More details on the correspondence between photon polarization
and qubit states will be given in Sec.~\ref{sec:gen_qubit}.

Photons do not interact with each other to a measurable extent, and can not
readily be stored, so they are unsuitable for most types of quantum
computer, but have the advantage that they can be
transmitted over great distances down optical fibers, preserving their
polarization. These properties will be
useful for some quantum protocols to be discussed in Chapter \ref{ch:qkd}.

%% file: lin_alg8.tex
The theory of quantum mechanics is based on linear algebra which is a
pre-requisite for the course and is standard material available in many books.
In this chapter we summarize those topics in linear algebra which will be
needed for this class. The treatment is quick and is intended as a review for
students, assuming that they have seen the material before.

\section{Vectors}

\index{vector}
An abstract vector $\vec{v}$ can be represented in terms of
its $N$ components $v_i, (i=1, \cdots,N)$
\index{basis!vectors}
\begin{equation}
\vec{v} = \sum_{i=1}^N v_i \hat{e}_i ,
\end{equation}
with respect to a set of basis vectors $\hat{e}_i$, which form an orthonormal
set, i.e.
\begin{equation}
\vec{e}_i \cdot \vec{e}_j = \delta_{ij} ,
\end{equation}
where the left hand side is a scalar product
\begin{equation}
\vec{a}\cdot \vec{b} = \sum_{i=1}^N a_i\, b_i ,
\end{equation}
and $\delta_{ij}$ is the Kronecker delta function,
\begin{equation}
\delta_{ij} = \left\{
\begin{array}{ll}
1 & (i = j), \\
0 & (i \ne j),
\end{array}
\right.
.
\end{equation}
\index{normalization}\index{orthogonality}
We say that a vector $\vec{v}$ is normalized if $\vec{v}\cdot\vec{v} = 1$, and that two 
vectors $\vec{a}$ and $\vec{b}$ are orthogonal if $\vec{a}\cdot \vec{b} = 0$.
A set of
vectors is said to be orthonormal if each is normalized and every pair is
\index{orthonormal}orthogonal. 
The number of independent basis states required to represent any vector is
called the size of the ``vector space". It is denoted here by $N$.

The vector $\vec{v}$ can be represented in terms of its components
$v_i$ as a column vector
\begin{equation}
\vec{v} = 
\begin{pmatrix}
v_1 \\ v_2 \\ \vdots \\ v_N \\
\end{pmatrix}
,
\end{equation}
and its transpose as a row vector 
\begin{equation}
\vec{v}^T = 
\begin{pmatrix}
v_1 & v_2 & \cdots & v_N \\
\end{pmatrix}
.
\end{equation}

The length of a vector is given by 
\begin{equation}
|v| = \left(\sum_{i=1}^N v_i^2\right)^{1/2} = \left(\vec{v} \cdot \vec{v}\,\right)^{1/2} .
\end{equation}

\index{basis!change of}
One can represent a vector with respect to different orthonormal bases rotated with
respect to each other. If a vector has components $v_i'$ with respect to the
new basis, there is a linear relation between the old and new components,
\begin{equation}
v_i' = \sum_{j=1}^N M_{ij} v_j,
\end{equation}
where $M$ is an $N \times N$ matrix with elements $M_{ij}$. In order that $M$
describes a rotation (which preserves lengths of vectors and angles between
them), it is necessary that $M$ be an orthogonal matrix, i.e.
\begin{equation}
M^{-1} = M^T ,
\end{equation}
where $M^T$ is the transpose matrix, and $M^{-1}$ is the matrix inverse which
means that $M^{-1} M = M M^{-1} = \mathbbm{1}$ where $\mathbbm{1}$ is the
identity matrix.
An example of a rotation matrix for two-component vectors is
\begin{equation}
M =
\begin{pmatrix}
\cos\theta & \sin\theta \\
-\sin\theta & \cos\theta \\
\end{pmatrix} ,
\end{equation}
where $\theta$ is the rotation angle.

The scalar product of two vectors is independent of basis, so
\begin{equation}
\vec{a} \cdot \vec{b} = \sum_{i=1}^N a_i b_i = \sum_{i=1}^N a'_i b'_i .
\end{equation}
This is why $\vec{a} \cdot \vec{b}$ is called a \textit{scalar} product.

\section{Complex Vectors}
\index{vector!complex}
In quantum mechanics, we need complex vectors, i.e.~vectors with complex
coefficients. 
The main new feature compared with real vectors is a
slight difference in the definition of the scalar product,
(called, more generally, an inner product),
namely one takes the complex conjugate of the
left hand vector, i.e.
\begin{equation}
\vec{a} \cdot \vec{b} = \sum_{i=1}^N a^\star_i b_i .
\end{equation}
In terms of rules for matrix multiplication one can view the scalar
product as the matrix product of the complex conjugate of the transpose vector
(row vector) for $a$ with the vector (column vector) for $b$, i.e.
\begin{equation}
\vec{a} \cdot \vec{b} \equiv \left(a^{T}\right)^\star b =  
\begin{pmatrix}
a_1^\star & a_2^\star & \cdots & a_N^\star \\
\end{pmatrix}
\begin{pmatrix}
b_1 \\ b_2 \\ \vdots \\ b_N \\
\end{pmatrix} ,
\label{inner}
\end{equation}
in which
$\left(a^T\right)^\star$ is an $1 \times N$ dimensional matrix,
$b$ is an $N \times 1$ dimensional matrix,
and
$\left(a^{T}\right)^\star b$ denotes matrix multiplication with the result
being a single number (scalar).

The length of a complex vector,
called the norm $|a|$ from now on,
\index{norm}
is still the
square root of the scalar product of the vector with itself, i.e.
\begin{equation}
|a| = \left(\vec{a} \cdot \vec{a}\right)^{1/2} =  \left(\sum_{i=1}^N |a_i|^2\right)^{1/2}.
\end{equation}


\section{Matrices}
\index{matrix!multiplication of}
If $A$ and $B$ are matrices then the matrix
product $C = AB$ is given in terms of its elements by
\begin{equation}
C_{ij} = \sum_{k=1}^M A_{ik} \, B_{k j} .
\end{equation}
We assume here that
$A$ is of dimension $N \times M$ ($N$ rows and $M$ columns), in which case $B$ must
have $M$ rows. If $B$ has $P$ columns then $C$ is of dimension $N \times
P$.
As noted above, it will sometimes be useful to think
of a column vector as an $N \times 1$ dimensional matrix ($N$ rows and 1 column), and
a row vector as a $1 \times N$ dimensional matrix.
Apart from
vectors, the matrices in this course will be square (number of rows
equals number of columns).

Matrix multiplication has the property that the order of multiplication
matters in general.
We define the commutator of two matrices by
\begin{equation}
[A, B] \equiv AB - B A.
\end{equation}
If $[A, B] =0$ we say that $A$ and $B$ commute.
\index{matrices!commuting}
However, in general matrices
do not commute, i.e.~their commutator is non-zero.
Lack of commutation of matrices will have important consequences in quantum
\index{matrix!commutator}
mechanics.

Some important, special types of matrices are:
\begin{itemize}
\item Symmetric: $M^T = M$ ($M^T$ is the transpose, so $\left(M^T\right)_{ij} = M_{ji}$).
\item Orthogonal: $M^T = M^{-1}$ 
($M^{-1}$ is the matrix inverse defined by $M^{-1} M = M M^{-1} = \mathbbm{1}$, the identity matrix, which has one 
on the diagonal elements and zero elsewhere.)
\index{matrix!Hermitian}
\end{itemize}
In quantum mechanics we will deal with complex matrices, as well as complex
vectors.
In the case of complex matrices, one is usually interested in Hermitian matrices
rather than symmetric ones, and unitary matrices rather than orthogonal ones, where these are defined by:
\index{matrix!unitary}
\begin{itemize}
\item Hermitian: $M^\dagger = M$ ($M^\dagger$ is the adjoint,
the complex conjugate of the transpose so $M^\dagger = \left(M^T\right)^\star$).
\item Unitary: $M^\dagger = M^{-1}$.\\
Unitary matrices have the useful property that the rows
form orthonormal vectors, as do the columns. To determine if a matrix is unitary
it may be easier to do this check rather than compute the inverse.
\end{itemize}
Hermitian and unitary matrices play important roles in quantum
mechanics.

We shall also meet the 
\index{identity matrix}identity matrix, $I$, which has 1 on the diagonal
elements and 0 on the off-diagonal elements, i.e.
\begin{equation}
I =
\begin{pmatrix}
1 &         0 & \cdots & 0 \\
0         & 1  & \cdots & 0 \\
\vdots    &  \vdots   & \ddots & \vdots\\
0         & 0         & \cdots & 1
\end{pmatrix} .
\end{equation}

\section{Matrix Diagonalization}
\label{sec:mat_diag}
\index{diagonalization of matrices}

Let $A$ be $N \times N$ matrix and $\vec{x}$ an $N$-component vector. Then
if $A \vec{x}$ is proportional to $\vec{x}$ itself, i.e.~if
\begin{equation}
\begin{split}
A \vec{x} &= \lambda \vec{x} \quad \mathrm{or,\ in\ terms\ of\ elements}, \\
\sum_{j=1}^N A_{ij} x_j &= \lambda x_i ,
\end{split}
\label{evects}
\end{equation}
\index{eigenvalues and eigenvectors}
then we say that $\lambda$ is an eigenvalue and $\vec{x}$ the corresponding
eigenvector of $A$. There are $N$ eigenvalues which may not all be
distinct. If two or more eigenvalues are equal we say that they are degenerate.
We can always multiply an eigenvector by a constant and it remains an
eigenvector. In quantum mechanics we will need to choose this multiplicative
constant so the vector is ``normalized", i.e.~has unit length.
\index{normalization}

The eigenvalues are obtained from solving
\index{matrix!determinant}
\begin{equation}
\det (A - \lambda \mathbbm{1} ) = 0,
\end{equation}
where $\det$ is short for determinant. Expanding out the determinant gives an
$N$-th order polynomial equation for $\lambda$. One can then get the
eigenvectors by solving the linear equations in Eq.~\eqref{evects} for each
value of $\lambda$.

The eigenvalues and eigenvectors of Hermitian matrices have special
properties:
\index{matrix!Hermitian}
\begin{itemize}
\item The eigenvalues are all real.
\item Eigenvectors corresponding to unequal (non-degenerate) eigenvalues are orthogonal. For
eigenvectors corresponding to degenerate eigenvalues, one can form linear
combinations which are orthogonal.
\end{itemize}

Once one has the eigenvectors, a matrix 
$A$ can be ``diagonalized'' as follows\footnote{There are
some matrices with degenerate eigenvalues which have less than $N$ independent
eigenvectors. These can not be diagonalized. However, this situation does not occur for
Hermitian or unitary matrices, the two categories that are of principle
interest in quantum mechanics, and so we will ignore non-diagonalizable matrices in
this course.}:
\index{matrix!diagonalization}
\begin{equation}
D = S^{-1} A S,
\label{DA}
\end{equation}
\index{eigenvalues and eigenvectors}
where $D$ is a diagonal matrix with the eigenvalues of $A$ on the diagonal,
\begin{equation}
D = 
\begin{pmatrix}
\lambda_1 &         0 & \cdots & 0 \\
0         & \lambda_2 & \cdots & 0 \\
\vdots    &  \vdots   & \ddots & \vdots\\
0         & 0         & \cdots & \lambda_N 
\end{pmatrix} ,
\end{equation}
and the
matrix $S$, which effects the diagonalization, is constructed out of the
eigenvectors of $A$ as follows:
\begin{equation}
S = \begin{pmatrix}
\vec{e}^{\,(1)}, & \vec{e}^{\,(2)}, & \cdots , \vec{e}^{\,(N)} \\
\end{pmatrix} ,
\end{equation}
where $\vec{e}^{\,(i)}$ is the $i$-th eigenvector of $A$ written as a column
vector.

If $A$ is Hermitian then the eigenvectors orthogonal, so if we normalize
them, the matrix of eigenvectors $S$ is unitary, so let's call it $U$, i.e.
$U^{-1} = U^\dagger$. Hence a Hermitian matrix $A$
is diagonalized by the following transformation
\begin{equation}
D = U^\dagger A U .
\end{equation}

If we consider two $N \times N$ matrices $A$ and $B$, one can show that they
have the same eigenvectors if and only if the matrices commute, i.e.~if $[A,
B] \equiv AB - BA = 0$. This result will have important consequences in
quantum mechanics.
\index{matrix!commutator}

\section{Some Important $2\times 2$ matrices}
\label{sec:important}

\index{matrix!Hermitian}
In quantum computing we deal most 
frequently with $2 \times 2$ matrices because qubits have
two states. 
Important examples of $2 \times 2$ Hermitian matrices are the Pauli (spin) matrices
\index{Pauli matrices!$X$ matrix}
\index{X|see {Pauli $X$ matrix}}
\index{Pauli matrices!$Y$ matrix}
\index{Y|see {Pauli $Y$ matrix}}
\index{Pauli matrices!$Z$ matrix}
\index{Z|see {Pauli $Z$ matrix}}
\begin{equation}
X =
\begin{pmatrix}
0 & 1 \\
1 & 0 
\end{pmatrix}, \qquad
Y = 
\begin{pmatrix}
0 & -i \\
i & 0 
\end{pmatrix}, \qquad
Z =
\begin{pmatrix}
1 & 0 \\
0 & -1 
\end{pmatrix} ,
\label{pauli}
\end{equation}
(called $\sigma_x, \sigma_y$ and $\sigma_z$ in the physics literature).
Note that $Y$ is not symmetric, though we do recover the original matrix if we
take the complex conjugate as well as the transpose, so it \textit{is} Hermitian. 

Any $2\times 2$ matrix can be
expressed as a linear combination of the three Pauli matrices plus the
identity. To see this note that $X, Y,Z$ and $\mathbbm{1}$ are linearly
independent (i.e.~we can't write any one as a linear combination of the others).  Also
a general $2\times 2$ matrix
\begin{equation}
A = 
\begin{pmatrix}
t & u \\
v & w \\
\end{pmatrix}
\end{equation}
has 4 complex elements, and so a total of 8 real parameters.  If we write
\begin{equation}
A = a_0 \mathbbm{1} + a_x X + a_y Y + a_z Z 
\label{intro_A}
\end{equation}
then there are also 4 complex coefficients (8 real parameters). Hence there
are just the right number
of coefficients
to specify any $2 \times 2$ matrix, so Eq.~\eqref{intro_A} is a general
expression for a $2 \times 2$ matrix.

\index{eigenvalues and eigenvectors}
Let's determine the eigenvalues and eigenvectors of $X$. The eigenvalues
$\lambda$ are obtained from 
\begin{equation}
\begin{vmatrix}
0 - \lambda & 1 \\
1 & 0 - \lambda
\end{vmatrix} = 0,
\end{equation}
which gives $\lambda^2 - 1 = 0$ or $\lambda = \pm 1$. These are real, which
they must be since $X$ is Hermitian.

Let us now get the
eigenvectors. We denote the corresponding normalized eigenvectors by
$\vec{e}_{+1}$ and $\vec{e}_{-1}$ and indicate the coefficients by $a$ and $b$.
\begin{itemize}
\item $\lambda = +1$. \\
\begin{equation}
\begin{pmatrix}
0 & 1 \\
1 & 0 
\end{pmatrix} 
\begin{pmatrix}
a \\ b
\end{pmatrix}
= \begin{pmatrix}
a \\ b
\end{pmatrix} ,
\end{equation}
which gives the equations $b=a$ and $a = b$, which are the same.
\index{normalization}
To normalize the eigenvector, we take $a = b = 1/\sqrt{2}$, so
\begin{equation}
\vec{e}_{+1} = {1 \over \sqrt{2}} 
\begin{pmatrix}
1 \\ 1
\end{pmatrix} .
\label{X1}
\end{equation}

\item $\lambda = -1$. \\
\begin{equation}
\begin{pmatrix}
0 & 1 \\
1 & 0 
\end{pmatrix} 
\begin{pmatrix}
a \\ b
\end{pmatrix}
= -\begin{pmatrix}
a \\ b
\end{pmatrix}
\end{equation}
which gives the two equations $b=-a$ and $a = -b$ (which are equivalent).
The normalized eigenvector is therefore
\begin{equation}
\vec{e}_{-1} = {1 \over \sqrt{2}} 
\begin{pmatrix}
1 \\ -1 
\end{pmatrix} .
\label{Xm1}
\end{equation}
The eigenvectors $\vec{e}_{+1}$ and $\vec{e}_{-1}$ are orthogonal, as we know
they must be since $X$ is Hermitian.
\end{itemize}

Forming the matrix of normalized eigenvectors gives
\begin{equation}
U = {1 \over \sqrt{2}}
\begin{pmatrix}
1 & 1 \\ 1 & -1
\end{pmatrix}
\end{equation}
which is unitary as expected.\footnote{A unitary matrix has the property that
$U^{-1} = U^\dagger$. Here, in addition, it turns out that $U^{-1}$ 
is equal to $U$ itself. This is not necessary for $U$ to be unitary,
though many unitary matrices in this course will have this property.}

It is instructive for the student to show that the eigenvalues of $Y$ and $Z$
are also $\pm 1$ and to determine their eigenvectors.
The student should also be able to show that $X, Y$ and $Z$ are not only
Hermitian but also unitary.

Pauli matrices have the property that the commutator of the two of them is
proportional to the third one,~e.g.
\begin{equation}
[X, Y] = 2 i Z,
\end{equation}
and similarly $[Y, Z]=2i X$ and $[Z, X] = 2 i Y$.
\index{matrices!anti-commuting}
Furthermore, if we define the \textit{anti-commutator} of two matrices by 
\begin{equation}
\{A, B\} \equiv AB + BA, 
\end{equation}
then, interestingly, different Pauli matrices \textit{anti-commute},~e.g.
\begin{equation}
\{X, Y\} = 0,
\end{equation}
and similarly $\{Y, Z\} = \{Z, X\} = 0$.

Another $2 \times 2$ matrix which is very important in quantum computing is
\index{Hadamard matrix (gate)}
the Hadamard, defined by
\begin{equation}
H ={1 \over \sqrt{2}}\left(X + Z\right) = {1 \over \sqrt{2}}
\begin{pmatrix}
1 & 1 \\
1 & -1 \\
\end{pmatrix} .
\label{had:intro}
\end{equation}
The Hadamard also has eigenvalues $\pm 1$.

\section{Properties of Matrices}
\label{sec:mat_props}

Two properties of square matrices will be important: the trace, which is the
sum of the diagonal elements, and the determinant.
\index{matrix!trace}
\index{matrix!determinant}
It is left as an exercise for the student to show (i) that the trace is the sum of the
eigenvalues, and (ii) that the trace of a product of matrices is invariant
under a cyclic permutation of the matrices so, for example, $\Tr AB = \Tr BA$
even if $A$ and $B$ don't commute so $AB \ne BA$.
\index{matrix!commutator}

We will now show (iii) that the determinant is the product of the eigenvalues.
If we multiply Eq.~\eqref{DA} on the left by $S$ and on the right by
$S^{-1}$ we get
\begin{equation}
A = S D S^{-1} .
\label{AD}
\end{equation}
An important result of linear algebra, which is not as well known in the
scientific community as it should be, is that
determinant of a product of matrices is equal to the product of the
determinants, i.e.
\index{matrix!determinant}
\begin{equation}
\det \left(A B\right) = \det A \, \det B .
\end{equation}
Taking the determinant of both sides of Eq.~\eqref{AD} gives
\begin{align}
\det A &= \det S \det D \det S^{-1} \nonumber \\
&=  \det D \det S \det S^{-1} \nonumber \\
&=  \det D \det \left(S S^{-1}\right) \nonumber \\
&= \det D = \prod_{m=1}^N \lambda_m ,
\end{align}
which is the desired result.

\hrulefill
\section*{Problems}
\input{hw_ch2.tex}

%% file: hw_ch2.tex
\begin{problems}

\item
For the following matrix $A$ 
$$
A = 
\begin{pmatrix}
1 & 2 & i \\
-2 & 1 & 3  \\
-i & 1 & 0 \\
\end{pmatrix}
,
$$
determine $A^T$ and $A^\dagger$.
\item
Show whether the following matrices are Hermitian or unitary or both or neither.
$$
(a) \quad
A = {1 \over \sqrt{2}}
\begin{pmatrix}
1 & 1 \\
1 & -1 \\
\end{pmatrix}
, \quad (b) \quad
B = 
\begin{pmatrix}
7 & 3i \\
-3i & 4 \\
\end{pmatrix}
,\quad (c) \quad
C =  {1 \over 5}
\begin{pmatrix}
3 & 4 \\
-4 & 3 \\
\end{pmatrix}
$$
\textit{Note:} To decide if a matrix $A$ is unitary it is simpler to check if
$A^\dagger A = \mathbbm{1}$ (where $\mathbbm{1}$ is the identity matrix) than to check if $A^{-1}= A^\dagger$.

\item 
Find the eigenvalues and normalized eigenvectors of the following matrix
\begin{equation}
\begin{pmatrix}
1 & 0 & 1 \\
0 & 1 & 0 \\
1 & 0 & 1 \\
\end{pmatrix} .
\end{equation}
\textit{Note:} As a check you should verify that the sum of the eigenvalues you
find
is equal to the trace (sum of diagonal elements). You should also check that
the eigenvectors are orthogonal.

\item
Verify that the trace of a matrix is equal to the sum of its eigenvalues for
the following matrices:
$$
(a)  \quad
A = 
\begin{pmatrix}
0 & 1 \\
1 & 0 \\
\end{pmatrix}
, \quad
(b) \quad
B = 
\begin{pmatrix}
2 & 0 & 2 \\
0 & 4 & 4 \\
1 & 0 & 3 \\
\end{pmatrix}
\label{trace}
$$
\item
For the matrices in Qu.~\ref{trace} verify that the determinant is equal to the
product of the eigenvalues.

\item
Show that the eigenvalues of a Hermitian matrix are real. Show also that the
eigenvectors of a Hermitian matrix belonging to distinct eigenvalues are
orthogonal.

\item
\textit{Cyclic invariance of the trace}\\
 Show that the trace of a product of
matrices is invariant under a cyclic permutation of the matrices, e.g.
\begin{equation}
\Tr (A B C ) = \Tr (B C A) = \Tr (C A B) \, .
\end{equation}
Hence show that the trace of a matrix is equal to the sum of its eigenvalues.

\item
Show that the eigenvalues of a matrix whose square is the identity are $\pm
1$.

\item
Show that two matrices $A$ and $B$ have common eigenvectors only if they
commute, i.e. if
\begin{equation}
[A, B] \equiv AB - BA = 0 .
\end{equation}

\item
Consider the Pauli spin matrices
\begin{equation}
\sigma_x \equiv X = 
\begin{pmatrix}
0 & 1 \\
1 & 0 \\
\end{pmatrix},\quad
\sigma_y \equiv Y = 
\begin{pmatrix}
0 & -i \\
i & 0 \\
\end{pmatrix},\quad
\sigma_z \equiv Z = 
\begin{pmatrix}
1 & 0 \\
0 & -1 \\
\end{pmatrix},\quad
\end{equation}
\begin{enumerate}[label=(\roman*)]
\label{pauli_hw}
\item
Show that $\sigma_i^2 = \mathbbm{1}$ for $i=x, y, z$.
\item
Determine the eigenvalues of each of the matrices.
\item
Determine the commutators $[\sigma_i, \sigma_j] \equiv \sigma_i \sigma_j -
\sigma_j \sigma_i$
for all distinct pairs $i$ and $j$. Express your results in terms of Pauli
matrices.
\item
Determine the anti-commutators $\{\sigma_i, \sigma_j\} \equiv \sigma_i \sigma_j +
\sigma_j \sigma_i$
for all distinct pairs $i$ and $j$.
\end{enumerate}

\item
Consider $\vec{\sigma} = \hat{x} \sigma_x + \hat{y} \sigma_y + \hat{z}
\sigma_z$, where the $\hat{x}$ etc.~refer to unit vectors in the indicated
coordinate directions. Show that
$$
(\vec{a} \cdot \vec{\sigma}) 
(\vec{b} \cdot \vec{\sigma})  = (\vec{a}\cdot
\vec{b}) \mathbbm{1} + i (\vec{a} \times \vec{b}) \cdot
\vec{\sigma}\, .
$$
\textit{Note:} The answers to Qu.~\ref{pauli_hw} will be useful here.

\item
If $A = B C$ show that $A^\dagger = C^\dagger B^\dagger$ (note the reverse
order). Hence show that if $B$ and $C$ are Hermitian, then $A$ is
Hermitian only if $[B, C] = 0$, (i.e.~if $B$ commutes with $C$.)

\end{problems}

%% file: qu_intro8.tex
In this chapter we give an introduction to quantum mechanics. A good textbook
on the subject, at an undergraduate level, is Griffiths~\cite{griffiths:05}.

\section{Quantum States as Complex Vectors}

\index{quantum state}
In Chapter \ref{ch:lin_alg} we reviewed linear algebra, including vectors,
generalized to the case where the coefficients of the vectors are complex. 

\index{vector!complex}
We now describe the basic postulates of quantum mechanics. We will see that
the framework is precisely that of complex vectors. The notation, however, is quite
different and so, for the next few equations, we will show
both
a statement concerning quantum mechanics in quantum mechanics notation, \textit{and} the
corresponding statement for complex vectors in the standard notation of linear
algebra.

While the discussion which follows may seem very abstract don't forget that
quantum mechanics is arguably the most successful theory in all of physics,
with countless precise comparisons between theory and experiment, some to the most
exquisite accuracy\footnote{For example, experimental and theoretical values
for the magnetic moment of the electron agree to better than a part in a
\index{magnetic moment}
\textit{trillion}, see Eq.~\eqref{3} of
\url{https://www.mdpi.com/2218-2004/7/2/45/pdf}.}.

Now we get started with quantum mechanics:

\medskip
\noindent \textbf{Ansatz 1:}
The state of a quantum system is a complex vector (which we
shall often call a ``state vector'' or just a vector).
\index{state vector|see {quantum state}}
\index{quantum state}

\medskip
\noindent In quantum computing one uses 
the notation of Dirac,\index{Dirac notation} in which a quantum state is written as
$|\psi\rangle$.
\begin{equation}
\mathrm{QM\ state}: \quad |\psi\rangle, \quad \Longleftrightarrow \quad
\mathrm{complex\ vector}: \quad \vec{v} .
\end{equation}
In equations with the double arrow $\Longleftrightarrow$ in the middle, the
part to the left of the arrow is in the notation of quantum mechanics, and the
part to the right is the corresponding statement in standard linear algebra notation.
The state $|\psi\rangle$ can be expressed as a linear
combination of basis states $|n\rangle$, 
\begin{equation}
|\psi\rangle = \sum_{n=1}^{N} c_n |n\rangle, \quad \Longleftrightarrow \quad
\vec{v} = \sum_{n=1}^{N} v_n \hat{e}_n ,
\label{lin_sup}
\end{equation}
in which the $c_n$ are called ``amplitudes".

We can write the state as a column vector
\begin{equation}
|\psi\rangle = 
\begin{pmatrix}\
c_1 \\c_2 \\ \vdots \\ c_{N}
\end{pmatrix}
\quad \Longleftrightarrow \quad
\vec{v} = 
\begin{pmatrix}\
v_1 \\v_2 \\ \vdots \\ v_{N}
\end{pmatrix} .
\end{equation}

We also introduce the dual state vector, denoted by $\langle \psi|$.
This
corresponds to the complex conjugate of the transpose vector introduced in
Eq.~\eqref{inner} in the context of the scalar product of a complex vector.
In other words, if
$|\phi\rangle$ is represented as a column vector by
\begin{equation}
|\phi\rangle = \begin{pmatrix} d_1 \\ d_2 \\ \vdots \\ d_{N}  \end{pmatrix},
\end{equation}
then the corresponding dual vector, needed to define the scalar (inner) product,  is
\begin{equation}
\langle \phi| = 
\begin{pmatrix}
d_1^\star & d_2^\star \cdots & d^\star_{N} \\
\end{pmatrix}
\quad \Longleftrightarrow \quad
\left(v^T\right)^\star =
\begin{pmatrix}
v_1^\star & v_2^\star \cdots & v^\star_{N} \\
\end{pmatrix} ,
\end{equation}
i.e.~a row vector in which the coefficients are the complex conjugate of the
coefficients in the original column vector. We will need the dual vector, as
well as the state vector, to define scalar (inner) products. Dirac called the
state vector a ``ket" and the dual vector a ``bra", and this notation is still
commonly used. 

In terms of components, the basis state $|n\rangle$ is a column vector with
all elements equal to zero except for the $n$-th element which has value 1, and the
dual basis state $\langle n|$ is a row vector with the $n-$th element equal to
1 and other elements equal to 0, i.e.
\begin{equation}
|n\rangle = \qquad
(\mathrm{row\ } n)
\begin{pmatrix}
0 \\ 0 \\ \vdots \\ 1 \\ \vdots \\  0 \\ 0
\end{pmatrix}
, \qquad
\langle n| = 
\begin{pmatrix}
0, & 0, & \cdots & 1, & \cdots & 0   & 0
\end{pmatrix} ,
\label{basis_comps}
\end{equation}

\index{inner product}
The scalar product of two vectors is called the ``inner product" in a general
context and this nomenclature
will be used here from now on. In quantum mechanics, the inner product of a vector
$|\psi\rangle$ with vector $|\phi\rangle $ is written as $\langle\phi|\psi\rangle$.
\begin{equation}
\langle\phi|\psi\rangle = \sum_{n=1}^{N} d_n^\star c_n 
\quad \Longleftrightarrow \quad
\vec{a} \cdot \vec{b} = \sum_{n=1}^{N} a^\star_n b_n .
\end{equation}
From this definition it follows that
\begin{equation}
\langle\phi|\psi\rangle = \langle\psi|\phi\rangle^\star .
\end{equation}

\index{norm}
The length of a vector in quantum mechanics
is called the ``norm" and
written $\|\psi\|$. As with ordinary vectors, the
norm of a state vector in quantum mechanics is the square root of the inner
product with itself, i.e.
\begin{equation}
\|\psi\| = \langle\psi|\psi\rangle^{1/2} = \left(\sum_{n=1}^{N} |c_n|^2 \right)^{1/2} \quad
\Longleftrightarrow \quad 
|v| = \left(\vec{v} \cdot \vec{v}\right)^{1/2} =  \left(\sum_{n=1}^N
|v_n|^2\right)^{1/2}.
\end{equation}
As we shall see later, in quantum mechanics state vectors must have
unit norm. Such vectors are said to be normalized.

Orthogonality. Two state vectors are said to be orthogonal if their inner
product is zero:
\begin{equation}
\langle\phi|\psi\rangle = \langle\psi|\phi\rangle = 0, \quad
\Longleftrightarrow \quad \vec{a} \cdot \vec{b} = \vec{b} \cdot \vec{a} = 0.
\end{equation}

We choose basis states $|n\rangle$ which are orthonormal, i.e.~normalized and orthogonal,
\index{normalization}\index{orthonormal}
\begin{equation}
\langle n|m\rangle = \delta_{nm},
\quad \Longleftrightarrow \quad
\vec{e}_n \cdot \vec{e}_m = \delta_{nm} .
\label{ortho_normal}
\end{equation}

So far, in this chapter we have emphasized the correspondence between quantum
mechanical states and complex vectors. Now that we are familiar with this
correspondence, 
from now on we will describe the formulation of quantum mechanics using only
quantum mechanics notation.

It will be useful to rewrite Eq.~\eqref{lin_sup} for a linear superposition
in a different way. Starting with Eq.~\eqref{lin_sup},
\begin{equation}
|\psi\rangle = \sum_{n=1}^{N} c_n |n\rangle,
\label{lin_sup2}
\end{equation}
we take the inner product of both sides with the dual of one of the basis
states, $\langle m |$ say. Using the orthonormality property in
Eq.~\eqref{ortho_normal}
gives us the coefficient\footnote{For ordinary vectors the corresponding expression would be 
$v_n = \vec{e}_n \cdot \vec{v}$.}
for state $|n\rangle$,
\begin{equation}
c_n = \langle n | \psi\rangle ,
\label{an}
\end{equation}
so we can rewrite Eq.~\eqref{lin_sup2} as
\begin{equation}
|\psi\rangle = \sum_{n=1}^{N} |n\rangle \langle n | \psi\rangle.
\label{psi_dirac}
\end{equation}
We call $\langle n | \psi\rangle$ the \textit{amplitude} for the
\index{amplitude}state
$|\psi\rangle$ to be in basis state $|n\rangle$.
Equation \eqref{psi_dirac} shows us that 
\begin{equation}
\sum_{n=1}^{N} |n\rangle \langle n| = \mathbbm{1} ,
\label{complete}
\end{equation}
the identity
matrix. Equation \eqref{complete} is sometimes called a completeness relation.
\index{completeness relation}
A single term in this sum, $|n\rangle \langle n|$ is an $N
\times N$ matrix with all elements 0 except for the $n$-th diagonal element
which is equal to 1. Note that a little more generally, $|n\rangle \langle m|$ is a matrix
with all elements zero except for the $m$-th row and $n$-th column which is
equal to 1. 
This is to be contrasted with $\langle n | m \rangle$ which is a single number
equal to $\delta_{n m}$.


To make our discussion
more concrete consider the following example of a 2-state system, i.e. a
single qubit,
\index{qubit}
\begin{equation}
|\psi'\rangle = 
\begin{pmatrix}
1 \\ 2i 
\end{pmatrix} .
\end{equation}
This is not normalized because the norm is
\index{norm}
\begin{equation}
\|\psi'\| = \sqrt{ 1^2 + |2i|^2} = \sqrt{1 + 4} = \sqrt{5}.
\end{equation}
To get a valid quantum state it must be properly normalized so we divide by the norm. Hence
\index{normalization}
\begin{equation}
|\psi\rangle =  {1 \over \sqrt{5}}  |\psi'\rangle =   {1 \over \sqrt{5}}
\begin{pmatrix}
1 \\ 2i 
\end{pmatrix} 
\end{equation}
is a valid quantum state. To get the dual state vector we take the complex
conjugate of the transpose, so
\begin{equation}
\langle\psi| = {1\over \sqrt{5}} 
\begin{pmatrix}
1,  & -2 i  \\
\end{pmatrix} .
\end{equation}

Suppose we also have a second state,
\begin{equation}
|\phi\rangle =   {1 \over \sqrt{5}}
\begin{pmatrix}
2 \\ -i 
\end{pmatrix}  ,
\end{equation}
which we see is normalized because $\sqrt{2^2 +|-i|^2} = \sqrt{5}$.
What then is the inner product $\langle \psi|\phi\rangle$? We have
\begin{equation}
\langle \psi|\phi\rangle = { 1 \over 5} \begin{pmatrix}
1 , & -2 i  \\
\end{pmatrix}
\begin{pmatrix}
2 \\ -i 
\end{pmatrix}
= {1 \over 5}( 1\cdot 2 + (-2i)\cdot (-i) ) = 0,
\end{equation}
so $|\psi\rangle$ and $|\phi\rangle$ are actually orthogonal. In this example we
had to be careful with the factors of $i$ because a complex conjugate is taken
when we form the dual vector (which we need to get the inner product with
another vector).

To make sure we haven't forgotten it, let's reiterate (with a bit more math
jargon) the first Ansatz of
quantum mechanics which we stated at the beginning of this section:\\
\noindent \textbf{Ansatz 1:}
The state of a quantum system is a vector in a complex vector space
(technically a Hilbert space though we won't need that level of mathematical
sophistication here).

\section{Phases}
\index{phases}
At this point it is convenient to discus an important topic, namely
\textit{phases}. Suppose we have a 2-state system with complex amplitudes,
\index{amplitude}
which we write in polar form as
\begin{equation}
|\psi\rangle = r_0 e^{i\theta_0} |0\rangle +  r_1 e^{i\theta_1} |1\rangle ,
\end{equation}
where $r_0^2 + r_1^2 = 1$ for normalization. Let's take out the factor of
$e^{i\theta_0}$, so
\begin{equation}
|\psi\rangle = e^{i\theta_0} \left(\,r_0 |0\rangle +  r_1
e^{i(\theta_1-\theta_0)} |1\rangle\,\right) .
\end{equation}
We call $\theta_0$ the \textit{global} phase which turns out to have no
physical significance, while $\theta_1-\theta_0$ is the \textit{relative} phase
(of basis states $|1\rangle$ and $|0\rangle$)
which is important because it gives rise to interference. It is crucial to
understand the difference between global phase and relative phase. \textbf{States
which differ only in the overall phase are physically identical}. As we will see
in Sec.~\ref{sec:meas} the reason for this is that no measurement can distinguish
states which only differ by a global phase. By contrast, \textbf{states
which differ in a relative phase are physically distinct} because measurements
can distinguish between them.
\index{phase!global}\index{phase!relative}
\index{interference!quantum}

For example,
\begin{equation}
|\psi_1\rangle = {1\over \sqrt{2}}
\begin{pmatrix}
1 \\ -1
\end{pmatrix},\qquad
|\psi_2\rangle = {1\over \sqrt{2}}
\begin{pmatrix}
-1 \\ 1
\end{pmatrix},\qquad
\end{equation}
describe the same state because one is just the negative of the other. By
contrast,
\begin{equation}
|\psi'_1\rangle = {1\over \sqrt{2}}
\begin{pmatrix}
1 \\ 1
\end{pmatrix},\qquad
|\psi'_2\rangle = {1\over \sqrt{2}}
\begin{pmatrix}
1 \\ -1
\end{pmatrix},\qquad
\end{equation}
describe different states because the relative phase of $|1\rangle$ and
$|0 \rangle$ is different in the two cases
($0$ for $|\psi'_1\rangle$ and $\pi$ for $|\psi'_2\rangle$).

\section{Observables}
\index{observables}
\label{sec:observables}
How is all this abstract stuff about complex vectors related to the real
world, i.e.~to quantities that we can measure.

The answer is that an observable quantity will be an \textit{operator},
$\hat{O}$ say, acting on these vectors. The ``hat" symbol ``$\ \hat{\ }\ $'' indicates an
operator, though, for simplicity of notation, we will usually omit the hat when context makes clear
that we are
dealing with an operator. In terms of components, operators are
represented by matrices.

An operator acting on a state vector gives another state vector, so
\begin{equation}
\hat{O} |\psi\rangle = |\phi\rangle, 
\end{equation}
\index{linearity}
A crucial point is that operators in quantum mechanics are \textit{linear},
i.e.
\begin{equation}
\hat{O}\left(\, a |\psi\rangle + b |\phi\rangle \,\right) = a\, \hat{O}
|\psi\rangle + b\, \hat{O} |\phi\rangle ,
\end{equation}
so an operator acts separately on the different pieces of a superposition.

This brings us to the second Ansatz of quantum mechanics:

\medskip
\index{measurements}
\noindent \textbf{Ansatz 2:} Observables are represented by linear Hermitian
operators. The result of a measurement is
one of the eigenvalues of the corresponding operator
$\hat{O}$. After the measurement, the system is in the eigenstate
corresponding to the measured eigenvalue.

Note this means that in general measurements \textbf{change the state of the system}.
The only exception is if the system was in an eigenstate of the measurement
operator \textit{before} the measurement. 

\medskip
Why is it assumed that quantity which can be measured is represented by a
\textit{Hermitian} operator? The answer is that the eigenvalues of a
Hermitian operator (matrix) are guaranteed to be real, and we know that the
results of a measurement must be real.

We now discuss how to represent operators as a matrix using the
Dirac notation. We take 
orthonormal basis vectors $|n\rangle$ which have the property $\langle
m|n\rangle = \delta_{mn}$.
In terms of components, $|n\rangle$ will be a column vector with 
the $n$-th entry equal to 1 and all the others zero, as shown in
Fig.~\eqref{basis_comps}.
Consider the action of an operator $A$ on one of the
basis vectors $|n\rangle$. It will give a linear combination of the basis
vectors.
\begin{equation}
\begin{pmatrix}
A_{11} & A_{12} & \cdots & A_{1n} & \cdots & A_{1,N-1} & A_{1N} \\
A_{21} & A_{22} & \cdots & A_{2n} & \cdots & A_{2,N-1} & A_{2N} \\
\vdots & \vdots & \ddots & \vdots & \ddots & \vdots   & \vdots \\
A_{n1} & A_{n2} & \cdots & A_{nn} & \cdots & A_{n,N-1} & A_{nN} \\
\vdots & \vdots & \ddots & \vdots & \ddots & \vdots   & \vdots \\
A_{N-1,1} & A_{N-1,2} & \cdots & A_{N-1,n} & \cdots & A_{N-1,N-1} & A_{N-1,N} \\
A_{N1}    & A_{N2}    & \cdots & A_{Nn}    & \cdots & A_{N,N-1}   & A_{NN}
\end{pmatrix}
 \ \begin{pmatrix}
0 \\ 0 \\ \vdots \\ 1 \\ \vdots \\ 0 \\ 0
\end{pmatrix} = 
\begin{pmatrix}
c_1 \\ c_2 \\ \vdots \\ c_n \\ \vdots \\ c_{N-1} \\ c_N
\end{pmatrix} .
\label{Anm}
\end{equation}
We see that $c_k$ is equal to the element of $A$ on the $k$-th row and $n$-th
column, i.e.~$A_{kn}$. We can therefore write Eq.~\eqref{Anm} as
\begin{equation}
A |n\rangle = \sum_k A_{kn} |k\rangle .
\end{equation}
Acting on the left with the dual vector $\langle m|$ and using the
orthonormality of the basis vectors, we get
\begin{equation}
A_{mn} = \langle m|A|n \rangle ,
\end{equation}
which is the connection between the usual suffix notation for an element of
a matrix, $A_{mn}$, and the Dirac notation for the same thing, $\langle
m|A|n\rangle$. They both refer to the $m$-th row and $n$th column of the matrix
$A$.

Recall that the definition of the adjoint of a matrix is $A^\dagger =
\left(A^T\right)^\star$. Hence, in Dirac notation,
\begin{equation}
\langle m|A^\dagger|n \rangle = \langle n|A|m \rangle^\star.
\end{equation}
If $A$ is Hermitian then it is equal to its adjoint so
\begin{equation}
\langle m|A|n \rangle = \langle n|A|m \rangle^\star \qquad (\mathrm{for}\ A\ \mathrm{Hermitian}).
\end{equation}
Note that this states, in component form, that the transpose of a Hermitian
matrix is equal to its complex conjugate, which is precisely the definition of
a Hermitian matrix. 

To gain still more familiarity with the Dirac notation consider
$\langle\phi|A|\psi\rangle$. If we write this out in components in some basis,
then
$|\psi\rangle$ is a column vector, $A$ is a matrix and $\langle\phi|$ is a row
vector, i.e.~we have
\begin{equation}
\langle\phi|A|\psi\rangle = 
\begin{pmatrix}
\phi_1^\star& \phi_2^\star& \cdots & \phi_N^\star 
\end{pmatrix}
\begin{pmatrix}
A_{11} & A_{12} & \cdots & A_{1N} \\
A_{21} & A_{22} & \cdots & A_{2N} \\
\vdots & \vdots & \ddots & \vdots \\
A_{N1} & A_{N2} & \cdots & A_{NN} 
\end{pmatrix}
\begin{pmatrix}
\psi_1 \\ \psi_2 \\ \vdots  \\ \psi_N
\end{pmatrix} ,
\end{equation}
in an obvious notation.
The multiplication can be done either by acting with $A$ on $|\psi\rangle$
to get $|A\psi\rangle$ and then taking the inner product with $\langle \phi|$,
or by acting with $A$ to the left on $\langle\phi|$ and then taking the
inner product with $|\psi\rangle$. But what does acting with $A$ to the left on
$\langle\phi|$ mean? Let's suppose that
\begin{equation}
\langle \phi| A = \langle \mu| .
\label{2}
\end{equation}
Then  we have
\begin{equation}
\begin{pmatrix}
\phi_1^\star& \phi_2^\star& \cdots & \phi_N^\star 
\end{pmatrix}
\begin{pmatrix}
A_{11} & A_{12} & \cdots & A_{1N} \\
A_{21} & A_{22} & \cdots & A_{2N} \\
\vdots & \vdots & \ddots & \vdots \\
A_{N1} & A_{N2} & \cdots & A_{NN} 
\end{pmatrix} = 
\begin{pmatrix}
\mu_1^\star& \mu_2^\star& \cdots & \mu_N^\star 
\end{pmatrix} .
\end{equation}
Evaluating components gives
\begin{equation}
\mu_m^\star = \sum_k \phi_k^\star A_{km} .
\end{equation}
This can be rearranged as
\begin{equation}
\begin{split}
\mu_m &= \sum_k \phi_k A^\star_{km}  \\ 
&= \sum_k \left(A^T\right)^\star_{mk} \phi_k = \sum_k
A^\dagger_{mk} \phi_k ,
\end{split}
\end{equation}
or, for the vector as a whole
\begin{equation}
|\mu\rangle = A^\dagger |\phi\rangle. 
\label{1}
\end{equation}
which is equivalent to Eq.~\eqref{2}. 
Hence the action of $A$ acting to the left on $\langle \phi|$ can be written
as
\begin{equation}
\langle \phi| A = \langle \, A^\dagger \phi\,| .
\label{phiA}
\end{equation}

Summarizing, we see that in $\langle\phi|A|\psi\rangle$, the operator $A$ can
be considered to act either to the left or the right as follows:
\begin{equation}
\langle \phi| A|\psi\rangle = \langle\, A^\dagger \phi\,|\psi\rangle = \langle
\phi|\, A \psi\,\rangle .
\end{equation}
In quantum mechanics $A$ will commonly be a Hermitian operator (since
observables are represented by Hermitian operators) for which $A^\dagger = A$,
so $A$ acts equally to the right and to the left as follows:
\begin{equation}
\langle \phi| A|\psi\rangle = \langle\,A \phi\,|\psi\rangle = \langle
\phi|\,A \psi\,\rangle \qquad (\mathrm{for}\ A\ \mathrm{Hermitian}).
\end{equation}

\section{The Computational Basis and Change of Basis}

When dealing with standard vectors, we know that we can work with different
sets of bases rotated with respect to each other. In quantum mechanics, too,
it will be convenient to represent state vectors in terms of different bases,
transformed with respect to each other. 

The standard basis for a single \index{qubit} qubit comprises the states $|0\rangle$ and
$|1\rangle$ and in this basis the Pauli operator $Z$ is diagonal, see
Eq.~\eqref{pauli}. This basis 
is called the computational basis. \textbf{It is
the basis in which measurements are performed.} 
Since $Z$ is diagonal in this basis the
\index{computational basis}
eigenvectors of $Z$ are the basis vectors. For this reason the computational
basis is sometimes called the $Z$-basis.

Note that for state $|0\rangle$ the eigenvalue of $Z$ is $+1$ and for 
state $|1\rangle$ the eigenvalue of $Z$ is $-1$. One might have thought it
should be the other way round but this is the convention that has been
adopted. 

\index{basis!change of}
We will also need to consider other bases, one of the most common being the
$X$-basis, i.e.~the basis in which $X$ (see Eq.~\eqref{pauli}) is diagonal.
We showed in Sec.~\ref{sec:important}
that the eigenvalues of $X$ are $+1$ and $-1$, with 
corresponding eigenvectors, called $|+\rangle$ and $|-\rangle$ (sometimes
called
$|0_x\rangle$ and $|1_x\rangle$), given by
\begin{equation}
\begin{split}
|0_x\rangle &\equiv |+\rangle = {1\over \sqrt{2}}\left(|0\rangle + |1\rangle \right) \\
|1_x\rangle &\equiv |-\rangle = {1\over \sqrt{2}}\left(|0\rangle - |1\rangle \right) .
\end{split}
\label{ZX}
\end{equation}
Note that the states $|0\rangle$ and $|1\rangle$ without subscripts always refer to
the $Z$-basis.
From these results it follows that, in the $X$ basis, the Pauli $X$-matrix is
written as
\begin{align}
& \quad |+\rangle \ \ \, |-\rangle \nonumber \\
X  =
\begin{matrix}
\langle + | \\
\langle - | 
\end{matrix}
& \begin{pmatrix}
\ 1\ \  &  0\  \\
\ 0\ \  & -1\ \\
\end{pmatrix} ,
\end{align}
which looks just like the Pauli-$Z$ matrix in the $Z$
(computational)\index{computational basis} basis.

There is a linear relation between the new basis vectors and the old ones.
Denoting the old basis vectors by Latin letters, e.g.~$|n\rangle$, and the new basis
vectors by Greek letters, e.g.~$|\alpha\rangle$, we write
\begin{equation}
|\alpha\rangle = \sum_n U_{\alpha n} |n\rangle .
\label{Ualphan}
\end{equation}
The new basis vectors must be orthonormal, like the old set, and this constrains
the matrix of coefficients $U$ in a way that we will now determine. Writing
the equivalent of
Eq.~\eqref{Ualphan} in terms of row vectors and taking the complex conjugate,
we get the following transformation for the dual basis state vectors
\begin{equation}
\langle \beta| = \sum_k U^\star_{\beta k}  \langle k| .
\label{Unalpha}
\end{equation}
Taking the inner product of Eqs.~\eqref{Ualphan} and \eqref{Unalpha} gives
\begin{align}
\langle \beta| \alpha\rangle &= \sum_{n, k} U^\star_{\beta k} U_{\alpha n}
\langle k|n\rangle \nonumber \\
&= \sum_{n} U^\star_{\beta n} U_{\alpha n} \nonumber \\
&= \sum_{n} U_{\alpha n}\left(U^T\right)^\star_{n \beta}
= \sum_{n} U_{\alpha n}U^\dagger_{n \beta}  = \left(U U^\dagger\right)_{\alpha\beta},
\end{align}
where we used that $\langle k|n\rangle = \delta_{kn}$ to get the second line. 
However, $\langle\beta| \alpha\rangle = \delta_{\alpha\beta}$ and so we must have
$U U^\dagger = \mathbbm{1}$, the identity matrix. Thus the matrix of
coefficients which transforms from one basis to another as in
Eq.~\eqref{Ualphan} must be unitary.
\index{matrix!unitary}
\index{unitary transformation}

As an example, according to Eq.~\eqref{ZX} the transformation from the
$Z$-basis to the $X$-basis can be written as
\begin{equation}
\begin{pmatrix}
|+\rangle \\ |-\rangle 
\end{pmatrix}
= U 
\begin{pmatrix}
|0\rangle \\ |1\rangle 
\end{pmatrix} ,
\label{ZXU}
\end{equation}
where
\begin{equation}
U = {1 \over \sqrt{2}}
\begin{pmatrix}
1 & 1 \\
1 & -1 
\end{pmatrix} .
\label{UZX}
\end{equation}
We can verify that this matrix is unitary by evaluating its inverse and
checking that $U^{-1} = U^\dagger$, or, more
simply, by recalling that the rows of a unitary matrix are orthonormal vectors, and
the same for the columns. By inspection, this is the case here. The inverse
transformation is given by
\begin{equation}
\begin{pmatrix}
|0\rangle \\ |1\rangle 
\end{pmatrix}
= U^{-1} 
\begin{pmatrix}
|+\rangle \\ |-\rangle 
\end{pmatrix} .
\label{XZU}
\end{equation}
Noting that $U^{-1} = U^\dagger (= U\ \mathrm{here})$, we can write the
inverse transformation of Eq.~\eqref{ZX} as
\begin{equation}
\begin{split}
|0\rangle &= {1\over \sqrt{2}}\left(|+\rangle + |-\rangle \right) \\
|1\rangle &= {1\over \sqrt{2}}\left(|+\rangle - |-\rangle \right) .
\end{split}
\label{XZb}
\end{equation}
Consequently, a linear superposition in the $Z$-basis
\begin{equation}
|\psi\rangle = \alpha |0\rangle + \beta|1\rangle
\label{psiZ}
\end{equation}
can be written in the $X$-basis as 
\begin{equation}
|\psi\rangle = {1\over \sqrt{2}} (\alpha + \beta)\,|+\rangle +
{1\over \sqrt{2}} (\alpha - \beta)\,|-\rangle .
\label{psiX}
\end{equation}
I emphasize that Eqs.~\eqref{psiZ} and \eqref{psiX} are equivalent ways of
writing the same quantum state $|\psi\rangle$.

\section{Outer Product Notation}
\index{outer product}
\label{sec:outer}

For orthonormal basis vectors, we have
$\langle i|j\rangle = \delta_{ij}$.
As a further exercise in familiarization with the Dirac notation, consider
what we
mean if we write the vector and the dual
vector the other way round i.e.~$|i\rangle \langle j|$, which is called an
``outer product".  It is actually a
matrix.
By
sandwiching it on the left and right by basis states we see that it is a
matrix
whose entries are all zero except for the element in the $i$-th row and $j$-th
column which is 1. In other words
\begin{align}
& \qquad\qquad\quad\quad \mathrm{(col.\ } j) &   \nonumber\\
|i \rangle\langle j| = \mathrm{(row\ } i) &
\begin{pmatrix}
0      & 0      &0      & \cdots & 0      & \cdots & 0      \\
0      & 0      &0      & \cdots & 0      & \cdots & 0      \\
\vdots & \vdots &\vdots & \ddots & \vdots & \ddots & \vdots \\
0      & 0      &0      & \cdots & 1      & \cdots & 0      \\
\vdots & \vdots &\vdots & \ddots & \vdots & \ddots & \vdots \\
0      & 0      &0      & \cdots & 0      & \cdots & 0      \\
0      & 0      &0      & \cdots & 0      & \cdots & 0      \\
\end{pmatrix} .
\end{align}

If $j=i$, then we have a $1$ in the $i$-th diagonal element and $0$ everywhere
else.
This is a projection operator on to state $i$, so we denote it by $P_i$, i.e.
\begin{equation}
P_i = |i \rangle\langle i| .
\end{equation}
One
can see it
is a projection operator because, if it acts on an arbitrary state
$|\psi\rangle$, we have
\begin{equation}
P_i |\psi\rangle = |i \rangle\, \langle i|\psi\rangle,
\end{equation}
which is the amplitude $\langle i|\psi\rangle$ for $|\psi\rangle$ to be along
\index{amplitude}
$|i \rangle$, times the state $|i \rangle$.

Clearly $\sum_i P_i$ has $1$ on all the diagonal elements and is zero
otherwise, so it is the identity matrix, i.e.
\begin{equation}
\sum_i P_i \equiv \sum_i |i \rangle\langle i| = \mathbbm{1} ,
\end{equation}
which is also known as a
\index{completeness relation} completeness relation, see Eq.~\eqref{complete}.

\section{Functions of operators}
\index{functions of operators}

We will need to evaluate functions of operators.  For example what is $e^A$?
In this case there is a convergent series expansion which can be used to
evaluate the function;
\begin{equation}
e^A = 1 + A + {A^2 \over 2!} + {A^3 \over 3!} + \cdots .
\end{equation}
In some cases the infinite series can be evaluated in closed form. Consider
for example $e^{c X}$ where $c$ is 
a constant and 
$X$, the Pauli operator, is
given in Eq.~\eqref{pauli}. We have $X^2 = \mathbbm{1}$ and so $X^3 =
X^5 \cdots = X^{2n+1} \cdots = X$, while $X^2 = X^4 \cdots = X^{2n}  \cdots = \mathbbm{1}$. Hence
\begin{align}
e^{cX} &= \mathbbm{1} \left(1 + {c^2 \over 2! } + {c^4 \over 4!} + \cdots
\right) + X \left( c + {c^3 \over 3!} + {c^5 \over 5!} + \cdots \right),
\nonumber \\
&= \mathbbm{1}\,\cosh c + X\,\sinh c =
\begin{pmatrix}
\cosh c & \sinh c \\
\sinh c & \cosh c 
\end{pmatrix} .
\end{align}

More generally, we can evaluate a
function of an operator by diagonalizing it. Consider first 
a diagonal matrix,
\begin{equation}
D = 
\begin{pmatrix}
\lambda_1 &         0 & \cdots & 0 \\
0         & \lambda_2 & \cdots & 0 \\
\vdots    &  \vdots   & \ddots & \vdots\\
0         & 0         & \cdots & \lambda_N 
\end{pmatrix} .
\end{equation}
When multiplying $D$ by itself $n$ times, say, all that happens is each diagonal element is
multiplied by itself $n$ times. Hence if $f(D)$ is some function of $D$ which
can be represented by a series expansion, we have
\begin{equation}
f(D) = 
\begin{pmatrix}
f(\lambda_1) &         0 & \cdots & 0 \\
0         & f(\lambda_2) & \cdots & 0 \\
\vdots    &  \vdots   & \ddots & \vdots\\
0         & 0         & \cdots & f(\lambda_N)  
\label{fD}
\end{pmatrix} .
\end{equation}
If the function $f(x)$ for scalar argument $x$ does not have a series
expansion, we take Eq.~\eqref{fD} as the \textit{definition} of the matrix
function $f(D)$ for a diagonal matrix $D$.

In general, a matrix $A$ is not already in diagonal form.  However, we can
diagonalize it by a similarity transform, see Eq.~\eqref{AD}, which we repeat
here:
\begin{equation}
A = S D S^{-1},
\end{equation}
where $D$ is a diagonal matrix with the eigenvalues of $A$ on the diagonal.
Hence it follows that
\begin{align}
A^2 &= S D S^{-1} S D S^{-1} = S D^2 S^{-1}, \nonumber \\
A^3 &= S D S^{-1} S D S^{-1} S D S^{-1} = S D^3 S^{-1}, \quad \mathrm{and\ so} \nonumber \\
A^n &= S D^n S^{-1}, \quad \mathrm{and\ hence}\nonumber \\
f(A) &= S f(D) S^{-1} \nonumber \\
&= S 
\begin{pmatrix}
f(\lambda_1) &         0 & \cdots & 0 \\
0         & f(\lambda_2) & \cdots & 0 \\
\vdots    &  \vdots   & \ddots & \vdots\\
0         & 0         & \cdots & f(\lambda_N)  
\end{pmatrix}
S^{-1} ,
\end{align}
which is the desired expression showing how to construct a function of a
matrix from its eigenvalues and eigenvectors.

\section{Measurements}
\index{measurements}
\label{sec:meas}
Now we have to discuss in detail the vexed topic of measurement in quantum
mechanics. The reason for using the term ``vexed" will become clear later,
especially in Chapter \ref{ch:EPR} when we discuss a famous thought experiment of
Einstein, Podolsky and Rosen (EPR).\index{EPR}

In a measurement, our delicate quantum system is
brought into contact with a macroscopic experimental apparatus.
Measurement is an irreversible process and as such has a special status in
quantum mechanics. 


Assume that the Hermitian operator $A$ corresponding to the measured quantity
of interest has eigenvalues
$\lambda_n$ and normalized eigenvectors $|n\rangle$. Because $A$ is Hermitian
the eigenvalues are real. In addition, for a
Hermitian matrix of size $N$ there are $N$ orthogonal
eigenvectors which can therefore
be used as a basis. Hence, we can write the state of the system before
measurement, $|\psi\rangle$, as a linear superposition of the eigenvectors of
$A$, 
\index{superposition}
\begin{equation}
\begin{split}
|\psi\rangle &= \sum_{n=1}^N a_n |n\rangle . \\
&= \sum_{n=1}^N |n\rangle\, \langle n|\psi\rangle,
\end{split}
\end{equation}
where the last line is from Eq.~\eqref{psi_dirac}.

According to ansatz 2 in Sec.~\ref{sec:observables},
a measurement will give one of the eigenvalues, $\lambda_n$, but which one?
To answer this question, we need to add one more ingredient to our Ansatz 2, 
one which was first proposed by Born in
a footnote in a 1926 paper, and which
is therefore called the ``Born rule". \index{Born rule}
This states that the probability, $P(n)$, to get
eigenvalue $\lambda_n$
(and after the measurement to leave the system in eigenstate $|n\rangle$), is the
square of the modulus of the amplitude $a_n$, i.e.
\begin{equation}
P(n) = |a_n|^2  \equiv |\langle n|\psi\rangle|^2
\equiv \langle \psi|n\rangle \,\langle n|\psi\rangle ,
\label{prob}
\end{equation}
where we used Eq.~\eqref{an} and that $\langle\psi|n\rangle = \langle n|\psi\rangle^\star$.
Since probabilities must add up to 1, it follows that state vectors in quantum
mechanics must be normalized to unity, i.e. 
\begin{equation}
1 = \sum_n P(n) = \sum_n |a_n|^2 = \sum_n |\langle n|\psi\rangle|^2 = \sum_n
\langle \psi|n\rangle \,\langle n|\psi\rangle = \langle \psi|\psi\rangle .
\end{equation}
Note that the probability of a getting a particular measured value only depends on the square of the
modulus of the amplitude of the 
corresponding eigenstate. This means that the \textit{global} phase of a state has no physical
significance since no measurement can distinguish two states which differ
only by a global phase. 

However if two states differ in the \textit{relative} phase difference between basis states in a superposition
there are measurements which
can distinguish between them. For example,
$ |+\rangle = {1\over \sqrt{2}}(|0\rangle + |1\rangle)$ and
$ |-\rangle = {1\over \sqrt{2}}(|0\rangle - |1\rangle)$ are eigenstates of $X$
with eigenvalues $+1$ and $-1$ respectively, so a measurement of $X$ will give
different 
results ($+1$ for $|+\rangle$ and $-1$ for $|-\rangle$, with probability $1$ in
both cases).

We therefore have to complete our Ansatz 2 to include the probabilities of
different results:

\medskip
\noindent \textbf{Ansatz $2'$:} ``Observables are represented by a linear
Hermitian operator. Measurement of an
observable corresponding to a
(linear) Hermitian operator $\hat{O}$ gives one of the eigenvalues of
$\hat{O}$. The probability of getting an eigenvalue is the square of the modulus of the 
amplitude for the state of the system to be in the corresponding eigenstate of $\hat{O}$.
\index{amplitude}
After the measurement, the system is in this eigenstate."

\medskip
The fact that probabilities enter into
the results of measurements has led to a lot of ``vexed" discussion.
Your first reaction might be ``What's the fuss? After all, don't probabilities enter in classical
physics too? If one tosses a coin isn't the result randomly heads or
tails with equal probability?" Well, is it \textit{really} random? If one could measure
with sufficient precision the initial momentum and angular momentum of the
coin, and integrate the equations of motion for its trajectory, including the
effects of air resistance, to sufficient accuracy then one would be able to
compute, \textit{with certainty}, on which side it would land. The difficulty is that
the coin toss has great sensitivity to the initial conditions, which means that if one changes
the initial velocity by an immeasurably small amount the result changes.
In other words, \textit{for all practical purposes} (FAPP) a coin
toss \textit{is} random.  Nonetheless, from a fundamental point of view it is not, since it
is uniquely determined by the initial conditions. However, the situation in quantum
mechanics is different since, as far as we know, probabilities \textit{enter
in a fundamental way}. 

The most famous critic of probabilities being
part of a fundamental theory of physics was Einstein\index{Einstein, Albert}, who had many discussions
on the topic with Niels Bohr\index{Bohr, Niels}. As we shall see
in our study of the EPR \index{EPR}thought experiment
in Chapter \ref{ch:EPR},
despite Einstein's claim that
``God doesn't play dice with the universe", quantum mechanics has been
repeatedly vindicated. 

We have said that after a measurement the system is left in eigenstate
$|n\rangle$. Measurement therefore ``projects" the initial state
$|\psi\rangle$ on to $|n\rangle$. This is accomplished by the projection
operator
\begin{equation}
\hat{P}_n = |n\rangle\langle n| 
\end{equation}
so
\begin{equation}
\hat{P}_n |\psi\rangle = |n\rangle\langle n|\psi\rangle ,
\label{Pn}
\end{equation}
(no sum on $n$). The sum of the projection operators must add to the identity,
i.e.
\begin{equation}
\sum_n \hat{P}_n \equiv \sum_n  |n\rangle\langle n| = \mathbbm{1} .
\label{sumPn}
\end{equation}
The fact that $\sum_n  |n\rangle\langle n| $ can be replaced by the identity is
called a ``completeness" relation\index{completeness relation}.

Note that the state in Eq.~\eqref{Pn} is not normalized. If we continue to
follow the system after the measurement then we need to multiply the state by
$1/|\langle n|\psi\rangle|$, 
so it is again correctly normalized and the sum of probabilities of
results of a future measurement will add to unity.  We note that something similar is also
done in classical statistics. If we have a sequence of measurements, and
we know the result of the first one, then we can determine the ``conditional
probability'' of subsequent measurements, given the result of the
first measurement, and these conditional probabilities add to unity. In
effect, this is what is done by multiplying a state by a constant to get its
norm back to 1 after a measurement. The resulting state will give
the conditional
probabilities for a subsequent measurement given the result of the first
measurement. The fact that a quantum state is not normalized after a
measurement is often
referred to as ``wavefunction collapse"\footnote{For continuum problems, a
quantum state expressed as a function of position is referred to as a
wavefunction.}.


Let's give a simple example of a measurement. 
Consider one qubit in state
$|\psi\rangle = {1\over\sqrt{2}}(|0\rangle + |1\rangle)$ and measure
$Z$. The eigenstates of $Z$ are $|0\rangle$ and $|1\rangle$ with eigenvalues
$+1$ and $-1$ respectively. Hence the results of a measurement of $Z$ are
\begin{equation}
\begin{split}
+1, \qquad &\mathrm{prob.}\ \left({1 \over \sqrt{2}}\right)^2 = {1 \over 2},
\quad \mathrm{qubit\ is\ in\ state\ } |0\rangle \ \mathrm{after\ the \ measurement},\\
-1, \qquad &\mathrm{prob.}\ \left({1 \over \sqrt{2}}\right)^2 = {1 \over 2},
\quad \mathrm{qubit\ is\ in\ state\ } |1\rangle \ \mathrm{after\ the \ measurement}
.
\end{split}
\end{equation}

Now suppose that we measure $X$. The eigenstates of $X$ are shown in
Eqs.~\eqref{X1} and \eqref{Xm1} to be ${1\over\sqrt{2}}(|0\rangle \pm |1\rangle)$.
Hence $|\psi\rangle$ is the eigenstate with eigenvalue $+1$, so the result
of the measurement of $X$ is $+1$ with $100\%$ probability. Similarly a
measurement of $X$ on state ${1\over\sqrt{2}}(|0\rangle - |1\rangle)$ would
give $-1$ with probability $1$.

We see that if the initial state is an eigenstate of the operator being
measured, then the result will, with certainty, be the corresponding
eigenvalue, and the state will remain unchanged after the measurement. However, if
this is not the case, i.e.~if the initial state is in a superposition of
eigenstates of the measurement operator, then (i) the result of
the measurement will take one of several values with
appropriate probabilities, and (ii) the measurement changes the state, leaving it in the
eigenstate corresponding to the eigenvalue which is measured.

\section{Statistics of Measurements}

If we prepare many identical copies of the system and measure each of them
what can we say about the statistics of the measured values $\lambda_n$, the 
eigenvalues of $A$. First of all,
what would be the mean of the measurements $\langle A \rangle$? We have
\begin{align}
\langle A \rangle &= \sum_n P(n) \lambda_n \nonumber \\
&=  \sum_n |\langle n|\psi \rangle|^2 \lambda_n = \sum_n  \langle \psi|n\rangle \lambda_n  \langle n|\psi\rangle \nonumber \\
&= \sum_n  \langle \psi| A | n\rangle \langle  n|\psi\rangle \nonumber \\
&= \langle \psi| A| \psi\rangle \, .
\label{expect}
\end{align}
where we used Eq.~\eqref{prob} to get the second line, we used that $A|n\rangle
= \lambda_n |n\rangle $ to get the third line, and Eq.~\eqref{sumPn} to get
the last line. The final result, $\langle
\psi| A| \psi\rangle $ is called the ``expectation value'' of $A$ in state
$|\psi\rangle$.
\index{expectation value}

In addition to the average result we are also often interested in the \textit{scatter}
about the average. This is characterized by the standard deviation defined by
\begin{equation}
\Delta A = \Bigl\langle\, \left( A - \langle A \rangle\,\right)^2
\, \Bigr\rangle^{1/2} ,
\end{equation}
which is the root mean square deviation about the mean.  It can be expressed
in a slightly simpler form since 
\begin{align}
\bigl\langle\, \left( A - \langle A \rangle\right)^2 \, \bigr\rangle &=
\langle A^2 - 2 A \langle A\rangle  + \langle A\rangle^2 \rangle \nonumber \\
&= \langle A^2 \rangle - 2 \langle A\rangle^2 + \langle A\rangle^2
\nonumber \\
&= \langle A^2 \rangle - \langle A\rangle^2 ,
\end{align}
so
\begin{equation}
\Delta A = \left(\, \langle A^2 \rangle - \langle A\rangle^2\, \right)^{1/2}
\label{DeltaA}
\end{equation}
We will call $\Delta A$ the uncertainty in $A$.
\index{uncertainty}

Let's illustrate this with the example we considered just above, namely
$|\psi\rangle = {1\over\sqrt{2}}(|0\rangle + |1\rangle)$. If we measure $Z$ we
have
\begin{equation}
\langle Z \rangle  = \langle \psi|Z|\psi \rangle = {1 \over 2} 
\begin{pmatrix}
1 & 1
\end{pmatrix}
\begin{pmatrix}
1 & 0 \\
0 & -1
\end{pmatrix}
\begin{pmatrix}
1 \\ 1
\end{pmatrix}
 = 0.
\end{equation}
This agrees with our previous discussion where we found $+1$ and $-1$ with
equal probability. We also have $Z^2 = \mathbbm{1}$ and, since the average of 1 is
always 1,
\begin{equation}
\langle Z^2 \rangle = 1,
\end{equation}
so
\begin{equation}
\Delta Z = \left(\, \langle Z^2 \rangle - \langle Z\rangle^2\, \right)^{1/2} =
1.
\end{equation}
This is a reasonable characterization of the uncertainty in $Z$ since a
measurement gives $+1$ or $-1$ with equal probability. 

For a measurement of $X$ we already showed that $|\psi\rangle$ is an
eigenstate with eigenvalue $1$ and so the measured value is always 1. If we
use Eqs.~\eqref{expect} and \eqref{DeltaA}, we obtain $\langle
X\rangle = 1, \langle X^2 \rangle = 1$, and so $ \Delta X = 0$ as expected.

If we consider a superposition
\begin{equation}
|\psi\rangle = \alpha |0\rangle + \beta |1\rangle,
\end{equation}
the student should now be able to show that
\begin{align}
\langle Z \rangle &=  \langle \psi| Z | \psi\rangle = |\alpha|^2 - |\beta|^2  \\
\langle X \rangle &=  \langle \psi| X | \psi\rangle = \alpha^\star \beta +
\alpha \beta^\star. 
\end{align}

\section{Composite Systems}
\label{sec:composite}
\index{composite systems}
So far, we have described states of just a single qubit. How should we
describe states of the many qubits which we will need for a quantum computer? 
Suppose, as an example, we have two qubits $A$ and $B$. We can label the
states of qubit $A$ by $|0_A\rangle$ and $ |1_A\rangle$, and similarly the states of qubit
$B$ by $|0_B\rangle$ and $ |1_B\rangle$. A state of both qubits is written as a
\index{tensor product}
``tensor product'', also known as a ``direct product"\index{direct product|see
{tensor product}},
e.g.~$|0_A\rangle \otimes |1_B\rangle$, which in this
example indicates
that qubit $A$ is in state $|0\rangle$ and qubit $B$ is in state
$|1\rangle$. 

This notation is heavy so we will usually write the same state more compactly
as $|0_A\rangle |1_B\rangle$, or even more concisely as $|01\rangle$ provided a
specification of the order of the qubits has been given. 
In this notation, the four possible
states of two qubits are
\begin{equation}
|00\rangle, \quad |01\rangle, \quad |10\rangle, \quad |11\rangle.
\label{four}
\end{equation}
Note that the label of each state is a number in binary notation from $0$ to
$3$. This provides an even more compact notation, which is particularly
convenient when the number of qubits is large, namely $|x\rangle_2$, where
$x=0,1,2$ or $3$. It is necessary to indicate the number of qubits by a
subscript on the bracket to avoid ambiguity. For example just writing a state
as
$|2\rangle$ we wouldn't know if it is state $|10\rangle$ for 2 qubits, or
$|010\rangle$ for 3 qubits and so on. An exception to this will be states
$|0\rangle$ and $|1\rangle$ (without subscript) which always refer to the
1-qubit
basis states. The four states in Eq.~\eqref{four} can therefore
also be written as 
\begin{equation}
|0\rangle_2, \quad |1\rangle_2, \quad |2\rangle_2, \quad |3\rangle_2.
\end{equation}

Similarly for three qubits, we can specify the 8 possible states by
$|x\rangle_3$ where $x = 0, 1, \cdots, 7$, and for $n$ qubits the $2^n$ states
are indicated by $|x\rangle_n$, where $x = 0, 1, \cdots, 2^n -1 $. We see that
to use this convenient binary notation we need to label the states starting
from 0 rather than 1. The last state then has label $2^n-1$.  

We just pointed out that
an $n$-qubit basis state $|x_{n-1}\,x_{n-2}\, \cdots\, x_2\, x_1\, x_0\rangle$, where the
$x_i$ are the values of the qubits, can also be represented as $|x\rangle_n$
where $x$ is the $n$-bit integer whose bits are the $x_i$.
As an example of this for $5$ qubits we
have
\begin{equation}
|23\rangle_5 \equiv |10111\rangle. 
\label{10_4_a}
\end{equation}
A state vector with 5 qubits has $2^5 = 32$ components. A single basis state
has all components equal to $0$ except for one entry which is $1$. For the
above state
\begin{equation}
|23\rangle_4 = (\mathrm{row}\ 24) \longrightarrow
\begin{pmatrix}
0 \\ \vdots \\ 0 \\ 1 \\ 0 \\ \vdots \\ 0 \\
\end{pmatrix}.
\label{10_4_b}
\end{equation}
The state representing the integer $|23\rangle_5$ is actually on the twenty-fourth row because
the entry on the first row is for state $|0\rangle_5$ not $|1\rangle_5$.
Note that the zeroes in Eq.~\eqref{10_4_a} represent the states of the $N$ individual qubits, while the
zeroes in Eq.~\eqref{10_4_b} are coefficients of the $2^N$ basis states.

An important point is that
summing over the two values ($0$ and $1$) of $x_i$ for each bit $i$, is equivalent to summing over
all the $x$ values from $0$ to $2^n -1 $.

We need to be familiar with these ways of labeling multi-qubit states.

Next we discuss matrix representations of operators on multiple qubits, and we
take as an example, the case of two qubits. An operator acting on the space of
two qubits is a $4 \times 4$ matrix. We will write the four basis states as
$|00\rangle, \ |01\rangle, \ |10\rangle, \ |11\rangle$. Consider an operator
where $X$ acts on the first (left hand) qubit and the identity $\mathbbm{1}$ acts on the
second (right hand) qubit. The 2-qubit operator is a tensor product of the 1-qubit
operators, i.e.~$X \otimes \mathbbm{1}$. Its action on the four basis states
is as follows:
\begin{equation}
\begin{split}
X \otimes \mathbbm{1} |00\rangle &= |10\rangle , \\
X \otimes \mathbbm{1} |01\rangle &= |11\rangle , \\
X \otimes \mathbbm{1} |10\rangle &= |00\rangle , \\
X \otimes \mathbbm{1} |11\rangle &= |01\rangle ,
\end{split}
\end{equation}
so its matrix representation is
\begin{align}
& \quad |00\rangle \ \ |01\rangle \ \ |10\rangle \ \ |11\rangle \nonumber \\
X \otimes \mathbbm{1} = 
\begin{matrix}
\langle 00 | \\
\langle 01 | \\
\langle 10 | \\
\langle 11 | 
\end{matrix}
& \begin{pmatrix}
\ 0\quad  & \quad\  0\  & \quad 1\  & \quad 0\  \\
\ 0\quad  & \quad\  0\  & \quad 0\  & \quad 1\  \\
\ 1\quad  & \quad\  0\  & \quad 0\  & \quad 0\  \\
\ 0\quad  & \quad\  1\  & \quad 0\  & \quad 0\  \\
\end{pmatrix}
=
\begin{pmatrix}
0 & \mathbbm{1} \\
\mathbbm{1} & 0 
\end{pmatrix}
,
\label{comp1}
\end{align}
where in the last expression each entry is a $2 \times 2$ block. Note how this
block structure reflects the operators in the tensor product on the left of
the expression. The $2
\times 2$ block structure is that of $X$ (the left hand operator)
while each block is made up of
the identity $\mathbbm{1}$ (the right hand operator).
Similarly
\begin{align}
& \quad |00\rangle \ \ |01\rangle \ \ |10\rangle \ \ |11\rangle \nonumber \\
\mathbbm{1} \otimes X = 
\begin{matrix}
\langle 00 | \\
\langle 01 | \\
\langle 10 | \\
\langle 11 | 
\end{matrix}
& \begin{pmatrix}
\ 0\quad  & \quad\  1\  & \quad 0\  & \quad 0\  \\
\ 1\quad  & \quad\  0\  & \quad 0\  & \quad 0\  \\
\ 0\quad  & \quad\  0\  & \quad 0\  & \quad 1\  \\
\ 0\quad  & \quad\  0\  & \quad 1\  & \quad 0\  \\
\end{pmatrix}
=
\begin{pmatrix}
X & 0 \\
0 & X 
\end{pmatrix}
.
\label{comp2}
\end{align}

As another example consider $X \otimes Z$. We have
\begin{equation}
\begin{split}
X \otimes Z |00\rangle &= |10\rangle , \\
X \otimes Z |01\rangle &= -|11\rangle , \\
X \otimes Z |10\rangle &= |00\rangle , \\
X \otimes Z |11\rangle &= -|01\rangle ,
\end{split}
\end{equation}
so its matrix representation is
\begin{align}
& \quad \,|00\rangle \ \ \ |01\rangle \ \ \ |10\rangle \ \ \ |11\rangle \nonumber \\
X \otimes Z = 
\begin{matrix}
\langle 00 | \\
\langle 01 | \\
\langle 10 | \\
\langle 11 | 
\end{matrix}
& \begin{pmatrix}
\ 0\quad  & \quad\  0\  & \quad 1\  & \quad 0\  \\
\ 0\quad  & \quad\  0\  & \quad 0\  & \quad -1\  \\
\ 1\quad  & \quad\  0\  & \quad 0\  & \quad 0\  \\
\ 0\quad  & \quad\ -1\  & \quad 0\  & \quad 0\  \\
\end{pmatrix}
=
\begin{pmatrix}
0 & Z \\
Z & 0 
\end{pmatrix}
.
\label{comp3}
\end{align}
Again notice how the block structure in the last expression reflects the operators in the tensor
product. I emphasize that in Eqs.~\eqref{comp1}, \eqref{comp2} amd \eqref{comp3}, the left hand
operator on the LHS of the equation acts on the left hand basis state on the
RHS.

\section{Generalized Born Rule}
\label{sec:gen}

\index{generalized Born rule}
In Sec.~\ref{sec:meas} we gave the standard physics text book discussion of measurement
in quantum mechanics. For quantum computing we need to extend this to deal
with situations involving multiple qubits where we measure only \textit{some}
of the qubits and we need to know the state of the remaining qubits after the measurement. As
a simple example, suppose we have 2 qubits $A$ and $B$, in a state
\begin{equation}
|\psi\rangle = a_0 |00\rangle + a_1 |01\rangle + a_2 |10\rangle +
a_3|11\rangle,
\label{psi2}
\end{equation}
where the left qubit is $A$ and the right qubit is $B$. Because the state has
to be normalized we need $|a_0|^2 + |a_1|^2 + |a_2|^2 + |a_3|^2 = 1$.

If we measure $Z$ for both qubits the Born rule tells us that we find that qubit $A$ has value $+1$ and qubit
$B$ also has value $+1$ with probability $|a_0|^2$, and similarly for the other
possible results.
Suppose instead
we measure $Z$ only for qubit-$A$, the left qubit. We want to know what
are the possible measurement results, what are the probabilities of the
different results,
and, for each case, in what state is qubit $B$ after the measurement.

We will rewrite Eq.~\eqref{psi2}, grouping together all the terms where qubit $A$ is
$|0\rangle$ (more generally an eigenstate of the measurement operator acting on qubit $A$),
and all the terms where qubit $A$ is $|1\rangle$ (the other eigenstate).
The terms involving qubit $A$ in state $|0\rangle$ are $a_0 |00\rangle + a_1
|01\rangle$. We write this as
\begin{equation}
a_0 |00\rangle + a_1 |01\rangle = \alpha_0 |0_A\rangle\, \left(
{a_0 \over \alpha_0} |0_B\rangle + 
{a_1 \over \alpha_0} |1_B\rangle \right) = 
\alpha_0 |0_A\rangle |\phi_{0,B}\rangle,
\end{equation}
where
\begin{equation}
|\alpha_0|^2 = |a_0|^2 + |a_1|^2
\end{equation}
and
\begin{equation}
|\phi_{0,B}\rangle = {1 \over \alpha_0}\left(a_0 |0_B\rangle + a_1
|1_B\rangle \right),
\end{equation}
is a \textit{normalized} state for qubit $B$.
Similarly
\begin{equation}
a_2 |10\rangle + a_3 |11\rangle = \alpha_1 |1_A\rangle\, \left(
{a_2 \over \alpha_1} |0_B\rangle + 
{a_3 \over \alpha_1} |1_B\rangle \right) = 
\alpha_1 |1_A\rangle |\phi_{1,B}\rangle,
\end{equation}
where
\begin{equation}
|\alpha_1|^2 = |a_2|^2 + |a_3|^2
\end{equation}
and
\begin{equation}
|\phi_{1,B}\rangle = {1 \over \alpha_1}\left(a_2 |0_B\rangle + a_3
|1_B\rangle \right),
\end{equation}
is normalized.

Combining we get
\begin{equation}
|\psi\rangle = \alpha_0 |0_A\rangle |\phi_{0,B}\rangle + 
\alpha_1 |1_A\rangle |\phi_{1,B}\rangle ,
\label{psiAB}
\end{equation}
where we emphasize that all the states in this expression are normalized since
$|\alpha_0|^2 + |\alpha_1|^2 = 1$.

The inner product of $| \phi_0\rangle$ and $|\phi_1\rangle$ is
\begin{equation}
\langle \phi_{0, B}| \phi_{1, B} \rangle ={a_0^\star a_2 +a_1^\star a_3 \over
\sqrt{|a_0|^2 + |a_1|^2}\, \sqrt{|a_2|^2 + |a_3|^2} },
\end{equation}
and since
\begin{equation}
\langle 0_B | 0_B \rangle = \langle 1_B | 1_B \rangle = 1, \qquad
\langle 0_B | 1_B \rangle = \langle 1_B | 0_B \rangle = 0
\end{equation}
there is no reason for this to be zero in general.
Hence, while $|\phi_{0, B}\rangle$
and $|\phi_{1, B}\rangle$ are normalized, they are not necessarily orthogonal.

Hence the natural extension of the Born rule, called the ``generalized Born'' rule,
is that,
when the two qubits are in the state given in Eq.~\eqref{psiAB},
the possible results of the measurement of $Z$ on qubit $A$, are
\begin{equation}
\begin{split}
\mathrm{result}\ {+1},& \quad \mathrm{probability}\ |\alpha_0|^2\ (= |a_0|^2 + |a_1|^2), \quad \mathrm{final\
state} \ \ |0_A\rangle |\phi_{0,B}\rangle , \\
\mathrm{result}\ {-1},& \quad \mathrm{probability}\ |\alpha_1|^2\ (= |a_2|^2 + |a_3|^2), \quad \mathrm{final\
state} \ \ |1_A\rangle |\phi_{1,B}\rangle .
\end{split}
\label{measA}
\end{equation}
It is straightforward to generalize this result to an arbitrary situation in which
there are $n+m$ qubits, $n$ of which are measured and we want to know the
possible
final states of the remaining $m$ qubits after the measurement, and to
arbitrary measurement operators.

Note that in the above example, if we measure \textit{both} qubits, the
probabilities of getting $|0_A\rangle $ and $|1_A\rangle$ for qubit $A$ are the
same as those in Eq.~\eqref{measA} for when we just measured qubit $A$. The
difference is that if we also measure qubit $B$, then, after the measurement,
qubit $B$ is left in a computational basis state, whereas if $B$ is not
measured, it is left in a superposition state, $|\phi_{0,B}\rangle$ or
$|\phi_{1,B}\rangle$.

\section{The Uncertainty Principle}
\index{uncertainty principle}

Now we come to a key concept in quantum mechanics, the \textit{uncertainty principle}.
We shall see that some variables are incompatible with each other, which means
that one can not have definite values for both of them in \textit{any} state.
The important quantity to see if two operators, $A$ and $B$ say, are
compatible is their commutator
\index{matrix!commutator}
\begin{equation}
[A, B] \equiv AB - BA .
\end{equation}
If $[A, B]\ne 0$ then it is shown in linear algebra texts that $A$ and $B$
have different eigenvectors. We have already noted that we only get a definite
value for some operator if the state is in an eigenstate of that operator.
Hence, if $[A, B]\ne 0$, so $A$ and $B$ have different eigenvectors, there is no state which will
give a definite value for both of them.

As an example of a commutator consider $X$ and $Z$. We have
\begin{equation}
[Z, X] = 
\begin{pmatrix}
1 & 0 \\
0 & -1
\end{pmatrix}
\begin{pmatrix}
0 & 1\\
1 & 0
\end{pmatrix}
-
\begin{pmatrix}
0 & 1\\
1 & 0
\end{pmatrix}
\begin{pmatrix}
1 & 0 \\
0 & -1
\end{pmatrix}
= 
\begin{pmatrix}
0 & 1 \\
-1 & 0
\end{pmatrix}
-
\begin{pmatrix}
0 & -1 \\
1 & 0
\end{pmatrix}
= 2 \begin{pmatrix}
0 & 1 \\
-1 & 0
\end{pmatrix}
= 2 i Y ,
\end{equation}
where $Y$ is defined in Eq.~\eqref{pauli}. Since the commutator is non-zero it
is impossible to find a state which is simultaneous eigenstate of both $X$ and
$Z$ and so either $\Delta X$ or $\Delta Z$, or both, must be non-zero.

An important inequality involving the uncertainties $\Delta A$ and
$\Delta B$ of two operators in a state $|\psi\rangle$ is
\begin{equation}
\left(\Delta A\, \Delta B\right)_\psi \ge {1 \over 2} \left|\left\langle [A, B] \right\rangle_\psi
\right|,
\end{equation}
which is known as the Heisenberg uncertainty principle. We shall not prove
\index{uncertainty principle}
this result. The most famous
case of the uncertainty principle is for $A = x$, the position of a particle,
and $B=p$, its momentum, for which the commutator is a
\index{position}
\index{momentum}
\index{Planck's constant}
constant\footnote{$\hbar$ is Planck's constant divided by $2\pi$. It is of
paramount importance in physics but does not explicitly play a role in the theory of quantum
computation. \label{hbar}} $i \hbar$ so
\begin{equation}
\Delta x \Delta p \ge {\hbar \over 2}.
\end{equation}
However, this particular version of the uncertainty principle does not play a
role in quantum computing which is concerned with (discrete) 2-state systems,
rather than (continuous) trajectories of particles. 

\section{Time Evolution of Quantum States}
\index{time evolution}

So far, we have described fixed quantum states. Now we need to discuss how they
evolve with time. If the state at an initial time is $|\psi\rangle$ and the state
at a later time
is $|\psi'\rangle$, then, according to quantum mechanics, that there is a linear relation between
the two, so
\begin{equation}
|\psi'\rangle = U |\psi\rangle ,
\label{psi'}
\end{equation}
for some linear operator $U$. The normalization condition must be preserved so
$\langle \psi'|\psi'\rangle = \langle \psi|\psi\rangle = 1$. This provides a
constraint on the form of $U$ as we will now show. The equation corresponding
to Eq.~\eqref{psi'} for the dual vector $\langle \psi'|$ is
\begin{equation}
\langle \psi'| = \langle \psi| U^\dagger .
\label{psi}
\end{equation}
To see this compare Eqs.~\eqref{1} and \eqref{2} and note that
$\left(A^\dagger\right)^\dagger = A$. Combining Eqs.~\eqref{psi'} and
\eqref{psi} we find
\begin{equation}
\langle \psi'|\psi'\rangle = \langle \psi|U^\dagger U|\psi \rangle .
\end{equation}
Since we must have $\langle \psi'|\psi'\rangle = \langle \psi|\psi \rangle\ (=1)$ for
any initial state $|\psi\rangle$
it follows that
\begin{equation}
U^\dagger U = \mathbbm{1} ,
\end{equation}
so $U$ has to be unitary. 
\index{unitary transformation}

In quantum computing we change the state of the
qubits by a sequence of \textit{discrete} unitary transformations.  Note that for a
unitary operator $U^{-1} = U^\dagger$, and $U^\dagger$ is well defined, so
the inverse transformation,  which acts on the final state and
converts it to the initial state, exists.  Thus quantum transformations are
\textit{reversible}. The exception is measurement, in which the quantum system
is coupled to a macroscopic, external apparatus which leads to an
\index{reversible computation}
irreversible change.
As we shall see, standard classical gates which manipulate the bits in a
classical computer are irreversible. The necessity of doing reversible
operations in a quantum computer will be a major difference compared with a
classical computer.

In a quantum computer, as noted above, we act on the qubits with a series
of discrete unitary operations, but we should be aware
that these are implemented by acting with some operation
for a finite amount of time, see e.g.~Chs.~14--17 of the book by
LaPierre~\cite{lapierre:21}.

Microscopically, quantum states evolve
\textit{continuously} with time, and we will finish this chapter with a brief discussion
of continuous time evolution in quantum mechanics (even though it will not be needed in the 
rest of the course). Time evolution is determined by the
Hamiltonian (energy), $\mathcal{H}$ a Hermitian operator,
\index{Hamiltonian}
according to Ansatz 3: 

\index{Schr\"odinger's equation}
\textbf{Ansatz 3:} The time dependence of a state is given by Schr\"odinger's
equation
\begin{equation}
i\hbar {\partial \over \partial t}|\psi(t)\rangle =\mathcal{H}\, |\psi(t)\rangle .
\label{sch}
\end{equation}

Assuming that $\mathcal{H}$ does not
change with time, we can integrate  Eq.~\eqref{sch} to get
\begin{equation}
|\psi(t)\rangle = U(t) |\psi(0)\rangle,
\end{equation}
where
\begin{equation}
U(t) = e^{-i\mathcal{H} t / \hbar} .
\end{equation}
\index{matrix!Hermitian}
\index{matrix!unitary}
Since $\mathcal{H}$ is Hermitian we can show that $U$ is unitary by the
following argument.
To get the adjoint of $U$ we take its complex conjugate 
and replace any operators in the expression for $U$ by their adjoint. Since $\mathcal{H}$ is self-adjoint
(Hermitian) we have
\begin{equation}
U^\dagger(t) = e^{i\mathcal{H} t / \hbar},
\end{equation}
from which one sees that 
\begin{equation}
U^\dagger(t) U(t) = e^{i\mathcal{H} t / \hbar}
e^{-i\mathcal{H} t / \hbar} = e^{i(\mathcal{H} t - \mathcal{H} t)/\hbar} = \mathbbm{1},
\label{Ut}
\end{equation}
so $U$ is unitary as required. Note that if we have operators in exponentials
which don't commute, we can't manipulate them as we do with ordinary numbers.
For example $e^A e^B$ does \textit{not} equal $e^{A+B}$ unless $[A, B] = 0$. However,
here both $A$ and $B$ are proportional to $\mathcal{H}$ which commutes with
itself, so combining the exponentials as done in Eq.~\eqref{Ut} \textit{is} valid.

\hrulefill
\section*{Problems}
\input{hw_ch3.tex}

%% file: hw_ch3.tex
\begin{problems}
\item
Consider the following state vectors:
$$
|\psi \rangle =
\begin{pmatrix}
2 \\
3 i \\
\end{pmatrix}
, \quad |\phi \rangle =
\begin{pmatrix}
4 \\
5 i\\
\end{pmatrix}
.
$$
\begin{enumerate}[label=(\roman*)]
\item
Write down the dual vectors $\langle \psi |$ and $\langle \phi |$.
\item
Normalize $|\psi \rangle$ and $|\phi \rangle$.
\item
For the normalized states determined in the last part, compute the inner
product $\langle \phi | \psi \rangle .$
\end{enumerate}
\item
Which of the following pairs of of quantum states represent the same physical state?
(You must explain your results.)
\begin{enumerate}[label=(\roman*)]
\item 
$|0\rangle$ and $-|0\rangle$ .
\item 
${1 \over \sqrt{2}}(|0\rangle + |1\rangle)$ and ${1 \over \sqrt{2}}(|0\rangle -
|1\rangle$
\item 
${1 \over \sqrt{2}}(|0\rangle - |1\rangle)$ and ${1 \over \sqrt{2}}(|1\rangle -
|0\rangle$
\item 
${1 \over \sqrt{2}}(|0\rangle + e^{i\pi/4} |1\rangle)$ and
${1 \over \sqrt{2}}(e^{-i\pi/4}|0\rangle + |1\rangle$
\item 
${1 \over \sqrt{2}}(|0\rangle + i |1\rangle)$ and
${1 \over \sqrt{2}}(i|0\rangle + |1\rangle$
\end{enumerate}
For those cases where the two states are different, what  measured  quantity
would give different results? Show that the measured quantity acting on the
states does yield different results.
\textit{Hint:} Think Pauli matrices.

\item
\textit{Tensor products of matrices}\\
Note the block structure of the following tensor product:
\begin{align}
& \quad |00\rangle \ \,|01\rangle \ \,|10\rangle \ \,|11\rangle \nonumber \\
X \otimes Z  =
\begin{pmatrix}
(0) Z & (1) Z \\
(1) Z & (0) Z \\
\end{pmatrix} 
=
& \begin{pmatrix}
\ 0\  & \  0\  & \  1\  & \ 0\  \\
\ 0\  & \  0\  & \ 0\  & -1\  \\
\ 1\  & \  0\  & \ 0\  & \ 0\  \\
\ 0\  &  -1\  & \ 0\  & \ 0\  \\
\end{pmatrix}
.
\end{align}
Obtain, in a similar way, the tensor products for $Z \otimes X$ and $H \otimes H$
where $H$ is the Hadamard matrix.

\item 
Consider the  2-qubit state
\begin{equation}
|\psi\rangle = {1\over \sqrt{2}}( |0\rangle_1 |0\rangle_2 - |1\rangle_1
1\rangle_2 ).
\end{equation}
Determine the expectation values of $\langle Z_1 Z_2\rangle$ and $\langle X_1
X_2\rangle$ in
this state.

%


%
%

\item
Consider the state
\begin{equation}
|\psi\rangle = {1\over \sqrt{2}}( |0\rangle + |1\rangle ).
\end{equation}
Determine $\langle Z\rangle, \langle Z^2\rangle, \Delta Z, \langle X\rangle,
\langle X^2 \rangle, \Delta X, \langle Y\rangle, \langle Y^2 \rangle$ and $ \Delta
Y$.
Show that the uncertainty is only zero for those operators for which the state
is an eigenstate.

\item
Consider the Bell states which are defined by
\begin{equation}
|\beta_{xy}\rangle = {|0y\rangle + (-1)^x |1\overline{y}\rangle \over \sqrt{2}} ,
\end{equation}
where $\overline{y}$ is the complement of $y$, i.e. $\overline{y} = 1 - y$.
Show that
\begin{align}
Z \otimes Z |\beta_{xy}\rangle &= (-1)^y  |\beta_{xy}\rangle , \\
X \otimes X |\beta_{xy}\rangle &= (-1)^x  |\beta_{xy}\rangle , \\
Y \otimes Y |\beta_{xy}\rangle &=-(-1)^{x+y}  |\beta_{xy}\rangle .
\end{align}

\item
Show that the unitary operator which transforms the $Z$-basis (i.e.~the basis
in which $Z$ is diagonal) to the $X$-basis is
\begin{equation}
H = {1\over \sqrt{2}}
\begin{pmatrix}
1& 1 \\
1 & -1 \\
\end{pmatrix} .
\end{equation}
$H$ is called the Hadamard operator.



\item 

Consider a system with two qubits, $A$ and $B$, in state
$$
|\phi\rangle = {1 \over 3} \left(|0\rangle_A |0\rangle_B + \sqrt{3}\,
|1\rangle_A |0\rangle_B +\sqrt{5} |1\rangle_A |1\rangle_B \right) .
$$
\begin{enumerate}[label=(\roman*)]
\item
Show that the state is normalized.
\item
If measurements are made of both qubits, what are the possible results and
their probabilities?
\item
If a measurement is made only on qubit $A$ what are the possible resulting
states for qubit $B$ and what are their probabilities?\\
\textit{Note:} Make sure that the probabilities add up to 1.
\end{enumerate}

\end{problems}

%% file: qubits8.tex
\label{ch:gen_qubit}
\section{General qubit states}
\label{sec:gen_qubit}

%
%

As already discussed in Sec.~\ref{sec:important}, the following $2 \times 2$
matrices, called Pauli matrices, acting on the states of a single qubit
will be important in the rest of the course:
\index{Pauli matrices!$X$ matrix}
\index{Pauli matrices!$Y$ matrix}
\index{Pauli matrices!$Z$ matrix}
\begin{subequations}
\label{Pauli:sigma}
\begin{align}
X \equiv \sigma_x &= 
\begin{pmatrix}
0 & 1 \\
1 & 0 \\
\end{pmatrix} , \\
Y \equiv \sigma_y &= 
\begin{pmatrix}
0 & -i \\
i & 0 \\
\end{pmatrix} , \\
Z \equiv \sigma_z &= 
\begin{pmatrix}
1 & 0 \\
0 & -1 \\
\end{pmatrix} .
\end{align}
\end{subequations}
In the physics literature the notation used is $\sigma_x, \sigma_y$ and
$\sigma_z$, 
but in this course we shall use the quantum computing notation:
$X, Y$, and $Z$. As shown in Sec.~\ref{sec:mat_diag}, an arbitrary $2\times
2$ matrix can be written as a linear combination of the three Pauli matrices
plus the $2\times 2$ identity matrix.
\index{matrix!Hermitian}
These matrices are Hermitian, and have
eigenvalues $\pm 1$, see Sec.~\ref{sec:mat_diag}. 

If the qubit is the spin of an electron, then the eigenstate with $Z=1$ has
spin along the $+z$ direction, and analogously the eigenstate with $Y=1$ has
spin along the $+y$ direction, and the eigenstate with $X=1$ has
spin along the $+x$ direction. Also, the eigenstate with $Z=-1$ has the
spin pointing in the $-z$ direction, and analogously for $X=-1$ and $Y=-1$. 

How can we specify a \textit{general} state of a qubit? To see this,
we first ask how many
parameters do we need to specify a general state?
A qubit vector has two complex components making a total of
four. However, one of these can be eliminated because the state must be normalized,
and another can be eliminated because an overall phase is unimportant.
This leaves two parameters necessary to describe a general
qubit state.

We shall see that we can conveniently take these two parameters to be the
two angles which describe a direction in space in spherical
polar coordinates.
To see this we compute the eigenstates for the spin of an electron
aligned along a general direction
with polar angle $\theta$ and azimuthal angle $\phi$, which describe 
a unit vector $\hat{n}$ where
\begin{equation}
\hat{n} = (\sin \theta \, \cos\phi,\,\, \sin\theta\, \sin\phi,\,\,
\cos\theta),
\label{hatn}
\end{equation}
so $n_x = \sin \theta \, \cos\phi$ etc.
In other words we compute the eigenvalues and eigenvectors of
$\vec{\sigma}\cdot\hat{n}$. We have
\begin{equation}
\vec{\sigma}\cdot\hat{n} =
\begin{pmatrix}
n_z & n_x - i n_y \\
n_x + i n_y & -n_z \\
\end{pmatrix}
\label{sigma_n}
\end{equation}
so the eigenvalues are given by
\begin{equation}
\begin{vmatrix}
n_z - \lambda & n_x - i n_y \\
n_x + i n_y & -n_z - \lambda \\
\end{vmatrix} = 0.
\end{equation}
Expanding the determinant, and using that $n_x^2 + n_y^2 + n_z^2 = 1$, we find
the eigenvalues to be
\begin{equation}
\lambda = \pm 1.
\end{equation}
Thus, the eigenvalues are not only $\pm 1$ when measured along the Cartesian
directions, but take the same values along \textit{any} direction.

\begin{figure}
\begin{center}
\includegraphics[width=8cm]{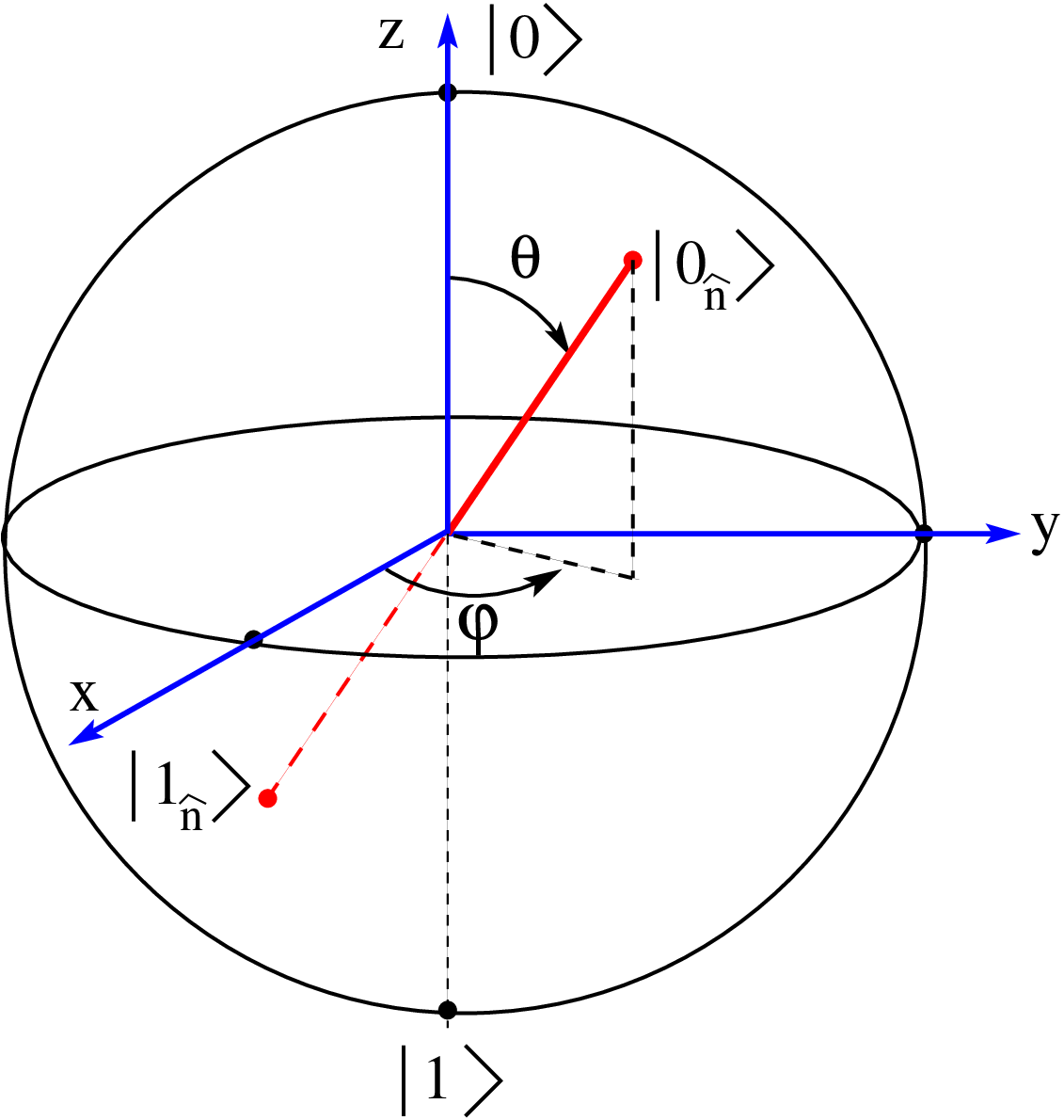}
\caption{The Bloch sphere.
\label{bloch}}
\end{center}
\end{figure}

Next we look at the eigenvectors. First the eigenvector for eigenvalue $+1$
is 
\begin{equation}
|0_{\hat{n}}\rangle =
\begin{pmatrix}
a \\ b
\end{pmatrix}
\end{equation}
where
\begin{equation}
\begin{pmatrix}
\cos\theta & \sin\theta\, e^{-i\phi} \\
\sin\theta\, e^{i\phi} & -\cos\theta \\
\end{pmatrix}
\begin{pmatrix}
a \\ b
\end{pmatrix}
=
\begin{pmatrix}
a \\ b
\end{pmatrix}
,
\end{equation}
where we used Eqs.~\eqref{hatn} and \eqref{sigma_n}.
Writing out the two equations we get
\begin{subequations}
\begin{align}
\sin\theta\, e^{-i\phi} \, b &= a (1 - \cos\theta), \\
\sin\theta\, e^{i\phi} \, a &= b (1 + \cos\theta). 
\end{align}
\end{subequations}
Both these equations are satisfied by
\begin{equation}
b\, \cos\smfrac{\theta}{2} = a\, e^{i\phi} \sin\smfrac{\theta}{2} ,
\end{equation}
in which we used that
\begin{equation}
\sin\theta = 2 \sin\theta/2\, \cos\theta/2,\quad \cos\theta = 2 \cos^2\theta/2
- 1 = 1 - 2 \sin^2\theta/2 .
\end{equation}
We require
the state to be normalized, i.e.~$|a|^2 + |b|^2 = 1$, so we get
\begin{equation}
|0_{\hat{n}}\rangle = 
\begin{pmatrix}
\cos\smfrac{\theta}{2} \\
e^{i\phi} \,\sin\smfrac{\theta}{2} \\
\end{pmatrix}  ,
\end{equation}
or equivalently, in Dirac notation,
\begin{subequations}
\label{01n}
\begin{equation}
|0_{\hat{n}}\rangle = \cos\smfrac{\theta}{2}\, |0\rangle + e^{i\phi}
\,\sin\smfrac{\theta}{2}\,|1\rangle \, .
\label{0n}
\end{equation}
This is the expression for a general qubit state. It depends on two parameters
$\theta$ and $\phi$ which can be taken to be the polar and azimuthal angles of
a point on a sphere, see Fig.~\ref{bloch}.

A similar calculation gives the eigenstate corresponding to eigenvalue $-1$ to
be
\begin{equation}
|1_{\hat{n}}\rangle = -\sin\smfrac{\theta}{2}\, |0\rangle + e^{i\phi}
\,\cos\smfrac{\theta}{2}\, |1\rangle \, .
\label{1n}
\end{equation}
\end{subequations}

It is straightforward to see that the states in Eqs.~\eqref{01n}
are normalized, i.e.
\begin{equation}
\langle 0_{\hat{n}}| 0_{\hat{n}}\rangle = 1,\qquad \langle 1_{\hat{n}}
|1_{\hat{n}}\rangle = 1,
\end{equation}
and are mutually orthogonal
\begin{equation}
\langle 0_{\hat{n}}| 1_{\hat{n}}\rangle = 0.
\end{equation}
Note that we can always multiply eigenstates by an arbitrary phase factor so
you might see expressions for these eigenstates which \textit{look} different
from Eqs.~\eqref{0n} and \eqref{1n},
but which are actually equivalent.

If we consider a point on a unit sphere (often called
the Bloch sphere)
with polar angles $\theta$ and $\phi$, then the eigenstate of
spin in that direction with eigenvalue $+1$ is given by Eq.~\eqref{0n}, see
Fig.~\ref{bloch}.\index{Bloch sphere}
Even if the qubit is not an electron spin, Eq.~\eqref{0n} provides a
\index{qubit!general state of}
convenient description of an arbitrary qubit state.

Similarly,
(apart from a possible unimportant overall phase factor)
the eigenstate with eigenvalue $-1$ is given by Eq.~\eqref{1n},
which corresponds to the antipodal point where $\theta \to \pi - \theta, \phi \to \phi + \pi$.

It is useful to consider four special cases of Eqs.~\eqref{01n}:
\begin{enumerate}[label=(\roman*)]
\item $(\theta = \phi = 0)$, the $z$ direction. Clearly $|0_{\hat{z}}\rangle =
|0\rangle$ and $|1_{\hat{z}}\rangle =
|1\rangle$ as required.
\item $(\theta = \pi/2, \phi = 0)$, the $x$ direction:
\begin{align}
|0_{\hat{x}}\rangle &= {1 \over \sqrt{2}}\left(\, |0\rangle + |1\rangle\,
\right) = |+\rangle ,\\
| 1_{\hat{x}}\rangle &= {1 \over \sqrt{2}}\left(\, -|0\rangle + |1\rangle\,
\right) = -|-\rangle .
\end{align}
These are the eigenstates of $X$ as expected. ($| 1_{\hat{x}}\rangle$ has the
opposite sign to the conventionally defined state $|-\rangle$, but the overall
sign of a state is of no importance.)
\item $(\theta $ arbitrary $\phi = 0)$, a direction  $\hat{n}$, in the $x$-$z$ plane at an
angle $\theta$ to the $z$ axis:
\begin{align}
|0_{\hat{n}}\rangle  &= \cos\smfrac{\theta}{2}\, |0\rangle +
\sin\smfrac{\theta}{2}\, |1\rangle  \qquad (\phi = 0), \\
|1_{\hat{n}}\rangle  &= -\sin\smfrac{\theta}{2}\, |0\rangle +
\cos\smfrac{\theta}{2}\, |1\rangle .
\end{align}

\item $(\theta = \pi/2, \phi = \pi/2)$, the $y$ direction:
\begin{align}
|0_{\hat{y}}\rangle &= {1 \over \sqrt{2}}\left(\, |0\rangle + i\,|1\rangle\,
\right) ,\\
|1_{\hat{y}}\rangle &= {1 \over \sqrt{2}}\left(\, -|0\rangle + i\,|1\rangle\,
\right) .
\end{align}
These are the eigenstates of $Y$ as expected.
\end{enumerate}

We mentioned in Sec.~\ref{sec:photons} that for certain 
quantum protocols photons make good
qubits, with the state of the qubit being characterized by its polarization
\index{photon!polarization}
(the direction and phase of the electric field).
Using Eqs.~\eqref{EBk}--\eqref{ph:diag} and \eqref{0n}, we find that
the electric field
of a photon propagating in the $\hat{z}$ direction, corresponding to
a qubit $|0_{\hat{n}}\rangle$ specified by angles $\theta$ and $\phi$,
is given by
\begin{equation}
\vec{E} = \Re\left[ E_0 \cos(\theta/2) e^{-i(k z-\omega t)}\, \hat{x} + 
E_0 \sin(\theta/2) e^{i\phi}\, e^{-i(k z - \omega t)}\, \hat{y} \right] ,
\end{equation}
where $\Re$ means real part,
so
\begin{equation}
\begin{split}
E_x &= E_0 \cos(\theta/2)\, \cos(\omega t - k z)  ,\\
E_y &= E_0 \sin(\theta/2)\, \cos(\omega t - k z -\phi).
\end{split}
\end{equation}

Hence one can create an arbitrary qubit state by an appropriate choice of
photon polarization.
The polarization states for a photon for each of the four
special cases given above are:
\begin{enumerate}[label=(\roman*)]
\item
$(\theta = \phi = 0)$, i.e. $|0_{\hat{z}}\rangle \equiv |0\rangle$.
Linearly polarized along $\hat{x}$. (The photon corresponding to
$|1 \rangle$ is polarized along $\hat{y}$.)
\item
$(\theta = \pi/2, \phi = 0)$, i.e. $|0_{\hat{x}}\rangle$.
Linearly polarized along a diagonal direction at $45^\circ$ to the $x$-axis.
(The photon corresponding to $|1_{\hat{x}}\rangle$ is polarized along the
other diagonal direction.)
\item
$(\theta\ \mathrm{arbitrary}, \phi=0)$.
Linearly polarized with the polarization direction at an angle $\theta/2$
to the $x$-axis. 
\item
$(\theta = \pi/2, \phi = \pi/2)$, i.e. $|0_{\hat{y}}\rangle$.
\index{photon!circular polarization}
Circularly polarized\footnote{Recall that $\cos(x -\pi/2)=\sin(x)$}
with the $\vec{E}$ vector rotating in a particular sense
as a function of time.
(The photon corresponding to $|1_{\hat{y}}\rangle$ is is circularly polarized
with the $\vec{E}$ vector rotating in the opposite sense.)
\end{enumerate}

\section{No-cloning theorem}
\index{no-cloning theorem}
A classical bit, $0$ or $1$, can be copied, i.e.~cloned. You just observe it
and create another one. With qubits, however, it turns out to be not possible to clone an
arbitrary, unknown state. This is called the ``no-cloning theorem''. It imposes an
important limitation on our ability to manipulate quantum states.  We now give
the simple derivation of this important result using proof by contradiction. 

Consider the general qubit state
\begin{equation}
|\psi\rangle = \alpha|0\rangle + \beta|1\rangle, \qquad \mathrm{where\ }
|\alpha|^2 + |\beta|^2 = 1. 
\end{equation}

We can't determine the state by measuring it because a measurement gives
$|0\rangle$ with probability $|\alpha|^2$ and $|1\rangle$  with probability
$|\beta|^2$, i.e.~it destroys the superposition.
\index{superposition}

Can we clone the state \textit{without} measuring it? If so, there must be a
unitary operator $U$ which acts on $|\psi\rangle$ and an ancilla qubit,
\index{ancilla qubits}
which is initialized to $|0\rangle$ say, and clones $|\psi\rangle$ as
follows:
\begin{equation}
U |\psi\rangle \, |0\rangle = |\psi\rangle \, |\psi\rangle.
\end{equation}
We shall see that no such operator can exist because operators in quantum
mechanics are \textit{linear}.

Suppose that
\begin{equation}
\begin{split}
U |\psi\rangle \, |0\rangle &= |\psi\rangle \, |\psi\rangle , \\
U |\phi\rangle \, |0\rangle &= |\phi\rangle \, |\phi\rangle  .
\end{split}
\end{equation}
Then, by linearity, 
\begin{equation}
U \left(\, \alpha|\psi\rangle +\beta|\phi\rangle\, \right) |0\rangle = \alpha|\psi\rangle \,
|\psi\rangle  + \beta|\phi\rangle \, |\phi\rangle \, .
\end{equation}
However, this is not a clone of $ \alpha|\psi\rangle +\beta|\phi\rangle$ which would be 
\begin{equation}
\left(\, \alpha|\psi\rangle +\beta|\phi\rangle\, \right) \, \left(\, \alpha|\psi\rangle
+\beta|\phi\rangle\, \right) = \alpha^2|\psi\rangle \, |\psi\rangle +
\alpha\beta|\psi\rangle \,
|\phi\rangle +
\alpha\beta|\phi\rangle \, |\psi\rangle + \beta^2|\phi\rangle \, |\phi\rangle  .
\end{equation}
There is an inconsistency so a unitary operator $U$ for cloning does not
exist.

The no-cloning theorem will be an important limitation when designing quantum
algorithms.

\section{Entanglement and Bell states}
\index{entanglement}

\index{Bell state}
A striking aspect of quantum states of more than one qubit,
which seems mysterious and plays a crucial role in quantum
algorithms, is called ``\textit{entanglement}''. Here we will illustrate this
concept for the simplest case of two qubits.

Let's suppose that the first qubit is in state $|\psi_1\rangle =\alpha_1|0\rangle + \beta_1
|1\rangle$ and the second qubit is in state $|\psi_2\rangle = \alpha_2|0\rangle + \beta_2
|1\rangle$. The state of the two-qubit system is the tensor product
\begin{equation}
|\psi_1\rangle \otimes |\psi_2\rangle = 
\begin{pmatrix}
\alpha_1 \\ \beta 1 \\
\end{pmatrix}
\otimes
\begin{pmatrix}
\alpha_2 \\ \beta 2 \\
\end{pmatrix}
=
\begin{pmatrix}
\alpha_1 \alpha_2 \\ \alpha_1 \beta_2\\ \beta_1 \alpha_2 \\ \beta_1 \beta_2 \\
\end{pmatrix}
.
\end{equation}
This is an example of what is called a \textit{product state} (it is also
sometimes called a
separable state).
\index{product state}

However, a general $2$-qubit state is \textit{not} a product state. It can be
written as
\begin{equation}
|\phi\rangle_2 = c_0 |00\rangle + c_1 |01\rangle +  c_2 |10\rangle +  c_3
|11\rangle ,
\end{equation}
or equivalently as
\begin{equation}
|\phi\rangle_2 = c_0 |0\rangle_2 + c_1 |1\rangle_2 +  c_2
|2\rangle_2 +  c_3 |3\rangle_2 = \sum_{x=0}^3 c_x |x\rangle_2 .
\end{equation}

The product state has 
\begin{equation}
c_0 = \alpha_1 \alpha_2, \qquad
c_1 = \alpha_1 \beta_2, \qquad
c_2 = \beta_1 \alpha_2, \qquad
c_3 = \beta_1 \beta_2, 
\end{equation}
and so satisfies
\begin{equation}
c_0 c_3 = c_1 c_2 .
\label{prod}
\end{equation}
This is the condition for a 2-qubit state to be a product state. States which do
not have this property are said to be \textit{entangled}.
\index{entanglement}

The most-studied
entangled states are so-called Bell states which involve two qubits. They are
named in
honor of the physicist John Bell\index{Bell, John}
whose inequalities (to be discussed later) demonstrated that the description of nature
provided by quantum mechanics is fundamentally different from the classical
description. The Bell states are defined by
\begin{subequations}
\label{bell_eq}
\begin{align}
|\beta_{00} \rangle = {1 \over\sqrt{2}} \left(\, |00\rangle + |11\rangle\, \right) , \\
|\beta_{01} \rangle = {1 \over\sqrt{2}} \left(\, |01\rangle + |10\rangle\, \right) , \\
|\beta_{10} \rangle = {1 \over\sqrt{2}} \left(\, |00\rangle - |11\rangle\, \right) , \\
|\beta_{11} \rangle = {1 \over\sqrt{2}} \left(\, |01\rangle - |10\rangle\, \right) .
\end{align}
\end{subequations}
These four equations can be combined as follows:
\begin{equation}
|\beta_{xy}\rangle = {1 \over \sqrt{2}} \left(\, |0y\rangle + (-1)^x |1
\overline{y}\rangle \, \right) \, ,
\label{bell_eqall}
\end{equation}
where $\overline{y}$ is the complement of $y$, i.e. $\overline{y} = 1 - y$.
The Bell states are clearly orthogonal and entangled. 

There are correlations between the qubits in the Bell states 
(quite generally between the qubits in entangled states).  For example, if we consider
$|\beta_{00}\rangle$ and do a
measurement on qubit 1, then a measurement of qubit 2 (if
performed) would find the same result with 100\% probability. We will discuss
quantum correlations in entangled states in some detail in Chapter \ref{ch:EPR} when
we investigate the claim of
Einstein-Podolsky-Rosen (EPR) \index{EPR} that quantum mechanics is incomplete.

\index{product state}
For the case of two qubits, Eq.~\eqref{prod} is a convenient way to test if a
state is a product state or entangled. In a more general case where we have,
say, $n= n_A + n_B$ qubits, we may want to know whether a partition of the
system into the two subsystems $A$, with $n_A$ qubits, and $B$,
with $n_B$ qubits, gives a product state, i.e.~if
\begin{equation}
|\psi\rangle_n = |\psi_A\rangle_{n_A} \otimes |\psi_B\rangle_{n_B} ,
\end{equation}
or whether the state is entangled with respect to this partition. In the
case with more than $n=2$ qubits, there
is no simple expression analogous to Eq.~\eqref{prod} for the $2^n$
coefficients $c_x, (x = 0, 1, \cdots, 2^n-1)$, which indicates a product
state. Instead, a systematic way to investigate whether such a state is
entangled or a product state is to use the density matrix, discussed in
Chapter \ref{ch:den_mat}.
\index{density matrix}

\begin{center}
{\Large\bf Appendix}
\end{center}

\begin{subappendices}
\section{Angular Momentum Eigenstates}
\label{sec:ang-mom}
\index{angular momentum eigenstates}
Physics students learn about quantum states which are eigenstates of angular
momentum. This appendix relates Bell states \index{Bell state}to spin 
angular momentum eigenstates of two
electrons. It is
intended for physics students and is
not essential reading for students of other disciplines. 

The spin of an electron $\vec{s}$ is given by
\begin{equation}
\vec{s} = {\hbar \over 2} \vec{\sigma},
\end{equation}
\index{Planck's constant}
where $\hbar$ is Planck's constant divided by $2 \pi$ and the Pauli operators
$\vec{\sigma}$ are defined in Eqs.~\eqref{Pauli:sigma}.

In general, spin angular momentum states, $|S, m\rangle$, are specified by two quantum
numbers $S$ and $m$. The total spin quantum number $S$ is defined by
\begin{equation}
S_x^2 + S_y^2 + S_z^2 = \hbar^2 S (S + 1) ,
\end{equation}
where $S_x$, for example, is the $x$-component of the total spin,
so $|S, m\rangle$ is an eigenvector of $\left(\vec{S}\right)^2$ with eigenvalue
$\hbar^2 S (S + 1)$.
The quantum number $m$ is defined such that $|S, m\rangle$ is an
also eigenstate of $S_z$ with eigenvalue $\hbar m$, where $m$ ranges from $-S$
to $S$ in integer steps (so there are $2S+1$ values of $m$
for a given $S$).  Thus the spin of an electron has $S=1/2$, and its
two basis states are $|S=1/2,\,m= 1/2\rangle$ and $|S=1/2,\,m=-1/2\rangle$,
which are often written as $|\uparrow \rangle$ and $|\downarrow \rangle$
respectively. The latter notation indicates that one thinks of these two
states as spin ``up" and spin ``down".
By convention, the correspondence between the basis states of the electron
spin in physics, $|\uparrow\rangle$ and $|\downarrow\rangle$, and the
computational basis states in quantum computer science, $|0\rangle$ and
$|1\rangle$, is taken to be
\begin{equation}
|\uparrow \rangle \equiv |0\rangle, \quad |\downarrow \rangle \equiv |1\rangle.
\label{up0}
\end{equation}

If we have two particles with total spin quantum numbers $S_1$ and $S_2$ then,
as shown in textbooks on quantum mechanics~\cite{griffiths:05},
the ``vector rule" for addition of angular momentum states that the total
spin quantum number of the combined system, $S_\mathrm{tot}$, takes integer
values between $S_1 + S_2$ and $|S_1 - S_2|$. Thus, two electrons can have
combined total spin quantum number $S_\mathrm{tot} = 1$ (for which
there are 3 values of $m_\mathrm{tot}$, namely $1, 0$ and $-1$, and
$S_\mathrm{tot} = 0$ (for which there is only one value of $m_\mathrm{tot}$,
namely $0$). These are called ``triplet" and ``singlet" states respectively.
Note that the total number of states works out right since there are
$2^2 = 4$ states
out of which $3$ have
$S_\mathrm{tot} = 1$ and 1 has $S_\mathrm{tot} = 0$, (i.e.~$2 \times 2 = 3 + 1$).

It is also shown in the quantum mechanics textbooks that the states of two
spin-$1/2$ particles with
specified values of $S_\mathrm{tot}$ and $m_\mathrm{tot}$ are given by
\begin{subequations}
\label{Stot}
\begin{align}
|S_\mathrm{tot} &= 1, m_\mathrm{tot}=1 \rangle & &= |\uparrow \uparrow \rangle & &\equiv |00 \rangle , \label{11}\\
|S_\mathrm{tot}& = 1, m_\mathrm{tot}=0 \rangle & &= {1\over \sqrt{2}}\left(\,
|\uparrow \downarrow \rangle + |\downarrow \uparrow \rangle\,\right) & &\equiv
{1 \over \sqrt{2}} \left(\, |01 \rangle + |10\rangle\, \right), \label{10} \\
|S_\mathrm{tot} &= 1, m_\mathrm{tot}=-1 \rangle & &= |\downarrow \downarrow \rangle & &\equiv |11 \rangle , 
\label{1-1} \\
|S_\mathrm{tot} &= 0, m_\mathrm{tot}=0 \rangle & &= {1\over \sqrt{2}}\left(\,
|\uparrow \downarrow \rangle - |\downarrow \uparrow \rangle\,\right) & &\equiv
{1 \over \sqrt{2}} \left(\, |01 \rangle - |10\rangle\, \right). \label{00}
\end{align}
\end{subequations}
Eqs.~\eqref{11}--\eqref{1-1} are the triplet states while Eq.~\eqref{00} is the singlet
state.

Comparing with Eqs.~\eqref{bell_eq} we see that
\begin{subequations}
\label{connect}
\begin{align}
&|S_\mathrm{tot} = 1, m_\mathrm{tot}=1 \rangle \ \ = {1\over \sqrt{2}}\left(\,
|\beta_{00}\rangle + |\beta_{10}\rangle \, \right) , \\
&|S_\mathrm{tot} = 1, m_\mathrm{tot}=0 \rangle\ \ =  |\beta_{01}\rangle, \\
&|S_\mathrm{tot} = 1, m_\mathrm{tot}=-1 \rangle =  {1\over \sqrt{2}}\left(\,
|\beta_{00}\rangle - |\beta_{10}\rangle \, \right) , \\
&|S_\mathrm{tot} = 0, m_\mathrm{tot}=0 \rangle \ \ = |\beta_{11}\rangle,
\end{align}
\end{subequations}
Equations \eqref{connect} connect Bell states and
angular momentum states, while Eqs.~\eqref{Stot} connect
computational basis states and angular momentum states.

In this chapter we have encountered three sets of states which can
describe 2 qubits:
\begin{itemize}
\item
the computational basis states $|x y\rangle$,
\item
the Bell states $|\beta_{xy}\rangle$, and
\item
the angular momentum states $|S_\mathrm{tot}, \, m_\mathrm{tot}\rangle$. 
\end{itemize}
Each of these forms a basis set. In quantum computing we generally use computational
basis states but sometimes the Bell basis will be useful. However, there does not seem
to be a use for angular momentum basis states in quantum computing. 

\end{subappendices}


%% file: density_matrix8.tex
\section{Introduction}
\index{density matrix}
The material in this chapter is not essential for the rest of the course and
so could be omitted if necessary.  It is, however, necessary for
advanced treatments of quantum error correction which go beyond the discussion in
Ch.~\ref{ch:err_corr}.

We will be interested in situations where 
a system is in contact with another, possibly much larger, system.
Let's call the system of interest \textit{sub}system $A$, and denote the other
system by \textit{sub}system $B$. We use the word ``\textit{sub}system" for $A$ and $B$,
since we now consider them as
the two parts of the combined $AB$ \textit{system}. We want to describe the
properties of subsystem $A$ without explicitly including the degrees of
freedom of subsystem $B$.
This is accomplished by the ``density matrix". Two situations where the
density matrix is useful are:
\begin{itemize}
\item
To determine whether a state is a product state or entangled with respect to a
\index{entanglement}
\index{product state}
partition of the system into two subsystems.
\item
Understanding and correcting errors in
quantum computers, where $A$ is the qubits of the 
computer and $B$ is the environment which inevitably couples to the
computational qubits. The environment is very complicated with a huge
(essentially infinite) number of degrees of freedom, so we cannot include it
explicitly and we need a description involving just the
degrees of freedom of $A$, in which the effects of the environment have been
averaged over in some sense. This description is 
provided by the density
matrix. In practice, approximations
will have to be made to determine it.
We will discuss the effects of the environment on a quantum computer
in the Chapter \ref{ch:err_corr}.
\end{itemize}

\noindent For further reading on the density matrix see Refs.~\cite{nielsen:00,vathsan:16,rieffel:14}.

\section{Definition of the Density Matrix}
\label{sec:dm}

To become familiar with the notation in a gentle way we first consider the
density matrix of a
system in a well-defined quantum state. This is not terribly useful in itself, and will
just be a rewriting of results we have already obtained, but doing this will help us
understand the much more useful case of the density matrix of a subsystem $A$, say,
coupled to another system $B$, such that
the total system $A\otimes B$ is in a well defined quantum state, but is
entangled with respect to the $A$-$B$ subdivision so neither $A$ nor $B$
are in a well defined state.

\subsection{Density matrix of a system in a well defined state}
\index{density matrix}

Consider, then, a quantum system in a well defined quantum state,
$|\psi\rangle$. We \textit{define} its density matrix $\rho$ by the outer
product
\index{outer product}
\begin{equation}
\rho = |\psi\rangle \langle \psi |  .
\label{rho_state}
\end{equation}
We will understand the reason for this definition as we go along.\footnote{In
standard linear algebra notation, $\rho_{nm} = c_n c_m^\star$ and $\rho_{mn} =
c_m c_n^\star = \left(\rho_{nm}\right)^\star$, so $\rho$ is Hermitian.\label{fnHerm}} The matrix
elements of $\rho$ are
\begin{equation}
\langle n | \rho | m \rangle  = \langle n | \psi \rangle \langle \psi |
m\rangle .
\end{equation}
Note that its diagonal elements are $|\langle n|\psi\rangle|^2$ which are the
probabilities, $P_n$,
of a measurement finding the system in state $|n\rangle$. Since probabilities
add up to one, the trace (sum of diagonal elements) must satisfy
\index{matrix!trace}
\begin{equation}
\Tr \rho = 1.
\end{equation}
This will turn out to be a general property of a density matrix.

We shall now show that expectation values of operators in state $|\psi\rangle$ can be
expressed in terms of $\rho$.  We showed in Eq.~\eqref{expect} that the expectation
value of an operator $\hat{O}$ is given by
\begin{equation}
\langle \hat{O} \rangle = \langle \psi|\hat{O}|\psi\rangle .
\end{equation}
This can be re-expressed in terms of the density matrix $\rho$ since
\begin{align}
\langle \psi|\hat{O}|\psi\rangle 
&= \sum_{m} \langle \psi|\hat{O}| m\rangle \langle m| \psi\rangle\nonumber \\
&= \sum_{m}  \langle m| \psi\rangle \langle \psi|\hat{O}| m\rangle  \nonumber \\
&= \sum_{m} \langle m| \rho\, \hat{O}| m\rangle , \nonumber \\
&= \Tr(\rho\, \hat{O}) ,
\label{Tr_rho_O}
\end{align}
where we used Eq.~\eqref{rho_state} and the completeness relation $\sum_n |n\rangle \langle n| =
\index{completeness relation}
\mathbbm{1}$, the identity.
Hence expectation values can be obtained directly from the density matrix.

I emphasize that this is a trivial example in which the density matrix is not
needed, but this discussion provides a useful starting point for the
general formulation of the density matrix described in the next subsection.

\subsection{Density matrix of a subsystem when the combined system is in a well defined state}
\label{rho_sub}
\index{density matrix}

In the rest of this chapter we consider a system composed of
two subsystems $A$ and $B$, such that the combined system is in a single
quantum state.  In general, subsystems $A$ and $B$ will be entangled, so
neither subsystem is in a well defined state. We will only be interested in
one of the subsystems, $A$ say, and would like a description in terms of just
the states of $A$.  This is where the density matrix becomes very useful. 

We will assume that subsystem $A$ has $n_A$ qubits, and subsystem $B$ has
$n_B$ qubits, so the number of states of each subsystem is given by
\begin{equation}
N_A = 2^{n_A}, \qquad N_B = 2^{n_B}.
\end{equation}

As in the previous subsection, the density matrix of the whole system is given
by
\begin{equation}
\rho^{AB} = |\psi_{AB} \rangle \langle \psi_{AB} |.
\label{rhoAB}
\end{equation}
This is a matrix involving the states of both $A$ and $B$. We
shall now show that information about averages of the $A$ degrees of freedom
can be obtained without explicitly considering the $B$ degrees of freedom from
the density matrix $\rho^A$ where\footnote{\label{fn1}As stated above,
in this chapter we assume that the combined
$AB$ system is in single quantum state. If, instead, the combined system is itself described
by a non-trivial density matrix $\rho^{AB}$, then the reduced density matrix for
subsystem $A$ is still given by $\rho_A = \Tr_B\, \rho^{AB}$
but $\rho^{AB}$ is no longer given by
Eq.~\eqref{rhoAB}.}
\begin{equation}
\rho^A = \Tr_B\, \rho^{AB} = \sum_{j_B=1}^{N_b}
\langle j_B |\psi_{AB} \rangle \langle
\psi_{AB} |j_B \rangle ,
\label{TrB}
\end{equation}
which is a matrix in the space of the states of $A$ only. Here $|j_B\rangle$ is a basis
state for subsystem $B$. We say that
we have ``traced out" the $B$ states. 

The state $|\psi_{AB}\rangle$ can be expressed in terms of basis states. In
the Dirac notation this has the rather cumbersome form
\begin{equation}
|\psi_{AB}\rangle = \sum_{i_A=1}^{N_A} \sum_{j_B=1}^{N_B}
|i_A\rangle |j_B\rangle \langle i_A| \langle
j_B| \psi_{AB}\rangle ,
\label{psi_AB_dirac}
\end{equation}
where  $|i_A\rangle$ is a basis 
state for subsystem $A$.
so I prefer to use here the standard matrix notation with indices, rather
than the Dirac notation, writing
Eq.~\eqref{psi_AB_dirac} as
\begin{equation}
|\psi_{AB}\rangle = \sum_{i_A=1}^{N_A} \sum_{j_B=1}^{N_B} c_{{i_A}{j_B}}\,
|i_A\rangle\,| j_B\rangle .
\label{psi_AB}
\end{equation}

From Eqs.~\eqref{TrB} and \eqref{psi_AB}, the matrix elements of $\rho^A$
are given in terms 
of the amplitudes $c_{{i_A}{j_B}}$ by\footnote{See footnote \ref{fnHerm} on page
\pageref{fnHerm}.}
\begin{equation}
\langle i_A| \rho^A |i'_A\rangle = \sum_{j_B} c_{{i_A}{j_B}}\, c^*_{i'_A{j_B}}. 
\label{rhoA}
\end{equation}
Because $|\psi_{AB}\rangle$ is normalized we have
\begin{equation}
\Tr \rho^A = \sum_{i_A, j_B} |c_{{i_A}{j_B}}|^2 = 1,
\label{norm}
\end{equation}
so the trace of the density matrix is always equal to $1$.
Also we see that
\begin{equation}
\langle i'| \rho^A |i\rangle = \langle i| \rho^A |i'\rangle^\star ,
\end{equation}
omitting for conciseness the label $A$ on $|i_A\rangle$ and  $|i'_A\rangle$ when
there is no ambiguity,
so $\rho^A$ is Hermitian.
As discussed earlier, this condition is a general property of density matrices. 

We want to compute the expectation value of some operator $\hat{O}_A$ acting
only on the $A$ degrees of freedom, i.e.
\begin{equation}
\begin{split}
\langle \hat{O}_A \rangle &= \langle \psi_{AB}|\hat{O}_A |\psi_{AB}\rangle  \\
&= \sum_{{i_A}, {i_A}'} \sum_{j_B, j_B'} \langle j'_B\, i'_A|\hat{O}_A |
i_A\, j_B\rangle\, c^*_{i'_A j'_B}\, c_{i_A j_B} \\
&= \sum_{i_A, i_A'} \sum_{j_B, j_B'}\, \langle j_B'|
j_B\rangle \langle i_A' |\hat{O}_A
| i_A\rangle \,\, c^*_{i'_A j'_B}\, c_{i_A j_B} \\
&= \sum_{i_A, i_A'} \sum_{j_B}  c_{i_A j_B}\, c^*_{i'_A j_B} \,\, \langle i_A' |\hat{O}_A|
i_A\rangle,
\end{split}
\end{equation}
where in the third line we used that $\hat{O}_A$ does not depend on the $B$ degrees of freedom,
and in the fourth line we used that $\langle j'_B | j_B\rangle= \delta_{j_B
j'_B}$. 

Hence, from Eq.~\eqref{rhoA}, 
\begin{equation}
\begin{split}
\langle \hat{O}_A \rangle &= \sum_{i, i'} \langle  i| \rho^A |i'\rangle \langle i'
|\hat{O}_A| i\rangle \\
&= \sum_i \langle  i| \rho^A \hat{O}_A| i\rangle \\
&= \mathrm{Tr}_A \left( \rho^A \hat{O}_A \right) ,
\end{split}
\label{trrhoO}
\end{equation}
which has the same form as Eq.~\eqref{Tr_rho_O}.
Thus we can compute averages of quantities involving subsystem $A$ from a
knowledge of the density matrix $\rho^A$, without needing to explicitly
consider subsystem $B$. All necessary information about $B$ is contained in the
density matrix $\rho^A$.
Note that $\rho^A$ is the same no matter what
quantity of system $A$
is to be calculated, and so it
only has to be calculated \textit{once}. 

One can equivalently trace out the degrees of freedom in $A$ to get the density
matrix for subsystem $B$, i.e.~$\rho^B = \mathrm{Tr}_A\, \rho^{AB}$, so
\begin{equation}
\langle j_B| \rho^B |j'_B\rangle = \sum_{i_A}  c_{i_A j_B}\, c^*_{i_A j'_B}. 
\end{equation}
%

As we shall see, it is useful to diagonalize the density matrix, obtaining its
eigenvalues $\lambda_\alpha$ and eigenvectors $|\phi_\alpha\rangle$. Since the sum of
the eigenvalues is equal to the trace we have, according to Eq.~\eqref{norm},
\begin{equation}
\sum_\alpha \lambda_\alpha = 1,
\end{equation}
which suggests that the eigenvalues can be interpreted as probabilities (since
probabilities also sum to 1).  We shall now see that this interpretation is
correct.

Let's consider Eq.~\eqref{trrhoO} in the basis where $\rho^A$ is diagonal. We have
\begin{equation}
\begin{split}
\langle \hat{O}_A \rangle &= \Tr \left( \rho^A \hat{O}_A \right) \\
&= \sum_\alpha \lambda_\alpha \langle \phi_\alpha| \hat{O}_A | \phi_\alpha \rangle .
\end{split}
\end{equation}
Thus we get the expectation value of $\hat{O}_A$ in state $|\psi_{AB}\rangle$
by (i) computing the expectation of $\hat{O}_A$ in state
$|\phi_\alpha\rangle$ (an eigenvector of $\rho^A$),
(ii) multiplying by $\lambda_\alpha$ (the corresponding eigenvalue of $\rho^A$), and (iii)
summing over $\alpha$. This clearly shows that $\lambda_\alpha$ should be
thought of as the probability that subsystem $A$ is in state $|\alpha\rangle$.
To emphasize this, from now on we will denote the eigenvalues of the density
matrix by $p_\alpha$.

Furthermore, if we consider Eq.~\eqref{trrhoO} in the basis $|m\rangle$ where $\hat{O}$ is
diagonal we get 
\begin{equation}
\begin{split}
\langle \hat{O}_A \rangle &= \Tr \left( \rho^A \hat{O}_A \right) \\
&= \sum_m \langle m|\rho|m \rangle \langle m| \hat{O}_A | m\rangle ,
\end{split}
\end{equation}
where $\langle m| \hat{O}_A | m\rangle$ is an eigenvalue of $\hat{O}_A$.
We interpret this to mean that the probability that a measurement of $\hat{O}$
yields eigenvalue $\langle m|\hat{O}|m \rangle$, leaving the system in
state $|m\rangle$, is the corresponding diagonal element of the density
matrix, $\langle m|\rho|m \rangle$.

To summarize, to determine the properties of a subsystem from the density
matrix when the state of
the whole system is in a single quantum state:
\begin{enumerate}
\item
We compute the elements of the density matrix according to Eq.~\eqref{rhoA}.
\item
The density matrix is Hermitian and so has real eigenvalues, $p_i$. The sum of the
eigenalues is equal to one, and the eigenvalues are interpreted as
probabilities.
\item
If the eigenstate corresponding to eigenvalue $p_i$ is denoted by $|u_i\rangle$ then the significance of 
the density matrix is that the subsystem can be thought of as being in state $|u_i\rangle$ with
probablity $p_i$.
\item
If measurements are made in some basis, then the probability that the measurement
finds the system in state
$|n\rangle$ is the corresponding diagonal element
of the density matrix, $\langle n|\rho|n \rangle$.
\item 
Expectation values of operators acting on the subsystem can be obtained from
Eq.~\eqref{trrhoO}.
\end{enumerate}

\section{Determining if a state is entangled}
\index{entanglement}
One use of the density matrix is that it gives a systematic prescription for
determining whether a state is a product state or entangled with respect to a
\index{product state}
partition into subsystems $A$ and $B$.  If it is not a product
state we say that it is a mixed state and is ``entangled" with respect to this
partition. We shall use the terms ``mixed state" and ``entangled state"
interchangeably.
\index{mixed state}

To see how the density matrix can determine if a state is a product state or
is entangled with respect to partition into $A$-$B$ subsystems, let's assume
initially that
$|\psi_{AB}\rangle$ is a product state, i.e.
\begin{equation}
|\psi_{AB}\rangle = |\phi\rangle_A\, |\mu\rangle_B .
\end{equation}
In this case subsystem $A$ is definitely in state $|\phi\rangle$, so the
eigenvalues of $\rho^A$ must be $p_1 = 1$ and $p_\alpha = 0$ for $\alpha\ne
1$. Also the eigenvector for the non-zero eigenvalue must be given by
$|\phi_1\rangle = |\phi\rangle$.

Hence, if the state of the combined $AB$ system is a product state then one of the
eigenvalues of the density matrix of $A$ (or of $B$) will be $1$ and the
others zero. Conversely if more than one of the eigenvalues of the density
matrix are positive (since they are probabilities they can only be positive or
zero) the state is mixed, i.e.~entangled. 

It is actually not necessary to diagonalize the density matrix to determine if
the state is a product state or entangled. Instead it is sufficient to take
its square.  To see this note that
\begin{equation}
\Tr \left(\rho^A\right)^2 = \sum_\alpha p_\alpha^2,
\end{equation}
where we used that the trace is the sum of the eigenvalues, see Sec.~\ref{sec:mat_props}, and
that the eigenvalues
of the square of a matrix are the square of the eigenvalues of that matrix. 
Since the $p_\alpha$ must lie between $0$ and $1$ and $\sum_\alpha p_\alpha =
1$, one can show that $\sum_\alpha p_\alpha^2 \le 1$, with the equality only
holding if one of the $p_\alpha$ is $1$ and the others zero. As an example,
consider the case of two states, for which the eigenvalues are $p$ and $1-p$ with
$0\le p \le 1$. Now
\begin{equation}
\Tr \left(\rho^A\right)^2 =
\sum_{\alpha=1}^2 p_\alpha^2 =
p^2 + (1-p)^2 = 1 -2 p + 2 p^2 = 1 - 2p(1-p).
\end{equation}
For $0 \le p \le 1$, we see that $0 \le 2 p (1-p) \le 1/2$ and is only zero for
$p=0$ and $1$. Consequently, $\Tr \left(\rho^A\right)^2 < 1$ unless $p=0$ or $1$.

Hence we have the following general criterion:
\begin{equation}
\mathrm{if\ }\Tr \left(\rho^A\right)^2 \left\{
\begin{array}{ll}
= 1, & \mathrm{then\ we\ have\ a\ product\ state}, \\
< 1, & \mathrm{then\ we\ have\ a\ \ mixed\ (entangled)\ state}, \\
\end{array}
\right.
\end{equation}
We emphasize again that $\Tr \rho^A = 1$ always.

Sometimes one defines the Von Neumann entanglement entropy by
\index{entanglement!entropy (Von Neuman)}
\begin{equation}
S(\rho^A) = - \Tr \rho^A \log \rho^A \ (= -\sum_\alpha p_\alpha \log p_\alpha).
\end{equation}
\index{product state}
It is easy to see that $S(\rho^A)= 0$ if the state is a product state since
$\lim_{x\to 0} (x \ln x) = 0$.  In the
opposite limit, of a maximally entangled state where $p_\alpha = 1/N_A$ for all
\index{maximally entangled state}
$\alpha$, one has $S(\rho^A)= \log N_A$. For the case where subsystem $A$ is a
single qubit, this gives  $S(\rho^A)= \log 2$.

\section{Some Simple Examples}
\label{sec:examples}
In this section we consider some simple examples where subsystems $A$ and $B$
each have just a single qubit.

\subsection{Example 1:}
We take
\begin{equation}
|\psi_{AB}\rangle = {1 \over 2}\left( |0_A 0_B\rangle + |0_A 1_B\rangle - |1_A
0_B\rangle - |1_A 1_B\rangle \right)
\end{equation}
\textit{Note:} We can see ``by inspection" that this is a product state
\begin{equation}
|\psi_{AB}\rangle =
{1 \over \sqrt{2}}\left(|0_A\rangle - |1_A\rangle \right) \otimes
{1 \over \sqrt{2}}\left(|0_B\rangle + |1_B\rangle\right) .
\label{separ}
\end{equation}
We shall now show how this result is obtained from the density matrices
$\rho^A$ and $\rho^B$. 

We have
\begin{equation}
c_{00} = c_{01} = \smfrac{1}{2},\qquad c_{10} = c_{11} = -\smfrac{1}{2},
\end{equation}
so, from Eq.~\eqref{rhoA},
\begin{equation}
\begin{split}
\rho^A_{00} &= c_{00} c_{00} + c_{01} c_{01} = \smfrac{1}{2} \\
\rho^A_{01} &= c_{00} c_{10} + c_{01} c_{11} = -\smfrac{1}{2} \\
\rho^A_{10} &= c_{10} c_{00} + c_{11} c_{01} = -\smfrac{1}{2} \\
\rho^A_{11} &= c_{10} c_{10} + c_{11} c_{11} = \smfrac{1}{2} ,
\end{split}
\end{equation}
and hence
\begin{equation}
\rho^A = {1 \over 2} \begin{pmatrix*}[r]
1 & -1 \\
-1 & 1
\end{pmatrix*} .
\end{equation}
The eigenvalues are given by 
\begin{equation}
\begin{vmatrix}
\smfrac{1}{2} - \lambda & -\smfrac{1}{2} \\
-\smfrac{1}{2} & \smfrac{1}{2} - \lambda  \\
\end{vmatrix}
 = 0
\end{equation}
so
\begin{equation}
\left(\lambda - \smfrac{1}{2}\right)^2 - \left(-\smfrac{1}{2}\right)^2 = 0
\end{equation}
which gives $\lambda = 1$ and $0$. Since only one eigenvalue is non-zero this is
a product state, as we saw above.

One easily finds that
\begin{equation}
\left(\rho^A\right)^2 = 
{1 \over 2} \begin{pmatrix*}[r]
1 & -1 \\
-1 & 1
\end{pmatrix*} ,
\end{equation}
and so $\mathrm{Tr} \left(\rho^A\right)^2 = 1$ as required for a product
state.

The eigenvector with eigenvalue $\lambda = 1$ is given by
\begin{equation}
{1 \over 2} \begin{pmatrix*}[r]
1 & -1 \\
-1 & 1
\end{pmatrix*}
\begin{pmatrix}
a \\ b \\
\end{pmatrix}
= \begin{pmatrix}
a \\ b \\
\end{pmatrix} .
\end{equation}
Both the resulting equations give $b = -a$ so the normalized eigenvector is
\begin{equation}
|\phi_{1, A}\rangle = {1\over \sqrt{2}}|\left(|0_A\rangle - |1_A\rangle \right) .
\end{equation}
Hence, with probability $1$, subsystem $A$ is in state $|\phi_{1}\rangle$, in
agreement with Eq.~\eqref{separ}.

One can repeat the same calculation for $\rho^B$. The results are
\begin{equation}
\begin{split}
\rho^B_{00} &= c_{00} c_{00} + c_{10} c_{10} = \smfrac{1}{2} \\
\rho^B_{01} &= c_{00} c_{01} + c_{10} c_{11} = \smfrac{1}{2} \\
\rho^B_{10} &= c_{01} c_{00} + c_{11} c_{10} = \smfrac{1}{2} \\
\rho^B_{11} &= c_{01} c_{01} + c_{11} c_{11} = \smfrac{1}{2} ,
\end{split}
\end{equation}
so
\begin{equation}
\rho^B = {1 \over 2} \begin{pmatrix*}[r]
1 & 1 \\
1 & 1
\end{pmatrix*}
\end{equation}
The eigenvalues are given by 
\begin{equation}
\begin{vmatrix}
\smfrac{1}{2} - \lambda & \smfrac{1}{2} \\
\smfrac{1}{2} & \smfrac{1}{2} - \lambda  \\
\end{vmatrix}
 = 0
\end{equation}
so
\begin{equation}
\left(\lambda - \smfrac{1}{2}\right)^2 - \left(\smfrac{1}{2}\right)^2 = 0
\end{equation}
which gives $\lambda = 1$ and $0$, the same as for $\rho^A$. It is true in
general that the non-zero eigenvalues of $\rho^A$ and $\rho^B$ must be equal,
provided that the combined $AB$ system is in a single quantum state. This is
discussed further in the more advanced material in Appendix \ref{app}

The eigenvector with eigenvalue $\lambda = 1$ is given by
\begin{equation}
{1 \over 2} \begin{pmatrix*}[r]
1 & 1 \\
1 & 1
\end{pmatrix*}
\begin{pmatrix}
a \\ b \\
\end{pmatrix}
= \begin{pmatrix}
a \\ b \\
\end{pmatrix} .
\end{equation}
Both the resulting equations give $b = a$ so the normalized eigenvector is
\begin{equation}
|\sigma_{1,B}\rangle = {1\over \sqrt{2}}|\left(|0_B\rangle + |1_B\rangle \right) .
\end{equation}
Hence, with probability $1$, subsystem $B$ is in state $|\sigma_{1}\rangle$, again in
agreement with Eq.~\eqref{separ}.

\subsection{Example 2:}
In this example we take one of the Bell states,\index{Bell state}
\begin{equation}
|\psi_{AB}\rangle = {1 \over \sqrt{2}}\left(|0_A 0_B\rangle + |1_A 1_B
\rangle\right),
\label{state2}
\end{equation}
which is clearly entangled.  Here we have
\begin{equation}
c_{00} = c_{11} = \smfrac{1}{\sqrt{2}},\qquad c_{10} = c_{01} = 0.
\end{equation}
Hence
\begin{equation}
\begin{split}
\rho^A_{00} &= c_{00} c_{00} + c_{01} c_{01} = \smfrac{1}{2} \\
\rho^A_{01} &= c_{00} c_{10} + c_{01} c_{11} = 0 \\
\rho^A_{10} &= c_{10} c_{00} + c_{11} c_{01} = 0 \\
\rho^A_{11} &= c_{10} c_{10} + c_{11} c_{11} = \smfrac{1}{2} ,
\end{split}
\end{equation}
so
\begin{equation}
\rho^A = {1 \over 2} \begin{pmatrix}
1 & 0 \\
0 & 1
\end{pmatrix}.
\end{equation}
This is already in diagonal form so we read off that the two eigenvalues are
both equal to $1/2$. Since more than one eigenvalue is positive, the state
is mixed. A density matrix like this, with all eigenvalues equal,
is \textit{maximally entangled}.
\index{maximally entangled state}
It is easy to see that the same eigenvalues are obtained from $\rho^B$.

Trivially 
\begin{equation}
\mathrm{Tr}\left(\rho^A\right)^2 = {1 \over 4} \mathrm{Tr}
\begin{pmatrix}
1 & 0 \\
0 & 1
\end{pmatrix}
= {1 \over 2} \ (< 1),
\end{equation}
which indicates, again, that Eq.~\eqref{state2} is a mixed state.

\subsection{Example 3:}
This example is slightly more complicated but it is useful to go through it in detail.
We take
\begin{equation}
|\psi_{AB}\rangle = {1 \over \sqrt{8}} \left(
|0_A 0_B\rangle + \sqrt{3}|0_A 1_B \rangle - \sqrt{3} |1_A 0_B \rangle -|1_A 1_B \rangle
\right),
\label{compl_state}
\end{equation}
so
\begin{equation}
c_{00} = \smfrac{1}{\sqrt{8}}, \qquad c_{01} = \sqrt{\smfrac{3}{8}}, \qquad c_{10} =
-\sqrt{\smfrac{3}{8}}, \qquad c_{11} = -\smfrac{1}{\sqrt{8}} .
\end{equation}
It follows that
\begin{equation}
\begin{split}
\rho^A_{00} &= c_{00} c_{00} + c_{01} c_{01} = \smfrac{1}{2} \\
\rho^A_{01} &= c_{00} c_{10} + c_{01} c_{11} = -\smfrac{\sqrt{3}}{4} \\
\rho^A_{10} &= c_{10} c_{00} + c_{11} c_{01} = -\smfrac{\sqrt{3}}{4} \\
\rho^A_{11} &= c_{10} c_{10} + c_{11} c_{11} = \smfrac{1}{2} ,
\end{split}
\end{equation}
so
\begin{equation}
\rho^A = {1 \over 4} \begin{pmatrix}
2 & -\sqrt{3} \\
-\sqrt{3} & 2
\end{pmatrix} .
\end{equation}
The eigenvalues are found to be
\begin{equation}
p_1 = {1\over 4}\left( 2 + \sqrt{3} \right), \qquad
p_2 = {1\over 4}\left( 2 - \sqrt{3} \right), 
\label{evals_3}
\end{equation}
with corresponding eigenvectors 
\begin{equation}
\begin{split}
|\phi_{1, A}\rangle &= {1 \over \sqrt{2}}\left(|0_A\rangle - |1_A\rangle \right) \\
|\phi_{2, A}\rangle &= {1 \over \sqrt{2}}\left(|0_A\rangle + |1_A\rangle \right) .
\end{split}
\end{equation}
Thus subsystem $A$ can be regarded as being in state $|\phi_1\rangle$ with
probability $p_1$ and in state $|\phi_2\rangle$ with
probability $p_2$.

It is straightforward to show that
\begin{equation}
\left(\rho^A\right)^2 = {1 \over 16}
\begin{pmatrix}
7 & - 4 \sqrt{3} \\
- 4 \sqrt{3} & 7 \\
\end{pmatrix}
\end{equation}
and so
\begin{equation}
\mathrm{Tr} \left(\rho^A\right)^2 ={7 \over 8} \ (< 1),
\end{equation}
in agreement with Eq.~\eqref{compl_state} being a mixed state.

Repeating the same arguments for $\rho^B$ gives
\begin{equation}
\begin{split}
\rho^B_{00} &= c_{00} c_{00} + c_{10} c_{10} = \smfrac{1}{2} \\
\rho^B_{01} &= c_{00} c_{01} + c_{10} c_{11} = \smfrac{\sqrt{3}}{4} \\
\rho^B_{10} &= c_{01} c_{00} + c_{11} c_{10} = \smfrac{\sqrt{3}}{4} \\
\rho^B_{11} &= c_{01} c_{01} + c_{11} c_{11} = \smfrac{1}{2} ,
\end{split}
\end{equation}
so
\begin{equation}
\rho^B = {1 \over 4} \begin{pmatrix}
2 & \sqrt{3} \\
\sqrt{3} & 2
\end{pmatrix} .
\end{equation}

The eigenvalues are found to be again given by Eq.~\eqref{evals_3} and
the corresponding eigenvectors are
\begin{equation}
\begin{split}
|\sigma_{1, B}\rangle &= {1 \over \sqrt{2}}\left(|0_B\rangle + |1_B\rangle \right) \\
|\sigma_{2, B}\rangle &= {1 \over \sqrt{2}}\left(-|0_B\rangle + |1_B\rangle \right) .
\end{split}
\end{equation}
Subsystem $B$ can therefore be regarded as being in state $|\sigma_1\rangle$ with
probability $p_1$ and in state $|\sigma_2\rangle$ with
probability $p_2$.

It is interesting to note that if we define
\begin{equation}
c_1 = {1 \over 2} \sqrt{2 + \sqrt{3}}, \qquad
c_2 = {1 \over 2} \sqrt{2 - \sqrt{3}}, 
\end{equation}
so
\begin{equation}
p_1 = c_1^2, \qquad p_2 = c_2^2,
\end{equation}
then a bit of algebra\footnote{Note that $\sqrt{2+\sqrt{3}} - \sqrt{2 - \sqrt{3}} =
\sqrt{2}$ and $\sqrt{2+\sqrt{3}} + \sqrt{2 - \sqrt{3}} =
\sqrt{6}$, which are proved by squaring both sides.} shows that
\begin{equation}
|\psi_{AB} \rangle = c_1 |\phi_{1, A}\rangle \otimes |\sigma_{1, B}\rangle + 
c_2 |\phi_{2, A}\rangle \otimes |\sigma_{2, B}\rangle .
\label{schmidt}
\end{equation}
\index{Schmidt!decomposition}
This is an example of Schmidt decomposition which is described in the more
advanced material in Appendix \ref{app}. The coefficients
$c_1$ and $c_2$ are known as Schmidt
coefficients.

According to Eq.~\eqref{schmidt} we can decompose $|\psi_{AB}\rangle$ in the
following way: with probability $p_1 =  c_1^2$ subsystem $A$ is in state
$|\psi_{1, A}\rangle$ and subsystem $B$ is in state
$|\sigma_{1, B}\rangle$, and with probability $p_2 =  c_2^2$ $(=1-p_1)$
subsystem $A$ is in state
$|\psi_{2, A}\rangle$ and subsystem $B$ is in state
$|\sigma_{2, B}\rangle$. In this way one can see why the non-zero eigenvalues of the two
subsystem density matrices $\rho^A$ and $\rho^B$ must be equal, namely for
both matrices the eigenvalues are given by
$c_1^2$ and $c_2^2$.

\section{Systems not in a single quantum state}
\label{sec:non_ortho}
An additional application for the density matrix is for systems which are not
described by a single quantum state. An example would be to characterize the
behavior of a stream of particles (electrons, say) which are polarized
in different directions. We need to average over the different spin
orientations using standard classical statistics.

Suppose for example that a fraction $p$ of the electrons are polarized in the
+$z$ direction, i.e are in state $|0\rangle$, and a fraction $1-p$ are in the
-$z$ direction, i.e are in state $|1\rangle$.
The density matrix for particles in state $|0\rangle$ is
\begin{equation}
|0\rangle\langle 0| = 
\begin{pmatrix}
1 & 0 \\
0 & 0 \\
\end{pmatrix},
\end{equation}
and that for state $|1\rangle$ is
\begin{equation}
|1\rangle\langle 1| = 
\begin{pmatrix}
0 & 0 \\
0 & 1 \\
\end{pmatrix}.
\end{equation}
The density matrix of the stream of electrons is therefore
\begin{equation}
\rho = p |0\rangle\langle 0| + (1-p) |1\rangle\langle 1| =
\begin{pmatrix}
p & 0 \\
0 & 1-p \\
\end{pmatrix}.
\end{equation}

For a less trivial example, consider the case that a fraction $p$ of the
electrons are in state $|0\rangle$ (polarized in the +$z$ direction) while
fraction $1-p$ are polarized in state $|+\rangle = {1\over \sqrt{2}}(|0\rangle
+ |1\rangle)$ (polarized in the +$x$ direction). The density matrix can then
be conveniently written as
\begin{equation}
\rho = p |0\rangle\langle 0| + (1-p) |+\rangle\langle +|
.\label{zx}
\end{equation}

\index{non-orthogonal states}
Note that states $|0\rangle$ and $|+\rangle$ are not orthogonal. If we rewrite
Eq.~\eqref{zx} in terms of orthogonal states, (for example computational basis
states) it becomes more complicated. To do this we
note that
\begin{equation}
|+\rangle\langle +| =  {1 \over 2}
\begin{pmatrix}
1 & 1 \\
1 & 1 
\end{pmatrix},
\end{equation}
so the density matrix for the beam of electrons can be written in the
computational basis as
\begin{equation}
\rho = 
\begin{pmatrix}
(1+p)/2 & (1-p)/2 \\
(1-p)/2 & (1-p)/2 
\end{pmatrix}.
\end{equation}
The eigenvalues of $\rho$ are 
\begin{equation}
\lambda_{\pm } = {1 \over 2}\left[ 1 \pm \sqrt{1 - 2p + 2 p^2}\right],
\label{lam+-}
\end{equation}
while the eigenvectors are
\begingroup
\renewcommand*{\arraystretch}{1.5}
\begin{equation}
|\psi_{\pm}\rangle =C_{\pm} 
\begin{pmatrix}
{\displaystyle p \pm \sqrt{1 - 2 p + 2 p^2} \over \displaystyle 1-p} \\ 1
\end{pmatrix}
,
\end{equation}
\endgroup
where the $C_{\pm} $ are normalization factors which are sufficiently messy that I prefer to not
write them down.

Here we have given a description of the density matrix in terms of
non-orthogonal states.\footnote{The statistical properties of measurements on
a system are
completely determined by its density matrix. However, this example shows
that the \textit{interpretation} of the density
matrix in terms of the system being in different states with various
probabilities is not unique if one allows for non-orthogonal states. Sometimes,
as in this example, it may be simpler to use non-orthogonal states.}
Note that the factors of $p$
and $(1-p)$ in Eq.~\eqref{zx} are \textit{not} the eigenvalues. These have to be
determined in an orthogonal basis and are given by
Eq.~\eqref{lam+-}.

\section{Conclusions}
We have seen that the density matrix is useful when studying the properties of a
system composed of two subsystems $A$ and $B$. More precisely, it can be used
to:
\begin{itemize}
\item
Determine the properties of one of the subsystems $A$ without explicitly having to
include the degrees of freedom of the other subsystem $B$. This is
particularly useful if $B$ contains a very large number of degrees of
freedom. An example of a large ``subsystem" is the environment, with which,
unfortunately, the qubits of a
quantum computer unavoidably interact.
\item
If the combined $AB$ system is in a single state, the properties of the subsystem density
\index{product state}
matrices tell us whether that state is a product state with respect to the
$A$-$B$ partition or whether, on the other hand, it is a mixed state in which the two 
subsystems are entangled.
\end{itemize}

\hrulefill
\section*{Problems}
\input{hw_ch5.tex}
\begin{center}
{\Large\bf Appendices}
\end{center}

\begin{subappendices}
\section{Schmidt Decomposition}
\index{Schmidt!decomposition}
\label{app}
(This is more advanced material which is not required for the course.)

It can be shown~\cite{nielsen:00} that a state $|\psi_{AB}\rangle$ can be
written as
\begin{equation}
|\psi_{AB}\rangle = \sum_\alpha c_\alpha\, |\phi_{\alpha, A}\rangle\,\otimes \,
|\sigma_{\alpha, B}\rangle,
\label{schmidt_decom}
\end{equation}
where the number of terms is less than or equal to the smaller of $N_A$ and
$N_B$, and the $|\phi_\alpha\rangle$ are mutually orthogonal as are the
$|\sigma_\alpha\rangle$.
The $c_\alpha$ are real, non-negative numbers called Schmidt coefficients.
\index{Schmidt!coefficients}
It is always possible to make the $c_\alpha$ real and non-negative
because the phases of $|\phi_{\alpha, A}\rangle$ and $|\sigma_{\alpha, B}\rangle$
can be chosen independently. 
The sum in Eq.~\eqref{schmidt_decom} is known as a Schmidt decomposition.
An example of a Schmidt decomposition is shown in Eq.~\eqref{schmidt}.

Using the definition of $\rho_A$ given in Eq.~\eqref{TrB} and working in the
$|\phi_\alpha\rangle$ basis for the states of $A$ and the $|\sigma_\alpha\rangle$ basis for
the states of $B$, one has
\begin{equation}
\begin{split}
\rho^A &= \mathrm{Tr}_B\, |\psi_{AB}\rangle \langle \psi_{AB}| \\
&= \sum_\alpha c_\alpha^2\, |\phi_{\alpha, A}\rangle\,\, \langle \phi_{\alpha, A}|.
\end{split}
\end{equation}
This shows that $\rho_A$ has non-zero eigenvalues $p_\alpha = c_\alpha^2$ with
corresponding eigenvectors $ |\phi_{\alpha}\rangle_A$. 

Similarly one has
\begin{equation}
\begin{split}
\rho^B &= \mathrm{Tr}_A\, |\psi_{AB}\rangle \langle \psi_{AB}| \\
&= \sum_\alpha c_\alpha^2\, |\sigma_{\alpha, B}\rangle\,\, {}\langle \sigma_{\alpha, B}|,
\end{split}
\end{equation}
which shows that $\rho_B$ has non-zero eigenvalues $p_\alpha = c_\alpha^2$
(the same as for $\rho^A$) with corresponding eigenvectors $
|\sigma_{\alpha}\rangle_B$. The number of non-zero Schmidt
coefficients (the $c_\alpha$) is called the Schmidt number (or Schmidt rank).
\index{Schmidt!number (rank)}
If the Schmidt
\index{product state}
number is $1$ the state is a product state, while if it is greater than 1, the
state is entangled (mixed).

\section{Change in the density matrix under a unitary transformation}
\label{appBB}

\index{unitary transformation}
If qubit $A$ (more generally subsystem $A$) is acted by a unitary
transformation $U^{A}$ then we show now that the density matrix for subsystem $A$
changes from $\rho^A$ to $\rho^{'A}$ where:
\begin{equation}
\rho^{'A} = U^{A} \rho^A \left(U^{A}\right)^\dagger.
\label{rhop}
\end{equation}
To see this, note that $|\psi_{AB}\rangle$ in Eq.~\eqref{psi_AB} goes to
$|\psi'_{AB}\rangle$ where
\begin{equation}
|\psi'_{AB}\rangle =
\sum_{i, j} c'_{ij} |i_A\rangle
 \otimes |j_B \rangle
\end{equation}
in which 
\begin{equation}
c'_{ij} = \sum_{k} U^A_{ik}\, c_{kj} 
\end{equation}
describes the change in amplitudes produced by the action of $U^A$.
Note that the second index $j$ on the amplitude $c_{ij}$ refers to subsystem
$B$ and is not changed.
Hence
\begin{equation}
\begin{split}
\rho^{'A}_{i ,i'} &= \sum_{j} c'_{ij} c^{'*}_{i'j} \\
&= \sum_{j, k_1, k_2} U^A_{i k_1} c_{k_1j} \, U^{A*}_{i'k_2} c^*_{k_2j} \\
&= \sum_{k_1, k_2} U^A_{ik_1}\left(\sum_j c_{k_1j} c^*_{k_2j}\right) U^{A*}_{i' k_2} \\
&= \sum_{k_1, k_2} U^A_{i k_1} \rho^A_{k_1, k_2}  \left(U^{A}_{k_2
i'}\right)^\dagger \\
&= \left(U^A \rho^A \left(U^{A}\right)^\dagger \right)_{i,i'},
\end{split}
\end{equation}
so we obtain Eq.~\eqref{rhop}.


Note that the most general operation that can be applied to the combined $AB$ system is a
unitary transformation acting on the \textit{whole system}, not just on subsystem $A$.
One can show that if one performs such a
\textit{general} unitary operation on the combined system, and then recomputes the
density matrix of subsystem $A$, the new density matrix is \textit{not in
general
related to the old one by a unitary transformation}. This is how 
irreversible processes can occur in a subsystem which is coupled to 
the environment. A
more detailed discussion of this is beyond the scope of the course but the
interested student is referred to the more advanced texts such as Nielsen and Chuang~\cite{nielsen:00} and
Rieffel and Polak~\cite{rieffel:14}.

\end{subappendices}

%% file: hw_ch5.tex
\begin{problems}

\item
Show that the following state is separable (i.e.~is a product state):
\begin{equation}
{1 \over 2} \left( |00\rangle - |01\rangle - |1 0\rangle + |1 1 \rangle \right)\, ,
\label{sep}
\end{equation}
and, by inspection, write the state in separable form.

\item
\begin{enumerate}[label=(\roman*)]
\item
Compute the reduced density matrix for qubit $A$ for the 2-qubit state
\begin{equation}
{1 \over \sqrt{2}} \left( |0_A 1_B\rangle - |1_A 0_B\rangle \right).
\end{equation}
How can you deduce from this density matrix that the state is entangled?
\item
Compute the reduced density matrix for the left-hand qubit in the state in
Eq.~\eqref{sep}.\\
How can you deduce from this density matrix that the state is a product
state?
\end{enumerate}

\item
\label{qu5}
Find the reduced density matrices for each subsystem for the state
\begin{equation}
|\psi_{AB}\rangle = {1 \over 2 \sqrt{2}} \left(\, |0_A 0_B\rangle + \sqrt{3}
|0_A 1_B\rangle + \sqrt{3}|1_A 0_B\rangle + |1_A 1_B\rangle\, \right)
\end{equation}
Determine the eigenvalues of each density matrix, $\rho^A$ and $\rho^B$,
and hence deduce if the state is separable or
entangled.\\
\textit{Note:} There is general theorem which states that if a system in a single state is
decomposed into subsystems $A$ and $B$ then $\rho^A$ has the same non-zero
eigenvalues as $\rho^B$.

\item
Using your result for $\rho^A$, the density matrix of subsystem $A$,
in question \ref{qu5} compute $\langle X_A
\rangle $ and $\langle Z_A\rangle$, where the average is for state
$|\psi_{AB}\rangle $.

\end{problems}

%% file: EPR8.tex
\section{Introduction}
In classical physics, objects have definite properties irrespective of whether
we measure them or not. This is called \textit{objective reality}.
\index{objective reality}
A measurement can be done in a sufficiently delicate way that it just \textit{reveals} a property which
\textit{already existed}. 

However, this is not the case in quantum mechanics. To see this, suppose that a qubit is initially
in the state
\begin{equation}
|\psi\rangle = {1\over\sqrt{2}}\left(|0\rangle +|1\rangle \right).
\label{super_EPR}
\end{equation}
If we measure the qubit (i.e.~measure $Z$)
the Born rule states\index{Born rule}
that we get $|0\rangle$ (i.e.~eigenvalue $+1$) with probability $\smfrac{1}{2}$ and $|1\rangle$
(i.e.~eigenvalue $-1$) with probability
$\smfrac{1}{2}$. However, we can not infer from this that, \textit{before} the
measurement, the qubit was in state $|0\rangle$ with probability $\smfrac{1}{2}$
and $|1\rangle$ with probability
$\smfrac{1}{2}$, for this leads to a contradiction as we will now see.

If we apply the Hadamard operator,
\begin{equation}
H = {1 \over \sqrt{2}}
\begin{pmatrix}
1 & 1 \\
1 & -1 \\
\end{pmatrix}
\end{equation}
to $|\psi\rangle $ we get 
\begin{equation}
H |\psi\rangle =|0\rangle .
\label{0}
\end{equation}
Hence, according to the Born rule if we measure a qubit in state
$H|\psi\rangle$, i.e.~after applying the Hadamard, we get $|0\rangle$ with probability 1. 


However, suppose we assume that, \textit{before} the
measurement, the qubit in state
$|\psi\rangle$ corresponds to being in state $|0\rangle$ with probability
$\smfrac{1}{2}$
and $|1\rangle$ with probability
$\smfrac{1}{2}$,
then the action of $H$ on $|\psi\rangle$ produces either
$\smfrac{1}{\sqrt{2}}(|0\rangle + |1\rangle )$ or 
$\smfrac{1}{\sqrt{2}}(|0\rangle - |1\rangle )$, again with equal probability, 
so a subsequent measurement of the qubit would give $|0\rangle$ or $|1\rangle$, again with equal
probability. This is in contradiction to Eq.~\eqref{0}, which states that
actually the
measurement would give $|0\rangle$ with probability 1. Hence we can \textit{not}
assume that state $|\psi\rangle$ in Eq.~\eqref{super_EPR} corresponds to its being in
$|0\rangle$ with probability
$\smfrac{1}{2}$
and $|1\rangle$ with probability
$\smfrac{1}{2}$ \textit{before} the measurement, even though this is the
\textit{result}
of the measurement. In other words the
description of the world provided by
quantum mechanics does not have objective reality.


One person who did not like
that quantum mechanics describes a world without objective reality
(and that quantum mechanics involves probabilities at a fundamental level),
was
Albert Einstein\index{Einstein, Albert}\footnote{He
reputedly claimed to Niels Bohr\index{Bohr, Niels} that ``God does
not play dice with the universe". Bohr's reported reply, which may be
apocryphal, was ``Albert, you shouldn't tell God what to do".}. In 1935 he
wrote a famous paper with Podolsky and Rosen (now called
EPR),\index{EPR}\index{Einstein, Podolsky, Rosen|see{EPR}}
in which they simply \textit{asserted} that
nature has the property of objective reality. According to this picture of the
world, the reason that, in general, measurements do not
give a definite answer but give different results with various probabilities,
is that quantum mechanics, as we have it, is \textit{incomplete}. Rather, there is a
deeper level of structure, which we don't have access to at present, with extra,
\index{hidden variable theories}
hidden, variables, such that if we could access those variables, the
measurement would be deterministic and would just reveal the state of the
system which existed previously, i.e.~we would have objective reality. The fact that measurements on a quantum state do not give a unique result is, in this picture, because the hidden variables have different values when the different measurements are done. 

The classical, EPR picture is called \textit{local realism}:
\begin{enumerate}
\index{local realism}
\item \textbf{Realism}. The measured values of each particle are objectively real. They
have definite values before measurement and irrespective of whether or not a
measurement is made.
\item \textbf{Locality}. A measurement of $A$ does not affect $B$
instantaneously. More precisely, the measurement of $A$ has no effect
on $B$ if $A$ and $B$ are spatially
separated, i.e.~$|\vec{r}_A - \vec{r}_B| > c t$ where $t$ is the time between
measurements and $c$ is the speed of light. This is just special relativity,
\index{special relativity}
one of Einstein's greatest insights. 
\end{enumerate}


\section{An EPR Experiment}
\index{EPR}
In this chapter we will describe an
experiment in which quantum mechanics gives different results from
\textit{any} local realistic theory. Such experiments have been done and found
to be in agreement with quantum mechanics and in disagreement with local
realism.

EPR examined a thought experiment with entangled particles. We shall consider
a simpler version of the EPR thought experiment due to Bohm\index{Bohm, David}. For this
experiment we will derive a condition (an inequality) which any theory with
local realism must have, but which is violated by quantum mechanics.  This
is one of many inequalities of a similar nature, initially
discovered by John Bell. \index{Bell, John} Hence they are known as Bell's
inequalities.\index{Bell's theorem}

\begin{figure}[tbh]
\begin{center}
\includegraphics[width=13cm]{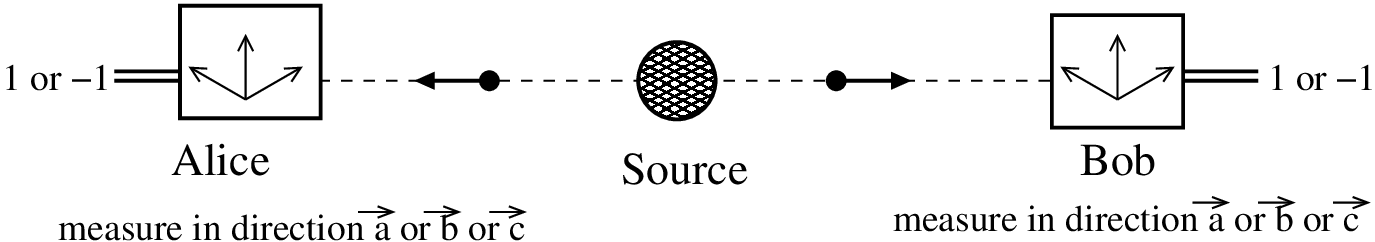}
\caption{
Sketch of the experimental setup for the version of the EPR experiment
discussed in the text. The source emits pairs of qubits (in practice photons) in the state
$|\psi\rangle = {1\over\sqrt{2}}(|01\rangle - |10\rangle)$ given in 
Eq.~\eqref{epr:psi}. For each pair Alice and Bob decide independently and randomly
which of the three non-orthogonal directions,
$\vec{a}, \vec{b}$ or $\vec{c}$ to measure along. The
result in each case is $+1$ or $-1$. The double lines indicate that the result of the measurement is a \textit{classical} bit.
\label{eprb}
}
\end{center}
\end{figure}

We suppose that an experimenter prepares pairs of 2-state particles
(qubits) in the following entangled Bell state\index{Bell state}
\begin{equation}
|\psi\rangle = {1\over \sqrt{2}}\left(|01 \rangle - |10\rangle \right).
\label{epr:psi}
\end{equation}
In experiments the qubits will be photons.
He sends one particle of the pair to Alice and the other, in the opposite direction, to
Bob, see Fig.~\ref{eprb}. He then repeats this for many pairs.
Suppose that Alice and Bob measure the particles in the computational ($Z$) basis. If Alice
measures $|0\rangle$ (for which the eigenvalue of $Z$ is $+1$) then Bob must measure
the opposite, i.e. $|1\rangle$ (for which the eigenvalue of $Z$ is $-1$).

Now consider a general basis. As discussed in Chapter \ref{ch:qubits} a general
qubit state $|0_{\hat{n}}\rangle$ is characterized by two parameters, $\theta$ and $\phi$, which are
the polar and azimuthal angles of a point in direction $\hat{n}$ on the unit sphere,
known, in this
context, as the Bloch sphere,\index{Bloch sphere} see Fig.~\ref{bloch}. The state on the antipodal point
on the sphere is denoted by $|1_{\hat{n}}\rangle$. 
The connection between $|0_{\hat{n}}\rangle$ and
$|1_{\hat{n}}\rangle$ and the basis states in the computational basis, $|0\rangle$ and
$|1\rangle$,
is given by Eqs.~\eqref{01n}.


It is shown in Eq.~\eqref{psi_n} in Appendix \ref{appa}, that
Eq.~\eqref{epr:psi} can equivalently be written as
\begin{equation}
|\psi\rangle = {1\over \sqrt{2}}\left(|0_{\hat{n}}\, 1_{\hat{n}}\,\rangle -
|1_{\hat{n}}\, 0_{\hat{n}}\, \rangle \right),
\label{psin}
\end{equation}
ignoring an overall phase, for \textit{any} direction $\hat{n}$.
Hence the state in Eq.~\eqref{epr:psi} has the interesting property that Alice and Bob
will always get opposite results as long as they measure in the same
basis\footnote{Note: when we say ``measure the qubit in the $\hat{n}$ basis'' we mean
measure $\vec{\sigma}\cdot\hat{n}$, where, as discussed in
Sec.~\ref{sec:mat_diag}, the
$\sigma_\alpha, (\alpha= x, y, z)$ are just another notation for the Pauli
operators
$X, Y$ and $Z$.\label{epr:fn1}} \textit{no
matter what that basis is}.

The results of the measurements of Alice and Bob are therefore strongly
correlated.  Of course, one can also have correlations between experimental
results in classical systems. However, we will show below that the quantum correlations
in entangled states like that in Eq,~\eqref{psin} 
are different from classical correlations.

In the experiment that we will consider, Alice and Bob each choose to
measure in one of three\footnote{In the simpler setup of two directions, one
finds that there is no incompatibility between quantum mechanics and local
realism. Three directions is the minimum needed to derive an inequality which
is violated by quantum mechanics.} distinct, non-orthogonal directions
$\vec{a}, \vec{b}$ and $\vec{c}$. Every time they receive a particle they
separately choose at random one of these three directions and record whether
they get $+1$ or $-1$

The timing of the measurements is
important. \textbf{They must be done in a causally disconnected manner} so
information about the direction that Alice,
for example, has chosen can not have reached Bob when he
makes his measurement, and vice versa.

The setup is sketched in Fig.~\ref{eprb}.

\section{Bells' Inequality}
\index{Bell's inequality|see{Bell's theorem}}\index{Bell's theorem}
If Alice and Bob choose the same
direction we know that they will get opposite results. 
Next consider in some
detail what happens when Alice and Bob do not choose the same direction.

Firstly let us see what happens in a \textbf{classical picture with objective reality}. 

The qubits then
have a well defined state prior to the measurement. The reason that we don't
always get the same result for measurements along a given direction must be
that the qubit pairs are not all emitted in the same state each time. Rather,
each possible result of the measurements corresponds to a particular type of
initial state.
There are three directions,
for each of which Alice and Bob get one of two possible results. Let's
first consider the results that Alice might get.
For each of the three directions she gets one of two possible results, $\pm 1$.
With objective reality, the result of the measurement is pre-ordained before
the measurement takes place, it just depends on the state of the photon.
Furthermore, even though only one measurement direction is used for each
photon, assuming
objective reality it makes sense to talk about the results that Alice \textit{would}
have got if she had measured in one of the other directions. For example, there are photons
where Alice would find $+1, +1, +1$ in the three directions.  Let call
these photons type 1.  Since there are $2^3 = 8$ possible results for the three
directions, 
there are eight possible types of photon, as far as Alice 
is concerned.

Now we incorporate Bob's results with those of Alice. Assuming that Bob's
results are not affected by the measurement direction chosen by Alice,
which is the case if the measurements are done in a causally disconnected
manner, Bob's
results are also determined only by the state of his photon when emitted by the
source.
In this case, if he and Alice measure in the
same direction we know that they must get opposite results\footnote{The
state $|\psi\rangle$ in Eq.~\eqref{epr:psi}, is
known as a ``spin singlet" state in the physics literature and has zero total
spin angular momentum. Assuming that the initial state of the
source, before the
qubits are omitted, has zero angular momentum, then conservation of angular
momentum \textit{requires} that the qubits be emitted in state $|\psi\rangle$ and therefore
that Alice and Bob must get opposite results if they measure in the same
direction.}.
For example if Alice receives a photon which would give $+1, +1, +1$ in the
three directions, then Bob's photon would give $-1, -1, -1$.
Hence, including
both Alice
and Bob's results, there are still only eight possible types of photon pair that
we need consider, and these are shown in Table \ref{tab1}.
For the $i$-th type, $N_i$ pairs will be generated where
\begin{equation}
N = \sum_{i=1}^8 N_i,
\end{equation}
is the total number of pairs.


\begin{table}
\begin{center}
\begin{tabular}{|c|ccc|ccc|}
\hline
 & \multicolumn{3}{c|} {Alice} & \multicolumn{3}{c|} {Bob} \\
\hline
Population & $\vec{a}$ & $\vec{b}$ & $\vec{c}$ & $\vec{a}$ & $\vec{b}$ &
$\vec{c}$ \\
\hline
$N_1$      & + & + & + & $-$ & $-$ & $-$ \\
$N_2$      & + & + & $-$ & $-$ & $-$ & + \\
$N_3$      & + & $-$ & + & $-$ & + & $-$ \\
$N_4$      & + & $-$ & $-$ & $-$ & + & + \\
$N_5$      & $-$ & + & + & + & $-$ & $-$ \\
$N_6$      & $-$ & + & $-$ & + & $-$ & + \\
$N_7$      & $-$ & $-$ & + & + & + & $-$ \\
$N_8$      & $-$ & $-$ & $-$ & + & + & + \\
\hline
\end{tabular}
\caption{The eight types of qubit pairs give different results
when measured along the $\vec{a}, \vec{b}$ and $\vec{c}$ directions. Note that
Alice and Bob get opposite results if they measure in the same direction, so
Bob's side of the table is precisely the opposite of Alice's. Hence there are
$2^3$ possible sets of outcomes.
\label{tab1}
}
\end{center}
\end{table}

Let us discuss next some examples taken from
Table \ref{tab1}.  For a qubit pair in population 4, Alice will get $+1$ if she
measures in direction $\vec{a}$, and Bob will get $+1$ if he measures in
direction $\vec{b}$. Similarly for population 7, Alice will get $-1$ if she
measures in direction $\vec{a}$ and Bob will get $+1$ if he measures in
direction $\vec{b}$. In all cases, if Alice and Bob measure in the same
direction they get opposite results.


We now make some simple observations. (Each observation is simple but one needs to
focus to follow the thread of the argument to the end.)  Clearly $N_i \ge 0$, so it must be true
that
\begin{equation}
{N_3 + N_4 \over N} \le {N_2 + N_4 \over N} + {N_3 + N_7 \over N},
\label{ineq1}
\end{equation}
since $N_2$ and $N_7$, which can not be negative, have been added on the RHS.

\begin{table}
\begin{center}
\begin{tabular}{|c|c|c|}
\hline
 & \multicolumn{1}{c|} {Alice} & \multicolumn{1}{c|} {Bob} \\
\hline
Population & $\vec{a}$ & $\vec{b}$ \\
\hline
$N_1$      & +   & $-$  \\
$N_2$      & +   & $-$  \\
$N_3$      & +   & +    \\
$N_4$      & +   & +    \\
$N_5$      & $-$ & $-$  \\
$N_6$      & $-$ & $-$  \\
$N_7$      & $-$ & +    \\
$N_8$      & $-$ & +    \\
\hline
\end{tabular}
\caption{The columns of Table \ref{tab1} for the case 
when Alice measures along $\vec{a}$ and Bob
along $\vec{b}$.
\label{tab2}
}
\end{center}
\end{table}

\begin{itemize}
\item $\boldsymbol{(N_3, N_4)}$
Let's suppose that Alice measures along $\vec{a}$ and Bob along $\vec{b}$. The
appropriate columns of Table \ref{tab1} are collected in Table \ref{tab2} for
clarity.
According to Table
\ref{tab2} only for populations $3$ and $4$ would Alice and Bob both get $+1$.
None of the other populations give this. Hence, among the
times that Alice measures along $\vec{a}$ and Bob along $\vec{b}$, the
probability that they both get $+1$ is $(N_3 + N_4)/N$. Let's call this 
$P(+\vec{a}; +\vec{b})$, in which the first argument refers to Alice and the
second to Bob, i.e.
\begin{equation}
{N_3 + N_4 \over N} = P(+\vec{a}; +\vec{b}).
\label{34}
\end{equation}

\item $\boldsymbol{(N_2, N_4)}$. Following similar arguments, only for populations $2$ and $4$ would Alice
get $+1$ measuring along $\vec{a}$ and Bob get $+1$ measuring along $\vec{c}$.
Hence
\begin{equation}
{N_2 + N_4 \over N} = P(+\vec{a}; +\vec{c}).
\label{24}
\end{equation}

\item $\boldsymbol{(N_3, N_7)}$. Similarly, only for populations $3$ and $7$ would Alice
get $+1$ measuring along $\vec{c}$ and Bob get $+1$ measuring along $\vec{b}$.
Hence
\begin{equation}
{N_3 + N_7 \over N} = P(+\vec{c}; +\vec{b}).
\label{37}
\end{equation}
\end{itemize}

Combining Eqs.~\eqref{ineq1}--\eqref{37}, we have\footnote{Recall what we mean
by these probabilities. $P(+\vec{a}; +\vec{b})$, for example, means that,
\textit{out
of the times when Alice measures along $\vec{a}$ and Bob measures along
$\vec{b}$}, this is the probability that they both get $+1$. The sum of the
probabilities for the different measurement results for these fixed directions
must add to $1$, i.e.~$P(+\vec{a}; +\vec{b}) + P(+\vec{a}; -\vec{b}) +
P(-\vec{a}; +\vec{b}) + P(-\vec{a}; -\vec{b}) = 1$. }
\begin{equation}
P(+\vec{a}; +\vec{b}) \le P(+\vec{a}; +\vec{c}) + P(+\vec{c}; +\vec{b}) \, .
\label{ineq2}
\end{equation}
In the simple case that all the populations are equal, each probability is $1/4$ so the
inequality is trivially satisfied. Equation \eqref{ineq2} is an example of a
Bell's inequality\index{Bell's theorem}. \textbf{It is satisfied by any theory with local realism.}
Note that there is nothing sophisticated about this Bell's inequality; it is
just bookkeeping. I emphasize that
Eq.~\eqref{ineq2} has nothing to do with quantum mechanics.  In fact, we will
now see that it is \textit{violated} by quantum mechanics for a broad range of
measurement directions $\vec{a}, \vec{b}, \vec{c}$.

We therefore now consider what \textbf{quantum mechanics} has to say. 

The 2-qubit state generated by the
source is
given by Eq.~\eqref{psin} for any direction $\hat{n}$, where
$|0_{\hat{n}}\rangle$ and $|1_{\hat{n}}\rangle$ are given by Eqs.~\eqref{epr:01n}.
We take the $\theta = \phi = 0$ direction to be that of $\vec{a}$,
so we write
\begin{equation}
|\psi\rangle = {1 \over \sqrt{2}} \left( |0_{\vec{a}}\rangle_1
|1_{\vec{a}}\rangle_2  - |1_{\vec{a}}\rangle_1  |0_{\vec{a}}\rangle_2\, \right),
\label{psib}
\end{equation}
where we indicate on the RHS which qubit is meant (1 for Alice's and 2 for Bob's).

We now compute $P(+\vec{a}; +\vec{c})$ according to quantum mechanics.
We need the probability amplitude for the state in Eq.~\eqref{psib} to have
eigenvalue $+1$ along $\vec{a}$ for Alice and eigenvalue $+1$ along $\vec{c}$
for Bob, i.e.~$|0_{\vec{a}}\rangle_1 |0_{\vec{c}}\rangle_2$.
Hence, to get $ P(+\vec{a}; +\vec{c})$ we compute first the amplitude
\begin{equation}
\big(\,\, _1\langle 0_{\vec{a}}|\,\, _2\langle 0_{\vec{c}}|\,\,\big)\, | \psi\rangle = {1 \over \sqrt{2}} \Big(\,
\langle 0_{\vec{a}} | 0_{\vec{a}}\rangle_1 \langle 0_{\vec{c}} | 1_{\vec{a}}\rangle_2 -
\langle 0_{\vec{a}} | 1_{\vec{a}}\rangle_1 \langle 0_{\vec{c}} | 0_{\vec{a}}\rangle_2 
\,\Big),
\end{equation}
Now $\langle 0_{\vec{a}} | 0_{\vec{a}}\rangle = 1$ and $\langle
0_{\vec{a}} | 1_{\vec{a}}\rangle = 0$, so
\begin{equation}
\langle 0_{\vec{a}}\, 0_{\vec{c}} | \psi\rangle =
{1 \over \sqrt{2}} \langle 0_{\vec{c}} | 1_{\vec{a}}\rangle .
\end{equation}
If $\vec{c}$ is at angles $(\theta_{ac}, \phi_{ac})$ relative to $\vec{a}$,
then, according to Eq.~\eqref{epr:0n},
\begin{equation}
{1 \over \sqrt{2}} \langle 0_{\vec{c}} | 1_{\vec{a}}\rangle  =
{1 \over \sqrt{2}} e^{i\phi_{ac}} \sin{\theta_{ac}\over 2} ,
\end{equation}
so
\begin{equation}
P(+\vec{a}; +\vec{c}) =
\left| \langle 0_{\vec{a}}\, 0_{\vec{c}} | \psi\rangle \right|^2 
= {1 \over 2} \left|e^{i\phi_{ac}} \sin{\theta_{ac}\over 2} \right|^2 
= {1\over 2} \sin^2\left({\theta_{ac}\over 2}\right) .
\label{Pac}
\end{equation}

We recall that out of the times when Alice measures along $\vec{a}$ and Bob measures along $\vec{c}$,
this is the probability that they both get $+1$. For these same directions
there are three other possibilities. It is straightforward to check that
$P(-\vec{a}; -\vec{c}) = P(+\vec{a}; +\vec{c}) $, and a calculation shows that
\begin{equation}
P(+\vec{a}; -\vec{c}) = P(-\vec{a}; +\vec{c}) = {1\over 2}
\cos^2\left({\theta_{ac}\over 2}\right) .
\end{equation}
Hence the sum of the probabilites for the four different ($\pm 1$) results
when Alice measures along $\vec{a}$ and Bob measures along $\vec{c}$ adds up to
$1$, i.e.
\begin{equation}
P(+\vec{a}; +\vec{c}) + P(+\vec{a}; -\vec{c}) +  P(-\vec{a}; +\vec{c}) +
P(-\vec{a}; -\vec{c}) = 1,
\end{equation}
as required. 

A further check on 
Eq.~\eqref{Pac} is that it predicts $P(+\vec{a}; +\vec{c})\to 0$ if $\vec{a}$ and $\vec{c}$ are 
in the same direction. This result is correct because when Alice and Bob
measure in the same direction they must get \textit{different} results because
of the nature of $|\psi\rangle$, see Eq.~\eqref{psin}.

Similarly one has
\begin{align}
P(+\vec{a}; +\vec{b}) &= {1\over 2} \sin^2\left({\theta_{ab}\over 2}\right), \\
P(+\vec{c}; +\vec{b}) &= {1\over 2} \sin^2\left({\theta_{cb}\over 2}\right).
\end{align}
Hence Bell's inequality,\index{Bell's theorem} Eq.~\eqref{ineq2}, when applied to quantum
mechanics, gives
\begin{equation}
\sin^2\left({\theta_{ab}\over 2}\right) \le
\sin^2\left({\theta_{ac}\over 2}\right) +
\sin^2\left({\theta_{cb}\over 2}\right) .
\label{ineq3}
\end{equation}
As we shall now see, it is easy to find cases where this is violated.

\begin{figure}[htb]
\begin{center}
\includegraphics[width=7.5cm]{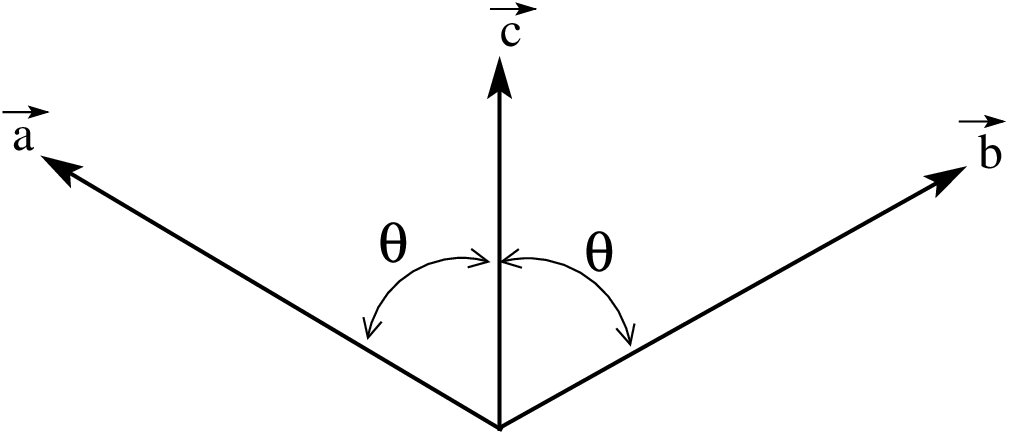}
\caption{
A possible choice of directions for which the Bell's inequality in
Eq.~\eqref{ineq3} is violated.
\label{angles}
}
\end{center}
\end{figure}

\begin{figure}[htb]
\begin{center}
\includegraphics[width=7.5cm]{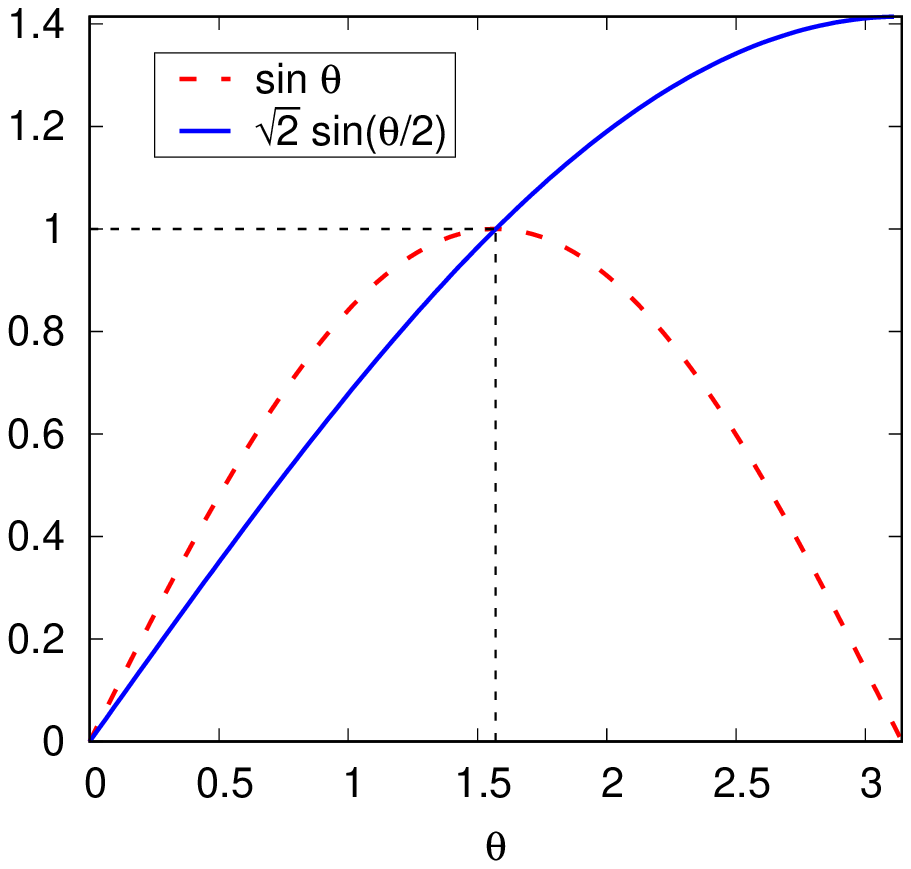}
\caption{
A graph showing that the inequality in Eq.~\eqref{sin_ineq} is violated for $0 <
\theta < \pi /2$.
\label{sins}
}
\end{center}
\end{figure}

Consider the situation in Fig.~\ref{angles} where
$\theta_{ac} = \theta_{cb} = \theta$, so $\theta_{ab}= 2\theta$, and take $\theta = \pi/3$. We have
\begin{equation}
\sin^2\left({\theta_{ac}\over 2}\right) = \sin^2\left({\theta_{cb}\over 2}\right) =
\sin^2\left({\theta\over 2}\right) = \sin^2({\pi\over 6})
= {1\over 4},
\end{equation}
and
\begin{equation}
\sin^2\left({\theta_{ab}\over 2}\right) = \sin^2 \theta =
\sin^2\left({\pi\over 3}\right) = {3 \over 4}. 
\end{equation}
Hence the LHS of Eq.~\eqref{ineq3} is $3/4$ while the RHS is $1/2$ so the
inequality is violated. For general $\theta$ in Fig.~\ref{angles}, the inequality in Eq.~\eqref{ineq3}
can be written
\begin{equation}
\sin \theta \le \sqrt{2}\sin\left({\theta \over 2}\right) ,
\label{sin_ineq}
\end{equation}
which is violated for the broad range $0 < \theta < \pi/2$, as shown
graphically in Fig.~\ref{sins}.

\section{Summary}

\textbf{We have seen that
quantum mechanics violates Bell's inequalities}.\index{Bell's theorem}
These inequalities
are satisfied by any
theory with local realism. Experiments along the lines of that sketched in Fig.~\ref{eprb} have been done, using
polarized photons.
\index{photon}
\index{photon!polarization}
\textbf{These experiments agree with quantum mechanics and disagree with local
realism.} See \url{https://physics.aps.org/articles/v8/123} for a brief
discussion of these experiments. Among the different experiments
there are variations in the initial state of
the entangled qubits and in which Bell's inequality is being tested, but they are
all equivalent. The more sophisticated experiments 
choose (randomly) the polarization directions while the photons are in flight.
This makes it impossible for the emitted photons to be affected by
the chosen orientations of the polarizers.
Similarly, the polarizer directions are set at times such
that
that information about the direction of one polarizer has not had time to reach the other
polarizer when it performs its measurement.
(Note that information can not
travel faster than the speed of light.)
Features of the experiment like these are 
necessary to show that no \textit{local} hidden variable theory can explain
the data. 
\index{hidden variable theories}

\label{no_infl}
Bell's inequalities characterize quantum correlations between \emph{two}
entangled qubits, which are different from classical correlations.  Very
recently non-classical correlations, \emph{distinct} from those of Bell, have
been found in experiments with three sources of pairs of entangled photons and
three detectors in the shape of a triangle, see
\url{https://physics.aps.org/articles/v12/106}.  Thus the study of
non-classical correlations in quantum mechanics, stimulated by EPR in the
1930s, made precise by Bell\index{Bell, John} in the 1960s, and studied experimentally since the 1970s, remains an active field up to the present day.

\index{non-local theories}
Although the experimentally found violations of Bell's inequalities rule out
local theories with objective reality, they do not, in principle, rule out
\textit{non-local}\footnote{The term non-local refers to information propagating faster than
the speed of light.} theories with objective reality. However, these would
violate special relativity. Hence
very few physicists think that a non-local theory of quantum mechanics
will turn out to be the correct
theory of nature.

\index{instantaneous transfer of information}
In an EPR-like \index{EPR} experiment the entangled state changes when one
qubit is measured. We can ask whether any information is instantaneously
transmitted to the other qubit at the moment of measurement. Since the two
qubits in an entangled state are correlated, naively one
might imagine that this occurs. If so, special
relativity, one of the cornerstones of modern physics, would be violated.
Fortunately, no information is transmitted at the moment of
measurement, as we show in Appendix \ref{sec:causal}, so special relativity
\textit{is} preserved.
\index{special relativity}

To conclude, we see that quantum mechanics is \textbf{strange}:
\begin{itemize}
\item Unlike in classical physics, probabilities enter in a
\textit{fundamental} way.
\index{objective reality}
\item Unlike in classical physics, we do not have objective reality. Reality
is an \textit{emergent} concept for bigger systems when we go over to a
description in terms of classical physics.
\end{itemize}

Many physicists feel uncomfortable with these aspects of quantum mechanics,
and hope that a better insight will emerge.  But, in the 90 years since the
EPR\index{EPR} paper this has not happened, so we will probably have to continue living with
the strange world of quantum mechanics as we now understand it.

Can we use the differences between the strange quantum world and the
familiar classical world to do more efficient computation, at least for some
problems? This question will be the focus of the rest of the course.

\hrulefill
\section*{Problems}
\input{hw_ch6.tex}

\begin{center}
{\Large\bf Appendices}
\end{center}

\begin{subappendices}
\section{The 2022 Physics Nobel Prize}
\label{sec:nobel}
\index{Nobel Prize, 2022}
In 2022 the Royal Swedish Academy of Sciences awarded the Nobel Prize in
Physics to Alain Aspect, John F.~Clauser and Anton Zeilinger 
\begin{quotation}
\noindent``for
experiments with entangled photons, establishing the violation of Bell
inequalities and pioneering quantum information science".
\index{Bell's theorem}
\end{quotation}
The announcement can
be seen at\\
\href{https://www.nobelprize.org/uploads/2022/10/advanced-physicsprize2022.pdf}
{\texttt{\textcolor{blue}{https://www.nobelprize.org/uploads/2022/10/advanced-physicsprize2022.pdf}}}

John Bell\index{Bell, John} himself died unexpectedly of a cerebral hemorrhage in 1990.
Apparently he had been nominated for the Nobel prize that year. Whether or not
he would have received it then, he would certainly have received it at some
point had he not died prematurely.

Here is a summary of the contributions made by the three awardees gleaned from
the Nobel announcement.

\noindent\textbf{John Clauser}\\
\index{Clauser, John}
The first experiment to test Bell inequalities was
performed by Stuart Freedman (now deceased) and Clauser, who found a violation
of a version of the Bell inequality proposed earlier by John Clauser, Michael Horne, Abner
Shimony and Richard Holt (CHSH). The results agreed well with quantum
mechanics.

\noindent\textbf{Alain Aspect}\\
\index{Aspect, Alain}
An assumption in the Bell inequalities is that the two observers, Alice and
Bob, make random choices of what to measure, independent of each other. For
this to be true, one must make sure that Alice cannot send a message to Bob
about which polarization direction she will measure which Bob receives before
he chooses his polarization direction.\index{photon!polarization} In other words, Alice will not
influence Bob's choices,
see the discussion at the top of p.~\pageref{no_infl}.
Assuming that special relativity is correct, this
locality condition amounts to making sure that such a message would have to
travel with a speed greater than that of light.
Alain Aspect was the first to
design an experiment to overcome this locality loophole. \index{locality
loophole} Aspect ensured the
independence of Alice and Bob by using polarization settings that changed
randomly during the time of flight of the photons between the detectors. His
results agreed well with quantum mechanics and violated the relevant Bell
inequality.

\noindent\textbf{Anton Zeilinger}\\
\index{Zeilinger, Anton}
We have discussed in this course that an unknown, arbitrary quantum states can not be copied,
i.e. cloned. However, as we will see in Chapter \ref{ch:qkd}, it is possible,
using entanglement, to ``teleport"\index{teleportation} an arbitrary state from one position to
another, as long as the original copy is destroyed. Zeilinger's group was one
of the first to demonstrate teleportation. It has been possible to create
entanglement over very large distances, and a Chinese group, in collaboration
with Zeilinger, was able to distribute entanglement between China and
Australia using a satellite. 

The locality loophole in experiments to test Bell inequalities was mentioned above in the
context of Aspect's work. This loophole was \textit{largely} eliminated by Aspect, but  in his
experiment the distance between the polarizers was too small to allow for
\textit{truly} random settings. Later, Zeilinger's group was able to test the
inequality under \textit{strict} local conditions with the observers separated by no
less than 400~m.

Another loophole is the ``detection loophole", \index{detection loophole} which arises
because no detector has 100\% efficiency, so a quantum skeptic could argue that
the lost photons might conspire to give a fake violation of a Bell inequality.
\index{Bell's theorem}
While this possibility seems unlikely it is important to rule it out. The
detection loophole was first closed in an experiment using trapped ions rather
than photons. However, in these systems one could not close the locality
loophole. It was only relatively recently, in the years 2015-17 that several
groups, including that of Zeilinger, managed to simultaneously close both the
locality and detection loopholes. 

\section{Information does not propagate faster than the speed of light}
\label{sec:causal}

Consider a pair of entangled qubits $A$ and $B$ which are widely separated.
If a measurement is done on qubit 
$B$ then the state of the system changes, the final state depending on the
result of the measurement. This change in state happens instantaneously. Does
this mean that \textit{information} is transmitted instantaneously to qubit $A$? If so,
this would violate special relativity.  We shall now see that this is not the
case, no information is transferred at the moment of measurement, and
therefore quantum
mechanics does not violate special relativity. 

\index{density matrix}
Since qubits $A$ and $B$ are entangled we have to describe qubit $A$ by a
density matrix. To see its form we separate out the parts of the entangled
state corresponding to $B$ being in state $|0\rangle$ and $B$ being in state $|1\rangle$. Referring 
to our discussion of the generalized Born rule in Sec.~\ref{sec:gen}, we
\index{generalized Born rule}
write the state of the two qubits \textit{before} the measurement as
\begin{equation}
|\psi_{AB}\rangle = \alpha |\psi_{0, A}\rangle |0_B\rangle +
\beta |\psi_{1, A}\rangle |1_B\rangle,
\label{psi_before}
\end{equation}
where $|\psi_{0, A}\rangle$ and $|\psi_{1, A}\rangle$ are normalized 
(but not, in general, orthogonal) states of
qubit $A$, and
$|\alpha|^2 + |\beta|^2 = 1$. As stated in Eq.~\eqref{rhoAB} in Sec.~\ref{sec:dm},
the density matrix of
the two qubits in a well-defined quantum state is
\begin{align}
\rho^{AB} &= |\psi_{AB}\rangle \langle \psi_{AB}| \nonumber \\
&= \left(\, \alpha |\psi_{0, A}\rangle |0_B\rangle +
\beta |\psi_{1, A}\rangle |1_B\rangle\,\right)\,
\left(\, \alpha^* \langle \psi_{0, A}| \langle 0_B| + \beta^* \langle \psi_{1, A}|
\langle 1_B|\, ]\right) ,
\end{align}
and the density matrix of qubit $A$ alone is
\begin{align}
\rho^A &= \Tr_B\, \rho^{AB} \nonumber \\
&= \langle 0_B| \rho^{AB} |0_B \rangle
+ \langle 1_B| \rho^{AB} |1_B \rangle \nonumber \\
&= |\alpha|^2 |\psi_{0, A}\rangle \langle \psi_{0, A}| 
+ |\beta|^2 |\psi_{1, A}\rangle \langle \psi_{1, A}| .
\label{rhoA_before}
\end{align}
Note that Eq.~\eqref{rhoA_before} is a representation of the density matrix in terms of 
the non-orthogonal
states $|\psi_{0, A}\rangle$ and $|\psi_{1, A}\rangle$.
Another example involving non-orthogonal states was described in Sec.~\ref{sec:non_ortho}.
\index{non-orthogonal states}
As discussed in Sec.~\ref{rho_sub}, Eq.~\eqref{rhoA_before} implies that that 
qubit-$A$ is in state $|\psi_{0, A}\rangle$ with probability $|\alpha|^2$ and
is in state $|\psi_{0, B}\rangle$ with probability $|\beta|^2$.

\index{density matrix}
Now consider the situation \textit{after} the measurement on qubit $B$. According to
the generalized Born rule discussed in Sec.~\ref{sec:gen}, for the state of
the combined $AB$ system in Eq.~\eqref{psi_before},
there is
probability $|\alpha|^2$ that qubit $B$ is measured to be in state $|0_B\rangle$ while qubit $A$ is
left in state $|\psi_{0, A}\rangle$, and there is probability $|\beta|^2$ that qubit $B$
is measured to be in state $|1_B\rangle$ while qubit $A$ is left in
in state $|\psi_{1, A}\rangle$.
For qubit $A$
this situation
is exactly the same as
we found \textit{before} the measurement,
see Eq.~\eqref{rhoA_before}.

Hence the density matrix for qubit $A$, which determines the probabilities of
results of subsequent measurements on $A$, is \textit{unchanged} by the
measurement of the distant qubit $B$, even though the two qubits are
entangled. Thus, although our \textit{description} of the state of the two
qubits does change instantaneously at the moment of measurement,
\textit{information} is not propagated instantaneously by the measurement and
so special relativity is satisfied.

\section{The spin-singlet state is isotropic}
\label{appa}

Equations~\eqref{01n} of Chapter \ref{ch:qubits} show that the eigenstate of
spin in a direction specified by polar angles $(\theta,\phi)$ with eigenvalue
$+1$ is given by
\begin{subequations}
\label{epr:01n}
\begin{equation}
|0_{\hat{n}}\rangle = \cos\smfrac{\theta}{2}\, |0\rangle + e^{i\phi}
\,\sin\smfrac{\theta}{2}\,|1\rangle \, ,
\label{epr:0n}
\end{equation}
see Fig.~\ref{bloch}.
We also showed that the eigenstate corresponding to eigenvalue $-1$ is
\begin{equation}
|1_{\hat{n}}\rangle = -\sin\smfrac{\theta}{2}\, |0\rangle + e^{i\phi}
\,\cos\smfrac{\theta}{2}\, |1\rangle \, ,
\label{epr:1n}
\end{equation}
\end{subequations}
which is the antipodal point where $\theta \to \pi - \theta, \phi \to \phi +
\pi$, see Eq.~\eqref{1n}.

From Eqs.~\eqref{epr:0n} and \eqref{epr:1n}, we see that the
unitary matrix which transforms from the $Z$ basis to the $\hat{n}$ basis
is
\begin{equation}
U = 
\begin{pmatrix}
\cos{\theta\over 2} & e^{i\phi}\, \sin{\theta\over 2} \\
-\sin{\theta\over 2} & e^{i\phi}\, \cos{\theta\over 2}
\end{pmatrix}.
\end{equation}
The inverse transformation is given by $U^{-1}$, but since $U$ is unitary we have
\begin{equation}
U^{-1} = U^\dagger \equiv \left(U^T\right)^\star =
\begin{pmatrix}
\cos{\theta\over 2} & -\sin{\theta\over 2} \\
e^{-i\phi}\,\sin{\theta\over2} & e^{-i\phi}\, \cos{\theta\over 2}
\end{pmatrix},
\label{Udagger}
\end{equation}
so
\begin{subequations}
\begin{align}
|0\rangle &= \cos\smfrac{\theta}{2}\, |0_{\hat{n}}\rangle - 
\sin\smfrac{\theta}{2} \,|1_{\hat{n}}\rangle \\
|1\rangle &= e^{-i\phi} \sin\smfrac{\theta}{2} \, |0_{\hat{n}}\rangle + e^{-i\phi}
\,\cos\smfrac{\theta}{2}\, |1_{\hat{n}}\rangle \, .
\end{align}
\end{subequations}

Hence the entangled \index{Bell state} Bell state $|\psi\rangle$ in
Eq.~\eqref{epr:psi},
(which is called the spin-singlet state in the physics literature) can be
written in the $\hat{n}$ basis as
\index{spin-singlet state}
\begin{align}
|\psi\rangle &= {1\over\sqrt2}  \left( |0 1\rangle - |1 0\rangle \right)
\label{psi_Z}\\
\begin{split}
&= {1\over\sqrt2}  \Big[
\left(\,\cos\smfrac{\theta}{2}\, |0_{\hat{n}}\rangle_1\!-\!\sin\smfrac{\theta}{2}\,|1_{\hat{n}}\rangle_1\,\right)
\left(\,e^{-i\phi}\sin\smfrac{\theta}{2} \,|0_{\hat{n}}\rangle_2\!+\!e^{-i\phi} \cos\smfrac{\theta}{2}\,|1_{\hat{n}}\rangle_2\,\right) -\\
&\qquad \left(e^{-i\phi}\sin\smfrac{\theta}{2} \,|0_{\hat{n}}\rangle_1\!+\!e^{-i\phi} \cos\smfrac{\theta}{2}\,|1_{\hat{n}}\rangle_1\,\right)
\left(\,\cos\smfrac{\theta}{2}\, |0_{\hat{n}}\rangle_2\!-\!\sin\smfrac{\theta}{2}\,|1_{\hat{n}}\rangle_2\,\right)
\Big]
\end{split} \\
&={e^{-i\phi}\over\sqrt2}\left( |0_{\hat{n}} 1_{\hat{n}}\rangle - |1_{\hat{n}} 0_{\hat{n}}\rangle \right) \, ,
\label{psi_n}
\end{align}
where, in the middle expression, we indicated by a subscript,
e.g.~$|\cdots\rangle_1$, whether the state is that of the first or second qubit.
Apart from the unimportant overall phase factor of\footnote{Note that
$e^{-i\phi}$ is just the determinant of the transformation matrix from the
computational basis to the $\hat{n}$ basis given in Eq.~\eqref{Udagger}. Quite
generally, if the ``singlet" state $|\psi\rangle$ in Eq.~\eqref{psi_Z} is
acted on by a unitary transformation $V$ then one can show that $V |\psi\rangle = \det V
|\psi\rangle$.
\index{unitary transformation}
\index{matrix!determinant}
Since $V$ is unitary its determinant can only be a pure phase.} $e^{-i\phi}$,
Eq.~\eqref{psi_n} is 
the same form that the state takes in the computational ($Z$) basis,
Eq.~\eqref{psi_Z}. Hence if two qubits in the entangled Bell state in
Eq.~\eqref{epr:psi} are observed in the same basis (see
footnote \ref{epr:fn1} on page \pageref{epr:fn1}), no matter which one, the
results of the two measurements will always be opposite, one giving $+1$ and
the other $-1$.

\end{subappendices}


%% file: hw_ch6.tex
\begin{problems}
\item
We showed in this chapter 
that, out of the times that Alice measures along $\vec{a}$ and Bob along
$\vec{c}$, the probability that they both get $+1$ is given by
$P(+\vec{a};+\vec{c}) = {1 \over 2} \sin^2(\theta_{ac}/2)$, where
$\theta_{ac}$ is the angle between the directions $\vec{a}$ and $\vec{c}$.

Perform a similar calculation to compute the probability that Alice gets $-1$
and Bob gets $+1$, which we call $P(-\vec{a};+\vec{c})$.

\item
We showed in this chapter
that the so-called ``singlet" state,
\begin{equation}
|\beta_{11}\rangle = {1\over \sqrt{2}} \left (|01\rangle - |10 \rangle\right)
,
\end{equation}
which is one of the Bell states, has the same form, apart from an unimportant
overall phase factor, in all bases.  Show that the same result is \emph{not}
true for the Bell state
\begin{equation}
|\beta_{01}\rangle = {1\over \sqrt{2}} \left (|01\rangle + |10 \rangle\right)
.
\end{equation}
\emph{Note}: like $|\beta_{01}\rangle$, the other two Bell states,
$|\beta_{10}\rangle$ and $|\beta_{00}\rangle$, also have a different form in
other bases.
\end{problems}

%% file: gates8.tex
Now, finally, we get to computation!

\bigskip
The elementary circuit elements which acts on the data in a computer are
called gates. In this chapter we will first discuss classical gates and then
go on to describe quantum gates.

\section{Classical Gates}
\label{sec:class}
\index{classical gates}

Data in a classical digital computer is in the form of bits, $x$, which take
values 0 or 1. The only operation involving a single classical bit, i.e.~the
only 1-bit classical gate, is NOT which takes 0 to 1 and vice versa.\index{NOT
gate}

Of particular interest are 2 bit gates, the most common ones being
\index{AND gate}\index{OR gate}\index{XOR gate} 
\begin{equation}
\begin{split}
\mathrm{AND}  \quad\quad\quad &
\begin{array}{l | c}
\mathrm{In} & \mathrm{Out}\\
\hline
00 & 0 \\
01 & 0 \\
10 & 0 \\
11 & 1 \\
 & \\
\end{array}
\qquad \qquad x \land y \\
\mathrm{OR}  \quad\quad\quad &
\begin{array}{l | c}
\mathrm{In} & \mathrm{Out}\\
\hline
00 & 0 \\
01 & 1 \\
10 & 1 \\
11 & 1 \\
 & \\
\end{array}
\qquad \qquad x \lor y \\
\mathrm{XOR}  \quad\quad\quad &
\begin{array}{l | c}
\mathrm{In} & \mathrm{Out}\\
\hline
00 & 0 \\
01 & 1 \\
10 & 1 \\
11 & 0 \\
 & \\
\end{array}
\qquad \qquad x \oplus y \\
\end{split}
\end{equation}

These have two input bits and one output bit. For the AND gate the result is 0
unless both inputs are 1.  For the OR gate the result is 0 unless one or both
of the inputs are 1.  The XOR gate only differs from the OR gate in giving zero 
if \textit{both} the inputs are 1. 

Note that AND gives the same results as multiplication of the bits $x y$. The
XOR operation is equivalent to addition of the bits modulo 2, i.e.~$x + y \,
(\!\!\!\mod 2)$.
To see this, note that the modulo operation gives the remainder after integer division.
For example, since $13 = (5 \times 2) + 3$ we have $13 \,(\! \mod 5) = 3$.
Referring to the XOR gate consider the case $x=y=1$, so we have $1 + 1 \,(\!\mod 2) = 0$, which
is the value of XOR in this case. It is trivial to see that XOR is also addition
modulo 2 for the other values of $x$ and $y$. For convenience of notation $x + y
\,(\!\mod 2)$ is written as $x \oplus y$. 

One can show that the AND, NOT and OR gates form a \textbf{universal set}
which means that any logical operation on a arbitrary number of bits on 
a classical computer can be
expressed in terms of these gates. Thus, classically, we only need 1-bit and
2-bit gates to perform any operation.

However, we cannot directly take over gates like AND, OR and XOR to a quantum
computer for the following reason.  A gate in a quantum computer will be
implemented by a unitary operator acting on a small number of qubits. A
unitary operator has the property that $U^{-1} = U^\dagger$. Now $U^{-1}$
performs the inverse operation, and since $U^\dagger$ is well defined the
inverse operation must exist. \textbf{Thus, quantum gates must be reversible.}

However, AND, OR and XOR can not be reversible because they have a different
number of outputs and inputs.  Suppose, for example, we know that the output
from an OR gate is 1, and want to know what is the input. We can't say
because there are three possible inputs, $01$, $10$ and $11$, which give this
output.

Thus, a major change in going from classical to quantum computing will be
having to deal with \textit{reversible} computation.  Next we will consider 
reversible \textit{classical} computation before doing the quantum case. 
\index{reversible computation}

Clearly a necessary condition for a gate to be reversible is that it has the
same number of input and output bits.  The 1-bit NOT gate has one input and
one output, and is reversible since acting twice gives back the original bit,
i.e. $(\mathrm{NOT})^2$ = IDENTITY, so (NOT)$^{-1}$ = NOT, i.e.~NOT is its own
inverse. 

We will now consider a reversible, classical, 2-bit gate, the quantum analog
of which will play an important role in quantum computing. This is the
controlled-NOT, or CNOT gate. It is similar to XOR except that it has a second
output bit, which is equal to one of the input bits, i.e.~this bit is unchanged on output.
As we shall see, this simple modification, namely keeping one of the input
bits as part of the output, suffices to make the CNOT gate a reversible
version of XOR.\index{CNOT classical gate}

One way of representing the action of CNOT is
\begin{equation}
\begin{pmatrix}
x \\ y 
\end{pmatrix}
\longrightarrow 
\begin{pmatrix}
x \\ x\oplus   y 
\end{pmatrix} .
\end{equation}

\index{control qubit}
\index{target qubit}
The first (upper) bit is called the control bit. This is unchanged by the
action of CNOT. The second (lower) bit is called the target bit, and the effect of
the XOR operation $x \oplus y$ is to flip $y$ if $x=1$ and to leave $y$ alone
if $x = 0$.  Hence, as far as the target bit is concerned, the gate is indeed a controlled NOT, since
the NOT acts if $x$, the control bit, is 1, and does not act if $x=0$. The
truth table is as follows:
\begin{equation}
\begin{array}{l l | l l }
x & y & x'  &y' \\
\hline
0 & 0 & 0 & 0 \\
0 & 1 & 0 & 1 \\
1 & 0 & 1 & 1 \\
1 & 1 & 1 & 0 \\
\end{array} .
\end{equation}

\begin{figure}[htb]
\begin{center}
\includegraphics[width=5cm]{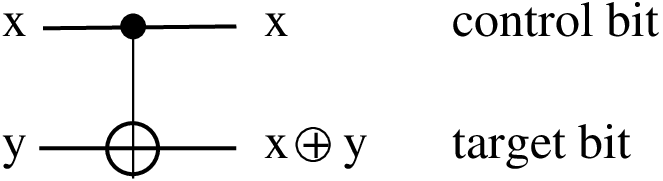}
\caption{
The CNOT gate. The input is on the left and the output on the right.
\label{CNOT2}
}
\end{center}
\end{figure}

It is useful to represent the CNOT gate by a diagram, as shown in
Fig.~\ref{CNOT2}. The input is on the left and the output on the right.
The upper line is the control bit, and has value $x$ on
input, while the lower line is the target bit and has value $y$ on input.
On output, the control qubit is unchanged and the target qubit is
the exclusive or (XOR) of $x$ and $y$.

It is easy to see that CNOT is reversible since, if we act twice, we get back the original input because
\begin{equation}
\begin{pmatrix}
x \\ y 
\end{pmatrix}
\mathrel{\stackrel{\makebox[0pt]{\mbox{\normalfont\tiny CNOT}}}{\longrightarrow}}
\begin{pmatrix}
x \\ x\oplus   y 
\end{pmatrix}
\mathrel{\stackrel{\makebox[0pt]{\mbox{\normalfont\tiny CNOT}}}{\longrightarrow}}
\begin{pmatrix}
x \\
x \oplus   x \oplus y
\end{pmatrix}
= 
\begin{pmatrix}
x \\ y 
\end{pmatrix} .
\end{equation}
The last line follows because $x \oplus x = 0 $ since $0 + 0 = 0$ and $1 + 1 =
0\, (\,\mathrm{mod}\,2)$. Thus CNOT is its own inverse. Hence, as mentioned
earlier, it can therefore be
regarded as a reversible version of XOR. 

Note that to be reversible it is not required that the inverse operator is the same as the original operator, only that the inverse operator exists.  However, it turns out that
\textit{most} quantum gates we consider will be their own inverse. 

\begin{figure}[htb]
\begin{center}
\includegraphics[width=3cm]{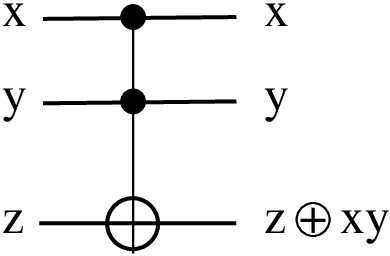}
\caption{
The Toffoli gate. This has two control bits $x$ and $y$ and one target bit $z$.
On output the control bits are unchanged and the target bit is flipped if
both control bits are 1, so $z \to z \oplus xy$.
\label{toffoli}
}
\end{center}
\end{figure}

\begin{figure}[htb]
\begin{center}
\includegraphics[width=7cm]{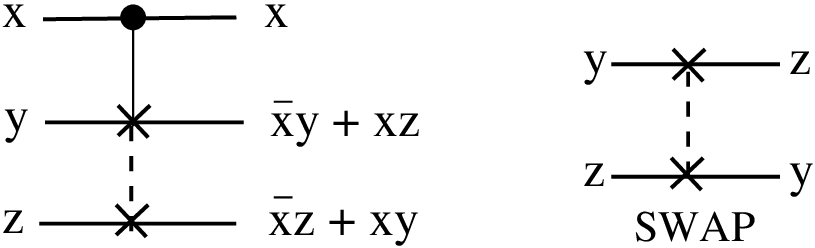}
\caption{
Left: the Fredkin gate. This is a controlled-swap gate. If the upper (control)
bit is 1 then the two lower (target) bits are swapped, and otherwise the
target bits are unchanged. $\overline{x} \equiv 1-x$ is the complement of
$x$.
Right: the elemental SWAP gate.
\label{fredkin}
}
\end{center}
\index{swap gate}
\end{figure}

\index{universal set of gates}
We mentioned above that the 1-bit (NOT) gate and a set of irreversible 2-bit gates (AND and OR) together
form universal set for a classical computer, which means that any logical operation on an arbitrary
number of bits can be constructed out of these gates. The question we now ask
is whether 1-bit and 2-bit \textit{reversible} classical gates are universal. The answer
is no. Classically one also needs a 3-bit gate such as the Toffoli gate shown
\label{Toffoli gate}\index{Toffoli gate}
in Fig.~\ref{toffoli} or the Fredkin gate\index{Fredkin gate} shown in Fig.~\ref{fredkin}.

Amazingly we shall see that 3-qubit gates are \textit{not} needed quantum mechanically.
In fact
it is possible build the Toffoli gate, for example, out of
1-qubit and 2-qubit gates, and you will go through how to do this in homework
question~\ref{qu:toffoli} in Ch.~\ref{ch:bv}.
We shall see that quantum mechanics allows for a big range of 1-qubit gates,
whereas we have already noted that classically the only 1-bit gate is NOT. It is this
wide range of possibilities for 1-qubit gates that allows us 
to construct a quantum mechanical
Toffoli gate out of 1-qubit and 2-qubit gates, whereas no such construction is
possible using classical gates.

\section{Quantum Circuits and Gates}
\begin{figure}[htb]
\begin{center}
\includegraphics[width=6cm]{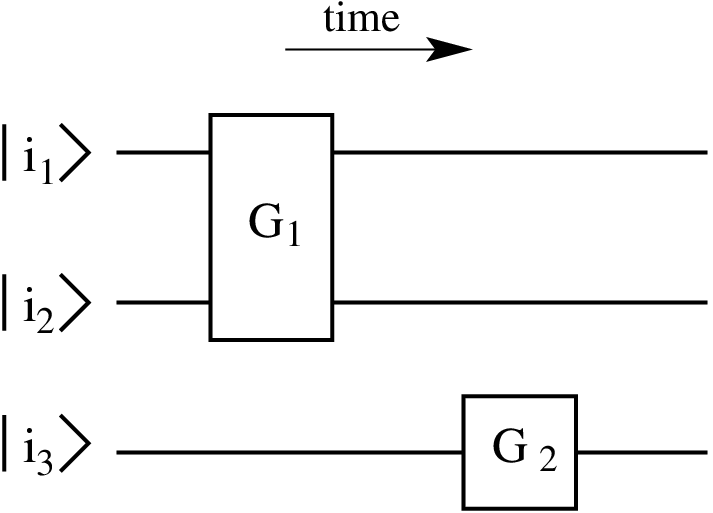}
\caption{A schematic circuit with three qubits and two gates. Time runs from left to
right. The initial state of the qubits is
$|i_1\rangle\otimes|i_2\rangle \otimes |i_3\rangle$.
\label{gen_circuit}
}
\end{center}
\end{figure}

\index{quantum gates}
Following David Deutsch\index{Deutsch, David} we represent the action of quantum gates by a circuit.
\index{circuit model}
The circuit
comprises a set of qubits in some initial state, acted on by gates and ending up in
a final state. Each qubit is represented by a line in the circuit diagram and
time runs from left to right, see e.g.~Fig.~\ref{gen_circuit}.

\index{register}
Sometimes we will indicate a set of $n$ qubits (called a register) compactly by a single line with
a slash through it as follows:
\hbox{\vspace{1.0cm}\includegraphics[width=2.5cm]{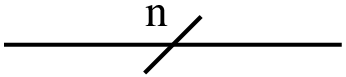}\, .}

\index{no-cloning theorem}
Quantum circuits have the following properties:
\begin{itemize}
\item There are no loops, because qubits can't go back in time.
\item Lines can't splay out (fan out) because of the no-cloning theorem.
\item Similarly lines can't merge.
\item Gates and circuits are \textit{linear}. We evaluate the effect of the
circuit on an initial state which is a computational basis state. However, if
the initial qubits are in a superposition of computational basis states, then
the final state of the qubits, after the circuit has acted, is easily
computed since it is
the corresponding linear superposition of outputs for each of the
computational basis state inputs.\index{linearity}
\end{itemize}

Circuits have several gates acting in succession on a qubit and it is important to understand the order in which they act. Unfortunately, this can be confusing. By convention, in diagrams time is from left to right, so in the diagram
\begin{center}
\includegraphics[width=5cm]{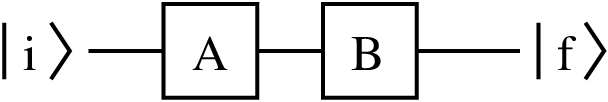}
\end{center}
\noindent $A$ (the leftmost gate) acts first on state $|i\rangle$, and then $B$
act, leaving the qubit in the final state $|f\rangle$. However, when writing
operator expressions, these work from right to left, so, the above diagram
corresponds to
\begin{equation}
|f\rangle = B A |i\rangle ,
\end{equation}
in which $A$ is on the \textit{right}. You simply have to get used to this
reversal of order when going from circuit diagrams to operator expressions.

Now we describe some commonly used quantum gates, 
recalling that quantum gates must be reversible and so are unitary operators. 

Firstly we consider 1-qubit gates.
\begin{itemize}
\index{quantum NOT gate|see{$X$ matrix}}
\index{bit-flip gate|see{$X$ matrix}}
\index{$\sigma_x$|see{Pauli $X$ matrix}}
\index{$\sigma_y$|see{Pauli $Y$ matrix}}
\index{$\sigma_z$|see{Pauli $Z$ matrix}}
\item NOT, i.e.~bit-flip (corresponds to the Pauli $X$ matrix)
\index{Pauli matrices!$X$ matrix}
\begin{equation}
\begin{array}{l}
X|0\rangle = |1\rangle \\
X|1\rangle = |0\rangle \\
\end{array}, \qquad X = 
\begin{pmatrix}
0 & 1 \\
1 & 0 \\
\end{pmatrix} , \qquad\mathrm{so\ } X
\begin{pmatrix} \alpha \\ \beta \end{pmatrix} = 
\begin{pmatrix} \beta \\ \alpha \end{pmatrix} .
\end{equation}

\index{phase-flip gate|see{Pauli $Z$ matrix}}\index{Pauli matrices!$Z$ matrix}
\item Phase flip (corresponds to the Pauli $Z$ matrix)
\begin{equation}
\begin{array}{l}
Z|0\rangle = |0\rangle \\
Z|1\rangle = -|1\rangle \\
\end{array}, \qquad Z = 
\begin{pmatrix}
1 & 0 \\
0 & -1 \\
\end{pmatrix} , \qquad\mathrm{so\ } Z
\begin{pmatrix} \alpha \\ \beta \end{pmatrix} = 
\begin{pmatrix} \alpha \\ -\beta \end{pmatrix} .
\end{equation}
In the physics literature $X$ and $Z$ are called Pauli spin matrices. 
There is also a third Pauli spin matrix, $Y$, where
\index{Pauli matrices!$Y$ matrix}
\begin{equation}
\begin{array}{l}
Y|0\rangle = -i|1\rangle \\
Y|1\rangle = i|0\rangle \\
\end{array},\quad \ \ 
Y = i X Z = 
\begin{pmatrix}
0 & -i \\
i & 0 \\
\end{pmatrix} , \qquad\mathrm{so\ } Y
\begin{pmatrix} \alpha \\ \beta \end{pmatrix} =  i
\begin{pmatrix} -\beta \\ \alpha \end{pmatrix},
\end{equation}
which corresponds to a combined bit- and phase-flip. The Pauli $Y$-matrix
will only appear again when we do quantum error correction in Chapter \ref{ch:err_corr}.

\index{Hadamard matrix (gate)}
\item Hadamard\\
\noindent The Hadamard gate $H$ will be very important.
\begin{equation}
H = {1 \over \sqrt{2}} (X + Z) = {1 \over \sqrt{2}}
\begin{pmatrix}
1 & 1 \\
1 & -1 \\
\end{pmatrix} .
\end{equation}
Note that $H^2 = \mathbbm{1}$, and similarly $X^2 = Y^2 = Z^2 = \mathbbm{1}$.

Now a matrix which squares to the identity has eigenvalues $\pm 1$. To see this note that if $\vec{x}$ is an eigenvector of $A$ with eigenvalue $\lambda$ then 
\begin{equation}
A^2 \vec{x} = A \left(A \vec{x}\right) = A \lambda \vec{x}  = \lambda A \vec{x} = \lambda^2 \vec{x} .
\end{equation}
But if $A^2 = \mathbbm{1}$ then it follows that $\lambda^2 = 1$ and so $\lambda = \pm 1$.

We need to become familiar with the action of $H$ on computational
basis\index{computational basis}
states. This is:
\begin{equation}
\begin{split}
H|0\rangle = & |+\rangle \equiv {1 \over \sqrt{2}} \left(|0\rangle + |1\rangle\right) \\
H|1\rangle = & |-\rangle \equiv {1 \over \sqrt{2}} \left(|0\rangle - |1\rangle\right) .
\end{split}
\end{equation}
Combining these two equations, the action of $H$ on a computational basis state $|x\rangle$ is seen to be
\begin{equation}
H |x\rangle = {1\over \sqrt{2}} \left(\, |0\rangle + (-1)^x |1\rangle \, \right) ,
\label{Had}
\end{equation}
for both values of $x$, namely 0 and 1. 
\end{itemize}

\index{superposition}
\noindent A crucial point is that these gates are linear, and so they act in the same way on a superposition.  For example:
\begin{equation}
H \left[\, \alpha|0\rangle + \beta|1\rangle\,\right] = {\alpha \over \sqrt{2}} \left( |0\rangle + |1\rangle \right) +
 {\beta \over \sqrt{2}}\left(|0\rangle - |1\rangle\right) = \left({\alpha + \beta \over \sqrt{2}}\right)|0\rangle + 
 \left({\alpha - \beta \over \sqrt{2}}\right)|1\rangle .
\end{equation}



\index{measurement gates}
We also need to consider measurement gates, in which a classical measurement
of a qubit takes place. The basis in which measurements are 
made is called the computational basis. The Pauli spin matrices are for the
computational basis and since
the Pauli $Z$ is
diagonal we also call the computational basis the $Z$-basis.

The result of the measurement is a classical bit. In
the circuit diagrams we indicate a classical bit by a double line, and so a
measurement gate is indicated as follows:
\begin{center}
\includegraphics[width=3.5cm]{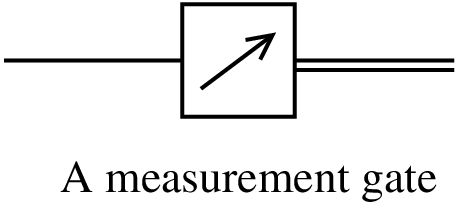}
\end{center}

The measurement apparatus acting on a qubit determines
the value of $Z$ for that qubit, obtaining either $+1$, in which case the qubit is left in
state $|0\rangle$, or $-1$, in which case the qubit is left in state
$|1\rangle$. If one wants to measure the value of
some other quantity one needs to perform
an appropriate unitary transformation first. For example, to determine the
value of $X$ one acts with a Hadamard before the measurement, since the
Hadamard converts the $X$-basis to the $Z$-basis and vice-versa.
In other words a state $\alpha|+\rangle +\beta|-\rangle$ 
becomes $\alpha|0\rangle + \beta|1\rangle$ after the Hadamard, and so a
measurement gives $|0\rangle$ with probability $|\alpha|^2$ and 
$|1\rangle$ with probability $|\beta|^2$. These are the probabilities 
one \textit{would} have of
measuring $|+\rangle$ and $|-\rangle$ respectively (before the Hadamard acted)
if one could measure $X$ directly.

Note, however, that this procedure leaves the qubit in an eigenstate of $Z$ which is
a problem if
we want to continue
to use the qubit after the measurement,
because then the qubit should be left in the eigenstate of the
measurement operator. It turns out that this can be done by
coupling the qubit to another ``ancilla" qubit and measuring the ancilla, as
explained in Fig.~\ref{stabilizer} below.

\begin{figure}[htb]
\begin{center}
\includegraphics[width=14cm]{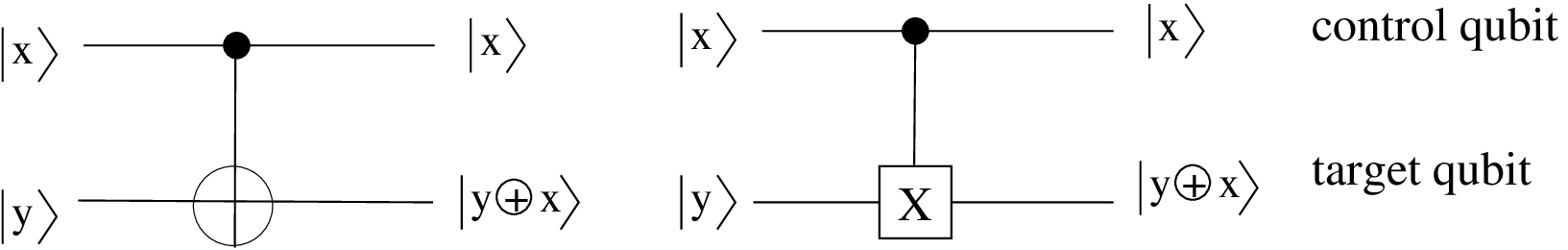}
\caption{Two equivalent ways of drawing a CNOT gate. The right hand way
makes clear that the NOT
operation is performed by the Pauli $X$ operator. If the initial state of the qubits
(in each case on the left of the diagram)
is a computational basis state, then the action of the quantum CNOT gate is the
same as that of the classical CNOT gate shown in Fig.~\ref{CNOT2}. The upper line
represents the control qubit and the lower line the target qubit.  Note that
the target qubit is only changed if the control qubit is $|1\rangle$.
\label{cnot_qu}
}
\end{center}
\end{figure}

Next we consider 2-qubit gates, the most important of which by far is the CNOT. We already
met the classical CNOT gate in Fig.~\ref{CNOT2}. In the quantum case, if
initially the qubits are in a computational basis state, then the action of the
CNOT is the same as classically. Since the NOT function is implemented by the
Pauli $X$ operator, so the CNOT operation
can equivalently be thought of as Ctrl-$X$, we indicate explicitly the action of $X$
in the circuit representation of the quantum CNOT gate
shown in 
Fig.~\ref{cnot_qu}.
\index{CNOT quantum gate}

The CNOT gate has the matrix representation
\begin{align}
& \quad |00\rangle \ |01\rangle \ |10\rangle \ |11\rangle \nonumber \\
U_\mathrm{CNOT} = 
\begin{matrix}
\langle 00 | \\
\langle 01 | \\
\langle 10 | \\
\langle 11 | 
\end{matrix}
& \begin{pmatrix}
\ 1\quad  & \quad\  0\  & \quad 0\  & \quad 0\  \\
\ 0\quad  & \quad\  1\  & \quad 0\  & \quad 0\  \\
\ 0\quad  & \quad\  0\  & \quad 0\  & \quad 1\  \\
\ 0\quad  & \quad\  0\  & \quad 1\  & \quad 0\  \\
\end{pmatrix}
.
\end{align}
\index{tensor product}
In this tensor product the control qubit is the one to the left. The target qubit (to
the right) is flipped if the control qubit is 1 (so, relative to
\index{control qubit}
\index{target qubit}
the identity matrix, columns 3 and
4 are interchanged). We can also write $U_\mathrm{CNOT}$ in terms of $2 \times 2$ blocks as
follows
\begin{equation}
U_{CNOT} = 
\begin{pmatrix}
\mathbbm{1} & 0 \\
0 & X \\
\end{pmatrix} .
\end{equation}

\begin{figure}[htb]
\begin{center}
\includegraphics[width=8cm]{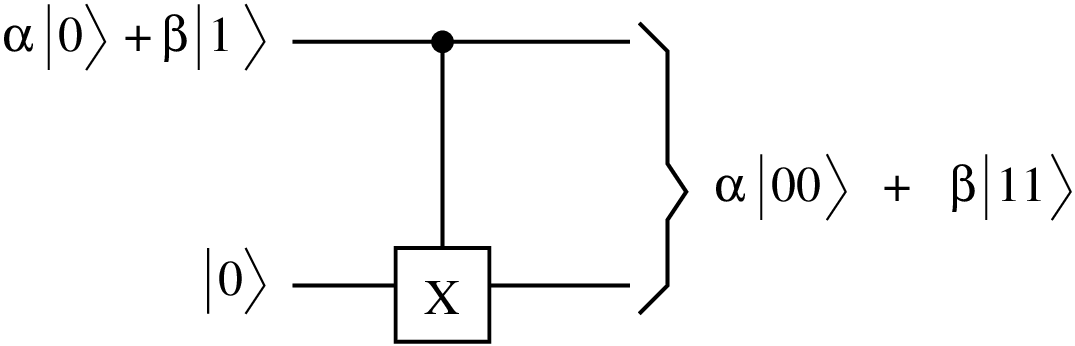}
\caption{The action of the CNOT gate when the upper (control) qubit is
initially in a
superposition $\alpha|0\rangle + \beta|1\rangle$, and the lower (target) qubit
is initially $|0\rangle$. By linearity, the final
state is $\alpha$ times the result of inputting $|0\rangle$ in the control
qubit plus $\beta$ times
the result of inputting $|1\rangle$, i.e.~$\alpha|00\rangle +
\beta|11\rangle$. We see that the final state is \textit{entangled}.
\label{cnot_super}
}
\end{center}
\end{figure}

The quantum aspect appears if we input (on the left) a linear
combination of basis states. Suppose, for example, we set the target (lower) qubit to
$|0\rangle$. Then if the control qubit is initially $|0\rangle$ the final
state of the 2-qubit system is $|00\rangle$, because the target qubit is
not flipped and stays as $|0\rangle$ (we take the control qubit to be the
left one). If the control qubit is initially $|1\rangle$ then the final state
of the 2-qubit system is $|11\rangle$ because the target qubit \textit{is}
flipped from $|0\rangle$ to $|1\rangle$. Hence, by linearity, if the initial
state of the control qubit is the superposition $\alpha|0\rangle +
\beta|1\rangle$, then the final state of the 2-qubit system is
$\alpha|00\rangle + \beta|11\rangle$, see Fig.~\ref{cnot_super}. Note that the
CNOT gate has entangled the control and target qubits. \textbf{Using a CNOT gate is the
standard way of entangling qubits in quantum computing.}

Because of entanglement, if we measure one of the qubits then the state of
the other qubit after the measurement depends on the result of the
measurement. For example, if we measure the upper qubit and find that the
result is $|0\rangle$ then the lower qubit will be left in state $|0\rangle$,
whereas if the result is $|1\rangle$ the lower qubit will be left in state $|1\rangle$.

Note that if $\alpha = 0$ (so $\beta = 1$ since $|\alpha|^2 + |\beta|^2 = 1$)
or $\alpha = 1 \ (\beta = 0)$, the final state is a clone of the initial state of the
control qubit. However, for a general input state, the final state of the two
qubits,
$\alpha|00\rangle + \beta|11\rangle$, is not a clone of the initial state of
the control qubit which
would be $(\alpha|0\rangle + \beta |1\rangle) \otimes (\alpha|0\rangle + \beta
|1\rangle) = \alpha^2 |00\rangle + \alpha\beta (|01\rangle + |10\rangle)
+ \beta^2|11\rangle$. Hence there is no violation of the no-cloning theorem
\index{no-cloning theorem}
which states that a \textit{general}, unknown quantum state can not be cloned.

\begin{figure}[htb]
\begin{center}
\includegraphics[width=6cm]{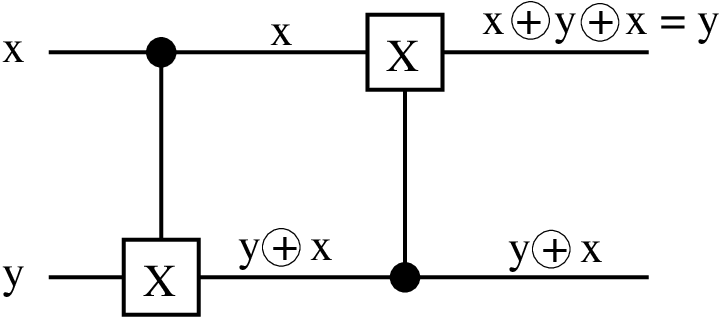}
\caption{A circuit with 2 CNOTs in opposite orientations. Note that
$x\oplus y\oplus x= y$ because $x \oplus x= 0$. Note that time goes from left
to right in circuit diagrams. 
\label{CU_example}
}
\end{center}
\end{figure}

To gain some more familiarity with the important CNOT gate consider the
circuit diagram in Fig.~\ref{CU_example} in which there are two CNOTs with opposite
orientations. Feeding in computational basis states $|x\rangle|y\rangle$ into the
inital state on the left, the figure shows the states at each stage. The net effect of the
second (right hand) CNOT is to convert the upper qubit from $x$ to $y$, which
is perhaps a bit surprising. 

In this course, we will specify the action of a gate by its action on an initial
computational basis state. If we denote a qubit by a Latin letter,
e.g.~$|x\rangle$, we mean that this is a computational basis state and $x$ takes values 0
or 1. General quantum states, i.e.~superpositions of computational basis
states, will be indicated by Greek letters, e.g.~$|\psi\rangle$. 

As already mentioned above, we do not need 3-qubit gates for quantum
computing. More precisely, the statement is that one can generate an arbitrary
unitary transformation (to a specified level of accuracy) on an arbitrary
number of qubits, using only CNOT and single-qubit gates. I do not prove this
result but refer interested
students to a more advanced text~\cite{nielsen:00}. It is fortunate that we
don't need 3-qubit gates given the difficulty of making quantum circuits.

\begin{figure}[htb]
\begin{center}
\includegraphics[width=8cm]{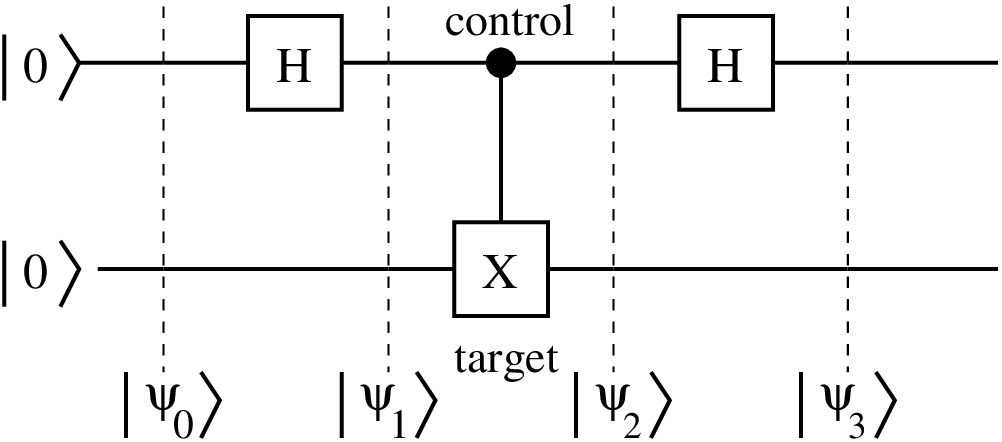}
\caption{The initial state of both qubits is $|0\rangle$. What is the final
state $|\psi_3\rangle$? Equation \eqref{ent} gives the state of the two qubits at each stage.
The end result is that the two qubits are entangled and, in contrast to what
one might have thought, the control (upper) qubit has a non-zero amplitude to be
flipped relative to its initial state, i.e.~to be in state $|1\rangle$.
\label{entangle}
}
\end{center}
\end{figure}

It is useful to mention here that one has to be careful when dealing with
superpositions, and one's initial intuition as to the final result may be
incorrect. As an example, consider the circuit in Fig.~\ref{entangle}.
Since $H^2 = \mathbbm{1}$ and the CNOT gate doesn't change the control (upper) qubit, 
one might think that the final state of the control qubit would be the same as
the initial state, i.e.~$|0\rangle$. However this is not correct because the
control and target qubits become entangled. Let's go through each stage of the
circuit using the notation for successive states indicated
in Fig.~\ref{entangle}, and taking the left-hand qubit
in the formulae to be the control qubit:
\index{entanglement}
\begin{equation}
\begin{split}
|\psi_0\rangle &= |00\rangle \\
|\psi_1\rangle &= {1\over \sqrt{2}} \left(\, |00\rangle + |10\rangle\, \right)\\
|\psi_2\rangle &= {1\over \sqrt{2}} \left(\, |00\rangle + |11\rangle\, \right) \\
|\psi_3\rangle &= {1 \over 2} \left(\, |00\rangle + |10\rangle + |01\rangle - |11\rangle\, \right) \\
               &= {1\over\sqrt{2}} 
	       \left[\, |0\rangle_c \otimes \left({|0\rangle_t + |1\rangle_t \over \sqrt{2}}\right)  +
	       |1\rangle_c \otimes \left({|0\rangle_t - |1\rangle_t \over \sqrt{2}}\right)\, \right] ,
\label{ent}
\end{split}
\end{equation}
where in the last expression we indicate explicitly which qubit is the control
qubit 
(``$c$''), and which the target qubit (``$t$''). We see that, contrary to what
one might have initially guessed, there is an amplitude for the final state of the
control qubit to be 
$|1\rangle$ because of its entanglement with the target qubit.

\section{A circuit to measure operators which have eigenvalues $\boldsymbol{\pm 1}$.}

We have noted that the Pauli operators $X, Y$ and $Z$, and the Hadamard
operator have eigenvalues $\pm 1$. Later in the course, when we consider the
important topic of quantum error correction, we will encounter
combinations of these operators on different qubits which also have $\pm 1$
eigenvalues. We will now describe a convenient way of measuring such
operators.
Let us denote such an operator by $U$. It will have matrix elements given by
\begin{equation}
U = \begin{pmatrix}
u_{00} & u_{01}\\
u_{10} & u_{11}\\
\end{pmatrix}
\end{equation}
and
eigenvalue $+1$ with eigenvector $|\psi_+\rangle$ and an eigenvalue $-1$ with 
eigenvector $|\psi_-\rangle$, i.e.
\begin{equation}
U|\psi_+\rangle = |\psi_+\rangle, \quad U|\psi_-\rangle = -|\psi_-\rangle.
\label{ctrl-U}
\end{equation}

We would like to
investigate the qubit (or qubits) to determine which
eigenstate of $U$ it is in, or, if it is in a linear superposition, to project
by measurement
on to one of the eigenstates, and know which one.

\begin{figure}[htb]
\begin{center}
\includegraphics[width=9cm]{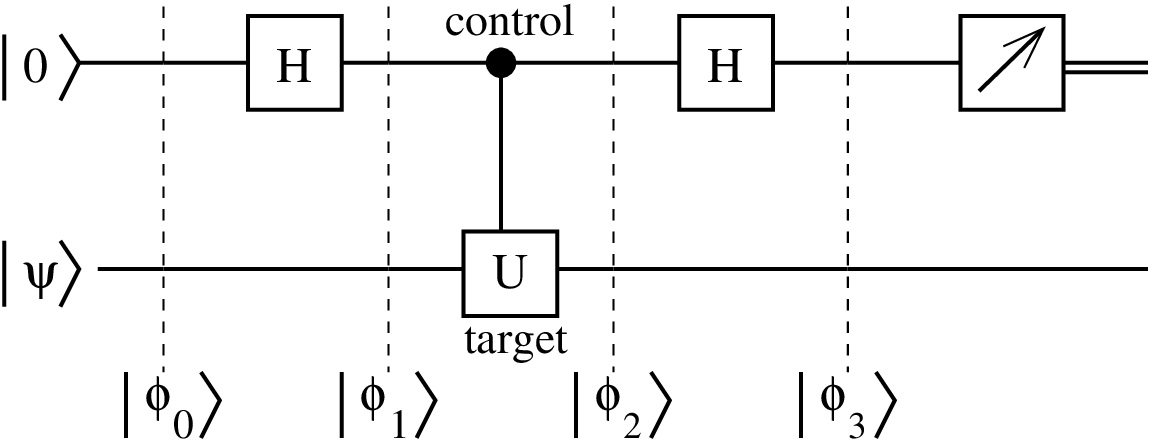}
\caption{A circuit with a control-$U$ gate in which the control (upper) qubit
is surrounded by Hadamards. $U$ is an operator with eigenvalues $\pm1$ and
corresponding eigenvectors $|\psi_+\rangle$ and $|\psi_-\rangle$.
As shown in the text, if a measurement of the upper qubit gives $|0\rangle$
then
the lower qubit will be in state $|\psi_+\rangle$,
and if the measurement gives $|1\rangle$ then the
lower qubit will be in state $|\psi_-\rangle$.
The states $|\phi_i\rangle\, (i=0, 1, 2, 3)$ are described in the text.
\label{stabilizer}
}
\end{center}
\end{figure}

\index{control-$U$ gate}
A convenient way is to use the circuit shown in Fig.~\ref{stabilizer}, which
has a control-$U$ gate\footnote{Apart from the absence of the final measurement
gate, Fig.~\ref{entangle} is a special case of Fig.~\ref{stabilizer} with $U = X$.}. The matrix
representation of control-$U$ is
\begin{align}
& \quad \ |00\rangle \ \ \ |01\rangle \ \ \ |10\rangle \ \ \ \ |11\rangle \nonumber \\
\mathrm{control}\!-\!U = 
\begin{matrix}
\langle 00 | \\
\langle 01 | \\
\langle 10 | \\
\langle 11 | 
\end{matrix}
& \begin{pmatrix}
\ 1\quad  & \quad\  0\  & \quad 0\  & \quad 0\  \\
\ 0\quad  & \quad\  1\  & \quad 0\  & \quad 0\  \\
\ 0\quad  & \quad\  0\  & \quad u_{00}\  & \quad u_{01}\  \\
\ 0\quad  & \quad\  0\  & \quad u_{10}\  & \quad u_{11}\  \\
\end{pmatrix}
=
\begin{pmatrix}
\mathbbm{1} & 0 \\
0 &  U \\
\end{pmatrix}
,
\end{align}
where the last expression is written in terms of $2 \times 2$ blocks.
If the control qubit is 1 then $U$ acts on the target qubit according to
Eq.~\eqref{ctrl-U}, while
if the control qubit is 0 then the
target qubit is unchanged.

The lower (target) qubit is initially in state $|\psi\rangle$, which can 
be written as a linear combination of the two eigenvectors
\begin{equation}
|\psi\rangle = \alpha_+ |\psi_+\rangle + \alpha_- |\psi_-\rangle \, ,
\end{equation}
and so, including the upper (control) qubit
which is initially in state $|0\rangle$, the
initial state of the circuit (on the left of Fig.~\ref{stabilizer}) is
\begin{subequations}
\begin{equation}
|\phi_0\rangle = \alpha_+ |0\, \psi_+\rangle + \alpha_- |0\, \psi_-\rangle \,
.
\end{equation}
In labeling the states, we put the state of the control qubit to the left and that of the
target qubit to the right.
After the first Hadamard on the upper qubit the state is
\begin{equation}
|\phi_1\rangle = {\alpha_+\over \sqrt{2}}\left(\,
|0\, \psi_+\rangle + |1\, \psi_+\rangle\,\right) +
{\alpha_-\over \sqrt{2}}\left( \,|0\, \psi_-\rangle +  |1\,
\psi_-\rangle\,\right)  \, .
\end{equation}
The effect of the control-$U$ gate on the target qubit is given by
Eq.~\eqref{ctrl-U} when the control qubit is 1 and has no effect if the
control qubit is 0. Hence, after the control-$U$ gate, the state is
\begin{equation}
|\phi_2\rangle = {\alpha_+\over \sqrt{2}}\left(\,
|0\, \psi_+\rangle + |1\, \psi_+\rangle\,\right) +
{\alpha_-\over \sqrt{2}}\left(\, |0\, \psi_-\rangle -  |1\, \psi_-\rangle\,
\right)  \, .
\end{equation}
Applying the righthand Hadamard in Fig.~\ref{stabilizer} to the upper (control) qubit we get
\begin{equation}
|\phi_3\rangle = \alpha_+ |0\, \psi_+\rangle + \alpha_- |1\, \psi_-\rangle \,
.
\label{gates:phi3}
\end{equation}
\label{gates:phi}
\end{subequations}
Hence if a measurement of the control (upper) qubit gives $|0\rangle$
(which it does with
probability $|\alpha_+|^2$) the target (lower) qubit will be in state $|\psi_+\rangle$,
and if the measurement gives $|1\rangle$ (for which the probability is $|\alpha_-|^2$) the
lower qubit
will be in state $|\psi_-\rangle$. We see that measuring the control qubit
projects the target qubit onto an eigenstate of $U$ and 
tells us which one. 

Note that we measure the state of the target qubit \textit{indirectly}. We
use the control qubit as an ancilla, and a measurement of the ancilla is used
to determine which eigenstate of $U$ the target qubit is in.
Could we not measure the state of the target qubit \textit{directly}? Since
measurements are always done in the $Z$-basis, one would have to:
\begin{enumerate}
\item
Act with a unitary operator\footnote{For the case of $U=X$, $S_U\, (=
S^\dagger_U\ \mathrm{here})$ is given by
Eq.~\eqref{UZX} (where it is called $U$, sorry for the confusing notation).} $S_U$ which converts the $U$-basis (where the
basis states are $|\psi_+\rangle$ and $|\psi_-\rangle$) to the $Z$-basis
(where the basis states are $|0 \rangle$ and $|1\rangle$). 
\item
Measure in the $Z$-basis.
\item
Convert back to the $U$-basis by acting with the inverse transformation
$S^\dagger_U$.
\end{enumerate}
If the measurement determines that the qubit is in state $|0\rangle$, then after
acting with $S^\dagger_U$ the qubit will be in state $|\psi_+\rangle$, and
similarly if the measurement gets $|1\rangle$, the final state will be 
$|\psi_-\rangle$, as required. However, it is complicated to construct gates which
implement the transformations $S_U$ and $S^\dagger_U$, so in practice one uses
the indirect method of coupling to an ancilla and
and measuring the ancilla, as shown in Fig.~\ref{stabilizer}. We will develop this
idea in detail in Chapter \ref{chap:error}, when we discuss quantum error correction. In that situation
each ancilla is coupled to several qubits and the operator $U$
involves a product of Pauli operators on those qubits.


We will return to the circuit in Fig.~\ref{stabilizer} in Chapter \ref{ch:err_corr} 
when we discuss quantum error correction.

\hrulefill
\section*{Problems}
\input{hw_ch7.tex}

%% file: hw_ch7.tex
\begin{problems}
\item
Show that the $n$-qubit Hadamard gate acts as
\begin{equation}
H^{\otimes n} |x\rangle_n = {1 \over\sqrt{2^n}} \sum_{y=0}^{2^n-1} (-1)^{x\cdot
y} |y\rangle ,
\end{equation}
where $x \cdot y$ is the bitwise inner product of $x$ and $y$ with modulo 2
addition:
\begin{equation}
x\cdot y = x_0 y_0 \oplus x_1 y_1 \oplus \ldots \oplus x_{n-1} y_{n-1} \, .
\end{equation}

\item
Show the following identities:
\begin{align*}
H X H &= Z \\
H Z H &= X \\
H Y H &= -Y ,
\end{align*}
where $H$ is the Hadamard matrix. 

\item
The ``SWAP" gate $S$ interchanges the two inputs. It is defined by
\begin{equation}
S|x y \rangle = |y x \rangle.
\end{equation}
\begin{enumerate}[label=(\roman*)]
\item
Give the matrix representing this state.
\item
Show that it is equivalent to three CNOT gates as
\begin{equation}
S_{12} = C_{12} C_{21} C_{12} \, ,
\end{equation}
where, in $C_{ij}$, $i$ refers to the control bit and $j$ to the target bit. 
\end{enumerate}

\item
Verify the following circuit identities:

\begin{center}
\includegraphics[width=11cm]{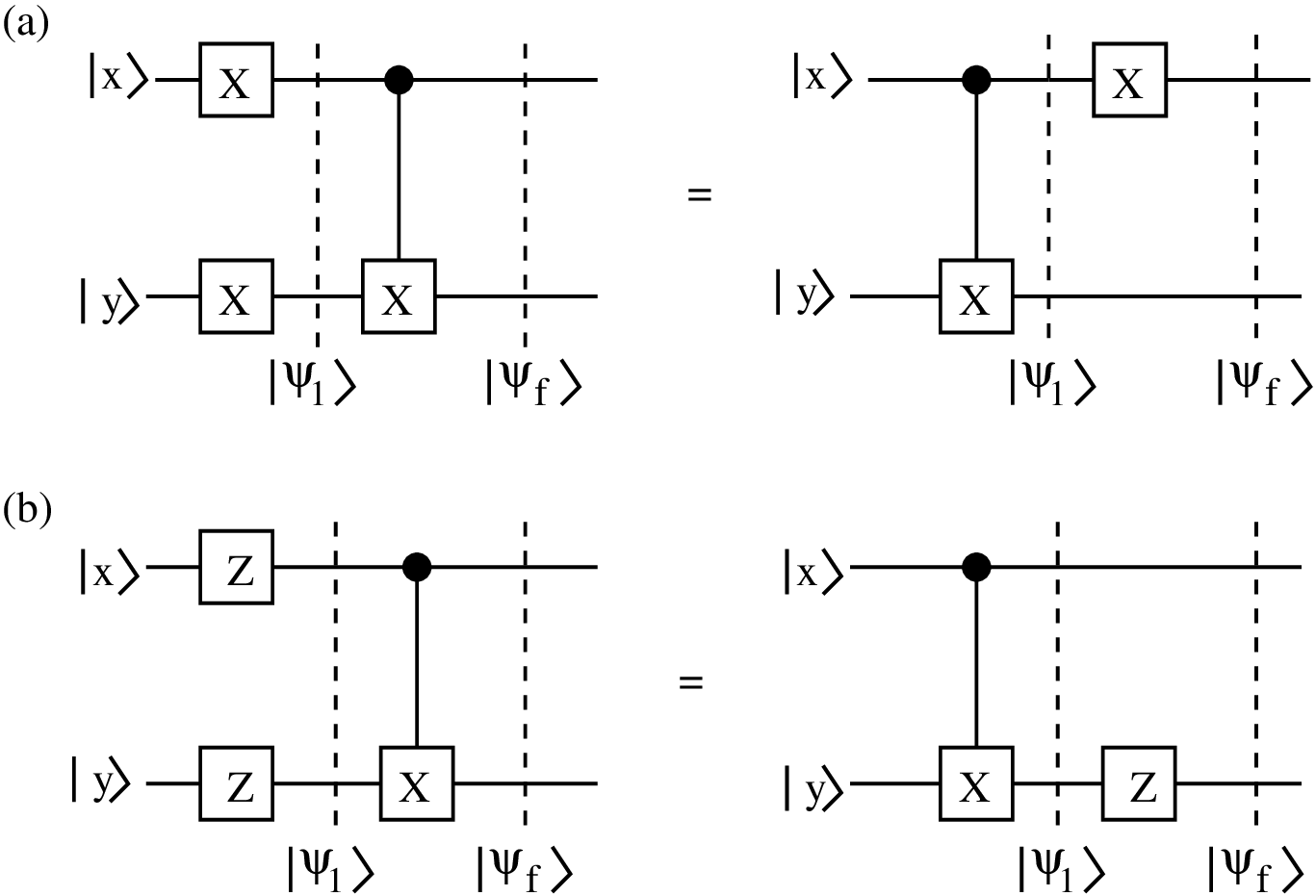}
\end{center}

\textit{Note:} Control-$X$ is another way of writing the CNOT gate.\\
\textit{Hint:} Consider arbitrary initial computational basis states
$|x\rangle$ and $|y\rangle$, and determine, for all four figures, the
intermediate states $|\psi_1\rangle$ and the final states $|\psi_f\rangle$.
Show that
the final state $|\psi_f\rangle$ is the same for the left-hand and the
corresponding right-hand figures.


\item
Consider a CNOT gate in which the target qubit is $|0\rangle$. Show that it
clones the control qubit if the control qubit is a computational
basis state, $|x\rangle$, where $x=0$ or $1$, but does not clone it if the control
qubit is a linear superposition of computational basis states. \\
\textit{Note:} This is in agreement with the no-cloning theorem which states
that one can not clone an \textit{arbitrary} unknown quantum state.

\item
The notion of controlled (i.e.~conditional) gate can
be generalized to an arbitrary single-qubit operation $U$ as follows
\begin{equation}
U_{CU} |x\rangle|y\rangle = |x\rangle U^x |y\rangle,
\end{equation}
where $x$ and $y$ are 0 or 1. Here $|x\rangle$ is the control qubit, and $|y\rangle$
is the target qubit. If $x=0$ then $U$ does not act because $U^x =
\mathbbm{1}$, whereas $U$ does act on the target qubit if $x=1$. The matrix
representation of this gate is
\begin{equation}
U_{CU} = 
\begin{pmatrix}
\mathbbm{1} & 0 \\
0 & U \\
\end{pmatrix},
\end{equation}
where $\mathbbm{1}$ and $U$ represent $2 \times 2$ blocks. The circuit
diagram is as follows:

\begin{center}
\includegraphics[width=4.5cm]{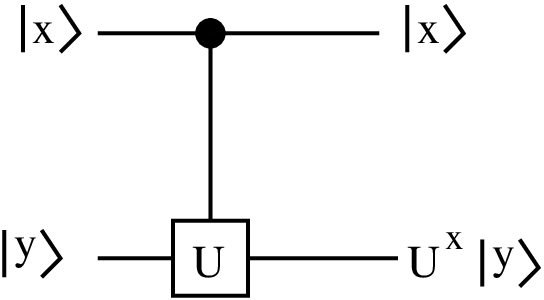}
\end{center}

In most of the examples that we will discuss, it turns out that $U^2
=\mathbbm{1}$ and so, as shown earlier, the eigenvalues are $\pm 1$. The
operator $U$ in the
circuit below has this property.

\begin{center}
\includegraphics[width=8.5cm]{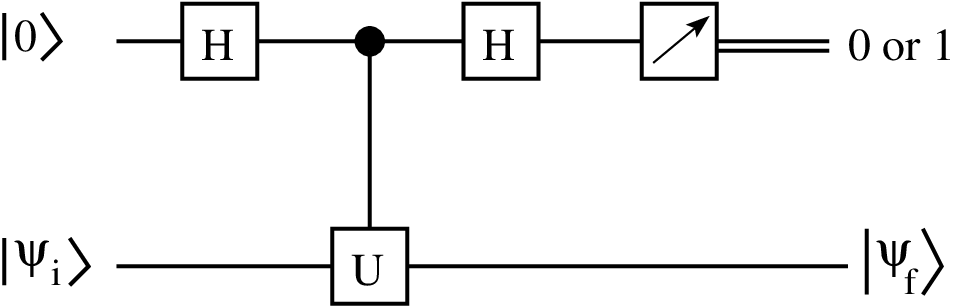}
\end{center}

Now we add a measurement of the control qubit as shown in the figure above.
The box with an arrow indicates a measurement.  The double line to the right
indicates that the result of the measurement is a classical bit, 0 or 1. \\

Show
that if the measurement of the upper (control) qubit
finds $|0\rangle$ then the lower (target) qubit ends up a state
$|\psi_f\rangle$ which is the eigenstate of $U$ with eigenvalue $+1$, whereas
if the measurement of the upper (control) qubit finds $|1\rangle$ then the lower (target)
qubit ends up in the eigenstate of $U$ with eigenvalue $-1$.

\textit{Note:}
We say that this circuit measures the operator $U$. It will
play an important role when we study quantum error correction. 

\end{problems}

%% file: bell8.tex
In this chapter we describe two simple quantum circuits.

Entangled states play an important role in quantum computing. The most-studied
entangled states are so-called Bell states
which involve two qubits, which we discussed in Ch.~\ref{ch:gen_qubit}.
\index{Bell state}
As a reminder, the Bell states are defined by
\begin{subequations}
\label{bell:bell_eq}
\begin{align}
|\beta_{00} \rangle = {1 \over\sqrt{2}} \left(\, |00\rangle + |11\rangle\, \right) , \\
|\beta_{01} \rangle = {1 \over\sqrt{2}} \left(\, |01\rangle + |10\rangle\, \right) , \\
|\beta_{10} \rangle = {1 \over\sqrt{2}} \left(\, |00\rangle - |11\rangle\, \right) , \\
|\beta_{11} \rangle = {1 \over\sqrt{2}} \left(\, |01\rangle - |10\rangle\, \right) .
\end{align}
\end{subequations}
These four equations can be combined as follows:
\begin{equation}
|\beta_{xy}\rangle = {1 \over \sqrt{2}} \left(\, |0y\rangle + (-1)^x |1
\overline{y}\rangle \, \right) \, ,
\label{bell:bell_eqall}
\end{equation}
where $\overline{y}$ is the complement of $y$, i.e. $\overline{y} = 1 - y$.
Note that the Bell states form a basis for two qubits, as do the computational states $|x\rangle_2$.

\begin{figure}[htb]
\begin{center}
\includegraphics[width=6cm]{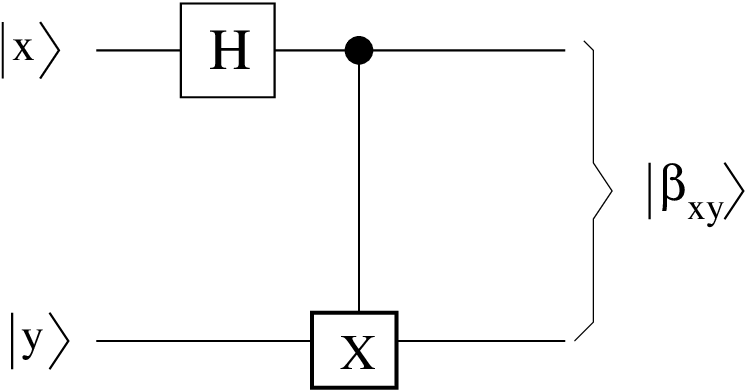}
\caption{
Circuit to create the Bell states defined by Eqs.~\eqref{bell:bell_eq}. In the CNOT
(Ctrl-$X$) gate, the upper qubit $|x\rangle$
is the control qubit and the lower qubit $|y\rangle$
is the target qubit. 
\label{bell}
}
\end{center}
\end{figure}

The Bell states are clearly entangled.  They can be created out of two
(unentangled) qubits in computational basis states $|x y\rangle$ by the
circuit shown in Fig.~\ref{bell}. I emphasize that time in circuit diagrams
runs from left to right whereas if we write the circuit in terms of
operators, they act \textit{from right to left}. Hence in Fig.~\ref{bell}, the
Hadmard on the upper qubit acts first and then the CNOT
(Ctrl-$X$) gate acts on the two qubits. In terms of operators, Fig.~\ref{bell} would be
written as
\begin{equation}
|\beta_{xy}\rangle =  \textrm{Ctrl-}\!X\, H^{(A)}| x y \rangle
\end{equation}
where $A$ refers to the leftmost qubit in the state $|xy\rangle$.

To see that Fig.~\ref{bell} generates a Bell state, note that, according to
Eq.~\eqref{Had}, after the Hadamard the
state is
\begin{equation}
|x y \rangle \rightarrow {1 \over \sqrt{2}} \left(\, 
|0 y\rangle + (-1)^x |1 y\rangle\, \right) \, .
\end{equation}
The effect of the CNOT gate is to flip $y$ in the second term (since $x=1$
there)
and so we get
Eq.~\eqref{bell:bell_eqall}.

\begin{figure}[htb]
\begin{center}
\includegraphics[width=8cm]{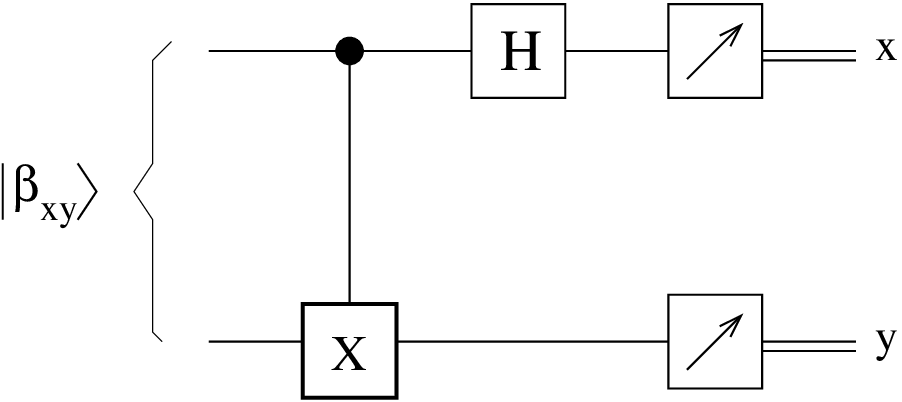}
\caption{
Circuit for Bell measurements. This will be used later in the course when we
discuss teleportation.
\label{bell_meas}
}
\end{center}
\end{figure}

The circuit in Fig.~\ref{bell} converts the computational basis to the Bell
basis. The reverse of this circuit can be used to convert the Bell basis back
to the computational basis as shown in Fig.~\ref{bell_meas}.  The measured
values of $x$ and $y$ tell us which Bell state we started with.  This is
called a \textit{Bell Measurement}. \index{Bell measurement}To see that this works note that after the
CNOT gate the state of the two qubits in Fig.~\ref{bell_meas}
is\footnote{The reason that
$\overline{y}$ in the Bell state, Eq.~\eqref{bell:bell_eqall},
changes to $y$ in the second term in Eq.~\eqref{phi} is because the control
qubit is $x=1$, so
the $y$ (target) qubit is flipped.}
\begin{equation}
{1 \over \sqrt{2}}\left[\, |0y\rangle + (-1)^x |1 y\rangle\, \right] ,
\label{phi}
\end{equation}
which is separable and so can be written as
\begin{equation}
{1 \over \sqrt{2}}\left[\, |0\rangle + (-1)^x |1\rangle\, \right] \otimes
|y\rangle .
\end{equation}
\index{control qubit}
\index{target qubit}
Recall that the left-hand qubit is the upper (control) qubit in
Fig.~\ref{bell_meas} and the right hand qubit is the lower (target) qubit. 
Acting with the Hadamard has the effect 
\begin{equation}
H {1 \over \sqrt{2}}\left[\, |0\rangle + (-1)^x |1\rangle\, \right] =
|x\rangle \, ,
\end{equation}
so the final state in Fig.~\ref{bell_meas} is $|x y\rangle$ as desired.

Note that the Bell states $|\beta_{xy}\rangle$ provide a basis for two qubits,
see Appendix \ref{sec:ang-mom} in Chapter \ref{ch:qubits},
since they are normalized and mutually orthogonal.
Consequently, if the state inputted into the Bell measurement circuit in
Fig.~\ref{bell_meas} is not a single Bell state, but rather a linear
combination,
\begin{equation}
|\psi_\mathrm{in}\rangle = \sum_{x,y=0}^1 \alpha_{xy} |\beta_{xy}\rangle,
\end{equation}
with $\sum_{x,y} |\alpha_{xy}|^2 = 1$, then the probability that the
measurements obtain a particular set of values for $x$ and $y$ is $|\alpha_{xy}|^2$.

%% file: functions8.tex
\section{An elementary quantum function}
\index{quantum functions}

So far we have met elementary 1-qubit gates such as $X, Z$ and $H$, and
2-qubit controlled gates where one qubit (the target) is acted on by one of
the one-qubit gates
(the $X$-gate for the case of CNOT) if the other qubit (the control) takes
the value 1. We extended this to the case of the Ctrl-$U$ gate where $U$ is an arbitrary
1-qubit operator with eigenvalues $\pm 1$.

In this chapter we discuss a general class of quantum functions, so we
need to learn how to define functions in a quantum
computer where they are determined by unitary transformations which are
\textit{reversible}?\index{reversible computation}

Let us first consider the simplest case, where the argument of the function,
$x$, is a single bit, and the result of the function, $f(x)$, is also a single bit. In other words,
$x$ takes only the values 0 and 1, and the same for $f(x)$.
We need to have a qubit for $x$ and an \textit{additional} qubit\footnote{We need
to have two qubits in both the initial and final states
in order that the function is reversible, just as we needed
two qubits in the final state as well as the initial state
to make the CNOT gate which is a reversible generalization of the XOR gate, 
see Chapter \ref{ch:gates}.} which contains information on the function $f(x)$.

The function $f(x)$ will be implemented by a unitary operator $U_f$ acting on two
qubits such that
\begin{equation}
U_f |x\rangle |y\rangle = |x\rangle |f(x) \oplus y\rangle .
\end{equation}
Note the similarity with the CNOT gate, which is precisely of this form with
$f(x) = x$. It is easy to see that $U_f^2 = \mathbbm{1}$ since
\begin{equation}
U^2_f |x\rangle |y\rangle = U_f |x\rangle |f(x) \oplus y\rangle
= |x\rangle |f(x) \oplus f(x) \oplus y\rangle = |x\rangle |y\rangle
\label{U2I}
\end{equation}
since, as discussed earlier in the course, $f(x) \oplus f(x) = 0$. Hence $U_f$
has an inverse, which is $U_f$ itself.

\begin{figure}[htb]
\begin{center}
\includegraphics[width=7cm]{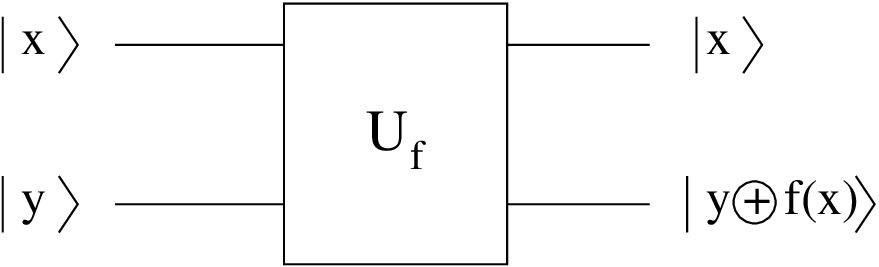}
\caption{Schematic diagram of a unitary transformation $U_f$ for a function
$f(x)$ in which both the argument $x$ and the function just take two values, 0
and 1.
\label{U11}
}
\end{center}
\end{figure}

The corresponding circuit diagram is shown in Fig.~\ref{U11}

Note that the function $U_f$ acts on \textit{both} qubits. It is distinct from
the 1-qubit gate $U$ described in Chapter \ref{ch:functions}.

For a general function, the range of inputs can be represented by $n$ bits, say, and the
range of outputs by $m$ bits. Thus we need a total of $n+m$ qubits both in the
initial state and final state. The unitary transformation is
\begin{equation}
U_f |x\rangle_n |y\rangle_m = |x\rangle_n |f(x) \oplus y\rangle_m ,
\label{Uf_nm}
\end{equation}
where the modulo 2 addition, indicated by $\oplus$, applies separately to each
of the $m$ bits of $f(x)$ and $y$. As an example, for $4$ qubits, if $f(x) =
0101$ and $y = 1100$ then $f(x) \oplus y= 1001$.

\begin{figure}[htb]
\begin{center}
\includegraphics[width=8cm]{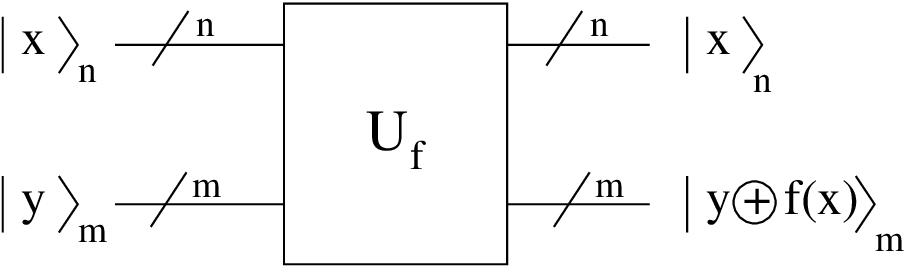}
\caption{Schematic diagram of a general unitary transformation $U_f$ for an
$n$-bit input $x$ and an $m$-bit output $f(x)$.
The
upper register in the figure has $n$ qubits and contains the input value $x$.
The lower
register has $m$ qubits and, in the final state on the right, 
contains information about the function
value $f(x)$. The registers are shown as single lines.
To ensure the
transformation is reversible there are $n+m$ qubits in both the initial state
(to the left) and final state (to the right). 
\label{Unm}
}
\end{center}
\end{figure}

The proof that $U_f$ is its own inverse is
the same as that in Eq.~\eqref{U2I}. The circuit diagram
corresponding to Eq.~\eqref{Uf_nm} is shown in
Fig.~\ref{Unm}.

\index{register!input}
\index{register!output}
One sometimes calls the upper register in Fig.~\ref{Unm} the ``input"
register, because it contains the input, $x$, and the
lower register the ``output register" because it contains information on the
function $f(x)$. However, since both registers are present in the
initial state (on the left) and the final state (on the right) this
terminology can be confusing.

Note that if $y=0$ the lower register contains precisely the function $f(x)$.

\section{Quantum Parallelism}
\index{quantum parallelism}
\label{sec:qp}
Things get interesting if we feed in a superposition. We can generate a
uniform superposition by acting with Hadamards on $|0\rangle_n$. Note that for
\index{superposition}
\index{Hadamard matrix (gate)}
one qubit
\begin{equation}
H|0\rangle = {1\over \sqrt{2}} \left(|0\rangle + |1\rangle\right) ,
\end{equation}
and similarly applying a Hadamard to each of two qubits
\begin{equation}
\begin{split}
H|0\rangle \otimes H |0\rangle &= {1\over 2}
\left(|0\rangle + |1\rangle\right)  \otimes \left(|0\rangle + |1\rangle\right)  \\
&=
{1 \over 2} \left( |00\rangle + |01\rangle + 
|10\rangle + |11\rangle \right) \\
&= {1 \over 2} \left( |0\rangle_2 + |1\rangle_2
+ |2\rangle_2 + |3\rangle_2 \right) = {1 \over 2} \sum_{x=0}^3 |x\rangle_2 .
\end{split}
\end{equation}
Generalizing we have
\begin{equation}
H^{\otimes n} |0\rangle_n = {1 \over 2^{n/2}} \sum_{x=0}^{2^n-1} |x\rangle_n .
\label{fun:super}
\end{equation}

\begin{figure}[htb]
\begin{center}
\includegraphics[width=6.5cm]{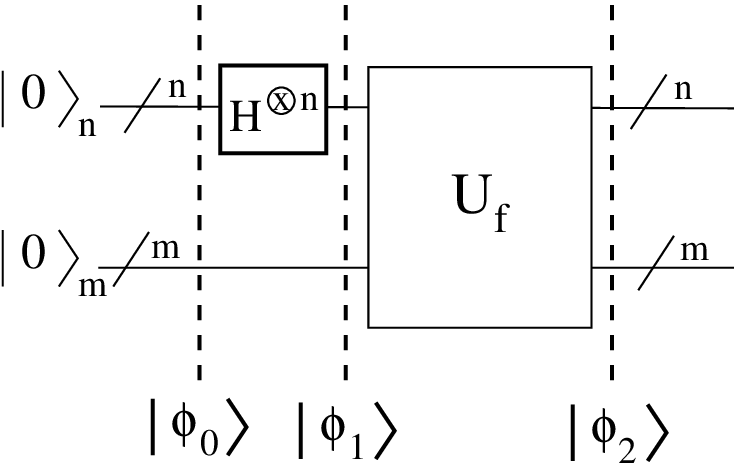}
\caption{Because of the Hadamards, the input to $U_f$ is now the uniform
superposition of all computational basis states in Eq.~\eqref{fun:super}. The
output from $U_f$ is given by Eq.~\eqref{final_super}.
\label{Unm_super}
}
\end{center}
\end{figure}

Now lets consider the circuit shown in Fig.~\ref{Unm_super}.
The initial state is
\begin{equation}
|\phi_0\rangle = 
|0\rangle_n |0\rangle_m ,
\end{equation}
so the state fed into the unitary operator $U_f$ is the superposition
\begin{equation}
|\phi_1\rangle = 
{1 \over 2^{n/2}} \sum_{x=0}^{2^n-1} |x\rangle_n |0\rangle_m .
\end{equation}
Noting that the lower register is initialized to $|0\rangle$, then
by linearity, according to Eq.~\eqref{Uf_nm}, the final state must be
\begin{equation}
|\phi_2\rangle = {1 \over 2^{n/2}} \sum_{x=0}^{2^n-1} |x\rangle_n |f(x)\rangle_m .
\label{final_super}
\end{equation}

This is an astonishing result. The final state contains the function values
for \textit{all} $2^n$ possible values of the input $x$. They have been
evaluated in parallel, a feature of quantum mechanics called, naturally
enough, ``quantum parallelism".
\index{quantum parallelism}
For $n=100$ we have $2^{100} \simeq 10^{30}$ function
evaluations in parallel.

A speedup of $10^{30}$ seems to good to be true, and, unfortunately, it is. What's the catch?
The catch is that the only way one can access the information contained in the
state is to do a measurement of the lower register.
This does not give $10^{30}$ results but just
one result, the value of $f(x$) for a single value of $x$. The probabilities
of the different results are the square of the amplitudes (which are all equal
here so there is a probability $1/2^n$ of getting the value of $f(x)$ for each
of the $2^n$ possible values of $x$). So, it seems that we have achieved
nothing. We have found the value of the function for one value of its
argument, which we could have got much more easily on a classical computer.
However, for some problems, one can gain enough useful information to get a
``quantum speedup" by doing clever pre-processing \textit{before} the
measurement, in order to reduce the number of possible measurement outcomes
(sometimes to just one.) How to achieve this in practice will occupy us for
most of the rest of the course. 

Philosophers, and some physicists, debate whether one can really state that all
$2^n$ values of the function have been evaluated since one can not observe
them. Most physicists would argue that  
the only ``real" quantities are those that can be observed, and, in particular, the 
quantum mechanical state itself is not real. Rather it is a device from which
one compute the results of measurements. 
From this point of view, it is not valid to
claim that all $2^n$ values of the function have actually been evaluated.

Now we have done enough preliminaries to study our first quantum algorithm!
This will be described in the next chapter.

%% file: deutsch8.tex
\section{Introduction}

\index{Deutsch's algorithm}
We now turn to our first algorithm, due to\index{Deutsch, David} David Deutsch
\cite{deutsch:85} which is generally felt to have started
the field of quantum computing.

As we shall see the problem is very trivial. It concerns functions which takes a 1-qubit argument
and give a 1-qubit output.  The problem is clearly contrived and is of
no practical interest. However, it does show a quantum speedup, and this
arises from the same features of quantum circuits, namely \textit{quantum parallelism}
and \textit{interference}, used in more sophisticated and useful quantum
algorithms such as that of Shor.
\index{interference!quantum}
\index{quantum parallelism}

Since the input takes one of two values, 0 and 1, as does the output, there
are only four distinct functions as shown in the table.

\begin{table}[htb]
\begin{center}
\begin{tabular}{|l|c|c|}
\hline\hline
 & $x = 0$ & $x = 1$ \\
\hline\hline
$\quad f_1$ & 0 & 0 \\
$\quad f_2$ & 0 & 1 \\
$\quad f_3$ & 1 & 0 \\
$\quad f_4$ & 1 & 1 \\
\hline\hline
\end{tabular}
\caption{The four functions which have a 1-qubit input and a 1-qubit output.
\label{deu:tab1}}
\end{center}
\end{table}

You see that $f_1$ and $f_4$ gave the same result for each input, they are
\textit{constant}. On the other hand, $f_2$ and $f_3$ give \textit{different} results
for the two inputs.  This is analogous to a coin toss. The two values of $x$
correspond to the two physical sides of the coin, the upper and the lower sides. The
function values correspond to what is represented 
on those sides, heads or tails. If
the two sides of the coin give different results (one heads
and the other tails), corresponding to a non-constant function, the coin is honest.
From now on we shall use the term ``balanced", rather than ``non-constant", to
indicate
a function which gives different results for $x=0$
and $x=1$.
However, if the two sides of the coin give the same result (both heads
or both tails), corresponding to a constant function, the coin is dishonest
since the person tossing the coin knows what the result will be. 

\index{black box}
We are given a ``black box''\footnote{The term ``black box" implies that the only information 
we can get about the function is by 
evaluating it for different inputs. We can't open up the box to see what is
inside. A black box function is often called an ``oracle".} 
function $f(x)$ and we want to learn about it. Of course we
could just feed in $x=0$ and $x=1$ and observe the results.  Suppose, however,
we only want to know whether the function is constant (satisfied by $f_1$ and
$f_4$) or balanced (satisfied by $f_2$ and $f_3$).  On a classical
computer the only thing to do is to evaluate the function for both values
of $x$ and compare them, i.e. we need to make \textit{two} calls to the
function. However, we shall see that we can answer this question on a quantum
computer with only \textit{one} call to the function. We get less information
than classically, because we don't determine the individual values of $f(0)$ and $f(1)$,
but we do determine whether or not $f$ is constant. 
Hence Deutsch's problem may be thought of as determining whether a coin to be
tossed is honest or not \textit{with just one toss of the coin}. 

\begin{figure}[htb]
\begin{center}
\includegraphics[width=6cm]{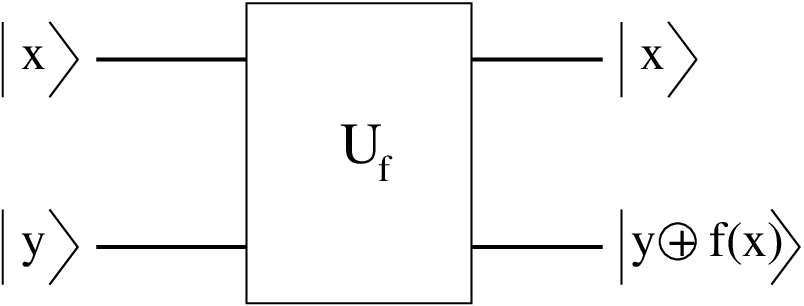}
\caption{
The blackbox routine $U_f$ for a function $f(x)$ which takes a 1-qubit input
$x$ and computes a 1-qubit function $f(x)$. Here $x$ and $y$ are computational
basis states $|0\rangle$ or $|1\rangle$. However, to gain a quantum speedup, we
will input superpositions, generated by Hadamard gates, as shown in
Fig.~\ref{Uf2}. We obtain the result of inputting a superposition from the
results of inputting computational basis states by using
linearity.
Recall that time runs from left to right in circuit diagrams.
\label{Uf}
}
\end{center}
\end{figure}

As we discussed in Chapter \ref{ch:functions},
a quantum function $f$ is implemented by a unitary operator $U_f$
as shown in Fig.~\ref{Uf}.

\begin{figure}[htb]
\begin{center}
\includegraphics[width=9cm]{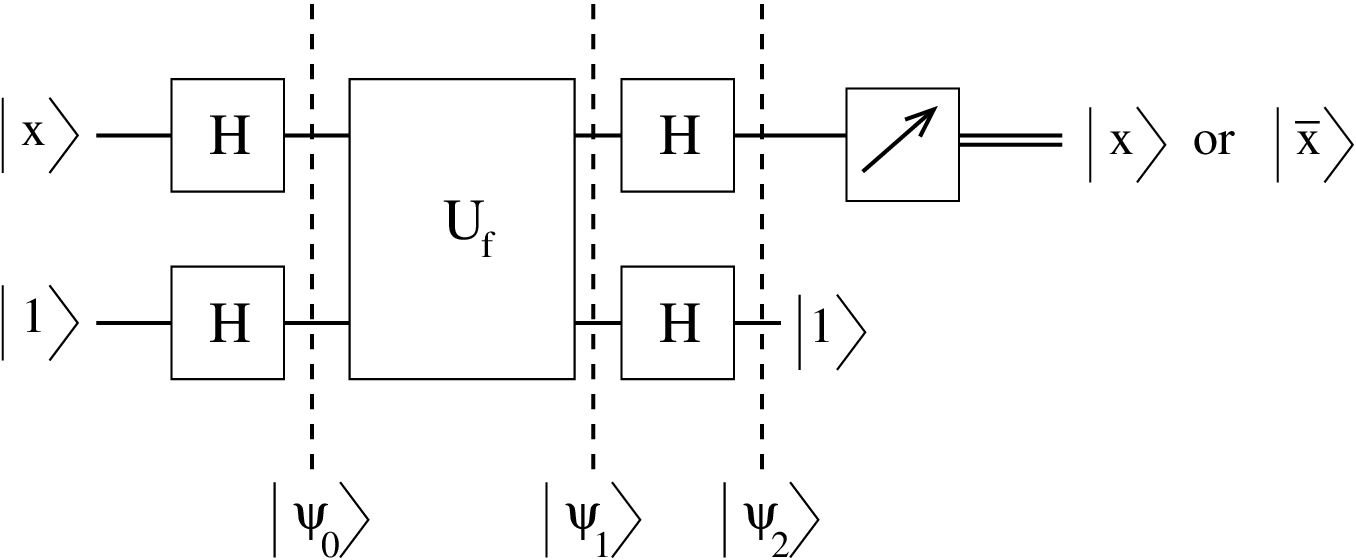}
\caption{
Circuit for Deutsch's algorithm.  The initial state (on left) can have either
$|0\rangle$  or $|1\rangle$ as the
upper (input) qubit but must have $|1\rangle$ as the lower (output) qubit. Hadamard
gates are applied to both qubits both before and after the function $U_f$
(which we assume to be an unknown black box). In the final state the lower
qubit is unchanged at $|1\rangle$. A measurement is made of the final value
(on right)
of the upper qubit. 
If the upper qubit is unchanged then the function is constant, whereas if 
it has flipped, the function is balanced.
\label{Uf2} }
\end{center}
\end{figure}

In order to take advantage of quantum parallelism we insert Hadamard
gates before the black box function $U_f$ on both the upper (input) and
lower (output) qubits,
and to take advantage of quantum interference of the results we will also put
Hadamards on both qubits after $U_f$ has acted\footnote{This is actually an
improved version of Deutsch's original algorithm.
The improved version works every time,
whereas the original version only worked half the time.}, see Fig.~\ref{Uf2}. We
initialize the
lower qubit to be
$|1\rangle$. The upper qubit could be initialized to either $|0\rangle$ or
$|1\rangle$ but it is essential to initialize the lower qubit to $|1\rangle$
as we shall see.

We will initialize the upper qubit to be $x=0$ but the conclusion would be same if
we set $x=1$.
Recalling that 
\begin{equation}
H|0\rangle = {1 \over\sqrt{2}}(|0\rangle + |1\rangle) , \quad 
H|1\rangle = {1 \over\sqrt{2}}(|0\rangle - |1\rangle) ,
\end{equation}
we find that after the first Hadamards the state in Fig.~\ref{Uf2} is
\begin{align}
|\psi_0\rangle &= {1\over 2}((|0\rangle_u + |1\rangle_u) \otimes (|0\rangle_l -
|1\rangle_l) \, , \nonumber \\
&= {1 \over 2}|0\rangle_u \otimes (|0\rangle_l - |1\rangle_l) +
{1 \over 2}|1\rangle_u \otimes (|0\rangle_l - |1\rangle_l) \, ,
\end{align}
where, in the tensor product, \index{tensor product} the 
the upper qubit (labeled ``u") is to the left and the lower qubit (labeled
``l") is to the right.

The function $U_f$ is then applied.  Recall from Fig.~\ref{Uf} that
if the state of the upper qubit is $x$, then the final
state of the lower
qubit is $f(x)$ if its initial state is zero, and the complement
$\overline{f(x)}$ if its initial state is one, i.e.
\begin{align}
|x\rangle\, |y\rangle &\to |x\rangle\,|y\oplus f(x)\rangle ,\ \mathrm{so} \\
|0\rangle\, |0\rangle &\to |0\rangle\,|f(0) \rangle , \nonumber \\
|0\rangle\, |1\rangle &\to |0\rangle\,|\overline{f(0)} \rangle , \nonumber \\
|1\rangle\, |0\rangle &\to |1\rangle\,|f(1) \rangle , \nonumber \\
|1\rangle\, |1\rangle &\to |1\rangle\,|\overline{f(1)} \rangle , \nonumber
\end{align}
Hence, after $U_f$ has been applied,
the state is 
\begin{equation}
|\psi_1\rangle = {1\over 2} |0\rangle_u \otimes (\,|f(0)\rangle_l - |\overline{f(0)}\rangle_l\, )
+ {1 \over 2} |1\rangle_u \otimes (\,|f(1)\rangle_l - |\overline{f(1)}\rangle_l \, )
\label{psi1}
\end{equation}
It is helpful to note that
\begin{align}
|f(x)\rangle_l - |\overline{f(x)}\rangle_l &= 
\left\{
\begin{array}{l}
|0\rangle_l - |1\rangle_l \quad \text{if\ } f(x) = 0, \\
|1\rangle_l - |0\rangle_l \quad \text{if\ } f(x) = 1, \\
\end{array}
\right. \nonumber \\
&=
(-1)^{f(x)} (\, |0\rangle_l - |1\rangle_l\,) \, .
\label{pkb}
\end{align}
Hence whether or not $f(x) = 0$ or $f(x) = 1$ just changes the overall sign of the
state. To get this effect it was necessary to prepare 
the lower qubit in state $|1\rangle$ rather than
$|0\rangle$. Vathsan~\cite{vathsan:16} calls Eq.~\eqref{pkb} ``\textit{phase kickback}".
\index{phase kickback}
Consequently we can write $|\psi_1\rangle$ as
\begin{equation}
|\psi_1\rangle = {(-1)^{f(0)} |0\rangle_u +
(-1)^{f(1)} |1 \rangle_u \over \sqrt{2}} \otimes {|0\rangle_l - |1\rangle_l
\over \sqrt{2}}\, .
\label{phase_kb}
\end{equation}
Now we run both qubits through Hadamards (those to the right of $U$ in
Fig.~\ref{Uf2}). It is easy to see that action on the
lower qubit (right hand one in the tensor product) is to convert
${1\over\sqrt{2}}(|0\rangle_l - |1\rangle_l)$ back to $|1\rangle_l$. The action of $H$ on
the upper qubit is to give
\begin{equation}
 {1 \over 2} \left[\, (-1)^{f(0)} (|0\rangle_u + |1\rangle_u) + 
(-1)^{f(1)} (|0 \rangle_u - |1\rangle_u) \, \right]
\end{equation}
which can be written as
\begin{equation}
{1\over 2} |0\rangle_u \left[\, (-1)^{f(0)} + (-1)^{f(1)} \, \right]
+  {1\over 2} |1\rangle_u \left[\, (-1)^{f(0)} - (-1)^{f(1)} \, \right] \, .
\end{equation}
Clearly this is $\pm |0\rangle_u$ if $f(0) = f(1)$ (where the plus sign is for
$f(0) = f(1) = 0$ and the minus sign for $f(0) = f(1) = 1$), and is $\pm |1\rangle_u$ if
$f(0) \ne f(1)$ (where the sign depends on whether $f(0)=1, f(1) = 0$ or vice
versa). Hence the state  to the right of the Hadamards in
Fig.~\ref{Uf2} is
\begin{equation}
|\psi_2\rangle = \left\{
\begin{array}{l}
\pm |0\rangle_u \otimes |1\rangle_l \quad \text{if\ } f(1) = f(0)\, , \\
\pm |1\rangle_u \otimes |1\rangle_l \quad \text{if\ } f(1) \ne f(0)\, . \\
\end{array}
\right.
\label{deu:psi2}
\end{equation}
Consequently, if a measurement of the upper qubit in Fig.~\ref{Uf2} (left in the tensor product) 
finds that it is unchanged at $x=0$
from its value in the initial state then $f(0)=f(1)$, whereas if it is flipped to $x=1$ then
$f(0) \ne f(1)$.
It does not matter whether we
initialize the upper qubit to be $0$ or $1$, the conclusion is the same.
Namely, if
the upper qubit is unchanged then the
function is constant, whereas if it is flipped the function is balanced.

We do this with \textit{one} call to the function so we have
achieved a ``quantum speedup" of 2, which is admittedly not spectacular but it
is interesting that we get any speedup at all. We will get more impressive speedups
in later algorithms.

If we could
measure the sign of the state we could determine the values of $f(0)$ and
$f(1)$ separately but the sign of the state (more generally its phase) has no
measurable effect and can not be determined.

\index{superposition}
A crucial role has been played by the Hadamards.  Those which act before $U$
is called generate a superposition state with both inputs $x=0$ and $1$
present. Looking at Eq.~\eqref{psi1} it ``seems'' that $U$ has computed
$f(x)$ for both values of $x$ with just one call to it.
This is ``\textit{quantum parallelism}".
If we do a measurement directly after
the application of $U$ we only get one value. However, for certain problems
like this one, if we do some
additional post-processing (in this case acting with Hadamards again), we can use
``\textit{quantum interference}" between the different pieces in the
superposition to set to zero the probability of getting certain results (in
this case all possible results bar one are suppressed). Consequently it is
possible to
get useful information (in this case whether the function is
constant or not) when the measurement is \textit{subsequently} done.  

Note that the Deutsch algorithm is not probabilistic: \textbf{it succeeds with
probability 1.} This shows that quantum algorithms don't necessarily have to
be probabilistic (though many are).  In this case, quantum interference
transforms the state to be measured into
an eigenstate of the computational basis. As we
know, if we measure an eigenstate we always get the same answer (the
eigenvalue) and there is no uncertainty.

\begin{center}
{\Large\bf Appendices}
\end{center}

\begin{figure}[htb!]
\begin{center}
\includegraphics[width=13cm]{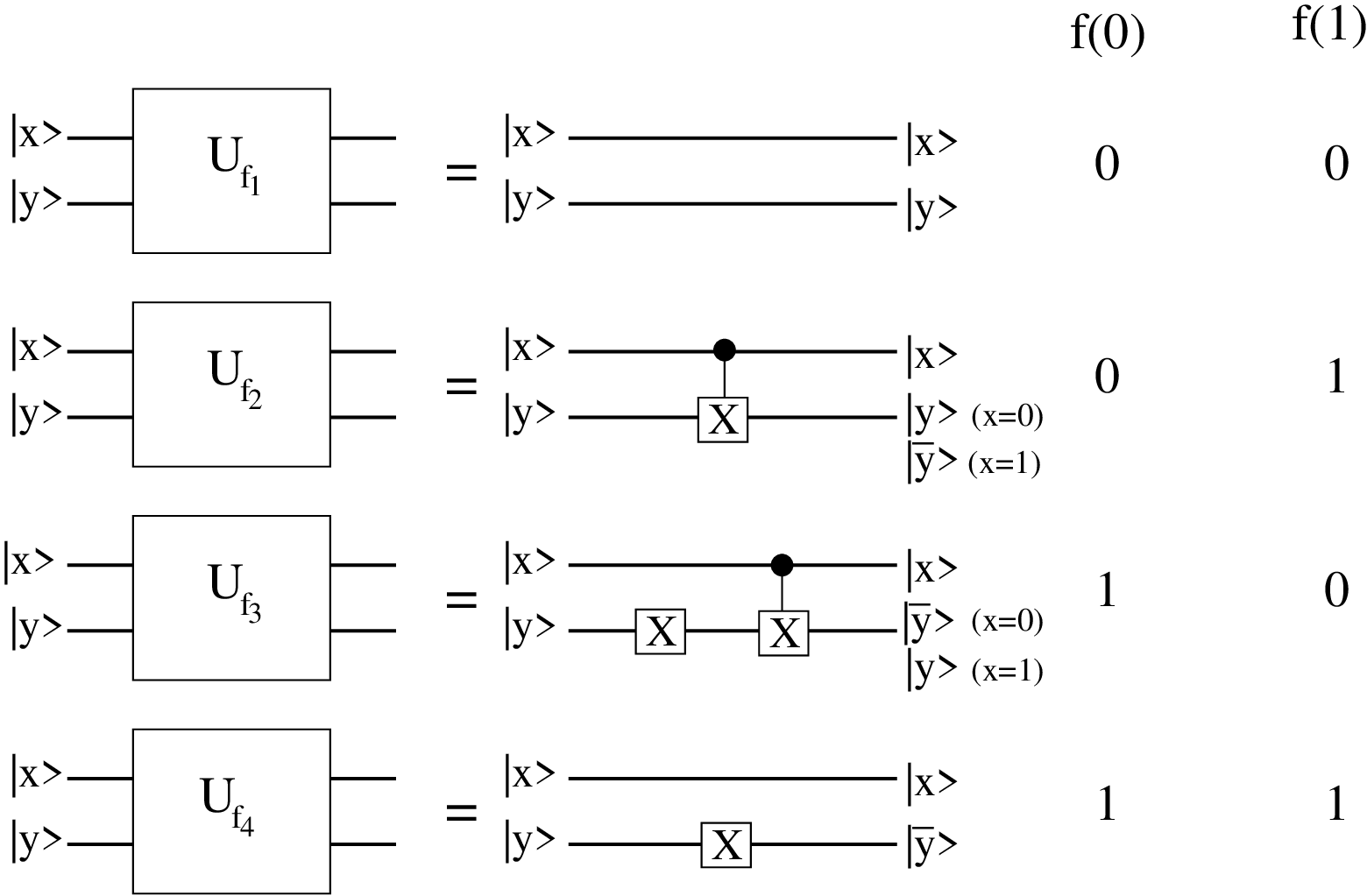}
\caption{
Circuit diagrams for each of the four functions $f_1, \cdots,f_4$ in Table
\ref{deu:tab1}. As seen in Fig.~\ref{Uf}, the function flips the lower (output) qubit
if the result of the function is $1$ but leaves it alone if the function gives
$0$.
If $f(x) =0$, $y$ is unchanged no matter what the value of $x$, but
if $f(x)=1$ then $y$ is flipped and becomes $\overline{y}$, the complement.
Note that $x$ is always unchanged.
For example,
with $f_1$ (top diagram), nothing happens. For $f_4$, $y$ is always flipped which
is done with the $X$ gate on the lower qubit.
For $f_2$, $y$ is only flipped if $x=1$ which is done by
the CNOT gate as shown. For $f_3$, $y$ is only flipped if $x=0$ which can be
accomplished by the extra $X$ gate on the $y$-qubit. 
\label{Uf_4}
}
\end{center}
\end{figure}
\begin{subappendices}
\section{An alternative derivation}
\label{app:1}

This appendix is based on Mermin~\cite{mermin:07}.

To familiarize ourselves with quantum circuits we will
obtain Eq.~\eqref{deu:psi2} in a different way by explicitly writing down circuits
for the four functions $f_1$ to $f_4$, see Fig.~2.1 of
Mermin~\cite{mermin:07}. Noting that the function flips the lower (output) qubit
if the result of the function is $1$ but leaves it alone if the function gives
$0$, we can represent the four functions in Table~\ref{deu:tab1} by the circuits shown
in Fig.~\ref{Uf_4}.

\begin{figure}[tbh!]
\begin{center}
\includegraphics[width=14cm]{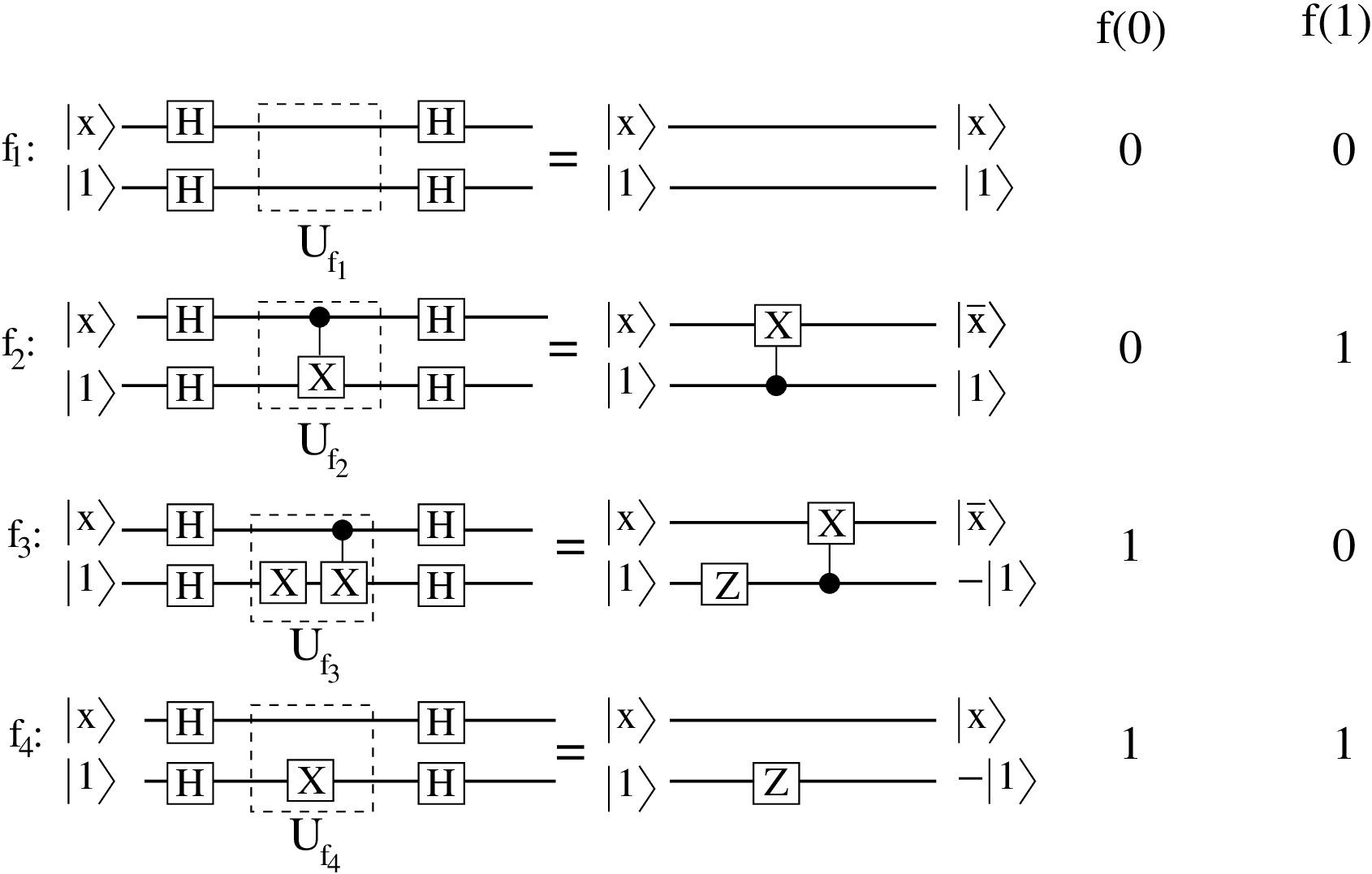}
\caption{
The circuits for the four functions $f_1,\cdots,f_4$ given in Fig.~\ref{Uf_4}
when sandwiched between Hadamards in order to perform the Deutsch algorithm.
The upper qubit is initialized in either of computational basis states,
$|x\rangle$ with $x=0$ or $1$, while the lower qubit is initialized to be
$|1\rangle$. The derivations of the equivalent circuits shown are given in the
text. One sees that the upper qubit is flipped for those functions which are
balanced, and is not flipped for the constant functions.
\label{Uf_4H}
}
\end{center}
\end{figure}

Explanations of why each circuit is equivalent to the
corresponding function are given in the figure caption.
We sandwich each of these functions between Hadamards to carry out the
Deutsch algorithm, as shown in Fig.~\ref{Uf2}, and prepare the qubits in the
initial state $|x\rangle \otimes |1\rangle$.  The results are shown in
Fig.~\ref{Uf_4H}.

We now explain each of the diagrams in this
figure.

\begin{itemize}
\item $\mathbf{f_1}$:\\
This follows simply because $U_{f_1}$ makes no change, see Fig.~\ref{Uf_4},
and $H^2 = \mathbbm{1}$ (the
identity), see Fig.~\ref{idents}(a) in Appendix \ref{app:2},
so the final qubits are the same as the initial qubits, $|x\rangle
\otimes |1\rangle$, see Fig.~\ref{Uf_4H}. In particular $x$ is unchanged indicating, correctly that
the function is constant.
\item $\mathbf{f_2}$:\\
The function $U_{f_2}$ has a CNOT gate in which the upper qubit is the
control and the lower qubit is the target, see Fig.~\ref{Uf_4}.  The result of sandwiching a CNOT
between Hadamards is, perhaps surprisingly, to interchange the role of the
target and control qubits.
This is shown in Appendix \ref{app:2}, see Fig.~\ref{idents}(f). Hence we see
that $x$ is flipped because the lower qubit is set to $|1\rangle$, see
Fig.~\ref{Uf_4H}. This is
correct because the function is balanced.
\item $\mathbf{f_3}$:\\
The 
circuit for $U_{f_3}$ is shown in
Fig.~\ref{Uf_4}. Noting that $H^2 = \mathbbm{1}$, one can insert two Hadamards
between the two $X$ gates in the circuit for $U_{f_3}$ in Fig.~\ref{Uf_4}.
As we noted for $\mathbf{f_2}$, the effect of putting Hadamards on either side of the CNOT
gate is to
interchange the role of the
target and control qubits.
In addition, we have $H X H = Z$, see Fig.~\ref{idents}(b) in the
Appendix \ref{app:2}. Hence $x$ is flipped and there is a sign change, see
Fig.~\ref{Uf_4H}. We
can't measure the sign change but the fact that $x$ is flipped correctly
indicates that the function is balanced.
\item $\mathbf{f_4}$:\\
The function $U_{f_4}$ has an $X$ gate on the lower qubit, see Fig.~\ref{Uf_4}, and 
again we have $H X H = Z$. Hence $x$ remains unchanged and there is a sign
change, see
Fig.~\ref{Uf_4H}. Again we cannot measure the sign change and the fact that $x$ is not
flipped indicates correctly that the function is constant.
\end{itemize}

\section{Derivation of some useful identities in quantum circuits}
\begin{figure}[tbh!]
\begin{center}
\includegraphics[width=10cm]{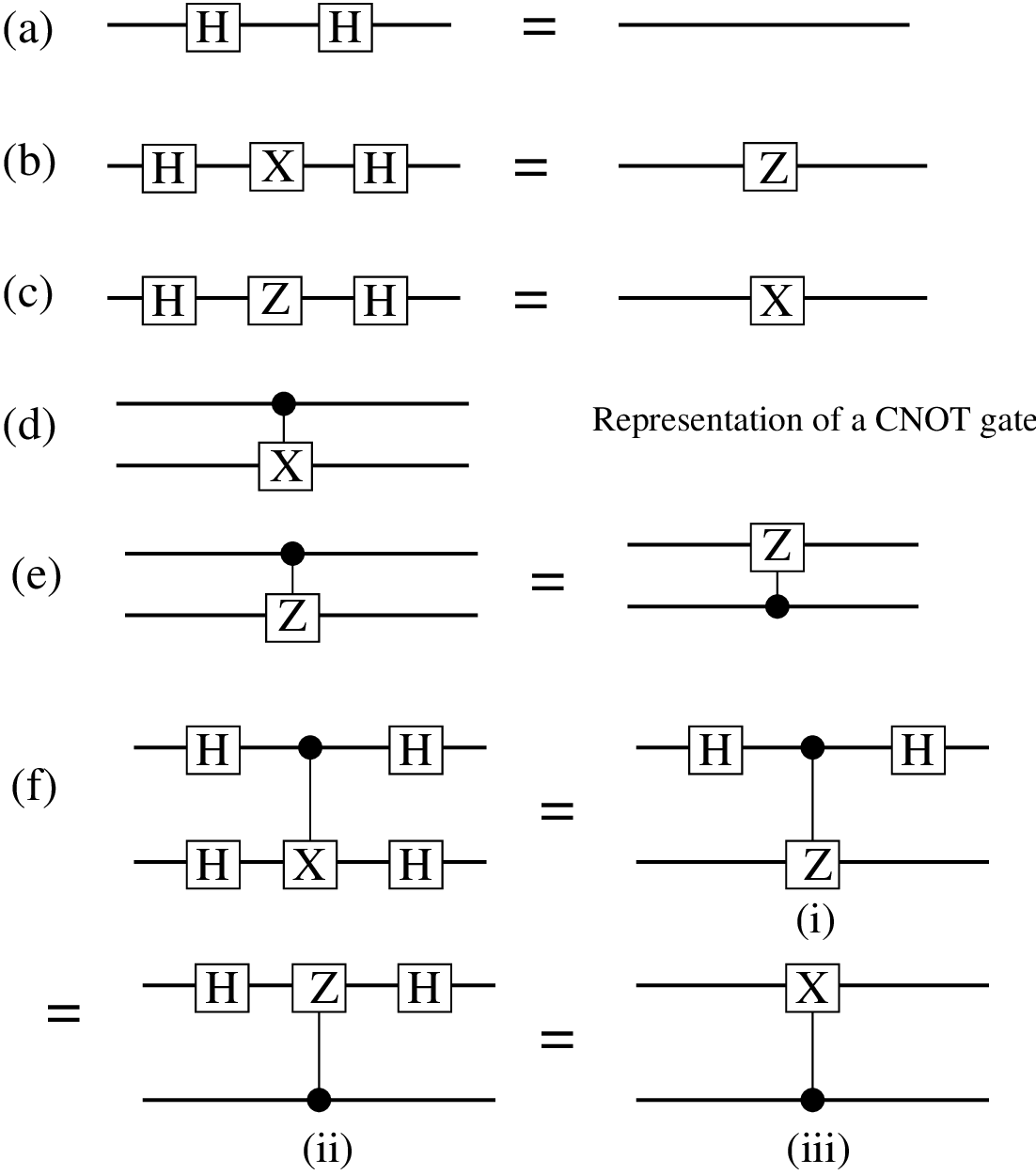}
\caption{
Some useful identities in quantum circuits. Of particular note is identity
(f) which shows that putting Hadamards around a CNOT gate is equivalent to a
CNOT gate without Hadamards, but with the control and target qubits
interchanged. These circuit identities can equivalently, but less conveniently,
be viewed as matrix identities. Identities (a)--(c) involve a single qubit and so
involve $2 \times 2$ matrices, while identities (d)--(f) involve two
qubits and so correspond to $4 \times 4$ matrices.
\label{idents}
}
\end{center}
\end{figure}

\index{circuit identities}
\label{app:2}
We have
\begin{equation}
X = 
\begin{pmatrix}
0 & 1 \\
1 & 0 \\
\end{pmatrix}
, \quad Z = 
\begin{pmatrix}
1 & 0 \\
0 & -1 \\
\end{pmatrix}
, \quad H = {1 \over \sqrt{2}}
\begin{pmatrix}
1 & 1 \\
1 & -1 \\
\end{pmatrix}
.
\end{equation}
By direct calculation it is easy to see that $X^2 = \mathbbm{1}, Z^2 =
\mathbbm{1}$, and
\begin{equation}
H^2 = \mathbbm{1}\, ,
\label{H2}
\end{equation}
where $\mathbbm{1}$ is the identity
\begin{equation}
\mathbbm{1} = 
\begin{pmatrix}
1 & 0 \\
0 & 1 \\
\end{pmatrix}
.
\end{equation}
Equation \eqref{H2} is represented graphically by Fig.~\ref{idents}(a)
Also by direct calculation, we have $X H = H Z$. Hence multiplying on the left
by $H$ gives
\begin{equation}
H X H = Z \, ,
\end{equation}
see Fig.~\ref{idents}(b) for a graphical illustration,
and multiplying on the right by $H$ gives
\begin{equation}
H Z H = X \, ,
\end{equation}
which is illustrated graphically in Fig.~\ref{idents}(c)

The NOT part of the CNOT gate is performed by the X operator.  Hence we
represent the CNOT gate as a control-X gate as in
Fig.~\ref{idents}(d).  We will also meet the control-Z gate, in which the
target
qubit is acted upon by $Z$ if the control qubit is 1, and otherwise the target
qubit is unchanged.  As with the control-X gate, there is no change in the
control qubit. With a bit of thought, we see that the
only effect of the control-Z gate is to change the overall sign of
the state if both the target and control qubits are one.  Thus the distinction
between target and control is non-existent, so control and target qubits can
be interchanged in a control-$Z$ gate, see Fig.~\ref{idents}(e).

Now consider a CNOT (control-$X$) gate sandwiched between Hadamards as shown
in Fig.~\ref{idents}(f). Consider the target (lower) qubit. If the control
qubit does not act on it, the target qubit is just acted on by the two Hadamards which is
equivalent to the
identity, see Fig.~\ref{idents}(a). If the control qubit does act on the
target qubit, the target qubit is acted on by the succession of gates $HXH$
which is equivalent to $Z$, see Fig.~\ref{idents}(b).  Both these
possibilities are taken care of by the equivalent circuit in
Fig.~\ref{idents}(f)(i), which is control-$Z$ gate. As illustrated in
Fig.~\ref{idents}(e), the target and control qubits in a control-$Z$ gate can
be interchanged so Fig.~\ref{idents}(f)(i) is equivalent to
Fig.~\ref{idents}(f)(ii). Now the target qubit is the upper one, and has the
sequence of gates $H\,$Ctrl-$Z H$ acting on it. Similar to the argument that 
showed Fig.~\ref{idents}(f) is equivalent to Fig.~\ref{idents}(f)(i),
this is equivalent to Ctrl-$X$
because of the identities in Fig.~\ref{idents}(a)
and Fig.~\ref{idents}(c). Hence Fig.~\ref{idents}(f) is equivalent to
Fig.~\ref{idents}(f)(iii).

So we see that a CNOT surrounded by Hadamards is equivalent to a CNOT gate
without Hadamards but with the control and target qubits interchanged, a quite
surprising result.

One could also derive this result by multiplying $4 \times 4$ matrices which
is more tedious.  However, for completeness we will do it here.
The CNOT gate has the matrix representation
\begin{align}
& \quad |00\rangle \ \ \, |01\rangle \ \ \, |10\rangle \ \ \, |11\rangle \\
U_{CNOT} = 
& \begin{pmatrix}
\ 1\quad  & \quad\  0\  & \quad 0\  & \quad 0\  \\
\ 0\quad  & \quad\  1\  & \quad 0\  & \quad 0\  \\
\ 0\quad  & \quad\  0\  & \quad 0\  & \quad 1\  \\
\ 0\quad  & \quad\  0\  & \quad 1\  & \quad 0\  \\
\end{pmatrix}
.
\end{align}
\index{tensor product}
\index{control qubit}
\index{target qubit}
In this tensor product the control qubit is to the left. The target qubit (to
the right) is flipped if the control qubit (to the left) is 1 (so, relative to
the identity matrix, columns 3 and
4 are interchanged). In a CNOT gate with target and control qubits swapped, the left hand qubit
is flipped if the right hand qubit is 1 (so columns 2 and 4 are interchanged).
Hence we have
\begin{align}
& \quad |00\rangle \ \ \, |01\rangle \ \ \, |10\rangle \ \ \, |11\rangle \\
U_{CNOT\_SWAP} =
& \begin{pmatrix}
\ 1\quad  & \quad\  0\  & \quad 0\  & \quad 0\  \\
\ 0\quad  & \quad\  0\  & \quad 0\  & \quad 1\  \\
\ 0\quad  & \quad\  0\  & \quad 1\  & \quad 0\  \\
\ 0\quad  & \quad\  1\  & \quad 0\  & \quad 0\  \\
\end{pmatrix}
.
\end{align}

The tensor product $H^{\otimes 2}$ is given by
\begin{equation}
H^{\otimes 2} = {1 \over \sqrt{2}}
\begin{pmatrix}
H & H \\
H & -H \\
\end{pmatrix}
= {1 \over 2} 
\begin{pmatrix*}[r]
1 & 1 & 1 & 1 \\
1 & -1 & 1 & -1 \\
1 & 1 & -1 & -1 \\
1 & -1 & -1 & 1 \\
\end{pmatrix*}
\end{equation}
One can check by working out the matrix multiplication that
\begin{equation}
U_{CNOT\_SWAP} = H^{\otimes 2}\, U_{CNOT}\, H^{\otimes 2} \, ,
\end{equation}
in agreement with Fig.~\ref{idents}(f).
This is a bit tedious so I used \textit{Mathematica}. It is more
straightforward to use the circuit identities shown in Fig.~\ref{idents}.
\end{subappendices}

%% file: bv8.tex
\section{The Algorithm}
\label{sec:bv1}
Like the Deutsch algorithm, the Bernstein-Vazirani\index{Bernstein-Vazirani algorithm}
algorithm finds information
\index{black box}
about a black box function, but has a bigger speedup. It is very similar to
the Deutsch-Josza algorithm 
which is set as a homework problem.

The black box takes as input an $n$-bit integer $x$ and returns as output a single bit
containing the value $a \cdot x$ where $a$ is an $n$-bit constant and
the dot indicates a bitwise inner product with modulo 2 addition:
\index{bitwise inner product}
\begin{equation}
f(x) = a \cdot x \equiv a_0 x_0 \oplus a_1 x_1 \oplus \cdots \oplus a_{n-1} x_{n-1}
\, .
\label{dot}
\end{equation}
The problem is to determine $a$. 

Let's make sure that we understand the ``dot".
We have $a_i x_i = 0$ or $1$, where the value $1$ only occurs if both $a_i$
and $x_i$ are equal to $1$. Hence
\begin{align}
a \cdot x &= a_0 x_0 \oplus a_1 x_1 \oplus \cdots \oplus a_{n-1} x_{n-1} \\
&= \left\{
\begin{array}{ll}
1 & \text{if\ an\ odd\ number\ of\ terms\ is\ 1} \\
0 & \text{if\ an\ even\ number\ of\ terms\ is\ 1} 
\end{array}
\right.
\end{align}
For example for $n=4$, if the bits of $a$ are $1101$ and the bits of $x$ are
$1110$ (recall that the zeroth bit is the least significant, i.e.~the
rightmost one) then\footnote{One can either do the mod 2 operation after each
addition or add up in the normal way and apply the mod 2 operation at the end.
In either case, the result is $0$ if an even number of terms in the sum are $1$, and $1$ 
if an odd number of terms are $1$.}
\begin{equation}
a \cdot x = (1 \times 0) +  (0 \times 1) + (1 \times 1) + (1 \times 1) \!\!\! \mod 2 = 0 + 0
+ 1 + 1 \!\!\! \mod 2 = 2 \!\!\! \mod 2 = 0.
\end{equation}
Hence, for these values of $a$ and $x$, $f(x) = 0$. If we take $x = 1000$ then
$f(x) = 0 + 0 + 0 + 1 \!\!\! \mod 2 = 1$.

Classically we can only determine the bits of $a$ one at a time. The $k$-th bit of $a$ can
be determined by feeding in $x = 2^k$.
To see this, consider the binary representations of $a$ and $x$:
\begin{equation}
\begin{split}
a &= a_0 + a_1 2^1 + \cdots + a_k 2^k + \cdots + a_{n-1} 2^{n-1} \, , \\
x &= x_0 + x_1 2^1 + \cdots + x_k 2^k + \cdots + x_{n-1} 2^{n-1} \, .
\end{split}
\end{equation}
Hence if $x = 2^k$ then $x_k =1$ while, for $l\ne k, x_l = 0$, so $a\cdot x = a_k$.
Consequently $f(2^k) = a_k$.
We have to do this for each bit, $k = 0, 1, 2, \cdots , n-1$, so it requires
$n$ calls of the function.

We will see that the quantum algorithm succeeds in determining $a$ with just \textit{one} call!

\begin{figure}[htb]
\begin{center}
\includegraphics[width=13cm]{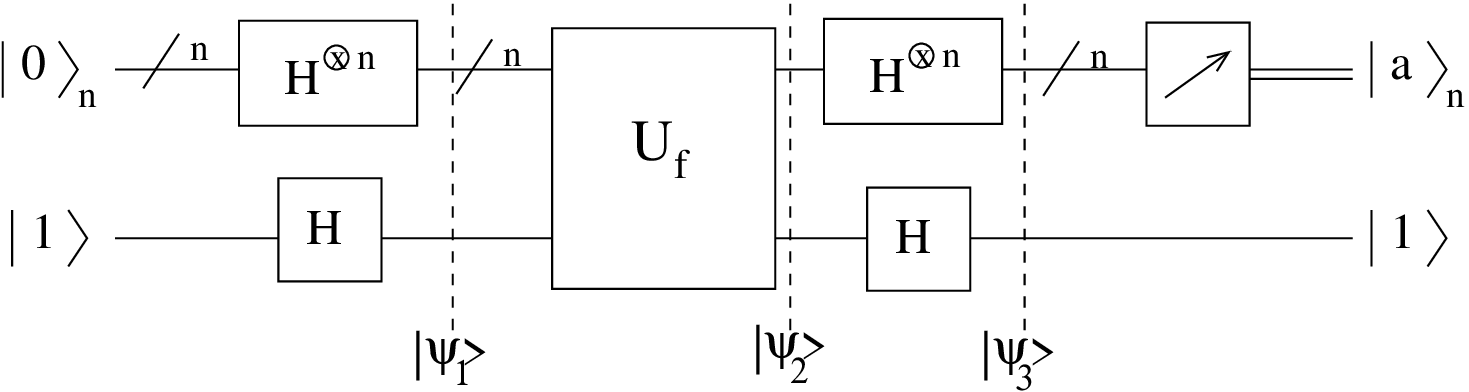}
\caption{Circuit diagram for the Bernstein-Vazirani algorithm. In the final
state the upper (input) register contains $|a\rangle$ while the
lower (output) qubit reverts to its initial state $|1\rangle$. The
desired value of $a$ can therefore be read off by measuring the upper
register.
\label{bv}
}
\end{center}
\end{figure}

A schematic diagram of a general reversible unitary transformation
which takes an $n$-bit input $x$ in the upper register
and generates an $m$-bit output $f(x)$ in the lower register is shown in
Fig.~\ref{Unm}. 
For the Bernstein-Vazirani Algorithm there are $n$ qubits in the
upper register but only $1$ qubit in the lower register.
In addition, the
unitary $U_f$ is surrounded by Hadamards, as shown in
Fig.~\ref{bv}. The upper register is set to $|0\rangle_n$ and the lower qubit
to $|1\rangle$. This is the same circuit as for the Deutsch-Josza algorithm,
see problem \ref{qu:dj}.

\index{superposition}
Acting with $H$ on $|0\rangle$ gives an equal linear superposition of the two
basis states. Similarly acting with $H^{\otimes n}$ on $|0\rangle_n$ gives an
equal superposition of the $2^n$ basis states.
Hence,
including the lower register, the state inputted to
$U_f$ is
\begin{equation}
|\psi_1 \rangle = H^{\otimes n} |0\rangle_n\, \otimes\, H |1\rangle =
{1 \over \sqrt{2^n}} \sum_{x=0}^{2^{n}-1} |x\rangle_n \,
\otimes\, {|0\rangle - |1\rangle \over \sqrt{2}} \, .
\end{equation}
For each term in the superposition, the function $U_f$ acts in the same way as
for the Deutsch algorithm described in Chapter \ref{ch:deutsch}.
The lower qubit is
flipped if $f(x) =1$, which is 
the same as changing the sign of
the state.
If $f(x) = 0$ there is no change.
Hence
each term in the superposition acquires a factor of $(-1)^{f(x)}$, so the
state of the system immediately after the action of $U_f$ is
\begin{equation}
|\psi_2\rangle = {1 \over \sqrt{2^n}} \sum_{x=0}^{2^{n}-1} (-1)^{f(x)} |x\rangle_n \,
\otimes\, {\left(\,|0\rangle - |1\rangle \,\right)\over \sqrt{2}} =
{1 \over \sqrt{2^n}} \sum_{x=0}^{2^{n}-1} (-1)^{a \cdot x} |x\rangle_n \, \otimes\,
{\left(\,|0\rangle - |1\rangle \,\right)\over \sqrt{2}}\, .
\label{afterU}
\end{equation}

Next consider the effect of the Hadamards acting after $U_f$. The action on
the lower qubit is to convert $(\,|0\rangle - |1\rangle\,)/\sqrt{2}$ to
$|1\rangle$.  
However, the effect of $H^{\otimes n}$ acting on an arbitrary
computational basis state $|x\rangle_n$ needs more thought.
Consider first just one qubit. Then
\begin{equation}
H |x\rangle = {1\over \sqrt{2}} \left(\, |0\rangle + (-1)^x |1\rangle \,
\right) = {{1\over \sqrt{2}} \sum_{y=0}^1} (-1)^{x y} |y\rangle \, .
\label{Ho1}
\end{equation}
Hence the effect of applying $H^{\otimes n}$ on an $n$-qubit computational
basis state is
\begin{align}
H^{\otimes n} |x \rangle_n &= {1\over 2^{n/2}}\,\sum_{y_{n-1}=0}^1 \cdots \sum_{y_1=0}^1
\sum_{y_0=0}^1 (-1)^{\sum_{j=0}^{n-1} x_j y_j} |y_{n-1}\rangle \cdots
|y_1\rangle |y_0\rangle \, , \nonumber \\
&= {1 \over 2^{n/2}} \sum_{y=0}^{2^n - 1} (-1)^{x \cdot y} |y \rangle_n \, ,
\label{Hon}
\end{align}
where $x \cdot y$ is the bitwise inner product with modulo 2 addition defined in Eq.~\eqref{dot},
and we have used the fact that we only need to know whether
$\sum_{j=0}^{n-1} x_j y_j$ is even
or odd.
In particular, for $x = a$, we have
\begin{equation}
H^{\otimes n} |a \rangle_n = {1 \over 2^{n/2}} \sum_{y=0}^{2^n - 1}
(-1)^{a \cdot y} |y \rangle_n \, ,
\end{equation}
so Eq.~\eqref{afterU} can be written as
\begin{equation}
|\psi_2\rangle = H^{\otimes n} |a\rangle_n \otimes H |1\rangle. 
\end{equation}
Since $H^2 = \mathbbm{1}$, acting with the final Hadamards in Fig.~\ref{bv}
gives the simple result
\begin{equation}
|\psi_3\rangle = |a\rangle_n \otimes |1\rangle .
\end{equation}
Consequently a measurement of the upper register in Fig.~\ref{bv} gives $a$,
\textit{with probability one},
even though we made just 
\textit{one} call to the function.

Since a classical computation of $a$ requires $n$ function calls, we have
obtained a ``\textit{quantum speedup}" of $n$. Note that the procedure is analogous to
Deutsch's algorithm. The first set of Hadamards generates a superposition
of inputs to the gate $U_f$ which ``evaluates"\footnote{To understand the
reason for the quotation marks see the discussion at the end of
Sec.~\ref{sec:qp}.
\index{superposition}
}
the function for all $2^n$
inputs using \textit{quantum parallelism}, and then the second set of
Hadamards destroys all the outputs apart from $a$, using quantum
\textit{interference}.\index{interference!quantum}\index{quantum parallelism}

\section{An Alternative Derivation}

\begin{figure}[htb]
\begin{center}
\includegraphics[width=8cm]{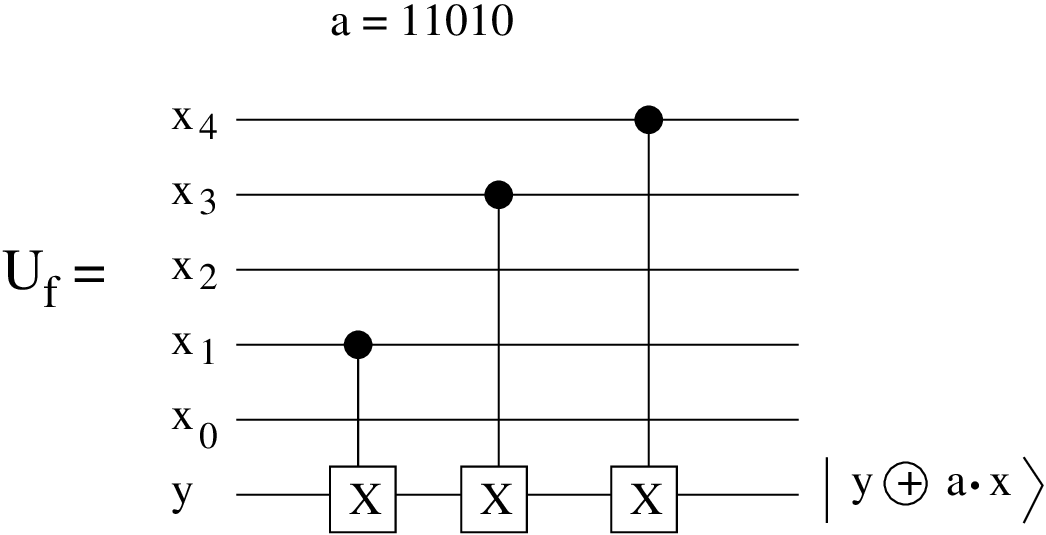}
\caption{A circuit diagram for $n=5$ to implement the function $f(x) = a \cdot x$ with
$a=11010$, i.e.~$f(x) = x_1 \oplus x_3 \oplus x_4$. The circuit
flips
the output
qubit, the lowest one, initialized to $y$, whenever $x_1 \oplus x_3 \oplus x_4 =1.$
(Note that flipping $y$ is equivalent to adding 1 to y mod 2.)
Hence the final value of the output qubit is $y \oplus (a \cdot x)
= y \oplus x_1 \oplus x_3 \oplus x_4$ as
required.
\label{ax}
}
\end{center}
\end{figure}

Following Mermin~\cite{mermin:07} and Vathsan~\cite{vathsan:16} it is useful to
give an alternative derivation of
how the circuit in Fig.~\ref{bv} works, by giving an
\textit{explicit} construction of the black box $U_f$.  It is convenient to illustrate by
a specific example. We take $n=5$ and $a = 11010$ so $a_0 = 0, a_1 = 1, a_2 =
0, a_3 = 1, a_4 = 1$ (recall we read the bits from right, the least significant,
to left, the most significant). The function $a \cdot x$ can be implemented by the gates
shown in Fig.~\ref{ax} since here $a \cdot x = x_1 \oplus x_3 \oplus x_4$ so
the lower register is $y \oplus a \cdot x = y \oplus x_1 \oplus x_3 \oplus
x_4$.

\begin{figure}[htb]
\begin{center}
\includegraphics[width=12.5cm]{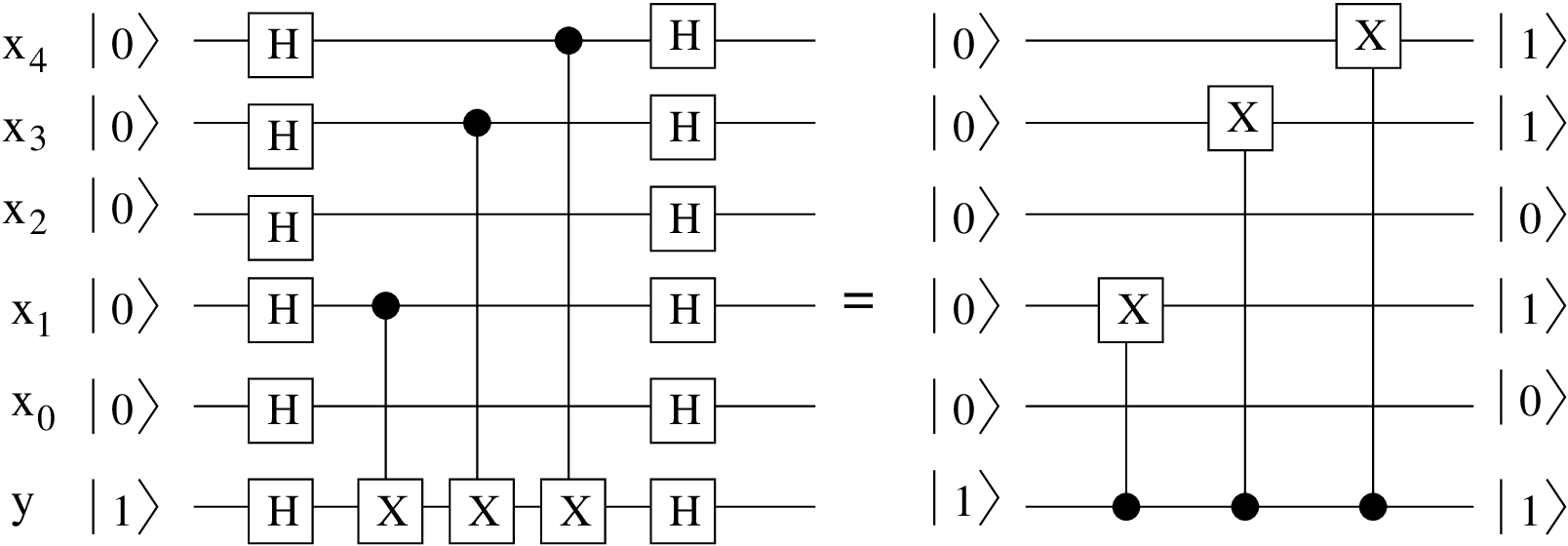}
\caption{Sandwiching the circuit for $U_f$ in Fig.~\ref{ax} between Hadamards,
and realizing that the effect of the Hadamards is to interchange the control
and target qubits in the CNOT (control-$X$) gates, we see immediately that
the final state of the upper (input) register contains $a = 11010$.
\label{bv2}
}
\end{center}
\end{figure}

To incorporate $U_f$ into the Bernstein-Vazirani algorithm, we
sandwich it in between Hadamards, see Fig.~\ref{bv}, and note that the Hadamards
interchange
\index{control qubit}
\index{target qubit}
control and target qubits in the CNOT (control-$X$) gates,
see Fig.~\ref{idents}(f) in Chapter \ref{ch:deutsch}.
As before, the initial upper register is $|0\rangle_n$ and the lower register
is $|1\rangle$.  We see immediately from Fig.~\ref{bv2} that $a$ is
\textit{directly} imprinted in the final state of the input register.  There does not
\textit{appear} to be any parallelism and interference. 
\index{interference!quantum}

Hence these two explanations of the Bernstein-Vazirani algorithm are quite
different. To quote Mermin~\cite{mermin:07}:
\begin{quotation}
\noindent ``The first applies $U_f$ to the quantum superposition of all possible inputs
and then applies operations which leads to perfect destructive interference of
\index{interference!quantum}
all states in the superposition except for the one in which the upper (input) register
is in the state $|a\rangle$. The second suggests a specific mechanism for
representing the subroutine that executes $U_f$ and then shows that
sandwiching such a mechanism between Hadamards \textit{automatically} (my
italics) imprints $a$ on the upper register. Interestingly, quantum mechanics
appears in the second method only because it allows the reversal of the
control and target qubits of a CNOT operation solely by means of 1-qubit
(Hadamard) gates."
\end{quotation}
(I have used the conventional spelling of ``qubit" rather
than Mermin's idiosyncratic ``Qbit".)
\index{control qubit}
\index{target qubit}

\hrulefill
\section*{Problems}
\input{hw_ch11.tex}

%% file: hw_ch11.tex
\begin{problems}

\item
\label{qu:dj}
\textit{The Deutsch-Josza Algorithm}\\
\index{Deutsch-Josza Algorithm}
This is an extension of the Deutsch algorithm discussed in class.  Recall that
in Deutsch's algorithm the input is one bit and the output is also one bit.  In the
Deutsch-Josza algorithm, the output is still one bit but the input has $n$
bits, so there are $2^n$ distinct inputs.  We are told that \textit{either}
the function is ``constant'' (in which case the function outputs the same value
for all $2^n$ inputs) \textit{or} is ``balanced" (in which case an equal
number of inputs give the results 1 and 0). Clearly this is a very
artificially constructed problem but it will be our first quantum algorithm
with more than a one-bit input.  Note that it is \textit{precisely} the
Deutsch algorithm for $n=1$.

The circuit for the Deutsch-Josza is almost identical to that for the Deutsch
algorithm except that the upper qubit in the Deutsch algorithm
(sometimes called the ``input" qubit) is
replaced by a $n$-qubit register. The circuit is shown in the figure below.

\begin{center}
\includegraphics[width=8cm]{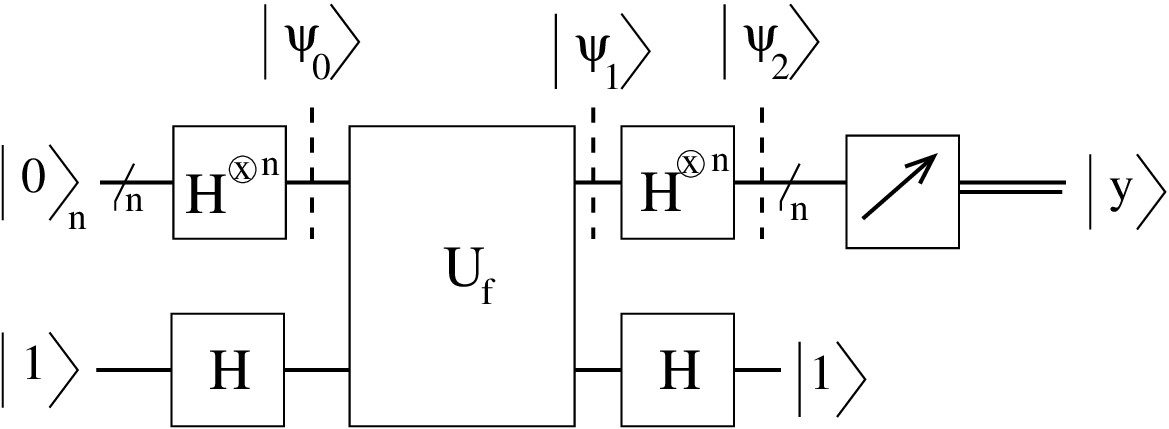}
\end{center}

The function $U_f$ acts as follows on computational basis states $|x\rangle_n$
and $|z\rangle$:
\begin{equation}
U_f |x\rangle_n |z\rangle = |x\rangle_n |z \oplus f(x)\rangle \, ,
\end{equation}
where $x$ is an $n$-bit integer, $|x\rangle$ is the state of the n-qubit
upper register in the figure, $z$ and $f(x)$ are 1-bit integers, and $|z\rangle$ is the lower
qubit in the figure.

As in the Deutsch algorithm, the lower qubit is initialized to $|1\rangle$.
In the Deutsch algorithm, the upper qubit is initialized to $|0\rangle$. Here
the single qubit is replaced by an $n$-qubit register
which, by analogy, is initialized to $|0\rangle_n$.

\begin{enumerate}[label=(\roman*)]
\item
Show that
\begin{equation}
|\psi_0\rangle_n = H^{\otimes n}|0\rangle_n = {1\over \sqrt{2^n}}
\sum_{x=0}^{2^n-1} |x\rangle_n \, ,
\label{p0}
\end{equation}
so the input to the function $U_f$ is the uniform superposition of all $2^n$
basis states. 

\item
Show that after the action of $U_f$ the state of the upper register is
\begin{equation}
|\psi_1\rangle_n = 
{1 \over \sqrt{2^n}} \sum_{x=0}^{2^n-1} (-1)^{f(x)} |x \rangle_n .
\end{equation}
The fact that the value of $f(x)$ only changes the overall sign of the state is called
the \textit{phase kickback} trick by Vathsan.
\item
Show that after the action of the second set of Hadamards on the $n$-qubit register,
the state of that register is
\begin{equation}
|\psi_2\rangle_n = H^{\otimes n}|\psi_1\rangle_n =
{1 \over 2^n} \sum_{x, y=0}^{2^n-1} (-1)^{[f(x) + x \cdot y]}
|y\rangle_n ,
\label{p2}
\end{equation}
where $x\cdot y$ is the bitwise inner product of $x$ and $y$ with modulo 2 addition:
\begin{equation}
x\cdot y = x_0 y_0 \oplus x_1 y_1 \oplus \ldots \oplus x_{n-1} y_{n-1} \, .
\end{equation}
\item
The upper register is then measured, and an $n$-bit integer $y$ is obtained.
Show that if  the function is a constant then $y=0$
with probability 1.
Show also that if the function is balanced then one must get a non-zero
value of $y$.  
Hence the Deutsch-Josza algorithm succeeds with
\textit{just one} function call.
\item
How does this compare with a classical approach? The only thing one can do
classically is
keep computing $f(x)$ for different values of $x$ and seeing if one gets more
than one value for the output.
If the function is balanced, one would probably get different
outputs quite quickly.  If the function is constant one would need to evaluate
half the inputs (plus 1), i.e.~$2^{n-1}+1$, to be 100\% sure that the function is not balanced.
This is exponentially (in $n$) worse than the quantum algorithm. 

However, this is arguably not fair.  We may well be content to establish that the
function is constant with some high probability\footnote{For later quantum
algorithms we will only be able to solve the problem with high probability.
Since we have to give up 100\% certainty in the quantum case, we we should not insist on 100\%
certainty here from the classical algorithm.}, a bit less than one. If the
function is constant, how many function calls would you need classically to
rule out the possibility that it is balanced with a probability of error of no
more than (i) $10^{-3}$ and (ii) $10^{-6}$. \\
\textit{Note:} For simplicity, assume that the number of function calls is much
less than $2^{n/2}$, the number of values of $x$ which give the same result if
the function is balanced.
\end{enumerate}

\item
\label{qu2}
Consider the Deutsch-Josza algorithm for $n=2$, and assume a constant function
$f(x) = 0$ for all $x$, i.e.~$x=0, 1, 2, 3$. Compute explicitly the state of the
system at each stage and show that you get the state $|y\rangle = |00\rangle$
in the upper register at the end. \\
\textit{Hint:} Evaluate explicitly Eqs.~\eqref{p0}--\eqref{p2} for this situation.

\item
Consider again the Deutsch-Josza algorithm for $n=2$ but this time assume that
$f(00) = f(01) = 0, f(10) = f(11) = 1$ (a balanced fiunction). Determine the final state of the
upper register and show that this implies the function is balanced, as indeed
it is. \\
\textit{Hint}: See the hint for Qu.~\ref{qu2}.

\item
\label{qu:toffoli}
\index{Toffoli gate}
\textit{The Toffoli Gate}.\\
We stated in Sec.~\ref{sec:class} that for \textit{classical} reversible computation we
need three-bit gates, such as the Toffoli gate, in addition to 1-bit and
2-bit gates, to be able to perform universal computation. However, three qubit
gates are, fortunately, not needed in quantum computation 
because the appropriate three-bit gates can be constructed out of
1-qubit and 2-qubit gates.

Here we consider the quantum Toffoli gate, which is a control-control-NOT (C-C-NOT)
gate:

\begin{center}
\includegraphics[width=3.5cm]{toffoli.eps}
\end{center}

The target qubit $z$ is flipped if both the control qubits, $x$ and
$y$, are 1 and is otherwise unchanged.

\begin{enumerate}[label=(\roman*)]
\item
Consider the following circuit for an arbitrary unitary operator $V$:

\begin{center}
\includegraphics[width=8cm]{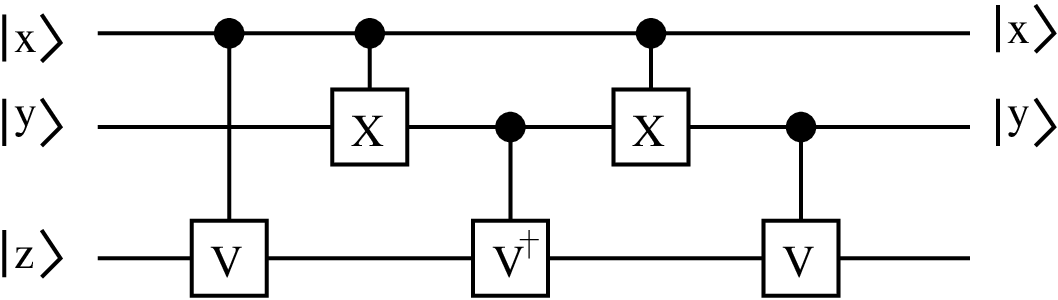}
\end{center}

Show that it acts with $V^2$ on $|z\rangle$ if both $x$ and $y$ are
1 and otherwise does nothing.\\
\textit{Hint:} One possible way of approaching this question (though not the
only way) is to consider separately what happens
for the four
possible input values of the control qubits $x$ and $y$, namely 00, 01, 10, and 11.\\
Another, more elegant, way is to note that the effect of
a Ctrl-$V$ gate in which $|z\rangle$ is the target and $|x\rangle$ is the
contol is
$V\, |x\rangle|z\rangle \longrightarrow |x\rangle V^x|z\rangle.$

\item
Now take $V$ to be the following 1-qubit gate:
\begin{equation}
V = (1 - i) {(\mathbbm{1} + i \, X) \over 2} \, . 
\end{equation}
Show that $V^\dagger V = \mathbbm{1}$, and hence $V$ is unitary. Show 
also that $V^2= X$ and hence the above circuit is a quantum Toffoli gate.\\
\textit{Note:} One sometimes says that $V$ is the ``square root of $X$''.

\end{enumerate}

\end{problems}

%% file: simon8.tex
\index{Simon's algorithm}
So far we have studied Deutsch's algorithm in Chapter \ref{ch:deutsch}
\index{Deutsch's algorithm}
which gave a quantum
speedup of a factor of 2, and the Bernstein-Vazirani algorithm in Chapter
\ref{ch:bv},\index{Bernstein-Vazirani algorithm}
which gave a speedup of
$n$, where $n$ is the size of the problem. Next we consider a problem, due to
Daniel Simon, which gives an \textit{exponential} speedup in $n$. Like the
previous algorithms it has an artificial character and is not of practical
use, but it has features in common with the vastly more useful algorithm of
Shor for factoring integers, which we shall spend a substantial amount of time
on in the next few chapters.
Like Shor's algorithm, Simon's is of a probabilistic nature.

In Simon's problem we are given a black box function which takes an $n$-bit input
\index{black box}
and has the property that
\begin{equation}
f(x \oplus a) = f(x),
\label{fxa}
\end{equation}
where $a$ is a non-zero $n$-bit integer and
$\oplus$ means bitwise addition modulo 2.
\index{bitwise addition}
Note that each bit is treated separately, so if the integer $x$ is represented
in binary notation by bits $x_{n-1}x_{n-2}\cdots x_1 x_0$, and similarly for
$a$ then $x \oplus a$ is an integer $y$ with binary representation $y_{n-1}
y_{n-2} \cdots y_1 y_0$ where $y_j = x_j \oplus a_j$.

Adding $a$ twice to $x$ (modulo 2) gives
back $x$, i.e.
\begin{equation}
x \oplus a \oplus a = x 
\end{equation}
since adding 
a bit to itself gives $0\,(\mod 2)$ irrespective of whether that bit is 0 or
1. Hence 
\begin{equation}
f(x) = f(x\oplus a) = f(x \oplus a \oplus a) 
\label{fxaa}
\end{equation}
and so on, so $f(x)$
is \textit{periodic}, with period $a$, under bitwise mod 2 addition. 
We are told that for every $x$ there is only one other input to the function, $x \oplus a$,
which gives the same output, so there are $2^{n-1}$ distinct values of $f$.
Hence we assume that we can represent $f$ by $n-1$ qubits. 
An example of a function with the desired property is shown in Table \ref{simon:tab1}.

The problem is to determine the period $a$ with the least number of function
calls.

If we input different values of $x$ and find a repeated output, 
i.e.~if $f(x_i) = f(x_j)$, then
$x_j = x_i \oplus a$. If we add $x_i$ to both sides (bitwise addition
modulo 2) we get
\begin{equation}
a = x_i \oplus x_j \, .
\label{xixj}
\end{equation}
so we obtain $a$ if we can find two values of $x$ which give the same
function value.

\begin{table}[htb]
\begin{center}
\begin{tabular}{|c|cccccccc|}
\hline
$x$ & 0 & 1 & 2 & 3 & 4 & 5 & 6 & 7 \\
\hline\hline
$f(x)$ & 3 & 2 & 2 & 3 & 0 & 1 & 1 & 0 \\
\hline
\end{tabular}
\caption{
An example with $n=3$ bits
of the type of function that is considered in Simon's algorithm.
The function
satisfies $f(x) = f(x \oplus a)$ for some non-zero $a$.
To determine $a$ we look for repetitions.  An example is $f(4) =
f(7) = 0$.  Hence, according to Eq.~\eqref{xixj},
$a = 4 \oplus 7 = 100 \oplus 111 = 011 = 3$.  The other repetitions
satisfy this same condition as you can check. 
\label{simon:tab1}
}
\end{center}
\end{table}


\textbf{Classically} this problem is hard, by which we mean that the number of function calls grows
\textit{exponentially} with $n$. All one can do is call the function with
different
values of $x$ until one finds a repeated output, i.e.~$f(x_i) = f(x_j)$,
which gives us $a$ from Eq.~\eqref{xixj}.
After $m$ calls to the function we have compared $m(m-1)/2$ pairs.  For a
reasonable chance of success we need $\smfrac{1}{2}m(m-1) \sim 2^n$, so
$m = O(2^{n/2})$, i.e.~exponential in the number of bits $n$.

\begin{figure}[htb]
\begin{center}
\includegraphics[width=14cm]{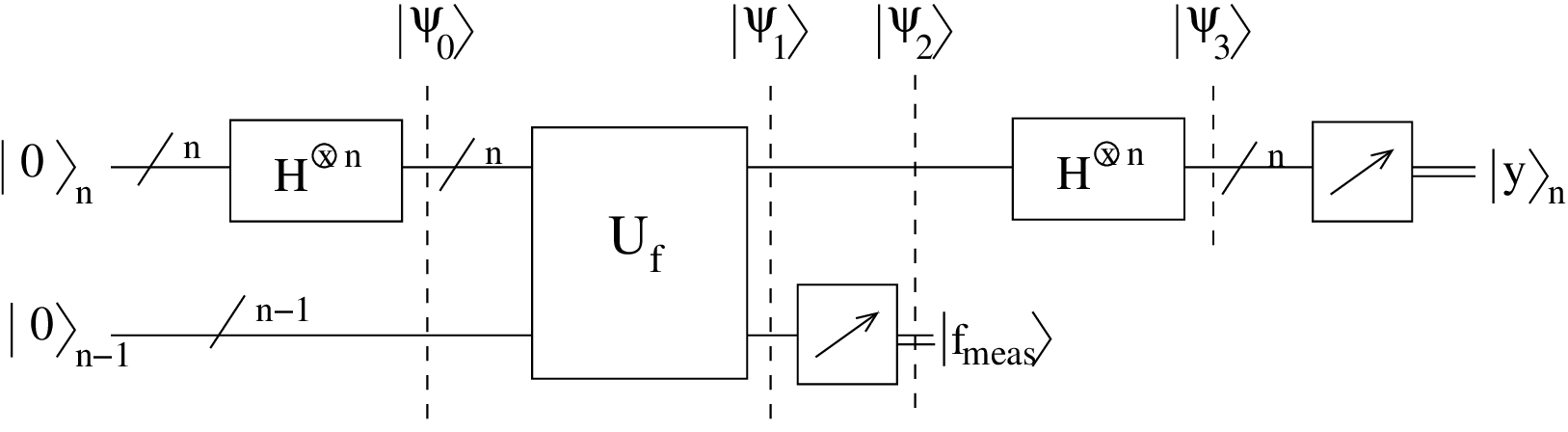}
\caption{Circuit diagram for Simon's algorithm.  The upper register has $n$
qubits and contains the $x$ values, while the lower register has $n-1$ qubits
and contains the values of the function $f(x)$.
\label{simon}
}
\end{center}
\end{figure}

The circuit to solve this problem \textbf{quantum mechanically}
is similar to that in the Bernstein-Vazirani\index{Bernstein-Vazirani
algorithm} algorithm
except that the lower register has enough qubits to contain the function
values, i.e.~$n-1$. 
Also the phase
kickback \index{phase kickback}is not used, so the lower register is initialized to $|0\rangle_{n-1}$
rather than $|1\rangle$ and we do not have Hadamards on the lower register. A
final difference is that we measure first on the lower register rather than the
upper one.  The circuit diagram is shown in Fig.~\ref{simon}.

After the first Hadamards in the upper register the state of the system is
\begin{equation} 
|\psi_0\rangle = {1 \over 2^{n/2}} \sum_{x=0}^{2^n-1} |x\rangle_n
\otimes |0\rangle_{n-1} \, .
\end{equation}
The function call makes the transformation $|x\rangle_n \otimes |y\rangle_{n-1}\to
|x\rangle_n \otimes |y\oplus f(x)\rangle_{n-1}$, see Fig.~\ref{bv} in Chapter
\ref{ch:bv}.
Here $y=0$ so,
after the function call the state becomes
\begin{equation} 
|\psi_1\rangle = {1 \over 2^{n/2}} \sum_{x=0}^{2^n-1} |x\rangle_n
\otimes |\,f(x)\, \rangle_{n-1} \, .
\end{equation}
A measurement is then done on the lower register which will record some value of
the function, $f_\mathrm{meas}$ say. All values are equally probable.
There are two values of $x$ which give function value $f_\mathrm{meas}$, and we denote
them by
$x_\mathrm{meas}$ and $x_\mathrm{meas}\oplus
a$. Hence, immediately after the measurement, the state of the system is
\begin{equation}
|\psi_2\rangle = {|x_\mathrm{meas}\rangle_n + |x_\mathrm{meas}\oplus a \rangle_n \over \sqrt{2}}\,
\otimes |\,f_\mathrm{meas}\, \rangle_{n-1} \, .
\label{bv:psi2}
\end{equation}

If we were now to measure the upper register, we would get \textit{either} $x_\mathrm{meas}$ or
$x_\mathrm{meas}\oplus a$. At first glance, this might seem like progress since we
appear to be halfway there. If we
could just get the other number,
we would have $a$. However there
is no way to get both. If we could clone the state several times and measure each clone
then, with high probability, we would be able to determine both of them. However,
the no-cloning theorem says that we can't clone an arbitrary, unknown state.
\index{no-cloning theorem}
Also, repeating the whole procedure doesn't help because, with high probability, we
would get a different function value, $\tilde{f}_\mathrm{meas}$, and one of a different pair of $x$-values,
$\tilde{x}_\mathrm{meas}$ or $\tilde{x}_\mathrm{meas} \oplus
a$, from which again we would not be able to extract $a$.

As in Deutsch's algorithm and the Bernstein-Vazirani algorithm, we must
do some more processing \textit{before} the final measurement. 
As we showed in Eq.~\eqref{Hon} in Chapter \ref{ch:bv} on the Bernstein-Vazirani 
algorithm,
the effect of
Hadamards on $n$-qubit register which is 
in a computational basis state $|x\rangle_n$, is given by
\begin{equation}
H^{\otimes n} |x\rangle_n ={1 \over 2^{n/2}} \sum_{y=0}^{2^n - 1} (-1)^{x
\cdot y} |y \rangle_n \, ,
\label{Hn}
\end{equation}
where $x \cdot y$ is the bitwise inner product modulo 2,
\begin{equation}
x \cdot y \equiv x_0 y_0 \oplus x_1 y_1 \oplus \cdots \oplus x_{n-1}y_{n-1}
\,\, \mod 2 ,
\end{equation}
discussed in Sec.~\ref{sec:bv1}.
Hence, applying Hadamards to the $n$-qubit upper register in state $|\psi_2\rangle$
in Eq.~\eqref{bv:psi2}, 
the state of that register becomes
\begin{equation}
|\psi_3\rangle_n = {1 \over \sqrt{2}}{1 \over 2^{n/2}} \sum_{y=0}^{2^n-1} 
\left[\, (-1)^{x_\mathrm{meas}\cdot y} + (-1)^{(x_\mathrm{meas}\oplus a) \cdot y} \, \right] |y\rangle_n\, .
\label{psi3}
\end{equation}
Now\footnote{This is the mod 2 version of the usual distributive rule for
addition and multiplication: $a \times (b + c) = (a \times b) + (a \times c)$.} 
$(x\oplus a) \cdot y = (x\cdot y) \oplus (a \cdot y)$
so we can write
\begin{equation}
|\psi_3\rangle_n = {1 \over \sqrt{2}}{1 \over 2^{n/2}} \sum_{y=0}^{2^n-1} 
(-1)^{x_\mathrm{meas}\cdot y} \left[\,1 + (-1)^{a \cdot y} \, \right] |y\rangle_n\, .
\label{psi3p}
\end{equation}
Noting that
$a\cdot y = 0 $ or $1$ we see that if $a\cdot y = 1$ then
the two terms in Eq.~\eqref{psi3p} cancel. Hence the
only terms with a non-zero amplitude are those with $a \cdot y = 0$. All
values of $y$ which satisfy this condition are equally probable. Note that the
condition does not depend on the value of $x_\mathrm{meas}$.

A measurement on the upper register then gives, with equal probability,
\textit{one} value of $y$ with $a \cdot y= 0$. This is a linear equation for
the $a_i$, the bits of $a$, i.e.
\begin{equation}
a_0 y_0 + a_1 y_1  + \cdots + a_{n-1} y_{n-1} = 0.
\end{equation}
If we can find $n$ such equations for the
$a_i$ \textit{which are linearly independent}, we can
obtain the solution. Hence we have to repeat the procedure, each time
determining the $y_i$.  As discussed in Appendix G of Mermin~\cite{mermin:07}
one needs to run the algorithm \textit{a little more} than $n$ times because
the set of equations one gets for the $a_i$ are not necessarily linearly
independent. The result is that if one runs $n + p$ times, then the probability
of getting $n$ linearly independent equations (and hence the solution for the
$a_i$) is is greater than
\begin{equation}
1 - {1 \over 2^{p+1}} \, .
\end{equation}
Hence there is less than one chance in a million of failure if one calls the
function $n+ 20$ times. A crucial point in this expression is that the number
of calls beyond $n$ needed to find a solution with some high probability
\textit{does not depend on} $n$.

The occurrence of probability, and some arcane mathematical arguments to prove
that one does get the solution with high probability within the specified
number of runs, is characteristic of several quantum algorithms including
Shor's. 

In the case of Simon's problem, the classical algorithm takes of order
$2^{n/2}$ function calls whereas the quantum algorithm finds the answer with
high probability with little more than $n$ calls\footnote{In the interests of
full disclosure I should state that one also needs to solve $n$ linear
equations on a classical computer, which takes of order $n^3$ steps. A
algorithm which takes a time proportional to a power of the problem size $n$
is said to be \textit{polynomial}. Since classical hardware is cheap
it is not clear if one should include this time using a
classical computer in the computational cost of Simon's algorithm.
However, since $n^3$ is polynomial,
even if one does include this time
the comparison is still between a polynomial quantum (+classical)
algorithm and an exponential purely classical algorithm, which is still an
exponential
speedup, see footnote \ref{fn3}.}. This is an \textit{exponential} speedup\footnote{An algorithm which takes a time proportional to a power of the problem size is said to have \textit{polynomial complexity}, while if the time increases exponentially with size (or exponentially with a power of the size) it is said to have \textit{exponential complexity}. If one algorithm has polynomial complexity and another has exponential complexity then the former is said to have an exponential speedup compared with the latter.\label{fn3}}.
\index{complexity!polynomial}
\index{complexity!exponential}

Finally a few words of anticipation for Shor's algorithm\index{Shor's factoring algorithm}
which we will do next.
Simon's problem \index{Simon's algorithm}
considers a function which is periodic under bitwise modulo 2 addition,
i.e.~$f(x\oplus a) = f(x)$.
Shor's
algorithm investigates functions which are periodic under \textit{ordinary}
addition: $f(x+a) = f(x)$, which is much more useful.
In Simon's problem, the action of the $n$-Hadamards in Eq.~\eqref{Hn} can be written
\begin{equation}
H^{\otimes n} |x\rangle_n ={1 \over 2^{n/2}} \sum_{y=0}^{2^n - 1}
e^{i\pi x \cdot y} |y \rangle_n \, ,
\label{Hnb}
\end{equation}
Since $x\cdot y$ is the bitwise inner product modulo 2, it only takes values 0 and 1, so
the phases in the complex
exponential are just $0$ and $\pi$. The core of Shor's algorithm is a
quantum Fourier transform (QFT), \index{QFT} where an essential difference from Eq.~\eqref{Hnb} is
that the bitwise inner product is replaced by ordinary multiplication. Hence
the QFT generates many different phases, with the result that, unlike Simon's
algorithm, it cannot, in general, be
constructed entirely out of 1-qubit gates. Fortunately, it \textit{can} be constructed
entirely out of 1- and 2-qubit gates.  
All this and more will be discussed in
Chapter \ref{ch:shor}.

\hrulefill
\section*{Problems}
\input{hw_ch12.tex}

%% file: hw_ch12.tex
\begin{problems}
\item
Consider Simon's problem, i.e.~we have a function $f(x)$, where $x$ has $n$
bits and $f$ has $(n-1)$ bits such that $f(x) = f(x \oplus a)$ where $a \ne
0$.
The quantum algorithm obtains values for $x$ such that $a \cdot x$ = 0. From
these linear equations for $x$ one deduces $a$.

Consider the case of $n=4$. You are given that \textit{some} of
the values of $x$ for which $x \cdot a= 0$ are
\begin{verbatim}
x =    3   (0011)
x =    4   (0100)
x =    7   (0111)
x =    9   (1001)
\end{verbatim}
\begin{enumerate}[label=(\roman*)]
\item
Using only this information, determine $a$.
\item
For this value of $a$, show that $a\cdot x = 1$ for $x = 1$ and $2$. \\
(Hence $x = 1$ and $x=2$ would not appear as possible results.)
\end{enumerate}
\end{problems}

%% file: rsa8.tex
\newcommand{\Mathematica}{\textit{Mathematica}}

\index{RSA encryption}
\index{Shor's factoring algorithm}
Shor's famous quantum algorithm, to be discussed in detail in
Chapter~\ref{ch:shor},
factors
large integers much more efficiently than
any known classical algorithm. Factoring is not just of interest to
mathematicians, however, because the difficulty of factoring is at the heart
of the popular RSA method 
of encrypting sensitive information sent via the
internet (or some other public channel). While RSA is not the only method use
to encrypt information, my understanding is that some version of Shor's
algorithm can be used to crack other currently used encryption methods such as
Diffie-Hellman\index{Diffie-Hellman encryption}.
RSA stands for the names of its inventors, Rivest, Shamir and
Adleman.

This chapter is a \LaTeX\ copy of a \Mathematica\ notebook, the original of which
is available at\\
\noindent \url{https://young.physics.ucsc.edu/150/rsa.nb}. 
In it, the RSA
algorithm is implemented, parameters are chosen, and random messages are
generated. These are encrypted, the encrypted messages are decrypted, and a
check is made that the original message is recovered. It you have
\Mathematica\ you can run the notebook version and verify that the RSA algorithm works.

Suppose that Bob wants to receive a message from Alice on the internet (a
public channel). Anything sent on a public channel can be intercepted by
others. How can Bob and Alice agree on a coding scheme and then send each
other coded messages which can be decoded by the other person but not by
anyone ``sniffing" on the internet? This has to be accomplished by only sending
messages down the public channel.

We will now describe the RSA encryption scheme for doing this. It uses a result
of number theory which we will quote but not prove.
To receive the message from Alice, Bob picks two large prime numbers $p$ and
$q$,
and sends to Alice, on the public channel, their product 
\begin{equation}
N = p q ,
\end{equation}
but not $p$ and $q$ separately. $N$ is taken to be large enough, typically a few
thousand bits, that it cannot be factored on a classical computer.
You might ask how can one choose the large prime numbers $p$ and $q$. If one
selects a large integer $N$ at random it can be shown that the probability that
it is prime is about $1/\ln N$. Hence, even if $N$ has, say, 400 digits (around
1000 bits) you only have to take test a few hundred to a thousand random
integers to typically find a prime number. But can one efficiently test if a
number is prime? It turns out that one can, even though, if the number is
found to be not prime, there is no known efficient classical algorithm to
determine the prime factors. The website
\url{http://mathworld.wolfram.com/PrimalityTest.html} explains how the test for
primality is done in \Mathematica.

Bob also sends a large ``encoding number" $c$ which has no factors in common with
$(p-1)(q-1)$. If there
\index{GCD}
are no factors in common then the greatest common divisor (GCD) is 1. The GCD of
two integers is easily determined by Euclid's algorithm discussed in
Sec.~\ref{sec:euclid}.
\index{Euclidean algorithm}
According to Appendix J of Mermin~\cite{mermin:07}, the
probability that two large random integers have no common factors is greater
that $1/2$, so it is not difficult to find a suitable value for $c$.

\fbox{Hence the public key (available to everyone) is $N$ and $c$.}
\index{public key}

Since Bob knows both $p$ and $q$, and hence $(p-1)(q-1)$, he can also determine the
encoding number $d$ such that
\begin{equation}
c\, d = 1 (\mathrm{mod}\ (p-1)(q-1)).
\label{cd}
\end{equation}
Let us remind ourselves of
this mod function. The value of $a\ \mathrm{mod}\ b$ is the result after one
subtracts (or adds) the appropriate multiple of $b$ to $a$ to get a value which
lies in the range $0$ to $b-1$. If $a$ is positive, things are simple,
one subtracts a multiple of $b$
(possibly 0) so the mod function is just the remainder after integer
division. Hence, for example, $9\ \mathrm{mod}\ 5 = 4$ because $9/5 = 1$ remainder $4$. If $a$ is
negative one has to add a multiple of $b$, so, for example, $(-13)\ \mathrm{mod}\ 5 = 2$
(since $-13 + (3\times 5) = 2$).

The above equation, $c\, d = 1\ \mathrm{mod}\ $(something), looks strange at first. If $c$ is
an integer we would normally think that its inverse should be a fraction less than $1$. However,
here $d$ is also an integer, and the product of two integers
\textit{can} give 1 if we use modular arithmetic. For example if $c=5$ and $d=3$ then $cd=15$,
and $cd\ \mathrm{mod}\ 7=1$ (since $15 = (7\times 2) + 1$).

The algorithm for computing $d$ in Eq.~\eqref{cd} is efficient and an
extension of Euclid's algorithm. It is given in Appendix \ref{sec:ext_euclid}
\index{Euclidean algorithm!extension}
and in Appendix J of 
Mermin~\cite{mermin:07}. It turns out that $d$ is unique. Hence Alice,
and anyone else sniffing on the public channel, knows $N$ and $c$ (but not $p$, and
$q$, and hence not $d$).

\fbox{The private key (known only to Bob) is $p$ and $q$ (and hence $d$).}
\index{private key}

Alice breaks up her message into chunks each containing a number of bits
less than the number of bits of the integer $N$. Each chunk is then a binary
number less than $N$. Let's denote by $a$ the numerical value of one chunk.
\begin{center}
$a$ is the original message.
\end{center}

Using the values of $N$ and $c$ that Bob has sent, Alice computes
\begin{equation}
b = a^c \ (\mathrm{mod}\ N)\quad \mathrm{the\ encoded\ message}.
\label{bac}
\end{equation}
The encoded message $b$ is another large integer, and is sent down the
public channel from Alice to Bob.

Bob knows not only $c$ and $N$, but also the
value of $d$. Here number theory kicks in and shows that the original
(unencoded) message $a$ is given by
\begin{equation}
a = b^d \ (\mathrm{mod}\ N)\quad \mathrm{(the\ original\ message\ is\ recovered)}.
\label{abd}
\end{equation}
For a proof of this result see the book by Mermin~\cite{mermin:07}.
Note the symmetry between the encoding formula, Eq.~\eqref{bac} and the decoding
formula, Eq.~\eqref{abd}, with $c$ and $d$ related by Eq.~\eqref{cd}.

Bob can compute the original message $a$ because
he knows $d$, but anyone
sniffing on the public channel does not know $d$.
However, if a third person, traditionally called Eve, listening on the public
channel, could factor $N$ (which is sent down the public channel) into its
factors $p$ and $q$, she would then have $(p-1)(q-1)$ and, since $c$ is also sent down
the public channel, she could determine $d$ where $c\, d = 1 \ (\mathrm{mod}\ (p-1)(q-1))$ using
the extension of the Euclid algorithm mentioned above.
Hence she could
find the original unencrypted message $a$ from Eq.~\eqref{abd}.

Let's do a simple example. We will take
\begin{equation}
p=7, \ q = 13, \ \mathrm{so}\ N = 91.
\end{equation}

For the encoding integer we take $c = 11$, which has no factors in common with
$(p-1)(q-1) = 6\times 12 = 72$. As shown in Appendix \ref{sec:ext_euclid}, using the
extended Euclid algorithm
one finds that $d = 59$. (Let's verify this: $c d = 11 \times 59 = 649 =
(9 \times 72) + 1$ so $cd\! \mod (p-1)(q-1) = 1$, as desired.)
The \Mathematica\ code below sets these values, checks that $p$ and $q$ are prime
while $N$ is not, and that $c d = 1 \ (\mathrm{mod}\ (p-1)(q-1))$. (Note: in
\Mathematica\ commands I use $n$ rather than $N$ because $N$ has a special meaning in
\Mathematica.) The code then generates a message $a$ by computing a random
integer between $0$ and $N-1$, and next computes the encoded message $b$ from 
$b = a^c \ (\mathrm{mod}\  N )$. It then computes $b^d \ (\mathrm{mod}\ N)$
and checks that it gives back the
original message $a$.
If you have \Mathematica\ you can run the code several times (each time a
different random value for the message $a$ will be generated) and see that the
original message is always returned.
\begin{verbatim}
In[1]:=  p=7; q=13; c=11; d=59; n=p*q
Out[1]=  91
\end{verbatim}
We check that $p$ and $q$ are prime. The Mathematica command $\mathrm{PrimeQ}[p]$ returns
``True" if $p$ is prime and ``False" if it is not, and the command
$\mathrm{Mod[x, y]}$ means $x\ \mathrm{mod}\ y$.
\begin{verbatim}
In[2]:=  PrimeQ[p]
Out[2]=  True
In[3]:=  PrimeQ[q]
Out[3]=  True
In[4]:=  PrimeQ[n]
Out[4]=  False
\end{verbatim}
We check that $cd = 1\ \mathrm{mod}\ ((p-1)(q-1))$.
\begin{verbatim}
In[5]:=  Mod[c * d, (p-1)(q-1)]
Out[5]=  1
\end{verbatim}
We generate a random message, using the command \textbf{Random[Integer, $n-1$]}
which generates a random integer between 0 and $n-1$.
\begin{verbatim}
In[6]:= mess = Random[Integer, n - 1]
Out[6]=  51
\end{verbatim}
We compute the encoded message.
\begin{verbatim}
In[7]:= encodedmess = Mod[mess^c, n]
Out[7]=  25
\end{verbatim}
We decode the encoded message and check that we recover the original message.
\begin{verbatim}
In[8]:= recoveredmess = Mod[encodedmess^d, n]
Out[8]=  51
In[9]:= recoveredmess == mess
Out[9]=  True
\end{verbatim}
Hence the message was successfully decoded.

\hrulefill
\section*{Problems}
\input{hw_ch13.tex}

\begin{center}
{\Large\bf Appendices}
\end{center}

\begin{subappendices}

\section{The Euclidean Algorithm}
\index{Euclidean algorithm}
\index{GCD}
\label{sec:euclid}
We want to efficiently find the Greatest Common Divisor (GCD) of two integers.
This is the largest factor that they have in common. As a simple example, the
GCD of 24 and 9 is 3 since $24 = 2^3 \times 3$ and $9 = 3^2$.

Suppose we want the GCD of two numbers $a_0$ and $b_0$ with $a_0 > b_0$. We proceed
iteratively. At each stage, the new value of $a$ is equal to the old value of $b$,
and the new value of $b$ is equal to the remainder when the old value of $a$ is
divided by the old value of $b$, i.e.
\begin{equation}
\begin{split}
a_{n+1} &= b_n \\
b_{n+1} &= a_n - [a_n /b_n]b_n \quad \textrm{which\ is\ the\ same\ as\ } b_{n+1} =
a_n \ \mathrm{mod}\  b_n ,
\end{split}
\label{anbn}
\end{equation}
where $[ \cdots ]$ means the integer part of the quantity in brackets.

Assuming that $b_n < a_n$ and using Eq.~\eqref{anbn} to get $a_{n+1}$ and $b_{n+1}$,
one finds
(i) $b_{n+1} < b_n$ since
the largest value that a number can have $\mathrm{mod}\ b_n$ is $b_n - 1$,
(ii) $b_n = a_{n+1}$ so combined with (i) we have
$b_{n+1} < a_{n+1}$ and (iii) $a_{n+1} = b_n < a_n$.
Hence $a_n$ and $b_n$:

\vspace{3mm}
(a) decrease at successive iterations, and 

\vspace{3mm}
(b) maintain the inequality $a_n > b_n$.

\vspace{3mm}

Note too that $a_n$ and $b_n$ have the same common factors as $a_0$ and $b_0$,
because $a_{n+1}$ and $b_{n+1}$ are linear combinations of the values at the
previous stage, $a_n$ and $b_n$, and so any common factor is preserved.
Eventually we get to a stage where all the factors of $b$ which are not
present in $a$ have been removed from $b_n$, so $b_n$ is the GCD and $a_n$ is an integer
times $b_n$. This means that $b_n$ divides $a_n$ exactly so $b_{n+1} = 0$. 
At this point the procedure stops, and the previous value of $b$, i.e.
$b_n$, is the GCD.
As an example we take $a_0 = 24, b_0 = 9$,

\begin{center}
\begin{tabular}{r|rr l}
$n$ & $a_n$ & $b_n$  \\
\hline
0 & 24 & 9 & (the initial values)\\
1 & 9 & 6  & (since $24 = 2\times 9 + 6$) \\
2 & 6 & 3  & (since $9= 6 \times 1 + 3$) \\
3 & 3 & 0 & (since $6 = 3 \times 2 + 0$) .
\end{tabular}
\end{center}

Hence the GCD of 24 and 9 is $b_2 \ (=3)$, which is correct.

\section{Extension of the Euclidean Algorithm to find an inverse modulo an
integer}
\label{sec:ext_euclid}
\index{Euclidean algorithm!extension}

Given $a$ and $c$ which have no common factors, and $a > c$, we want to find
$d$ where 
\begin{equation}
c \, d = 1 \ \mathrm{mod}\ a .
\end{equation}

The greatest common divisor of $c$ and $a$ is 1 since, by assumption, they have no
common factors. We go through the Euclid algorithm
\begin{equation}
\begin{split}
a_{n+1} &= c_n \\
c_{n+1} &= a_n - [a_n /c_n]c_n 
\end{split}
\end{equation}
until we get to the stage where $c_n =1 $, the greatest common divisor. 
One can then obtain $d$ by working backwards through the iterations. This is
best shown by an example. We take $p = 7, q = 13$, as in example above, so
we have $a = (p-1)(q-1) = 72$ and hence we initialize $a_0 = 72$. We also take $c =
11$ (again as in the example) which has no factors in common with $a$, and so
initialize $c_0 = 11$. Hence the Euclid algorithm proceeds as follows

\begin{center}
\begin{tabular}{r |rr l}
$n$\quad & $a_n$ & $c_n$ & \\
\hline
\ 0\quad &\ 72\ &\ 11\ &\qquad  $a_0 = a,\ \, c_0 = c$ (the initial values) \\
\ 1\ &\ 11\ &\ 6\  & \qquad $a_1 = c_0,\ c_1 = a_0 - 6 c_0 = 6$ \\
\ 2\ &\ 6\ &\  5\  & \qquad $a_2 = c_1,\ c_2 = a_1 - c_1 = 5$ \\
\ 3\ &\ 5\ &\  1\  & \qquad $a_3 = c_2,\ c_3 = a_2 - c_2 = 1\ (c_3=1$ so we
stop).
\end{tabular}
\end{center}
Hence working backwards, 
\begin{equation}
1 = a_2 - c_2 = c_1 - (a_1 - c_1) = 2c_1 - a_1 = 2 (a_0 - 6 c_0) - c_0 = 2a_0
- 13 c_0\ (= 2a - 13 c).
\end{equation}
We want to take this $(\mathrm{mod}\ a)$. Now $2 a \ (\mathrm{mod}\ a) = 0$.
Since
$-13c$ is negative we need to make it positive by adding $a\, c$
(which is zero $(\mathrm{mod}\ a)$). Hence
\begin{equation}
1 =  2a - 13 c \ (\mathrm{mod}\ a) = - 13 c \ (\mathrm{mod}\ a) = (- 13+a)  c \
 (\mathrm{mod}\ a) = 59 c\ (\mathrm{mod}\ a),
\end{equation}
where we used that $a=72$ to get the last equality.
Hence $d = 59$ as stated in the above example.

\end{subappendices}

%% file: hw_ch13.tex
\begin{problems}
\item
Consider the RSA scheme for encryption with $p=11, q = 3$ so $N = p q = 33$.
For the encoding integer take $c=3$ which has no factors in common with
$(p-1)(q-1) = 20$. 
\begin{enumerate}[label=(\roman*)]
\item
Using the extended Euclid algorithm, find the decoding number $d$ which
satisfies $c  d = 1 \!\! \mod (p-1)(q-1)$.
\item
Assume that the original message $m$ is represented by the integer $7$.
Compute the encoded message $m'$ given by
\begin{equation}
m^\prime= m^c \!\! \mod N.
\end{equation}
\item
Compute $(m')^d \!\! \mod N$, and show that you recover the original message
$m$.
\end{enumerate}

\end{problems}

%% file: period8.tex
In this chapter, we explain how finding the period of a certain
\index{period finding algorithm}
function will enable us to factor integers. We will also illustrate the technique
with a simple example. This  will probably
seem a strange approach for factoring, and is
not the preferred method on a classical computer, but it is the method used by
Shor in his quantum algorithm. 

We take two large primes $p$ and $q$ and form the product
\begin{equation}
N = p\, q \, .
\end{equation}
The goal is to find the factors $p$ and $q$ given only the product $N$. This
is a problem which is hard classically. For applications in cryptography $p$ and $q$ may
have around 600 digits (around 2000 bits) so $n$, the number of bits of $N$,
will be several thousand.

We proceed by choosing a random integer $a$ less than $N$ which has no factors in common
with $N$. Whether or not $a$ and $N$ have a common factor can be determined
efficiently using Euclid's algorithm, which was described in
Sec.~\ref{sec:euclid}.
\index{Euclidean algorithm}
In the very unlikely event that $a$ and $N$ do
have a common factor we have found a factor of $N$ and the problem is solved.
Otherwise we compute the following function
\begin{equation}
f(x) \equiv a^x\ (\!\!\!\! \mod N\, )
\end{equation}
for $x = 1, 2, \cdots$, so $f(0) = 1, f(1) = a, \cdots$.
As stated, $a$ and $N$ have no common factors, and for 
this case one can show that
eventually we will get $f(x)=1$
for some value, $x=r$
say, so
\begin{equation}
a^r \equiv 1 \ (\!\!\!\! \mod N ) \, .
\label{r}
\end{equation}
The function then repeats since
\begin{equation}
f(x+r) \equiv a^{x + r}\ (\!\!\!\! \mod N \,)  \equiv a^x \ (\!\!\!\! \mod N \,)
\times \ a^r \ (\!\!\!\! \mod N \,) \equiv a^x\ (\!\!\!\! \mod N \,)
= f(x) \, ,
\end{equation}
using Eq.~\eqref{r}.  Hence $r$ is the period of the function. 

\begin{figure}[htb]
\begin{center}
\includegraphics[width=0.6\columnwidth]{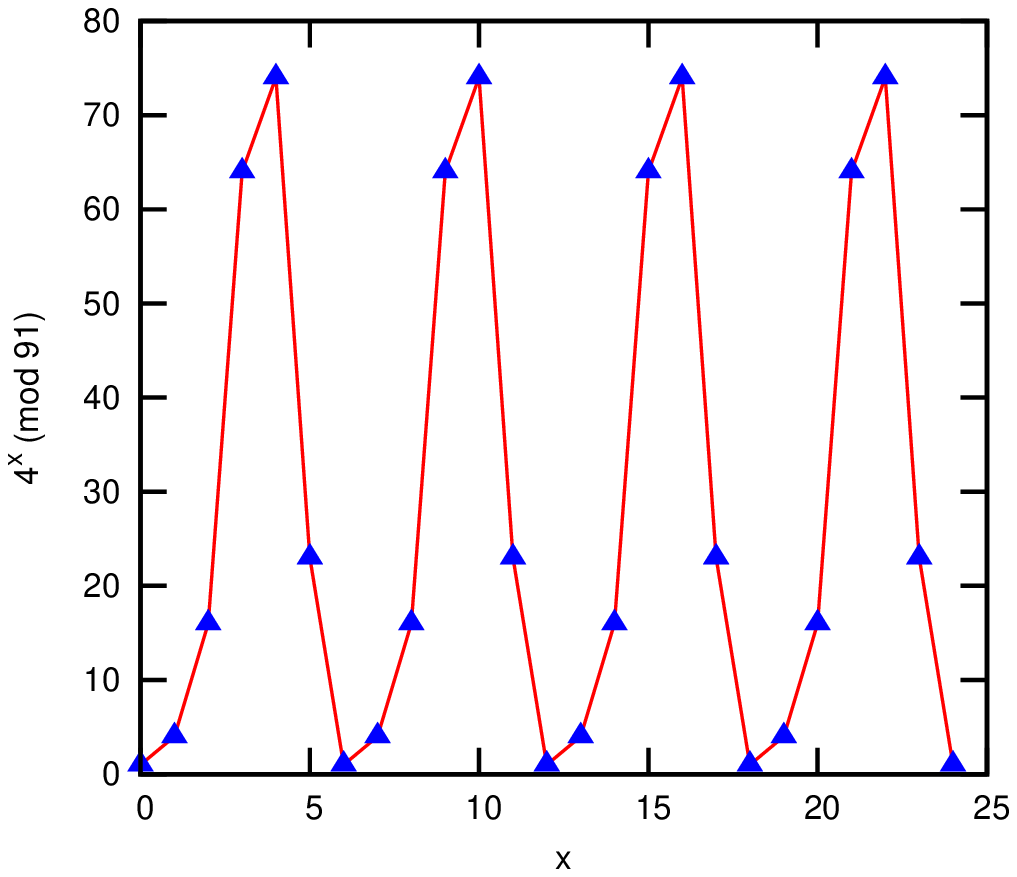}
\caption{
The function $f(x) \equiv 4^x\ (\!\!\!\!\mod 91\,)$. The
period is seen by inspection to equal
6.
\label{fx}
}
\end{center}
\end{figure}

We we illustrate with a simple example, 
\begin{equation}
N = p\,q = 91, \quad \mathrm{with\ factors}\  p = 13, q = 7.
\end{equation}
We also take $a = 4$, which has no factors in common with $91$. We plot $f(x)
\equiv 4^x\ (\!\!\!\!\mod 91\,)$ in Fig.~\ref{fx}.
The periodic nature is clear, and the period
is found to equal 6 by inspection. Let's make sure we understand how this
figure is obtained by working out the values of
$4^x \ (\!\!\!\!\mod 91\,)$ for $x=1, 2, \cdots,6$.
\begin{subequations}
\begin{align}
x = 1,\quad 4^x &= 4 \, ,  \\
x = 2,\quad 4^x &=16 \, ,  \\
x = 3,\quad 4^x &=64\, , \label{fx3} \\
x = 4,\quad 4^x &= 64 \times 4 = 256 = 2 \times 91 +
74 \equiv 74 \ (\!\!\!\!\mod 91\,)\, ,  \\
x = 5,\quad 4^x &\equiv 74 \times 4 = 296 = 3 \times 91 + 
23 \equiv 23 \ (\!\!\!\!\mod 91\,) \, ,  \\
x = 6,\quad 4^x &\equiv 23 \times 4  =92 =  91 + 
1 \equiv 1 \ (\!\!\!\!\mod 91\,) \, .   \label{fx6}
\end{align}
\label{fxvals}
\end{subequations}
In the above equations the symbol $\equiv$ means equivalent to $(\!\!\!\!\mod 91)$.
   
\begin{figure}[htb]
\begin{center}
\includegraphics[width=0.6\columnwidth]{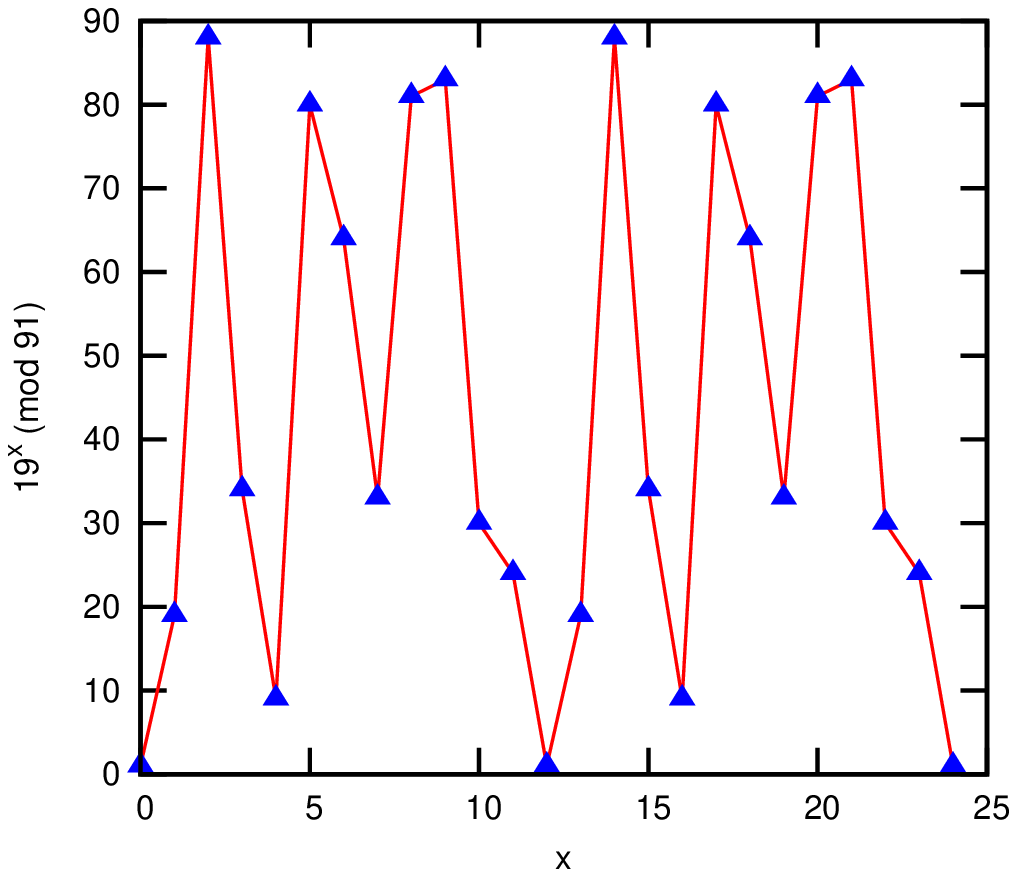}
\caption{
The function $f(x) \equiv 19^x \ (\!\!\!\!\mod 91\,)$.
The period is seen by inspection to equal
12.
\label{fx2}
}
\end{center}
\end{figure}

The plot in Fig.~\ref{fx} seems to have a fairly regular
behavior, but such smooth behavior is exceptional and occurs here only because of the
particularly simple choice of
parameters.  Figure~\ref{fx2} shows a plot for the same value of $N$ but
with $a = 19$.  This is a much
more random looking figure, as is typical.  In this case the period is $r=12$.
The apparently random shape of
$f(x)$ means that one can not estimate the period by taking a few nearby
values of $x$ and extrapolating.

Having found the period we now need to be lucky in two respects:
\begin{enumerate}
\item The period $r$ must be even. This means that $r/2$ is an integer and
so is $a^{r/2}$. Hence we can write
\begin{equation}
0 \equiv a^r - 1 \equiv (a^{r/2}-1)(a^{r/2}+1) \ (\!\!\!\!\mod\ p\, q\,) \, .
\label{br0}
\end{equation}

\item We need that
\begin{equation}
a^{r/2}+1 \not\equiv 0 \ (\!\!\!\!\mod p\, q\,) \, .
\label{lucky2}
\end{equation}
It is automatically true that $a^{r/2}-1 \not\equiv 0 \ (\!\!\!\!\mod p\, q\,)$ because,
by defintion,
$x=r$ is the smallest power for which $a^x-1 \equiv 0 \ (\!\!\!\!\mod
p\, q\,)$. Hence, if Eq.~\eqref{lucky2} is true,
neither $a^{r/2}+1$ nor $a^{r/2}-1$ is divisible by $N = p\,q$
but, according to Eq.~\eqref{br0},
their product is, i.e.~$\left(a^{r/2} + 1\right)\left(a^{r/2}-1\right) = \text{const.}\ p q$.
Since $p$ and $q$ are primes (and neither ~$\left(a^{r/2} + 1\right)$ nor
$\left(a^{r/2}-1\right)$ are multiples of $p q$), this is only possible if
$a^{r/2}+1$ is a multiple of one of the factors, $p$ say, i.e.~$a^{r/2}+1= C
p$, and $a^{r/2}-1$ is
a multiple of the other one $q$, i.e.~$a^{r/2}-1 = C' q$
($C$ and $C'$ are constants).
Consequently $p$ is the greatest common
divisor of $N \,(= p\, q)$ and $a^{r/2}+1 \,(= C p)$,
and $q$ is the greatest common divisor of $N \,(= p\, q)$ and
$a^{r/2}-1 \,(= C'q)$. We can therefore find $p$ and $q$ using the Euclidean algorithm
mentioned earlier. 
\end{enumerate}

What are the odds that we will be doubly lucky in this way?  According to
Appendix M in
Mermin~\cite{mermin:07} the probability is greater than 0.5
for large $N$. If one \textit{is} unlucky one tries a different choice for $a$. Since
the probability of success is quite high at each attempt,
one does not have to repeat the process very many times to succeed with very
high probability.

Back to our example.  For $N=91, a=4$ we found $r = 6$. Indeed we are lucky!
This is even. Also $a^{r/2} +1= 65 \not\equiv 0 \ (\!\!\!\!\mod 91\,)$. So we
are doubly lucky! However, this is not remarkable. 
As noted above the probability of this double luck is
greater than 0.5 (at least for large $N$).

Hence one of the factors is the greatest common divisor (GCD) of $91$ and $a^{r/2} +
\index{GCD}\index{greatest common divisor|see {GCD}}
1 = 65$. The other factor is the greatest common divisor of $91$ and $a^{r/2}
-1 = 63$. 

Applying Euclid's algorithm, described in
\index{Euclidean algorithm}
Sec.~\ref{sec:euclid},
to 
$f_0=91, g_0=65$:
\begin{align}
f_1 &= 65, \nonumber \\
g_1 &=  91 - [91/65]\, 65 = 91 - 65 = 26, \nonumber \\
f_2 &=  26, \nonumber \\
g_2 &=  65 - [65/26]\, 26 = 65 - 52 = 13, \nonumber \\
f_3 &=  13, \nonumber \\
g_3 &=  26 - [26 / 13]\, 13 = 26 - 26 = 0.
\end{align}
Hence the GCD is $g_2 = 13$, which is indeed one of the factors of $91$. The
other factor of 91 is $91/13 = 7$. One could also obtain this last result by
using the same process to obtain the GCD of $63$ and $91$.

Period finding is a rather indirect method for factoring integers and is not
the most efficient one on a classical computer because of the amount of work
in computing $a^x (\!\!\!\!\mod N\,)$ for all $x$ from 1 to $r$ where $r$ is
of order $N$. However Shor realized that it
lends itself to a very efficient implementation on a quantum computer. Part
of Shor's algorithm, which we will discuss in Chapter \ref{ch:shor}, uses
quantum parallelism to compute all needed values of $a^x
(\!\!\!\!\mod N\,)$ with a time that only increases as a power of 
$n$ rather than exponentially in $n$, where we recall that the number to be factored, $N$, has $n$ bits.

%% file: FFT8.tex
\section{Introduction}
\index{Fourier Transform}
\index{Fast Fourier Transform|see{FFT}}\index{FFT}
The standard Fourier Transform concerns a \textit{continuous} function, $x(t)$
say. For descriptive purposes it will be convenient to think of $t$ as time, but this is not essential.
In the Fourier transform we
decompose $x(t)$ into its components at different ``frequencies" $\omega$ as
follows:
\begin{equation}
y(\omega) = {1 \over \sqrt{2\pi}} \int_{-\infty}^\infty e^{i \omega t}\, x(t)
\, d t \, .
\label{ftcont}
\end{equation}
If $x(t)$ comprises oscillations at a frequency $\omega_0$, say,
(i.e.~has a period $T$ equal to $2 \pi / \omega_0$), so $x(t) \sim
e^{-i\omega_0 t}$,
then $y(\omega)$ will be sharply peaked at $\omega = \omega_0$ (or equivalently
at $\omega = 2 \pi/ T$). Note the inverse relation between the period $T$ and
the position of the peak in the Fourier Transform. The larger the period, the
smaller the value of $\omega$ at the peak. 

As an example, if $x(t) = \cos\omega_0 t = {1\over 2}\left(e^{i \omega_0 t} +
e^{-i\omega_0 t}\right)$ then $y(\omega)$ has sharp ``delta
function" peaks at $\omega = \pm \omega_0$. A completely different situation
is when $x(t)$ is random (i.e.~white noise) in which case $y(\omega)$ is a constant
(at least for $|\omega|$ less than a cut-off value $\omega_c$.)

There is also an inverse Fourier transform, 
\begin{equation}
x(t) = {1 \over \sqrt{2\pi}} \int_{-\infty}^\infty e^{-i \omega t}\, y(\omega)
\, d \omega \, ,
\label{iftcont}
\end{equation}
which has almost the same form as the original (forward) transform, apart from
the sign of $i$ in the exponential.  It is shown in standard mathematics texts
that substituting for $y(\omega)$ from Eq.~\eqref{ftcont} into
the RHS of Eq.~\eqref{iftcont} does give back $x(t)$ for a wide class of
functions $x(t)$.

This chapter is concerned with the discrete analog of Eqs.~\eqref{ftcont} and
\eqref{iftcont} in which the data $x_m$ is at a set of $N$ equally spaced ``times",
and the Fourier transform $y_k$ is at a set of $N$ equally spaced
``frequencies". In addition, in the discrete Fourier Transform, the data
only covers a finite range, whereas the data in the original, continuous
Fourier Transform extends to $\pm \infty$.

%

\section{The Discrete Fourier Transform}
\label{sec:FT}
If we have a set of $N$ data points $x_m\, (m=0, 1, \cdots, N -1)$,
the discrete Fourier transform (FT) is a set of
$N$ new values $y_k$ given by
\begin{equation}
y_k = {1 \over \sqrt{N}} \sum_{m=0}^{N-1} \exp(2 \pi i\,k m/ N)\,  x_m \, ,
\label{ft}
\end{equation}
evaluated for $k = 0, 1, \cdots, N-1$. We don't need to consider $k$ values
outside this range because $y_{k+N} = y_k$ (so the $y_k$ are periodic
with period $N$). Equation \eqref{ft} corresponds to
a discretized and finite-range version of Eq.~\eqref{ftcont} with $m$
corresponding to $t$ and $2 \pi k / N$ corresponding to $\omega$.
If $x_m$ is a periodic function of $m$ with period
$T$, i.e. $x_m \sim e^{-2\pi i m /T}$,
then $y_k$ will be peaked for $k$ around $N/T$ since the terms in
Eq.~\eqref{ft}
then add up in phase. This corresponds, in the
continuous Fourier Transform, to a peak for $\omega$ at around $2\pi/T$.

The inverse Fourier
transform has almost the same form; one just needs to take the complex
conjugate of the exponential, i.e.
\begin{equation}
x_m = {1 \over \sqrt{N}} \sum_{k=0}^{N-1} \exp(-2 \pi i \, k m / N) \, y_k \, ,
\qquad (m = 0, 1, \cdots, N-1).
\label{ift}
\end{equation}
To see this we substitute Eq.~\eqref{ft} into Eq.~\eqref{ift} so
\begin{align}
x_m &= {1 \over \sqrt{N}} \sum_{k=0}^{N-1} \exp(-2 \pi i\, k m / N) \,
{1 \over \sqrt{N}} \sum_{l=0}^{N-1} \exp(2 \pi i \, k l / N)\,  x_l \nonumber
\\
&= {1 \over N} \sum_{l=0}^{N-1} x_l \left[
\sum_{k=0}^{N-1} \exp(2 \pi i\, k (l-m) / N)
\right]  \nonumber \\
&= {1 \over N} \sum_{l=0}^{N-1} x_l \left[
{1 - \exp(2 \pi i \, (l-m) ) \over 1 - \exp(2 \pi i \, (l-m)/N )}
\right], \label{ftcheck}
\end{align}
where, in the last expression, we summed up the geometric series.
The numerator in the brackets is always zero. The denominator is only zero if
$l = m$. Hence, as long as $ l \ne m$ the sum is zero.
However, if $l =m$ we get $0/0$, which is undefined, and so, to get the
answer, we either
evaluate it as the limit $l \to m$ or go back the start and put $l=m$ from
the beginning. In either method one finds that the term in rectangular
brackets is equal to $N$ for $l=m$.. Hence the RHS of
Eq.~\eqref{ftcheck} is $x_m$, showing that the inverse transform in
Eq.~\eqref{ift} does give back the original dataset $x_m$ as claimed. 

Note that $x_{m+N} = x_m$, so the $x$-values obtained from the inverse Fourier
transform are actually a periodic repetition of the original data
(i.e.~the $x_m$
for $m=0,\cdots,N-1$) with period $N$.

The discrete Fourier transform can be conveniently written as
\begin{equation}
y_k = {1 \over \sqrt{N}} \sum_{m=0}^{N-1} \omega^{k m}  x_m \, ,
\qquad (k = 0, 1, \cdots, N-1),
\label{FT1}
\end{equation}
where
\begin{equation}
\label{omega}
\omega = \exp(2 \pi i / N) \, ,
\end{equation}
is the $N$-th root of unity.

For example, for $N=4$ we have $\omega = i$, and so
\begin{equation}
\vec{y} = U \vec{x} \, ,
\end{equation}
where the matrix of coefficients is
\begin{equation}
U =  {1 \over 2}
\begin{pmatrix}
1 & 1 & 1 & 1 \\
1 & i  & i^2 & i^3 \\
1 & i^2 & i^4 & i^6 \\
1 & i^3 & i^6 & i^9 
\end{pmatrix}
= {1 \over 2}
\begin{pmatrix*}[r]
1 & 1 & 1 & 1 \\
1 & i & -1 & -i \\
1 & -1 & 1 & -1 \\
1 & -i & -1 & i 
\end{pmatrix*} .
\label{DFT_N4}
\end{equation}

To determine the discrete FT, each application of Eq.~\eqref{FT1} requires $N$ additions
and $N$ multiplications for each of the $N$ values of $k$, so the operation count
is $O(N^2)$. 

In the appendices of this chapter we describe the fast Fourier transform (FFT)
which is a much more efficient way to calculate a discrete Fourier transform.
We don't need the FFT for this course, but I include a description of it here
in the appendices partly to stimulate students' interest in it (since it is a
gem of computer science), and partly because it bears a strong resemblance to
Shor's quantum Fourier transform (QFT),\index{QFT} see Chapter \ref{ch:qft},
which is the heart of his factoring algorithm.  We shall show this connection
in the appendices of Chapter \ref{ch:qft}.

The Fast Fourier Transform (FFT) requires
an operation count of only $N \log_2 N$ compared with $N^2$ which is needed for a
straightforward evaluation of Eq.~\eqref{FT1} for all $k$.
This reduction (which is considerable for large $N$) is possible because
$\omega^n$ is a periodic function of $n$ with period $N$ and so
$\omega^{k m}$ takes only $N$ distinct values, even though $k\,m$ runs
over $O(N^2)$ values. Incredibly, as we shall see, the QFT does the discrete
FT with only of order $\left(\log_2 N\right)^2$ operations. 

The FFT is discussed in the appendices which now follow. As mentioned above,
this material is not
required for the rest of the course and can be omitted.


\begin{center}
{\Large \bf Appendices}
\end{center}

\begin{subappendices}

\section{The Fast Fourier Transform; an example with $N = 8$}

We will understand the Fast Fourier Transform (FFT) by first working out in
detail a simple example.
The number of data points $N$ must be a power of $2$. If it's not a power of 2
then one pads the data with zeroes to make it so.
We will take $n=3$, i.e.~$N=8$. 
Written out explicitly, the discrete Fourier Transform for $N = 8$ data points is
\begin{subequations}
\label{FT}
\begin{align}
y_0 &=  \smfrac{1}{\sqrt{8}}\left( \,x_0 + \ \ \,\,x_1 + \ \ \,\,x_2 + \ \
\,\,x_3 +\ \ \,\,\,
x_4 + \ \ \,\,x_5 + \ \ \,\,\,x_6 + \ \ \,\,x_7\, \right)\, ,  \\
y_1 &= \smfrac{1}{\sqrt{8}} \left(\, x_0 + \omega\, x_1 + \omega^2 x_2 + \omega^3 x_3 +
\omega^4 x_4 + \omega^5 x_5 + \omega^6 x_6 + \omega^7 x_7\, \right)\, ,  \label{y1}\\
y_2 &= \smfrac{1}{\sqrt{8}} \left(\, x_0 + \omega^2 x_1 + \omega^4 x_2 + \omega^6 x_3 +
\ \ x_4 + \omega^2 x_5 + \omega^4 x_6 + \omega^6 x_7 \, \right) \, , \\
y_3 &= \smfrac{1}{\sqrt{8}} \left(\, x_0 + \omega^3 x_1 + \omega^6 x_2 + \omega\, x_3 +
\omega^4 x_4 + \omega^7 x_5 + \omega^2 x_6 + \omega^5 x_7 \, \right) \, , \\
y_4 &= \smfrac{1}{\sqrt{8}} \left(\, x_0 + \omega^4 x_1 + \ \ \,\,x_2 + \omega^4 x_3 +
\ \ \,\,x_4 + \omega^4 x_5 + \ \ \,\,x_6 + \omega^4 x_7 \, \right) \, , \\
y_5 &= \smfrac{1}{\sqrt{8}} \left(\, x_0 + \omega^5 x_1 + \omega^2 x_2 + \omega^7 x_3 +
\omega^4 x_4 + \omega\, x_5 + \omega^6 x_6 + \omega^3 x_7 \, \right) \, , \\
y_6 &= \smfrac{1}{\sqrt{8}} \left(\, x_0 + \omega^6 x_1 + \omega^4 x_2 + \omega^2 x_3 +
\ \ \,\,x_4 + \omega^6 x_5 + \omega^4 x_6 + \omega^2 x_7 \, \right) \, , \\
y_7 &= \smfrac{1}{\sqrt{8}} \left(\, x_0 + \omega^7 x_1 + \omega^6 x_2 + \omega^5 x_3 +
\omega^4 x_4 + \omega^3 x_5 + \omega^2 x_6 + \omega\, x_7 \, \right) \, ,
\end{align}
\end{subequations}
where the $x_j$ are the original data, the $y_j$ are the Fourier transformed data,
\begin{equation}
\omega = \exp(2 \pi i / 8) = {1 \over \sqrt{2}} (1 + i) \, ,
\end{equation}
and we note that
\begin{equation}
\omega^8 = 1 \ = \omega^0 \, ,
\end{equation}
so we have reduced all the powers of $\omega$ to be between $0$ and
$7\ (=N-1)$. We also note that
\begin{equation}
\omega^2 = i, \ \omega^4 = -1 \, .
\label{powers}
\end{equation}

To evaluate Eqs.~\eqref{FT} efficiently the FFT proceeds recursively. We
firstly define Fourier transforms of length 2:
\begin{subequations}
\label{step1}
\begin{align}
u_0 &= \smfrac{1}{\sqrt{2}} (x_0 + x_4) \qquad  = \smfrac{1}{\sqrt{2}} (x_0 + \omega^{4 k} x_4)\ (k = 0) \, ,\\
u_1 &= \smfrac{1}{\sqrt{2}} (x_1 + x_5) \qquad  = \smfrac{1}{\sqrt{2}} (x_1 + \omega^{4 k} x_5)\ (k = 0) \, ,\\
u_2 &= \smfrac{1}{\sqrt{2}} (x_2 + x_6) \qquad  = \smfrac{1}{\sqrt{2}} (x_2 + \omega^{4 k} x_6)\ (k = 0) \, ,\\
u_3 &= \smfrac{1}{\sqrt{2}} (x_3 + x_7) \qquad  = \smfrac{1}{\sqrt{2}} (x_3 + \omega^{4 k} x_7)\ (k = 0) \, ,\\
u_4 &= \smfrac{1}{\sqrt{2}} (x_0 - x_4) \qquad = \smfrac{1}{\sqrt{2}} (x_0 + \omega^{4 k} x_4)\ (k = 1) \, ,\label{4}\\
u_5 &= \smfrac{1}{\sqrt{2}} (x_1 - x_5) \qquad = \smfrac{1}{\sqrt{2}} (x_1 + \omega^{4 k} x_5)\ (k = 1) \, ,\label{6}\\
u_6 &= \smfrac{1}{\sqrt{2}} (x_2 - x_6) \qquad = \smfrac{1}{\sqrt{2}} (x_2 + \omega^{4 k} x_6)\ (k = 1) \, ,\label{5}\\
u_7 &= \smfrac{1}{\sqrt{2}} (x_3 - x_7)\qquad = \smfrac{1}{\sqrt{2}} (x_3 + \omega^{4 k} x_7)\ (k = 1) \, .\label{7}
\end{align}
\end{subequations}

\begin{figure}[thb!]
\begin{center}
\includegraphics[width=9cm]{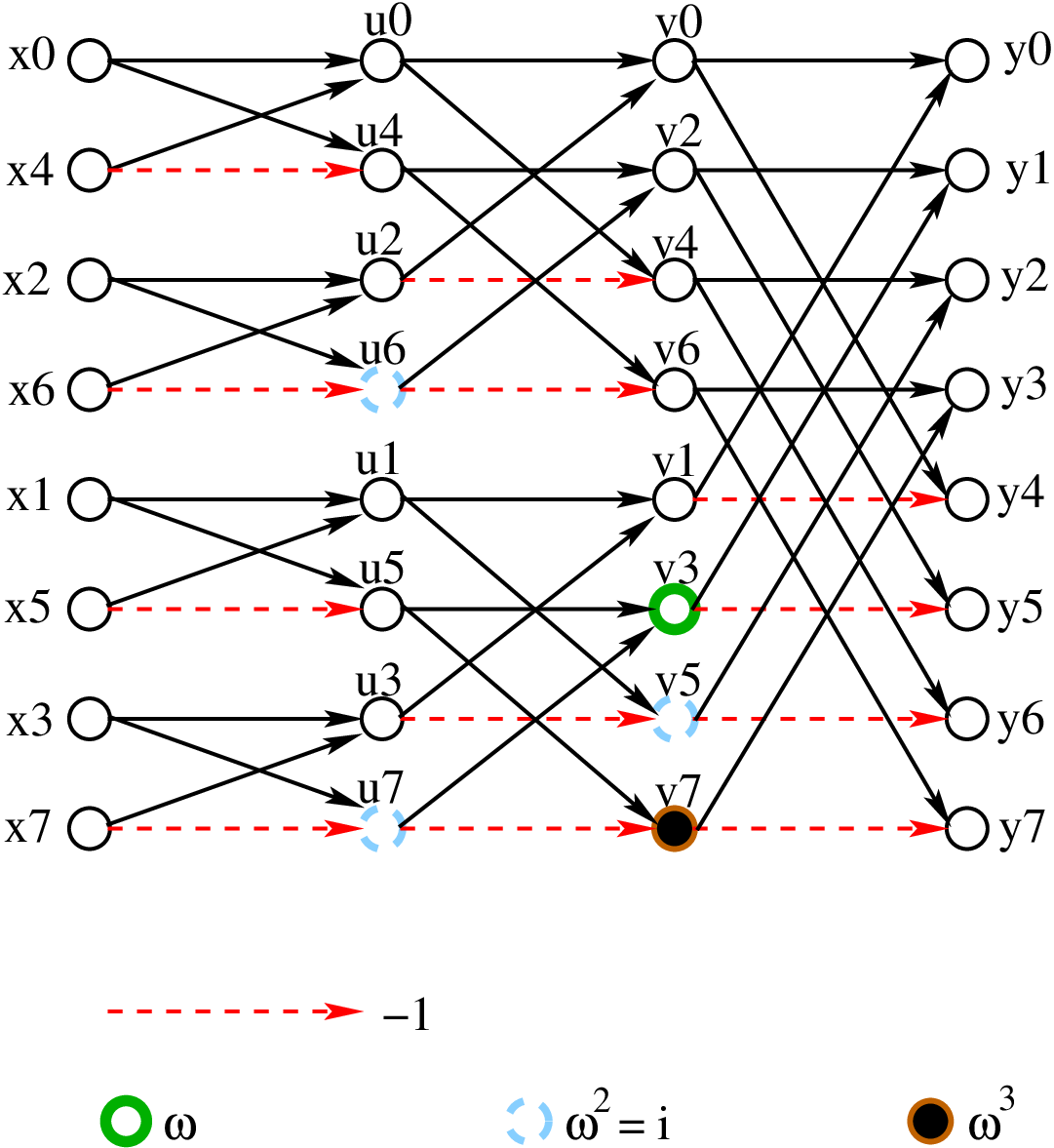}
\caption{A graphical representation of Eqs.~\eqref{step1}--\eqref{step3},
which is the FFT for $N = 8\, (=2^n\ \text{with}\ n=3)$. The original data are
the $x_j$ and the Fourier transformed data are the $y_j$. The
dashed (red) lines have a factor of $-1$ and the solid lines have
a factor of $1$.  The thick (green) circle transmits a factor of
$\omega$ to the right, the dashed (blue) circles transmit a factor of $\omega^2\ (=i)$
to the right, and the (brown) filled-in circle transmits a factor of $\omega^3$ to the
right. In Sec.~\ref{sec:general} we will change to a notation applicable for
general $n$, as follows: $y_j \equiv x^{(0)}_j, v_j \equiv x^{(1)}_j, u_j
\equiv
x^{(2)}_j$, and $x_j = x^{(3)}_j$.
(Adapted from R.~Vathsan \textit{Introduction to Quantum Physics and
Information Processing}.)
\label{fig:FFT8}
}
\end{center}
\end{figure}

Pairs of quantities in Eqs.~\eqref{step1} are combined into Fourier Transforms of length 4:
\begin{subequations}
\label{step2}
\begin{align}
v_0 &= \smfrac{1}{\sqrt{2}}(u_0 + u_2)   \qquad\, = \smfrac{1}{\sqrt{2}}(u_0 + \omega^{2 k} u_2) \ (k = 0) \, ,\\
v_1 &= \smfrac{1}{\sqrt{2}}(u_1 + u_3)   \qquad\, = \smfrac{1}{\sqrt{2}}(u_1 + \omega^{2 k} u_3) \ (k = 0) \, , \\
v_2 &= \smfrac{1}{\sqrt{2}}(u_4 + iu_6)  \qquad = \smfrac{1}{\sqrt{2}}(u_4 + \omega^{2 k} u_6) \ (k = 1) \, , \label{FFT:2}\\
v_3 &= \smfrac{1}{\sqrt{2}}(u_5 + iu_7)  \qquad = \smfrac{1}{\sqrt{2}}(u_5 + \omega^{2 k} u_7) \ (k = 1) \, , \label{3}\\
v_4 &= \smfrac{1}{\sqrt{2}}(u_0 - u_2)   \qquad\, = \smfrac{1}{\sqrt{2}}(u_0 + \omega^{2 k} u_2) \ (k = 2) \, ,\\
v_5 &= \smfrac{1}{\sqrt{2}}(u_1 - u_3)   \qquad\, = \smfrac{1}{\sqrt{2}}(u_1 + \omega^{2 k} u_3) \ (k = 2) \, , \\
v_6 &= \smfrac{1}{\sqrt{2}}(u_4 - i u_6) \qquad = \smfrac{1}{\sqrt{2}}(u_4 + \omega^{2 k} u_6) \ (k = 3) \, ,\\
v_7 &= \smfrac{1}{\sqrt{2}}(u_5 - i u_7) \qquad = \smfrac{1}{\sqrt{2}}(u_5 + \omega^{2 k} u_7) \ (k = 3) \, ,
\end{align}
\end{subequations}
and finally pairs of quantities in Eqs.~\eqref{step2} are combined to form
the Fourier Transform in Eqs.~\eqref{FT}:
\begin{subequations}
\label{step3}
\begin{align}
y_0 &= \smfrac{1}{\sqrt{2}}(v_0 + v_1)   \qquad\ \ \,= \smfrac{1}{\sqrt{2}}(v_0 + \omega^{k} v_1) \ (k = 0) \, ,\\
y_1 &= \smfrac{1}{\sqrt{2}}(v_2 + \omega\, v_3) \qquad = \smfrac{1}{\sqrt{2}}(v_2 + \omega^{k} v_3) \ (k = 1) \, , \label{FFT:1}\\
y_2 &= \smfrac{1}{\sqrt{2}}(v_4 + i v_5)   \qquad\  = \smfrac{1}{\sqrt{2}}(v_4 + \omega^{k} v_5) \ (k = 2) \, ,\\
y_3 &= \smfrac{1}{\sqrt{2}}(v_6 + \omega^3 v_7) \qquad\!\! = \smfrac{1}{\sqrt{2}}(v_6 + \omega^{k} v_7) \ (k = 3) \, ,\\
y_4 &= \smfrac{1}{\sqrt{2}}(v_0 - v_1)   \qquad\ \ \, = \smfrac{1}{\sqrt{2}}(v_0 + \omega^{k} v_1) \ (k = 4) \, ,\\
y_5 &= \smfrac{1}{\sqrt{2}}(v_2 - \omega v_3) \qquad \, = \smfrac{1}{\sqrt{2}}(v_2 + \omega^{k} v_3) \ (k = 5) \, ,\\
y_6 &= \smfrac{1}{\sqrt{2}}(v_4 - i v_5)   \qquad\ \, = \smfrac{1}{\sqrt{2}}(v_4 + \omega^{k} v_5) \ (k = 6) \, ,\\
y_7 &= \smfrac{1}{\sqrt{2}}(v_6 - \omega^3 v_7) \qquad\!\! = \smfrac{1}{\sqrt{2}}(v_6 + \omega^{k} v_7) \ (k = 7) \, ,
\end{align}
\end{subequations}
Equations \eqref{step1}--\eqref{step3} are represented graphically by
Fig.~\ref{fig:FFT8}.

We see that the FFT, specified by
Eqs.~\eqref{step1}--\eqref{step3}, requires $8 \times 3\ (= N \log_2 N\
\text{for}\ N=8)$ additions and multiplications, whereas a direct evaluation
of the FT according to Eq.~\eqref{FT} takes $8 \times 8 \ (= N^2)$ additions
and multiplications.  For large $N$, the speedup factor, $N / \log_2 N$, in
using the FFT rather than direct evaluation of the FT is considerable.

Let's check that this works by evaluating $y_1$. We have
\begin{subequations}
\label{check}
\begin{align}
y_1 &= \smfrac{1}{\sqrt{2}}\left(v_2 + \omega\, v_3\right)\, , \label{a}\\
&= \smfrac{1}{2}\left(\,u_4 + i u_6 + \omega (u_5 + i u_7)\,\right) = \smfrac{1}{2} \left(\,u_4 + \omega^2 u_6 +
\omega\, u_5 + \omega^3 u_7\,\right) \, , \label{b} \\ 
&= \smfrac{1}{\sqrt{8}}\left(\,x_0 - x_4 + \omega^2(x_2 - x_6) + \omega(x_1 - x_5) +
\omega^3(x_3 - x_7)\,\right) \, , \label{c} \\
&= \smfrac{1}{\sqrt{8}}\left(\,x_0 + \omega \, x_1 + \omega^2 x_2 + \omega^3 x_3 + \omega^4 x_4 + \omega^5
x_5 + \omega^6 x_6 + \omega^7 x_7\,\right) \, ,
\label{d}
\end{align}
\end{subequations}
which agrees with Eq.~\eqref{y1}. We have used Eq.~\eqref{FFT:1} to get
Eq.~\eqref{a}, Eqs.~\eqref{FFT:2} and \eqref{3} to get Eq.~\eqref{b}, and
Eqs.~\eqref{4}, \eqref{5}, \eqref{6} and \eqref{7} to get
Eq.~\eqref{c}. Equation \eqref{d} is the same as Eq.~\eqref{c} with powers of
$\omega$ written out explicitly using Eq.~\eqref{powers}. 

It is instructive to write the linear transformations in Eqs.~\eqref{FT},
\eqref{step1}, \eqref{step2} and \eqref{step3} in matrix form.
Equation \eqref{FT} is written in matrix formulation as 
\begin{equation}
\vec{y} = U \vec{x} \, ,
\end{equation}
where 
\begin{equation}
U = {1 \over \sqrt{8}}
\begin{pmatrix}
1     & 1        & 1        & 1        & 1        & 1         &  1       &  1        \\
1     & \omega   & \omega^2 & \omega^3 & \omega^4 & \omega^5  & \omega^6 &  \omega^7 \\
1     & \omega^2 & \omega^4 & \omega^6 & 1        & \omega^2  & \omega^4 &  \omega^6 \\
1     & \omega^3 & \omega^6 & \omega   & \omega^4 & \omega^7  & \omega^2 &  \omega^5 \\
1     & \omega^4 & 1        & \omega^4 & 1        & \omega^4  & 1        &  \omega^4 \\
1     & \omega^5 & \omega^2 & \omega^7 & \omega^4 & \omega    & \omega^6 &  \omega^3 \\
1     & \omega^6 & \omega^4 & \omega^2 & 1        & \omega^6  & \omega^4 &  \omega^2 \\
1     & \omega^7 & \omega^6 & \omega^5 & \omega^4 & \omega^3  & \omega^2 &  \omega  
\end{pmatrix}
\, .
\end{equation}
Equation \eqref{step1} in matrix form is
\begin{equation}
\vec{u} = D \vec{x} \, ,
\end{equation}
where
\begin{equation}
D = {1 \over \sqrt{2}}
\begin{pmatrix}
1     & 0        & 0        & 0        & 1        & 0         &  0       &  0        \\
0     & 1        & 0        & 0        & 0        & 1         & 0        &  0        \\
0     & 0        & 1        & 0        & 0        & 0         & 1        &  0        \\
0     & 0        & 0        & 1        & 0        & 0         & 0        &  1        \\
1     & 0        & 0        & 0        & \omega^4 & 0         & 0        &  0        \\
0     & 1        & 0        & 0        & 0        & \omega^4  & 0        &  0        \\
0     & 0        & 1        & 0        & 0        & 0         & \omega^4 &  0        \\
0     & 0        & 0        & 1        & 0        & 0         & 0        &  \omega^4 
\end{pmatrix}
\, .
\end{equation}
Equation \eqref{step2} in matrix form is
\begin{equation}
\vec{v} = E \vec{u} \, ,
\end{equation}
where
\begin{equation}
E = {1 \over \sqrt{2}}
\begin{pmatrix}
1     & 0        & 1        & 0        & 0        & 0         & 0        & 0        \\
0     & 1        & 0        & 1        & 0        & 0         & 0        & 0        \\
0     & 0        & 0        & 0        & 1        & 0         & \omega^2 & 0        \\
0     & 0        & 0        & 0        & 0        & 1         & 0        & \omega^2 \\
1     & 0        & \omega^4 & 0        & 0        & 0         & 0        & 0        \\
0     & 1        & 0        & \omega^4 & 0        & 0         & 0        & 0        \\
0     & 0        & 0        & 0        & 1        & 0         & \omega^6 & 0        \\
0     & 0        & 0        & 0        & 0        & 1         & 0        & \omega^6 
\end{pmatrix}
\, .
\end{equation}
Equation \eqref{step3} in matrix form is
\begin{equation}
\vec{y} = F \vec{v} \, ,
\end{equation}
where
\begin{equation}
F = {1 \over \sqrt{2}}
\begin{pmatrix}
1     & 1        & 0        & 0        & 0        & 0         & 0        & 0        \\
0     & 0        & 1        & \omega   & 0        & 0         & 0        & 0        \\
0     & 0        & 0        & 0        & 1        & \omega^2  & 0        & 0        \\
0     & 0        & 0        & 0        & 0        & 0         & 1        & \omega^3 \\
1     & \omega^4 & 0        & 0        & 0        & 0         & 0        & 0        \\
0     & 0        & 1        & \omega^5 & 0        & 0         & 0        & 0        \\
0     & 0        & 0        & 0        & 1        & \omega^6  & 0        & 0        \\
0     & 0        & 0        & 0        & 0        & 0         & 1        & \omega^7 
\end{pmatrix}
\, .
\end{equation}
Notice that $D, E$ and $F$, which describe the FFT,
are very sparse, they have only two entries in
each row and column, so they can be multiplied very efficiently, whereas the
matrix $U$, which describes the original Fourier transform, is dense.
With some tedious matrix manipulations one can verify that
\begin{equation}
U = F\, E\, D \, ,
\end{equation}
as required. (I used \textit{Mathematica}.)

\section{Beyond $N = 8$}

Now we discuss how we obtained Eqs.~\eqref{step1}--\eqref{step3}.
For a general value $n$, with $N=2^n$, the discrete FT is defined by
\begin{equation}
y_k = {1 \over \sqrt{N}} \sum_{m=0}^{N-1} \omega^{k m} x_m \, , \qquad (k= 0, 1, \cdots, N-1)
\label{FT_N}
\end{equation}
with $\omega$ given by Eq.~\eqref{omega}.
We can break Eq.~\eqref{FT_N} into even and odd terms as follows:
\begin{align}
y_k &= {1 \over\sqrt{N}}\left[ \sum_{m=0}^{N/2-1} \omega^{2 km} x_{2 m} +
\sum_{m=0}^{N/2-1} \omega^{
k (2m+1)} x_{2 m + 1} \right] \, , \nonumber\\
&= {1 \over\sqrt{2}}\left[ \sqrt{2\over N} \sum_{m=0}^{N/2-1} (\omega^2)^{km} x_{2 m} +
\omega^k \sqrt{2\over N} \sum_{m=0}^{N/2-1}
(\omega^2)^{km} x_{2m + 1} \right]\, , \qquad (k= 0, 1, \cdots, N-1) \, . \label{recurse}
\end{align}
Noting that $\omega^2$ is the complex exponential factor analogous to Eq.~\eqref{omega}
which figures in a Fourier
Transform with $N/2$ points, we see that the first term in Eq.~\eqref{recurse}
is a FT for the $N/2$ even points and the second term is the FT for the $N/2$
odd points. 
We can write Eq.~\eqref{recurse} as 
\begin{equation}
\boxed{
y_k =\smfrac{1}{\sqrt{2}}\left[v_{2k} + \omega^k v_{2k+1}\right] \, , \quad (k= 0, 1, \cdots, N-1)\, , }
\label{ykv}
\end{equation}
where
\begin{subequations}
\begin{align}
v_{2k} &=\sqrt{2\over N} \sum_{m=0}^{N/2-1} (\omega^2)^{km} x_{2 m} \, , \\
v_{2k+1} &= \sqrt{2 \over N} \sum_{m=0}^{N/2-1} (\omega^2)^{km} x_{2 m+1}\, ,\quad  (k= 0, 1,
\cdots,N-1) \, .
\end{align}
\label{v}
\end{subequations}
Here $k$ runs over the range $0, 1, \cdots,N-1$ so the indices on the $v_j$ in
Eqs.~\eqref{v} run from
0 to $2 N - 1$.  However, since $\omega^N = 1$, see Eq.~\eqref{omega}, it follows
from the definition of
the $v_j$ in Eq,~\eqref{v} that $v_{j_+N} = v_j$. Hence the index $j$, of
the $v_j$
is to be evaluated modulo $N$. This applies in an obvious way to other
quantities as well, such as the $u_j$, and, in Sec.~\ref{sec:general}, to the lower
index on the $x^{(\ell)}_j$.

For $N=8$ please check that Eq.~\eqref{ykv} corresponds
to our Eqs.~\eqref{step3} for $k = 0, 1, 2,
\cdots 7$ and that, according to Eqs.~\eqref{v},
the expressions for the $v_k$ in terms of the original data $x_m$
are
\begin{subequations}
\begin{align}
v_0 &= \smfrac{1}{2} \sum_{m=0}^{3} x_{2 m} \, ,
\qquad v_2 = \smfrac{1}{2} \sum_{m=0}^{3} (\omega^2)^m x_{2 m}\, ,
\qquad v_4 = \smfrac{1}{2} \sum_{m=0}^{3} (\omega^2)^{2m} x_{2 m}\, ,
\qquad v_6 = \smfrac{1}{2} \sum_{m=0}^{3} (\omega^2)^{3m} x_{2 m} \, ,\\
v_1 &= \smfrac{1}{2} \sum_{m=0}^{3} x_{2m + 1}\, ,
\quad v_3 = \smfrac{1}{2} \sum_{m=0}^{3} (\omega^2)^{m} x_{2m + 1}\, ,
\quad v_5 = \smfrac{1}{2} \sum_{m=0}^{3} (\omega^2)^{2m} x_{2m + 1}\, ,
\quad v_7 = \smfrac{1}{2} \sum_{m=0}^{3} (\omega^2)^{3n} x_{2m + 1}\, ,
\end{align}
\end{subequations}
so $v_0, v_2, v_4 $ and $v_6$ are the FT of the 4 even points for $k = 0, 1,
2$ and $3$ respectively, while $v_1, v_3, v_5$ and $v_7$ are the FT of the 4
odd points for  $k = 0, 1,
2$ and $3$ respectively.

We can again separate each of Eqs.~\eqref{v} into even and odd terms by
analogy with Eq.~\eqref{recurse}. We have
\begin{subequations}
\begin{align}
v_{2k} &= \sqrt{2 \over N} \left[\sum_{m=0}^{N/4-1} (\omega^4)^{km} x_{4 m} + 
\left(\omega^2\right)^k \sum_{m=0}^{N/4-1} (\omega^4)^{km} x_{4 m + 2} \right] \, ,\\
v_{2k+1} &= \sqrt{2 \over N} \left[\sum_{m=0}^{N/4-1} (\omega^4)^{km} x_{4 m+1} + 
\left(\omega^2\right)^k \sum_{m=0}^{N/4-1} (\omega^4)^{km} x_{4 m + 3} \right] \, .
\end{align}
\label{recurse2}
\end{subequations}
We can write these equations as
\begin{subequations}
\begin{align}
v_{2k} &= \smfrac{1}{\sqrt{2}} \left[ u_{4k} +
\left(\omega^2\right)^k u_{4k+2}\right]\, ,\\
v_{2k+1} &= \smfrac{1}{\sqrt{2}} \left[ u_{4k+1} +
\left(\omega^2\right)^k u_{4k+3}\right] \, , \quad(k = 0, 1,
\cdots, N/2 - 1)\, ,
\end{align}
\label{v2k}
\end{subequations}
where
\begin{subequations}
\begin{align}
u_{4k} &= \sqrt{4 \over N} \sum_{m=0}^{N/4-1} (\omega^4)^{km} x_{4 m}\, , 
\qquad\quad u_{4k+1} = \sqrt{4 \over N} \sum_{m=0}^{N/4-1} (\omega^4)^{km} x_{4 m+1}\, , \\
u_{4k+2} &= \sqrt{4 \over N} \sum_{m=0}^{N/4-1} (\omega^4)^{km} x_{4 m+2}\, , 
\qquad u_{4k+3} = \sqrt{4 \over N} \sum_{m=0}^{N/4-1} (\omega^4)^{km} x_{4 m+3}\, .
\end{align}
\label{u4k}
\end{subequations}
Note that the two equations in Eqs.~\eqref{v2k} can be combined as
\begin{equation}
\boxed{
v_{2k+p} = \smfrac{1}{\sqrt{2}}\left[ u_{4k+p} + \left(\omega^2\right)^k u_{4k+p+2}\right]\, , \quad
(p=0,1),(k=0, 1, \cdots, N/2-1)\, .}
\label{v2kp}
\end{equation}
Again, the index $j$ on the $u_j$ is to be evaluated modulo $N$.

For $N=8$ please check that Eq.~\eqref{v2kp} corresponds to our Eqs.~\eqref{step2} for
$p=0, 1$, and $k=0, 1, 2$ and $3$,
and that, according to Eqs.~\eqref{u4k}, the explicit expressions for the $u_j$ are
\begin{subequations}
\begin{align}
u_0 &= \smfrac{1}{\sqrt{2}}\sum_{m=0}^{1} x_{4 m} = \smfrac{1}{\sqrt{2}}(x_0 + x_4) \, , \qquad\quad\quad\quad
u_1 = \smfrac{1}{\sqrt{2}}\sum_{m=0}^{1} x_{4 m+1} = \smfrac{1}{\sqrt{2}}(x_1 + x_5)\, , \\
u_2 &= \smfrac{1}{\sqrt{2}}\sum_{m=0}^{1} x_{4 m + 2} = \smfrac{1}{\sqrt{2}}(x_2 + x_6)\, ,\qquad\quad\quad
u_3 = \smfrac{1}{\sqrt{2}}\sum_{m=0}^{1} x_{4 m + 3} = \smfrac{1}{\sqrt{2}}(x_3 + x_7) \, , \\
u_4 &= \smfrac{1}{\sqrt{2}}\sum_{m=0}^{1} (\omega^4)^{m} x_{4 m} = \smfrac{1}{\sqrt{2}}(x_0 - x_4) \, , \quad\quad
u_5 = \smfrac{1}{\sqrt{2}}\sum_{m=0}^{1} (\omega^4)^{m} x_{4 m + 1} = \smfrac{1}{\sqrt{2}}(x_1 - x_5) \, , \\
u_6 &= \smfrac{1}{\sqrt{2}}\sum_{m=0}^{1} (\omega^4)^{m} x_{4 m+2} = \smfrac{1}{\sqrt{2}}(x_2 - x_6) \, , \quad
u_7 = \smfrac{1}{\sqrt{2}}\sum_{m=0}^{1} (\omega^4)^{n} x_{4 m + 3} = \smfrac{1}{\sqrt{2}}(x_3 - x_7) \, .
\end{align}
\label{u}
\end{subequations}
Equations \eqref{u} agree with the expressions in Eq.~\eqref{step1}.
They can be written as a single equation as
\begin{equation}
\boxed{
u_{4k + p} = \smfrac{1}{\sqrt{2}} [x_{p} + (-1)^k x_{p+4}] \, , \quad(p=0, 1, 2, 3), (k=0,1) \, .}
\label{ukp}
\end{equation}

Thus we have seen that the FFT for $N=8\, (=2^n\ \text{with}\ n=3)$, which is
written out explicitly in Eqs.~\eqref{step1}--\eqref{step3}, corresponds to
firstly doing the Fourier transforms of length $2$ in Eq.~\eqref{ukp},
followed by two applications of the iterative procedure, the first shown in
Eq.~\eqref{v2kp} for length 4 and the second shown in Eq.~\eqref{ykv} for
length 8.

\section{The General Case}
\label{sec:general}

So far we have unsystematically labeled the results at each stage of
iteration by a different symbol, $x\to u \to v \to y$, see Fig.~\ref{fig:FFT8}.
When writing a code applicable for $N=2^n$ data points for arbitrary $n$,
one would use a common symbol but add a second index, so 
\begin{subequations}
\begin{align}
x_j &\equiv x^{(n)}_j\, , \\
& \vdots \nonumber \\
u_j &\equiv x^{(2)}_j \, , \\
v_j &\equiv x^{(1)}_j \, , \\
y_j &\equiv x^{(0)}_j\, .
\end{align}
\end{subequations}
Note that since $\omega = \exp(2\pi i / 2^n)$ we have
\begin{equation}
\omega^{2^n} = \exp(2 \pi i) = 1, \quad \omega^{2^{n-1}} = \exp(\pi i ) = -1.
\end{equation}

The $\ell$-th iteration, analogous to Eqs.~\eqref{v2kp}, \eqref{ykv} and
\eqref{ukp}
is
\begin{equation}
\boxed{
x^{(\ell-1)}_{2^{\ell-1}k + p} = \smfrac{1}{\sqrt{2}} \left[x^{(\ell)}_{2^{\ell} k + p} +
(\omega^{2^{\ell-1}})^k x^{(\ell)}_{2^{\ell} k + p + 2^{\ell-1}}\right]\, ,}
\label{iterate}
\end{equation}
with
\begin{equation}
p = 0, 1, \cdots, 2^{\ell-1}-1, \quad k = 0, 1, \cdots, 2^{n - \ell + 1} - 1 \, .
\end{equation}
Sorry that the notation is messy but I can't
see how to improve it; one just has to keep track of the indices and the
powers of $\omega$.
Recall that the lower index $j$ on the $x^{(\ell)}_j$
is to be evaluated modulo $2^n$. 

Let's see how this works.
\begin{itemize}
\item
We start with $\ell = n$, for which $x^{(\ell)}_j \equiv x_j$, the original data
points.\\
Equation \eqref{iterate} is then
\begin{equation}
x^{(n-1)}_{2^{n-1}k + p} = \smfrac{1}{\sqrt{2}}\left[x_{p} + (-1)^k x_{p + 2^{n-1}}\right] \, , \quad
(p=0,1,\cdots,2^{n-1}-1),\,(k=0,1)\, .
\end{equation}
For $n=3\, (N=8)$ this corresponds to Eq.~\eqref{ukp} with
$x^{(n-1)}_j \equiv u_j$.
\item
We then iterate Eq.~\eqref{iterate} for $\ell = n-1, n-2, \cdots, 2, 1$.\\
At the
next to the last iteration, $\ell = 2$, we have
\begin{equation}
x^{(1)}_{2k + p} = \smfrac{1}{\sqrt{2}}\left[x^{(2)}_{4 k + p} +
(\omega^{2})^k x^{(2)}_{4 k + p + 2}\right]\, , \quad
(p = 0, 1), \, (k = 0, 1, \cdots, 2^{n - 1} - 1) \, ,
\end{equation}
which corresponds to Eq.~\eqref{v2kp} with, $x^{(1)}_j \equiv v_j , x^{(2)}_j \equiv u_j$.
At the last iteration, $\ell = 1$, we obtain
\begin{equation}
y_{k} = \smfrac{1}{\sqrt{2}}\left[x^{(1)}_{2 k} + \omega^k x^{(1)}_{2k + 1}\right], \qquad (k=0, 1, 2,
\cdots, 2^n - 1) \, , 
\end{equation}
which is Eq.~\eqref{ykv}. (Recall that $x^{(0)}_j \equiv y_j$, the Fourier
transformed data, and $x^{(1)}_j \equiv v_j$.)
\end{itemize}

Note that the iterations are
evaluated in reverse, starting with $\ell = n$ and working down to $\ell = 1$. 


\end{subappendices}

%% file: QFT-FFT-all8.tex
\section{Introduction}
This chapter introduces the quantum Fourier transform (QFT), which is at the
heart of Shor's algorithm for period finding, and hence for factoring.
Shor's algorithm will be discussed in Chapter \ref{ch:shor}.
The appendices of this chapter make a detailed comparison with the (classical) Fast Fourier
Transform(FFT)\index{FFT}. The FFT is not part of the course so if you
are not interested in this comparison 
you can ignore the
appendices.

The QFT can be defined as follows. Starting with $n$ qubits in a single computational
basis state $|x\rangle_n$, where $x$ is an $n$-bit integer,
one generates the following superposition:
\begin{equation}
|x\rangle_n 
\mathrel{\stackrel{\makebox[0pt]{\mbox{\normalfont\tiny QFT}}}{\longrightarrow}}
|\psi_x \rangle_n ={1 \over 2^{n/2}} \sum_{y=0}^{2^n-1}
\exp[2 \pi i x y / 2^n]
|y\rangle_n 
\label{QFT_n}
\end{equation}
where $y$ is also an $n$-bit integer. The real power of the QFT arises, of
course, because it acts \textit{in parallel} if one inputs a superposition
$\sum_{x=0}^{2^n-1} a_x |x\rangle_n$, i.e.
\begin{equation}
\sum_{x=0}^{2^n-1} a_x |x\rangle_n 
\mathrel{\stackrel{\makebox[0pt]{\mbox{\normalfont\tiny QFT}}}{\longrightarrow}}
{1 \over 2^{n/2}} \sum_{x=0}^{2^n-1} a_x\, \sum_{y=0}^{2^n-1} \exp[2 \pi i x y / 2^n] |y\rangle_n 
= {1 \over 2^{n/2}} \sum_{y=0}^{2^n-1} \left[ \sum_{x=0}^{2^n-1}\, a_x\, \exp[2 \pi i x y / 2^n]
\right] |y\rangle_n .
\label{QFT_n_super}
\end{equation}
The circuit to perform the QFT, the derivation of which is the main topic of
this chapter and which is shown below in Fig.~\ref{had_gen_n},
takes no more time to act on the superposition in
Eq.~\eqref{QFT_n_super} than on the single basis state in
Eq.~\eqref{QFT_n}. This is where the power of the QFT lies.

Note that the effect of the QFT acting on a superposition, 
given in Eq.~\eqref{QFT_n_super}, can be written as
\begin{equation}
\sum_{x=0}^{2^n-1} a_x |x\rangle_n 
\mathrel{\stackrel{\makebox[0pt]{\mbox{\normalfont\tiny
QFT}}}{\longrightarrow}}
\sum_{y=0}^{2^n-1} a'_y |y\rangle_n ,
\end{equation}
where the transformed amplitudes $a'_y$ are related to the original amplitudes
$a_x$ by
\begin{equation}
a'_y = {1 \over 2^{n/2}} \sum_{x=0}^{2^n-1}\, \exp[2 \pi i x y / 2^n]\, a_x
\end{equation}
which is a discrete Fourier transform on the \textit{amplitudes}. This
transformation of the amplitudes is a
useful alternative way of defining a QFT, and is equivalent to Eq.~\eqref{QFT_n}.

\section{The trivial case of the QFT with one qubit}
\begin{figure}[htb!]
\begin{center}
\includegraphics[width=6cm]{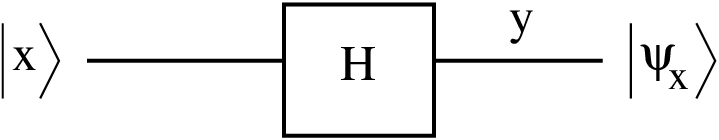}
\caption{The QFT for one qubit. The output from the Hadamard gate is $|\psi_{x}\rangle =
{1\over \sqrt{2}}(|y\!\!=\!0\rangle +
(-1)^{x} |y\!\!= \!1\rangle)$. This can be expressed as
${1 \over \sqrt{2}}\sum_{y=0}^1  (-1)^{x y}
|y\rangle = {1 \over \sqrt{2}}\sum_{y=0}^1 \exp[2 \pi i y x /2]|y_1\rangle$,
which is the QFT.
Recall that $x$ takes a fixed value 0 or 1.
\label{had}
}
\end{center}
\end{figure}

For the trivial case of one qubit, the QFT is 
\begin{equation}
|\psi_x\rangle = {1 \over \sqrt{2}} \sum_{y=0}^1  \exp[2 \pi i x y / 2]\, |y\rangle =
{1 \over \sqrt{2}} \sum_{y=0}^1 (-1)^{xy} |y\rangle .
\label{QFT0}
\end{equation}
This transformation is precisely that of the Hadamard gate, see Eq.~\eqref{Ho1}.
Graphically it is shown in Fig.~\ref{had}.

\section{QFT with two qubits}
\label{sec:qft}

\index{QFT} \index{quantum Fourier transform|see{QFT}}

%

The Quantum Fourier Transform (QFT) in Eq.~\eqref{QFT_n} for $n=2$ qubits is 
\index{superposition}
\begin{equation}
|\psi_x \rangle_2 = 
{1 \over 2} \sum_{y=0}^3 \exp\left[2 \pi i x y / 2^2\right] \, |y\rangle_2,
\label{QFT1}
\end{equation}
where $|x\rangle_2 \equiv |x_1 x_0\rangle$ and
$|y\rangle_2 \equiv |y_1 y_0\rangle$.
The $|\psi_x\rangle_2$
form a basis just as the $|x\rangle_2$ form a basis because one can show that
they are orthonormal, i.e.
\begin{equation}
_2\langle \psi_{x}| \psi_{x'} \rangle_2 = \delta_{{x},{x'}}\, .
\end{equation}

Noting that $y = y_0 + 2 y_1$ and $x= x_0 + 2 x_1$
we can simplify the argument of the exponential:
\begin{equation}
{2\pi i x y \over 2^2} = {2\pi i (x_0 + 2 x_1) (y_0 + 2y_1) \over 2^2}
= 2\pi i \left\{ y_0\left({x_0 \over 4} + {x_1 \over 2}\right) + y_1 \left(
{x_0 \over 2} + x_1\right) \right\}  \, .
\label{xy}
\end{equation}
Now $\exp(2 \pi i y_1 x_1) = 1$ so the factor $y_1 x_1$ above can  be neglected.
Hence Eq.~\eqref{QFT1} becomes
\begin{equation}
|\psi_x \rangle_2 =
\left( {1 \over \sqrt{2}}\sum_{y_0=0}^1 \exp\left[2\pi i y_0\left({x_0\over 4}
+ {x_1 \over 2}\right)\right] \right) \, \left({1 \over \sqrt{2}} \sum_{y_1=0}^1
\exp\left[2 \pi i y_1 {x_0 \over 2}\right] \right) |y_1 y_0 \rangle.
\label{QFT2}
\end{equation}

Next we will explain how to perform the operations in Eq.~\eqref{QFT2} using quantum
gates.

According to Eq.~\eqref{QFT0}, the second factor on the RHS of Eq.~\eqref{QFT2},
including the sum over $y_1$, is generated by
the Hadamard gate shown in Fig.~\ref{had}.

What about the first factor on the RHS of Eq.~\eqref{QFT2} which involves $y_0$? There
are two pieces in the exponential.  The factor involving $2\pi i y_0 x_1/2$,
including the sum over $y_0$, can
be dealt with by a Hadamard, similar to Fig.~\ref{had} but with the left hand
qubit being $x_1$ and the right hand qubit being labeled by $y_0$.  However,
the piece involving $2\pi i y_0 x_0/4$ is different.  It induces a phase
shift of $e^{i \pi / 2}$ for $y_0 = 1$ provided that $x_0$ is also 1. This
requires a controlled phase gate.  We define a phase gate $R_d$
by\footnote{This is the definition of $R_d$ that I
find most convenient. Some other authors adopt a
slightly different definition with  a factor of $e^{2\pi i / 2^d}$ instead of $e^{\pi i /
2^d}$.}
\begin{equation}
R_d = 
\begin{pmatrix}
1 & 0 \\
0 & e^{\pi i / 2^d} 
\end{pmatrix}
\, . \label{Rd}
\end{equation}
Acting on $|0\rangle, R_d$ makes no change, while acting on $|1\rangle\, R_d$
changes the phase by $\pi/ 2^d$.
Note that $R_0$ is just the $Z$ gate. Here we need $R_1$.

\begin{figure}[htb!]
\begin{center}
\includegraphics[width=8cm]{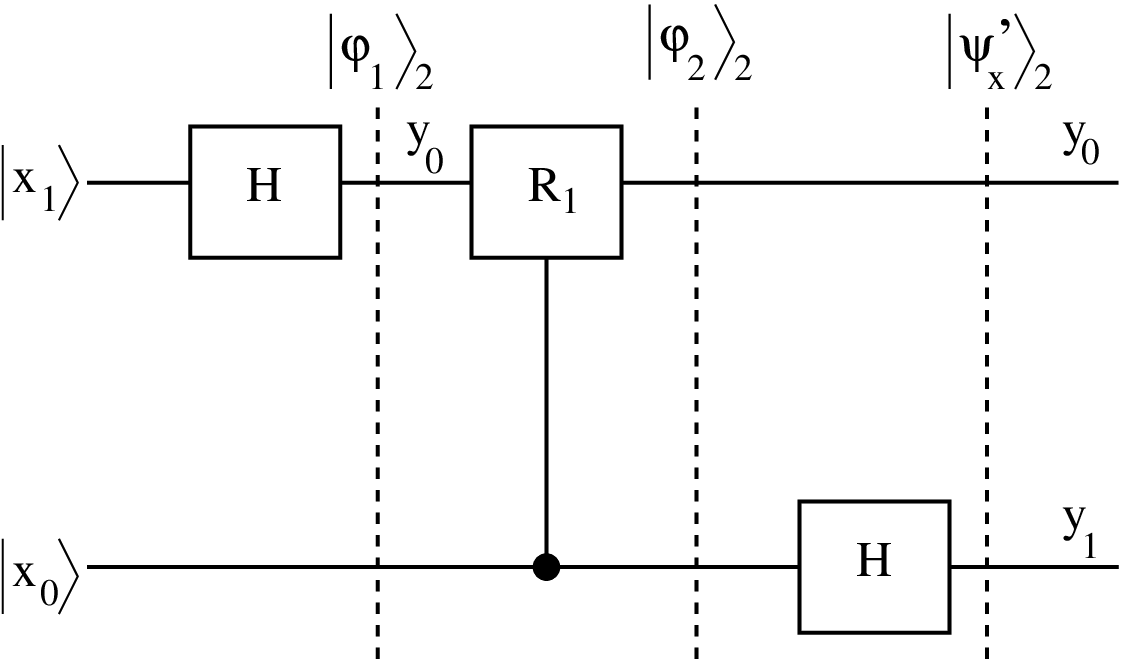}
\caption{The initial state
on the left is the single quantum state  $|x\rangle_2 \equiv |x_1 x_0\rangle$ 
in the computational basis. The
final state on the right is the superposition
$|\psi'_x\rangle_2 = (1/2)\sum_{y=0}^3 \exp(2 \pi i x y / 2^2) |y_0y_1\rangle$, which is
almost  $|\psi_x\rangle_2$,
the QFT of $|x\rangle_2 \equiv |x_1 x_0\rangle$ given in Eq.~\eqref{QFT2}, except
that the
order of the bits in the final state is the reverse of what it should be according to Eq.~\eqref{QFT2}.  This can be corrected by a swap gate as shown in
Fig.~\ref{had_R1_b}. Note the controlled-$R_1$ phase gate. This acts if the
control qubit, $x_0$, is 1, and changes the phase of the state if the target
qubit, $y_0$, is also equal to 1. The
general phase  gate $R_d$ is defined in Eq.~\eqref{Rd}.
\label{had_R1}
}
\end{center}
\end{figure}

Hence the exponential in the first term on the RHS of Eq.~\eqref{QFT2} can be
generated by a Hadamard followed by a controlled $R_1$ gate as shown for the
top qubit in Fig.~\ref{had_R1}, in which the $R_1$ gate on the upper qubit
is controlled by the lower
qubit, $x_0$.  Including the Hadamard on the lower qubit, Fig.~\ref{had_R1} generates
both factors on the RHS of Eq.~\eqref{QFT2}.

\begin{figure}[htb!]
\begin{center}
\includegraphics[width=10cm]{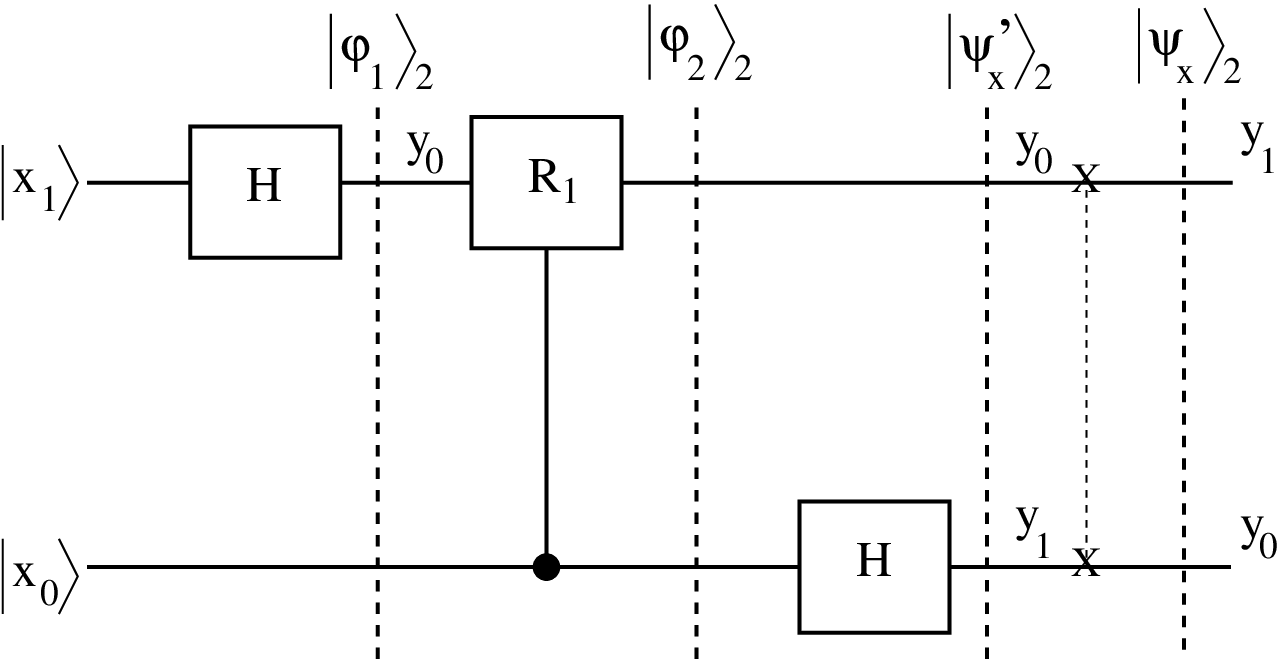}
\caption{The same as Fig.~\ref{had_R1} but with the addition of a swap gate on
the right (the dashed line with crosses at the ends). The final state is now
precisely
$|\psi_x\rangle_2 = (1/2)\sum_{y=0}^3 \exp(2 \pi i x y / 2^2) |y_1y_0\rangle$, the QFT given in Eq.~\eqref{QFT2}.
\label{had_R1_b}
}
\end{center}
\end{figure}

To make sure we understand we understand what is happening in the circuit
in Fig.~\ref{had_R1} we now write down the state at each of
the steps shown in the figure. The initial state is
\begin{subequations}
\begin{equation}
|x\rangle_2 = |x_1 x_0\rangle.
\end{equation}
After the first Hadamard the state is
\begin{equation}
|\phi_1\rangle_2 = {1 \over \sqrt{2}} \sum_{y_0=0}^1 e^{2\pi i y_0x_1/2} 
|y_0 x_0\rangle.
\end{equation}
After the controlled-$R_1$ gate we have
\begin{equation}
|\phi_2\rangle_2 = {1 \over \sqrt{2}} \sum_{y_0=0}^1
e^{2\pi i y_0x_1/2}  \,
e^{2\pi i y_0x_0/4} 
|y_0 x_0\rangle .
\end{equation}
The final state after the Hadamard on the lower qubit is therefore
\begin{equation}
|\psi'_x\rangle_2 = \left({1 \over \sqrt{2}} \sum_{y_0=0}^1
e^{2\pi i y_0x_1/2}  \,
e^{2\pi i y_0x_0/4} \right) \, \left(
{1 \over \sqrt{2}}
\sum_{y_1=0}^1 e^{2\pi i y_1x_0/2} \right)
|y_0 y_1\rangle .
\end{equation}
\end{subequations}
$|\psi'_x\rangle$ is almost the desired QFT in Eq.~\eqref{QFT2}, except that
the order of
the qubits on in the final state on the right has been reversed.  This can be
compensated for by adding a swap gate on the right as shown in
Fig.~\ref{had_R1_b}.

In terms of operators the circuit in Fig.~\ref{had_R1_b} corresponds to
\begin{equation}
\text{QFT}_2 = (\text{SWAP})\,(I \otimes H)\, (\text{Ctrl-}R_1) (H \otimes I) ,
\label{mat_prod}
\end{equation}
where in the tensor product the left operator refers to the upper qubit in the
figure. We
recall that for operators we read from right to left (the opposite of
circuit diagrams).

The $4 \times 4$ matrices for each piece in this operator product are
\begin{align}
\text{SWAP}\ &=  
\begin{pmatrix}
1 & 0 & 0 & 0 \\
0 & 0 & 1 & 0 \\
0 & 1 & 0 & 0 \\
0 & 0 & 0 & 1 
\end{pmatrix},\\
I \otimes H &= {1 \over \sqrt{2}}
\begin{pmatrix*}[r]
1 & 1  & 0 & 0 \\
1 & -1 & 0 & 0 \\
0 & 0  & 1 & 1 \\
0 & 0  & 1 & -1 
\end{pmatrix*}, \\
\text{Ctrl-}R_1 & =
\begin{pmatrix}
1 & 0 & 0 & 0 \\
0 & 1 & 0 & 0 \\
0 & 0 & 1 & 0 \\
0 & 0 & 0 & i 
\end{pmatrix} ,\\
H \otimes I &= {1 \over \sqrt{2}}
\begin{pmatrix*}[r]
1 & 0  & 1  & 0 \\
0 & 1  & 0  & 1 \\
1 & 0  & -1 & 0 \\
0 & 1  & 0  & -1 
\end{pmatrix*} .
\end{align}
For a dicsusion of how to construct matrices for a direct product of operators
on two qubits see Sec.~\ref{sec:composite}.
Multiplying the above matrices in the order specified in Eq.~\eqref{mat_prod} one can
verify that one correctly obtains the Fourier Transform for $N=4$ states
given in Eq.~\eqref{DFT_N4}.

This confirms that Fig.~\ref{had_R1_b} displays the circuit to
implement the QFT for 2 qubits.
I emphasize
that initially (on the left) the qubits are in a single computational
basis state, $|x_0\rangle$ and $|x_1\rangle$, whereas in the final state (on
the right) there is a sum over the states $y_0$ and $y_1$ (the sum being
generated by the Hadamards).


\section{QFT with three or more qubits}
\label{sec:3orm}

We next
do another special case, this time with
$n=3$ qubits. After this, we will be able to see the structure of the circuit for
\textit{general} $n$.

The QFT analogous to Eq.~\eqref{QFT2} is
\begin{align}
|\psi_x \rangle_3 &={1 \over 2^{3/2}} \sum_{y=0}^7 \exp[2 \pi i x y / 2^3]
|y\rangle_3 \\
\begin{split}
= \left( {1 \over \sqrt{2}}\sum_{y_0=0}^1 \exp\left[2\pi i y_0\left({x_0\over 8}
+ {x_1 \over 4} + {x_2 \over 2}\right)\right] \right) \, \left({1 \over \sqrt{2}} \sum_{y_1=0}^1
\exp\left[2 \pi i y_1\left({x_0\over 4} + {x_1 \over2}\right)\right] \right) \\
\times \left({1 \over \sqrt{2}} \sum_{y_2=0}^1 \exp\left[2 \pi i y_2{x_0 \over
2}\right]\right)
|y_2 y_1 y_0 \rangle,
\end{split}
\label{QFT3}
\end{align}
where we have again replaced factors of $\exp(2 \pi i\times \,\mathrm{integer})$ by
unity.

Note that the terms in the exponential are of the form
\begin{equation}
2 \pi i x_j y_k {2^j 2^k \over 2^n} 
\label{xjyk}
\end{equation}
where $k$ runs from $0$ to $n-1$ and $j$ runs from $0$ to $n - j - 1$.

\begin{figure}[tbh!]
\begin{center}
\includegraphics[width=12cm]{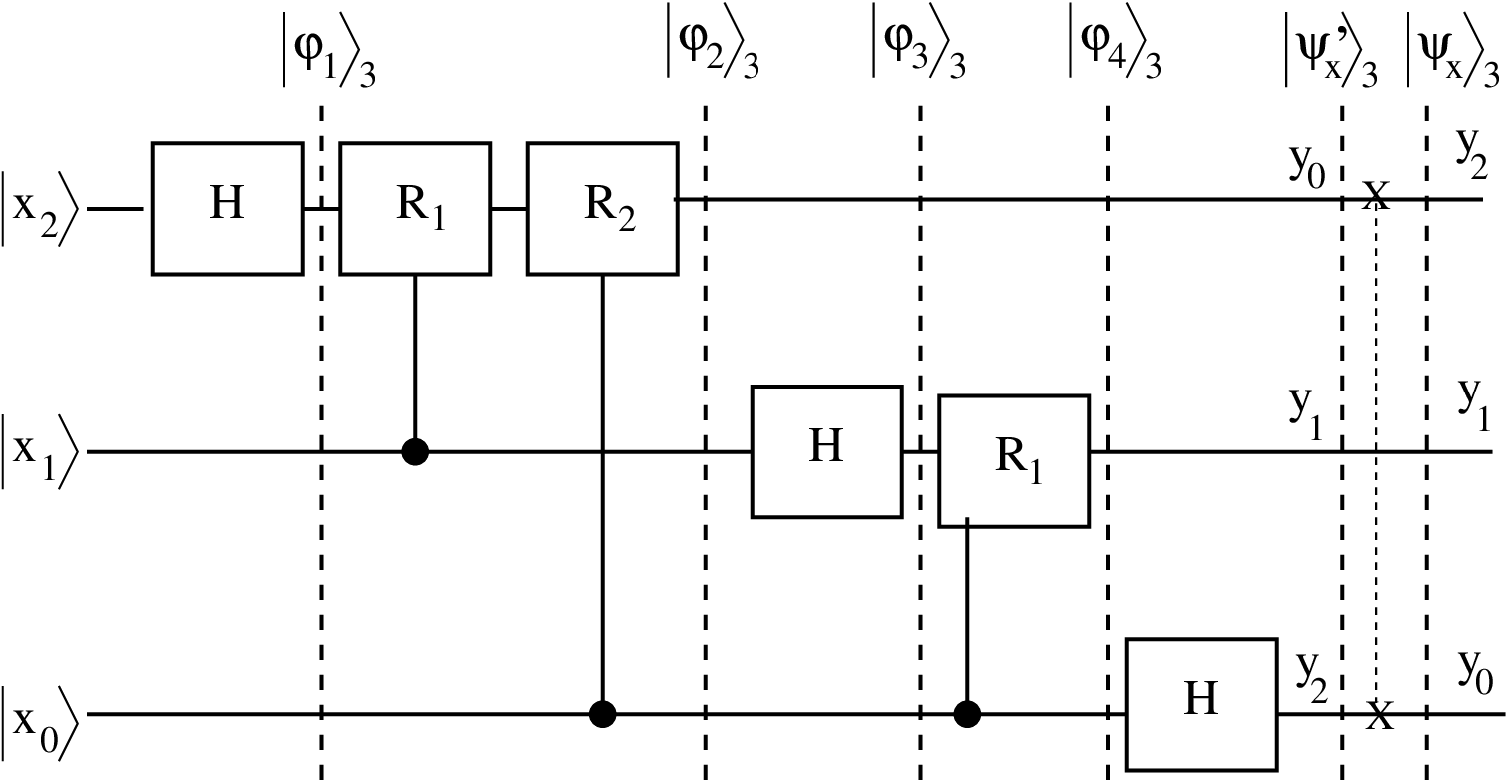}
\caption{
Circuit diagram for performing the QFT with $n=3$ qubits. It generates the
transformation shown in Eq.~\eqref{QFT3}. The initial state is $|x\rangle_3 =
|x_2 x_1 x_0\rangle$ and the subsequent states are given in Eqs.~\eqref{states_3}.
The phase gates, $R_d$, are defined
in Eq.~\eqref{Rd}. The dashed line with crosses at the ends indicates a swap
gate between qubits 0 and 2.  This serves to reverse the order of the qubits.
\label{had_R1_R2}
}
\end{center}
\end{figure}

Following along the lines in the previous section, the circuit diagram which
will perform this is shown in Fig.~\ref{had_R1_R2}. 
To make sure we understand this circuit we will write down the
state at each stage indicated on the figure. (Although these expressions look
rather complicated is useful to make the effort to understand them.)
The initial state is
\begin{subequations}
\label{states_3}
\begin{equation}
|x\rangle_3 = |x_2 x_1 x_0\rangle , \hspace{12cm}
\end{equation}
and the subsequent states, labeled in Fig.~\ref{had_R1_R2}, are
\begin{align}
|\phi_1\rangle_3 = 
&\left( {1 \over \sqrt{2}}\sum_{y_0=0}^1 \exp\left[2\pi i y_0\left(
{x_2 \over 2}\right)\right] \right) \, 
|y_0 x_1 x_0 \rangle , \\
|\phi_2\rangle_3 = 
&\left( {1 \over \sqrt{2}}\sum_{y_0=0}^1 \exp\left[2\pi i y_0\left({x_0\over 8}
+ {x_1 \over 4} + {x_2 \over 2}\right)\right] \right) \, 
|y_0 x_1 x_0 \rangle , \\
|\phi_3\rangle_3 = 
&\left( {1 \over \sqrt{2}}\sum_{y_0=0}^1 \exp\left[2\pi i y_0\left({x_0\over 8}
+ {x_1 \over 4} + {x_2 \over 2}\right)\right] \right) \, \left({1 \over \sqrt{2}} \sum_{y_1=0}^1
\exp\left[2 \pi i y_1\left({x_1 \over2}\right)\right] \right)
|y_0 y_1 x_0 \rangle , 
\end{align}
\begin{equation}
\begin{split}
|\phi_4 \rangle_3 =
\left( {1 \over \sqrt{2}}\sum_{y_0=0}^1 \exp\left[2\pi i y_0\left({x_0\over 8}
+ {x_1 \over 4} + {x_2 \over 2}\right)\right] \right) \, \left({1 \over \sqrt{2}} \sum_{y_1=0}^1
\exp\left[2 \pi i y_1\left({x_0\over 4} + {x_1 \over2}\right)\right] \right)
\\
|y_0 y_1 x_0 \rangle ,
\end{split}
\end{equation}
\begin{equation}
\begin{split}
|\psi'_x \rangle_3 = 
\left( {1 \over \sqrt{2}}\sum_{y_0=0}^1 \exp\left[2\pi i y_0\left({x_0\over 8}
+ {x_1 \over 4} + {x_2 \over 2}\right)\right] \right) \, \left({1 \over \sqrt{2}} \sum_{y_1=0}^1
\exp\left[2 \pi i y_1\left({x_0\over 4} + {x_1 \over2}\right)\right] \right) \\
\times \left({1 \over \sqrt{2}} \sum_{y_2=0}^1 \exp\left[2 \pi i y_2{x_0 \over
2}\right]\right)
|y_0 y_1 y_2 \rangle.
\end{split}
\end{equation}
\end{subequations}
$|\psi'_x\rangle$ is almost the desired QFT in Eq.~\eqref{QFT3}, except that
the order of
the qubits on in the final state on the right has been reversed.  This can be compensated for by adding a swap
gate between qubits 1 and 3. Hence $|\psi_x\rangle$ in the figure is the
desired QFT for 3 qubits given in Eq.~\eqref{QFT3}.

Intuitively, the reason that for the reverse order of the qubits in the final
state before the swaps, is the following.  The Hadamards generate the
superpositions, i.e.~the sums over the $y_j$. They also produce the factors
in the exponential involving $2 \pi i / 2$.
From
the straightforward generalization of Eq.~\eqref{QFT3} to arbitrary $n$,
see Eq.~\eqref{xjyk}
it follows that the factors generated by the Hadamards are $(2 \pi
i / 2)\sum_{j=0}^{n-1} x_j y_{n-j-1}$. Here $x_j$ is the label of the $j$-th
physical
qubit in its initial state, and $y_{n-j-1}$ is the dummy label for the state of
the same physical qubit in its final state.  Because it is
the label $y_{n-j-1}$ (rather than $y_j$) which occurs on the same physical qubit as $x_j$,
the qubits in the final state are in reverse order.

\begin{figure}[tbh!]
\begin{center}
\includegraphics[width=14cm]{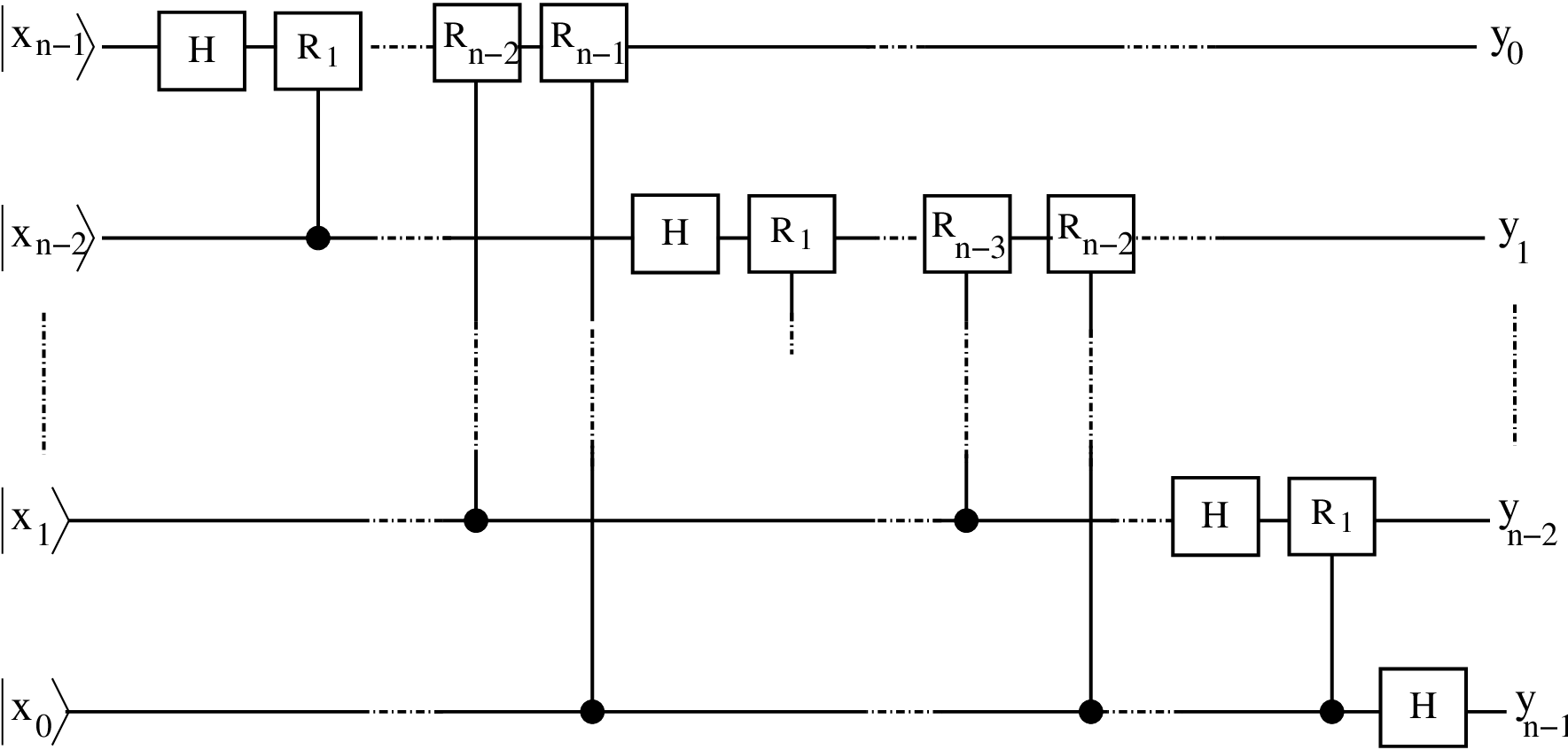}
\caption{
Circuit diagram for performing the QFT with an arbitrary number of qubits.
For clarity the final swaps are not shown, so the input states
on the left, $x_i$, and the output states on the right, $y_i$, are in opposite
order. Note that the controlled phase gate between qubits $x_i$ and $x_j$ is
$R_{|i-j|}$, which makes the structure fairly simple. The state inputted 
on the left is a single computational basis state $|x\rangle_n$, and if we
add the final swaps, the
state outputted on the right is the superposition in Eq.~\eqref{QFT_n}. 
\label{had_gen_n}
}
\end{center}
\end{figure}

Comparing with the case for two
qubits shown in Fig.~\ref{had_R1_b}, and that for three qubits in Fig.~\ref{had_R1_R2},
the generalization to an arbitrary number of
qubits
can be deduced and is shown in Fig.~\ref{had_gen_n}.
Each $x_j$ is acted on by a Hadamard followed by controlled phase gates in
which the control is provided by the $x_i$ for all $i$ less than $j$.
Note that the
controlled phase gate between qubits $x_i$ and $x_j$ is
$R_{|i-j|}$, which makes the structure fairly simple.

For an $n$-qubit QFT 
one needs $n$ Hadamard gates.  The number of
controlled phase gates is $1 + 2 +\cdots + n-1 = n(n-1)/2$. Also $[n/2]$
swaps are required, where $[k]$ denotes the largest integer less than or equal to $k$.
The circuit therefore provides an algorithm for performing
the QFT in $O(n^2)$ steps. By contrast the FFT requires $O(n 2^n)$ steps which
is exponentially greater. Note that the number of terms in the Fourier Series
is $N = 2^n$. 

However, we cannot obtain the $2^n$ Fourier
amplitudes from the QFT since a measurement will just give one of the
basis states with a probability
proportional to the square of the absolute value of its Fourier amplitude.
However, the QFT does give useful information if the input state is a linear
combination $\sum_x a_x |x\rangle$,
see Eq.~\eqref{QFT_n_super}, in which the $a_x$ are \textit{periodic} in $x$ with
some period $r$.  As we shall see in Chapter \ref{ch:shor}
the Fourier amplitudes are then strongly
peaked at values of $y$ which are
multiples of $2^n / r$, so there is a high probability that a measurement of
$y$ will give a value
which is equal or close to a multiple of $2^n / r$. As we shall also see in
Chapter \ref{ch:shor}, from this information
one can then deduce the period $r$ with high probability.  Hence the QFT is very useful
for period finding.

As we saw in Chapter \ref{ch:period},
period finding\index{period finding algorithm}
can be used to factor integers. If one could factor large integers, one would
be able to
decode messages sent down the internet which have been encoded with
the standard RSA encryption method. \index{RSA encryption} We discussed RSA encryption in 
Chapter \ref{ch:rsa}.


Another application of the QFT is to estimate the phase of the eigenvalues of
a unitary matrix.  This is discussed in section \ref{sec:phase_est}.

\section{The Phase Estimation Algorithm}
\label{sec:phase_est}
\index{phase estimation}
This section gives an application of the quantum Fourier transform.

The eigenvalue of a unitary operator $U$ must be a pure phase, i.e.~$\lambda =
e^{i\theta}$. The reason is that $U$ preserves the norm of states, so if
$|\psi'\rangle = U |\psi\rangle$, we have
\begin{equation}
\langle \psi'| \psi'\rangle =
\langle U \psi| U  \psi\rangle =
\langle \psi| U^\dagger U | \psi\rangle =
\langle
\psi | \psi\rangle = 1,
\end{equation}
since $U^\dagger U = \mathbbm{1}$ and we used Eq.~\eqref{phiA}.
If $|\psi\rangle$ is an eigenstate of $U$, i.e.~$|\psi'\rangle = \lambda
|\psi\rangle$ this last equation becomes
\begin{equation}
1 = \langle \psi'| \psi'\rangle =
\langle \lambda \psi| \lambda \psi\rangle =
\langle \psi| \lambda^\star \lambda | \psi\rangle =
|\lambda|^2 \langle \psi | \psi\rangle = |\lambda|^2,
\end{equation}
so $|\lambda|^2 = 1$, and hence $\lambda = e^{i\theta}$ for some $\theta$.

The objective of this section is to determine an eigenvalue of a unitary
matrix, which is equivalent to determining its (complex) phase (since, as we
just showed, its modulus is 1). Hence this problem is called ``phase
estimation".

Let us write 
\begin{equation}
\theta = 2 \pi \phi
\end{equation}
so $0\le \phi < 1$.
The result for the phase $\phi$ will be encoded as an integer (formed
from the values of the measured qubits) and
let's suppose we want to determine $\phi$ correct to $n$ bits of precision.
The procedure is to compute an $n$-bit integer $\phi'$, related to $\phi$ and
$\theta$ by
\begin{equation}
\phi' = 2 ^n \phi, \qquad
\mathrm{so}\ \ \theta = 2 \pi {\phi' \over 2^n}.
\label{theta_phi}
\end{equation}
The possible values of $\phi'$ are  $0, 1, 2, \cdots, 2^n - 1$.

\begin{figure}[htb]
\begin{center}
\includegraphics[width=9cm]{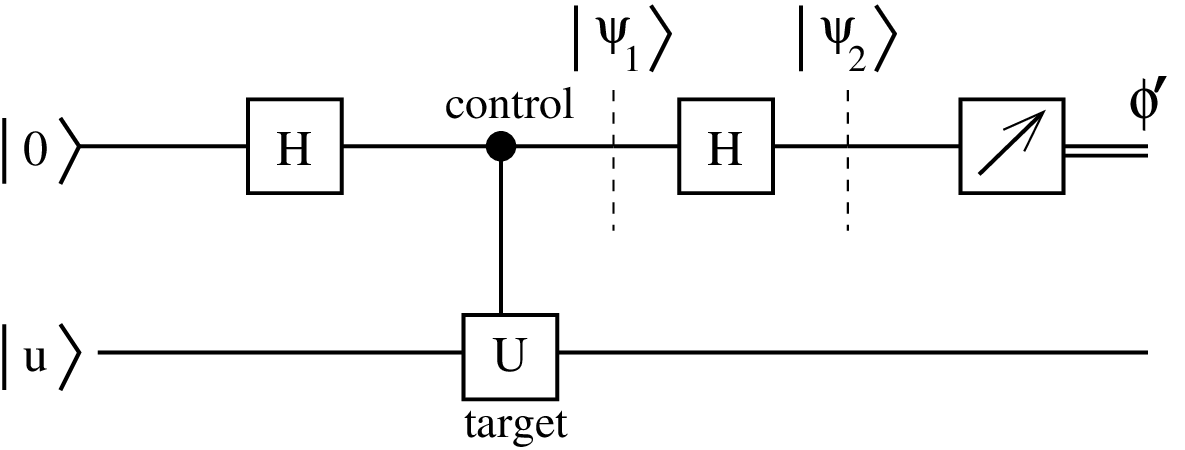}
\caption{The circuit for phase estimation for 1 bit of precision.
\label{phase1}
}
\end{center}
\end{figure}

We start with a simple example in which we only require 1 bit accuracy, so $\phi' =
0$ or $1$. We will see
that circuit in Fig.~\ref{phase1} does the trick.
Figure \ref{phase1} is essentially the same as Fig.~\ref{stabilizer} in Chapter \ref{ch:gates}. Here we
assume that $|u\rangle$ is an eigenstate of $U$ with eigenvalue $\exp(2 \pi i
\phi'/2)$. Following the discussion after Fig.~\ref{stabilizer} we find that 
\begin{align}
|\psi_1\rangle &= {1\over \sqrt{2}}\left(|0\rangle + e^{2 \pi i
\phi'/2}|1\rangle \right) , \nonumber \\
|\psi_2\rangle &= {1\over 2}\left[\, \left(1 + e^{2 \pi i \phi'/2} \right) |0\rangle
+ \left(1 - e^{2 \pi i \phi'/2} \right)|1\rangle  \, \right]
\end{align}
We see that if $\phi' = 0$ the measurement of the upper qubit gives $|0\rangle$ and
if $\phi' = 1$, the measurement of the upper qubit gives $|1\rangle$. Hence a
measurement of the upper qubit in Fig.~\ref{phase1} determines the phase to
one bit of precision.

We note that the right hand Hadamard on the upper qubit in Fig.~\ref{phase1} is just the QFT
for 1 qubit, see Fig.~\ref{had}. In fact, one can obtain $\phi'$ to an
arbitrary accuracy of
$n$-bits by using the $n$-bit QFT (strictly speaking the inverse
QFT).

\begin{figure}[htb]
\begin{center}
\includegraphics[width=12cm]{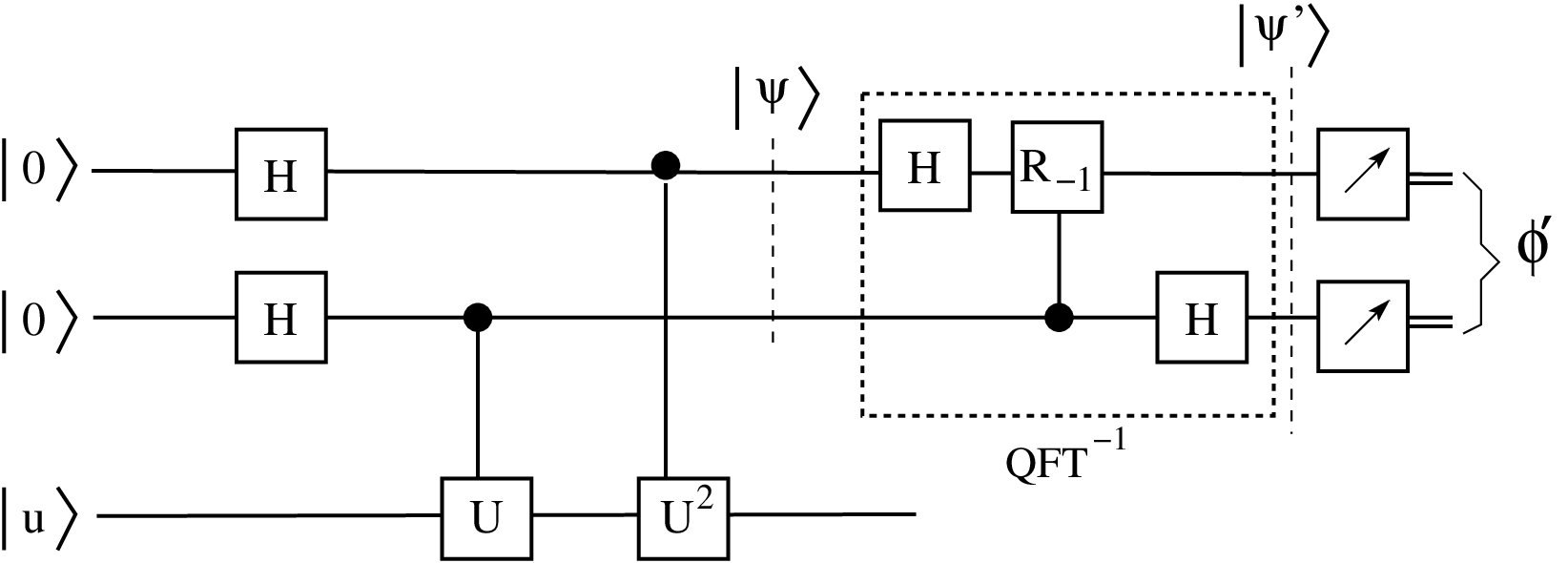}
\caption{The circuit for phase estimation for two bits of precision.
In their final state the two upper qubits contain the two bits of $\phi'$,
which is related to the phase $\theta$ by Eq.~\eqref{theta_phi}.
\label{phase2}
}
\end{center}
\end{figure}

To see this we proceed gently by considering the circuit in Fig.~\ref{phase2}
which is for two qubits. Both of the upper qubits are acted on by a Hadamard,
after which one of them is the control for a control-$U$ gate and the other is the
control for a control-$U^2$ gate. The state $|\psi\rangle$ is given by
\begin{align}
|\psi\rangle &= {1 \over \sqrt{2}}\left(\, |0\rangle + e^{2 \pi i \phi'/2^2} |1\rangle\,\right)
{1 \over \sqrt{2}} \left(\, |0\rangle + e^{4 \pi i \phi'/2^2} |1\rangle\,\right) \nonumber \\
&= {1 \over 2}\left(\, |00\rangle + e^{2 \pi i \phi'/2^2} |01\rangle
+ e^{4 \pi i \phi'/2^2} |01\rangle + e^{6 \pi i \phi'/2^2} |11\rangle\,\right) \nonumber \\
&= {1 \over 2}\left(\, |0\rangle_2 + e^{2 \pi i \phi'/2^2} |1\rangle_2
+ e^{4 \pi i \phi'/2^2} |2\rangle_2 + e^{6 \pi i \phi'/2^2} |3\rangle_2\,\right) \nonumber \\
&= {1 \over 2} \sum_{k=0}^3 e^{2 \pi i k \phi'/2^2} |k\rangle_2  .
\end{align}
This is just the  QFT of $|\phi'\rangle$ which can be
undone by an inverse QFT, i.e.
\begin{equation}
|k\rangle_2 \to {1 \over 2} \sum_{y=0}^2 e^{-2 \pi i y k / 2^3} 
|y\rangle_2 ,
\end{equation}
since, after the inverse QFT, the state of the system $|\psi'\rangle$ is given by
\begin{align}
|\psi'\rangle  & =
{1 \over 2} \sum_{k=0}^2 e^{2 \pi i k \phi' / 2^2}\ 
{1 \over 2} \sum_{y=0}^2 e^{-2 \pi i y k / 2^2} 
|y\rangle_2 \nonumber \\
& = {1 \over 2^2} \sum_{y=0}^2 \left[\, \sum_{k=0}^3 e^{2 \pi i (\phi'-y) k / 2^2} \, \right]|y\rangle_2
\nonumber \\
&= {1 \over 2^2}  \sum_{y=0}^2 2^2\, \delta_{y, \phi'}\,|y\rangle_2 \nonumber \\
&= |\phi'\rangle_2.
\label{phip}
\end{align}

In terms of gates, what is the difference between the quantum Fourier transform and its
inverse? For the quantum Fourier transform we use phase gates $R_d$,
defined by Eq.~\eqref{Rd}, which increase the phase of basis state $|1\rangle$ by $\pi/2^d$ and leave the 
phase of basis state $|0\rangle$ unchanged. In the inverse transform these are
replaced by gates, which we label $R_{-d}$, which \textit{decrease} the phase of
basis state $|1\rangle$ by $\pi/2^d$ and leave the phase of basis state
$|0\rangle$ unchanged, i.e. 
\begin{equation}
R_{-d} = 
\begin{pmatrix}
1 & 0 \\
0 & e^{-\pi i / 2^d} 
\end{pmatrix}
\, . \label{Rmd}
\end{equation}

The gates which perform the inverse quantum Fourier transform for two qubits are indicated in
Fig.~\ref{phase2}.
According to Eq.~\eqref{phip},
the final measurement in Fig.~\ref{phase2}, after the inverse quantum Fourier transform 
has been done, gives the $2$-bit integer $\phi'$ from which
the eigenvalue is given by $\lambda =e^{2 \pi i \phi' / 2^2}$.

\begin{figure}[htb]
\begin{center}
\includegraphics[width=12cm]{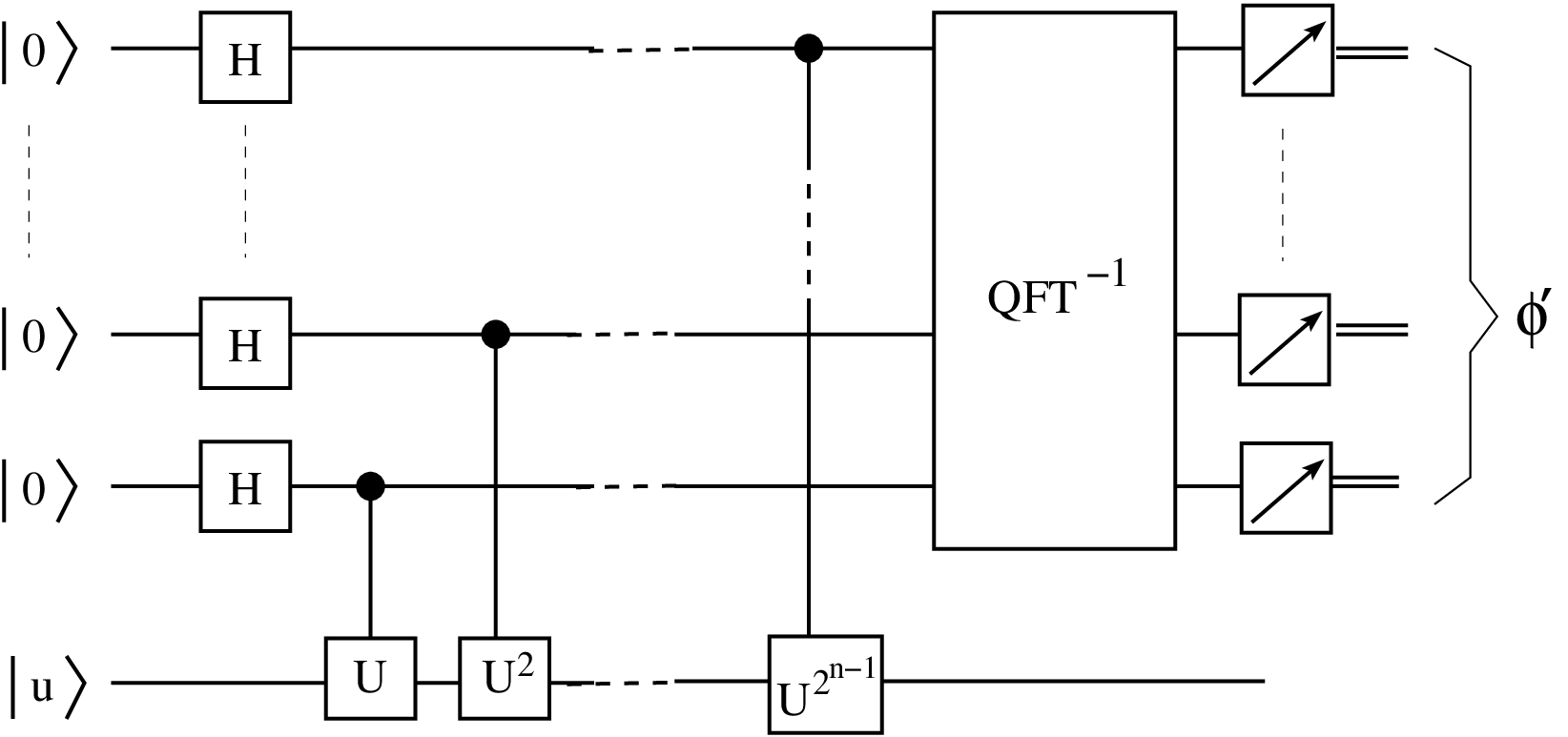}
\caption{The circuit for phase estimation. The values of the $n$ measured qubits 
form the binary representation
integer $\phi'$, related to an eigenvalue $\lambda = e^{i\theta}$ of the unitary operator $U$
by $\theta = 2 \pi \phi'/2^n$.
\label{phase3}
}
\end{center}
\end{figure}

This generalizes to the case of $n$ bits of precision. We need $n$ qubits
to act as control-$U$, control-$U^2$, control-$U^4,\cdots$, control-$U^{2^{n-1}}$ gates
on the qubit containing $|u\rangle$.
After the control-$U^{2^l}$ gates, for $l=0, 1, \cdots, n-1$,
have acted, we run the qubits through the
inverse Fourier transform to get $|\phi'\rangle_n$, from which $\theta = 2 \pi
\phi'/2^n$. The circuit is shown in Fig.~\ref{phase3}.

What happens if $|u\rangle$ is not a single eigenstate of $U$ as we have been assuming up
to now, but a superposition?
After the inverse QFT, the
state of the $n$ qubits will be a superposition of
computational basis states $|\phi'\rangle$
for \textit{each} of the eigenvalues present in the decomposition of $|\psi\rangle_n$ into
its eigenstates. Measurement will then project on to the value
of $\phi'$ corresponding to \textit{one} of the eigenvalues.

\hrulefill
\section*{Problems}
\input{hw_ch16.tex}

\begin{center}
{\Large\bf Appendices}
\end{center}

\begin{subappendices}
\input{app-QFT-FFT4.tex}

%% file: hw_ch16.tex
\begin{problems}
\item
We stated (without proof) that any quantum gate can be made out
of single qubit gates and the CNOT gate (i.e.~the only gate needed with more
than one qubit is CNOT). Here we illustrate this for the controlled phase gate
used in Shor's quantum Fourier transform.

The (uncontrolled) phase gate (acting on one qubit) has the matrix
representation
\begin{equation}
R(\theta) = 
\begin{pmatrix}
1 & 0 \\
0 & e^{i \theta} \\
\end{pmatrix} ,
\end{equation}
so the phase is changed by $\theta$ if the qubit is in state $|1\rangle$ and is
unchanged if the qubit is in state $|0\rangle$.

Now we want this gate to be controlled by a control qubit such that the gate
will only act on the target qubit
if the control qubit is in state $|1\rangle$. We want to find
out how to do this using 1-qubit gates (including $R(\theta)$) and the CNOT
(Ctrl-$X$) 2-qubit gate.
Note that the $4 \times 4 $ matrix representation for the controlled phase
gate is
\begin{align}
& \quad |00\rangle \ \ \ |01\rangle \ \ \ |10\rangle \ \ \,|11\rangle
\nonumber \\
\begin{matrix}
\langle 00 | \\
\langle 01 | \\
\langle 10 | \\
\langle 11 | 
\end{matrix}
& \begin{pmatrix}
\ 1\  & \quad\  0\  & \quad 0\  & \ 0\  \\
\ 0\  & \quad\  1\  & \quad 0\  & \ 0\  \\
\ 0\  & \quad\  0\  & \quad 1\  & \ 0\  \\
\ 0\  & \quad\  0\  & \quad 0\  & \ e^{i\theta}\  \\
\end{pmatrix}
,
\label{m1}
\end{align}
where the control qubit is to the left and the target qubit is to the right.

Show that the following circuit almost generates a controlled-phase gate:

\begin{center}
\includegraphics[width=7cm]{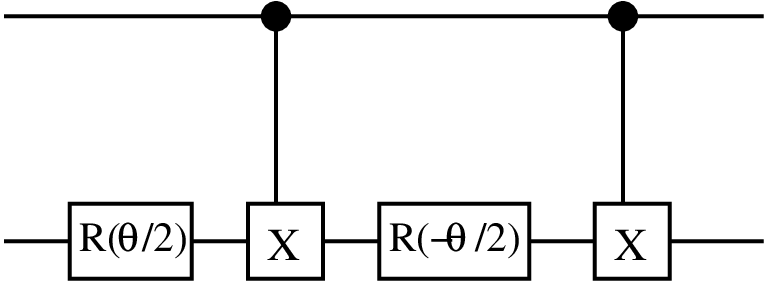}
\end{center}

In particular, write the $4 \times 4$ matrix representation of this circuit.
From this you should be able show that if the control (upper) qubit is 0 then the qubits are
unchanged (as required) but that if the control qubit is 1 (so the gate is
activated) then the relative phase between the $|1\rangle$ and $|0\rangle$
states of the target (lower) qubit is $\theta$ as required, but the overall phase
of these two states is not correct\footnote{Note that what I mean by
this overall phase is the common phase of
the two states of the target qubit when the control qubit is $|1\rangle$.
The existence of this phase means that there is an error in
the \textit{relative} phase between the two states when the control qubit is $|1\rangle$
and those when the control qubit is $|0\rangle$ compared with what is expected
in Eq.~\eqref{m1}}.

Show that this phase can be corrected by adding another $R(\theta/2)$
gate on the \textit{control} qubit after the other gates have acted (i.e.~at the right).

\textit{Note:} If we have two or more qubits we often find it convenient to
associate the global phase (or the sign in simple cases) with just \textit{one} of the
qubits. However, you should appreciate that this is simply a manner of speaking;
the global phase is a property of the whole state.

%

\item
Consider a function $f(x)$ which is periodic with period $N$.
We are given a unitary operator $U_y$ that performs
the transformation
\begin{equation}
U_y|f(x)\rangle = |f(x+y) \rangle.
\end{equation}
Show that the state
\begin{equation}
|\tilde{f}(k)\rangle = { 1 \over \sqrt{N}} \sum_{x=0}^{N-1} e^{-2\pi i k x / N}
|f(x)\rangle
\end{equation}
is an eigenvector of $U_y$. Calculate the corresponding eigenvalue.

\item
We have defined the quantum Fourier transform (QFT) in terms of a transformation of
basis states
\begin{equation}
|x\rangle 
\mathrel{\stackrel{\makebox[0pt]{\mbox{\normalfont\tiny QFT}}}{\longrightarrow}}
{1 \over \sqrt{2^n}} \sum_{y=0}^{2^n - 1} \exp[2 \pi i x y / 2^n ]
|y \rangle \, .
\end{equation}
\begin{enumerate}[label=(\roman*)]
\item
If we consider a superposition
\begin{equation}
|\psi\rangle = \sum_{x=0}^{2^n - 1} c_x |x\rangle ,
\end{equation}
show that one can regard the QFT as a transformation of the coefficients
\begin{equation}
|\psi\rangle  
\mathrel{\stackrel{\makebox[0pt]{\mbox{\normalfont\tiny QFT}}}{\longrightarrow}}
\sum_{y=0}^{2^n - 1} \tilde{c}_y |y\rangle 
\end{equation}
where
\begin{equation}
\tilde{c}_y = {1 \over \sqrt{2^n}} \sum_{x=0}^{2^n - 1} \exp[2 \pi i x y / 2^n ]\, c_x .
\end{equation}
\item
Now suppose we shift the basis states by $a$, say, in the sense that we define a new state
\begin{equation}
|\psi'\rangle = \sum_{x=0}^{2^n - 1} c_x |x + a\rangle .
\end{equation}
Show that, after the quantum Fourier transform
\begin{equation}
|\psi'\rangle 
\mathrel{\stackrel{\makebox[0pt]{\mbox{\normalfont\tiny QFT}}}{\longrightarrow}}
\sum_{y=0}^{2^n - 1} \tilde{c}'_y |y\rangle 
\end{equation}
where
\begin{equation}
\tilde{c}'_y = e^{2\pi i a y / 2^n} \tilde{c}_y
\end{equation}
This is called the ``shift-invariance" property of the Fourier transform.
\end{enumerate}

\item
\begin{enumerate}[label=(\roman*)]
\item
Write down the $4 \times 4$ matrix for the Fourier transform for $N = 4$ (2
qubits).
\item
Consider the circuit diagram below, 

\begin{center}
\includegraphics[width=6cm]{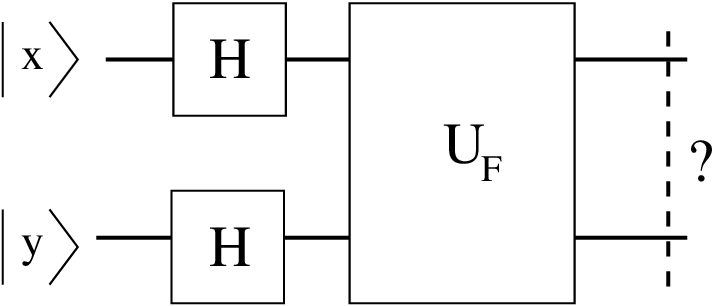}
\end{center}

where $F$ indicates the quantum Fourier transform.

What is the final state if $x = y= 0$? 

\item
What is the final state for the other possible values of $x$ and $y$?

\end{enumerate}

\item

The circuit for the quantum Fourier transform with 4 qubits is shown in the
figure below. (The final swap gates are omitted). 

\begin{center}
\includegraphics[width=12cm]{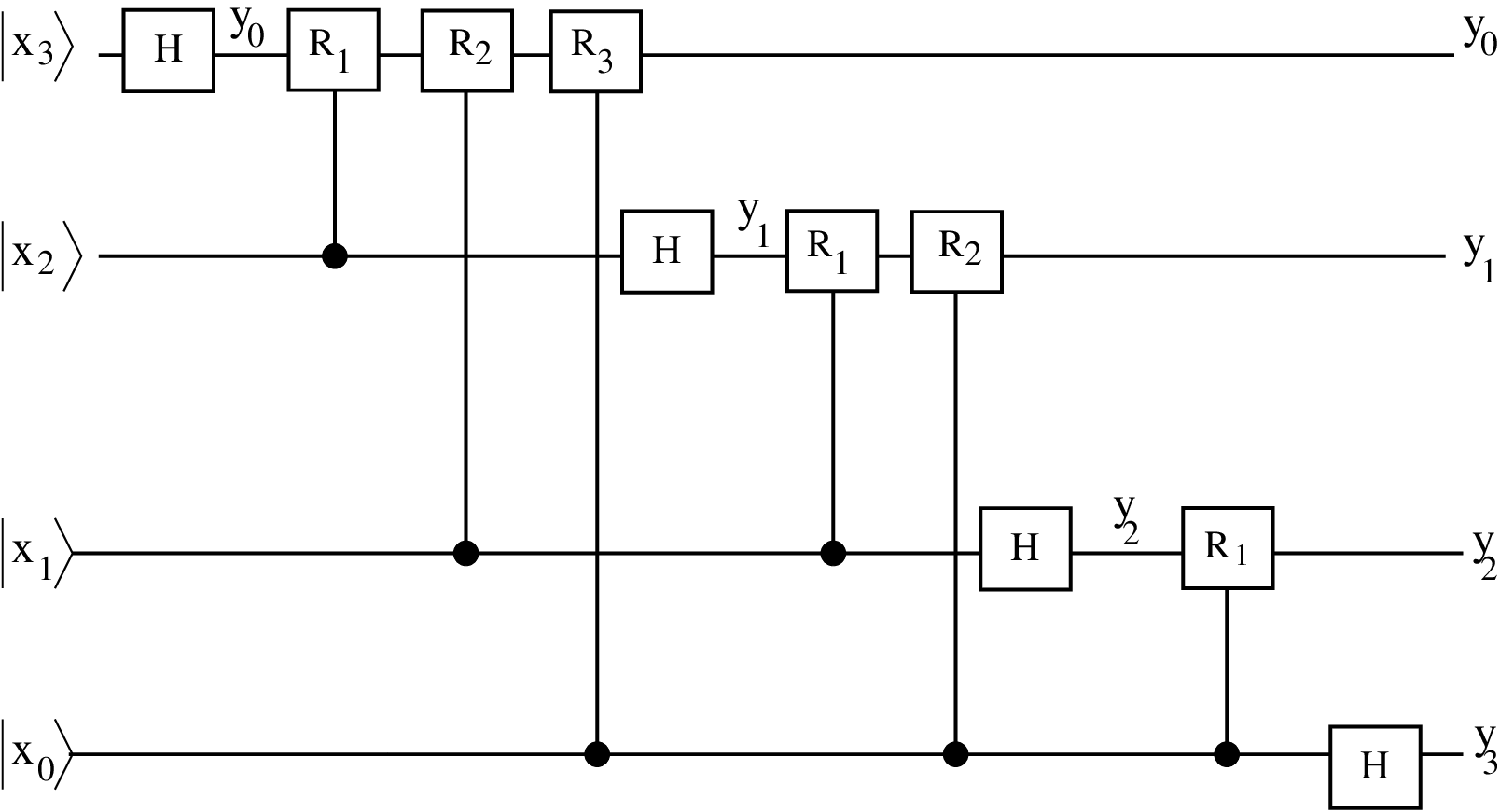}
\end{center}

Write down the following states of the system:
\begin{enumerate}[label=(\roman*)]
\item
Immediately after the $R_3$ gate on the top qubit.
\item
Immediately after the $R_2$ gate on the next to top qubit.
\item
Immediately after the $R_1$ gate on the second from top qubit.
\item
The final state at the right.
\end{enumerate}

\item \textit{Phase Estimation Algorithm}

We showed in Sec.~\ref{sec:phase_est} that the eigenvalues of a unitary matric are a pure phase,
i.e.~are of the form $e^{i\theta}$ for some phase $\theta$. We also showed
that to determine the
phase with a quantum algorithm
one takes out a factor of $2 \pi$ and writes $\theta = 2 \pi \phi$ where
$0\le \phi < 1$. To determine $\phi$ to $n$ bits of precision one then
writes $\phi = \phi'/2^n$ where $\phi'$ is an integer in the range from $0$
to $2^n-1$. The quantum algorithm, discussed in Sec.~\ref{sec:phase_est}, determines $\phi'$.

Here we consider the following two unitary matrices:
\begin{align}
& \mathrm{(a)}\quad
X = 
\begin{pmatrix}
0 & 1 \\
1 & 0 
\end{pmatrix}
, \\
&\mathrm{(b)}\quad
R_1 = 
\begin{pmatrix}
1 & 0 \\
0 & e^{i\pi/2}  
\end{pmatrix}
.
\end{align}

For each matrix determine how many bits $n$ you need to evaluate the
eigenvalues, and draw the quantum circuit for each case. Explain how each
circuit works.

\end{problems}

%% file: app-QFT-FFT4.tex
%

\section{Comparison between FFT and the QFT for $N=4$}
\index{FFT}
In this and the subsequent appendix in this chapter, we describe the
connection between the QFT and FFT. This material is not necessary for the
rest of the course and can be skipped.

We start by considering the simplest case of $2$ qubits, i.e.~$N=4$.
The FFT for $N=4$ is
\begin{subequations}
\label{qft:FT}
\begin{align}
y_0 &=  \smfrac{1}{2}\left( \,x_0 + x_1 + x_2 + x_3 \right) \, , \\
y_1 &= \smfrac{1}{2} \left(\, x_0 + i x_1 + i^2 x_2 + i^3 x_3\right) \, , \label{qft:y1}\\
y_2 &= \smfrac{1}{2} \left(\, x_0 + i^2 x_1 + x_2 + i^2 x_3 \, \right) \, , \\
y_3 &= \smfrac{1}{2} \left(\, x_0 + i^3 x_1 + i^2 x_2 + i x_3 \, \right) \, ,
\end{align}
\end{subequations}
where the $x_j$ are the original data, the $y_j$ are the Fourier transformed data, and
we have used that
\begin{equation}
\exp(2 \pi i / 4) = i \, .
\end{equation}

To evaluate Eqs.~\eqref{qft:FT} efficiently, the FFT proceeds recursively. We
firstly define Fourier transforms of length 2:
\begin{subequations}
\label{qft:step1}
\begin{align}
u_0 &= \smfrac{1}{\sqrt{2}} (x_0 + x_2) \qquad  = \smfrac{1}{\sqrt{2}} (x_0 + i^{2 k} x_2)\ (k = 0) \, ,\\
u_1 &= \smfrac{1}{\sqrt{2}} (x_1 + x_3) \qquad  = \smfrac{1}{\sqrt{2}} (x_1 + i^{2 k} x_3)\ (k = 0) \, ,\\
u_2 &= \smfrac{1}{\sqrt{2}} (x_0 - x_2) \qquad  = \smfrac{1}{\sqrt{2}} (x_0 + i^{2 k} x_2)\ (k = 1) \, ,\\
u_3 &= \smfrac{1}{\sqrt{2}} (x_1 - x_3) \qquad  = \smfrac{1}{\sqrt{2}} (x_1 + i^{2 k} x_3)\ (k = 1) \, ,\\
\end{align}
\end{subequations}

Pairs of quantities in Eqs.~\eqref{qft:step1} are combined to form
the Fourier Transform in Eqs.~\eqref{qft:FT}:
\begin{subequations}
\label{qft:step3}
\begin{align}
y_0 &= \smfrac{1}{\sqrt{2}}(u_0 + u_1)   \qquad\ \ \,= \smfrac{1}{\sqrt{2}}(u_0 + i^{k} u_1) \ (k = 0) \, ,\\
y_1 &= \smfrac{1}{\sqrt{2}}(u_2 + i\, u_3) \qquad = \smfrac{1}{\sqrt{2}}(u_2 + i^{k} u_3) \ (k = 1) \, , \\
y_2 &= \smfrac{1}{\sqrt{2}}(u_0 - u_1)   \qquad\  = \smfrac{1}{\sqrt{2}}(u_0 + i^{k} u_1) \ (k = 2) \, ,\\
y_3 &= \smfrac{1}{\sqrt{2}}(u_2 - i u_3) \qquad\!\! = \smfrac{1}{\sqrt{2}}(u_2 + i^{k} u_3) \ (k = 3) \, ,\\
\end{align}
\end{subequations}

Let's check that this works by evaluating $y_1$. We have
\begin{subequations}
\label{qft:check}
\begin{align}
y_1 &= \smfrac{1}{\sqrt{2}}\left(u_2 + i\, u_3\right)\, ,
\\
&= \smfrac{1}{2}\left(\,x_0 - x_2 + i (x_1 - x_3)\,\right) =
\smfrac{1}{2} \left(\,u_0 + i x_1 + i^2 x_2 + i^3 x_3\right) 
\, ,
\\ 
\end{align}
\end{subequations}
which agrees with Eq.~\eqref{qft:y1}.

It is instructive to write the linear transformations in Eqs.~\eqref{qft:FT},
\eqref{qft:step1}, and \eqref{qft:step3} in matrix form.
Equation \eqref{qft:FT} is written in matrix formulation as 
\begin{equation}
\vec{y} = U \vec{x} \, ,
\end{equation}
where 
\begin{equation}
U = {1 \over 2}
\begin{pmatrix}
1     & 1        & 1        & 1   \\
1     & i        & i^2      & i^3 \\
1     & i^2      & 1        & i^2 \\
1     & i^3      & i^2      & i   \\
\end{pmatrix}
\, . \label{U}
\end{equation}
Equation \eqref{qft:step1} in matrix form is
\begin{equation}
\vec{u} = U_1 \vec{x} \, ,
\end{equation}
where
\begin{equation}
U_1 = {1 \over \sqrt{2}}
\begin{pmatrix}
1     & 0        & 1        & 0     \\
0     & 1        & 0        & 1     \\
1     & 0        & i^2      & 0     \\
0     & 1        & 0        & i^2   \\
\end{pmatrix}
\, . \label{D}
\end{equation}
Equation \eqref{qft:step3} in matrix form is
\begin{equation}
\vec{y} = U_2 \vec{u} \, ,
\end{equation}
where
\begin{equation}
U_2 = {1 \over \sqrt{2}}
\begin{pmatrix}
1     & 1        & 0        & 0    \\
0     & 0        & 1        & i    \\
1     & i^2      & 0        & 0    \\
0     & 0        & 1        & i^3   \\
\end{pmatrix}
\, . \label{F}
\end{equation}
With some matrix manipulations one can verify that
\begin{equation}
U = U_2\, U_1 \, ,
\label{UFD}
\end{equation}
as required. (I used \textit{Mathematica}.)

\index{Comparison between FFT and QFT}


We will now show that there is a close connection between the
FFT and the QFT, and in particular that the transformations $U_1$ and
$U_2$ correspond to different parts of the diagram in Fig.~\ref{had_R1_b}.

The swap gate interchanges states $|01\rangle$ and $|10\rangle$,
so it has the matrix representation
\begin{equation}
S = 
\begin{pmatrix}
1 & 0 & 0 & 0 \\
0 & 0 & 1 & 0 \\
0 & 1 & 0 & 0 \\
0 & 0 & 0 & 1
\end{pmatrix}
\, . 
\label{S}
\end{equation}
The Hadamard gate acting on the lower qubit of Fig.~\ref{had_R1_b} was shown
in Eq.~\eqref{H2}. Including now also the (unchanged) upper qubit, the matrix
representation of the transformation induced by this gate is
\begin{equation}
H_l = {1 \over \sqrt{2}}
\begin{pmatrix}
1 & 1 & 0 & 0 \\
1 & -1 & 0 & 0 \\
0 & 0 & 1 & 1 \\
0 & 0 & 1 & -1
\end{pmatrix}
\, . 
\label{H}
\end{equation}
The Hadamard on the upper qubit has a similar representation, except that the
two qubits have been interchanged, i.e. 
\begin{equation}
H_u =
{1 \over \sqrt{2}}
\begin{pmatrix}
1 & 0 & 1 & 0 \\
0 & 1 & 0 & 1 \\
1 & 0 &-1 & 0 \\
0 & 1 & 0 &-1
\end{pmatrix}
\, . 
\end{equation}
The controlled $R_1$ phase gate gives a multiplicative factor of $i$ if $y_0$
and $x_0$ are both 1, i.e.~state $|3\rangle$. Hence
\begin{equation}
R_1 = 
\begin{pmatrix}
1 & 0 & 0 & 0 \\
0 & 1 & 0 & 0 \\
0 & 0 & 1 & 0 \\
0 & 0 & 0 & i
\end{pmatrix}
\, .  \label{R2}
\end{equation}

The total effect of the quantum circuit in Fig.~\ref{had_R1_b}, reading from
left to right on the circuit, is given by the
matrix product $S H_l R_1 H_u$. Note that one reads from
from right to left in a product
of operators because the operators act on the right. It can be confusing that
the direction of time in the circuit diagram is opposite to that in an
expression of operators. 
Multiplying out these matrices using \textit{Mathematica} one gets
the expected result,
\begin{equation}
S H_l R_1 H_u  = U\, ,
\label{RHS}
\end{equation}
where $U$ is the Fourier transform, shown in Eq.~\eqref{U}. 
Recall that $S$ is the swap, $H_l$ is the Hadamard on the lower qubit,
$R_1$ is the controlled phase gate, and $H_u$ is the
Hadamard on the upper qubit.
Hence the gates in
the quantum circuit in Fig.~\ref{had_R1_b} do indeed affect a Fourier
transform for 2 qubits.

\begin{figure}
\begin{center}
\includegraphics[width=10cm]{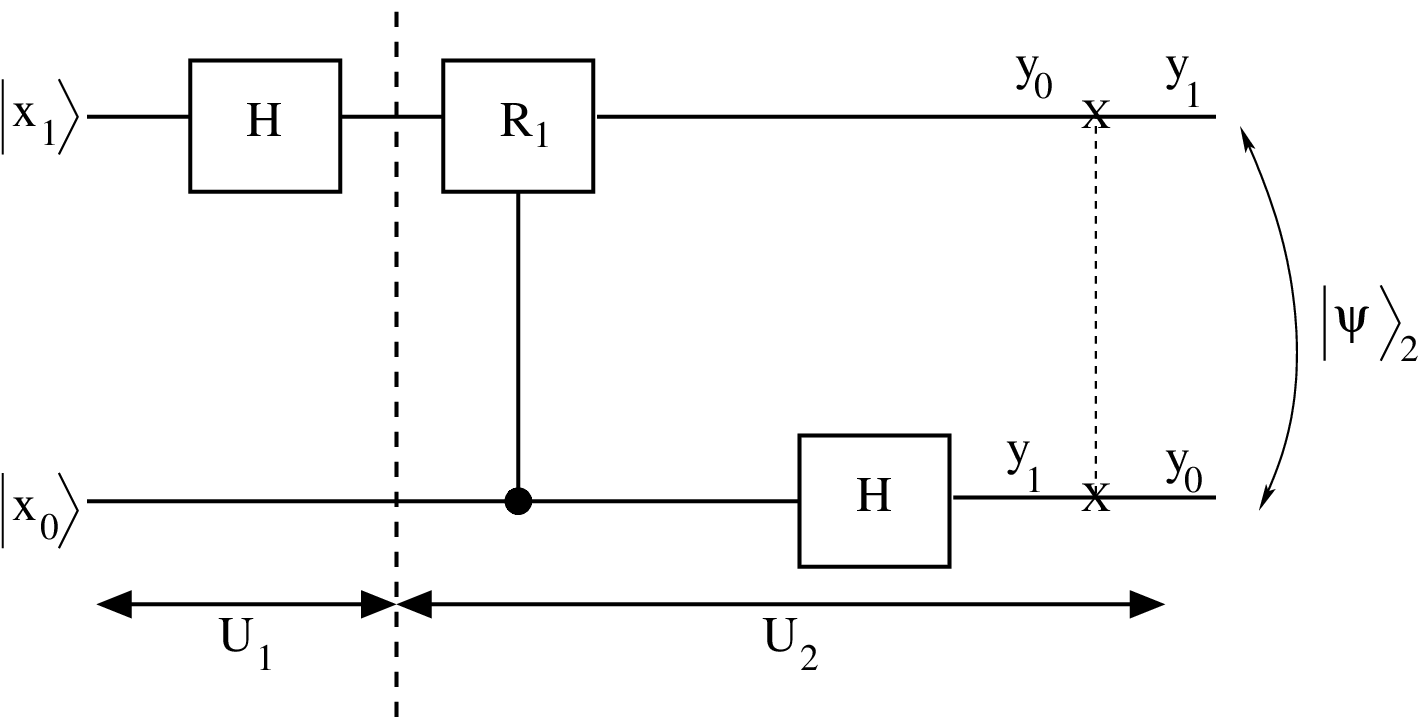}
\caption{The same as Fig.~\ref{had_R1_b} but also showing the correspondence with the
breakup of the FFT into two operations $U = U_2 U_1$, see
Eqs.~\eqref{FD}.
\label{had_R1_c}
}
\end{center}
\end{figure}

In the FFT we decomposed $U$ into a product of two sparse matrices, $U = U_2 U_1$,
see Eq.~\eqref{UFD}.
We can also make a connection between the individual matrices $U_2$ and $U_1$ of
the FFT and the individual matrices $S, H_l, H_u$ and $R_1$ of the QFT. One
finds 
\begin{subequations}
\begin{align}
U_1 &= H_u \, , \\
U_2 & = S  H_l R_1 \, .
\end{align}
\label{FD}
\end{subequations}
The first is obtained by inspection and the second I checked with
\textit{Mathematica}.  Hence the first operation $U_1$ in the FFT for $N=4$
corresponds, in the QFT, to the Hadamard on the upper qubit in
Fig.~\ref{had_R1_b}, while the second operation $U_2$ in the FFT corresponds to
the remaining operations in the QFT: the controlled phase gate on the upper
qubit, the Hadamard on the lower qubit, and the swap. This breakup is shown in
Fig.~\ref{had_R1_c}.

To conclude this section, we have seen that for $2$ qubits there is close connection between
the breakup used in the FFT and that used in the QFT. This should not be a
surprise.  In the FFT we iteratively divide the FT into two FTs of half the
length, while in the QFT we have a binary representation of the states and
treat each bit in turn, so clearly these are related. For $N = 4$, this
connection is expressed in Eqs.~\eqref{FD}.

\section{Comparison of the FFT and QFT for $N = 8$ and
generalization to larger $N$}
\index{Comparison between FFT and QFT}

\begin{figure}[tbh]
\begin{center}
\includegraphics[width=12cm]{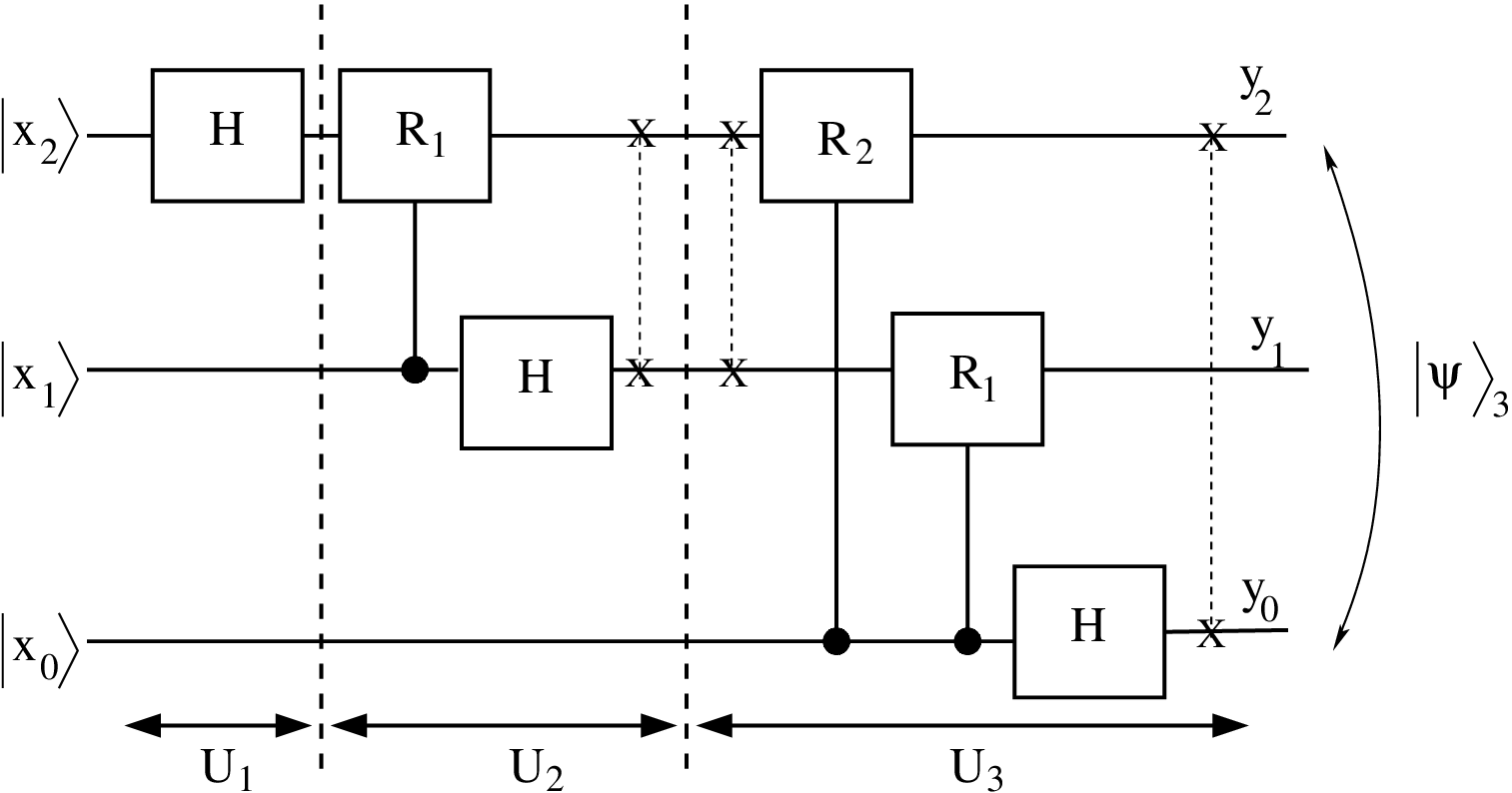}
\caption{
Like Fig.~\ref{had_R1_R2} except that the $R_2$ gate has been moved to the
right of the Hadamard on the middle qubit
(which has no effect) and that a pair of reversals of the order of qubits
1 and 2 have
been added (which also has no effect). The reversal is accomplished by a
swap gate. Note that the final reversal of the order of all three qubits (on the right of the diagram)
is also accomplished by a single swap gate. The correspondence with the
breakup of
the FFT ($U = U_3 U_2 U_1$) is indicated, see Eqs.~\eqref{U123}. To see this
correspondence it is necessary to include the pair of reversals of the order
of qubits 1 and 2. 
\label{had_R1_R2_b}
}
\end{center}
\end{figure}

In this appendix we show how the breakup of the FFT for $3$ qubits, i.e.~$N=8$ is related to
the circuit for the QFT.  Our final result will be Fig.~\ref{had_R1_R2_b}, which is the
analog of Fig.~\ref{had_R1_c} for $N=4$.

As shown in Chapter \ref{ch:fft},
the FFT for $N=8$ can be written as
\begin{equation}
U^{(8)} = U_3^{(8)}\, U_2^{(8)}\, U_1^{(8)}
\label{UFED}
\end{equation}
where
\begin{equation}
U^{(8)} = {1 \over \sqrt{8}}
\begin{pmatrix}
1     & 1        & 1        & 1        & 1        & 1         &  1       &  1        \\
1     & \omega   & \omega^2 & \omega^3 & \omega^4 & \omega^5  & \omega^6 &  \omega^7 \\
1     & \omega^2 & \omega^4 & \omega^6 & 1        & \omega^2  & \omega^4 &  \omega^6 \\
1     & \omega^3 & \omega^6 & \omega   & \omega^4 & \omega^7  & \omega^2 &  \omega^5 \\
1     & \omega^4 & 1        & \omega^4 & 1        & \omega^4  & 1        &  \omega^4 \\
1     & \omega^5 & \omega^2 & \omega^7 & \omega^4 & \omega    & \omega^6 &  \omega^3 \\
1     & \omega^6 & \omega^4 & \omega^2 & 1        & \omega^6  & \omega^4 &  \omega^2 \\
1     & \omega^7 & \omega^6 & \omega^5 & \omega^4 & \omega^3  & \omega^2 &  \omega  
\end{pmatrix}
\, ,
\end{equation}
\begin{equation}
U_1^{(8)} = {1 \over \sqrt{2}}
\begin{pmatrix}
1     & 0        & 0        & 0        & 1        & 0         &  0       &  0        \\
0     & 1        & 0        & 0        & 0        & 1         & 0        &  0        \\
0     & 0        & 1        & 0        & 0        & 0         & 1        &  0        \\
0     & 0        & 0        & 1        & 0        & 0         & 0        &  1        \\
1     & 0        & 0        & 0        & \omega^4 & 0         & 0        &  0        \\
0     & 1        & 0        & 0        & 0        & \omega^4  & 0        &  0        \\
0     & 0        & 1        & 0        & 0        & 0         & \omega^4 &  0        \\
0     & 0        & 0        & 1        & 0        & 0         & 0        &  \omega^4 
\end{pmatrix}
\, ,
\end{equation}
\begin{equation}
U_2^{(8)} = {1 \over \sqrt{2}}
\begin{pmatrix}
1     & 0        & 1        & 0        & 0        & 0         & 0        & 0        \\
0     & 1        & 0        & 1        & 0        & 0         & 0        & 0        \\
0     & 0        & 0        & 0        & 1        & 0         & \omega^2 & 0        \\
0     & 0        & 0        & 0        & 0        & 1         & 0        & \omega^2 \\
1     & 0        & \omega^4 & 0        & 0        & 0         & 0        & 0        \\
0     & 1        & 0        & \omega^4 & 0        & 0         & 0        & 0        \\
0     & 0        & 0        & 0        & 1        & 0         & \omega^6 & 0        \\
0     & 0        & 0        & 0        & 0        & 1         & 0        & \omega^6 
\end{pmatrix}
\, ,
\end{equation}
and 
\begin{equation}
U_3^{(8)} = {1 \over \sqrt{2}}
\begin{pmatrix}
1     & 1        & 0        & 0        & 0        & 0         & 0        & 0        \\
0     & 0        & 1        & \omega   & 0        & 0         & 0        & 0        \\
0     & 0        & 0        & 0        & 1        & \omega^2  & 0        & 0        \\
0     & 0        & 0        & 0        & 0        & 0         & 1        & \omega^3 \\
1     & \omega^4 & 0        & 0        & 0        & 0         & 0        & 0        \\
0     & 0        & 1        & \omega^5 & 0        & 0         & 0        & 0        \\
0     & 0        & 0        & 0        & 1        & \omega^6  & 0        & 0        \\
0     & 0        & 0        & 0        & 0        & 0         & 1        & \omega^7 
\end{pmatrix}
\, .
\end{equation}
One can verify by doing the matrix multiplication (using
\textit{Mathematica} helps) that Eq.~\eqref{UFED} is
satisfied.

One can see from Fig.~\ref{had_R1_R2} that the QFT can be written
as\footnote{Recall that we work from right to left
in operator equations like Eq.~\eqref{U8QFT}
but from left to right in circuit diagrams such as Fig.~\ref{had_R1_R2}.}
\begin{equation}
U^{(8)} = S_{02}^{(8)}\, H_{l}^{(8)}\, R^{(8)}_{1,m} \, H_{m}^{(8)} \,
R^{(8)}_{2,u} \, R^{(8)}_{1,u}\, H_{u}^{(8)} 
\, ,
\label{U8QFT}
\end{equation}
in a fairly obvious notation, where
\begin{equation}
S_{02}^{(8)} = 
\begin{pmatrix}
1 & 0 & 0 & 0 & 0 & 0 & 0 & 0 \\
0 & 0 & 0 & 0 & 1 & 0 & 0 & 0 \\
0 & 0 & 1 & 0 & 0 & 0 & 0 & 0 \\
0 & 0 & 0 & 0 & 0 & 0 & 1 & 0 \\
0 & 1 & 0 & 0 & 0 & 0 & 0 & 0 \\
0 & 0 & 0 & 0 & 0 & 1 & 0 & 0 \\
0 & 0 & 0 & 1 & 0 & 0 & 0 & 0 \\
0 & 0 & 0 & 0 & 0 & 0 & 0 & 1 
\end{pmatrix}
\, , \label{S8}
\end{equation}
\begin{equation}
H^{(8)}_l = {1 \over \sqrt{2}}
\begin{pmatrix}
1 & 1 & 0 & 0 & 0 & 0 & 0 & 0 \\
1 &-1 & 0 & 0 & 0 & 0 & 0 & 0 \\
0 & 0 & 1 & 1 & 0 & 0 & 0 & 0 \\
0 & 0 & 1 &-1 & 0 & 0 & 0 & 0 \\
0 & 0 & 0 & 0 & 1 & 1 & 0 & 0 \\
0 & 0 & 0 & 0 & 1 &-1 & 0 & 0 \\
0 & 0 & 0 & 0 & 0 & 0 & 1 & 1 \\
0 & 0 & 0 & 0 & 0 & 0 & 1 &-1 
\end{pmatrix}
\, ,
\end{equation}
\begin{equation}
R^{(8)}_{1,m} = 
\begin{pmatrix}
1 & 0 & 0 & 0 & 0 & 0 & 0 & 0 \\
0 & 1 & 0 & 0 & 0 & 0 & 0 & 0 \\
0 & 0 & 1 & 0 & 0 & 0 & 0 & 0 \\
0 & 0 & 0 & i & 0 & 0 & 0 & 0 \\
0 & 1 & 0 & 0 & 1 & 0 & 0 & 0 \\
0 & 0 & 0 & 0 & 0 & 1 & 0 & 0 \\
0 & 0 & 0 & 0 & 0 & 0 & 1 & 0 \\
0 & 0 & 0 & 0 & 0 & 0 & 0 & i 
\end{pmatrix}
\, ,
\end{equation}
\begin{equation}
H^{(8)}_m = {1 \over \sqrt{2}}
\begin{pmatrix}
1 & 0 & 1 & 0 & 0 & 0 & 0 & 0 \\
0 & 1 & 0 & 1 & 0 & 0 & 0 & 0 \\
1 & 0 &-1 & 0 & 0 & 0 & 0 & 0 \\
0 & 1 & 0 &-1 & 0 & 0 & 0 & 0 \\
0 & 0 & 0 & 0 & 1 & 0 & 1 & 0 \\
0 & 0 & 0 & 0 & 0 & 1 & 0 & 1 \\
0 & 0 & 0 & 0 & 1 & 0 &-1 & 0 \\
0 & 0 & 0 & 0 & 0 & 1 & 0 &-1 
\end{pmatrix}
\, ,
\end{equation}
\begin{equation}
R^{(8)}_{2,u} = 
\begin{pmatrix}
1 & 0 & 0 & 0 & 0 & 0 & 0 & 0 \\
0 & 1 & 0 & 0 & 0 & 0 & 0 & 0 \\
0 & 0 & 1 & 0 & 0 & 0 & 0 & 0 \\
0 & 0 & 0 & 1 & 0 & 0 & 0 & 0 \\
0 & 1 & 0 & 0 & 1 & 0 & 0 & 0 \\
0 & 0 & 0 & 0 & 0 & \omega & 0 & 0 \\
0 & 0 & 0 & 0 & 0 & 0 & 1 & 0 \\
0 & 0 & 0 & 0 & 0 & 0 & 0 & \omega 
\end{pmatrix}
\, ,
\end{equation}
\begin{equation}
R^{(8)}_{1, u} = 
\begin{pmatrix}
1 & 0 & 0 & 0 & 0 & 0 & 0 & 0 \\
0 & 1 & 0 & 0 & 0 & 0 & 0 & 0 \\
0 & 0 & 1 & 0 & 0 & 0 & 0 & 0 \\
0 & 0 & 0 & 1 & 0 & 0 & 0 & 0 \\
0 & 0 & 0 & 0 & 1 & 0 & 0 & 0 \\
0 & 0 & 0 & 0 & 0 & 1 & 0 & 0 \\
0 & 0 & 0 & 0 & 0 & 0 & i & 0 \\
0 & 0 & 0 & 0 & 0 & 0 & 0 & i 
\end{pmatrix}
\, ,
\end{equation}
\begin{equation}
H^{(8)}_u = {1 \over \sqrt{2}}
\begin{pmatrix}
1 & 0 & 0 & 0 & 1 & 0 & 0 & 0 \\
0 & 1 & 0 & 0 & 0 & 1 & 0 & 0 \\
0 & 0 & 1 & 0 & 0 & 0 & 1 & 0 \\
0 & 0 & 0 & 1 & 0 & 0 & 0 & 1 \\
1 & 0 & 0 & 0 &-1 & 0 & 0 & 0 \\
0 & 1 & 0 & 0 & 0 &-1 & 0 & 0 \\
0 & 0 & 1 & 0 & 1 & 0 &-1 & 0 \\
0 & 0 & 0 & 1 & 0 & 1 & 0 &-1 
\end{pmatrix}
\, ,
\end{equation}
One may verify Eq.~\eqref{U8QFT} using \textit{Mathematica}. Note that
$S_{02}$ swaps qubits 0 and 2, as required to reverse the order of the
qubits.

Can we make a connection between the individual matrices,
$U^{(8)}_1, U^{(8)}_2$, and $U^{(8)}_3$,
in the FFT,
Eq.~\eqref{UFED}, and the individual matrices,
$S_{02}^{(8)}, H_{l}^{(8)}, R^{(8)}_{1,m}, H_{m}^{(8)},
R^{(8)}_{2,u}, R^{(8)}_{1,u},$ and $ H_{u}^{(8)}$, in the QFT,
Eq.~\eqref{U8QFT}?

One immediately sees that $U_1^{(8)} = H_{u}^{(8)}$. However to make a
connection between the other parts of the FFT, $U_2^{(8)} $ and $U_3^{(8)} $,
we introduce the swap operator between qubits 1 and 2:
\begin{equation}
S_{12}^{(8)} = 
\begin{pmatrix}
1 & 0 & 0 & 0 & 0 & 0 & 0 & 0 \\
0 & 1 & 0 & 0 & 0 & 0 & 0 & 0 \\
0 & 0 & 0 & 0 & 1 & 0 & 0 & 0 \\
0 & 0 & 0 & 0 & 0 & 1 & 0 & 0 \\
0 & 0 & 1 & 0 & 0 & 0 & 0 & 0 \\
0 & 0 & 0 & 1 & 0 & 0 & 0 & 0 \\
0 & 0 & 0 & 0 & 0 & 0 & 1 & 0 \\
0 & 0 & 0 & 0 & 0 & 0 & 0 & 1 
\end{pmatrix}
\, , \label{S12}
\end{equation}

We also need to realize that we can move the $R_2$ gate in
Fig.~\ref{had_R1_R2} to the
right as long as it does not cross the Hadamard on the lowest
qubit (since this is the control qubit). Hence we can also write
Eq.~\eqref{U8QFT}
as
\begin{equation}
U^{(8)} = S_{02}^{(8)}\, H_{l}^{(8)}\,
R^{(8)}_{1,m} \,
R^{(8)}_{2,u} \,
H_{m}^{(8)} \,
R^{(8)}_{1,u}\,
H_{u}^{(8)} \, ,
\label{U8QFTb}
\end{equation}
where we have moved $R^{(8)}_{2,u}$ to the \textit{left}.
We then find that
\begin{subequations}
\begin{align}
U_1^{(8)} & = H_{u}^{(8)}\, , \\
U_2^{(8)} & = S_{12}^{(8)}\, H_{m}^{(8)}\, R^{(8)}_{1,u} , \\
U_3^{(8)} & = S^{(8)}_{02}\, H_{l}^{(8)}\, R^{(8)}_{1,m} \, R^{(8)}_{2,u}\,
S_{12}^{(8)}\, , 
\end{align}
\label{U123}
\end{subequations}
which agrees with Eqs.~\eqref{U8QFTb} and \eqref{UFED} since
$\left(S_{12}^{(8)}\right)^2$ is the
identity (swapping twice makes no change). This breakup is shown in
Fig.~\ref{had_R1_R2_b}. Apart from the reversals of qubit order, the
correspondence between the QFT and the FFT is straightforward to see.

\begin{figure}[tbh!]
\begin{center}
\includegraphics[width=14cm]{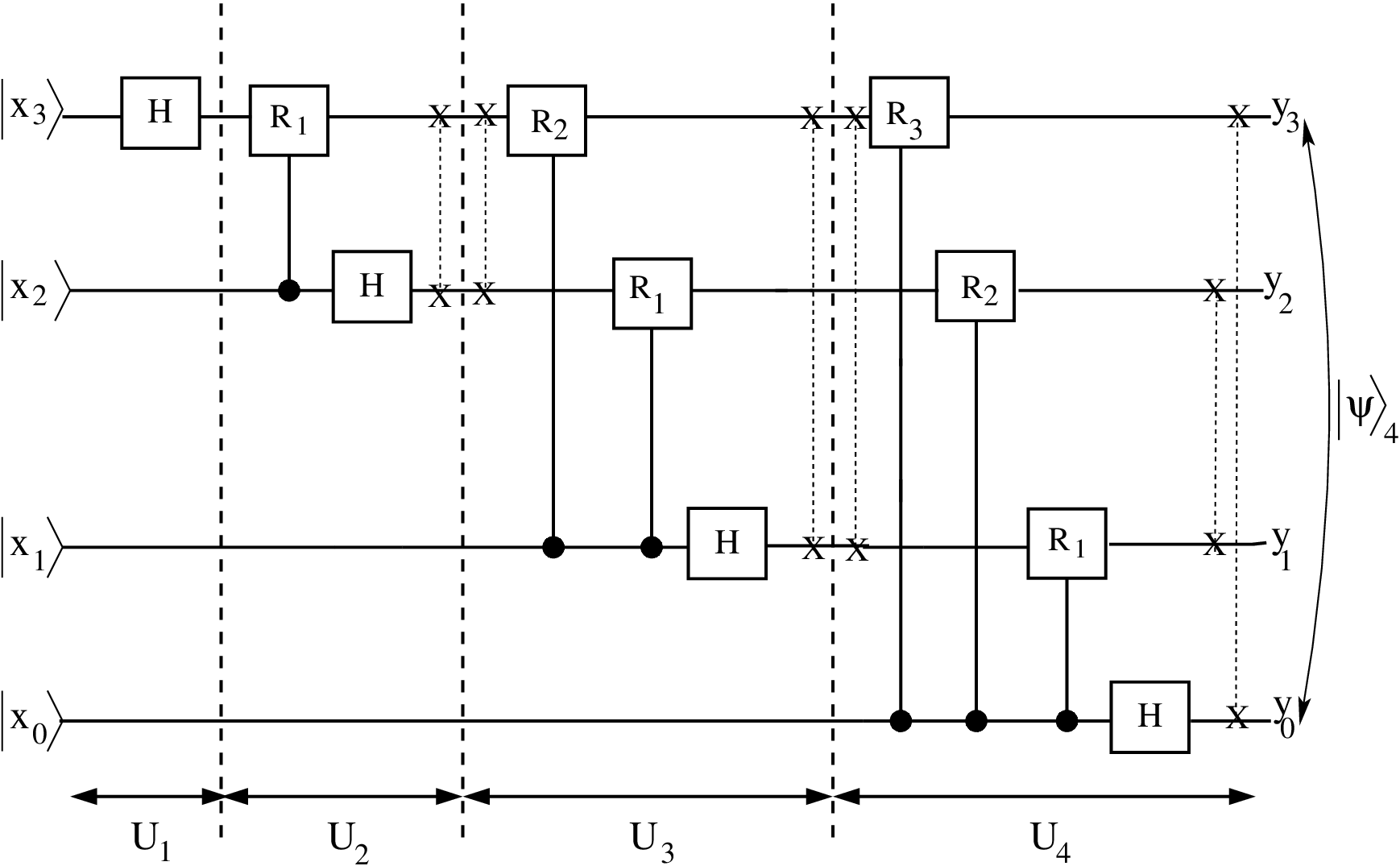}
\caption{
The generalization of Figs.~\ref{had_R1_R2_b} and \ref{had_R1_c} to the case of four
qubits.
The correspondence with the breakup of
the FFT ($U = U_4 U_3 U_2 U_1$) is indicated.
\label{had_4bit}
}
\end{center}
\end{figure}

Following the structure of Fig.~\ref{had_R1_c} for two qubits, and
Fig.~\ref{had_R1_R2_b} for three qubits the generalization to four qubits is
shown in Fig.~\ref{had_4bit}. The correspondence with the FFT is clear, the
only complication being that, in order to show the
correspondence, pairs of reversals of the order of the qubits
(which cancel each other out) have
to be introduced, with one reversal being in one stage of the QFT and the other
reversal in the next stage of the QFT. Reading Fig.~\ref{had_4bit} from left to
right, the first reversal pair reverses qubits 2 and 3 (which needs a single swap gate between qubits
2 and 3), the next reversal reverses
qubits 1, 2 and 3 (which only needs a single swap gate between qubits 1 and 3),
and the last reversal (not a pair because this is the last
one so there is no additional stage to compensate it) reverses all 4 qubits (which needs
two swap gates, one between qubits 0 and 3 and the other between qubits 1 and
2).

To conclude, we see that there is a close parallel between the breakup of the FFT and
circuit of the QFT.  The details are slightly complicated because one needs
reversals of the order of the qubits to make the correspondence precise.  Note that Fig.~1 in
\url{https://arxiv.org/pdf/1005.3730.pdf} is related to the results
presented here. 
\end{subappendices}

%% file: shor8.tex
\index{Shor's factoring algorithm}
\index{Shor, Peter}
\vspace{-0.8cm}
\begin{quotation}
\noindent \textit{When computers we build become quantum, \\
Then spies of all factions will want 'em. \\
Our codes will all fail, \\
They'll hack our email, \\
But crypto that's quantum will daunt 'em.}
\end{quotation}
This is a slightly edited version of a limerick by Peter and Jennifer Shor.
(The original version is printed in the book by Nielsen and
Chuang~\cite{nielsen:00}.)
Continuing in a literary vein, on p.~453 of Nielsen and Chuang is a very
well-crafted (Shakespearean) sonnet by Daniel Gottesman on quantum error correction.
It seems that quantum computing brings out latent literary qualities in
some scientists who work on it, but unfortunately not for me!

\section{Introduction}
Consider an integer $N$ composed of two prime factors $p$ and $q$, i.e.~$N = p\, q$.
In Chapter \ref{ch:period} we
showed how to determine 
the factors of $N$ from the period $r$ of the function 
\begin{equation}
f(x) \equiv a^x \,(\!\! \mod N\,) \, ,
\label{shor:fx}
\end{equation}
where $a$ is some number less than
$N$ and which has no factors in common with $N$. Since $a^0 = 1$, the period is the
smallest value $x=r$
such that
\begin{equation}
a^r \,(\!\! \mod N\,)  = 1.
\label{ar1}
\end{equation}

In 1994 Peter Shor~\cite{shor:94} developed a famous quantum algorithm for period finding
which is much more efficient for factoring large integers than 
any known algorithm running on a classical computer. The ability to
factor a large integer can be used to decode messages sent down a public channel
(such as the internet) which have been encrypted with the RSA
\index{RSA encryption} scheme. The
first four lines of the above limerick refer to this\footnote{The last line of
the limerick refers to quantum key distribution (QKD)\index{quantum key distribution}
which will be discussed
in Chapter \ref{ch:qkd}.}.
The RSA
encryption scheme is described in Chapter \ref{ch:rsa}.

Here we describe in detail Shor's algorithm
to determine the period of the function $f(x)$ in Eq.~\eqref{shor:fx}.  Useful
references are~\cite{mermin:07,nielsen:00,vathsan:16}. There is also a helpful
YouTube video at \url{https://www.youtube.com/watch?v=lvTqbM5Dq4Q},
which is less technical than the present discussion. 

We denote by $n_0$ the number of bits needed
to contain $N$, so $N$ is comparable to $2^{n_0}$.
In cryptography, $N$ may have of
order 600 digits (so $n_0 \sim 2000$ bits). 

\section{Modular Exponentiation}
\label{sec:mod_exp}

In Shor's algorithm the period is found by a Quantum Fourier transform of the
function in Eq.~\eqref{shor:fx} evaluated for $x = 0, 1, 2, \cdots , 2^n - 1$. What
do we take for $n$?  Now the period may be comparable to $N$ and, according
to Mermin~\cite{mermin:07}, in general
we need at least $N$ periods in the data, i.e.~$2^n > N^2$, and so
set $n=2n_0$.  We will see why the
doubling of the number of qubits is necessary in Sec.~\ref{generic}. Hence,
if $n_0 \sim 2000$ we have $n \sim 4000$.

It would seem to be a formidable (nay, impossible) task to calculate
$a^x \,(\!\!\mod N\,)$  for a value of $x$ of order $2^{4000}$.
However, it can be done as follows.
First compute $a, a^2, a^4, \cdots a^{2^n}
\,(\!\!\mod N\,)$ by successively squaring. This only takes $n$
multiplications and so can be done on a
classical or quantum computer. Let the binary expansion of $x$ be
\begin{equation}
x= x_{n-1} x_{n-2} \cdots x_2 x_1 x_0. 
\end{equation}
Then we have
\begin{equation}
a^x = \prod_{j=0}^{n-1}\left(a^{2^j}\right)^{x_j} \, .
\label{shor:ax}
\end{equation}
For example for $n=4$, $x=10$, the binary expansion of $x$ is $1010$ (note the
least significant bit is to the right) so
\begin{equation}
a^{10} = \left(a^8\right)^1
\left(a^4\right)^0 \left(a^2\right)^1 \left(a^1\right)^0 .
\end{equation}
The use of Eq.~\eqref{shor:ax} to compute $a^x$ for 
huge values of $x$ is called ``\textit{modular exponentiation}". 
\index{modular exponentiation}

We can compute $a^x$ using modular exponentiation on a classical or quantum computer as 
follows. We start with the value for
$x \equiv x_{n-1}x_{n-2}\cdots x_2x_1x_0$
in the input register and $1\, \equiv 000\cdots 001$ in the output
register.  We also need an additional work register with $n_0$ qubits,
whose contents we will denote by $w$, with initial value $w=a$.
The following steps compute $a^x\,(\!\!\mod
N\,)$ using Eq.~\eqref{shor:ax}: 
\begin{itemize}
\item (a)\ Multiply the output register by $w$ if $x_0 = 1$.
\item (b)\ Replace $w$ by its square $w \to w^2$.
\item (a') Repeat (a) but for $x_1$.
\item (b') Repeat (b)
\item Continue repeating (a) (with successive bits of $x$) and (b).
\end{itemize}

On a classical computer, the computation has to be performed separately
for each $x$,
whereas on a quantum computer,
as we shall see, Eq.~\eqref{shor:ax} can be computed efficiently for \textit{all}
$x$ between $1$ and $2^n-1$ using quantum parallelism.
\index{quantum parallelism}

\begin{figure}[htb!]
\begin{center}
\includegraphics[width=8cm]{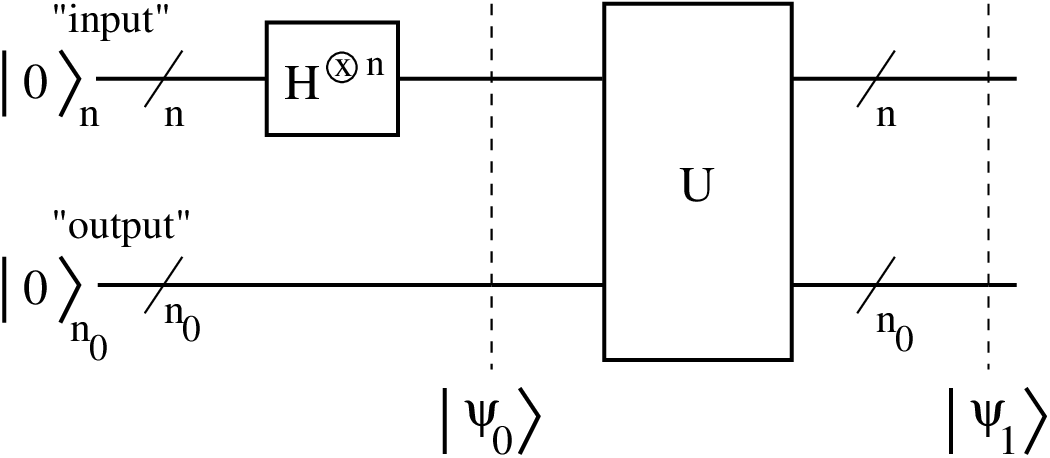}
\caption{
Schematic circuit diagram for performing the modular exponentiation. The
workings of the black box $U$ are described in the text. The state entering $U$
is $|\psi_0\rangle = {1 \over 2^{n/2}}\sum_{x=0}^{2^n - 1} |x\rangle_n \,
|0\rangle_{n_0}$ and the state exiting from $U$ is $|\psi_1\rangle = {1 \over
2^{n/2}}\sum_{x=0}^{2^n - 1} |x\rangle_n \,
|f(x) \rangle_{n_0}$.
\label{mod_exp}
}
\end{center}
\end{figure}

A schematic circuit diagram for doing modular exponentiation on a quantum
computer is shown in
Fig.~\ref{mod_exp}. There are $n$ upper or ``input" qubits, which contain the
values of $x$, and $n_0$ lower or
``output" qubits, which contain the function values $f(x)$. As discussed above,
we usually take $n=2 n_0$.
We will call
the \textit{set} of input qubits the ``input register", and similarly denote the
output qubits as the ``output register".
The notation ``input" and ``output", though often used, can be rather confusing since
both input and output registers are present in the initial state
(left
edge of the circuit diagram in Fig.~\ref{mod_exp}) and in the final state (right edge
of the circuit), so from now on we will refer to these registers as ``upper" and ``lower". 

Both the upper and lower registers are initialized to $|0\rangle$. The qubits
in the upper register
\index{Hadamard matrix (gate)}
are each run through a Hadamard gate.
As shown in Sec.~\ref{sec:qp},
Hadamards acting on $n$ qubits gives
the symmetric sum of all $2^n$ basis states. Hence before entering into the
box $U$ shown in Fig.~\ref{mod_exp}, the state of the system is
\begin{equation}
|\psi_0\rangle = {1 \over 2^{n/2}}\sum_{x=0}^{2^n - 1} |x\rangle_n \,
|0\rangle_{n_0} \, .
\label{psi0}
\end{equation}
On exiting the box $U$, the state of the system has the
values of $f(x)$ in the lower register, i.e.
\begin{equation}
|\psi_1\rangle = {1 \over 2^{n/2}}\sum_{x=0}^{2^n - 1} |x\rangle_n \,
|f(x) \rangle_{n_0} \, .
\label{shor:psi1}
\end{equation}
Note that, in general,
if the lower qubits were initialized to $|y\rangle_{n_0}$, then
after the function acted they would be in state $|y\oplus f(x)\rangle_{n_0}$, but here
$y=0$. 

How many operations does this require?  If we consider (b) we need to do $n$
squares of an $n_0$-bit number. Multiplying two $n_0$ bit numbers in the
simplest way\footnote{As mentioned in Nielsen and Chuang~\cite{nielsen:00},
there are more sophisticated methods of multiplying
$n$-bit numbers which only take 
$O(n \, (\ln n)\,(\ln\ln n))$ operations rather than $O(n^2)$.
This gives a total operation count for modular exponentiation of $O(n^2 \ln
n\, \ln\ln n)$, hardly more than $O(n^2)$. \label{mult}}
takes $O(n_0^2)$ operations.  Since $n = 2n_0$ we see that the
operation count for (b) is $O(n^3)$.  The operation count for (a) is similar,
so the total operation count for modular exponentiation is $O(n^3)$. 

On a classical computer one would have to perform these calculations
sequentially for $x = 1, 2, \cdots, r$, where the period $r$ is of order $N$
where $N$ is of order $2^n/2$, but on a quantum computer they are are done in
parallel using quantum superposition. Hence a quantum computer performs the modular
exponentiation part of Shor's algorithm exponentially faster than a classical computer.

\section{Quantum Fourier Transform (QFT)}
\index{QFT}
\label{qft}

\begin{figure}[htb!]
\begin{center}
\includegraphics[width=12cm]{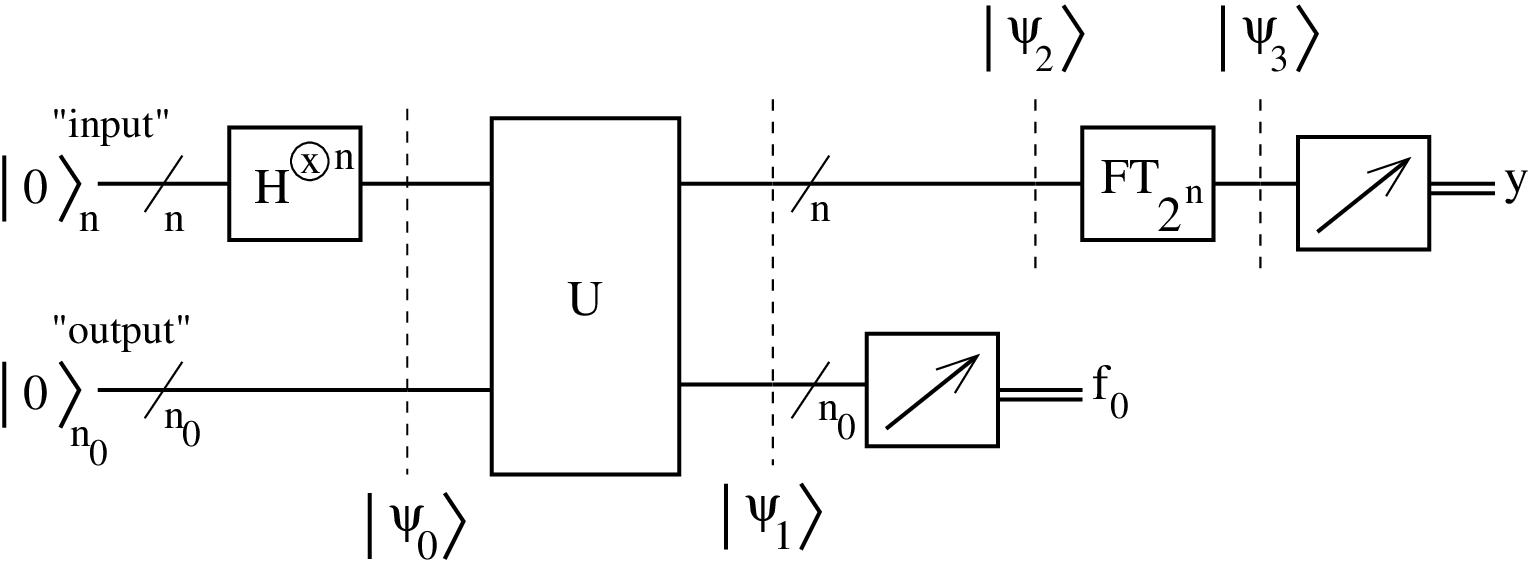}
\caption{
Schematic circuit diagram for Shor's algorithm for period finding on a quantum
computer.
The black box $U$ does the modular exponentiation as described in the 
\index{black box}
text, see also Fig.~\ref{mod_exp}.
The state inputted to $U$ is given by $|\psi_0\rangle$ in Eq.~\eqref{psi0}
and the state outputted from $U$ is given by $|\psi_1\rangle$ in Eq.~\eqref{shor:psi1}.
A measurement (indicated by the box with the arrow) is 
performed on the lower register, giving some value $f_0$.
The double lines indicate that the measurement gives
classical bits which take values 0 \textit{or} 1.  The state of the
upper
register is then given by $|\psi_2\rangle$ in Eq.~\eqref{shor:psi2}, the equally
weighted superposition of all values of $x$ for which $f(x) = f_0$.
The $n$ qubits in the upper register then go through the quantum
Fourier transform the result of which is given by $|\psi_3\rangle$ in Eq.~\eqref{shor:psi3}.
A measurement of the upper qubits then gives a result $y$ which is close to an
integer multiple of $2^n/r$, where $r$ is the period, as discussed in the text.
\label{per_find}
}
\end{center}
\end{figure}

Now that the state of the qubits contains $f(x)$ for all $x$ from $0$ to
$2^n-1$, how do we determine the period $r$?
A schematic of the full circuit for doing this is shown in
Fig.~\ref{per_find}.

\begin{figure}[htb!]
\begin{center}
\includegraphics[width=11cm]{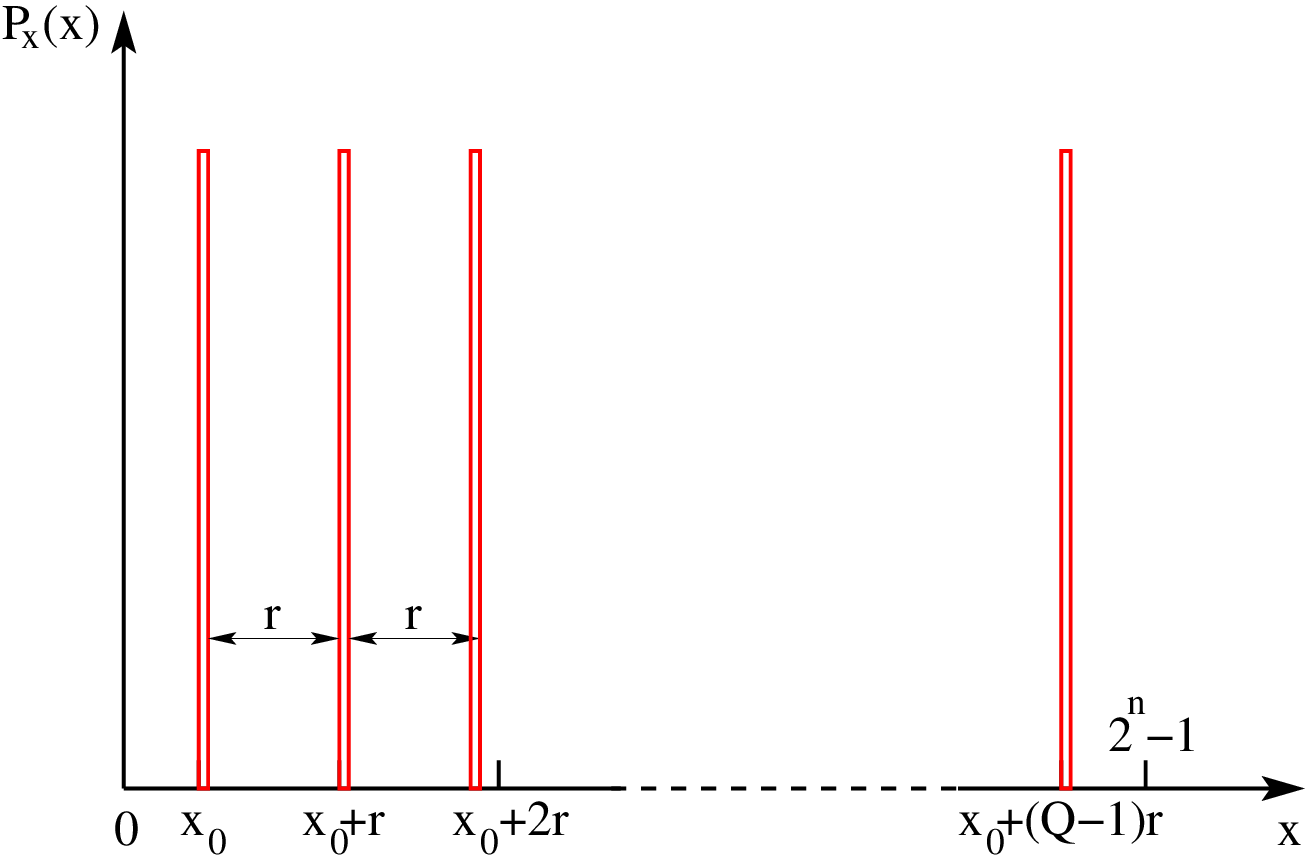}
\caption{
The probability of getting state $x$ in the upper register if a measurement
were performed 
\textit{before} doing the Quantum Fourier Transform. There are $Q$ delta
functions, where $Q = [2^n/r]$, each with weight $1/Q$ separated by $r$, the period. The values of
$x$ where these delta functions appear,
$x_0 + k r, k=0, 1, \cdots,Q-1$, are those values for which $f(x) = f_0$
the result obtained from the measurement of the lower register. A measurement
would get a value for $x_0 + k r$ for some $k$ but since we don't know $x_0$
this is no help in determining the period $r$. Hence measuring the upper
register at this point is not useful.
We need to Fourier transform the state of the upper register
before measuring it, in order to determine the period. 
\label{pxx}
}
\end{center}
\end{figure}

The first (left) part of the algorithm
is the modular exponentiation also
shown in Fig.~\ref{mod_exp}. 
A measurement is then made of the result in the (lower
register from the modular exponentiation routine $U$. This measurement is indicated by the
lower box with an arrow in Fig.~\ref{per_find}.
The measurement will yield some value for $f(x)$, say $f_0$. According to the extended Born hypothesis,
the upper register will then
contain a superposition of those basis states for which $f(x)$ = $f_0$. Since
$f(x)$ is periodic with period $r$, the possible values of $x$ are of the form
$x_0 + k r$, so,
after the measurement on the lower register,
the state of the upper register becomes
\begin{equation}
\label{shor:psi2}
|\psi_2\rangle = {1 \over \sqrt{Q}}\, \sum_{k=0}^{Q-1}
|x_0 + k r\rangle_n \, . 
\end{equation}
Here $0 \le x_0 \le
r-1$, $x_0 + (Q-1) r \le 2^n-1$, $f(x_0+k r) = f_0$, and the number states in the sum is
\begin{equation}
Q = \left[ 2^n \over r\right] \, ,
\label{Q}
\end{equation}
where $[\cdots]$ denotes the integer part. Thus $P_x(x)$, the probability of
of measuring state $|x\rangle$ in the upper register, consists of $Q$ delta functions at positions $x_0 + k r, k = 0, 1,
\cdots, Q-1$, see Fig.~\ref{pxx}.

\begin{figure}[htb!]
\begin{center}
\includegraphics[width=13cm]{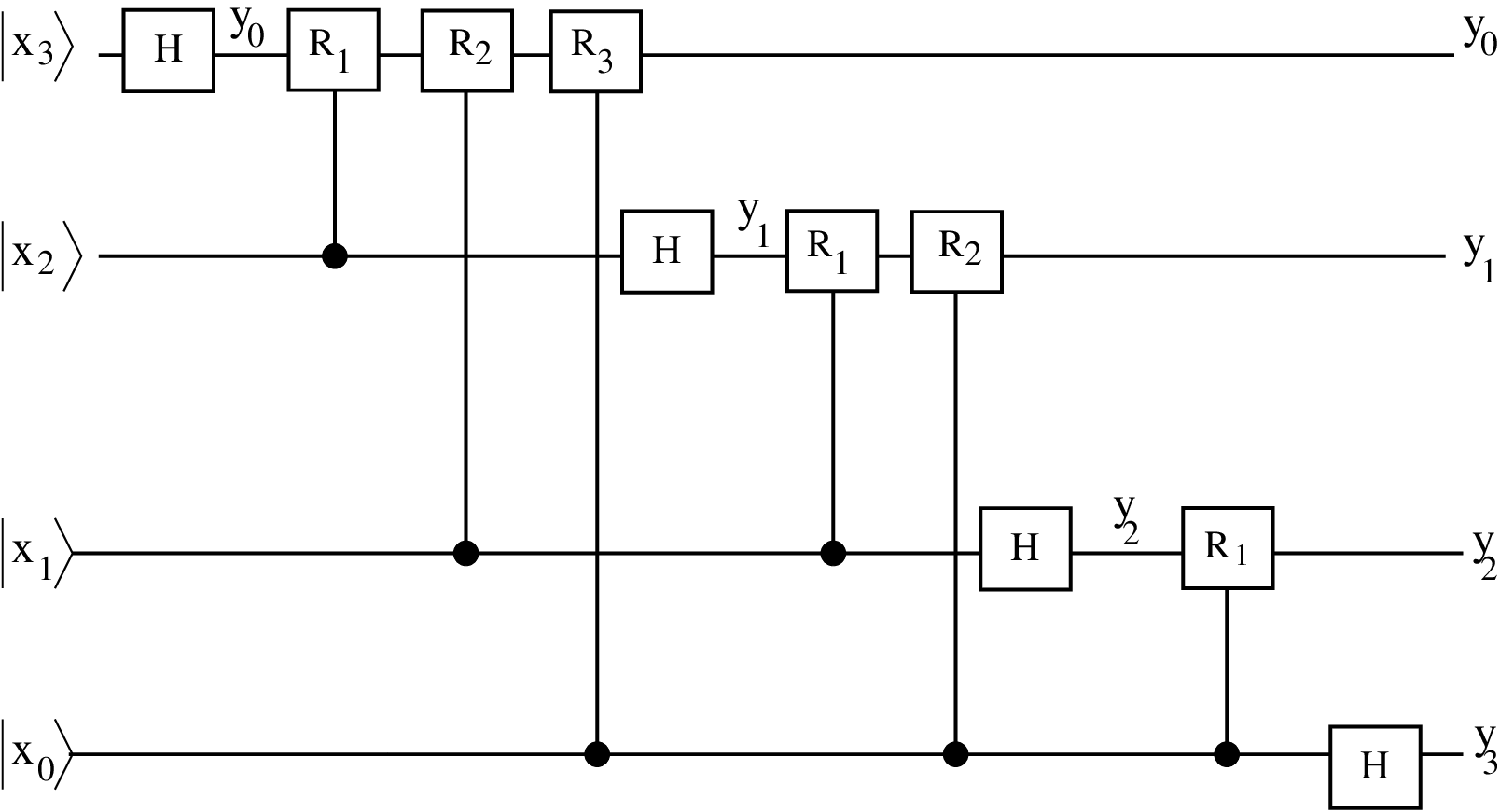}
\caption{
The circuit for the quantum Fourier transform for $n=4$ qubits.  The
controlled phase gates act on the target qubit according to
Eq.~\eqref{Rd} if the control qubit is 1 and otherwise does nothing.  The
final swap gates to reverse the order of the qubits outputted 
on the right are not included here. Note that the controlled phase gate
between qubits $x_i$ and $x_j$ is $R_{|i-j|}$ which makes the structure of
the circuit quite simple to understand.
\label{had_4bit_b}
}
\end{center}
\end{figure}

If we were to measure $|\psi_2\rangle$ we would just get one value of $x_0 + k
r$, which, because of the dependence on the unknown quantity
$x_0$, does not give any information from which we might be able to determine
the period $r$.
In order to extract information on $r$, we have
perform a quantum Fourier transform on the states in Eq.~\eqref{shor:psi2}
before measuring. This gives
\begin{equation}
|\psi_3\rangle 
= \sum_{y=0}^{2^n-1} \left({1 \over \sqrt{2^n Q}}\,  \sum_{k=0}^{Q-1} 
e^{2\pi i (x_0 + k r) y / 2^n} |y\rangle_n \right) \, .
\label{shor:psi3}
\end{equation}
The quantum circuit which performs the quantum Fourier transform is described
in Chapter \ref{ch:qft}.
An example for $n=4$ qubits is shown in
Fig.~\ref{had_4bit_b}. The controlled phase gates act on the
target qubit according to Eq.~\eqref{Rd}
if the control qubit is 1, and otherwise do nothing.
Like the controlled $Z$ gate, the controlled phase gate is
symmetric between the control and target qubits (the phase is changed only if
both qubits are $|1\rangle$), so the control and target qubits can be
exchanged.  We will use this in Appendix \ref{shor:app:1}
when we see how to actually eliminate these
2-qubit gates.

Generalizing the diagram in Fig.~\ref{had_4bit_b} to the case of $n$ qubits we
see that controlled phase gates $R_d$ are required for $d = 1, 2, \cdots,
n-1$.  Hence, in total, we need $n$ Hadamard gates and $1 + 2 + \cdots + n-1 =
n(n-1)/2$ controlled phase gates.  However, as discussed in Sec.~3.9 of
Mermin~\cite{mermin:07}, and in Appendix \ref{app:B}, it is both impossible to
contruct gates giving a phase change which is exponentially small in $n$, and
also not necessary to do this to obtain the QFT with the required precision.
Mermin shows that one only needs controlled phase gates $R_d$ for $d < \log_2
(\text{const.}\, n)$, where the constant Mermin gives is large but independent
of $n$. Thus the number of controlled phase gates needed \textit{in practice}
is of order $n \log_2 n$ which is considerably less than $O(n^2)$ if $n$ is
several thousand. 

In fact we can eliminate the 2-qubit controlled phase gates by measuring each
qubit immediately after the gates of the QFT have acted on it, rather than
after completion of the QFT. This is discussed in Appendix \ref{shor:app:1}.

After the quantum Fourier transform we measure the upper (input) register in
Fig.~\ref{per_find}, obtaining a
value for $y$.  The probability of getting a particular state $y$ is given by
the square of the absolute value of the amplitude of $|y\rangle$ in
Eq.~\eqref{shor:psi3}, i.e.
\begin{equation}
P(y) = {1 \over 2^n Q} \, \left| \sum_{k=0}^{Q-1} e^{2 \pi i k r y/ 2^n}
\right|^2 \, .
\label{Py}
\end{equation}
Note that the dependence on $x_0$, which was troublesome before doing the
Fourier transform, and appears just as a
phase factor after the Fourier transform, Eq.~\eqref{shor:psi3}, now drops out
completely
when we take the square of the absolute value to get the probabilities in
Eq.~\eqref{Py}. 

\begin{figure}[htb!]
\begin{center}
\includegraphics[width=11cm]{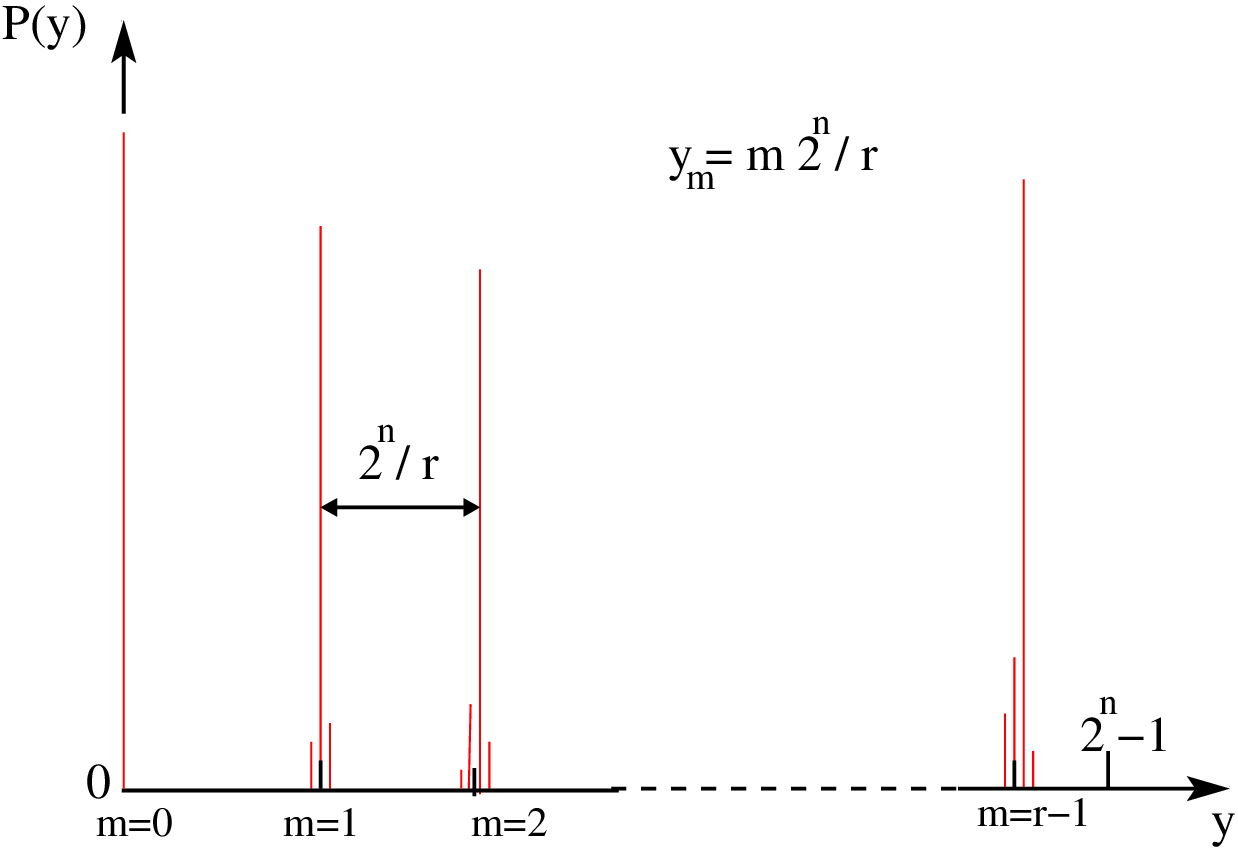}
\caption{
A sketch of the probability of getting state $y$ in the input register \textit{after}
the Quantum Fourier Transform. There are $r$ peaks
at $y_m =  m\, 2^n / r$ for $m=0,1,2,\cdots,r-1$. Note that $[2^n/r] = Q$ so
the \textit{separation} between the peaks in $P(y)$ is no more than 1 away
from $Q$,
the \textit{number} of peaks in the distribution $P_x(x)$ for
the state \textit{before} the quantum
Fourier transform, see Fig.~\ref{pxx}. 
Precise values of $P(y)$ will be calculated in Sec.~\ref{generic} for a
particular case and the resulting values of $P(y)$ will
be shown in Fig.~\ref{fig1}. 
\label{py}
}
\end{center}
\end{figure}

If $y$ could take real values, the exponentials would add up precisely in phase
(and so there would be a peak in the probability for $y$), when $y r / 2^n$ is
an integer, i.e.~for $y=y_m$ where
\begin{equation}
y_m = m\, {2^n \over r},
\label{ym}
\end{equation}
in which $m$ is an integer. Note that there are $r$ values of $m$, from 0 to
$r-1$ since $y$ runs over a range of $2^n$ values.
We emphasize that $y_m$ is not an integer in general,
but the measured values of $y$ \textit{are} are integers, so there will be
peaks in $P(y)$ at integer values \textit{close to} the $y_m$ in
Eq.~\eqref{ym}, see the sketch in Fig.~\ref{py}. Precise values of $P(y)$ for
a particular case will be calculated in Sec.~\ref{generic}.  
Hence there is a high probability that we will obtain an
integer $y$ close to an integral multiple of $2^n / r$.

To summarize this part, $P(y)$ has $r$ peaks separated by $2^n / r$. We recall
that $r$ is the period, which is what we want to compute. 

\section{A special case: 
the period $\mathbf{r}$ is a power of 2.}
\label{sec:power2}

In some special cases the period $r$ will be a power of 2, in which case an integer
number of periods fits \textit{exactly} into the range of $x$-values $(2^n)$. An example
discussed by Mermin~\cite{mermin:07} is if both $p$ and $q$ are both primes of
the form $2^\ell + 1$ (e.g.~the commonly studied case of $N=p q = 15$).
In this situation we will not need
$n$ to be as big as $2 n_0$ (where $n_0$ is the number of bits needed to
contain $N$). Rather, we will see that we just need $2^n$ to be big enough to
contain some integer number\footnote{Even one period is sufficient,
i.e.~$2^n=r$.}
of periods for us to exactly determine an integer
multiple of $2^n/r$.  Since the period might be as large as $N$, when $r$ is a power of 2 we need
\begin{equation}
\begin{split}
2^n &= \mathrm{const.}\ 2^{n_0} \quad \mathrm{rather\ than}  \\
2^n &= 2^{2 n_0} \quad \mathrm{in\ the\ general\ case.}
\end{split}
\end{equation}
Here we go through this special case because the mathematics is simpler than
the general case which we will study in the next section.

First of all we check for $N = 15$ that 
the period \textit{is} a power of 2 as stated above.
Let's take $a = 7$ which has no factors in
common with $15$:
\begin{subequations}
\begin{align}
x = 1,\quad a^x &= 7 \, ,  \\
x = 2,\quad a^x &= 7\times 7 = 49 \equiv 4 \ (\!\!\!\!\mod 15\,) \, ,  \\
x = 3,\quad a^x &\equiv 7 \times 4=28 \equiv 13 \ (\!\!\!\!\mod 15\,) \, ,  \\
x = 4,\quad a^x &\equiv 7 \times 13 =91 \equiv 1 \ (\!\!\!\!\mod 15\,) \, ,  
\end{align}
\label{fxvals15}
\end{subequations}
so the period is $r=4$, i.e.~a power of 2 as claimed.

Now, we perform the sum in Eq.~\eqref{Py}.  Since $r$ is a power of 2 here, and
$2^n \ge r$, it follows that 
$2^n / r$ is an integer, so $Q$, the number of terms in the sum in
Eq.~\eqref{Py}, is given
\textit{exactly} by
\begin{equation}
Q = {2^n \over r }\, .
\label{Qnr}
\end{equation}
From Eq.~\eqref{Qnr}, we see that Eq.~\eqref{Py} becomes
\begin{equation}
P(y) = {1 \over r} \left| {1 \over Q} \sum_{k=0}^{Q-1} e^{2 \pi i k y / Q}
\right|^2\, . \label{Py3}
\end{equation}

Firstly suppose that $y= m Q$ for integer $m$. It is trivial to see that all
the exponentials in Eq.~\eqref{Py3} are unity so
\begin{equation}
P(y= m Q) = {1\over r}.
\end{equation}
Note that there are $r$ distinct values of $m$, $m=0, 1, 2, \cdots, r-1$ since
$y$ runs over a range of $2^n$ values and $Q = 2^n/r$, see Eq.~\eqref{Qnr}. Hence
the sum of the probabilities for the set of values $y=m Q$ is unity. Since the
\textit{total}
probability must be unity there can be
no probability for other values of $y$, as we will now verify.

\begin{figure}[htb!]
\begin{center}
\includegraphics[width=10cm]{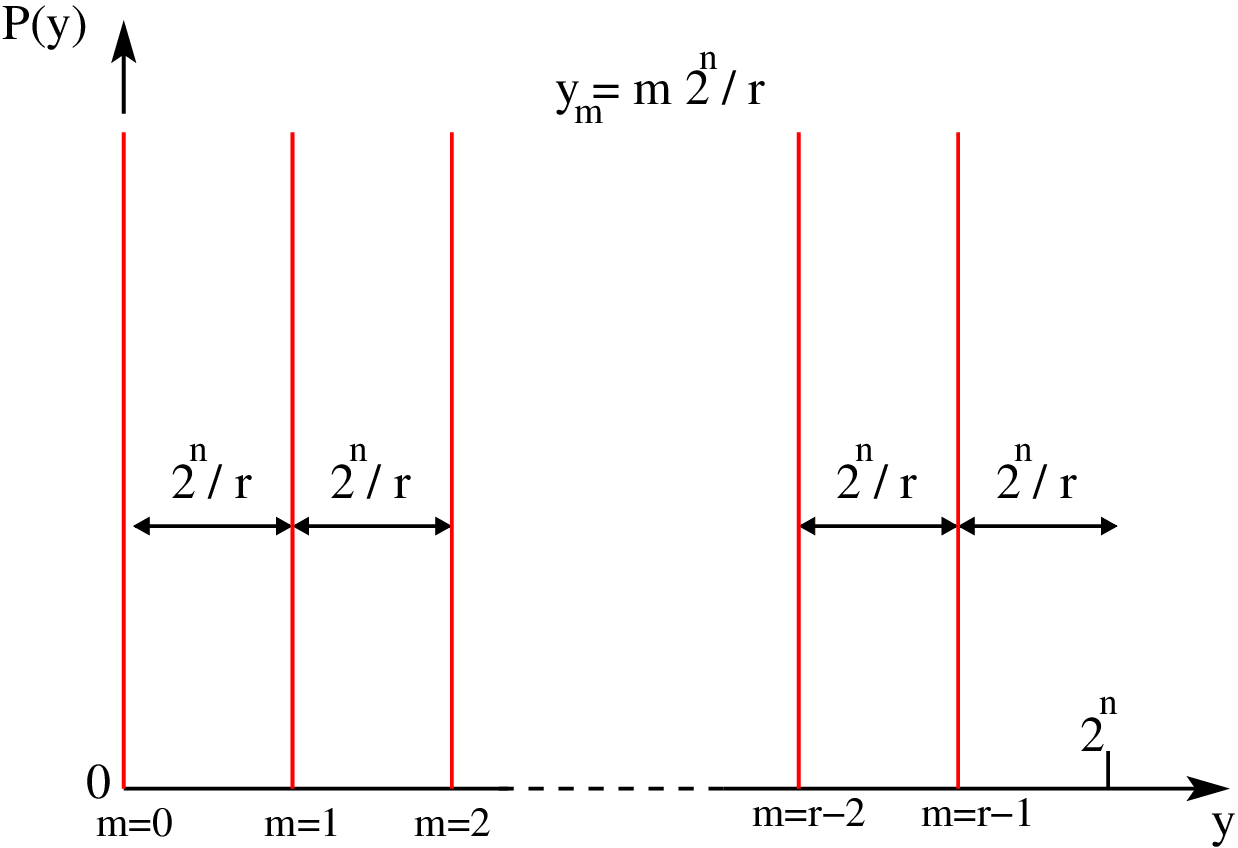}
\caption{
The probability of getting state $y$ in the input register after
the Quantum Fourier Transform for the special case where $r$ is a power of $2$
so there are an exact number of periods in the interval $2^n$. There are 
$r$ delta functions of equal weight at exactly $ y_m = m\,2^n / r$, for
$m = 0, 1, \cdots, Q-1$.
\label{py2}
}
\end{center}
\end{figure}

The sum in Eq.~\eqref{Py3} is a geometric series, which
can be summed to give
\begin{equation}
\sum_{k=0}^{Q-1} e^{2 \pi i k y / Q} = 
{1 - e^{2 \pi i y} \over 1 - e^{2 \pi i y/Q}} \, .
\end{equation}
The numerator is 
zero for all $y$ (recall that $y$ is an integer), but 
for $y \ne m Q$ the denominator is non-zero, so
\begin{equation}
P(y \ne m Q) = 0 \, ,
\end{equation}
as required.
Thus, with probability $1$, the measured value of $y$ is an integer multiple
of $2^n/r$. This is shown in Fig.~\ref{py2}. Superficially, this may look
similar to the situation before the QFT shown in Fig.~\ref{pxx}. The
difference is that the unknown quantity $x_0$ does not appear in Fig.~\ref{py2}.
Rather, the delta functions occur at positions  $y_m$ where $y_m/2^n = m / r$ from which one
can determine $r$.

Notice the reciprocal relation between the period $r$ in the original data in
Fig.~\ref{pxx} and the period in the Fourier transformed data which is the
size of the dataset, $2^n$, \textit{divided} by $r$. To use terminology from
sound waves and frequencies, quite generally, if the original dataset is a
periodic function of ``time" with period $r$,
the Fourier transform will have a peak at the ``fundamental
frequency", $2^n/r$, and in addition can have peaks at ``higher
harmonics" ($m\, 2^n/r$ for $m>1$). It can also have a component
at zero ``frequency" ($y=0$) if the average of the original data is
non-zero. The special nature of the original dataset here (equally weighted, uniformly spaced
delta functions, see Fig.~\ref{pxx}), leads to a Fourier transform which also
comprises
\textit{equally} weighted, uniformly spaced delta functions.

Let us give a simple example so we can see in detail
how to extract the period $r$ from this knowledge. We take our previous
example of $N=15, a=7$, for which we found in Eq.~\eqref{fxvals15} that
the period is $r=4$. 
This means that $7^4 \equiv 1 \mod 15$. We will 
assume that we have $n=5$ qubits, so $2^n = 32$.
The only possible results of a measurement of $y$ are an integer
multiple of $Q = 2^n / r\, (=8)$, so here we have
$y = 0, 8, 16$ or $24$, each with equal probability $1/4$, see Table~\ref{ec:table2}.

\begin{table}[ht]
\begin{center}
\begin{tabular}{|c|c|c|c|}
\hline
$y$ & $m$ & $\dfrac{y}{2^n} \left(= \dfrac{m_0}{r_0}\right)$ & $c =\dfrac{r}{r_0} $ \\
\hline \hline
0  &  0     & 0     & -- \\
8  &  1     & 1/4   & 1  \\
16 &  2     & 1/2   & 2  \\
24 &  3     & 3/4   & 1  \\
\hline
\end{tabular}
\caption{The possible results of a measurement of $y$ for 
the case of $N=15, a=7, 2^n = 32$ for which $r=4$. The value of $y$ gives us
the fraction $y/2^n$, which is also equal to $m/r$
for some $m$. However, any common factor, $c$, is divided out,
so we write $y/2^n$ as $m_0/r_0$ with $m = c m_0, r = c r_0$.
Hence we obtain $r_0$ (and $m_0$), but not $c$. We determine $c$ by
computing the function $a^{c r_0}$ for $c= 1, 2, \cdots$ until we get the value 1. There is
a probability $1/2$ that $c = 1$ works, and it is extremely unlikely that a
large value of $c$ will be needed. The values of $c$ in this example are shown
in the last column.
\label{ec:table2}
}
\end{center}
\end{table}

From the measurement of $y$ we determine 
the  fraction $y/2^n$, which is also equal to $m/r$
for some $m$. However, any common factor $c$
is divided out. We therefore write $y/2^n$ as $m_0/r_0$ with $m = c m_0, r = c r_0$.
The values of $c$ in this example are shown in the last column of Table~\ref{ec:table2}.
In general, to determine $r= c r_0$
we compute the function
$a^{c r_0} \mod N$ for the first few values of $c=1, 2, \cdots$ and
see\label{shor:fn3} for what value of $c$ we obtain 1, the result if $c r_0 = r$,
see Eq.~\eqref{ar1}. The common ratio $c$ is unlikely to be large.  For
example if $m$ is odd, which occurs with probability $1/2$, then $c$ must
equal $1$.
Similarly there is probability $1/4$ that $m$ is even but not a multiple of
$4$ in which case $c$ cannot be greater than $2$. Proceeding in this vein we
see that it is very unlikely that $c$ is large.
In the rare case that the common ratio $c$ \textit{is} large, we would stop after
the first few values of $c$ and restart the quantum computation (the steps
shown in Fig.~\ref{per_find}).

In Table \ref{ec:table2} we see that the value
$y=0$ does not give useful information but, since the number of possible
results is equal to $r$ and each result is equally probable, the probability
of getting $y=0$ is small if the period $r$ is large (the situation if
one needs a quantum computer).

In this section, 
we have seen that in the rare situation that the period is a power of $2$,
the measurement of $y$ gives an integer multiple of $2^n / r$
\textit{with probability one}. Hence $y/2^n = m/r$ with integer $m$
\textit{exactly}.
%
However, in the general case, which we discuss in the next
section,
the measurement of $y$ will give, with a probability which is high
but less than one, a value such that $y/2^n$ is \textit{close to} (but not
equal to) $m / r$.  The continued fraction method in Appendix \ref{app:cf} is then
needed to determine $m/r$. For the continued fraction method to work it turns
out that we need
to have at least $N$ periods in the range of values of $x$,
and so we will take $n = 2 n_0$.

\section{The general case: the period is not a power of 2.}
\label{generic}

We now evaluate the
sum in Eq.~\eqref{Py} for the general case when $r$ is not a power of 2 so
we do not have an exact integer number of periods in the range 
of $x$-values, $2^n$, over which $f(x)$ is calculated. As discussed after
Eq.~\eqref{Py}, $P(y)$ has $r$ peaks, where each peak is in the vicinity of one of the values of $y_m = m 2^n / r$
where $m = 0, 1, 2, \cdots, r-1$.
We set
\begin{align}
y &= y_m + \delta_m \, , \nonumber \\
&= m{2^n \over r} + \delta_m \, .
\label{yy0}
\end{align}
We assume that $\delta_m$ is small, so we are close to the $m$-th peak,
but $2^n, r$ and $m$ are large, since we only need the quantum algorithm when
these numbers are large. (Recall that $y$, the measured value is an integer,
whereas $y_m$ and $\delta_m$ are not.) 

\begin{figure}[htb!]
\begin{center}
\includegraphics[width=9cm]{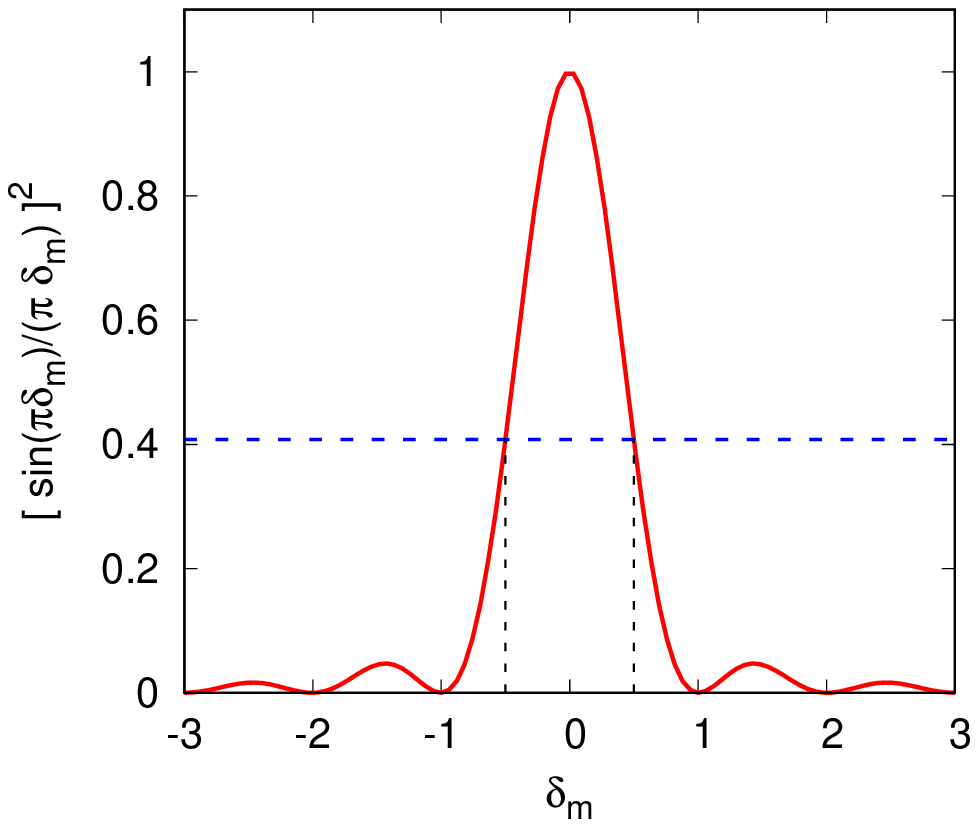}
\caption{
A plot of the function in Eq.~\eqref{Py2}, neglecting the factor of $1/r$
where $r$ is the number of peaks. The area under the curve is 1. The result of
a measurement will be one of a series
uniformly spaced possible values of $\delta_m$ separated by 1. For example,
if $y_m + 0.3$ is an integer, the possible measured values of $\delta_m$
would be $\cdots,-1.7, -0.7, 0.3, 1.3,\cdots$.
One of these values for $\delta_m$ must be within $1/2$ of
0 and the figure shows that the probability for this is greater than
$4/\pi^2$, the dashed horizontal line.  (The dashed vertical lines are at
$\delta_m = \pm 1/2$).
An example of real data is shown in Fig.~\ref{fig2}.
\label{sin2ox2}
}
\end{center}
\end{figure}

Equation~\eqref{Py} involves a geometric series which can be summed as
follows:
\begin{align}
\sum_{k=0}^{Q-1} e^{2 \pi i k r y/ 2^n} &= 
\sum_{k=0}^{Q-1} e^{2 \pi i k m} e^{2 \pi i k r \delta_m/ 2^n} \, , \nonumber \\
&= \sum_{k=0}^{Q-1} e^{2 \pi i k r \delta_m/ 2^n} \, , \nonumber  \\
&= {1 - e^{2 \pi i Q r \delta_m/ 2^n} \over 1 - e^{2 \pi i r \delta_m/ 2^n}}
\, ,\nonumber \\
&= {e^{\pi i Q r \delta_m/ 2^n} \sin\left(\pi Q r \delta_m/ 2^n\right) \over
e^{\pi i r \delta_m/ 2^n} \sin\left(\pi r \delta_m/ 2^n\right)} \, .
\label{sum_gen}
\end{align}
where we used that $\sin x = \smfrac{1}{2i}(e^{ix} - e^{-ix})$.
Inserting Eq.~\eqref{sum_gen} into Eq.~\eqref{Py} the phase factors drop out and we get
\begin{equation}
P(y) = {1 \over 2^n Q}\, {\sin^2\left(\pi Q r \delta_m/ 2^n\right) \over
\sin^2\left(\pi r \delta_m/ 2^n\right)} \, .
\end{equation}
Now $Q$ is within an integer of $2^n / r$ and $Q$ is also large so so we can replace
$Q r / 2^n$ by 1 with negligible error.  Also $r / 2^n$ is very small, since
we take $n$ to be big enough that there are many periods within the range of
$x$ computed, so the sine in the denominator 
can be replaced by its argument.  Hence, to a good approximation,
\begin{equation}
\boxed{
P(y) = {1 \over r} \left({\sin \pi\delta_m \over \pi \delta_m}\right)^2 \, ,}
\label{Py2}
\end{equation}
for $y$ in the vicinity of $y_m$.  Recall that the relation between $\delta_m$
and $y$ is given in Eq.~\eqref{yy0}. The function in Eq.~\eqref{Py2} is
plotted in Fig.~\ref{sin2ox2}.  The area under the curve is 1, and most of the
weight is in the peak centered at 0.

To find the period we would like to get the integer $y$ which is
\textit{closest} to $m\, 2^n / r$ for some integer $m$
i.e.~$|\delta_m| < 1/2$. Writing $\pi\delta_m = x$, this corresponds to $|x| < \pi/2$, and in this region
\begin{equation}
{\sin x \over x} > {2 \over \pi} \, ,
\end{equation}
so, according to Eq.~\eqref{Py2}, the probability of getting the nearest
integer to $y_m$ is greater than 
\begin{equation}
{1 \over r}\, {4 \over \pi^2} \simeq {0.40 \over r} ,
\end{equation}
see Fig.~\ref{sin2ox2}. 
There are $r$ distinct values\footnote{One of these is for $m=0$ which
doesn't give useful information but since we are interested in situations
where $r$ is large, the difference between $r$ and $r-1$ is negligible.} of
$m$ so
the total probability of getting the closest integer to \textit{one} of the
$y_m$ is greater than 40\%.\footnote{In fact, according to Mermin~\cite{mermin:07},
Appendix L, when $N$
is the product of two primes (as we have here) the period is not only less
than $N$ but less than $N/2$. As a result, still using $n = 2 n_0$ qubits in
the input register, the algorithm will provide a result for $r$ not only if
the measured value of $y$ is the closest integer to $m\, 2^n / r$, but also if
it is the second, third or fourth closest. This increases the probability of a
successful run to about $0.9$.}

So, with fairly high probability, we 
have obtained the nearest integer to $m\,2^n / r$
for some integer $m$ (which we don't know).  How can we determine $r$
from this information? We need some post-processing which will be done on a
\textit{classical} computer. 

In deriving Eq.~\eqref{Py2} we just needed that 
the range of $x$ studied contains many periods, i.e.~$2^n \gg r$. Since $r$ can
not be bigger than $N$ we needed $2^n \gg N$. However, to actually
extract $r$ we need a stronger condition, $2^n > N^2$, as we shall now see.

We assume now that we have been successful and 
found a $y$ which is within 1/2 of $2^n m/ r$. Dividing by $2^n$ we
have
\begin{equation}
\left| {y \over 2^n} - {m \over r}\right| < {1 \over 2^{n+1}} \, ,
\label{smalldiff}
\end{equation}
so $y/2^n$, our estimate for $m/r$, is off by no more than $1/(2\cdot 2^n)$. 

The value of
\index{continued fraction}$m/r$ can then be obtained using continued
fractions.
A continued fraction representation of a number $x$ has the form
\index{continued fraction}
\begin{equation}
x = c_0 + \cfrac{1}{c_1 + \cfrac{1}{c_2 + \cfrac{1}{c_3 + \cdots}}} 
\, ,
\end{equation}
where the $c_i$ are integers known as the continued fraction coefficients.
If we stop after a certain number of iterations and ignore the remainder we
have a ``partial sum'', which is an approximation for $x$.
If $x$ is a rational number (ratio of
two integers) the continued fraction will eventually terminate. If $x$ is
irrational (like $\pi$) the continued fraction will go on for ever.
More details about continued fractions are given in
Appendix \ref{app:cf}.

The crucial result of continued fractions which we need 
is theorem A4.16 in Appendix 4 of
Ref.~\cite{nielsen:00}, which states that if 
\begin{equation}
\left|{y \over 2^n}  - {m \over r} \right| < {1 \over 2 r^2} 
\end{equation}
then $m/r$ is one of the partial sums in the continued fraction representation
of $y /2^n $. Here $r < N \sim 2^{n_0} = 2^{n/2}$ so we see from
Eq.~\eqref{smalldiff} that the theorem applies\footnote{It is at this point
that we need the data to contain at least $N$ periods.}. Hence $m/r$ will appear as one of
the partial sums in the continued fraction representation of $y/2^n$. Since $r
< N$ this must be a partial sum with denominator less than $N$. 
Successive partial sums get more and more accurate, so we want the one
with the \textit{largest denominator} less than\footnote{If we have two
approximants for $y/2^n$, $p/q$ and $p'/q'$ say, then
$$\left|{p\over q} - {p' \over q'}\right| = {|p q' - p' q| \over 2 q q'} > {1
\over N^2}$$ (since $q$ and $q'$ are less than $N$) unless the two approximants are equal,
so there is at most one approximant with denominator less than $N$ which
satisfies Eq.~\eqref{smalldiff}. Since successive approximants give better
approximations, the unique partial fraction that we want must be the one with the
\textit{largest} denominator less than $N$.}
$N$.

As we already noted for the special case when $r$ is a power of 2 (Sec.~\ref{sec:power2}),
if $m$ and $r$ have a
common factor, $c$ say, then the continued fraction representation will divide
this out and give $m_0 / r_0$ where $m_0 = m/c, r_0 = r/c$. Thus we actually
get $r_0$ which is a divisor of $r$.
However, we may be lucky and still get $r$ straight away. 
As shown in Appendix J of Mermin~\cite{mermin:07}, the probability that two
large numbers chosen at random have no common factors is greater than $1/2$.
Thus, with probability greater than $1/2$, we get $r$ directly. We can check
if $r_0$ is the period $r$ by computing, on a classical computer, $a^{r_0}
(\!\!\mod N\,)$ and seeing if we get 1. If we do not, we would try simple
multiples, $r=2 r_0, 3 r_0, 4r_0, \cdots$, since it is very unlikely that the
common factor is large. If we \textit{are} very unlucky, and the common factor
\textit{is} large, we could start again from the beginning, get another value
for $m/r$ and hence get another value for $r_0$, and compute $a^{r_0}
(\!\!\mod N\,)$. If this is not 1, then again we try $r=2 r_0, 3 r_0, 4r_0,
\cdots$. There is also a chance that the measured value of $y$ is not close
enough to one of the $y_m$ to get the period from continued fractions.  Again,
if this happens we need to repeat the whole procedure. However, we will not
have to repeat very many times because the probability of success in one run
is quite high.

The probabilistic nature of Shor's algorithm, with the resultant need to run
the algorithm several times (usually not very many), is a quite common feature
of quantum algorithms. 

\section{An example}
\label{ex}
The last section was probably hard going, so we will try to clarify things by
discussing a simple example.
Consider the following, which was also discussed in Chapter \ref{ch:period},
$N = 91, a = 4$.
As shown in Eq.~\eqref{fxvals}, the period is $r=6$.
Since the period is not a power of 2 this is a general example, as discussed in
the previous two sections.

\begin{table}[ht]
\begin{center}
\begin{tabular}{|c|D{.}{.}{2}|r|c|}
\hline
order $(m)$ & \text{peak\ position\ } (y_m=m\, 2^n /r) &
nearest  integer & $P(\text{nearest\ int.})$  \\
\hline \hline
0 &   0     & 0     \qquad\qquad& 0.167 \\
1 & 2730.67 & 2731  \qquad\qquad& 0.114 \\
2 & 5461.33 & 5461  \qquad\qquad& 0.114 \\
3 & 8192    & 8192  \qquad\qquad& 0.167 \\
4 & 10922.67& 10923 \qquad\qquad& 0.114 \\
5 & 13653.33& 13653 \qquad\qquad& 0.114 \\
\hline
\end{tabular}
\caption{The peak positions in the Fourier transform for the example discussed
in this chapter. The output is at integer values of $y$ and the nearest
integers to the peaks are shown along with the probability at those nearest
integer values, computed numerically from Eq.~\eqref{Py}.
Neglecting the zeroth order peak at $y = 0$, which doesn't
give useful information, the sum of the other probabilities at the nearest
integers is $0.623$, so we have a greater than 60\% probability of obtaining
the nearest integer to a non-zero multiple of $2^n/ r$, from which one can
deduce $r$ using continued fractions, as discussed in the text and Appendix
\ref{app:cf}. 
\label{ec:table1}}
\end{center}
\end{table}

One needs $n_0 = 7$ bits to represent $N$ so we take
$n= 2 n_0 = 14$. Hence
\begin{equation}
{2^n \over r} = 2730.67 \, 
\end{equation}
so 
\begin{equation}
Q = 2730\, .
\end{equation}
Hence there are $2730$ (and two thirds) periods in our data. As discussed in
Mermin~\cite{mermin:07} and Sec.~\ref{generic}
we need at least $N\,(=91)$ periods so $2730$ is something of an
overkill.
The peaks in the
Fourier transform, which are at integers next to multiples of $2^n/r$ as
discussed above, are shown in Table~\ref{ec:table1}.

\begin{figure}[htb!]
\begin{center}
\includegraphics[width=0.6\columnwidth]{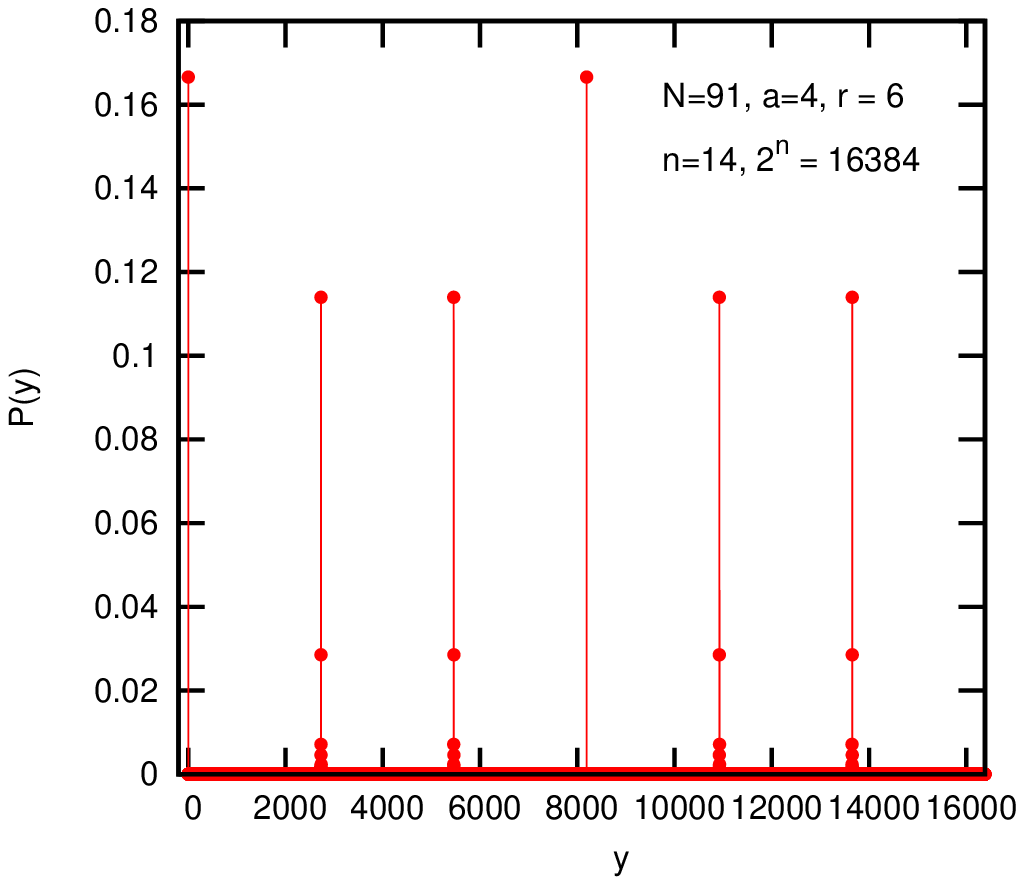}
\caption{
Probabilities for the different components of the Fourier transformed state
for the example studied with $N=91, a=4$ for which the period is $r=6$. These
are computed numerically from Eq.~\eqref{Py}.
There are six sharp peaks near $y_m = m\, 2^n / r$, for $m=0, 1, \cdots,5$.
The one at $y=0\, (m=0)$ doesn't give useful information.
However, the probability of hitting the highest point of one of the other five
peaks, i.e.~the nearest integer to a non-zero multiple of $2^n/r$, is greater
than 60\%, see Table \ref{ec:table1}. If, as is likely, the measurement gives one of these
results, it can then be used to determine the period $r$, as discussed in the
text and
Appendix \ref{app:cf}. 
A blowup of the $m=2$ peak is shown in
Fig.~\ref{fig2}.
\label{fig1}
}
\end{center}
\end{figure}

\begin{figure}[htb!]
\begin{center}
\includegraphics[width=0.6\columnwidth]{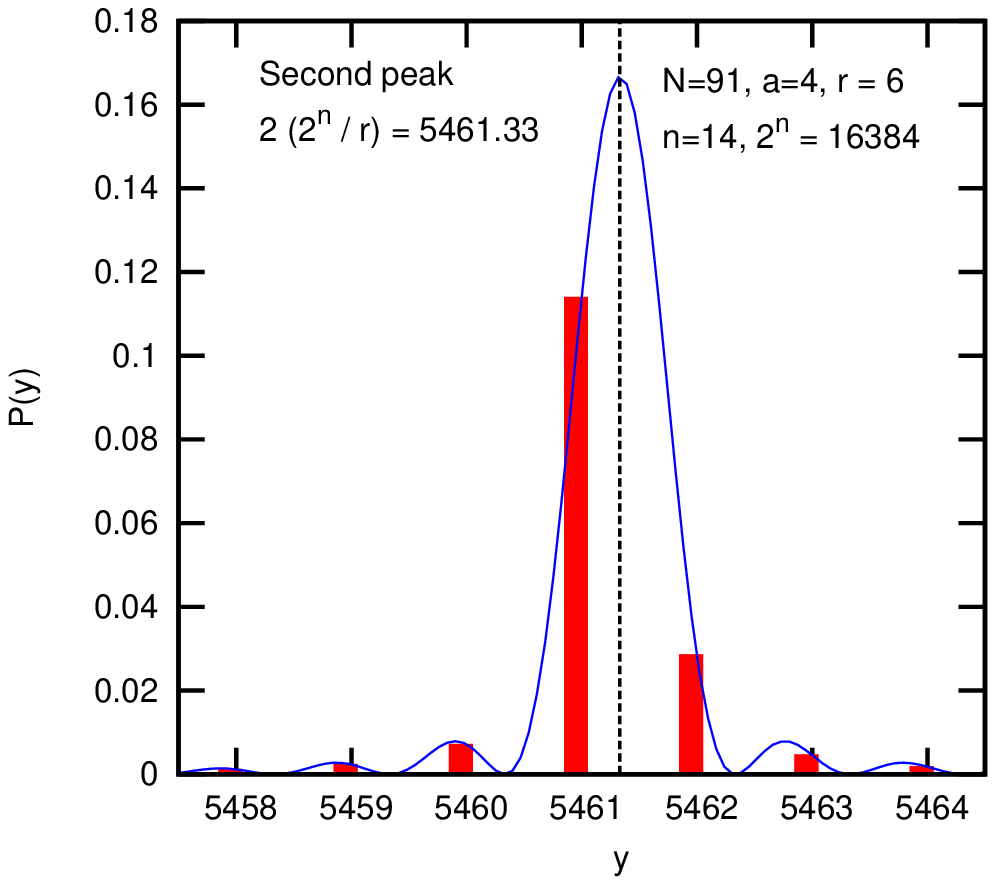}
\caption{
A blowup of the region around the $m=2$ peak in Fig.~\ref{fig1} (see also Table
\ref{ec:table1}). The histogram is obtained from numerical evaluation of
Eq.~\eqref{Py}. The
probability is dominated by the biggest bar, which is at $y=5461$ the
nearest integer to $y_2 = 2 \times (2^n / r) = 5461.33$ (indicated by the vertical
dashed line). According to Eq.~\eqref{sum}, 
the sum of the weights in the histogram is $1/r\ (=1/6\ \mathrm{here})$.
The solid curve is the expression shown in Eq.~\eqref{Py2},
with $y\ (= y_m + \delta_m)$ considered to be a continuous variable. 
\label{fig2}
}
\end{center}
\end{figure}

I have evaluated $P(y)$ numerically from Eq.~\eqref{Py} and the results are
shown in Fig.~\ref{fig1}. There are $r=6$ peaks at values close to $y_m = m\, 2^2 / r$.
There is a trivial peak at exactly $y
= 0\, (m=0)$ but this can not give any useful information about the period $r$. 
For the other $5$ peaks, the peaks are not, in general, centered at
exactly integer values, so the possible observed (integer) values of $y$ are a set of
discrete values around each peak, as shown in the histogram in Fig.~\ref{fig2}
which blows up the region around the $m=2$ peak.

As discussed in Sec.~\ref{generic}, the sum in Eq.~\eqref{Py} can be
evaluated, and is given, to a good approximation, by Eq.~\eqref{Py2} in the
region of the $m$-th peak, where $y$ is given by Eq.~\eqref{yy0}, and $y_m$,
given by Eq.~\eqref{ym}, indicates the peak position.
(Recall that $y$ itself is an integer.)  The function in Eq.~\eqref{Py2} is
plotted for continuous $y$ as the solid curve in Fig.~\ref{fig2}. When
evaluated at integer $y$, it agrees very well with the values numerically
computed from Eq.~\eqref{Py} which are shown as the histogram in
Fig.~\ref{fig2}.

Note that $\delta_m$ in Eq.~\eqref{Py2} is defined in Eq.~\eqref{yy0} and can be
written as 
\begin{equation}
\delta_m = \epsilon + \ell
\label{delta}
\end{equation}
where $\ell$ is an integer and $|\epsilon|< 0.5$. Note too that
\begin{equation}
\sum_{\ell = -\infty}^\infty  \left(\sin (\pi( \epsilon + \ell)) \over 
\pi (\epsilon + \ell) \right)^2
= 1,
\label{sum}
\end{equation}
for arbitrary $\epsilon$ (recent versions of Mathematica know this).
Hence, according to Eqs.~\eqref{Py2}, \eqref{delta},
and \eqref{sum},
the weight
around each of the peaks in Fig.~\ref{fig1} is equal to $1/r\ (= 1/6\ \mathrm{here})$.
There are $r$ peaks so the total probability is $r \times (1/r) = 1$ as
required.
Referring to Fig.~\ref{fig2}, the weight in the largest bar is $0.114$
which is 68\% of $1/6$, the total weight in all the bars for this $(m=2)$ peak.

From Table~\ref{ec:table1} we see that the probability of getting the nearest
integer
to an integral multiple of $2^n / r$ is greater than 60\%.  Let's suppose we
get one of these.  In fact, lets suppose we get the large bar at $y=5461$ in
Fig.~\ref{fig2}. (Recall that Fig.~\ref{fig2} is a blowup of the $m=2$ peak in
Fig.~\ref{fig1}.) Given the measured value, $y=5461$,
we will now see how to determine the period $r$ using continued fractions.

We define
$x = y / 2^n$. This is close to $m / r$, where $r$, the period,
is what we want to determine.
Since $r$ is no greater than $N$, as discussed in Sec.~\ref{generic},
the best guess for $x$ is the partial sum
having the largest denominator less than $N$. As stated above we assume in
this example that the
measurement gave the value
$y=5461$, the highest histogram for the peak in Fig.~\ref{fig2}.
We therefore determine the continued fraction representation for
$x = 5461 / 16384$ (since $n=14$ we have $2^n = 16384$).
Since this is a rational fraction the continued fraction
terminates.

We use the methods of Appendix \ref{app:cf} to determine coefficients as follows.
We have $c_0 = [x] = 0$ (note: $[\cdots]$ means the integer part of what is in the brackets).
We subtract $c_0$ from $x$ and call the inverse of the remainder $x_1$, so
$x_1 = 16384/5461$. $c_1$ is the integer part of $x_1$ so $c_1 = 3$. Subtract
$c_1$ from $x_1$ and call the inverse of the remainder $x_2$. Since $x_1 - c_1
= 1/5416$, we have $x_2 = 5461$. Since this is an integer, the continued
fraction terminates at this point. Hence the coefficients are
\begin{equation}
c_0 = 0,\ c_1 = 3,\ c_2 = 5461,
\end{equation}
and the corresponding partial sums are
\begin{equation}
\begin{split}
c_0 &= 0 , \\
c_0 + \cfrac{1}{c_1} &= {1 \over 3} , \\
c_0 + \cfrac{1}{c_1+ \cfrac{1}{c_2}} &= {5416 \over 16384} .\\
\end{split}
\end{equation}
The last result has a denominator
bigger than  
$ N\ (=91)$ so we neglect it and conclude that\footnote{In this case, where
there are many more than $N$  periods in the intervals $2^n$, one gets the
right answer from the continued fraction
if the measurement gives one of the other nearby $y$ values.  For
example, if we get $y=5460$ (the third closest to the peak), the continued
fraction coefficients are $0, 3,$ and $1365$ which give the partial sums
$0, 1/3, 1365/4096$.
The last value has a
denominator greater than $N$, so we ignore it and take the previous partial
sum, again getting $m/r = 1/3$.}
\index{continued fraction}
\begin{equation}
{m \over r} = {1 \over 3}\, .
\end{equation}
It is possible that $m$ and $r$ have a common factor, i.e.~$m = k, r = 3k$ for
some integer $k$. 
We try some small values for $k$. Starting with $k=1$, so $r=3$, we compute $a^3
\ (\!\!\!\!\mod 91\,)$ and find that it is not 1, see Eq.~\eqref{fx3}.
However, we find that $k=2$ 
does work, since $a^6 \equiv 1
\ (\!\!\!\!\mod 91\,)$, see Eq.~\eqref{fx6}. Hence the period $r$ is equal to
$6$, the desired result. 

\section{Summary}
What is the operation count for Shor's period finding algorithm?
\index{Shor's factoring algorithm!operation count}

To factor an
integer with $n$ bits, the QFT requires, in principle,
$O(n^2)$ operations, as shown in section
\ref{qft}.  Note, however, as discussed there, in Appendix \ref{app:B}, and in
Mermin~\cite{mermin:07}, in practice one only needs of order
$n \log_2 n$ gates to perform the QFT to within the necessary precision.

The computation of the function values using modular
exponentiation takes $O(n^3)$ operations, as shown in section \ref{sec:mod_exp}
(but see footnote \ref{mult} on page \pageref{mult} which states that the
operation count is $O(n^2 \, \log n\, \log\log n)$,
not much more than $O(n^2)$, if one uses a sophisticated
method for multiplying two large numbers).

What about the continued
fraction part, which is, of course, done on a classical computer? Each
division of an $n$-bit number takes of order $n^2$ operations if the division
is done in a simple way. In fact, division can be rewritten as several
multiplications, see
\url{https://en.wikipedia.org/wiki/Division\_algorithm}, so the operation
count can be reduced to that for multiplication, i.e.~$O(n \, \log n\,
\log\log n)$. The depth of
the continued fraction where the denominator is $O(N)$ is
$O(\log N)$,
since the coefficients in the continued fraction multiply to get the
numerator and denominator. This is $O(n)$ since $N$ contains no more than
$n/2$ bits.  Hence the operation count for the
continued fraction post-processing is $O(n^3)$, but recall that this is done
on a classical computer. Again the count is not much more than $O(n^2)$ if one
uses a sophisticated 
method for dividing two large numbers.
Hence, the overall operation count of Shor's algorithm is\footnote{This can
be reduced to $O(n^2\, \log n \, \log\log n)$ using sophisticated methods for
multiplying and dividing large numbers.}
$O(n^3)$.

Shor's
algorithm for factoring integers therefore runs in polynomial time as a function of
$n$, the
number of bits in $N$. For comparison, no polynomial time classical algorithm
for factoring integers is known. The fastest classical algorithm at present,
the general number field sieve (GNFS),
takes a time
$\exp(\mathrm{const.}\, n^{1/3}\log^{2/3} n) $. It is currently not known whether 
there exists
a yet to be determined
polynomial time classical algorithm for factorization.

Even though the power of $n$ in the exponent of the GNFS algorithm is less than
one, it still much slower for large $n$ than Shor's polynomial-time algorithm.
Hence, if the considerable technical difficulties could be overcome, and a
quantum computer with a sufficiently large number of qubits built with the
error rate made sufficiently low, then such a device could decode encrypted
messages currently being sent down the internet which are currently
impossible to decode on a classical computer.

\hrulefill
\section*{Problems}
\input{hw_ch17.tex}
\begin{center}
{\Large \bf Appendices}
\end{center}

\begin{subappendices}

\section{Continued Fractions}
\label{app:cf}

\index{continued fraction}
Continued fractions are a convenient way of finding a simple rational
approximation to a number.

The continued fraction representation of a number $x$ is obtained
as follows. If there is an integer part of $x$ call
this $c_0$. Subtract $c_0$ from $x$ and call the inverse of the remainder
$x_1$, so
\begin{equation}
x = c_0 + \cfrac{1}{x_1} \, .
\end{equation}
Let the integer part of $x_1$ be $c_1$. Subtract $c_1$ from $x_1$ and call the
inverse of the remainder $x_2$ so $x_1 = c_1 + 1/x_2$.
Continuing in the same way for $c_2$ and $x_3$
etc.~we get
\begin{equation}
x = c_0 + \cfrac{1}{c_1 + \cfrac{1}{x_2}} 
= c_0 + \cfrac{1}{c_1 + \cfrac{1}{c_2 + \cfrac{1}{x_3}}} \cdots 
= c_0 + \cfrac{1}{c_1 + \cfrac{1}{c_2 + \cfrac{1}{c_3 + \cdots}}} 
\, .
\end{equation}

To evaluate continued fractions we start at the bottom. For example if we wish
to evaluate
\begin{equation}
x = \cfrac{1}{2 + \cfrac{1}{5 + \cfrac{1}{4}}}
\end{equation}
we determine first that
\begin{equation}
{5 + \cfrac{1}{4}} = {21 \over 4}
\end{equation}
and then that
\begin{equation}
2 + \cfrac{4}{21} = {46 \over 21}
\end{equation}
so 
\begin{equation}
x = {21 \over 46}.
\end{equation}

If we stop after a certain number of iterations and ignore the remainder we
have a ``partial sum'', which is an approximation for $x$. After each
iteration the approximation improves.
If $x$ is a rational number (ratio of
two integers) the continued fraction will eventually terminate. If $x$ is
irrational (like $\pi$) the continued fraction will go on for ever. 
The first few continued fraction
coefficients $c_i (i=0, 1, 2\cdots)$ for $\pi=3.141592654\dots$ are
\begin{equation}
3, 7, 15, 1, 292, 1, \cdots.
\label{cfpi}
\end{equation}
It is a property of continued fractions, which you can verify, that if a
relatively large coefficient appears at some point, stopping the continued
fraction at the previous coefficient gives an accurate approximation to the
number.  For, example, omitting 15 and subsequent coefficients in
Eq.~\eqref{cfpi} gives the well known approximation\footnote{A much more accurate result is
obtained by omitting $292$ and subsequent terms,
which gives a value $355/113 = 3.141592920\ldots$,
which has an error of a bit less than $3 \times 10^{-7}$. This
rational approximation to $\pi$ was apparently first obtained by a Chinese
mathematician Zu Chouygzhi about $1500$ years ago.}
\begin{equation}
\mathbf{3} + \cfrac{1}{\mathbf{7}} = 
{22 \over 7} = 3.14286\ldots\, ,
\end{equation}
which has an error of about $10^{-3}$
(the continued fraction coefficients are in bold).

In the present case we are interested in the continued fraction representation
of $y / 2^n$, which is a rational fraction so the continued fraction
will eventually terminate.  As discussed in the text, the value of $y / 2^n$
is close to $m / r$ where $r$ is no
bigger than $N$ ($N$ can be represented by $n_0$ qubits
with $n_0 = n / 2$). So we are interested in a
continued fraction \textit{approximation} to $y/2^n$ with a denominator no bigger than $N$. 
(Recall that $2^n = \left(2^{n_0}\right)^2$ which is greater than $N^2$.)

Consider the example described in this chapter which has $N=91, a=4$ and
$n=14$ so $2^n = 16384$.  The most probable results for $y$ are those in the
column labeled ``nearest integer'' in Table~\ref{ec:table1}.  Suppose the
measurement of $y$ gives the nearest integer for $m=5$, i.e. $13653$.
The continued fraction representation of $13653 / 16384$ is obtained as
follows:
\begin{equation}
\begin{split}
x &= {13653 \over 16384} , \\
c_0 &= \left[x\right] = 0, \qquad\ \, x_1 = \left(x - c_0 \right)^{-1}\ = {16384 \over 13653}\\
c_1 & = \left[x_1\right] = 1, \qquad x_2 =  \left(x_1 - c_1 \right)^{-1} = {13653\over 2731} \\
c_2 & = \left[x_2\right] = 4, \qquad x_3 =  \left(x_2 - c_2 \right)^{-1} = {2731 \over 2729} \\
c_3 & = \left[x_3\right] = 1, \qquad x_4 =  \left(x_3 - c_3 \right)^{-1} = {2729 \over 2 }\\
c_4 & = \left[x_4\right] = 1364, \ \, x_5 =  \left(x_4 - c_4 \right)^{-1} = 2 \\
c_5 & = \left[x_5\right] = 2,
\end{split}
\end{equation}
and the series terminates since $x_5$ is an integer.
Hence the exact continued fraction coefficients of $13653/16384$ are
\begin{equation}
0, 1, 4, 1, 1364, 2 \, .
\end{equation}
Successive partial sums are $0, 1, 4/5, 5/6, 6824/8189$ and $13653/16384$. We
want the partial sum with the largest denominator less than $N \ (=91)$, 
which is $5/6$.
This tells us, if $m$ and $r$ have no common factors, that $m= 5$ and
$r=6$.
\index{continued fraction}

We check if $r=6$ works by directly calculating $4^6 (\!\!\mod\, 91)$. We
find that it is equal to 1, see Eq.~\eqref{fx6}, so the period is indeed 6.
According to Appendix M in Mermin~\cite{mermin:07}
the probability of two large randomly chosen numbers not having a
common factor is greater than 1/2. If we are unlucky and the assumption of no
common factor does not work, then usually we would only have to try a few
values for the common factor i.e.~$2, 3, 4, \cdots$, before succeeding. If we
are \textit{really} unlucky, and the common factor is very large, we would
give up at some point, start again and get a different value for $y$. In the
related example studied in detail in Sec.~\ref{ex}, where the measurement 
gives the nearest integer to the second peak, the common factor
is 2.

\section{Eliminating the two-qubit gates}
\label{shor:app:1}
It is possible to replace the 2-qubit gates by 1-qubit gates which act or not
depending on the result of a measurement. This is important from a
technological point of view since 1-qubit gates are much easier to
implement than 2-qubit gates. The point is that we measure the final state of the
QFT anyway, and we will see that we can eliminate the 2-qubit gates by
measuring each qubit \textit{immediately after all the gates of the QFT have acted
on it} rather than waiting until the QFT is completed. We now see how to do
this. 

\begin{figure}[htb!]
\begin{center}
\includegraphics[width=13cm]{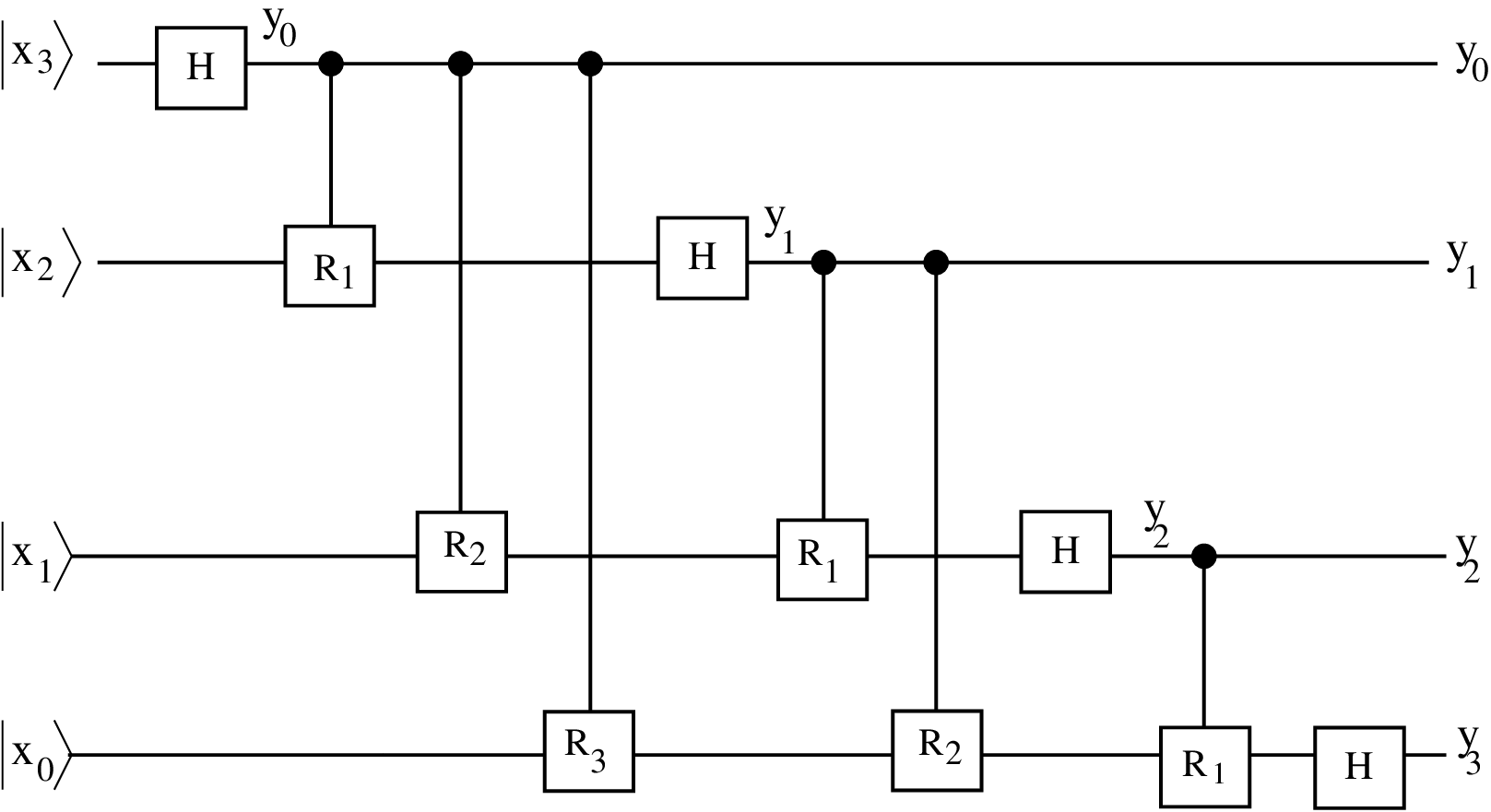}
\caption{
Circuit equivalent to Fig.~\ref{had_4bit_b} but with the target and control
qubits interchanged on the controlled phase gates.
\label{had_4bit_c}
}
\end{center}
\end{figure}

\index{control qubit}
\index{target qubit}
First of all we note that, similar to the control-$Z$ gate,
the target and control qubits in the controlled
phase gates can be interchanged. Hence Fig.~\ref{had_4bit_b} is equivalent to
Fig.~\ref{had_4bit_c}. 

\begin{figure}[htb!]
\begin{center}
\includegraphics[width=12.5cm]{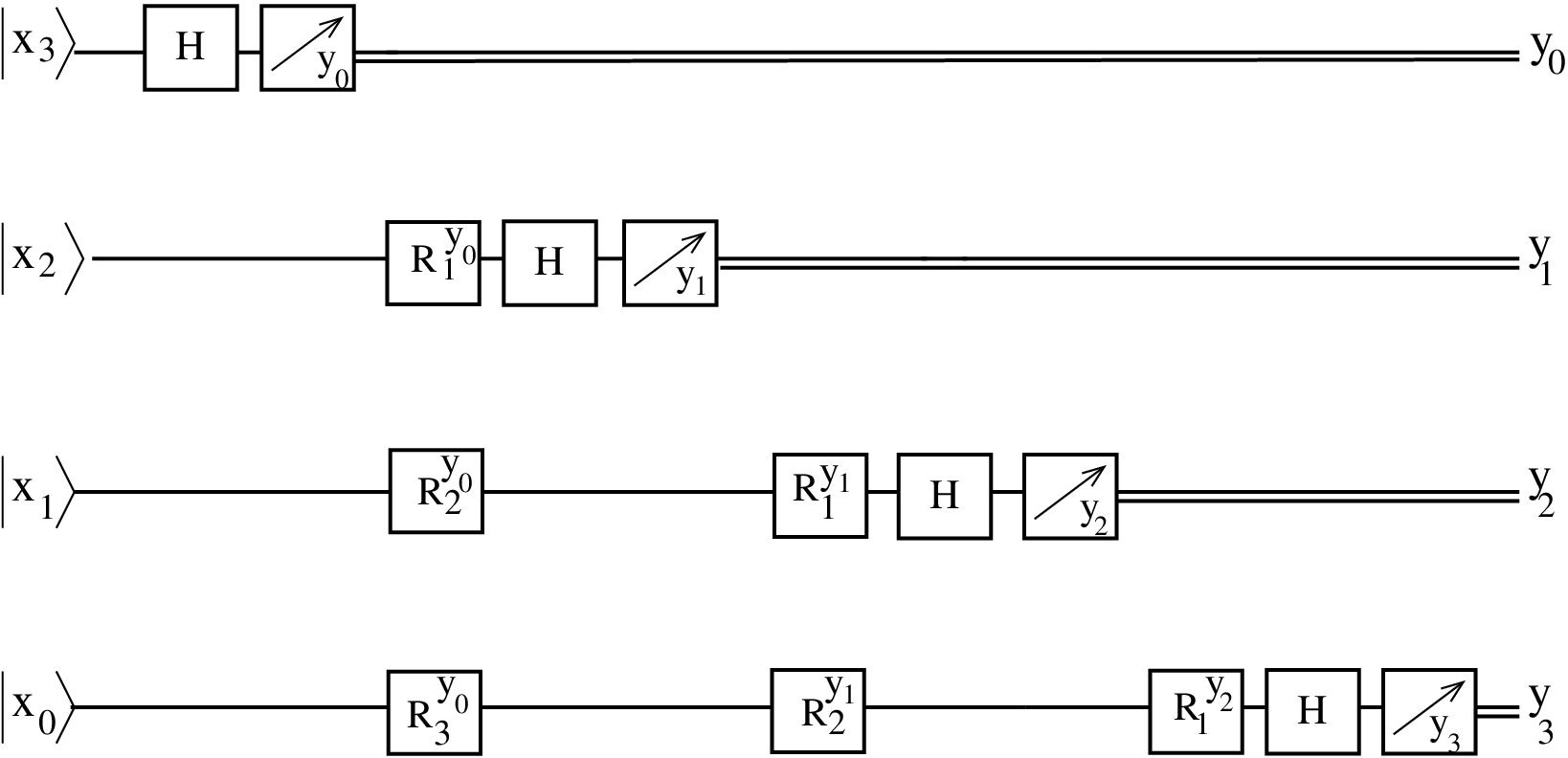}
\caption{
Circuit for the QFT with 4 qubits equivalent to Fig.~\ref{had_4bit_c} but in
which each qubit is measured immediately after the Hadamard gate.
Subsequent phase gates
(on qubits lower in the diagram) are controlled by classical circuits
(not shown) which
use the values of the already measured qubits. Note that $R_1^{y_0}$ means
$R_1$ to the power $y_0$. Since $y_0$ is 0 or 1 this gives $R_1$ if $y_0 = 1$
and $1$ if $y_0=0$. Hence we obtain the required control, but done by a classical
circuit rather than the 2-qubit controlled phase gates in
Fig.~\ref{had_4bit_c}.
\label{qft_nomeas}
}
\end{center}
\end{figure}

In Fig~\ref{had_4bit_c} we see that, for each qubit, once the phase gates and Hadamard 
have acted
the qubit doesn't change, so it could be measured at this point. (Recall that
time flows from left to right in circuit diagrams). Consider the top qubit
$x_3$ which, on output, is $y_0$. We can measure it immediately after the
Hadamard has acted, since it doesn't change after that. If the result is $y_0 = 1$ then the $R_1$ phase gate for $x_2$
is activated, as well as the $R_2$ phase gate for $x_1$ and the $R_3$ phase gate
for $x_0$.
However, if the result is $y_0 = 0$ then those phase gates are not activated.
Since $y_0$ has been measured, this control can be done by a
\textit{classical}
circuit, which is much simpler to implement than a 2-qubit quantum gate.  Similarly we
measure $x_2$, which is $y_1$ on output, immediately after its Hadamard. Hence
the $R_1$ gate on $x_1$ and the $R_2$ gate on $x_0$
can be activated classically if 
$y_1 = 1$. We can proceed in this way for the whole circuit,
measuring the qubit after the Hadamard, and using the result to phase change
other qubits, or not, using classical control. The circuit is shown
in Fig.~\ref{qft_nomeas}.

\section{Unimportance of Small Phase Errors}
\label{app:B}

The action of the controlled-phase gate is given by Eq.~\eqref{Rd} and the QFT
requires, in principle, these gates for $d = 1, 2, \cdots, n-1$. The total
number of controlled phase gates is therefore $1 + 2 + \cdots + n-1 = O(n^2)$.
However, it is clearly impossible to accurately construct a phase
gate for a phase  which is exponentially small in $n$ if $n$ is large.  For
factoring, $n$ would typically be several thousand.  

Fortunately it is not
necessary to include controlled phase gates with such small phase changes.
Mermin~\cite{mermin:07} shows that one can generate the closest integer to a
multiple of $2^n/ r$ within almost the same probability as when one includes
all gates (reduced by at most
1\%) if one neglects controlled phase gates with $d > d^\star =
\log_2 (C n)$, where the
constant $C$ is quite large ($500 \pi$) but independent of $n$.  Hence,
in practice, one
only needs of order $d^\star n$ controlled phase gates ($\sim n \log_2 n$) to
obtain the desired result, rather than $O(n^2)$ which would be needed if one
includes all the gates with $d$ up to $n$. Hence the size of the circuit does
not grow much faster than $n$ which is a huge improvement compared with
$O(n^2)$ if $n$ is several thousand.

\end{subappendices}

%% file: hw_ch17.tex
\begin{problems}

\item \textit{Continued Fractions}\\
Consider the Shor algorithm for $n=10$, so $2^n = 1024$. Recall that there are
peaks in the quantum Fourier transform
in the vicinity of $y_m = m 2^n /
r$, for integer $m$, where $r$ is the period that we wish to determine. Suppose we measure $y = 695$, which, with high
probability will be close to one of the peaks.
\begin{enumerate}[label=(\roman*)]
\item
Go through the continued fraction
calculation to determine the period.  
\item
Compare the resulting value of $y_m = m 2^n / r $ with the measured (integer) value of
$695$.
\end{enumerate}
\textit{Note:}
\begin{itemize}
\item
Recall you want the continued fraction with the largest denominator less than
$N$, the number being factored. Since we take $2^n$ to be comparable to $N^2$,
as discussed in class, you may assume that $N$ is no bigger than $50$. 
\item
Note that the period could, potentially, be a multiple of the denominator,
$r_0$, you
found in the continued fraction. In a real situation, this would be checked by
seeing if $a^{c r_0} \mod N = 1$, for $c=1, 2, \cdots$. Neglect this possibility
here and take the period $r$ to equal the denominator $r_0$.
\item
If you wish you may use a package such as Mathematica, or write
your own computer program, to help with evaluating the continued fraction.
\end{itemize}

\item 
Consider Shor's algorithm for determining the period $r$ of the function
\begin{equation}
f(x) = a^x \!\! \mod N,
\end{equation}
so $a^r \!\! \mod N = 1$. Recall that the register containing
the values of $f(x)$ is measured, and then the register containing the
$x$-values is acted on by a quantum Fourier transform (QFT). The values of $x$
range from $0$ to $2^n-1$.
We showed that if one then
measures the $n$-qubit register containing the $x$-values the probability of
getting the
(integer) value $y$ is given by
\begin{equation}
P(y) = {1 \over 2^n Q}\left| \sum_{k=0}^{Q-1} e^{2 \pi i r k y / 2^n}
\right|^2
\end{equation}
where, in general,
\begin{equation}
Q = \left[{2^n \over r}\right],
\end{equation}
in which $[x]$ means the integer part of $x$.

In this question we consider a simple case in which $r$ is a power of 2, so
here
$2^n/r$ \textit{is} precisely an integer.
\begin{enumerate}[label=(\roman*)]
\item
Show that the probability of getting $y = m Q$ for $m=0,
1, 2 \cdots, r-1$ is given by $P(y=mQ) = 1/r$, and that the probability of
getting
any other $y$-value is zero.
\item
Suppose $n=6$ (so $2^n = 64$) and the period is $r = 8$. What are the possible
values of $y$?
\item
We showed in class that $y/2^n = k / r$ for some integer $k$. For each of the
possible values of
$y$ from the previous part what are the values of $k/r$ (dividing out any
common factors)?\\
\textit{Example}: for $y = 48$ we have
$k/r = 3/4$.
\item
For each of the possible $y$ values there is still a little work to determine
the period $r$. Explain what you have to do for each possible value of $y$.\\
\textit{Example}: in the example in the last part, for $y=48$ we have $k/r =
3/4$.  Is the period equal to the denominator, i.e.~4?
How would one check this? If 4 is not
the period, what would one check next?
\end{enumerate}

\end{problems}

%% file: coherent8.tex
In the next chapter we shall discuss the effects of external noise on qubits.
This will require us to understand
the distinction between
a \textit{coherent} superposition of \textit{amplitudes} in quantum mechanics
and an
\textit{incoherent} (classical) addition of \textit{probabilities}. This is
the topic that we discuss here.

\section{Coherent Linear Superposition: $1$ qubit}
\label{coherent}
\index{coherence}
\index{superposition}

To illustrate coherent superposition,
consider one qubit in the following state
\begin{equation}
|\psi\rangle = \alpha |0\rangle + \beta|1\rangle ,
\label{coh:psi}
\end{equation}
where $|\alpha|^2 + |\beta|^2 = 1$. We denote $|\alpha^2|$ by $p$.
Evidently $|\psi\rangle$ is a linear superposition of
basis states $|0\rangle$ and $|1\rangle$. We say it is a \textit{coherent}
superposition because there is a well defined phase relationship between the
pieces in the superposition, which means that there can be \textit{interference} between these pieces in
subsequent operations.
\index{interference!quantum}

If we measure $|\psi\rangle$ in the computational basis we get
\begin{equation}
\begin{split}
&|0\rangle\ \mathrm{with\ probability\ } |\alpha|^2 = p, \\
&|1\rangle\ \mathrm{with\ probability\ } |\beta|^2 = 1-p .
\end{split}
\label{prob1}
\end{equation}

To show the effects of interference we apply a Hadamard gate, defined in
Eq.~\eqref{had:intro}, before
doing the measurement. The result is
\begin{equation}
|\psi'\rangle = H |\psi\rangle =
{\alpha \over\sqrt{2}} \left( |0\rangle +|1\rangle \right) +
{\beta \over\sqrt{2}} \left( |0\rangle -|1\rangle \right)  = 
\left({\alpha + \beta \over \sqrt{2}}\right) |0\rangle + 
\left({\alpha - \beta \over \sqrt{2}}\right) |1\rangle . 
\end{equation}
If we do a measurement in the computational basis 
\textit{after} applying the Hadamard, the results are
\begin{equation}
\begin{split}
&|0\rangle\ \mathrm{with\ probability\ } \smfrac{1}{2}|\alpha + \beta|^2 =
\smfrac{1}{2}\left(1+ \alpha\beta^* + \alpha^* \beta\right),  \\
&|1\rangle\ \mathrm{with\ probability\ } \smfrac{1}{2}|\alpha - \beta|^2 =
\smfrac{1}{2}\left(1- \alpha\beta^* - \alpha^* \beta\right).
\label{probs_cohere}
\end{split}
\end{equation}
The factor $\alpha\beta^* + \alpha^* \beta$ comes from \textit{interference}
between the two pieces in the linear combination of $|\psi\rangle$ in
Eq.~\eqref{coh:psi}. In particular, if $\alpha = \beta = \smfrac{1}{\sqrt{2}}$, so $p =
\smfrac{1}{2}$, we get
\begin{equation}
\begin{split}
&|0\rangle\ \mathrm{with\ probability\ } 1, \\
&|1\rangle\ \mathrm{with\ probability\ } 0 ,
\end{split}
\end{equation}
showing that there is \textit{zero} probability of getting state $|1\rangle$
in this case
if we measure after performing a Hadamard.
The vanishing probability of getting $|1\rangle$ is due to destructive
interference between the two pieces of the superposition in state
$|\psi\rangle$ in Eq.~\eqref{coh:psi}.

We emphasise that it is incorrect to claim that the state in Eq.~\eqref{coh:psi}
corresponds to the qubit being in state $|0\rangle$ with probability $|\alpha|^2$ and
in state $|1\rangle$ with probability $|\beta|^2$. Although this gives the
correct result if we measure without acting with the Hadamard it gives
incorrect results if we apply the Hadamard before measuring. The reason is
that,
after acting with the Hadamard gate, the system would be in state $H|0\rangle=
\smfrac{1}{2}(|0\rangle + |1\rangle)$ with probability  $|\alpha|^2$ and state
$H|1\rangle=
\smfrac{1}{2}(|0\rangle - |1\rangle)$ with probability  $|\beta|^2$. Adding
the probabilities and using $|\alpha|^2 + |\beta|^2=1$, we find that a
measurement would then give
\begin{equation}
\begin{split}
&|0\rangle\ \mathrm{with\ probability\ {1 \over 2}} 
,  \\
&|1\rangle\ \mathrm{with\ probability\ } {1 \over 2} 
, 
\label{probs_incohere3}
\end{split}
\end{equation}
which does not have the interference terms present in Eq.~\eqref{probs_cohere}.

\section{Incoherent (Classical) Addition of Probabilities}

\subsection{Example with 1 qubit}

An example of a situation with classical probabilities is measuring a single qubit in the
presence of external noise. Suppose the qubit starts out in state
$|\psi\rangle$ in Eq.~\eqref{coh:psi} but is then acted on by noise which
randomises the phases of the two
parts of the superposition.
After the noise has
acted for some time, we can write the state in terms of a global phase $\theta$
and a relative phase $\phi$ as
\begin{equation}
|\psi\rangle = e^{i\theta} \left( \alpha |0\rangle + e^{i \phi} \beta|1\rangle\right) .
\label{incoh:psi}
\end{equation}

Measuring $|\psi\rangle$ in the computational basis gives the same results as
without noise in Eq.~\eqref{prob1}.
However, there is a difference if we apply a
Hadamard before doing the measurement. After a Hadamard this state becomes
\begin{equation}
e^{i\theta}\left({\alpha + e^{i \phi} \beta \over \sqrt{2}}\right) |0\rangle + 
e^{i\theta} \left({\alpha -  e^{i \phi}\beta \over \sqrt{2}}\right) |1\rangle . 
\end{equation}
If we then measure, we will get state
$|0\rangle$ with probability
\begin{equation}
{1 \over 2} e^{i\theta}\, \left(\alpha + e^{i\phi} \beta\right)\, e^{-i\theta}\left(\alpha^\star +
e^{-i\phi}\beta^\star\right) = {1 \over 2} \left(\, |\alpha|^2 + |\beta|^2 + e^{-i\phi}
\alpha\beta^\star + e^{i\phi} \alpha^\star \beta \, \right) .
\end{equation}
The global phase $\theta$ drops out, of course,
but we still need to average over the relative phase $\phi$. After some time the external noise will have
completely randomized the phases so each value of $\phi$ will be equally
probable. Since $\int_0^{2\pi} e^{i\phi} \, d \phi = 0$
the interference terms disappear when we average over the relative phase,
so the probability of getting state
$|0\rangle$ is
$\smfrac{1}{2}\left(|\alpha|^2+ |\beta|^2\right) =
\smfrac{1}{2}$.
In other words measuring the qubit after acting with a Hadamard one finds
\begin{equation}
\begin{split}
&|0\rangle\ \mathrm{with\ probability\ } \smfrac{1}{2} , \ \mathrm{and \
similarly} \\
&|1\rangle\ \mathrm{with\ probability\ } \smfrac{1}{2}.
\label{probs_incohere2}
\end{split}
\end{equation}

The probabilities in Eq.~\eqref{probs_incohere2} differ from those in
Eq.~\eqref{probs_cohere}, which is for the case of a coherent
superposition, by the absence of the factors of
$\alpha\beta^* + \alpha^* \beta$ which came from interference. Interference
does not happen here because the phase relation between the $|0\rangle$ and
$|1\rangle$ parts of the
qubit state has been erased by incoherent noise. 

However Eq.~\eqref{probs_incohere2} is the same as Eq.~\eqref{probs_incohere3} where we assumed that the
qubit is in state $|0\rangle$ with probability $|\alpha|^2$ and in state
$|1\rangle$ with probability $|\beta|^2$ and added those probabilities as in
classical statistics. Thus we shall call the results like
Eq.~\eqref{probs_incohere2}, when noise has erased the
phase difference, an \textit{incoherent} (classical) average over \textit{probabilities},
as opposed to a result like Eq.~\eqref{probs_cohere} for the superposition state in Eq.~\eqref{coh:psi}, which is a
\textit{coherent} sum over \textit{amplitudes}.

\subsection{Example with 2 qubits}

As another example of the incoherent addition of probabilities, consider two qubits in the
following entangled state
\begin{equation}
|\psi_2\rangle = \alpha |0 0 \rangle + \beta |1 1 \rangle ,
\end{equation}
where we again denote $|\alpha|^2$ by $p$.
If $\alpha = \pm \beta = \smfrac{1}{\sqrt{2}}$ this is a Bell state.
\index{Bell state}Let us
write $|\psi_2\rangle$ more explicitly as
\begin{equation}
|\psi_2\rangle = \alpha |0_A\rangle \otimes |0_B\rangle +
\beta |1_A\rangle \otimes |1_B\rangle.
\label{coh:psi2}
\end{equation}
If we focus on qubit $A$, say, then state $|\psi_2\rangle$ looks rather
similar to the $1$-qubit state $|\psi\rangle$ in Eq.~\eqref{coh:psi}, in that
there is a piece where qubit $A$ is $|0\rangle$ with amplitude $\alpha$ and a
piece where qubit $A$ is $|1\rangle$ with amplitude $\beta$. However, for
$|\psi_2\rangle$, unlike for $|\psi\rangle$,
each of these pieces goes with a different state for qubit
$B$ (i.e.~$|\psi_2\rangle$ is entangled). Because of this entanglement, we will \textit{not} get
\index{entanglement}
interference between the pieces of $|\psi_2\rangle$ if we perform operations on
qubit $A$ followed by a measurement of that qubit, as we now show. 

%

If we measure qubit $A$ before doing any operation on it we get 
\begin{equation}
\begin{split}
&|0\rangle\ \mathrm{with\ probability\ } p \ (= |\alpha|^2) , \\
&|1\rangle\ \mathrm{with\ probability\ } 1-p  \ (= |\beta|^2) ,
\end{split}
\label{prob2}
\end{equation}
which is the same as for the other examples.

However,
if we apply a Hadamard the state is given by
\begin{equation}
\begin{split}
|\psi'_2\rangle &= H_A |\psi_2\rangle \\
&= \alpha \left(H_A |0_A\rangle\right) \otimes |0_B\rangle + 
\beta \left(H_A |1_A\rangle\right) \otimes |1_B\rangle \\ 
&= {\alpha \over \sqrt{2}}\left(|0_A0_B\rangle + |1_A0_B\rangle \right) +
{\beta \over \sqrt{2}}\left(|0_A1_B\rangle - |1_A1_B\rangle \right) \\
&= {1 \over \sqrt{2}}
\Bigl[\,
|0_A\rangle \otimes \left( \alpha |0_B\rangle + \beta|1_B\rangle \right)
+ |1_A\rangle \otimes \left( \alpha |0_B\rangle - \beta|1_B\rangle \right)
\, \Bigr]  \\
&= {1 \over \sqrt{2}} \Bigl[\,
|0_A\rangle\otimes|\phi_{0,B}\rangle +
|1_A\rangle\otimes|\phi_{1,B}\rangle
\, \Bigr] ,
\end{split}
\end{equation}
where
\begin{equation}
\begin{split}
|\phi_{0,B}\rangle & = \alpha |0_B\rangle + \beta|1_B\rangle \\ 
|\phi_{1,B}\rangle &= \alpha |0_B\rangle - \beta|1_B\rangle .
\end{split}
\end{equation}
According to the generalized Born hypothesis discussed in Sec.~\ref{sec:gen},
if one measures qubit $A$ \textit{after} acting with the Hadamard one finds
qubit $A$ is in state $|0\rangle$ with probability $1/2$, leaving qubit $B$
in state $|\phi_{0,B}\rangle$, and is
in state $|1\rangle$ with probability $1/2$, in which case qubit $B$ is left
in state $|\phi_{1,B}\rangle$.
Again, the probabilities 
differ from those in
Eq.~\eqref{probs_cohere}, which is for the case of a coherent
superposition, by the absence of the factors of
$\alpha\beta^* + \alpha^* \beta$ which came from interference.

One could also obtain these results
by computing the density matrix for qubit $A$, see Chapter
\ref{ch:den_mat}, particularly Example $2$ in Sec.~\ref{sec:examples}.

Intuitively, interference terms do not appear when the qubit being
investigated (qubit $A$ here) is entangled with another qubit because there is
then no well defined phase relation between the two parts of the superposition
($|0_A\rangle$ and $|1_A|\rangle$). 

\section{Summary}
For a coherent superposition, to compute probabilities one sums the amplitudes and
then squares, e.g.
\begin{equation}
{1\over 2}|\alpha + \beta|^2 ,
\end{equation}
while for an incoherent addition of probabilities, which happens when the relative phase is erased
by noise or by entanglement with other qubits, one squares and then sums, e.g.
\begin{equation}
{1\over 2} \left(|\alpha|^2 + |\beta|^2\right) .
\end{equation}

%% file: error_corr8.tex
\label{chap:error}
\section{Introduction}
Quantum error correction has developed into a huge topic, so here we will only
be able to describe the main ideas.

Error correction is essential for quantum computing, but appeared at
first to be impossible, for reasons that
we shall soon see. The field was transformed in 1995 by
Shor~\cite{shor:95} and Steane~\cite{steane:96} who showed that quantum error
correction \textit{is}
feasible.  Before Shor and Steane, the goal of building a useful
quantum computer seemed clearly unattainable. After those two papers,  while building a quantum
computer obviously posed enormous experimental challenges, it was not \textit{necessarily}
impossible. 

Some general references on quantum error correction are
Refs.~\cite{mermin:07,nielsen:00,vathsan:16,rieffel:14}.

Let us start by giving a simple discussion of classical error correction which
will motivate our study of quantum error correction. Classically,
error correction is not necessary for computation. This is because the hardware for one bit is
huge on an atomic scale and the states 0 and 1 are so different that the
probability of an unwanted flip is tiny. However, error correction is needed
classically for transmitting a signal over large distances where it attenuates
and can be corrupted by noise. 

To perform error correction one needs redundancy.
One simple way of doing classical error correction is to encode each
\textit{logical} bit by three \textit{physical} bits, i.e.
\begin{subequations}
\begin{align}
|0\rangle \rightarrow |\overline{0} \rangle &\equiv |0\rangle|0\rangle|0\rangle \equiv
|000\rangle \, , \\
|1\rangle \rightarrow |\overline{1} \rangle &\equiv |1\rangle|1\rangle|1\rangle \equiv
|111\rangle \, ,
\end{align}\label{01_three}
\end{subequations}
(for convenience we are using Dirac notation here even though these are classical bits for now.)
The sets of three bits, $|000\rangle$ and $|111\rangle$, are called \textit{codewords}.
One monitors the codewords to look for errors. If the bits in a codeword are
\index{codeword}
\index{majority rule}
not all the same one uses ``majority rule'' to correct. For example
\begin{equation}
\begin{split}
&|010\rangle\ \mathrm{is\ corrected\ to}\ |000\rangle \\
&|110\rangle\ \mathrm{is\ corrected\ to}\ |111\rangle .\\
\end{split}
\end{equation}
This works if no more than one bit is corrupted and so the error rate must be
sufficiently low that the probability of two or more bits in a codeword being
corrupted is negligible.

In quantum error correction one also uses multi-qubit codewords and
monitoring. However, there are several major differences compared with
classical error correction:
\begin{enumerate}
\item
\textit{Error correction is essential}.
Quantum computing requires error correction. This is because the
physical systems for a single qubit are very small, often on an atomic scale,
so any small outside interference can disrupt the quantum state.
\item
\textit{Measurement destroys quantum information}.
\label{point2}
In contrast to the classical case checking for errors is problematic.
Monitoring means measuring, and measuring a general quantum state alters it.
Thus it seems that any attempt at error correction must destroy
important quantum information.\label{destroy}
\item
\textit{More general types of error can occur}.
Bit flips are not the only possible errors. For example one can have phase
errors where ${1\over\sqrt{2}}(|0\rangle + |1\rangle) \to 
{1\over\sqrt{2}}(|0\rangle + e^{i \phi} |1\rangle)$.
\item
\textit{Errors are continuous}.
Unlike all-or-nothing bit flip errors for classical bits, errors in qubits can
grow continuously out of the uncorrupted state.
\end{enumerate}
One might imagine that point (\ref{point2}), in particular, would be fatal.
Amazingly this is not so as we shall see.

\section{Correcting bit flip errors}
\index{bit-flip code, 3-qubits}
We start our discussion of quantum error correction by considering how one
can correct for just bit flip errors. If the error rate is low we
might hope to correct them by tripling the number of qubits as in the classical
case, Eq.~\eqref{01_three}.

\begin{figure}[htb]
\begin{center}
\includegraphics[width=5cm]{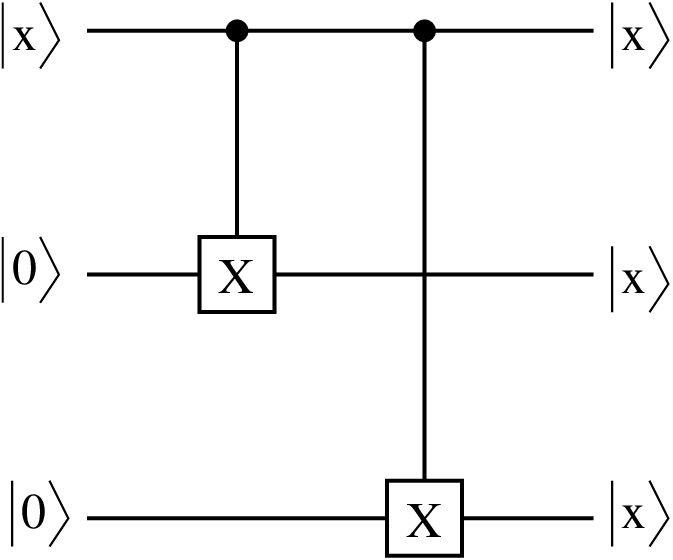}
\caption{
Circuit to encode the 3-qubit bit-flip code. Here $|x\rangle$ is $|0\rangle$ or
$|1\rangle$ in the
computational basis. The effect of this circuit on a linear combination of
$|0\rangle$
and $|1\rangle$ is shown in Fig.~\ref{triple_psi}.
\label{triple}
}
\end{center}
\end{figure}

The tripling of the qubits can be accomplished by the circuit in
Fig.~\ref{triple}. To see how this works suppose that the
input qubit, $|x\rangle$, is
$|0\rangle$. Then none of the Ctrl-X (CNOT) gates act on their target qubit so all
three qubits are $|0\rangle$ at the end (i.e.~on the right).
However, if the input qubit $|x\rangle$ is $|1\rangle$ then the Ctrl-X
gates act so all three qubits are 1 at the end.

\begin{figure}[htb]
\begin{center}
\includegraphics[width=10cm]{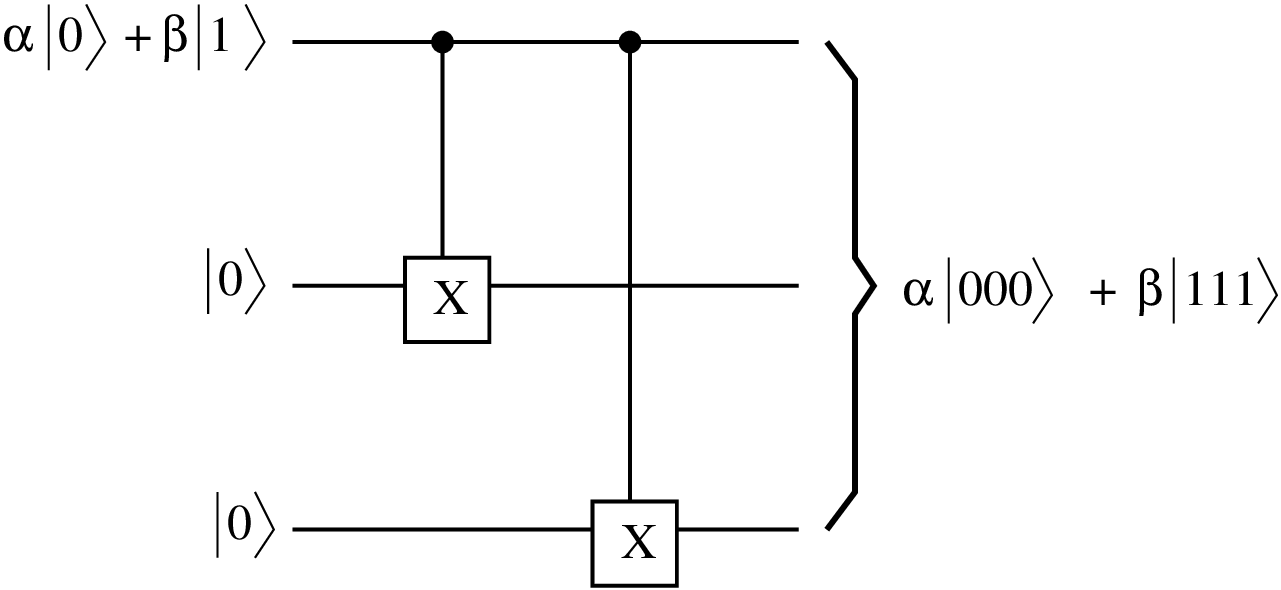}
\caption{
Circuit to encode the 3-qubit bit-flip code acting on a linear combination of
$|0\rangle$ and $|1\rangle$.
\label{triple_psi}
}
\end{center}
\end{figure}

By linearity a linear
combination of $|0\rangle$ and $|1\rangle$ is transformed as we want:
\begin{equation}
\alpha |0\rangle + \beta |1\rangle \to \alpha |000\rangle + \beta|111\rangle
\, ,
\label{bitflip}
\end{equation}
see Fig.~\ref{triple_psi}.
Note that this is not a clone of the input state which would be
\begin{equation}
\left(\,\alpha|0\rangle +\beta|1\rangle\,\right)^{\otimes 3} = 
\alpha^3 |000\rangle +
\alpha^2 \beta \left(\,|001\rangle + |010\rangle + |100\rangle\,\right) +
\alpha \beta^2 \left(\,|110\rangle + |101\rangle + |011\rangle\,\right) +
\beta^3 |111\rangle \, .
\end{equation}
We recall that cloning an arbitrary unknown state is impossible according to
the no-cloning theorem.
\index{no-cloning theorem}

\begin{figure}[htb]
\begin{center}
\includegraphics[width=7cm]{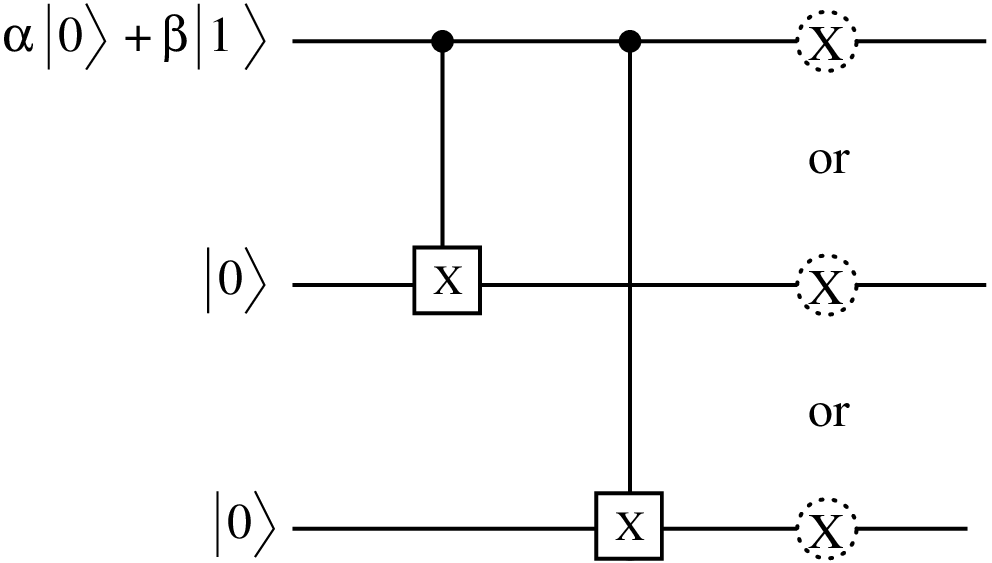}
\caption{Circuit indicating that at most one of the three bits generated by
the circuit in Fig.~\ref{triple_psi} has flipped due to an error.  The goal will be to
determine whether any have flipped, if so which one, and then correct the
error. 
\label{flipped1}
}
\end{center}
\end{figure}

Now we have to check if any of the three qubits generated by the circuit in
Fig.~\ref{triple_psi} are flipped, i.e.~if the situation is that shown in
Fig.~\ref{flipped1}. We assume that no more than one has been flipped, which
is a reasonable approximation if the
error rate is small.

We have therefore to consider one uncorrupted state and three corrupted states:
\begin{subequations}
\begin{align}
|\,\psi\,\rangle &= \alpha |000\rangle + \beta|111\rangle \, ,  \label{uncorr} \\
|\psi_1\rangle &= \alpha |100\rangle + \beta|011\rangle = X_1
|\,\psi\,\rangle \quad\quad
\mathrm{(qubit \ 1\ flipped)}\, , \\
|\psi_2\rangle &= \alpha |010\rangle + \beta|101\rangle  = X_2
|\,\psi\,\rangle \quad\quad
\mathrm{(qubit \ 2\ flipped)} \, ,\\
|\psi_3\rangle &= \alpha |001\rangle + \beta|110\rangle  = X_3
|\,\psi\,\rangle \quad\quad
\mathrm{(qubit \ 3\ flipped)} \, .
\end{align}
\label{syndrome}
\end{subequations}
\index{syndrome}
These four states are called the ``syndromes". Note that
we denote the left hand
qubit as the first qubit, the one to its right as the second qubit, and so on,
e.g.~$|x_1 x_2 x_3\rangle$.
Hence in Eq.~\eqref{syndrome} $|\psi_i\rangle$ refers to the state in which
qubit $i$ is flipped relative to the uncorrupted state $|\psi\rangle$.

Classically, to determine if one of the bits is flipped we just have to look
at them. However, quantum mechanically, if we measure $|\psi\rangle$, say, we
get $|000\rangle$ with probability $|\alpha|^2$ and $|111\rangle$ with
probability $|\beta|^2$, which destroys the coherent superposition.
\index{coherence}It
might
therefore seem that quantum error correction is impossible. 

\index{ancilla qubits}
Amazingly this is not so.  The secret is to couple the codeword qubits to 
ancillary qubits and measure only the ancillas.  This will give enough information
to determine which syndrome the state is in \textit{without
destroying the coherent superposition}.

\begin{figure}[htb]
\begin{center}
\includegraphics[width=15cm]{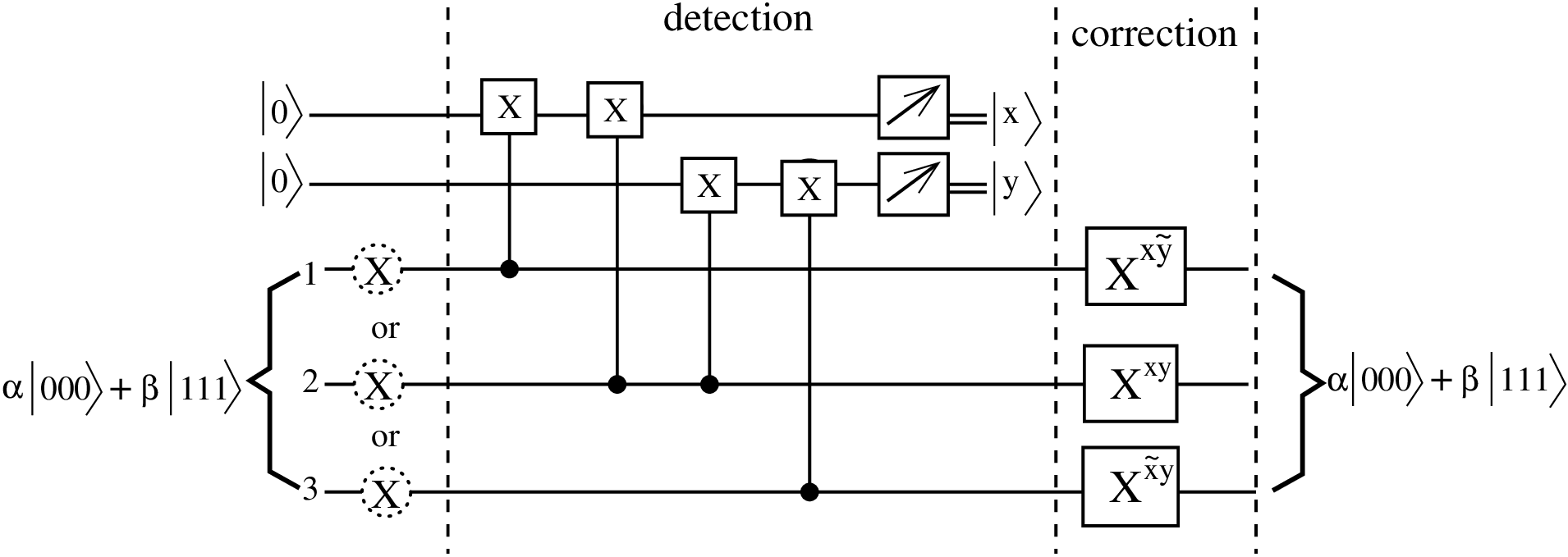}
\caption{Circuit to determine the syndrome for the 3-qubit bit-flip code, and
correct if necessary. A box with an arrow denotes a measurement. The double
lines indicate that the result of a measurement is a classical bit.
\label{flipped2}
}
\end{center}
\end{figure}

Here we need two ancillary qubits. The circuit including them is shown in
Fig.~\ref{flipped2}.  The three codeword qubits are at the bottom and the
ancillary
qubits are at the top. The ancillary qubits are measured and give values $x$
and $y$. We shall now see that each of the four possible pairs of values for
$x$ and $y$ corresponds to one of the syndrome states in Eq.~\eqref{syndrome}.

Both ancillas are targeted by two of the codeword qubits:

\vspace{3mm}
1st (upper) ancilla $(x)$ is targeted by codeword qubits 1 and 2.\\
\indent 2nd (lower) ancilla $(y)$ is targeted by codeword qubits 2 and 3.

\vspace{3mm}
Let's see what happens for the four syndrome states. 
\begin{itemize}
\item [$|\psi\rangle$] 
Codeword $|000\rangle$. No ancilla flipped so $x=0, y=0$.\\
\indent Codeword $|111\rangle$. Both ancillas are flipped twice so again $x=0, y=0$.\\
Note that the result of the measurement is the same for both the
$|000\rangle$ and $|111\rangle$ parts of the state $|\psi\rangle$. Hence the coherent
superposition of $|\psi\rangle$ is not destroyed by the measurement on the ancillas.
If the result of the measurement were different for the different parts of the
superposition, then only the piece corresponding to the measured value would
survive and the superposition would be broken.
\item [$|\psi_1\rangle$] 
Codeword $|100\rangle$. $x$ is flipped once, and $y$ is not flipped, so $x=1, y=0$.\\
\indent Codeword $|011\rangle$. $x$ is flipped once and $y$ is flipped twice so again
$x=1, y=0$.\\
Recall that the qubits are ordered such that qubit 1 is on the left.
\item [$|\psi_2\rangle$] 
Codeword $|010\rangle$. $x$ and $y$ are both flipped once so $x=1, y=1$.\\
\indent Codeword $|101\rangle$. $x$ and $y$ are both flipped once so again
$x=1, y=1$.
\item [$|\psi_3\rangle$] 
Codeword $|001\rangle$. $x$ is not flipped and $y$ is flipped once so $x=0, y=1$.\\
\indent Codeword $|110\rangle$. $x$ is flipped twice and $y$ is flipped once so
again $x=0, y=1$.
\end{itemize}
Hence we get the table of results shown in Table \ref{table1}. Note that in
all cases the coherent superposition of the syndrome state is not destroyed by the
measurement of the ancillas. 

\begin{table}[tbh]
\begin{center}
\begin{tabular}{|c|c|cc|}
\hline
syndrome  & bit flipped &\ x\ &\ y\ \\
\hline
$|\,\psi\,\rangle$   & none & 0 & 0 \\
$|\psi_1\rangle$ & 1 & 1 & 0 \\
$|\psi_2\rangle$ & 2 & 1 & 1 \\
$|\psi_3\rangle$ & 3 & 0 & 1 \\
\hline
\end{tabular}
\caption{Results of measurement of the ancillary qubits for the different
syndromes of the codeword qubits.
\label{table1}
}
\end{center}
\end{table}

Hence by measuring the auxiliary qubits we can determine which if any of the
codeword qubits have flipped and then apply a compensating flip if necessary. 
The $X$-gates which perform these
compensating flips are shown at the right of Fig.~\ref{flipped2}. 
For example the $X^{x\tilde{y}}$
gate on qubit 1
indicates that a flip is done by
acting with the $X$ operator on qubit 1 only if $x\tilde{y}=1$, i.e.~$x=1$
and $y = 0$, which corresponds to the
second entry in the Table \ref{table1} ($\tilde{y}$ means the complement of $y$).

We have assumed up to now that the state of the system has had
a qubit flipped with
probability one. However, as already noted, errors in quantum circuits can
arise continuously from zero, and we are concerned with the situation in
which the error rate is small (otherwise we can not error correct). 
Consider then, a more realistic scenario in which the state of the three
qubits in the codeword 
has a small amplitude to have any of the qubits flipped, i.e.~the state
of the codeword is given by
\begin{equation}
|\psi\rangle \rightarrow 
\left[ 1 + (\,  \epsilon_1 X_1 + \epsilon_2 X_2 +
\epsilon_3 X_3\,) \right]|\psi\rangle \, ,
\label{eps}
\end{equation}
where
$|\psi\rangle$ is given by Eq.~\eqref{uncorr}, 
the $\epsilon_i$ may be complex, $|\epsilon_i| \ll 1$,
we have only indicated terms to first order in the $\epsilon_i$, and ignored
corrections to the normalization which are second order in the $\epsilon_i$.

Hence, to first order in $\epsilon$, the state of the codeword qubit and ancilla qubits which is inputed to
the detection phase of the circuit in Fig.~\ref{flipped2} is
\begin{equation}
\left[ 1 + (\,  \epsilon_1 X_1 + \epsilon_2 X_2 +
\epsilon_3 X_3\,) \right]|\psi\rangle \otimes |00\rangle_A, ,
\end{equation}
where $|\cdots\rangle_A$ refers to the ancillas.
In the detection phase, the codeword qubits are entangled with 
ancillas in such a way that the state of the combined codeword-ancilla system,
just before the measuring gates in Fig.~\ref{flipped2}, is
\begin{equation}
|\psi\rangle |00\rangle_A +
\epsilon_1 X_1 |\psi\rangle |10\rangle_A +
\epsilon_2 X_2 |\psi\rangle |11\rangle_A +
\epsilon_3 X_3 |\psi\rangle |01\rangle_A,
\end{equation}
where $|\cdots\rangle_A$ refers to the ancillas. The ancillas are then
measured
with the possible results shown below
\begin{center}
\begin{tabular}{|c|c|c|c|}
\hline
measured ancillas & probability & resulting syndrome & operator to correct the
error \\
\hline\hline
$|00\rangle_A $ & $\simeq 1$   & $|\psi\rangle $ & none needed \\
$|10\rangle_A $ & $|\epsilon_1|^2$ & $X_1|\psi\rangle $ & $X_1$ \\
$|11\rangle_A $ & $|\epsilon_2|^2$ & $X_2|\psi\rangle $ & $X_2$ \\
$|01\rangle_A $ & $|\epsilon_3|^2$ & $X_3|\psi\rangle $ & $X_3$ \\
\hline
\end{tabular}
\end{center}


Since the
$\epsilon_i$ are small,
the \textit{probability} that a corrupted state is detected is small, so the
most probable situation is that projection is on to the uncorrupted state so
no correction is needed.
However, there is a small probability that the projection will be on to one of
the corrupted syndromes.
The corrupted syndromes differ \textit{substantially} from the uncorrupted
state. They are
further, in fact, from the uncorrupted state than the original state in Eq.~\eqref{eps}.
This might, at first, seem like a retrograde step but it is not because
the corrupted state is \textit{known precisely} so
it is possible to correct it
back to to the uncorrupted state.

To summarize this part,
quantum error correction is feasible,
even though errors arise continuously, because possibly corrupted states
are projected on to one of a \textit{discrete} set of states which can be corrected
if necessary. We will discuss this important point again in
Sec.~\ref{discrete} when
we consider how general errors arise.

It should be noted that in classical analog computers, where errors also arise
continuously, no such projection can be done, and hence error correction can
not be performed.  This is why we don't have classical analog computers. 

\begin{figure}[htb]
\begin{center}
\includegraphics[width=15cm]{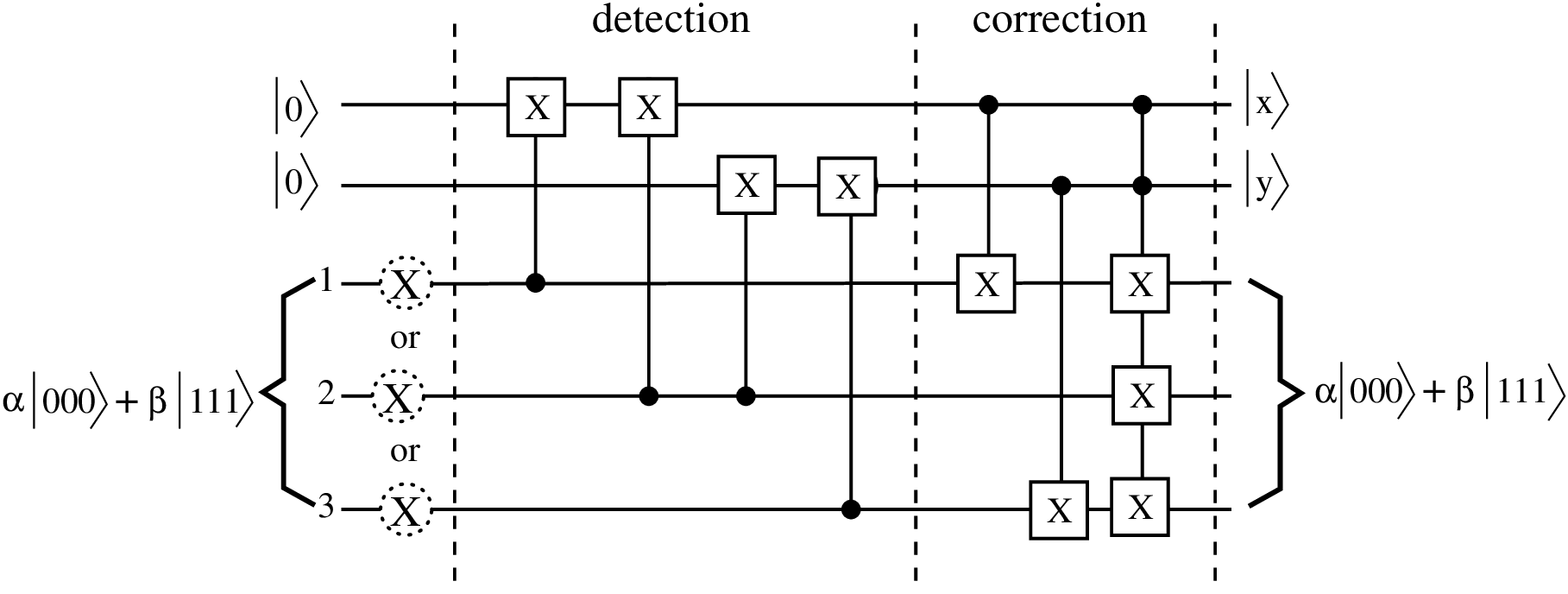}
\caption{Automation of the error correction procedure of Fig.~\ref{flipped2}.
The three controlled gates on the right have the same effect as the NOT
(i.e.~$X$) gates
on the right of Fig.~\ref{flipped2} which depend on the result of measurements of the $x$
and $y$ ancillary qubits. The rightmost gate, with two control qubits and three
target qubits, is discussed in the text. The values of the control bits $x$ and
$y$ at the end depend on which of the four syndromes is present (i.e.~which if any of the
$X$ gates on the left of the figure have acted) according to Table
\ref{table1}. Before this circuit can be used again, the ancillary qubits have
to be reinitialized to $0$.
\label{flipped3}
}
\end{center}
\end{figure}

\index{coherence}
Going back to the discussion of Fig.~\ref{flipped2},
one can avoid explicitly measuring the qubits and instead
correct any bit-flip error
coherently and automatically
by having the ancillas interact back on
the codeword qubits as shown in Fig.~\ref{flipped3}.  In that figure, the
rightmost three controlled gates have the same effect as the NOT (i.e.~$X$) gates
in the right of Fig.~\ref{flipped2} which depend on the result of measurements
of the $x$
and $y$ ancillary qubits. The rightmost gate in Fig.~\ref{flipped3} has two control qubits and three
target qubits.  This gate flips all the target qubits if \textit{both} control
qubits are 1.  It is a generalization of the Toffoli gate $T$ which has two
\index{Toffoli gate!generalized}
\index{control qubit}
\index{target qubit}
control qubits, and one target qubit which is flipped if both control qubits
are 1, i.e.~$T|x\rangle|y\rangle|z\rangle = |x\rangle|y\rangle|z \oplus
x\,y\rangle$.
If we denote by $T^*$ the rightmost gate in Fig.~\ref{flipped3}
then ~$T^*|x\rangle|y\rangle|z\rangle|u\rangle|v\rangle =
|x\rangle|y\rangle|z \oplus
x\,y\rangle |u  \oplus x\, y\rangle |v  \oplus x\, y\rangle$.
\index{Toffoli gate}
Note that this gate is
equivalent to three separate Toffoli gates, in which the two ancilla qubits are the controls, 
qubit 1 is the target for the first Toffoli, qubit 2 for the second Toffoli,
etc.
After the error on the computational bits has been corrected the
ancilla qubits have to be reinitialized to zero.

It is instructive to show for the different syndromes in
Eq.~\eqref{syndrome} that the circuits in Figs.~\ref{flipped2} and
\ref{flipped3} give the same result, i.e.~the end product is the uncorrupted state
$|\psi\rangle$. The results from the circuit of Fig.~\ref{flipped2} have
already been discussed above. For the circuit in Fig.~\ref{flipped3}
we just consider the case of $|\psi_2\rangle$
(so qubit 2 has been flipped), and we have $x=1,
y=1$ according to Table \ref{table1}. Consider the rightmost three gates in
Fig.~\ref{flipped3} (these are the ones that do the error correction).
For $x=1, y=1$, the rightmost gate is active and flips all
three codeword qubits. Hence,
between them, the rightmost three gates flip codeword qubit 1 twice, 
flip codeword qubit 2 once, and flip codeword qubit 3 twice.  The net result is that 
only codeword qubit 2 is flipped so we recover the
uncorrupted state $|\psi\rangle$. It is useful to check that the circuit in
Fig.~\ref{flipped3} also works to correct $|\psi_1\rangle$ and $|\psi_3\rangle$.

\section{Stabilizer formalism}
\label{sec:stabilizer}
\index{stabilizers}
In order to conveniently generalize the ideas in the previous section to
arbitrary errors  we need
to reformulate them. 

\index{matrix!Hermitian}
For reasons that will shortly become clear, consider the two
Hermitian\footnote{As discussed in Chapter \ref{ch:qu_intro}
it is an axiom of quantum mechanics that measurable quantities are represented
by Hermitian operators.}
operators $Z_1Z_2$ and
$Z_2 Z_3$. Because $Z_i^2 = \mathbbm{1}$ (the identity) and different $Z's$ commute we have
\begin{equation}
\left(Z_1Z_2\right)^2 = \mathbbm{1}, \quad \left(Z_2Z_3\right)^2 = \mathbbm{1}\, .
\end{equation}
An operator whose square is unity 
has eigenvalues equal to $\pm 1$, since acting twice with the operator on an
eigenvector gives the eigenvector, so the square of the eigenvalue is 1. 
We also we know that $Z_1Z_2$ and $Z_2 Z_3$ commute with each other
\textit{and hence have the same eigenvectors}. 

\begin{table}[tbh]
\begin{center}
\begin{tabular}{|c|c|cc|cc|}
\hline
syndrome  &  & $Z_1 Z_2$ & $Z_2 Z_3$ & \ x\ &\ y\ \\
\hline
$|\,\,\psi\,\rangle$   &   &  \,\,1 & \,1  & 0 & 0 \\
$|\psi_1\rangle$ & $X_1|\psi\rangle$ & -1 & \,\,1 & \,\,1 & 0 \\
$|\psi_2\rangle$ & $X_2|\psi\rangle$ & -1 &-1 & \,\,1 & \,\,1 \\
$|\psi_3\rangle$ & $X_3|\psi\rangle$ &  \,\,1 &-1 & 0 & \,\,1 \\
\hline
\end{tabular}
\caption{The eigenvalues of the stabilizers $Z_1Z_2$ and $Z_2Z_3$ for the four
syndromes for the 3-qubit bit-flip code, and a comparison with the
measurements of the ancillary qubits $x$ and $y$ used to measure them, see Fig.~\ref{flipped4}.
The uncorrupted state has eigenvalue $+1$ for both stabilizers.  This is an
important property that stabilizers must have in general.
Note that $Z_1Z_2=1$
corresponds to $x=0$, and $Z_1Z_2=-1$ corresponds to $x = 1$. There is a
similar connection between $Z_2 Z_3$ and $y$, so $Z_1 Z_2 =(-1)^x, Z_2 Z_3 =
(-1)^y$. The second column shows how the corrupted state is generated from the
uncorrupted state.
\label{table2}
}
\end{center}
\end{table}

One can verify that the syndrome states in Eq.~\eqref{syndrome} are eigenvectors of 
$Z_1Z_2$ and $Z_2 Z_3$ according to Table~\ref{table2}.
In general we use the
term ``stabilizers'' to denote operators like operators $Z_1Z_2$
and $Z_2Z_3$ whose $\pm 1$ eigenvalues distinguish the different syndromes. 
As we will see below, each of the stabilizers is measured by an ancilla qubit,
$|x\rangle$ for $Z_1Z_2$ and $|y\rangle$ for $Z_2 Z_3$, see Fig~\ref{flipped4} below.
The ancilla state $|x=0\rangle$ corresponds to $Z_1Z_2 = +1$, and
$|x=1\rangle$ corresponds to $Z_1Z_2 = -1$, or in other words,
$Z_1 Z_2 =(-1)^x$,  and similarly $Z_2 Z_3 =
(-1)^y$.

Below we will discuss the circuit with which we measure the stabilizers, but first we
show
a more straightforward way to determine whether the eigenvalue of a
stabilizer in a
syndrome is $+1$ or $-1$ than simply acting with the stabilizer on the syndrome.

We note first that the eigenvalue of
all the stabilizers is $+1$ in the uncorrupted syndrome $|\psi\rangle$. This
is an essential property that stabilizers must have.
Also
note that
the operators for the stabilizers will be built out of the single-qubit operators
$Z_i$ and $X_i$. For the 3-qubit, bit-flip code we
only have the $Z_i$ but the $X_i$ will also be needed to correct for general
errors. Furthermore the syndromes with a single qubit error are obtained 
by acting on the uncorrupted syndrome with the $X_i, Y_i$ and
$Z_i$ operators.\footnote{Recall that the Pauli operators
$X, Y$ and $Z$ are given by
$X \equiv\sigma^x =  \begin{pmatrix} 0 & 1 \\ 1 & 0 \end{pmatrix},
Y \equiv\sigma^y =  \begin{pmatrix} 0 & -i \\ i & 0 \end{pmatrix},
Z \equiv\sigma^z =  \begin{pmatrix} 1 & 0 \\ 0 & -1 \end{pmatrix}$ and so
$Y = i X Z$.}
Again,
for our simple example above, we only had the $X_i$, but the other operators
will also be used when we deal with general errors.

The Pauli operators, $X_i, Y_i, Z_i$, have the property that they commute for
different qubits $i$, whereas different operators on the same qubit anti-commute,
where the
anti-commutator of $A$ and $B$ is defined by $\{A, B\} \equiv AB + BA$.  Hence
we have, for example,
\begin{subequations}
\begin{align}
[X_i, Y_j] &\equiv X_i Y_j - Y_j X_i  = 0 \quad(i \ne j)\, , \\
\{X_i, Y_i\} &\equiv X_i Y_i + Y_i X_i  = 0 \, .
\label{anticomm}
\end{align}
\end{subequations}
(Verify the anti-commutation relations like Eq.~\eqref{anticomm} by explicitly
working out some cases.)

Consequently, if we consider a general stabilizer $A_\alpha$ and a syndrome state
$|\psi_\beta\rangle = B_\beta |\psi \rangle$ then $A_\alpha$ either commutes or
anti-commutes with $B_\beta$. Note that $B_\beta$ only involves a single
Pauli operator (which, in general, can be an $X$ or a $Y$ or a $Z$) whereas
$A_\alpha$ involves a product of Pauli operators, which, in the general case,
can be made up of $X$'s and $Z$'s.
%
We will now show that if 
$A_\alpha$ commutes with $B_\beta$ the eigenvalue of the stabilizer $A_\alpha$
in state $|\psi_\beta \rangle$ is +1 and if they anti-commute the eigenvalue
is $-1$. 

Firstly, if $A_\alpha$ commutes with $B_\beta$ then
\begin{equation}
A_\alpha |\psi_\beta \rangle = A_\alpha B_\beta |\psi \rangle
= B_\beta A_\alpha |\psi \rangle = 
B_\beta |\psi \rangle = |\psi_\beta \rangle \, ,
\label{+1}
\end{equation}
where we used that the eigenvalues of all the stabilizers $A_\alpha$
are $+1$ in the uncorrupted state $|\psi\rangle$ to get the third equality.
Hence the eigenvalue of $A_\alpha$ in state $|\psi_\beta\rangle$
is $+1$ if $A_\alpha$ commutes with $B_\beta$.
Similarly if $A_\alpha$ anti-commutes with
$B_\beta$ then
\begin{equation}
A_\alpha |\psi_\beta \rangle = A_\alpha B_\beta |\psi \rangle
= -B_\beta A_\alpha |\psi \rangle = 
-B_\beta |\psi \rangle = -|\psi_\beta \rangle \, ,
\label{-1}
\end{equation}
so the eigenvalue is $-1$.

We emphasize that the syndromes must be eigenstates of \textit{all} the stabilizers
which means that the \textit{stabilizers must commute with each other.}


Next we will see how to determine efficiently if a stabilizer commutes or anti-commutes with
the operator which generates a corrupted syndrome out of the uncorrupted state.

For the case of the 3-qubit, bit-flip code discussed so far the stabilizers
are
\begin{equation}
Z_1 Z_2\ \text{and}\ Z_2 Z_3\, ,
\end{equation}
\index{stabilizers!for 3-qubit code}
and the operators which generate the corrupted syndrome from the uncorrupted
state are
\begin{equation}
X_1, X_2 \ \text{and} \ X_3.
\end{equation}
As an example, we see that $X_1$ commutes with $Z_2 Z_3$ because there are no sites in
common, so the eigenvalue of $Z_2 Z_3$ for $|\psi_1\rangle$ must be $+1$ which
agrees with Table \ref{table2}. On the other hand $X_2$ has one site in common
with $Z_2 Z_3$ so
\begin{equation}
X_2\, Z_2 Z_3 = - Z_2 X_2 Z_3 = - Z_2 Z_3 X_2 \, ,
\end{equation}
and the operators anticommute, so the eigenvalue of $Z_2 Z_3$ for
$|\psi_2\rangle$ must be $-1$, which again agrees with Table \ref{table2}.

\begin{quotation}
\noindent The point is that every time we have to interchange the order of two
different operators acting on the same qubit we pick up a minus sign.
\end{quotation}

\noindent Hence it is straightforward to
deduce the overall sign.
Note that operators of the same type, e.g.~the $Z_i$,
always commute.

As a more complicated example, which occurs in a scheme for full
error correction, consider the stabilizer $Z_1 Z_3 X_4 X_5$. For the syndrome which
has been
corrupted by $Z_4$ the eigenvalue is $-1$, the minus sign coming from
interchanging the order of $X_4$ and $Z_4$. However, for the syndrome which
was
corrupted by $X_4$ the eigenvalue is $+1$ since, for the qubit in common,
(qubit $4$), both operators are $X$ and so commute. As another example, for the
syndrome
which was corrupted by $X_2$ the eigenvalue is $+1$, because $X_2$ and the
stabilizer commute since they have no qubits in common.

\index{stabilizers}
To summarize, in the stabilizer formalism we need to construct a 
set of Hermitian operators (the stabilizers) which have the
following properties:
\begin{enumerate}
\item
they square to the identity, (so the eigenvalues are $\pm 1$),
\item
they mutually commute (so they have the same eigenstates),
\item
the syndromes are eigenstates
\item
the uncorrupted syndrome has eigenvalue $+1$ for all stabilizers, and
\item
the set of $\pm 1$ eigenvalues of the
stabilizers
uniquely specifies the syndrome. Whether the eigenvalue is $+1$ or $-1$ is
easily determined from the commutation properties of the stabilizer with
respect to the operator which generates the corruption in the syndrome.
\end{enumerate}

In Sec.~\ref{shor} we will describe an example with full
error correction which has codewords with 9 qubits and needs 8 stabilizers.

\begin{figure}[htb]
\begin{center}
\includegraphics[width=9cm]{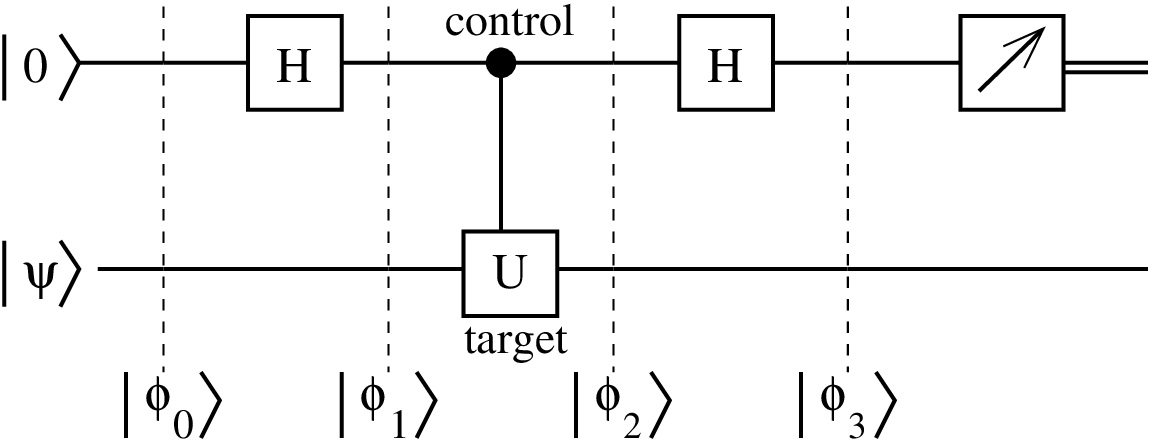}
\caption{A circuit with a control-$U$ gate in which the control (upper) qubit
is surrounded by Hadamards. $U$ is an operator with eigenvalues $\pm1$ and
corresponding eigenvectors $|\psi_+\rangle$ and $|\psi_-\rangle$.
As shown in the text, if a measurement of the upper qubit gives $|0\rangle$ then
the lower qubit will be in state $|\psi_+\rangle$,
and if the measurement gives $|1\rangle$ then the
lower qubit will be in state $|\psi_-\rangle$.
The states $|\phi_i\rangle\, (i=0, 1, 2, 3)$ are described in the text. Note
that this figure is identical to Fig.~\ref{stabilizer} and was discussed in Chapter
\ref{ch:gates}.
\label{ec:stabilizer}
}
\end{center}
\end{figure}

Next we describe the circuit which will measure the eigenvalues of the
stabilizers and hence determine which syndrome has
occurred. Consider the circuit in Fig.~\ref{ec:stabilizer} which includes a
control-$U$ gate in which the control qubit is sandwiched between Hadamards. 
Here $U$ is an operator, which, like the stabilizers, has eigenvalues $\pm1$. If the control qubit is 1 the
effect on the target qubit is
\begin{equation}
U|\psi_+\rangle = |\psi_+\rangle, \quad U|\psi_-\rangle = -|\psi_-\rangle,
\label{ec:ctrl-U}
\end{equation}
where $|\psi_+\rangle$ and $|\psi_-\rangle$ are the eigenvectors with
eigenvalue $+1$ and $-1$ respectively. If the control qubit is 0 then the
target qubit is unchanged. The initial state of the target qubit can be
written as a superposition of eigenstates, i.e.
\begin{equation}
|\psi\rangle = \alpha_+ |\psi_+ \rangle + \alpha_- |\psi_- \rangle .
\end{equation}

We discussed the circuit of Fig.~\ref{ec:stabilizer} in Chapter \ref{ch:gates}
and found that the states $|\phi_i\rangle, \, (i=0, 1, 2, 3)$ are given by
Eqs.~\eqref{gates:phi}. In particular, the final state $|\phi_3\rangle$,
before the measurement of the upper qubit, is given by 
\begin{equation}
|\phi_3\rangle = \alpha_+ |0\, \psi_+\rangle + \alpha_- |1\, \psi_-\rangle \, .
\end{equation}
Hence if a measurement of the upper qubit gives $|0\rangle$
(which it does with
probability $|\alpha_+|^2$) the lower qubit will be in state $|\psi_+\rangle$,
and if the measurement gives $|1\rangle$ (probability is $|\alpha_-|^2$) the lower qubit
will be in state $|\psi_-\rangle$. Hence we see that measuring the control qubit tells us
which eigenstate of $U$ the target qubit is in.

Stabilizers involve more than one codeword qubit so the gates we need will have
several target qubits. For the 3-qubit, bit-flip code, the circuit equivalent
to Fig.~\ref{flipped2} is shown in Fig.~\ref{flipped4}. We see that the $x$
ancilla is the control qubit for a control-$Z_1Z_2$ gate which is sandwiched
between Hadamards, and similarly the $y$
ancilla is the control qubit for a control-$Z_2Z_3$ gate. Hence
if $x=0$ the
state of the codeword
bits has $Z_1 Z_2= +1$, whereas if $x=1$ the state of the codeword
bits has $Z_1 Z_2= -1$. There is an analogous correspondence between $y$
and $Z_2 Z_3$.

\begin{figure}[htb]
\begin{center}
\includegraphics[width=15cm]{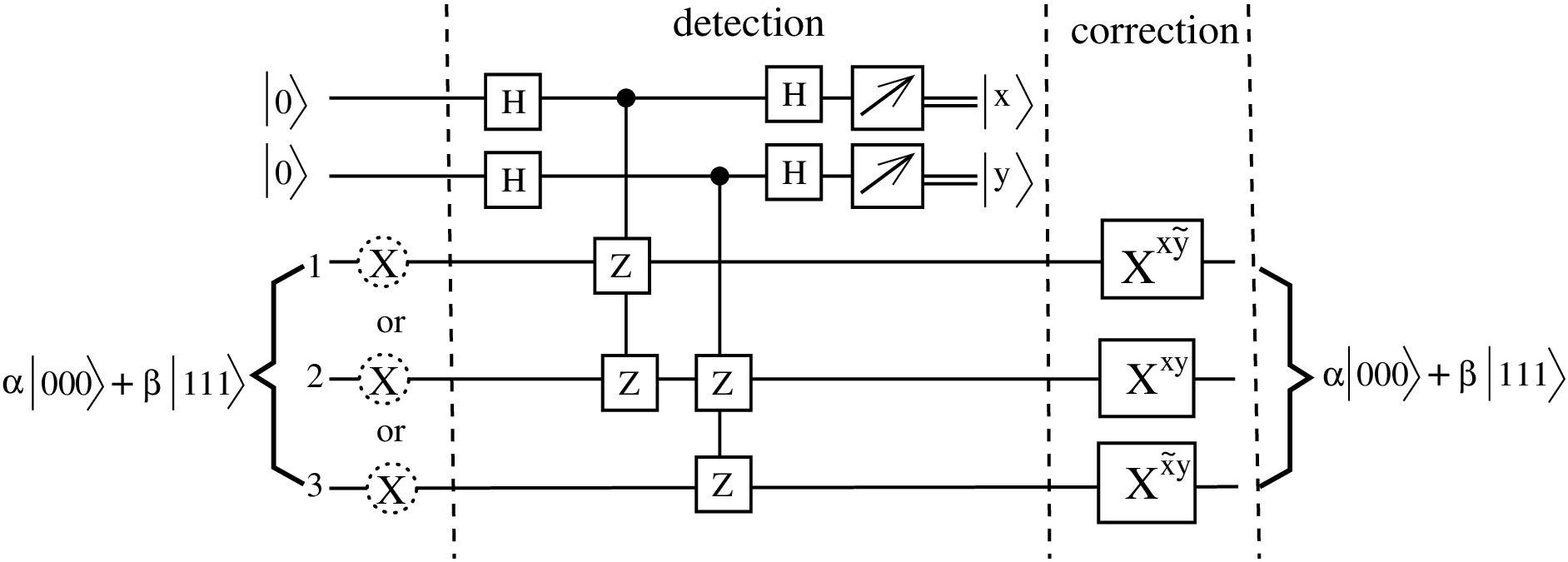}
\caption{Circuit equivalent to that in Fig.~\ref{flipped2} but in the
stabilizer formalism. In this circuit $x$ measures $Z_1 Z_2$, and $y$
measures $Z_2 Z_3$. In other words, if $x=0$ the state of the codeword
bits has $Z_1 Z_2= +1$, whereas if $x=1$ the state of the codeword
bits has $Z_1 Z_2= -1$, with an analogous correspondence between $y$
and $Z_2 Z_3$.
Note that $Z_1 Z_2$ and $Z_2 Z_3$ have
eigenvalues $\pm 1$ and commute with each other. 
\label{flipped4}
}
\end{center}
\end{figure}

\begin{figure}[htb]
\begin{center}
\includegraphics[width=10cm]{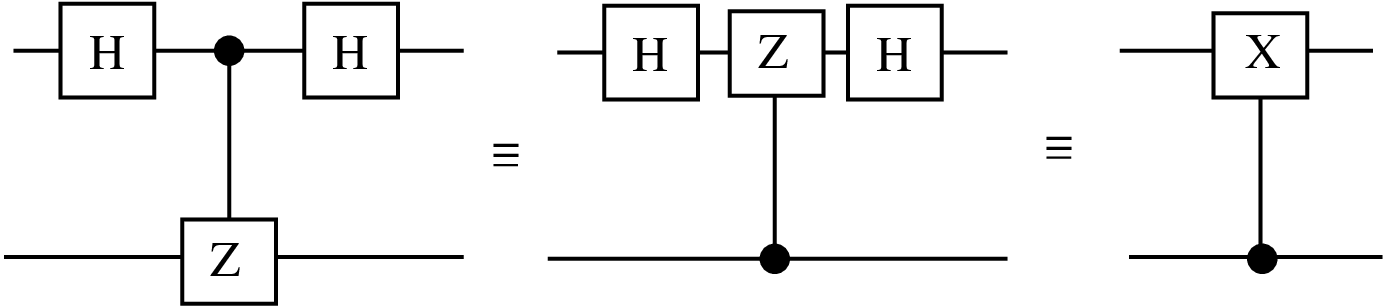}
\caption{The equalities in this figure are helpful to understand the
equivalence of Figs.~\ref{flipped2} and \ref{flipped4}. The
left-hand equality comes from the fact that the target and control qubits
can be exchanged in a control-$Z$ gate,
and the right-hand equality is because $HZH = X$ and $H^2 = \mathbbm{1}$.
\label{equiv}
}
\end{center}
\end{figure}

The equivalence of the circuits in Figs.~\ref{flipped2} and \ref{flipped4} can
also be understood from the simpler case of the equivalences shown
in Fig.~\ref{equiv} in which the
left-hand equality comes from the fact that the target and control qubits can
be exchanged in a control-$Z$ gate,\footnote{Because the only effect of the gate
is to change the sign of the state if both target and control qubits are 1.}
and the right-hand equality is because $HZH = X$ and $H^2 = \mathbbm{1}$ (the
identity).

The stabilizer formalism will be convenient when devising circuits for full
error correction rather than just correcting bit flips as we have done up to
now.

\section{Phase Flip Code}
\index{phase flip code}
Before discussing how to correct general errors, we will briefly mention another
special case, a phase flip, which has no classical equivalent since classical
bits don't have any property corresponding to phase. In this error model, with some
probability $p$, the relative phase 
of $|0\rangle$ and $|1\rangle$ is flipped so
\begin{equation}
|\psi\rangle = 
\alpha |0\rangle + \beta |1\rangle \ \rightarrow \  \alpha |0\rangle - \beta
|1\rangle\, .
\label{psiab}
\end{equation}
Phase flips are generated by the $Z$ operator since
\begin{equation}
\begin{pmatrix}
\alpha \\
\beta
\end{pmatrix}
\rightarrow Z
\begin{pmatrix}
\alpha \\
\beta
\end{pmatrix}
= \begin{pmatrix}
\alpha \\
-\beta
\end{pmatrix}
\quad (\text{computational\ basis}).
\end{equation}

The phase-flip error model can be turned into the already-studied bit-flip
model by transforming to the $\pm$ basis (also called the
$X$-basis because it is the basis in which $X$ is diagonal) where
\begin{equation}
|+\rangle = {1\over \sqrt{2}}\left(|0\rangle + |1\rangle\right), \quad
|-\rangle = {1\over \sqrt{2}}\left(|0\rangle - |1\rangle\right),
\end{equation}
One transforms between the $\pm$ basis and the computational basis
using Hadamards:
\begin{subequations}
\begin{align}
H |0\rangle &= |+\rangle, \quad H |1\rangle = |-\rangle, \\
H |+\rangle &= |0\rangle, \quad H |-\rangle = |1\rangle.
\end{align}
\end{subequations}
In the $\pm$ basis the roles of $X$ and $Z$ are interchanged since
\begin{subequations}
\label{XZ}
\begin{align}
X |0\rangle &= |1\rangle, \quad X |1\rangle = |0\rangle, \quad
Z |0\rangle = |0\rangle, \quad Z |1\rangle = -|1\rangle, \\
Z |+\rangle &= |-\rangle, \quad \!\!Z |-\rangle = |+\rangle, \quad\!\!\!
X |+\rangle = |+\rangle, \quad\!\!\! X |-\rangle = -|-\rangle. 
\end{align}
\end{subequations}
Thus we shall find in Sec.~\ref{shor} that stabilizers to detect phase errors
involve $X$ operators, as opposed to those used to
detect bit-flip errors which involve $Z$ operators (see Fig.~\ref{flipped4}).



\begin{figure}[htb]
\begin{center}
\includegraphics[width=10cm]{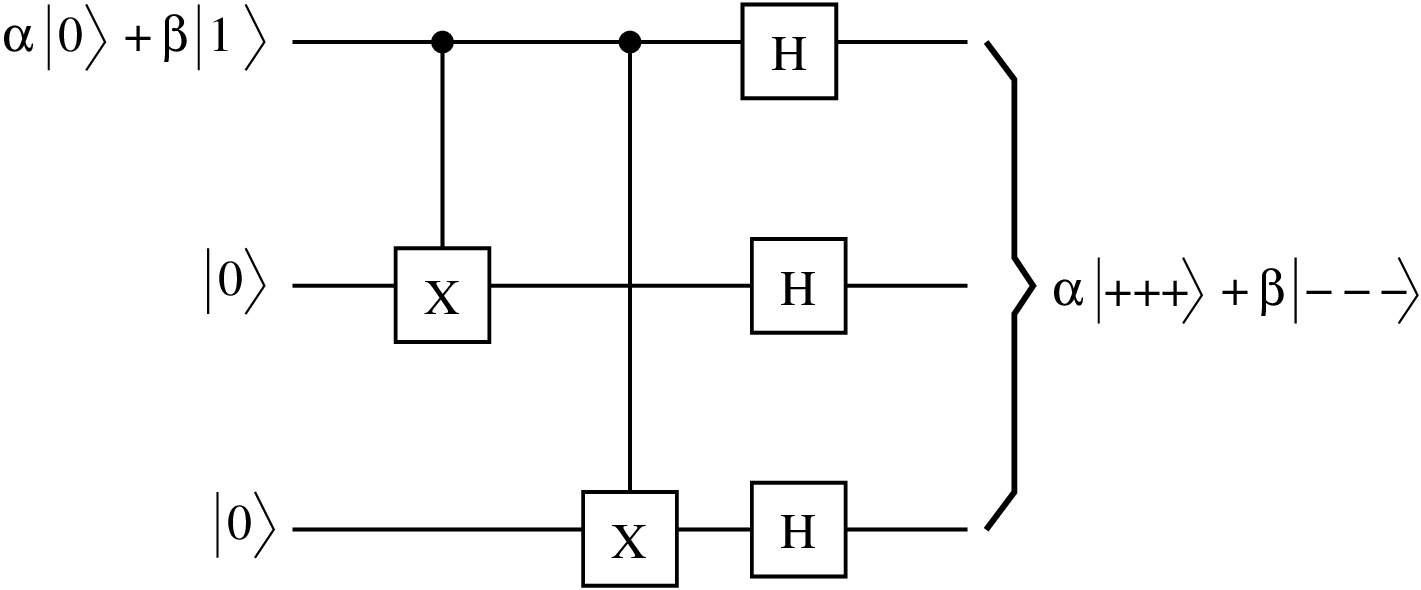}
\caption{
Encoding circuit for the 3-qubit phase-flip code.
\label{triple_psi2}
}
\end{center}
\end{figure}


The encoding circuit for the 3-qubit phase-flip code is obtained from that for
the 3-qubit bit-flip code in Fig.~\ref{triple_psi}
by adding Hadamards to the circuit, with
the result shown in
Fig.~\ref{triple_psi2}.  We shall use this circuit in Sec.~\ref{shor} as part of
the encoding circuit in Fig.~\ref{shor_encode} for a code (due to Shor) which
corrects general 1-qubit errors.



\section{General Errors and the Effects of the Environment}
\label{discrete}

In our discussion of errors we have so far implicitly assumed that the errors
occur because of some malfunction in the circuit. The state has underdone a
unitary transformation, but not exactly the right one.
Another, and very important,
source of error is interaction between the qubits and
the environment, which is unavoidable even though quantum computer
engineers work very hard to reduce it to a minimum. This can lead to errors
due to a \textit{non-unitary} change in the computational qubits (though the combined
\index{non-unitary transformation}
system of qubits plus environment undergoes unitary time development.)
In this section we include the effects of the environment and
also consider the most general type of single qubit error. The 
discussion below follows Mermin~\cite{mermin:07}.

Consider
a single qubit $|x\rangle$, and call the environment $|e\rangle$. Unlike
the state of the qubit, the state of the environment is in a
space of very many dimensions.  Ideally $|x\rangle$ evolves under the effects
of the gates only, independent of the environment. However, interactions with
the environment cannot be avoided which leads to a corruption of the qubit
and an entangling of the qubit with the environment. 

The most general such form of these effects is
\begin{subequations}
\begin{align}
|e\rangle \,|0\rangle &\rightarrow |e_0\rangle \,|0\rangle + |e_1\rangle
\,|1\rangle, \\
|e\rangle \,|1\rangle &\rightarrow |e_2\rangle \,|0\rangle + |e_3\rangle
\,|1\rangle, 
\end{align}
\label{decohere}
\end{subequations}
where $|e_i\rangle\, (i=0, \cdots,3)$ are possible final states of the
environment. The environment states are not normalized, and not orthogonal
either.  However, the two states on the right hand side of Eqs.~\eqref{decohere} must be
orthogonal since the time evolution of the combined qubit-environment system
is unitary. In other words
\begin{equation}
\langle e_2 |e_0\rangle + \langle e_3 |e_1\rangle
= 0\, .
\end{equation}
The corruption of the computation by the environment indicated in
Eq.~\eqref{decohere} is called ``decoherence".\index{decoherence}  It is the main source of
difficulty in building a practical quantum computer. 

\index{entanglement}
In previous sections we have neglected entanglement with the environment.
Rather, errors were assumed to occur because of mistakes made in the circuit
itself.  This corresponds to a special case of Eqs.~\eqref{decohere}, where all
the environment states are the same, apart from normalization,
i.e.~$|e_i\rangle = c_i |e\rangle$, for $i=0, \cdots, 3$. 

We are interested in the case where the probability of an error is small
(otherwise we would not be able to correct for it), i.e.
\begin{equation}
\langle e |e\rangle = 1, \quad
\langle e_0 |e_0\rangle \simeq 1, \quad
\langle e_3 |e_3\rangle \simeq 1, \quad
\langle e_1 |e_1\rangle \ll 1, \quad
\langle e_2 |e_2\rangle \ll 1. 
\end{equation}

Equations \eqref{decohere} can be combined into one as 
\begin{equation}
|e\rangle \,|x\rangle \rightarrow  \left\{
\left(\, {|e_0\rangle + |e_3\rangle \over 2} \, \right)\, \mathbbm{1} +
\left(\, {|e_0\rangle - |e_3\rangle \over 2} \, \right)\, Z +
\left(\, {|e_2\rangle + |e_1\rangle \over 2} \, \right)\, X +
\left(\, {|e_2\rangle - |e_1\rangle \over 2} \, \right)\, (iY)
\right\} \, |x\rangle ,
\label{IXYZ}
\end{equation}
where $x=0$ or $1$ and, as usual,\footnote{I prefer to write equations like
\eqref{IXYZ} in terms of $iY (=ZX)$ rather
than $Y$ to avoid having explicitly complex
elements in the matrices. Many texts on quantum computing write $ZX$ rather than $iY$. Note
that $iY \, (=ZX)$ is not Hermitian (though $Y$ is) but we do not need the
Hermitian property here. What we do need is that, $iY$, like $X, Y$ and $Z$, is
unitary.}
\begin{equation}
Z = 
\begin{pmatrix}
1 & 0 \\
0 & -1
\end{pmatrix}
, \quad
X = 
\begin{pmatrix}
0 & 1 \\
1 & 0
\end{pmatrix}
, \quad
i Y = Z X = 
\begin{pmatrix}
0 & 1 \\
-1 & 0
\end{pmatrix}
, \quad
\mathbbm{1} =
\begin{pmatrix}
1 & 0 \\
0 & 1
\end{pmatrix}
.
\label{decohere2}
\end{equation}
Please evaluate Eq.~\eqref{IXYZ} separately for $x=0$ and $1$ 
to verify that it is equivalent to Eqs.~\eqref{decohere}. There is nothing
special about these environment states so we can write
\begin{equation}
|e\rangle \,|x\rangle \rightarrow  \left(\,
|d\rangle \mathbbm{1} +
|a\rangle X +
|b\rangle (iY) +
|c\rangle Z
\, \right) |x\rangle .
\label{decohere3}
\end{equation}
Equation \eqref{decohere2} applies to both $x=0$ and $x=1$. Since time
evolution of the combined qubit-environment system follows quantum mechanics
and so is unitary and linear, it also applies to a linear superposition
\index{superposition}
$|\psi\rangle = \alpha |0\rangle + \beta |1\rangle$ so
\begin{equation}
|e\rangle \,|\psi\rangle \rightarrow  \left(\,
|d\rangle \mathbbm{1} +
|a\rangle X +
|b\rangle (iY) +
|c\rangle Z
\, \right) |\psi\rangle .
\label{decohere4}
\end{equation}

We see that the effects of the environment on the uncorrupted state
of a single qubit can be
expressed entirely in terms of the Pauli operators, $X, (iY)$ and $Z$. These are
characterized as follows:
\begin{itemize}
\item $X$ corresponds to a bit-flip error,
\item $Z$ corresponds to a phase-flip error, and
\item $iY (=ZX)$ corresponds to combined bit-flip and phase-flip errors.
\end{itemize}
Intuitively, the reason that the new state can be expressed in terms of the Pauli operators
and the identity, is that any $2 \times 2$ matrix can be written as a linear
combination of these operators, see Eq.~\eqref{intro_A}. 

We remind the reader that the environment states are not normalized, and so,
in the important case where 
the initial state is close to the final
state, we have
\begin{equation}
\langle a |a\rangle \ll 1, \quad
\langle b |b\rangle \ll 1, \quad
\langle c |c\rangle \ll 1,
\end{equation}
in Eq.~\eqref{decohere4},

We now extend this discussion to the situation where we have expanded a single
qubit into an $n$-qubit codeword which we write as $|\psi\rangle_n$. In this
course we just consider how to correct single-qubit errors, so we 
neglect the possibility that two or more of the qubits in the codeword are
corrupted. From Eq.~\eqref{decohere4},
we see that all single qubit errors are incorporated by
\begin{equation}
|e\rangle \,|\psi\rangle_n \rightarrow  \left(\,
|d\rangle \mathbbm{1} +
\sum_{k=1}^n |a_k\rangle X_k +
\sum_{k=1}^n |b_k\rangle (iY_k) +
\sum_{k=1}^n |c_k\rangle Z_k
\, \right) |\psi\rangle_n .
\label{decohere5}
\end{equation}

Based on Eq.~\eqref{decohere5}, single qubit quantum error correction 
involves the following steps:
\begin{itemize}
\item
Expand the logical qubit to an $n$-qubit codeword.
\item
Project the possibly corrupted state to \textit{one} of the $3n+1$ states (syndromes)
on the
right hand side of Eq.~\eqref{decohere5}, with information indicating
\textit{which} one.
\item
Correct, if necessary, the 1-qubit error by acting with the appropriate $X_k,
Y_k$ or $Z_k$.
\end{itemize}

\textbf{Please note the following important points:}
\begin{enumerate}
\item
The whole \textit{continuum} of errors can be represented by a
finite set of \textit{discrete} errors.  Errors emerge continuously from the
uncorrupted state by increasing from zero the size of the terms in
Eq.~\eqref{decohere5} involving $X_i, Y_i$ and $Z_i$, which are characterized by
$\langle a_i |a_i\rangle^{1/2},
\langle b_i |b_i\rangle^{1/2}$ and 
$\langle c_i |c_i\rangle^{1/2}$ respectively.
However, the projection is always to one of the $3n+1$
discrete states. If the amplitude of the error is small then, with high
probability, the projection will be to the uncorrupted state (which needs no
correction) but with small but non-zero probability the projection will be to one of the
$3n$ corrupted states (which do need correction). 
\item
An \textit{arbitrary} error on a single qubit will be corrected, not just
bit-flip ($X$), or phase-flip ($Z$), or combined bit- and phase-flip ($iY)$ errors
but also \textit{any combination of the errors on a single qubit}. For example, suppose that the $k$-th qubit has
been reinitialized to zero, i.e.
\begin{equation}
|0_k\rangle \to |0_k\rangle, \ |1_k\rangle \to |0_k\rangle. 
\end{equation}
The matrix which accomplishes this transformation is\footnote{The
reader will notice that the
transformation in Eqs.~\eqref{nu1}, which involves a \textit{linear
combination} of $X, iY$ and $Z$ on a single qubit, are not unitary. Now
the evolution of an isolated (closed) system \textit{is} unitary, However, qubits
are coupled to the environment. If we consider a
system coupled to the environment (called an open system), and subject the combined
system+environment to a
unitary transformation, and finally consider the behavior of just the system by
tracing out over the environment, the resulting transformation of the system is
not necessarily unitary~\cite{nielsen:00,rieffel:14}.
\label{fn6}}
\begin{equation}
\begin{pmatrix}
1 & 1 \\
0 & 0
\end{pmatrix}
\label{nu1}
\end{equation}
which can be written as 
\begin{equation}
{\mathbbm{1} + X_k + i Y_k + Z_k \over 2}  .
\end{equation}
This is an example of the result shown in Eq.~\eqref{decohere4} that a general error on a
single qubit can be expressed as a combination of a bit-flip ($X$) error, a
phase-flip ($Z$) error and a combined bit- and phase-flip ($iY)$ error. 
Hence the state of the codeword qubits and environment has been transformed as follows:
\begin{equation}
|e\rangle\, |\psi\rangle_n \to |e'\rangle\, |\psi'\rangle_n =  |e'\rangle \,
{1 \over 2}
\left(\mathbbm{1} + X_k + i Y_k + Z_k\right) |\psi\rangle_n .
\label{ec:psi'}
\end{equation}
The codeword qubits are now in a linear combination of four syndromes, corresponding to the four
terms in this equation. 
A general syndrome measuring circuit, such as the Shor 9-qubit
code discussed
in the next section, will detect these syndromes and obtain a \textit{unique} set
of values for the ancilla qubits for each of them. Hence, even for this
non-unitary error, measuring the ancillas
\index{non-unitary transformation}
will project on to one of the syndromes which can then be corrected if
necessary.

\index{entanglement}
\item
A full discussion of how the entanglement of qubits with the environment
generates errors and how they can subsequently
be corrected, requires a detailed treatment of
the density matrix, see Chapter \ref{ch:den_mat}.
This advanced material is discussed in Refs.~\cite{nielsen:00,rieffel:14} but
is beyond the scope of the present course.

\end{enumerate}

\section{Correcting Arbitrary Errors: the 9-qubit Shor code}
\label{shor}
\index{Shor's 9-qubit code}
In the section we discuss a code, due to Peter Shor~\cite{shor:95}, for correcting
arbitrary 1-qubit errors.
This code needs code words of nine qubits to represent one logical
qubit. It is not the most efficient code, there are others which use smaller
code words and so don't need as many physical qubits, but the structure of
Shor's code follows quite naturally from the discussion we have already given
of 1-qubit bit-flip, and 1-qubit phase-flip errors, so will discuss it here.  

Shor's algorithm includes both
bit-flip ($X$) and phase-flip ($Z$) codes, which
turns out to then automatically correct combined bit-flip, phase-flip
($iY$) errors. As discussed in the previous section, it then also corrects 
\textit{arbitrary} 1-qubit errors.

\begin{figure}[htb]
\begin{center}
\includegraphics[width=5.5cm]{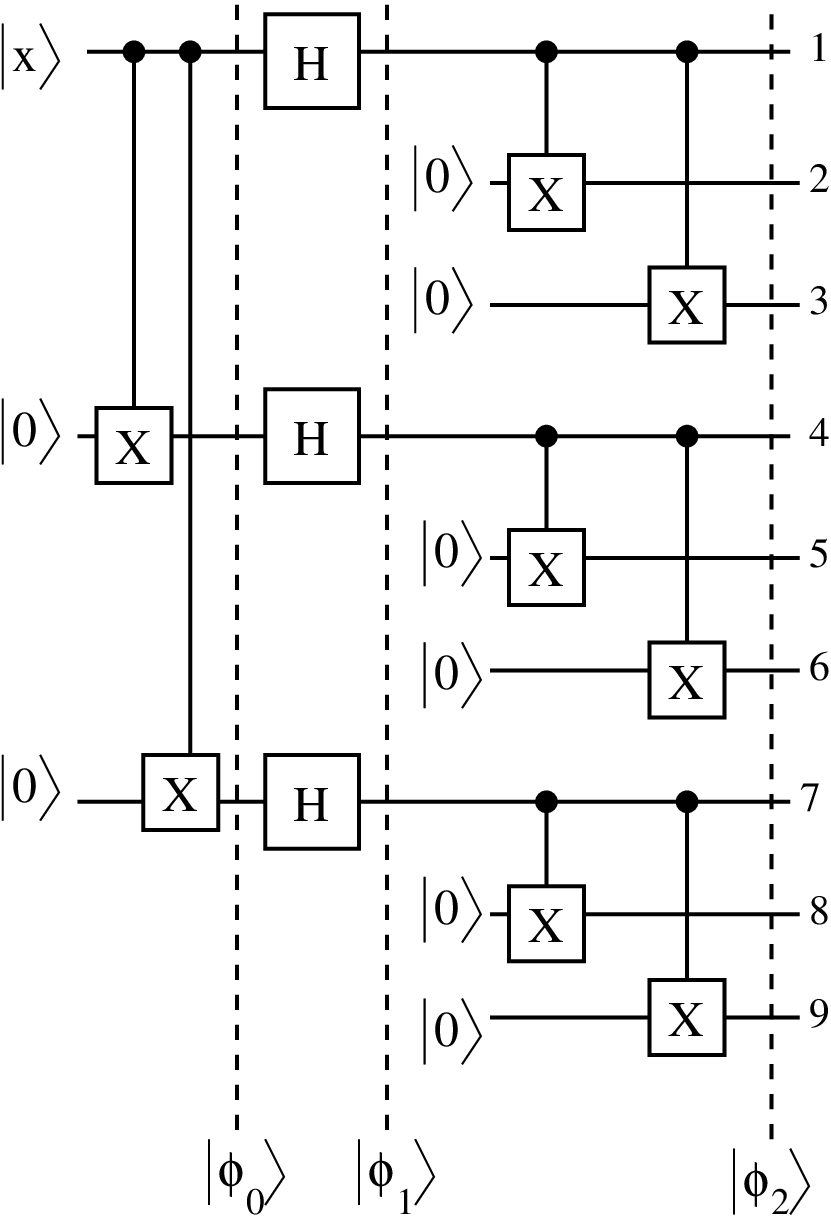}
\caption{
Encoding for the Shor 9-qubit code. If the initial state at the top left,
$|x\rangle$, is a computational basis state,
$|0 \rangle$ or $|1\rangle$, then
$|\phi_0\rangle = |xxx\rangle$ and $|\phi_1\rangle = 2^{-3/2}
(|0\rangle + (-1)^x|1\rangle)(|0\rangle + (-1)^x|1\rangle)(|0\rangle +
(-1)^x|1\rangle)$ since $H|x\rangle = 2^{-1/2}(|0\rangle + (-1)^x|1\rangle)$.
By comparison with Fig.~\ref{triple}, we see that $|\phi_2\rangle =
|\overline{x}\rangle$ given in Eq.~\eqref{xbar}.
Hence, if the initial state at the
top left is a linear combination $\alpha|0\rangle + \beta |1\rangle$ then, by
linearity, the final state at the right is $\alpha|\overline{0}\rangle
+ \beta |\overline{1}\rangle$. The numbers at the right are the labels of the
nine qubits. Note that this circuit 
is a concatenation of the encoding circuit for phase-flips
shown in Fig.~\ref{triple_psi2}, and that for bit-flips in
Fig.~\ref{triple}.
\label{shor_encode}
}
\end{center}
\end{figure}

We first encode for phase flip errors:
\begin{equation}
|0\rangle \rightarrow |+++ \rangle, \quad 
|1\rangle \rightarrow |--- \rangle,
\end{equation}
and then encode for bit-flip errors
\begin{equation}
|+\rangle  = {1\over \sqrt{2}}\left(\, |0\rangle + |1\rangle\,
\right) \rightarrow {1\over \sqrt{2}}\left(\, |000\rangle + |111\rangle\,
\right), \quad
|-\rangle  = {1\over \sqrt{2}}\left(\, |0\rangle - |1\rangle\,
\right) \rightarrow {1\over \sqrt{2}}\left(\, |000\rangle - |111\rangle\,
\right). 
\end{equation}
The final result is the 9-qubit encoding
\begin{subequations}
\begin{align}
|0\rangle &\rightarrow |\overline{0}\rangle = {1 \over2^{3/2}}
\left(\, |000\rangle + |111\rangle \, \right)\,
\left(\, |000\rangle + |111\rangle \, \right)\,
\left(\, |000\rangle + |111\rangle \right)\, ,\label{0bar} \\
|1\rangle &\rightarrow |\overline{1}\rangle = {1 \over2^{3/2}}
\left(\, |000\rangle - |111\rangle \, \right)\,
\left(\, |000\rangle - |111\rangle \, \right)\,
\left(\, |000\rangle - |111\rangle \right)\, . \label{1bar}
\end{align}
\label{shor_01bar}
\end{subequations}
These two equations can be combined as
\begin{equation}
|x\rangle \rightarrow |\overline{x}\rangle = {1 \over2^{3/2}}
\left(\, |000\rangle + (-1)^x |111\rangle \, \right)\,
\left(\, |000\rangle + (-1)^x |111\rangle \, \right)\,
\left(\, |000\rangle + (-1)^x |111\rangle \right)\, , \label{xbar}
\end{equation}
or more concisely as
\begin{equation}
|\overline{x}\rangle = {1 \over2^{3/2}}
\left(\, |000\rangle + (-1)^x |111\rangle \, \right)^{\otimes 3} .
\end{equation}
Such a code is called a \textit{concatenated} code. The circuit to achieve
this encoding is obtained by concatenating the phase flip and the bit flip
encodings as shown in Fig.~\ref{shor_encode}. 
Note the labeling of the qubits.
The qubits in each of the three blocks in Eq.~\eqref{shor_01bar}
have labels $123, 456$ and $789$.

\begin{figure}[htb!]
\begin{center}
\includegraphics[width=16cm]{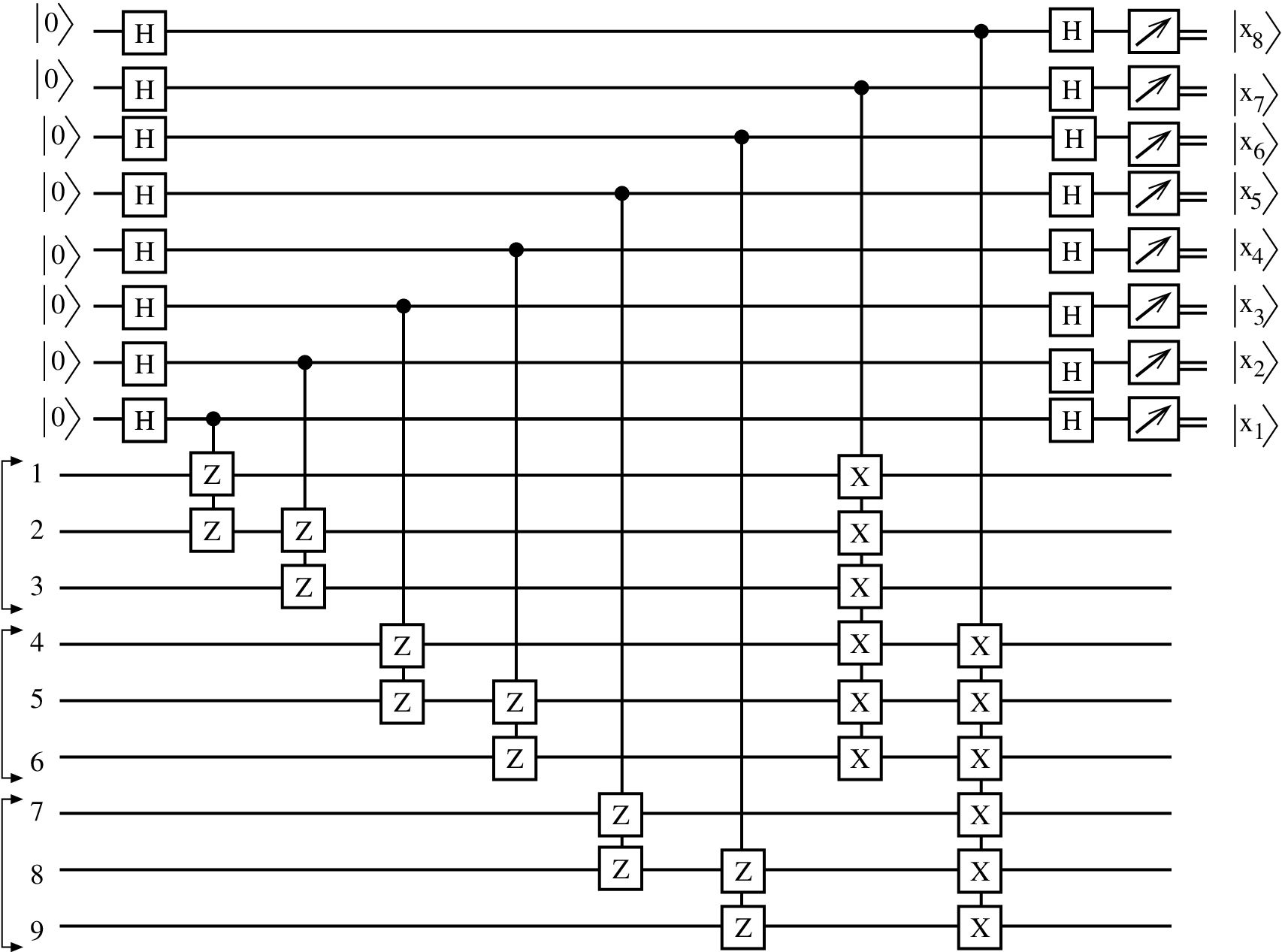}
\caption{
A circuit to measure the error syndrome for the Shor 9-qubit code.
The nine codeword qubits are at the bottom
and the eight ancillary qubits at the top. The ancillary qubits determine the
values of the eight, mutually commuting stabilizers in Eq.~\eqref{shor_stab},
$M_1 = Z_1 Z_2, M_2 = Z_2 Z_3, M_3 = Z_4 Z_5, M_4 = Z_5 Z_6, M_5 = Z_7 Z_8,
M_6 = Z_8 Z_9, M_7 = X_1 X_2 X_3 X_4 X_5 X_6$ and $M_8 = X_4 X_5 X_6 X_7
X_8 X_9$. The nine codeword qubits can be conveniently grouped into three
groups of three as indicated.
The measured value of the $i$-th ancilla $x_i \ (= 0 \ \textrm{or}\ 1)$, is
related to the 
value of the corresponding stabilizer
$M_i$ by $M_i = (-1)^{x_i}$. The measured values of the eight $x_i$
(or equivalently the $M_i$) determine which
syndrome in Eq.~\eqref{decohere6} has been projected out by the
measurement of the ancillas, as discussed in the text
and Table~\ref{table_shor}.
If one of the corrupted syndromes is found,
it can be corrected back to the uncorrupted state
by acting with the appropriate $X_i, Y_i$
or $Z_i$.
\label{shor_meas}
}
\end{center}
\end{figure}

The form of the 1-qubit corruption in Eq.~\eqref{decohere5} simplifies a
little here because if $|\psi\rangle$ is a linear combination of the codeword
states in Eq.~\eqref{shor_01bar} then
\begin{subequations}
\begin{align}
Z_1 |\psi\rangle = Z_2 |\psi\rangle =Z_3 |\psi\rangle, \\
Z_4 |\psi\rangle = Z_5 |\psi\rangle =Z_6 |\psi\rangle, \\
Z_7 |\psi\rangle = Z_8 |\psi\rangle =Z_9 |\psi\rangle.
\end{align}
\end{subequations}
The reason is that, for example,
changing the first of the $+$ signs in
Eq.~\eqref{0bar} into a $-$ sign, and the first $-$ sign in Eq.~\eqref{1bar}
into a $+$ sign, can be accomplished by acting with either $Z_1, Z_2$ or
$Z_3$.

Hence, the general form of a 1-qubit corruption contains only 22 independent
syndromes rather than $28 = (3\times 9) +1$:
\begin{equation}
|e\rangle \,|\psi\rangle \rightarrow  \left(\,
|d\rangle I +
|c\rangle Z_1 +
|c'\rangle Z_4 +
|c''\rangle Z_7 +
\sum_{i=1}^9 |a_i\rangle X_i +
\sum_{i=1}^9 |b_i\rangle iY_i 
\, \right) |\psi\rangle .
\label{decohere6}
\end{equation}

\index{stabilizers!for Shor's 9-qubit code}
The eight stabilizers which we use to diagnose the error are
\begin{align}
M_1 &= Z_1 Z_2, \quad M_2 = Z_2 Z_3, \quad M_3 = Z_4 Z_5, \quad
M_4 = Z_5 Z_6, \quad  M_5 = Z_7 Z_8, \quad M_6 = Z_8 Z_9, \nonumber \\
M_7 &= X_1 X_2 X_3 X_4 X_5 X_6, \quad  M_8 = X_4 X_5 X_6 X_7 X_8 X_9 \, .
\label{shor_stab}
\end{align}
Note that the nine qubits can conveniently be grouped into three blocks of
three, containing qubits $123, 456$ and $789$ respectively. $M_1$ and $M_2$ act entirely on
the first block, and do so in the same way as the stabilizers of the 3-qubit,
bit flip code shown in Fig.~\ref{flipped4}. Similarly $M_3$ and $M_4$ act on
the second block and $M_5$ and $M_6$ act on the third block. $M_7$ acts on all
qubits in blocks 1 and 2, while $M_7$ acts on all qubit in blocks 2 and 3.

The circuit for determining the syndrome eigenvalues is
shown in Fig.~\ref{shor_meas}.

We will now show that the $M_i$ have the desired properties:
\begin{itemize}
\item
They all square to unity (since each of the $Z$'s and $X$'s square to
unity and the $X$'s commute amongst each other as do the $Z$'s). Hence
their eigenvalues are $\pm 1$.
\item
They mutually commute. The six $Z$-stabilizers trivially commute with each other
as do the two $X$-stabilizers. Comparing the indices on the $Z$-stabilizers with
the $X$-stabilizers one sees that either they have none in common, in which case
this $X$-stabilizer and $Z$-stabilizer trivially commute, or they have two in
common, in which case there are two minus signs when one pulls one of the
stabilizers through the other, so the overall sign is positive and again the
$X$-stabilizer
and the $Z$-stabilizer commute).
\item
The eigenvalue of the uncorrupted codewords $|\overline{0}\rangle$ and
$|\overline{1}\rangle$ is $+1$ for all stabilizers.

This is trivially seen for
$M_1$--$M_6$ which involve pairs of $Z$ operators, since, for each pair, both
qubits are 0 or both are 1 in the codewords. Note that the pairs are entirely within the
blocks of three adjacent qubits in Eq.~\eqref{shor_01bar}, see
Fig.~\ref{shor_encode}.

Next consider $M_7$ and $M_8$ which involve a product of six $X$ operators, each
spanning two of the three blocks shown in Fig.~\ref{shor_encode}. For example,
$M_7$ is a product of the $X$ operators for the qubits in the first two
blocks. We have
\begin{align}
M_7 |\overline{0}\rangle = X_1 X_2 X_3 X_4 X_5 X_6
&{1 \over2^{3/2}}
\left(\, |000\rangle + |111\rangle \, \right)\,
\left(\, |000\rangle + |111\rangle \, \right)\,
\left(\, |000\rangle + |111\rangle \right) \nonumber \\
= &{1 \over2^{3/2}}
\left(\, |111\rangle + |000\rangle \, \right)\,
\left(\, |111\rangle + |000\rangle \, \right)\,
\left(\, |000\rangle + |111\rangle \right) \nonumber \\
=& {1 \over2^{3/2}}
\left(\, |000\rangle + |111\rangle \, \right)\,
\left(\, |000\rangle + |111\rangle \, \right)\,
\left(\, |000\rangle + |111\rangle \right) \nonumber \\
=& |\overline{0}\rangle,
\end{align}
and
\begin{align}
M_7 |\overline{1}\rangle = X_1 X_2 X_3 X_4 X_5 X_6
&{1 \over2^{3/2}}
\left(\, |000\rangle - |111\rangle \, \right)\,
\left(\, |000\rangle - |111\rangle \, \right)\,
\left(\, |000\rangle - |111\rangle \right) \nonumber \\
=& {1 \over2^{3/2}}
\left(\, |111\rangle - |000\rangle \, \right)\,
\left(\, |111\rangle - |000\rangle \, \right)\,
\left(\, |000\rangle - |111\rangle \right) \nonumber \\
=& {1 \over2^{3/2}}
\left(\, |000\rangle - |111\rangle \, \right)\,
\left(\, |000\rangle - |111\rangle \, \right)\,
\left(\, |000\rangle - |111\rangle \right) \nonumber \\
=& |\overline{1}\rangle ,
\end{align}
so $M_7$ has eigenvalue $+1$ for both uncorrupted codewords. The argument for
$M_8$ goes along the same lines.

\begin{table}[tbh]
\begin{center}
\begin{tabular}{|c| l l l l l l l l|}
\hline
Syndrome      & $M_1$ & $M_2$ & $M_3$ & $M_4$ & $M_5$ & $M_6$ & $M_7$ & $M_8$ \\
\hline\hline
$\mathbbm{1}$ & $+$   & $+$   & $+$   & $+$   & $+$   & $+$   & $+$   & $+$  \\
\hline
$X_1$         & $-$   & $+$   & $+$   & $+$   & $+$   & $+$   & $+$   & $+$  \\
$X_2$         & $-$   & $-$   & $+$   & $+$   & $+$   & $+$   & $+$   & $+$  \\
$X_3$         & $+$   & $-$   & $+$   & $+$   & $+$   & $+$   & $+$   & $+$  \\
$X_4$         & $+$   & $+$   & $-$   & $+$   & $+$   & $+$   & $+$   & $+$  \\
$X_5$         & $+$   & $+$   & $-$   & $-$   & $+$   & $+$   & $+$   & $+$  \\
$X_6$         & $+$   & $+$   & $+$   & $-$   & $+$   & $+$   & $+$   & $+$  \\
$X_7$         & $+$   & $+$   & $+$   & $+$   & $-$   & $+$   & $+$   & $+$  \\
$X_8$         & $+$   & $+$   & $+$   & $+$   & $-$   & $-$   & $+$   & $+$  \\
$X_9$         & $+$   & $+$   & $+$   & $+$   & $+$   & $-$   & $+$   & $+$  \\
\hline
$Y_1$         & $-$   & $+$   & $+$   & $+$   & $+$   & $+$   & $-$   & $+$  \\
$Y_2$         & $-$   & $-$   & $+$   & $+$   & $+$   & $+$   & $-$   & $+$  \\
$Y_3$         & $+$   & $-$   & $+$   & $+$   & $+$   & $+$   & $-$   & $+$  \\
$Y_4$         & $+$   & $+$   & $-$   & $+$   & $+$   & $+$   & $-$   & $-$  \\
$Y_5$         & $+$   & $+$   & $-$   & $-$   & $+$   & $+$   & $-$   & $-$  \\
$Y_6$         & $+$   & $+$   & $+$   & $-$   & $+$   & $+$   & $-$   & $-$  \\
$Y_7$         & $+$   & $+$   & $+$   & $+$   & $-$   & $+$   & $+$   & $-$  \\
$Y_8$         & $+$   & $+$   & $+$   & $+$   & $-$   & $-$   & $+$   & $-$  \\
$Y_9$         & $+$   & $+$   & $+$   & $+$   & $+$   & $-$   & $+$   & $-$  \\
\hline
$Z_1 \,(=Z_2 = Z_3)$  & $+$   & $+$   & $+$   & $+$   & $+$   & $+$   & $-$   & $+$  \\
$Z_4 \,(=Z_5 = Z_6)$  & $+$   & $+$   & $+$   & $+$   & $+$   & $+$   & $-$   & $-$  \\
$Z_7 \,(=Z_8 = Z_9)$  & $+$   & $+$   & $+$   & $+$   & $+$   & $+$   & $+$   & $-$  \\
\hline
\end{tabular}
\caption{The eigenvalues of the 8 stabilizers defined in Eq.~\eqref{shor_stab}
for the 22 syndromes of Shor's 9-qubit error correcting code. The left column
indicates which Pauli operator generates the syndrome from the uncorrupted
state. A $+$ sign
indicates eigenvalue $+1$ and a $-$ sign indicates eigenvalue $-1$. Each
stabilizer $M_i$ is measured by an ancilla qubit $x_i$, see
Fig.~\ref{shor_meas}, such that if $M_i = +1$ then $x_i = 0$ and if $M_i = -1$
then $x_i = 1$. An essential feature is that each of the 22 rows,
i.e.~syndromes, has a unique pattern of $+$ and $-$ signs.  Hence
the measured values of the $x_i$ indicate which syndrome has been projected
out by the measurement.
If this is one of the corrupted syndromes, the set of $x_i$ indicate which
Pauli operator generated the corruption, and the syndrome is then corrected by
applying the same Pauli operator. This works because the Pauli operators
square to the identity.
\label{table_shor}
}
\end{center}
\end{table}

\item
The $\pm 1$ eigenvalues of the stabilizers allow one to determine which of the
22 syndromes in Eq.~\eqref{decohere6} the system has projected on
to. Recalling the discussion in Sec.~\ref{sec:stabilizer}, the eigenvalue is $+1$ if the
stabilizer commutes with the operator which caused the 1-qubit corruption, and
is $-1$ if it anti-commutes. Each time two different operators on the same
qubit are pulled through each other to perform the commutation one generates a
minus sign. The operators which generate the corruption are the 21 $X_i, Y_i$ and
$Z_i$ in Eq.~\eqref{decohere6}.
A table of the eigenvalues of the stabilizers for all 22 syndromes is given in
Table~\ref{table_shor}.
\end{itemize}

Let's make sure that we understand how the syndrome-detection circuit in
Fig.~\ref{shor_meas} works. Firstly we remind the reader that if the measurement of an
auxiliary qubit, $x_i$ say, is $0$, then the value of the corresponding stabilizer
$M_i$ is $+1$, while if the measurement is $1$,
then the value of $M_i$ is $-1$. Thus we can say that $x_i$ measures $M_i$, see the discussion
of Fig.~\ref{ec:stabilizer} on page~\pageref{ec:stabilizer}. Next we discuss how
each of the stabilizers works.
\begin{itemize}
\item
We consider first $M_1$--$M_6$, the stabilizers involving $Z$ operators.\\
The ancilla qubits $x_1$ and $x_2$
measure $M_1 = Z_1 Z_2$ and $M_2 = Z_2 Z_3$ respectively, and so detect a bit-flip error in
the first group of three qubits in the 9-qubit encoding of
Eq.~\eqref{shor_01bar}, in exactly the same way as for the 3-qubit, bit-flip code shown in Fig.~\ref{flipped4}.
Similarly $x_3$ and $x_4$ detect a bit-flip error in the second group of three
qubits (qubits $4$--$6$), and $x_5$ and $x_6$ detect a bit-flip error in 
the third group of three qubits (qubits $7$--$9$). 
\item
Next we consider $M_7$ and $M_8$, the stabilizers involving $X$ operators.\\
The ancilla $x_7$ measures $M_7=X_1X_2X_3X_4X_5X_6$ and the ancilla $x_8$ measures
$M_8=X_4X_5X_6X_7X_8X_9$.
These detect phase flips. $M_7$ detects a phase flip in the first two groups of three qubits 
(qubits $1$--$6$) while $M_8$ detects a phase flip in the second and third groups of three qubits 
(qubits $4$--$9$).

\end{itemize}

We now
illustrate in more detail how Table~\ref{table_shor} was obtained by working through a few
cases. 
(Eigenvalues are taken to be $+1$ unless otherwise stated.)
\renewcommand{\labelenumi}{(\alph{enumi})}
\begin{enumerate}
\item  Syndrome $\boldsymbol{Z_2|\psi\rangle}$:
Clearly $Z_2$ commutes with all the $Z$-stabilizers. It anticommutes with $M_7$ (because
it has one qubit in common and $X$ and $Z$ anticommute)
and commutes with $M_8$ because it has no qubits
in common. Hence $M_7$ has eigenvalue $-1$ while all other stabilizers have
eigenvalue $+1$.
\item  Syndrome $\boldsymbol{Z_4|\psi\rangle}$:
Both $M_7$ and $M_8$ have eigenvalue $-1$ since they have one qubit in common
with $Z_4$ (and  $X$ and $Z$ anticommute).
\item  Syndrome $\boldsymbol{X_4|\psi\rangle}$:
Clearly $X_4$ commutes with both $X$-stabilizers. It anticommutes with $M_3$ because it
has one qubit in common (and $Z$ and $X$ anticommute). Hence $M_3$ has
eigenvalue $-1$.
\item  Syndrome $\boldsymbol{Y_5|\psi\rangle}$:
We note that $Y$ anticommutes with both $X$ and $Z$ so we have to consider all the
stabilizers. $Y_5$ has a qubit in common with $M_3, M_4, M_7$ and $ M_8$ so these
stabilizers have eigenvalue $-1$.

\end{enumerate}

Table \ref{table_shor} shows that each syndrome gives rise to a unique set of $+1$ and  $-1$
eigenvalues of the stabilizers as required.
Thus, measuring the eigenvalues of the eight stabilizers in Eq.~\eqref{shor_stab}
projects the corrupted state on to one of the
22 syndromes in Eq.~\eqref{decohere6}, and the set of eigenvalues determines which
one it is. One then applies an appropriate unitary
transformation to correct the state if necessary. Note that the Shor code is
\textit{explicitly}
designed to detect and correct bit-flip ($X$) and phase-flip ($Z$)
errors, but then \textit{automatically} detects and corrects combined bit-flip and
phase-flip ($ZX \equiv iY)$ errors.

Not only that, it also corrects \textit{arbitrary} errors on a single qubit,
which, as discussed in Sec.~\ref{discrete}, can be expressed as 
\textit{linear combinations} of bit-flip, phase-flip, and
combined bit- and phase-flip errors.  As an example consider the situation
mentioned in Eq.~\eqref{ec:psi'} in Sec.~\ref{discrete} in
which a qubit has been reset to $|0 \rangle$. This is an example of a
\textit{non-unitary}\footnote{In footnote \ref{fn6} we noted that,
\index{non-unitary transformation}
while a transformation of
the combined system+environment \textit{is} unitary, if the system is coupled
to the environment, then a unitary operation applied to system+environment followed by 
a trace over the
environment leaves the system in a new state which is not, in general, related by a
unitary transformation to its initial state.}
operation on the qubit.
Let's take it to be qubit 1
and indicate the codeword qubits by putting the first on the left, the last on the
right (we will use the same ordering below for the ancilla qubits). In other
words
\begin{equation}
|\psi\rangle = \alpha |\overline{0}\rangle + \beta |\overline{1}\rangle 
\end{equation}
has been transformed to
\begin{equation}
\begin{split}
|\psi'\rangle = 
& {\alpha \over2^{3/2}}
\left(\, |000\rangle + |011\rangle \, \right)\,
\left(\, |000\rangle + |111\rangle \, \right)\,
\left(\, |000\rangle + |111\rangle \right) + \\
& {\beta \over2^{3/2}}
\left(\, |000\rangle - |011\rangle \, \right)\,
\left(\, |000\rangle - |111\rangle \, \right)\,
\left(\, |000\rangle - |111\rangle \right) .
\end{split}
\label{reset}
\end{equation}
According to Eq.~\eqref{ec:psi'} this can be written as
\begin{equation}
|\psi'\rangle = 
{1 \over 2}
\left(\mathbbm{1} + X_1 + i Y_1 + Z_1\right) |\psi\rangle,
\label{psi'2}
\end{equation}
where
\begin{subequations}
\begin{align}
|\psi\rangle &=
\alpha\left(\, \ \ |000\rangle + |111\rangle\, \right)\, (\cdots)_+ \, (\cdots)_+ \, +
\beta \left(\, \ \ \, |000\rangle - |111\rangle\, \right)\, (\cdots)_- \, (\cdots)_- \, \\
X_1 |\psi\rangle &=
\alpha\left(\, \ \ |100\rangle + |011\rangle\, \right)\, (\cdots)_+ \, (\cdots)_+ \, +
\beta \left(\, \ \ \, |100\rangle - |011\rangle\, \right)\, (\cdots)_- \, (\cdots)_- \, \\
iY_1 |\psi\rangle &=
\alpha\left(\, -|100\rangle + |011\rangle\, \right)\, (\cdots)_+ \, (\cdots)_+ \, +
\beta \left(\, -|100\rangle - |011\rangle\, \right)\, (\cdots)_- \, (\cdots)_- \, \\
Z_1 |\psi\rangle &= 
\alpha\left(\, \ \ |000\rangle - |111\rangle\, \right)\, (\cdots)_+ \, (\cdots)_+ \, +
\beta \left(\, \ \ \, |000\rangle + |111\rangle\, \right)\, (\cdots)_- \, (\cdots)_- \, ,
\end{align}
\label{pieces}
\end{subequations}
in which
\begin{equation}
\begin{split}
(\cdots)_+ &\equiv \left(\, |000\rangle + |111\rangle \, \right) \\
(\cdots)_- &\equiv \left(\, |000\rangle - |111\rangle \, \right) .
\end{split}
\end{equation}
One can verify that adding Eqs.~\eqref{pieces} (and dividing by 2 according to
Eq.~\eqref{psi'2}) does indeed give Eq.~\eqref{reset}.

Equation \eqref{psi'2} is the input to the syndrome measurement circuit. According
to Table \ref{table_shor}, after the syndrome measurement circuit in Fig.~\ref{shor_encode}
has acted, the state of the system is
\begin{equation}
{1 \over 2} \left[\,\,
    |\psi\rangle\, |00000000\rangle_A + 
X_1|\psi\rangle\,\,  |10000000\rangle_A +
iY_1|\psi\rangle\,\,  |10000010\rangle_A +
Z_1|\psi\rangle\,\,   |00000010\rangle_A \,\, \right] ,
\end{equation}
where $|\cdots\rangle_A$ denotes the ancillas, which 
are ordered from 1 on the left to 8 on the right.
Measuring the ancillas gives one of the following results:
\begin{enumerate}[label=(\roman*)]
\item With probability $(1/2)^2 = 1/4$ the ancillas are measured to be
$|00000000\rangle_A$ and the computational qubits are in the uncorrupted state
$|\psi\rangle$. No correction is needed.
\item With probability $1/4$ the ancillas are measured to be
$|10000000\rangle_A$ and the computational qubits are in the corrupted state $X_1|\psi\rangle$. 
The error is corrected by acting with $X$ on qubit 1.
\item With probability $1/4$ the ancillas are measured to be
$|10000010\rangle_A$ and the computational qubits are in the corrupted state $i Y_1|\psi\rangle$. 
The error is corrected by acting with $Y$ on qubit 1.
\item With probability $1/4$ the ancillas are measured to be
$|00000010\rangle_A$ and the computational qubits are in the corrupted state $Z_1|\psi\rangle$. 
The error is corrected by acting with $Z$ on qubit 1.
\end{enumerate}


Thus, Shor's 9-qubit code, and other codes designed to correct both bit-flip
and phase-flip errors,
actually correct \textit{arbitrary} 1-qubit errors.  I
find this amazing.

\section{Other error-correcting codes}
The Shor code uses nine physical qubits to encode one logical qubit. What is
the minimum number of physical qubits needed to correct all 1-qubit errors? If
we encode using $n$ qubits the dimension of the space of states is $2^n$.  Now
the uncorrupted syndrome is a linear combination of $|\overline{0}\rangle$ and
$|\overline{1}\rangle$, i.e.~two basis states. Similarly each of the corrupted
syndromes is a linear combination of two basis states.
Hence $2^n$ must be sufficient to contain $3n+1$ mutually orthogonal 
2-d subspaces for the syndromes (the
1 is for the uncorrupted state and there are $n$ possible corruptions with
each of the $X, iY$ or $Z$ operators). Hence we need
\begin{equation}
2^n \ge 2 (3 n + 1) \, ,
\end{equation}
so the smallest value is $n=5$ which satisfies this condition as an equality. 

\index{5-qubit code}
There \textit{is} a 5-qubit code, but it turns out to be difficult to
construct the necessary gates.  A more popular choice is a 7-qubit code due to
Steane~\cite{steane:96}. The Shor code, which has 9-qubit codewords,
is now mainly of pedagogical interest.

\subsection{The 5-qubit code}
We now state, without much discussion, the codewords and stabilizers for the
5-qubit code. Further details are in Mermin~\cite{mermin:07}.

\begin{figure}[htb]
\begin{center}
\includegraphics[width=10cm]{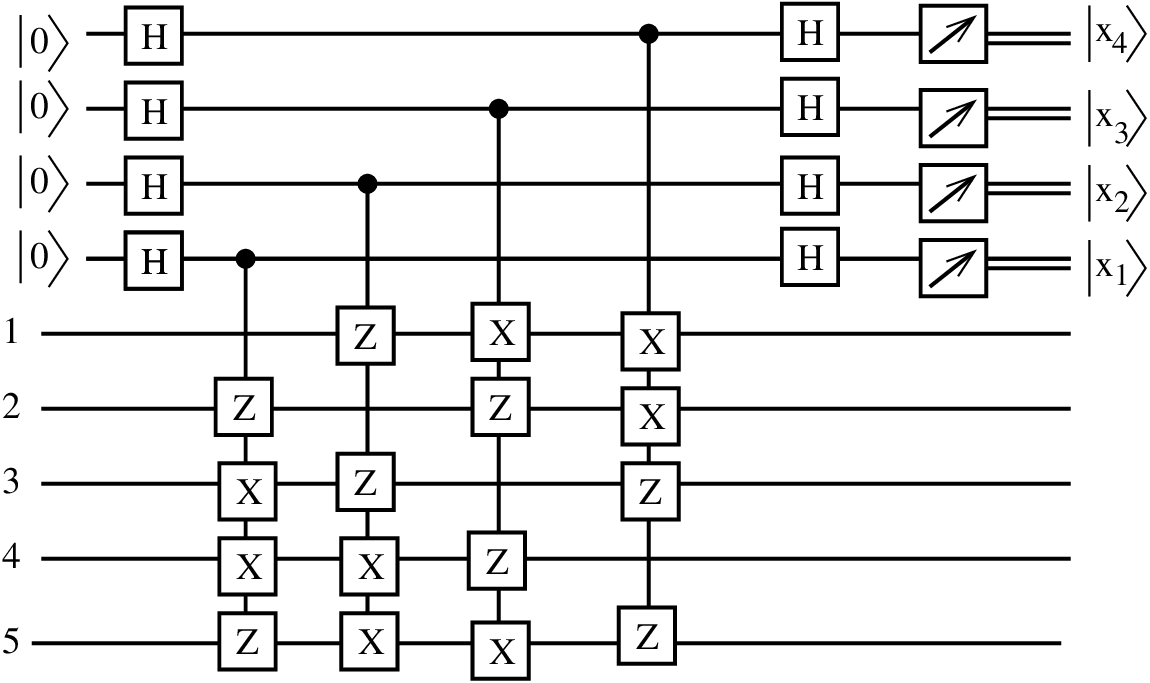}
\caption{
A circuit to measure the error syndrome for the 5-qubit code.
The five codeword qubits are at the bottom
and the four ancillary qubits at the top. The ancillary qubits determine the
values of the four, mutually commuting stabilizers in Eq.~\eqref{five_qubit_stab},
$M_1 = Z_2 X_3 X_4 Z_5 , 
M_2 = Z_3 X_4 X_5 Z_1 , 
M_3 = Z_4 X_5 X_1 Z_2 , 
M_4 = Z_5 X_1 X_2 Z_3 .$
\label{five_qubit_meas}
}
\end{center}
\end{figure}

For the 5-qubit code we have $(3\times 5) + 1 = 16$ mutually orthogonal,
two-dimensional subspaces, i.e.~16 syndromes. There are four stabilizers and, since they each
have two eigenvalues $(\pm 1)$, the number of distinct sets of eigenvalues
is $2^4 = 16$ which is just enough to distinguish the syndromes. These stabilizers are
\index{stabilizers!for 5-qubit code}
\begin{subequations}
\begin{align}
M_1 &= Z_2 X_3 X_4 Z_5 , \\
M_2 &= Z_3 X_4 X_5 Z_1 , \\
M_3 &= Z_4 X_5 X_1 Z_2 , \\
M_4 &= Z_5 X_1 X_2 Z_3 . 
\end{align}
\label{five_qubit_stab}
\end{subequations}
The circuit to measure the $M_i$ is shown in Fig.~\ref{five_qubit_meas}.

\begin{table}[tbh]
\begin{center}
\begin{tabular}{| c | c  | c | c | c |}
\hline
Syndrome & $M_1 = Z_2X_3X_4Z_5$ & $M_2 = Z_3X_4X_5Z_1$ & $M_3 = Z_4X_5X_1Z_2$ &
$M_4 = Z_5X_1X_2Z_3$ \\
\hline\hline
$ \mathbbm{1}$ &$+$&$+$&$+$&$+$\\
\hline
$X_1$ & $+$ & $-$ & $+$ & $+$ \\
$X_2$ & $-$ & $+$ & $-$ & $+$ \\
$X_3$ & $+$ & $-$ & $+$ & $-$ \\
$X_4$ & $+$ & $+$ & $-$ & $-$ \\
$X_5$ & $-$ & $+$ & $+$ & $-$ \\
\hline
$Y_1$ & $+$ & $-$ & $-$ & $-$ \\
$Y_2$ & $-$ & $+$ & $-$ & $-$ \\
$Y_3$ & $-$ & $-$ & $+$ & $-$ \\
$Y_4$ & $-$ & $-$ & $-$ & $+$ \\
$Y_5$ & $-$ & $-$ & $-$ & $-$ \\
\hline
$Z_1$ & $+$ & $+$ & $-$ & $-$ \\
$Z_2$ & $+$ & $+$ & $+$ & $-$ \\
$Z_3$ & $-$ & $+$ & $+$ & $+$ \\
$Z_4$ & $-$ & $-$ & $+$ & $+$ \\
$Z_5$ & $+$ & $-$ & $-$ & $+$ \\
\hline
\end{tabular}
\caption{The table shows
whether the four stabilizers $M_i$ for the 5-qubit error correcting code
commute ($+$) or anti-commute ($-$) with the 15
operators $X_i, Y_i$ and $Z_i,\, i = 1, 2, \cdots,5$ (which generate a
corruption of the codeword) as well as with the identity.
Each of the 16 rows has a unique pattern of $+$ and $-$ signs. A
$+$ sign corresponds to an eigenvalue $+1$ while a $-$ sign indicates an
eigenvalue $-1$. 
\label{table3}
}
\end{center}
\end{table}

The 5-qubit codewords are most conveniently expressed in terms of the $M_i$:
\begin{subequations}
\label{cw5qubit}
\begin{align}
|\overline{0}\rangle &= {1\over 4}(1 + M_1)(1 + M_2)(1 + M_3)(1 + M_4)|00000\rangle, \\
|\overline{1}\rangle &= {1\over 4}(1 + M_1)(1 + M_2)(1 + M_3)(1 + M_4)|11111\rangle. 
\end{align}
\end{subequations}
Note that $|\overline{0}\rangle$ is composed of the 16 basis states with an
even number of 1's, while $|\overline{1}\rangle$ is composed of the 16 basis
states with an odd number of 1's, so the two codewords are orthogonal. It is
not completely trivial to generate these codewords, see
Mermin~\cite{mermin:07} for details.

Furthermore the $M_i$ square to unity, are mutually commuting and each
has eigenvalue $+1$ for the uncorrupted codewords in Eq.~\eqref{cw5qubit}.
Each of them
commutes or anti-commutes with the $X_i, Y_i$ and $Z_i$ error operators, so
the 15 corrupted syndromes and the uncorrupted state are distinguished by the
set of $\pm 1$ eigenvalues of the $M$'s, as shown in Table \ref{table3}. 

\subsection{The Steane 7-qubit code}
\index{Steane's 7-qubit code}
Next I describe briefly the 7-qubit Steane code. 

\begin{figure}[tbh]
\begin{center}
\includegraphics[width=11cm]{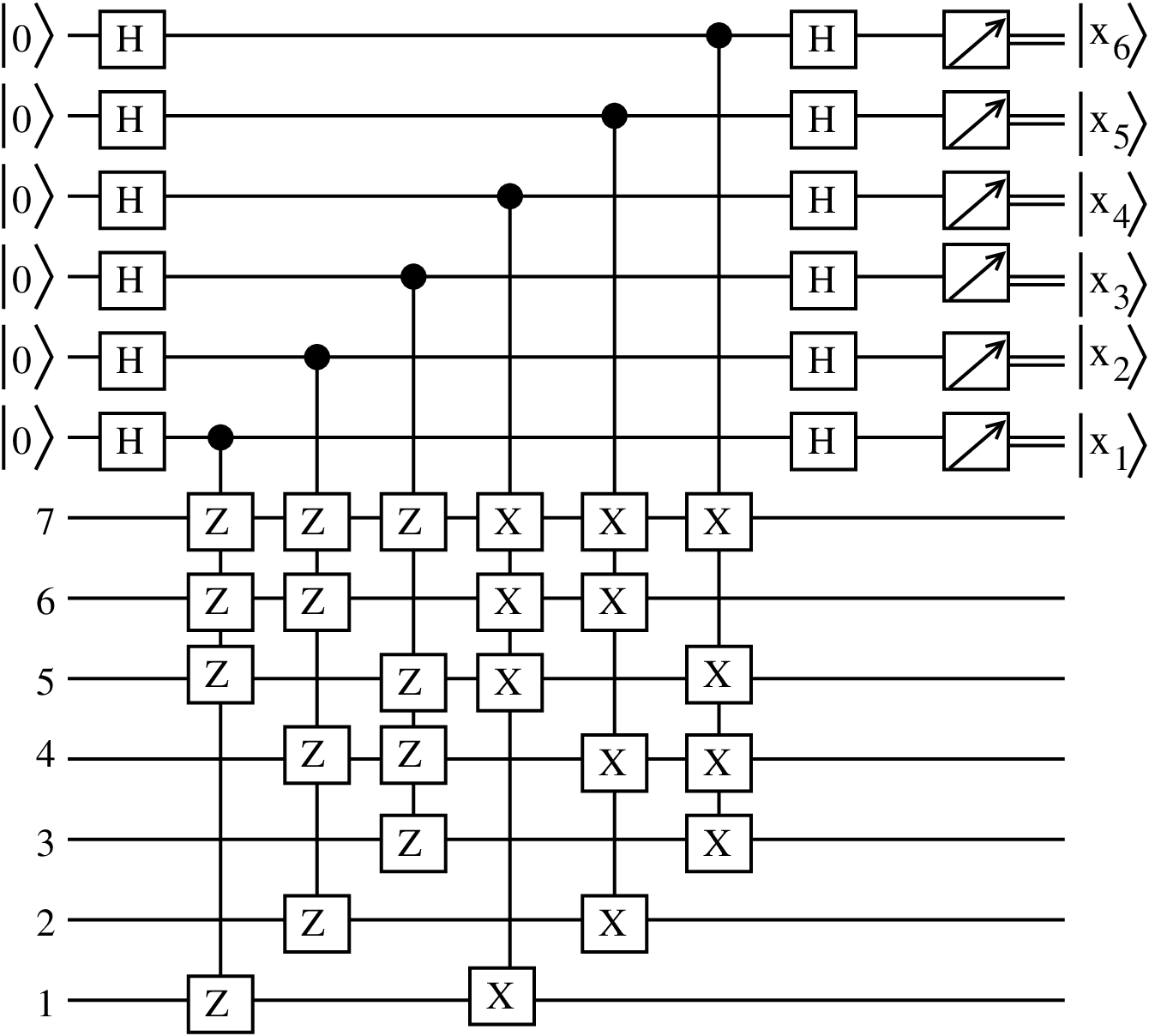}
\caption{The circuit of Steane's 7-qubit code to detect errors in the
computational qubits, (labeled 1--7 in the figure). There are also six ancilla
qubits (at the top) each of which is associated with one of the stabilizers as
follows: $N_1$-$N_3$ correspond to $x_1$-$x_3$ respectively, and  $M_1$-$M_3$ correspond to
$x_4$-$x_6$ respectively, in the usual way, e.g.~$N_1 = (-1)^{x_1}, M_1 =
(-1)^{x_4}$.
\label{steane}}
\end{center}
\end{figure}

There are 6 stabilizers
which are
\index{stabilizers!for Steane's 7-qubit code}
\begin{align}
M_1 &= X_1 X_5 X_6 X_7, \qquad N_1 = Z_1 Z_5 Z_6 Z_7, \nonumber \\
M_2 &= X_2 X_4 X_6 X_7, \qquad N_2 = Z_2 Z_4 Z_6 Z_7, \nonumber \\
M_3 &= X_3 X_4 X_5 X_7, \qquad N_3 = Z_3 Z_4 Z_5 Z_7. 
\end{align}
The circuit to detect errors is shown in Fig.~\ref{steane}.
The 7-qubit codewords are given by
\begin{align}
|\overline{0}\rangle &= {1 \over \sqrt{8}}(1 + M_1)(1 + M_2)(1 +
M_3)|0\rangle_7, \nonumber \\
|\overline{1}\rangle &= {1 \over \sqrt{8}}(1 + M_1)(1 + M_2)(1 +
M_3)\overline{X} |0\rangle_7, 
\label{steane_cw}
\end{align}
where
\begin{equation}
\overline{X} = X_1 X_2 X_3 X_4 X_5 X_6 X_7,
\label{oX}
\end{equation}
so
\begin{equation}
|1111111\rangle = \overline{X}|0000000\rangle .
\end{equation}

It is instructive for the student to show the following:
\begin{enumerate}
\item
The stabilizers mutually commute and square to the identity.
\item
The two states in Eq.~\eqref{steane_cw} are orthogonal.
\item
The two states in Eq.~\eqref{steane_cw} are normalized.\\
\textit{Hint:} You will need to use that the $M_i$ square to the identity, as
does
$\overline{X}$, and that $\overline{X}$ commutes with the $M_i$.
\item
The codewords $|\overline{0}\rangle$ and $|\overline{1}\rangle$ are
eigenstates of each of the stabilizers with eigenvalue $+1$.\\
\textit{Hint:} Note that $M_i(1+M_i) = 1 + M_i$ (why?), that the $N_j$
commute with $\overline{X}$ (explain why), and that $|0\rangle_7$ is an
eigenstate of the $N_i$ with eigenvalue $1$.
\end{enumerate}

\subsection{Surface Codes}
\index{surface codes}
A different approach to quantum error correction, but one that seems the most
promising, is to use ``surface codes" in which the physical qubits are arranged in a
square array and the values of the logical qubits are encoded in complicated
entangled states of the square array. Unfortunately, I have not been able to
find a simple introduction to this topic.

\section{Fault Tolerant Quantum Computing}
\index{fault tolerant quantum computing}
So far we have assumed that an error has occurred in some way and that we can
correct it by \textit{perfect} gates which do not introduce any further
errors.  This is, of course unreasonable since all aspects of quantum
computing can introduce errors: acting with gates, measurements, or simply
waiting.  Looking at the number of gates for Shor's 9-qubit syndrome-detection code in
Fig.~\ref{shor_meas} we might imagine that this circuit could introduce more
errors than it corrects. Of particular importance is that a circuit does not
spread an error initially in one qubit into multiple qubits which would then
be much harder to correct. A circuit which does not spread errors is said to
be ``fault tolerant".

An important result in quantum error correction is the ``threshold theorem"
which states that if the intrinsic error rate in an individual gate in a fault
tolerant circuit is less than a critical value $p_c$ then the overall error
rate in the circuit can be reduced to arbitrary low levels by quantum error
correction. This means that errors are being corrected faster than they are
being generated. However, since error correction requires duplication,
getting the error rate down to an acceptable
level will require that the number of
physical qubits is much greater than the number of logical qubits (those that
appear in the algorithm).

To see how one might reduce errors to an arbitrarily low level
suppose that the intrinsic error rate is $p$ and we have a fault tolerant
error correction scheme which corrects 1-qubit errors. 
This means that the error rate after error
correction is\footnote{The crucial point is that the new error rate is
proportional to the \textit{square} of the old error rate. I don't think it's obvious
that one can design a circuit with this property, but a detailed study
indicates that one can~\cite{nielsen:00,rieffel:14}. Unfortunately, I 
have not been able to find a simple explanation of this result.}
$c p^2$ for some constant $c$. If $p c < 1$ then we have
decreased the errors, so the threshold error rate is $p_c = 1/c$.

How can we go decrease the errors further?
Suppose the error correction procedure requires $n$ physical qubits for each logical
qubit, so, for example, $n=9$ for the Shor code and $n=7$ for the 7-qubit Steane code.
We can then take each of the $n$ qubits and error correct these with the
same code. This procedure is known as \textit{concatenation}. We then have
$n^2$ physical qubits and the error rate is $c (cp^2)^2 = c^{-1} (c p) ^{2^2}$.
Generalizing, if we concatenate
$l$ times, then the number of qubits is $n^l$ while the resulting error rate
is $c^{-1}(c p)^{2^l}$. Note that while the number of qubits increases
exponentially with the level of concatenation $l$, the error rate decreases
\textit{doubly} exponentially with $l$.  As an example, to get a feel for what
this means, consider the case $p=1/8, c= 2$, so $cp = 1/4$ and also suppose
that $n=7$ (corresponding to the Steane code). Then successive concatenations
give the numbers in Table \ref{tab:ft}.

\begin{table}[tbh]
\begin{center}
\begin{tabular}{|l|l|l|l|}
\hline\hline
no.~of concatenations ($l$)  & error rate (formula) & error rate (numeric) & no.~of qubits \\
\hline\hline
0 & $p$  & $1/2^3 =0.125$ & 1   \\
1 & $cp^2 = c^{-1}(cp)^2 $  & $1/2^5 = 0.03125$& $n \,\,\,(= 7)$ \\
2 & $c (cp^2)^2 = c^{-1}(cp)^{2^2} $  & $1/2^9 = 1.953 \times 10^{-3}$ &
$n^2 \,(=49)$ \\
3 & $c ( (cp^2)^2)^2 = c^{-1}(cp)^{2^3} $  &
$1/2^{17}= 7.629 \times 10^{-6}$ & $n^3 \,(=343)$ \\
4 & $c (c ( (cp^2)^2)^2)^2 = c^{-1}(cp)^{2^4} $  &
$1/2^{33}=1.164 \times 10^{-10}$ & $n^4 \,(=2401)$ \\
5 & $c(c (c ( (cp^2)^2)^2)^2)^2 = c^{-1}(cp)^{2^5} $  &
$1/2^{65} =2.711 \times 10^{-20}$ & $n^5 \,(=16807)$ \\
\hline\hline
\end{tabular}
\caption{Parameters for the concatenation of a fault tolerant circuit with an
(artificial) choice of parameters discussed in the text.
\label{tab:ft}
}
\end{center}
\end{table}

These numbers are not realistic. They correspond to a threshold value of $p_c
= 1/c = 1/2$ and any realistic circuit would have a much smaller value.
However, they do show, and this is the main point, that the error rate goes
down much faster than the number of physical qubits goes up. Of course, the
number of physical qubits per logical qubit will still have to be very large to get the
error rate down to an acceptable value for computation. 

Various calculations have estimated the threshold for 7-qubit Steane code at around
$10^{-5}$. To perform error correction one would need individual circuit
elements with an error rate significantly
less than this, which, to my knowledge, is not
\index{surface codes}
feasible at present. Surface codes, which were briefly mentioned above, are
estimated to have a higher threshold, of around $10^{-2}$, and it does seem
feasible to make gates with a lower error rate than this.  For example, at the
end of a very long and technical paper, Ref.~\cite{fowler:12} estimates that to
factor, using Shor's algorithm, an integer which is too large to be factored
on a classical computer (2000 bits), would require no less than
around $220 \times 10^6$
qubits with then state-of-the-art
superconducting qubits using quantum error correction with surface codes.
At present, quantum computers (using the ``gate" model of quantum computing
which is the topic of this course) have at most a few tens of qubits, so a
huge increase in scale will be required. However, who is to say that this
cannot happen in a few decades?  An example of a comparable increase in scale
which has \textit{already} happened is the number of transistors on a modern
chip compared with the number on early integrated circuits.

Thus, in my view, in the next
few years, we may see
quantum computers with a modest number of logical qubits which
perform error correction. However, quantum computers with error correction
\textit{having
enough logical qubits to outperform classical computers} for some \textit{useful}
problem such as integer factorization are for the distant future, if
ever.

\vspace{0.3cm}

\noindent\textit{Note.} This last remark may be too pessimistic; see the
comments at the end of Chapter \ref{ch:qu_s}.


\section{Summary of Quantum Error Correction}
\label{sec:summ_err_corr}
This chapter has been quite involved and it is easy to get lost in the
details.
I have therefore summarized the main ideas in this
section.

A logical qubit is represented by $n$ physical qubits. 
We consider codes that can correct errors in just one of those qubits.
The initial state is therefore assumed to be a superposition of the
uncorrupted state, with an amplitude close to 1, plus all possible single
qubit corruptions with small amplitude. Since each qubit can be corrupted with an $X, Y$ or $Z$
Pauli operator, there are usually $3 n$ corrupted states\footnote{The Shor
$9$-qubit code that we discussed in detail has fewer because some corruptions
give the same state.} and so there are usually $3n +1$ states in total in the superposition.
These are called syndromes.

Omitting to write the states of the environment for simplicity of notation,
the initial state is
\begin{equation}
|\psi\rangle \to \sum_{\alpha = 0}^{N_s -1} c_\alpha A_\alpha | \psi\rangle
\end{equation}
where $|\psi\rangle$ is the uncorrupted state, $\alpha=0$ represents the uncorrupted state
so $A_0 = I$ (the identity), the other
$A_\alpha$ are Pauli operators $X_i, Y_i$ or $Z_i (i = 1, 2, n)$, $N_s$ is the
number of syndromes (usually $N_s = 3n+1$), $c_0$ is the amplitude of the
uncorrupted state which is close to $1$ in magnitude, and the other $c_\alpha$
are much less than $1$ in magnitude.

In addition we have $m$ ancilla qubits. We denote a state of the ancillas by
$|x\rangle_A$ where $x$ is an $m$-bit integer whose binary representation is the state
of the ancilla qubits. Initially, the state of the ancillas
is
$|0\rangle_A$.

The error detection circuit entangles the $n$ codeword qubits with the $m$
ancilla qubits, so the final state of the combined codeword-ancilla system,
after the error detection circuit has acted, is
\begin{equation}
\sum_{\alpha = 0}^{N_s -1}  c_\alpha A_\alpha | \psi\rangle \otimes
|x_\alpha\rangle_A ,
\end{equation}
where each syndrome is associated with a \textit{distinct} state of the ancillas,
represented by the integer $x_\alpha$, with the unperturbed syndrome having
$x_0 = 0$.

A measurement is then made of the ancillas, whose state after the measurement is represented by the $m$-bit integer
$x_{\tilde{\alpha}}$ corresponding to one of the syndromes $\tilde{\alpha}$. The
codeword has then been projected on to the $\tilde{\alpha}$ syndrome,
i.e.~$A_{\tilde{\alpha}}|\psi\rangle$. From the measured $x_{\tilde{\alpha}}$ 
we know $\tilde{\alpha}$ (since each $x_\alpha$ specifies a unique syndrome
$\alpha$), and hence, if $\tilde{\alpha} \ne 0$ so there is an error, we can correct that
error by acting on the codeword qubits with\footnote{Recall that the $A_\alpha$ are Pauli operators
which square to unity.}
$A_{\tilde{\alpha}}$.
The codeword qubits are then in the uncorrupted state
$|\psi\rangle$, as required.

I thank Eleanor Rieffel for a helpful email exchange on quantum error
correction.

\hrulefill
\section*{Problems}
\input{hw_ch19.tex}




%% file: hw_ch19.tex
\begin{problems}

\item
\label{qu3qb}
Consider the 3-qubit, bit-flip code discussed in class, and in the lecture material. The circuit
is shown in Fig.~\ref{bitflip7}.
We
commented that this circuit works in the situation where a bit-flip error
builds up continuously from zero. Let us verify this. Consider  the corrupted state
\begin{equation}
|\psi' \rangle = \left[(1 -\epsilon^2/2)\mathbbm{1} + i \left(\, \epsilon_1 X_1 +
\epsilon_2 X_2 + \epsilon_3 X_3\, \right)\right]|\psi\rangle,
\label{psi'7}
\end{equation}
where $\epsilon_k \ll 1$ and $\epsilon^2 = \sum_{k=1}^3 \epsilon_k^2$ and
\begin{equation}
|\psi\rangle = \alpha|000\rangle + \beta|111\rangle
\end{equation}
is the uncorrupted state.
We will
work to first order in $\epsilon$ (the factor of $1-\epsilon^2/2$ is inserted
so that the normalization constant is 1 through order $\epsilon^2$).
$|\psi'\rangle$ is the initial state (on
the left) of the three computational qubits, labeled 1, 2 and 3, in
Fig.~\ref{bitflip7}. 

Determine the state of the
system (computational qubits plus ancillas) after the error detection circuit
has operated. 

Then consider the correction phase.
What are the possible results of the measurements of the
ancillas, what are the probabilities of these results, and what is the
resulting state of the computational qubits? \\
(You should conclude that the bit-flip error has been corrected for all
possible results of the measurement of the ancillas.)

\begin{figure}[tbh]
\begin{center}
\includegraphics[width=11cm]{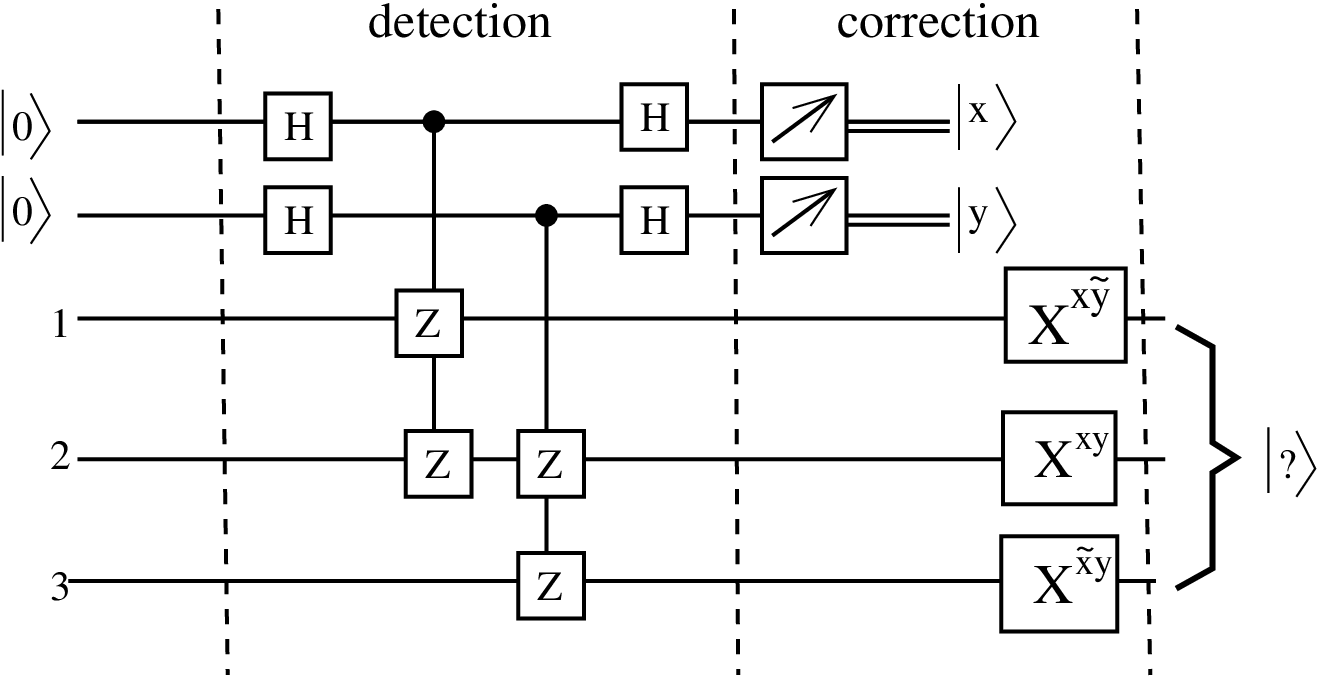}
\caption{Circuit for syndrome detection for the 3-qubit bit-flip code, and
for correction if necessary.
\label{bitflip7}}
\end{center}
\end{figure}

\item
\label{qu3}
In question \ref{qu3qb} we implicitly assumed that the time dependence of the computational qubits has
proceeded in a unitary manner including the point where the error has developed.
In other words the error is in the 
circuit itself. However, a very common cause of errors in a quantum computer
is that the qubits have an unwanted interaction with the environment.  The
environment becomes entangled with the qubits leading to ``decoherence", which
is the main difficulty in building a useful quantum computer.

Let us apply the same 3-qubit, bit-flip code shown in Fig.~\ref{bitflip7} to an
error model where the error comes from the prior interaction of the qubits
with the environment.


A system interacting with environment is not in a single quantum state but can
be represented as being 
in different quantum states with various probabilities\footnote{The correct
way to describe this is with the density matrix discussed in Chapter 5, but we
will not need the details of the density matrix here.}.
Let us suppose, then, that the system is described as follows (in which we
again only allow for single bit-flips):
\begin{align}
\mathrm{Probability:} P_0, \quad |\psi'\rangle &=
\alpha|000\rangle + \beta|111\rangle = |\psi\rangle \nonumber \\
\mathrm{Probability:} P_1, \quad |\psi'\rangle &=
\alpha|100\rangle + \beta|011\rangle \nonumber = X_1 |\psi\rangle \\
\mathrm{Probability:} P_2, \quad |\psi'\rangle &=
\alpha|010\rangle + \beta|101\rangle = X_2 |\psi\rangle \nonumber \\
\mathrm{Probability:} P_3, \quad |\psi'\rangle &=
\alpha|001\rangle + \beta|110\rangle = X_3 |\psi\rangle ,
\end{align}
where, of course, $\sum_{i=0}^3 P_i = 1$.  Note that these states are 
\textit{incoherent} in the sense that there is no interference between the
different states.  This is different from Eq.~\eqref{psi'7} where the different
pieces of the wave function have well defined relative phases (i.e.~the
superposition is \textit{coherent}) and so can 
potentially interfere.

Describe the result of acting with the ``detection" part of the circuit. 

Then
consider the ``correction" part and derive the possible results of the
measurements of the ancillas and their probabilities. Show that, like the case
of the coherent bit-flip error of Eq.~\eqref{psi'7} in Qu.~\ref{qu3qb}, the circuit succeeds in
correcting the error. 

\textit{Note:} The difference between questions \ref{qu3qb} and \ref{qu3} is
that in the former the corruption is due to a \emph{coherent} superposition of
1-qubit corrupted states, while in the latter it is due to an
\emph{incoherent} sum of 1-qubit corrupted states with various probabilities.
To answer Qu.~\ref{qu3} you have to discuss, \emph{for each of the states in the
incoherent sum},  what is the state of the ancillas and how the error
correction is done. \\ By doing both these two questions you see that error
correction works irrespective of whether the error is due to a coherent
addition of corrupted states (perhaps due to the gates not functioning
correctly) or to an incoherent addition of corrupted states due to the
computational qubits becoming entangled with the environment.

\item  \textit{Shor's 9-qubit code}

\begin{enumerate}[label=(\roman*)]
\item
We mentioned in class that it is necessary that the (uncorrupted) codewords
are eigenvectors
of all the stabilizers with eigenvalue $+1$. Show that this is the case for
stabilizers $M_1$ and $M_8$ of Shor's 9-qubit code.
\item
We all discussed in detail the table of $\pm 1$ eigenvalues for the
stabilizers acting on the 22 syndromes. Here is an extract from that table,
for the syndrome where there is a 1-qubit corruption due to $Y_4$ (i.e.~a
combined phase-flip and bit-flip acting on qubit 4).

\begin{center}
\begin{tabular}{|c| l l l l l l l l|}
\hline
Syndrome      & $M_1$ & $M_2$ & $M_3$ & $M_4$ & $M_5$ & $M_6$ & $M_7$ & $M_8$
\\
\hline\hline
$Y_4$         & $+$   & $+$   & $-$   & $+$   & $+$   & $+$   & $-$   & $-$ \\
\hline
\end{tabular} ,
\end{center}
\noindent Here ``$+$" means eigenvalue $+1$ and ``$-$" means eigenvalue $-1$.

Explain the sign of each of these $\pm 1$ eigenvalues.

\end{enumerate}
\item
\label{qu1}
As discussed in class, the four stabilizers for the 5-qubit
error correcting code are
\begin{subequations}
\begin{align}
M_1 &= Z_2 X_3 X_4 Z_5 , \\
M_2 &= Z_3 X_4 X_5 Z_1 , \\
M_3 &= Z_4 X_5 X_1 Z_2 , \\
M_4 &= Z_5 X_1 X_2 Z_3 . 
\end{align}
\end{subequations}
We also stated that the pattern of $+1$ and $-1$ eigenvalues for the
stabilizers among the 16 syndromes (1 uncorrupted and $3\times 5 = 15$
corrupted) are given by

\begin{center}
\begin{tabular}{|l|l|l|l|l|l|l|}
\hline
          & $X_1Y_1Z_1$ & $X_2Y_2Z_2$ & $X_3Y_3Z_3$ & $X_4Y_4Z_4$ &
$X_5Y_5Z_5$ & $\mathbbm{1}$ \\
\hline\hline
$M_1 = Z_2X_3X_4Z_5$ &$+++$  &$--+$  & $+--$  & $+--$ & $--+$ & $+$ \\
$M_2 = Z_3X_4X_5Z_1$ &$--+$  &$+++$  & $--+$  & $+--$ & $+--$ & $+$ \\
$M_3 = Z_4X_5X_1Z_2$ &$+--$  &$--+$  & $+++$  & $--+$ & $+--$ & $+$ \\
$M_4 = Z_5X_1X_2Z_3$ &$+--$  &$+--$  & $--+$  & $+++$ & $--+$ & $+$ \\
\hline
\end{tabular},
\end{center}

where the top row indicates which Pauli operator is used to generate the
corrupted state from the uncorrupted state.

\begin{enumerate}[label=(\roman*)]
\item Show that the stabilizers square to the identity.
\item Show that they are mutually commuting.
\item By considering the nature of the commutation of the stabilizer with the
relevant Pauli operator, explain the results in the table for the columns
$X_3, Y_4$ and $Z_5$.\\
\textit{Note:} You may assume without proof that the right-hand column is
correct, i.e.~the eigenvalues of all the stabilizers are $+1$ for the
uncorrupted state.
\end{enumerate}

\item
Using the expressions for the stabilizers of the 5-qubit code given in Qu.~\ref{qu1}, draw the 
circuit to detect 1-qubit errors in the 5-qubit code. 

\item \textit{(More challenging)}\\
Consider the 7-qubit Steane code. There are 6 stabilizers
which are 
\begin{align}
M_1 &= X_1 X_5 X_6 X_7, \qquad N_1 = Z_1 Z_5 Z_6 Z_7, \nonumber \\
M_2 &= X_2 X_4 X_6 X_7, \qquad N_2 = Z_2 Z_4 Z_6 Z_7, \nonumber \\
M_3 &= X_3 X_4 X_5 X_7, \qquad N_3 = Z_3 Z_4 Z_5 Z_7. 
\end{align}
The circuit to detect errors is shown in Fig.~\ref{steane2}.
The 7-qubit codewords are given by
\begin{align}
|\overline{0}\rangle &= {1 \over \sqrt{8}}(1 + M_1)(1 + M_2)(1 +
M_3)|0\rangle_7, \nonumber \\
|\overline{1}\rangle &= {1 \over \sqrt{8}}(1 + M_1)(1 + M_2)(1 +
M_3)\overline{X} |0\rangle_7, 
\label{steane_cw2}
\end{align}
where ``1" refers to the identity operator,
\begin{equation}
\overline{X} = X_1 X_2 X_3 X_4 X_5 X_6 X_7,
\label{oX2}
\end{equation}
so
\begin{equation}
\overline{X}|0000000\rangle .
=
|1111111\rangle
\end{equation}

\begin{figure}[tbh]
\begin{center}
\includegraphics[width=11cm]{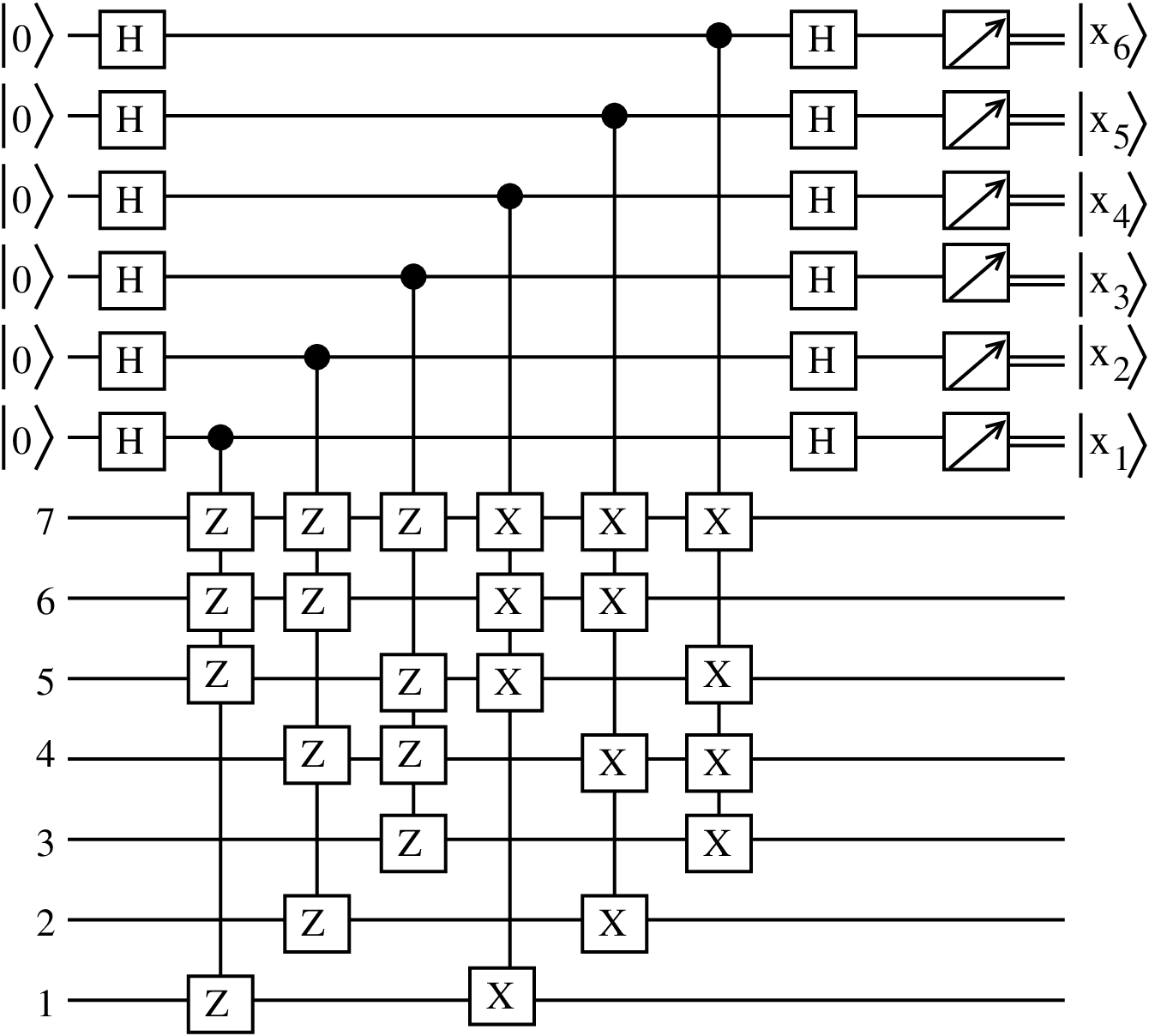}
\caption{The circuit of Steane's 7-qubit code to detect errors in the
computational qubits, (labeled 1--7 in the figure). There are also six ancilla
qubits (at the top) each of which is associated with one of the stabilizers as
follows: $N_1$-$N_3$ correspond to $x_1$-$x_3$, and  $M_1$-$M_3$ correspond to
$x_4$-$x_6$, in the usual way, e.g. $M_1 = (-1)^{x_1}$.
\label{steane2}}
\end{center}
\end{figure}

\begin{enumerate}[label=(\roman*)]
\item 
Show that the stabilizers mutually commute and square to the identity. 
\item
Show that the two states in Eq.~\eqref{steane_cw2} are orthogonal.
\item
Show that the two states in Eq.~\eqref{steane_cw2} are normalized.\\
\textit{Hint:} You will need to use that the $M_i$ square to the identity, as does 
$\overline{X}$, and that $\overline{X}$ commutes with the $M_i$.
\item
Show that the codewords $|\overline{0}\rangle$ and $|\overline{1}\rangle$ are
eigenstates of each of the stabilizers with eigenvalue $+1$.\\
\textit{Hint:} Note that $M_i(1+M_i) = 1 + M_i$ (why?), that the $N_j$
commute with $\overline{X}$ (explain why), and that $|0\rangle_7$ is an
eigenstate of the $N_i$ with eigenvalue $1$.
\end{enumerate}

\item \textit {(More challenging)}\\
Consider operators which act equally on all qubits in the 7 qubit code:
\begin{equation}
\overline{Z} = Z_1 Z_2 Z_3 Z_4 Z_5 Z_6 Z_7,\qquad
\overline{H} = H_1 H_2 H_3 H_4 H_5 H_6 H_7,
\end{equation}
and similarly $\overline{X}$ defined in Eq.~\eqref{oX2}.
\begin{enumerate}[label=(\roman*)]
\item Show that $\overline{X}$ implements the logical NOT gate (i.e.~logical
$X$) on the codewords, i.e.
\begin{equation}
\overline{X}|\overline{0}\rangle = |\overline{1}\rangle, \quad
\overline{X}|\overline{1}\rangle = |\overline{0}\rangle.
\end{equation}
\item Show that $\overline{Z}$ implements the logical 
$Z$ on the codewords, i.e.
\begin{equation}
\overline{Z}|\overline{0}\rangle = |\overline{0}\rangle, \quad
\overline{Z}|\overline{1}\rangle = -|\overline{1}\rangle.
\end{equation}
\item (Harder) Show that $\overline{H}$ implements the logical 
$H$ on the codewords, i.e.
\begin{equation}
\overline{H}|\overline{0}\rangle = {1\over \sqrt{2}} \left(\,
|\overline{0}\rangle + |\overline{1}\rangle \, \right), \quad
\overline{H}|\overline{1}\rangle = {1 \over\sqrt{2}} \left(\,
|\overline{0}\rangle - |\overline{1}\rangle\, \right) .
\end{equation}
\textit{Hints:}
\begin{itemize}
\item
We want to show that
\begin{equation}
\langle \overline{0}| \overline{H}|\overline{0}\rangle =
\langle \overline{1}| \overline{H}|\overline{0}\rangle =
\langle \overline{0}| \overline{H}|\overline{1}\rangle ={1 \over \sqrt{2}},
\quad
\langle \overline{1}| \overline{H}|\overline{1}\rangle = - {1 \over \sqrt{2}}.
\end{equation}
\item
Hence we need to calculate
\begin{multline}
\langle \overline{x}| \overline{H}|\overline{y}\rangle ={1 \over 8}\,
{}_7\langle 0|\overline{X}^x (1+M_1)(1+M_2)(1+M_3) \overline{H}(1 + M_1) \\
\times (1 + M_2)(1+M_3) \overline{X}^y |0\rangle_7 .
\label{xhbary}
\end{multline}
\item
Derive the results 
\begin{equation}
\overline{H} M_i = N_i \overline{H}, \quad M_i \overline{H} = \overline{H}
N_i,
\end{equation}
and use them to show that you can replace the $M_i$ in Eq.~\eqref{xhbary} by $N_i$.
\item
Show that each $N_i$ commutes with $\overline{X}$ and apply this result.
\item
Use that each $N_i$ acts as the identity on $|0\rangle_7$. 
\end{itemize}

\end{enumerate}
\textit{Note:} Having codeword gates that are tensor products of single qubit
gates is very helpful when designing circuits to
implement an error correcting code. A similar result also holds
for CNOT. In Steane's code the logical CNOT gate that takes $|\overline{x}\rangle
|\overline{y}\rangle $ to $|\overline{x}\rangle |\overline{x\oplus y}\rangle$,
is simply made up of CNOT gates applied to each of the seven pairs of qubits
in the two codewords.

The results in this question for Hadamards and CNOT
gates do not apply, for example, to the 5 qubit code of Qu.~\ref{qu1}. That
they do apply to
Steane's 7 qubit code is one of the reasons why this code is a popular choice.

\item
In the last question we showed that, for the 7-qubit Steane code, 
the logical $\overline{X}$ acting on the codewords
is implemented by $\prod_j X_j$, and the logical $\overline{Z}$ is implemented by
$\prod_j Z_j$. Show that the corresponding results for Shor's 9-qubit code do
not hold. Instead, show that one has, rather curiously, 
\begin{equation}
\prod_{j=1}^9 Z_j \equiv \overline{X}, \qquad
\prod_{j=1}^9 X_j\equiv \overline{Z} .
\end{equation}

\end{problems}

%% file: grover8.tex
\section{Introduction}

\index{Grover's search algorithm}
Grover's algorithm discussed in this chapter is of a  different type from
Shor's algorithm.  Whereas Shor's (and related algorithms like Simon's) depend
on a quantum Fourier transform (of some sort), Grover's algorithm involves
a different approach, \textit{amplitude amplification}.\index{amplitude amplification}

To motivate Grover's algorithm consider looking up someone in a phone directory. 
It is straightforward to lookup a person's phone number in a directory if one
is given the name, because names are in alphabetic order. To locate the name
systematically one would go to the midpoint of the list, see which half the
name is in, divide that half in two, again
see which half the number is in, and so on. One continues this procedure
until the size of the region containing the desired entry
is just one. For a directory
with $N$ entries, this 
\textit{bisection} method takes $\log_2 N$ operations (rounded up to the
nearest integer if $N$ is not a power of $2$) since one halves the range over
which the special entry could be at each stage. 

By contrast, suppose one is given the number and asked which person has that
number. Since the numbers are not ordered, all one can do is go through the
entries one at a time and see if each one has the desired name. On average
this would take $N/2$ operations before success was achieved. 

If $N$ is large this is a huge difference.  For example if $N= 10^6$ then
$\log_2 N \simeq 20$, to be compared with $N/2 = 5 \times 10^5$. Note that if the
$N$ possible values are represented by the configurations of $n$ qubits
then
\begin{equation}
N = 2^n .
\label{grover:Nn}
\end{equation}

The quantum search algorithm algorithm discussed here, due to Grover, is
often presented as such
a search of an unstructured database.\footnote{Though it is
doubtful it would ever be used in this way since it would
be a very extravagant use of
a precious resource to use qubits to store classical information.} Grover's
algorithm requires a quantum computer running a subroutine for which the input
is a number corresponding to an entry in the database, and which performs a
test to see if this is the special value being searched for. For
large $N$ it will determine the special value, with probability close to $1$,
by calling the subroutine only $(\pi/4)\sqrt{N}$ times.  This is a
\textit{quadratic} speedup compared with a classical computer.
While less spectacular than the exponential
speedup of Shor's algorithm\index{Shor's factoring algorithm}, it can potentially be applied to a wide variety of
problems\footnote{However, most applications of practical interest have some structure, whereas Grover is designed for problems with no structure. In most cases that Grover could potentially be applied, the structure of the problem allows an efficient classical algorithm which outperforms Grover. Thus it is debated whether the Grover algorithm would be of practical utility, even if one could overcome the
severe experimental difficulties of building a large quantum computer.}.


\section{The Black Box (Oracle)}
\index{oracle|see {black box}}
\index{black box}

To formulate the problem we consider $n$-bit integers, one of which, $a$, is
special. The goal is to find $a$.
We need a subroutine which outputs 1 if the
input value $x$ is equal to $a$ and outputs 0 otherwise, i.e.
\begin{equation}
\begin{split}
f(x) &= 0, \quad (x \ne a) , \\
f(a) & = 1 \, .
\end{split}
\end{equation}

As usual, the function will be determined from a unitary transformation acting
on an $n$-qubit ``input" register and an ``output" qubit which is flipped
or not flipped depending on whether $x$ is the special number $a$ or not:
\begin{equation}
U|x\rangle_n |y \rangle_1 = |x\rangle_n |y \oplus f(x) \rangle_1 \, .
\label{grover:U}
\end{equation}

\begin{figure}[htb!]
\begin{center}
\includegraphics[width=8cm]{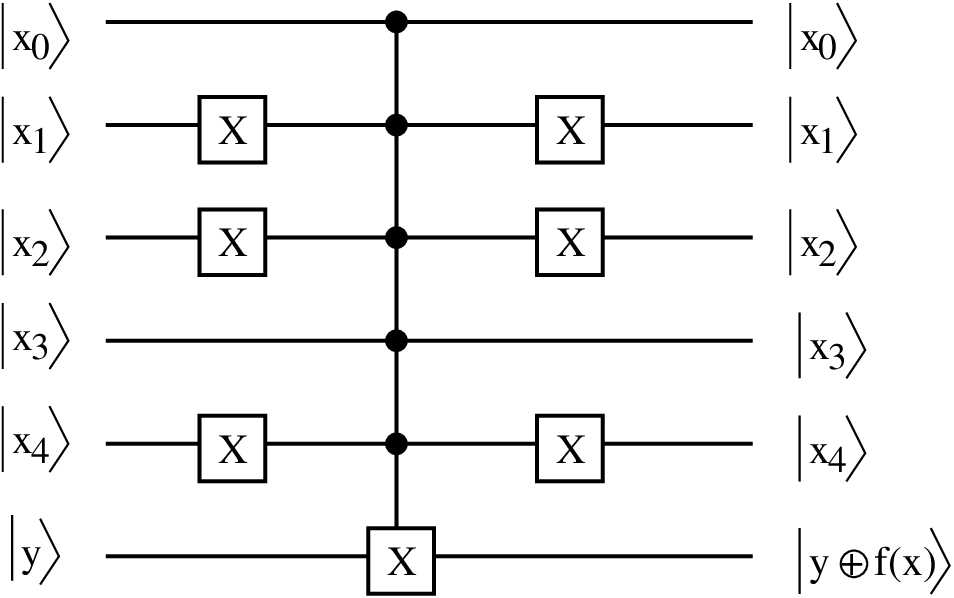}
\caption{
\label{grover_bb}
A black box circuit that executes the first part of a Grover iteration, 
Eq.~\eqref{grover:U}, in which $f(x) = 0$ if $x \ne a$ and $f(a) = 1$, for the case
of $n=5$ qubits and where
the special number $a$ is $01001$. The 6-qubit gate in the center
is a five-fold-controlled-NOT gate which acts to flip the target qubit $y$ 
only if all the control qubits are 1. The $X$ gates on the left
flip qubits $x_1, x_2$ and $x_4$. Hence the target qubit is flipped if
and only if $x_0=1,
x_1=0, x_2=0, x_3=1, x_4 = 0$, which are the bits of $a$.
The $X$-gates on the right flip back those qubits which
had previously
been flipped, thus leaving the ``input" register, the $\{|x_i\rangle\}$, unchanged.
The  lower ``output" 
qubit, which is initialized to $|y\rangle$, contains information on the function $f(x)$ in
its final state.
}
\end{center}
\end{figure}

\index{control qubit}
\index{target qubit}
A simple example of such a function for $n=5$ and $a = 01001$ is shown in
Fig.~\ref{grover_bb}. Recall that $x_0$ is the least significant (i.e.~right-hand) bit.
The target qubit is flipped only if all five of the control bits
are one, which requires $x_0=1, x_1=0, x_2=0, x_3=1, x_4 = 0$ (the 
bits of $a$). How to construct such a
five-fold-controlled-NOT gate out of 1-qubit and 2-qubit elementary gates is
discussed in Mermin~\cite{mermin:07} \S4.2. 

Such a black box function is called an oracle.  An oracle
\index{black box}
gives the output for the input values which are fed into it but one is not allowed
to ``open the box" and see how it is made.
Of course, for the implementation in Fig.~\ref{grover_bb} if you did look at
the workings of the circuit you would immediately determine the special value
$a$.
However, the implementation of the black box in Fig.~\ref{grover_bb}
is a simple example.  The Grover algorithm can also be applied in more
useful situations where the value of $f(x)$ is \textit{not} built in
explicitly but has to be calculated in a non-trivial way and so for
these cases ``opening the
box" wouldn't help to solve the problem. Examples are
discussed in Mermin~\cite{mermin:07} and Nielsen and Chuang~\cite{nielsen:00}.

It is useful to initially set
the ``output" qubit $y$ to be 1 and then apply a Hadamard gate
before applying $U$. The ``output" qubit is then
\begin{equation}
H |1\rangle = {1\over \sqrt{2}} \left(\, |0\rangle - |1\rangle\, \right)\, .
\end{equation}
If the result of $U$ is $f(x)=0$ then the ``output" qubit is unchanged. If the
result is $f(x) = 1$ then $|0\rangle \to |1\rangle$ and vice-versa, so the
``output" qubit changes sign. We already met this ``phase kickback" in our
discussion of the Deutsch algorithm, see Eq.~\eqref{phase_kb}.
Consequently
\begin{equation}
U \left(\, |x\rangle_n \otimes H |1\rangle_1\, \right) = 
(-1)^{f(x)} |x\rangle_n \otimes H |1\rangle_1 \, .
\end{equation}
We can associate the possible sign change with the ``input" register" in which
case the ``output" qubit remains unchanged. Hence, for simplicity, the ``output"
qubit will be
ignored in what follows. Thus we consider the following
unitary operator $\hat{O}$
acting only on the $n$-qubit ``input" register\footnote{We omit the subscript
$n$ on the states from now on since we will only be dealing with $n$-qubit
states.}:
\begin{equation}
\hat{O} |x\rangle = (-1)^{f(x)} |x\rangle = \left\{
\begin{array}{ll}
\ \ |x\rangle, &\  x \ne a, \\
-|a\rangle, &\  x = a. \\
\end{array}
\right.
\end{equation}

Since $U$, and hence $\hat{O}$, are linear, acting with $\hat{O}$
on a superposition changes
the sign of the component along $|a\rangle$ but leaves the component
perpendicular to $|a\rangle$ unchanged. Hence if
\begin{equation}
|\psi\rangle = \sum_x c_x |x\rangle, 
\end{equation}
then
\begin{equation}
|\psi'\rangle \equiv \hat{O} |\psi\rangle = \sum_{x\ne a} c_x  |x\rangle - c_a|a \rangle
= \sum_x  c_x  |x\rangle - 2 c_a|a \rangle
= |\psi\rangle - 2 |a\rangle \langle a| \psi\rangle
\label{Ohat}
\end{equation}
since $c_a = \langle a| \psi\rangle$. 
You should check that $\langle a|\psi'\rangle = -\langle
a|\psi\rangle\, (=-c_a)$ and, for $x \ne a$, that  $\langle x|\psi'\rangle =
\langle x|\psi\rangle \,(= c_x)$, as required.
You should also verify that $|\psi'\rangle$ is correctly normalized if $|\psi\rangle$ and
$|a\rangle$ are.

We initialize the $n$-qubit input register into a uniform superposition of all
basis states by acting with $n$ Hadamards on $|0\rangle$:
\begin{equation}
|\psi_0\rangle = H^{\otimes n}|0\rangle = {1 \over\sqrt{N}}
\sum_{x=0}^{N - 1} |x\rangle \, .
\label{grover:psi}
\end{equation}
We can also write $|\psi_0\rangle$ as
\begin{equation}
|\psi_0\rangle = {1 \over\sqrt{N}} |a\rangle + \sqrt{N-1\over N}|a_\perp
\rangle,
\label{a_aperp}
\end{equation}
where $|a_\perp\rangle$ is a normalized, \textit{uniform}
superposition of all basis states
perpendicular to $|a\rangle$, i.e.
\begin{equation}
|a_\perp \rangle = {1 \over \sqrt{N-1}}
\sum_{\begin{array}{c}\scriptstyle x=0\\
^{(x \ne a)}\end{array}}^{N-1}|x\rangle \, .
\label{aperp}
\end{equation}
\index{superposition}

\begin{figure}[htb!]
\begin{center}
\includegraphics[width=7.5cm]{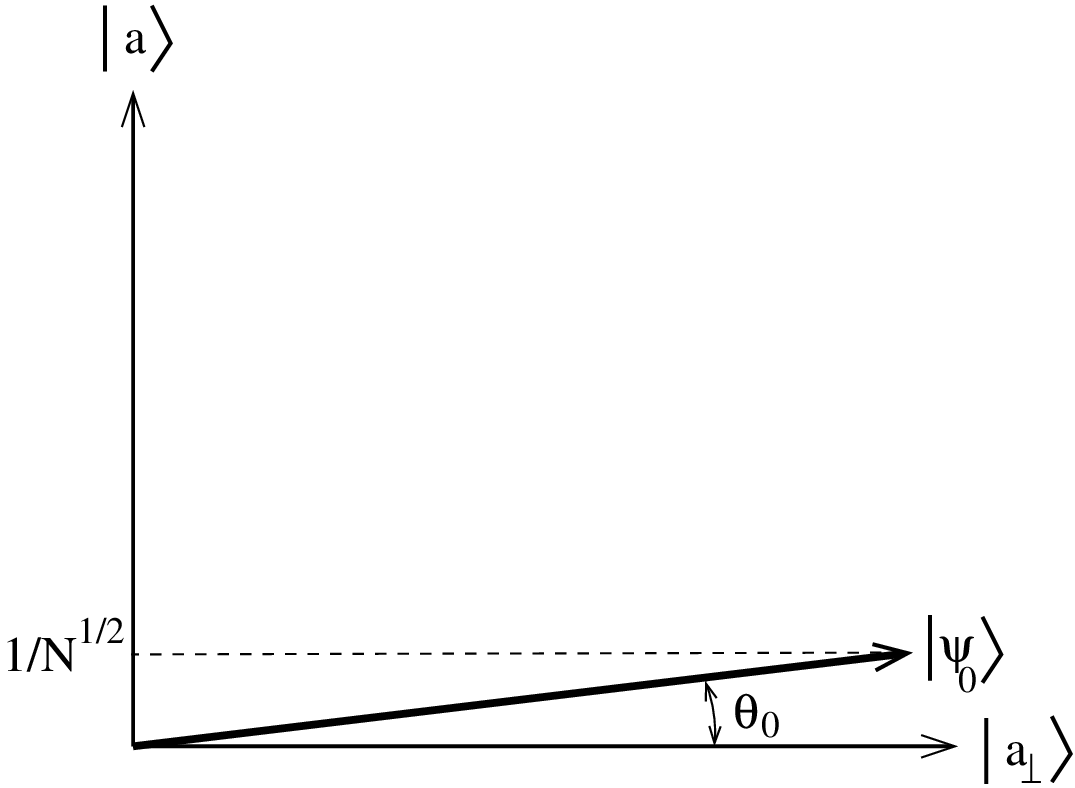}
\caption{
Projection of the $2^N$-dimensional space on to a 2-dimensional space
spanned by $|a\rangle$ and
$|a_\perp\rangle$, the latter being
a (normalized) equal linear combination of all basis
states except for $|a\rangle$ itself, see Eq.~\eqref{aperp}. The vector in
bold is the initial state  $|\psi_0\rangle$, an equal linear combination
of all basis states, see
Eq.~\eqref{grover:psi}. The vector $|\psi_0\rangle$ has a projection $1/\sqrt{N}$ on 
to $|a\rangle$, so $\sin\theta_0 = 1/\sqrt{N}$, where $\theta_0$ is the angle
between $|\psi_0\rangle$ and $|a_\perp\rangle$.
\label{grover1}
}
\end{center}
\end{figure}

We shall see that all the subsequent states generated during the Grover
algorithm can also be written as a linear
combination of $|a\rangle$ and $|a_\perp \rangle$.
These can be conveniently
drawn as
vectors in the 2-dimensional space spanned by these two basis vectors, see
Fig.~\ref{grover1}.

Hence $|\psi_0\rangle$ 
makes an angle $\theta_0$ with the $|a_\perp \rangle$ axis 
where $\sin\theta_0 = \langle a|\psi_0\rangle$, or
\begin{equation}
\sin\theta_0 = {1 \over \sqrt{N}} \, ,
\label{sintheta}
\end{equation}
so we can express $|\psi_0\rangle$ in Eq.~\eqref{a_aperp} as
\begin{equation}
|\psi_0\rangle = \sin\theta_0\, |a\rangle + \cos\theta_0 \, |a_\perp \rangle .
\label{psi0_theta}
\end{equation}
Note that $|\psi_0\rangle,  |a_\perp\rangle$ and $|a\rangle$ are all
normalized.

From Eq.~\eqref{psi0_theta} we see that if we
were to measure $|\psi_0\rangle$ now, we would get $|a\rangle$ with
probability $\sin^2\theta_0 \,(= 1/N)$, which is \textit{very small} for large $N$.
(Of course we can also see that the probability is $1/N$
directly from Eq.~\eqref{grover:psi}.)
The goal of the Grover algorithm is to iteratively
rotate the vector representing the
state of the input register
from its initial direction, that of $|\psi_0\rangle$ (which is close to
the $|a_\perp\rangle$ axis), to a direction close to the $|a\rangle$ axis,
because a measurement of it will then give $a$ with \textit{high probability}. This
is called \textit{amplitude amplification}.

%
%
%
%

\begin{figure}[htb!]
\begin{center}
\includegraphics[width=7.5cm]{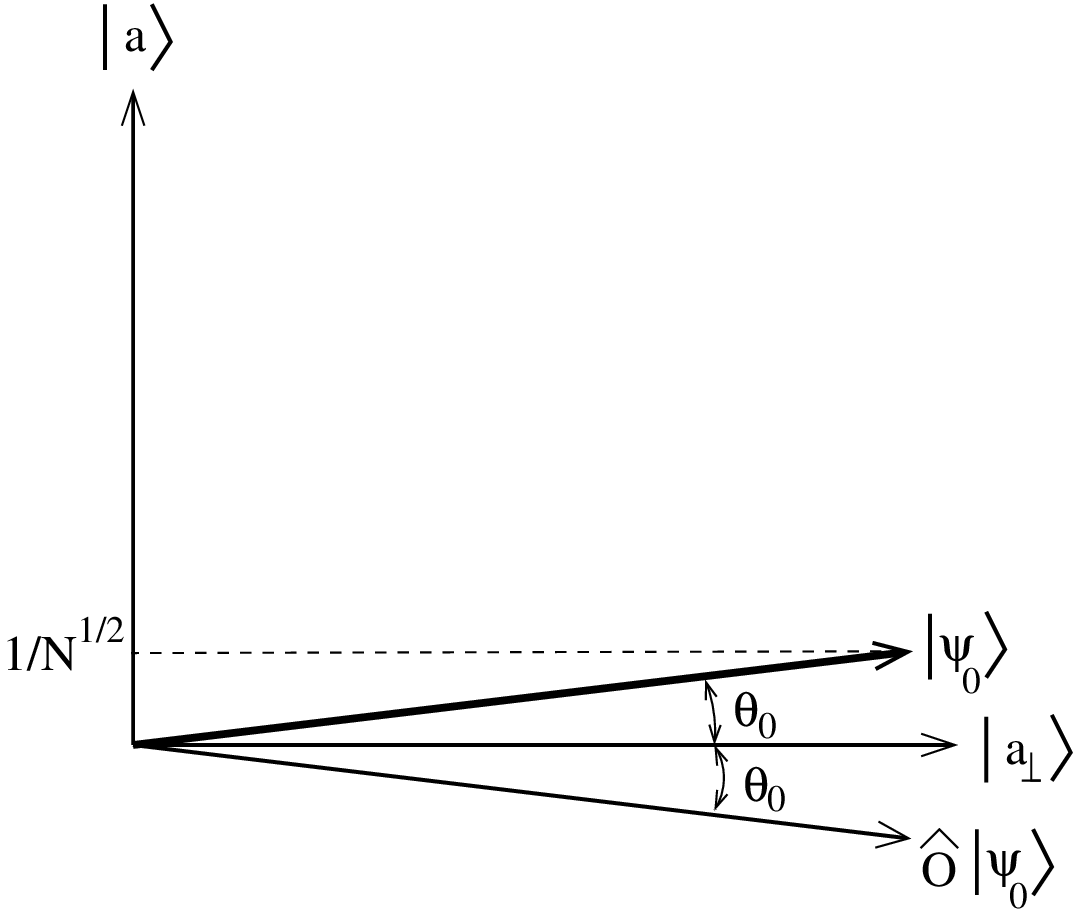}
\caption{
Figure showing that the action of the operator $\hat{O}$ is to reflect the 
state it is acting on, in this case
$|\psi_0\rangle$, about the $|a_\perp\rangle$ axis. 
\label{grover2}
}
\end{center}
\end{figure}

As shown in Eq.~\eqref{Ohat} the action of $\hat{O}$ is to invert the
component along $|a\rangle$ of the vector
it acts on,
while keeping the component perpendicular to
$|a\rangle$ unchanged. The net effect is to \textit{reflect} about the 
$|a_\perp\rangle$ axis. Figure~\ref{grover2} shows the effect of $\hat{O}$ on
the initial state $|\psi_0\rangle$.
To rotate the direction of the state towards the $|a\rangle$ axis we will need a
second unitary operation that is discussed in the next section.

\section{The second step of the Grover iteration}

\begin{figure}[htb!]
\begin{center}
\includegraphics[width=7.5cm]{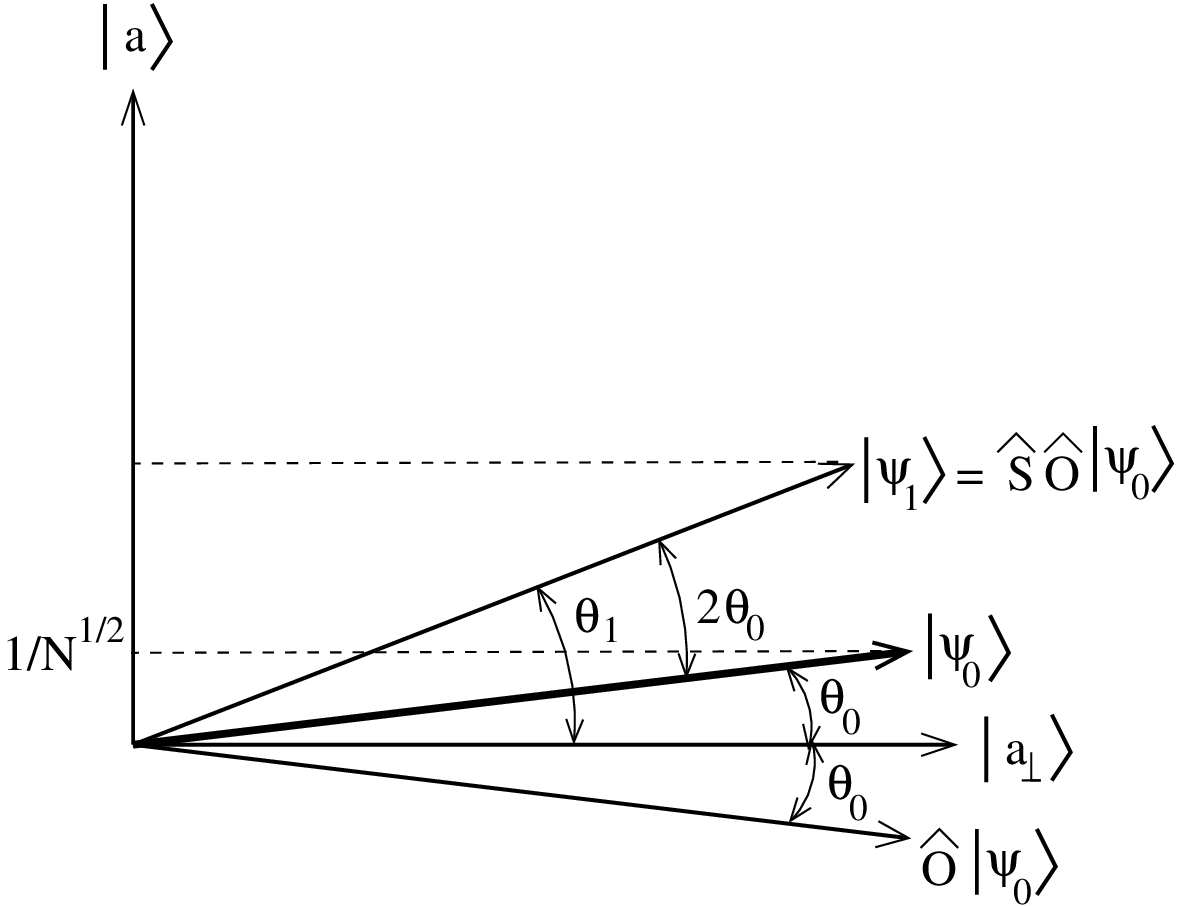}
\caption{
Figure showing that the action of the operator $\hat{S}$ is to reflect
the state
it is acting on, in this case $\hat{O}|\psi_0\rangle$, about the direction of 
$|\psi_0\rangle$ which is defined in Eq.~\eqref{grover:psi}.  The net result of the two
operations, $\hat{O}$ followed by $\hat{S}$, is to rotate the direction of
$|\psi_0\rangle$ by $2 \theta_0$ in an anti-clockwise direction. We will call the new state $|\psi_1\rangle$. It is at an angle $\theta_1 = \theta_0 + 2 \theta_0$ to the $|a_\perp\rangle$ axis.
\label{grover3}
}
\end{center}
\end{figure}

The second stage of a single Grover iteration is independent of the special
number $a$. It changes the sign of the component perpendicular to the initial
state $|\psi_0\rangle$ and keeps unchanged the component along $|\psi_0\rangle$.
Denoting this operation by $\hat{S}$ we have 
\begin{equation}
|\phi\rangle \to |\phi'\rangle = 
\hat{S} |\phi\rangle = 2 |\psi_0\rangle \langle \psi_0|\phi\rangle - |\phi\rangle,
\label{Shat}
\end{equation}
where $|\phi\rangle$ is an arbitrary state.  You should check that
$\langle \psi_0|\phi' \rangle  = \langle \psi_0|\phi \rangle$,
so the component along $|\psi_0\rangle$ is unchanged, and for a state
$|\mu\rangle$ which is orthogonal to $|\psi_0\rangle$,
$\langle \mu|\phi' \rangle  = -\langle \mu|\phi \rangle$,
showing that the component perpendicular to $|\psi_0\rangle$ has the sign
changed.
The net result is to reflect $|\phi\rangle$ about
the direction of the original $|\psi_0\rangle$.

Figure~\ref{grover3} shows the effects of
$\hat{S}$ acting on the state generated by $\hat{O}|\psi_0\rangle$. The combined
effect of $\hat{O}$ followed by $\hat{S}$ is to rotate the initial state
$|\psi_0\rangle$ by $2 \theta_0$ in an anti-clockwise direction, i.e.~$2\theta_0$
towards the desired direction of the $|a\rangle$ axis.  The combination of
these two operations is called a Grover iteration, implemented by the Grover
operator
\begin{equation}
\hat{G} = \hat{S} \hat{O}.
\label{Grover_op}
\end{equation}

The effect of the first Grover iteration, therefore, is to take the initial state $|\psi_0\rangle$ and rotate it anti-clockwise by $2\theta_0$. We will call the resulting state $|\psi_1\rangle$. It is at an angle $\theta_1$ to the $|a_\perp\rangle$ axis, where
\begin{equation}
\theta_1 = \theta_0 + 2 \theta_0 ,
\end{equation}
see Fig.~\ref{grover3}.

\section{Subsequent iterations}
\begin{figure}[htb!]
\begin{center}
\includegraphics[width=12cm]{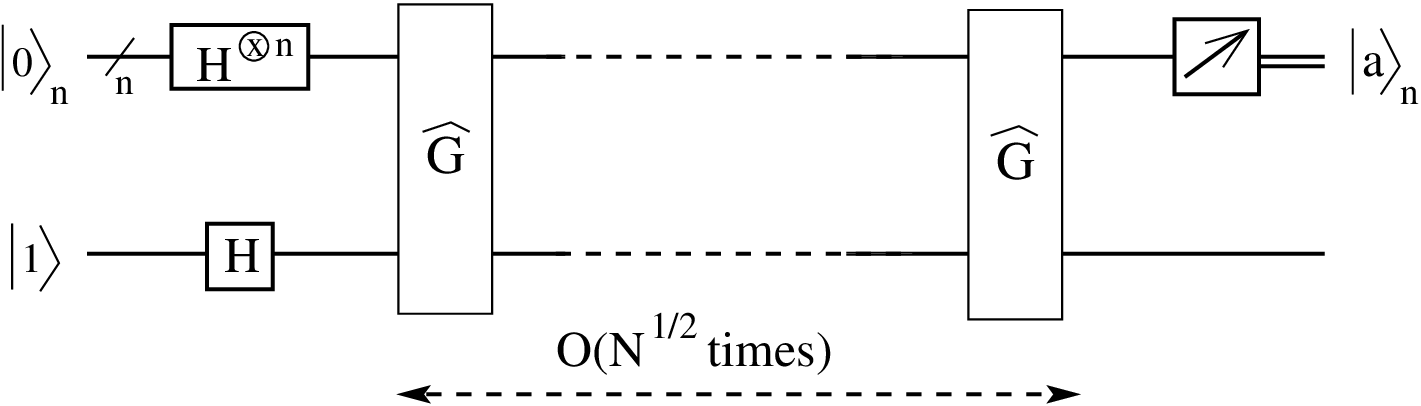}
\caption{
Circuit implementing the Grover algorithm. $\hat{G}$ is the Grover operator,
given by $\hat{G} = \hat{S} \hat{O}$ where $\hat{O}$ and $\hat{S}$ are
given by Eqs.~\eqref{Ohat} and \eqref{Shat} respectively. It
acts only on the $n$ input qubits (the upper
line). The output qubit (the lower line) remains unchanged by $\hat{G}$.
After $O(\sqrt{N})$ iterations of the
Grover operator, the result of a
measurement on the input qubits is the special value $a$ with high
probability.
\label{grover_circ}
}
\end{center}
\end{figure}

Subsequent Grover iterations perform the same two steps: $\hat{O}$ which
reflects about $|a_\perp\rangle$ followed by $\hat{S}$ which reflects about
$|\psi_0\rangle$.
The overall circuit implementing the Grover algorithm is shown in
Fig.~\ref{grover_circ}.

\begin{figure}[htb]
\begin{center}
\includegraphics[width=7.5cm]{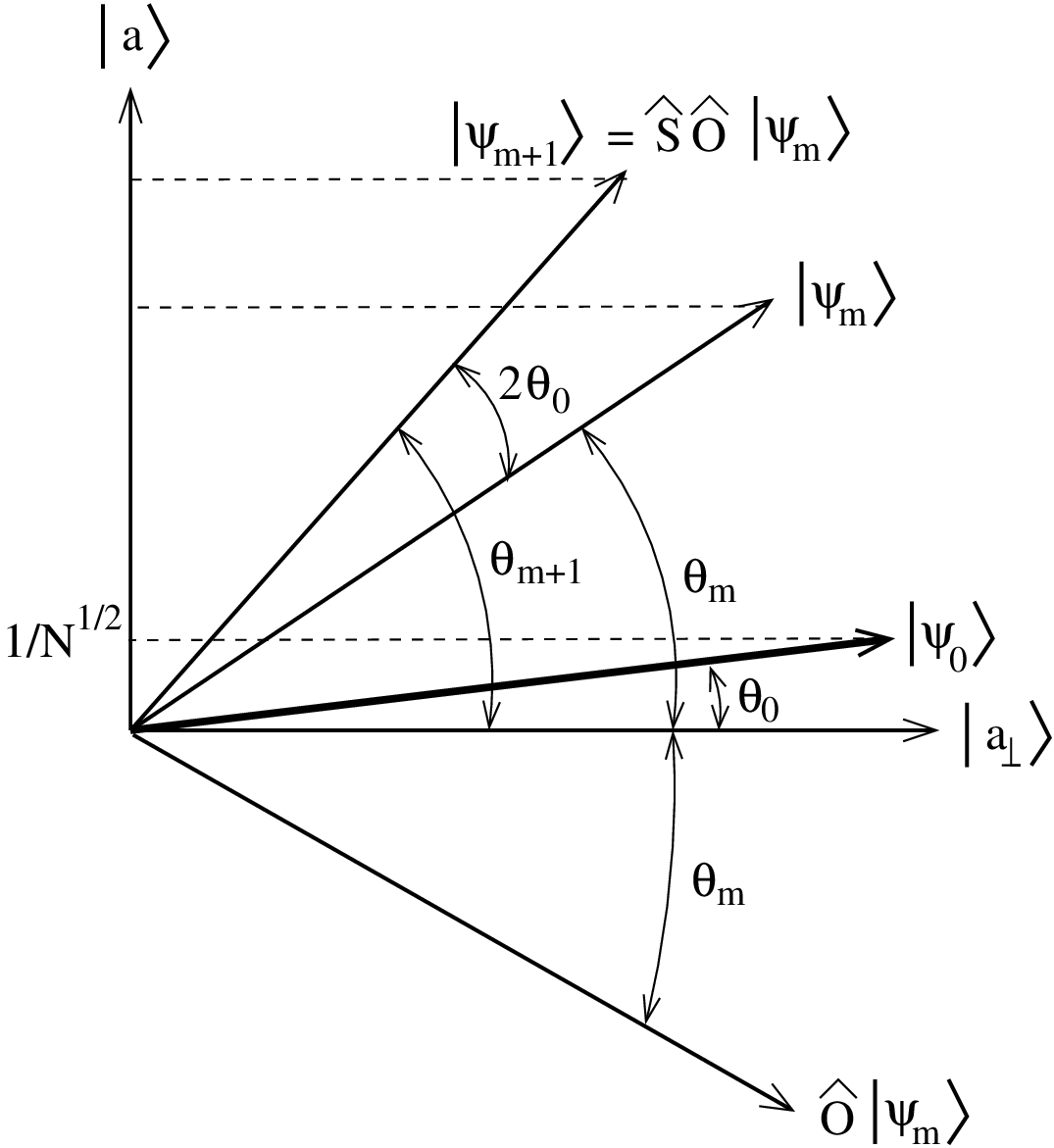}
\caption{
After the $m$-th iteration of the Grover algorithm, the state $|\psi_0\rangle$
has been rotated to $|\psi_m\rangle$, which makes an angle $\theta_m$ with the
$|a_\perp\rangle$ axis. At the next iteration of the Grover algorithm, 
firstly the
action of $\hat{O}$ reflects $|\psi_m\rangle$ about the $|a_\perp\rangle$
axis as shown. This is equivalent to a clockwise rotation by $2 \theta_m$
so $\hat{O} |\psi_m\rangle$ is at an angle $\theta_m$ below the
$|a_\perp\rangle$ axis. 
Secondly, the state $\hat{O} |\psi_m\rangle$ is acted on by
$\hat{S}$ which reflects about the direction of $|\psi_0\rangle$. This is
equivalent to an anti-clockwise
rotation by $2(\theta_m + \theta_0)$. The net effect of the two operations is
to rotate $|\psi_m\rangle$ by an angle $2\theta_0$ in an anti-clockwise
direction. Hence the new state $|\psi_{m+1}\rangle$ is at an angle
$\theta_{m+1} = \theta_m + 2 \theta_0$ to the $|a_\perp\rangle$ axis. The
amplitude for the state $|\psi_m\rangle$ to be $|a\rangle$ is the projection
on to the vertical axis, which increases with $m$ up to the point
where $\theta_m = \pi/2$.
\label{grover4}
}
\end{center}
\end{figure}

If $m$ iterations have already been done, so the current state is
$|\psi_m\rangle$, Fig.~\ref{grover4} shows the effect of doing an additional
iteration. The state $|\psi_m\rangle$ makes an angle $\theta_m$ with the $|a_\perp\rangle$
axis, so $\hat{O}$ rotates the direction by $2 \theta_m$ clockwise, while
$\hat{S}$ rotates it by $2(\theta_m + \theta_0)$ anti-clockwise. The net result is a
rotation by $2 \theta_0$ (independent of $\theta_m$) anti-clockwise, which 
is towards the desired direction, $|a\rangle$, i.e.
\begin{equation}
\theta_{m+1} = \theta_m + 2 \theta_0 ,
\end{equation}
which gives
\begin{equation}
\theta_m  = (2m+1) \theta_0
\label{theta_m}
\end{equation}

The relationship between $|\psi_m\rangle, |a\rangle$ and $|a_\perp\rangle$ is
\begin{equation}
|\psi_m\rangle = \cos\theta_m |a_\perp\rangle + \sin\theta_m |a\rangle.
\label{pmaaperp}
\end{equation}

According to Eq.~\eqref{pmaaperp}, 
the amplitude for $|\psi_m\rangle$ to be measured in state $|a\rangle$,
i.e.~$\langle a|\psi_m\rangle$, is $\sin\theta_m = \sin[(2m+1)\theta_0]$, the
projection on to the vertical axis in Fig.~\ref{grover4}. This
increases as $m$ increases up to the point where $\theta_m = \pi/2$ but then decreases.
One therefore takes the number of Grover iterations, $m$, to be such that
$\theta_m \simeq \pi/2$. From Eqs.~\eqref{theta_m} and \eqref{sintheta} we see that we need
\begin{equation}
\theta_m = (2m+1) \theta_0 = (2m+1) \sin^{-1}{1\over\sqrt{N}} = {\pi \over 2},
\end{equation}
which, for large $N$, gives
\begin{equation}
m = {\pi \over 4}\, \sqrt{N}.
\end{equation}
When $\theta_m \simeq \pi/2$  measuring the state gives $a$ with high
probability. 

We do not have to get the number of iterations precisely right. After $m$ iterations, the probability that a measurement gives $a$ is $\sin^2\theta_m = \sin^2[(2m+1)\theta_0]$. Any value of $\theta_m$ in the range
\begin{equation}
{\pi\over 4} < \theta_m <  {3 \pi \over 4}
\end{equation}
will get determine $a$ correctly with a probability greater than 1/2. For large $N$ this corresponds to
\begin{equation}
 {\pi \over 8}\, \sqrt{N} < m <  {3 \pi \over 8}\, \sqrt{N} .
\end{equation}
Note that the probability \textit{decreases} for $m > (\pi/4)\sqrt{N}$, unlike
many algorithms where increasing the number of iterations progressively
improves the probability of success.


The operation count of the Grover algorithm is $O(\sqrt{N})$ which is a
quadratic speedup compared with the $O(N)$ count on a classical computer. The
quantum speedup comes, of course, from quantum parallelism; all
\index{quantum parallelism}
$N=2^n$ values of
$f(x)$ are evaluated in parallel, so naively it looks as though we 
should
be able to get a speedup by a factor of $N$, i.e.~an operation count of $O(1)$.
However, if one measured directly 
after computing the function, one would
just get one value of $x$ and the corresponding $f(x)$, which is
no better than on
a classical computer.
It requires additional operations, in the form of 
the Grover operator $\hat{G}$ applied iteratively, 
to extract a speedup, which in this case only reduces the operation count to
$O(\sqrt{N})$ not $O(1)$.
One can show that the $O(\sqrt{N})$ operation count of the Grover algorithm is
optimal. An operation count of $O(1)$ is \textit{proved} to be impossible.

\section{Extensions}
\index{Grover's search algorithm!more than one special value}
\subsection{More than one special value}
\label{Mgt1}

In the standard implementation of the Grover algorithm it is assumed that
there is only one special value.  If there are $M$ solutions, $a_i, i = 1,
\cdots,M$
then, proceeding
along the lines of the derivation for one solution, one
finds~\cite{nielsen:00,mermin:07,vathsan:16,rieffel:14}:
\begin{enumerate}
\item
\index{superposition}
The states generated by the Grover algorithm can be written as a linear
combination of a uniform superposition of all the special states,
\begin{equation}
|a\rangle = {1\over \sqrt{M}}\sum_{x\,\in\,\{a_i\}} |x\rangle,
\end{equation}
and a uniform superposition of all the other states,
\begin{equation}
|a_\perp\rangle = {1\over \sqrt{N-M}}\sum_{x\,\not\in \,\{a_i\}} |x\rangle .
\end{equation}
We see that $|a\rangle$ and $|a_\perp\rangle$ are normalized.
\item
The initial state, $|\psi_0\rangle$, the uniform superposition of
\textit{all} states given in Eq.~\eqref{grover:psi}, can be written in terms of
$|a\rangle$ and $|a_\perp\rangle$ as
\begin{equation}
|\psi_0\rangle = \sqrt{M\over N}\, |a\rangle + \sqrt{N-M\over N}\, |a_\perp \rangle .
\label{psi_NM}
\end{equation}
Since $|a\rangle$ and $|a_\perp\rangle$ are normalized it follows that
$|\psi_0\rangle$ is also normalized.
Hence $|\psi_0\rangle$ makes an angle $\theta_0$ with the
$|a_\perp\rangle$ axis where $\sin\theta_0 = \langle a|\psi_0\rangle$, or
\begin{equation}
\sin\theta_0 = \sqrt{M \over N},
\label{sinthetaM}
\end{equation}
rather than Eq.~\eqref{sintheta}. Consequently we can write Eq.~\eqref{psi_NM} in terms of
$\theta_0$ in the
same way as for $M=1$, namely 
Eq.~\eqref{psi0_theta}.
\item
Subsequent iterations rotate the direction of the state by an angle $2 \theta_0$
towards the $|a\rangle$ axis and so, after $m$ iterations, the angle $\theta_m$ is given by
Eq.~\eqref{theta_m}, and the state $|\psi_m\rangle$ is given by Eq.~\eqref{pmaaperp}. Hence the effect of each 
Grover iteration, when expressed in terms of $\theta_0$, is the same as for $M=1$, 
and the only difference compared with $M=1$ is that 
$\theta_0$ is given by Eq.~\eqref{sinthetaM} rather than \eqref{sintheta}.
\item
Assuming $M \ll N$, then $\theta_m$ is approximately $\pi/2$
when
the number of iterations $m$ is given by
\begin{equation}
m = {\pi \over 4} \sqrt{N \over M} .
\end{equation}
After this number of iterations of the Grover operator, with high
probability a measurement of the state
will give \textit{one} of the special values $a_i$ with equal likelihood.
\end{enumerate}
The student is advised to check these steps.


\subsection{Quantum Counting}
\index{quantum counting algorithm}

The results of the previous subsection are only useful if we \textit{know}
in advance how many special values, $M$, there are.  If we have no
prior knowledge of $M$,
how can we determine it? We saw that the Grover operator $\hat{G}$ rotates vectors in the
$|a\rangle$--$|a_\perp\rangle$ plane by an angle $2 \theta_0$, where
$\theta_0$ is given by Eq.~\eqref{sinthetaM} and so depends on $M$. In other
words, in the space of $|a\rangle$ and $|a_\perp\rangle$, the Grover operator 
has the standard form of a rotation matrix
\begin{equation}
\hat{G} = \begin{pmatrix*}[r]
\cos 2\theta_0 & -\sin2\theta_0 \\
\sin 2\theta_0 & \cos 2\theta_0
\end{pmatrix*} .
\end{equation}

The eigenvalues of $\hat{G}$ are easily found to be $\exp(\pm 2 i \theta_0)$. (It is a general
property of unitary matrices that their
eigenvalues are a pure phase.) We already showed in
section~\ref{sec:phase_est} in Chapter \ref{ch:qft} 
that the phase of the eigenvalue of a unitary matrix can be
determined from the phase estimation algorithm using Shor's quantum Fourier
transform.
\index{phase estimation}


Consequently, we can determine $\theta_0$ (and hence $M$), and also get one of
the special values $a_i$, by combining the Quantum Fourier Transform 
with Grover's algorithm. In fact this ``quantum counting"
algorithm will even tell us whether or not a special value exists at all,
i.e.~whether or not $M = 0$. 
The interested student can find
more details in advanced texts such as Refs.~\cite{nielsen:00,rieffel:14}.


\hrulefill
\section*{Problems}
\input{hw_ch20.tex}

%% file: hw_ch20.tex
\begin{problems}

\item 
Consider the Grover algorithm in which you have to find one marked state out
of $N=4$ states.  Show that the algorithm succeeds with probability 1 after 1
iteration.

\item 
You have to find one marked state out of $N=2$ states. Classically, picking
one state at random has a probability of $1/2$ to succeed.  Show that the
Grover algorithm does not improve these odds.

\item 
Assume that there are $M$ marked states out of $N$. Fill in the details
of the derivation, sketched in Sec.~\ref{Mgt1},
of the required number of Grover iterations.
(Assume that $N$ is large.)

\end{problems}

%% file: qkd8.tex
There are several problems of interest where qubits can be considered one at a time,
without needing any qubit-qubit interactions. Photons are ideal qubits for this
because their interactions with each other are immeasurably weak, and they can be propagated
down optical fibres for a big distance with little attenuation while
preserving their polarization. You will recall from Sec.~\ref{sec:photons} that it is the polarization of
\index{photon!polarization}
the photon which characterizes the qubit, e.g.:
\index{photon}
\begin{equation}
\begin{split}
|0\rangle &\equiv\ |\leftrightarrow\,\rangle , \qquad \mathrm{(left\!-\!right)} \\
|1\rangle &\equiv\ |\updownarrow\,\rangle , \qquad \mathrm{(up\!-\!down)}  \\
|+\rangle = H|0\rangle = {1\over \sqrt{2}}(|0\rangle + |1\rangle)\ &\equiv\
|\rotatebox[origin=c]{45}{\Large$\leftrightarrow$}\rangle , \qquad
\mathrm{(one\ of\ the\ diagonals)} \\
|-\rangle =  H|1\rangle = {1\over \sqrt{2}}(|0\rangle - |1\rangle)\ &\equiv\
|\rotatebox[origin=c]{45}{\Large$\updownarrow$}\rangle , \qquad \mathrm{(the\ other\ diagonal)} .
\end{split}
\label{pol}
\end{equation}
The connection between the polarization of photons and qubit states was
described in more detail in Sec.~\ref{sec:gen_qubit}.

Several quantum protocols involving photons have been
successfully implemented. Here we will discuss applications 
to cryptography and ``teleportation", the latter being set
as a homework problem with lots of help. Some references
are~\cite{nielsen:00,vathsan:16,mermin:07}.

\section{Quantum Key Distribution}

\index{cryptography}
\index{QKD|see{quantum key distribution}}\index{quantum key distribution}
Cryptography is concerned with transmitting secret messages. There are two
main approaches:
\begin{itemize}
\item
\index{public key cryptography}
\textbf{Public Key}\\
An example is the RSA scheme\index{RSA encryption}
which we already met in Chapter \ref{ch:rsa} in the
context of Shor's algorithm for factoring integers. Let us briefly review the basic
idea. Suppose Bob wants to send a message to Alice. Alice sends her public key
\index{public key}
down an open channel to Bob who uses this to encrypt his message. Alice
\index{private key}
decodes the encrypted message using her private key. The private key is not
shared, only the public key.  Security depends on the difficulty of decoding
the message without the private key. In the case of RSA we recall that this
required factoring a large integer. 

\item \textbf{Private key (or symmetric key)}.
\index{private key cryptography}
(Note: public key encryption is not
symmetric between sender and receiver.)\\
Alice and Bob share a private key, which has been generated and shared in
advance. This must be as long as the message and, as we shall explain later,
can only be used once.  But how do Alice and Bob share the private key
securely?  Perhaps Alice could put it in a box and send it to Bob by FedEx. This is not 
convenient which is why internet transactions use
public key encryption instead.
\end{itemize}

We shall now see that quantum mechanics can help with securely sharing 
private keys, using what is called \textbf{Quantum Key Distribution} (QKD). 

The idea of QKD is to create a one-time codepad which Alice and Bob share. By
using quantum mechanics, Alice and Bob will be able to detect whether an
eavesdropper whom, following tradition,
we shall call Eve,
is trying to intercept their messages when they share the codepad.

The codepad is a shared random string of bits $R$, which must be at least as long as the
message. Alice encodes the message $M$ by bit-wise
XOR-ing it with the random string, i.e.
\begin{equation}
\mathrm{Alice:}\qquad M \longrightarrow M \oplus R \ (= M') .
\end{equation}
Bob decodes the encoded message $M'$ by also XOR-ing it with $R$, i.e.
\begin{equation}
\mathrm{Bob:}\qquad M' \longrightarrow M' \oplus R = M .
\end{equation}
This works because
$M \oplus R \oplus R = M$, as we have discussed several times before in the
course.

We now explain why this codepad can only be used once securely. Suppose we
send two encoded messages using the same codepad, i.e.
\begin{equation}
\begin{split}
M_1' &= M_1 \oplus R \\
M_2' &= M_2 \oplus R.
\end{split}
\end{equation}
Anyone intercepting the message can XOR the two messages with the result
\begin{equation}
M_1' \oplus M_2' = M_1 \oplus R \oplus M_2 \oplus R = M_1 \oplus M_2,
\end{equation}
so the random string has dropped out. The eavesdropper can then use standard methods
(e.g.~letter frequency) to decrypt. This is harder than for a single message
since one has to extract both messages, but may be feasible. Hence the
great security\footnote{If the bit string is truly random it is impossible to decrypt
the message without knowing the string.}
coming from using a random bit string has been lost.

How do Alice Bob know that their random bit string $R$ was not intercepted by
Eve as they were sharing it? This is where quantum mechanics comes into play.

\subsection{BB84 protocol}
\index{quantum key distribution!BB84 protocol}
We describe here the method proposed by Bennett and Brassard in 1984 (BB84).
Alice sends Bob a long string of photons. Each photon is in one of the four
polarization states in Eq.~\eqref{pol}. The polarization states corresponding
\index{photon!polarization}
to qubits $|0\rangle$ and $|1\rangle$ we will call $Z$-basis qubits (since this
is the basis in which $Z$ is diagonal). The
polarization states corresponding to $H|0\rangle = {1\over \sqrt{2}}
(|0\rangle + |1\rangle)$ and $H|1\rangle = {1\over \sqrt{2}}
(|0\rangle - |1\rangle)$ we will call $X$-basis qubits (since this 
is the basis in which $X$ is diagonal). To decide in which basis to send a photon
Alice generates a random integer taking values $0$  and $1$. If she gets $0$
she sends a
$Z$-basis photon, and if she gets $1$ she sends an $X$-basis photon. Within each basis-type 
there are two states, which Alice chooses by generating a second random integer,
again taking values $0$ and $1$. If she gets $0$ she sends
$|0\rangle$ if the $Z$-basis were chosen
and $H|0\rangle$ if the $X$-basis were chosen. If she gets $1$ for the second random 
number, she sends $|1\rangle$
or $H|1\rangle$, depending on whether the $Z$-basis or
$X$-basis was chosen. An example of a set of photons sent to Bob is
\begin{align}
\mathrm{basis}\qquad & Z \quad X \quad X \quad X \quad Z \quad Z \quad X \quad Z
\quad X \quad \cdots \nonumber \\
\mathrm{state}\qquad & \ 0 \quad \ 1 \quad \ 0 \quad \ 1 \quad \ 1 \quad \,0 \quad \ 1 \quad
\ 0 \quad \ 0 \quad \,\cdots
\end{align}

Bob receives these qubits and decides randomly whether to measure in the
$Z$-basis or the $X$-basis. Note that the photons are individually
identifiable by the sequence in which they arrive.

If the \textit{basis}
in which Alice sends a photon ($Z$ or $X$) is the same as that in which Bob measures it,
then the \textit{state} which Bob measures, $0$ or $1$, must be the same as the state that Alice sent.
However if the bases for sending and measuring are different, then Bob will
only find the same state as Alice about half the time. Alice tells Bob over an
insecure channel which photons were in the $Z$ basis and which in the $X$-basis,
but not the state.  Bob then tells Alice over an insecure channel for which of the
photons he measured in the same basis as she sent it in. They keep these and
discard the others (about 1/2 on average).

The onetime codepad is the set of random bits corresponding to the \textit{state} of the
qubits for which Alice and Bob measured in the same basis. Note that this
information was \textit{not} sent down the insecure channel, only the
\textit{basis} was sent. Recall that if Alice and Bob use the same basis they
must get the same state.

Let's complete the above example with a possible set of measurements that Bob
made.
\begin{align}
\mathrm{Alice} \quad \mathrm{basis}\qquad & Z \quad \fbox{X} \quad \fbox{X} \quad X \quad \fbox{Z} \quad Z \quad X \quad \fbox{Z}
\quad X \quad \cdots \nonumber \\
\mathrm{state}\qquad & \,0^\star \quad \fbox{1} \quad \ \fbox{0} \quad \ 1 \quad \,\fbox{1} \quad 0^\star \quad \,1 \quad
\,\fbox{0} \quad 0 \quad \cdots \nonumber \\
\ \nonumber \\
\mathrm{Bob} \quad \mathrm{basis}\qquad & X \quad \fbox{X} \quad \fbox{X} \quad Z \quad \fbox{Z} \quad X \quad Z \quad \fbox{Z}
\quad Z \quad \cdots \nonumber \\
\mathrm{state}\qquad & \,1^\star \quad \boxed{1} \quad \ \boxed{0} \quad \ 1 \quad \,\boxed{1} \quad 1^\star \quad \,1 \quad
\,\boxed{0} \quad 0 \quad \cdots
\end{align}
For the photons where Alice's and Bob's bases agree, the information is boxed.
For these photons, the state that Alice generated and that which Bob measured
agree. For the other photons, the states agree only half the time on average.
The cases where the states disagree are starred (in
this example, the states differ for 2 out of the 5 cases where the bases differ). 

The codepad which Alice and Bob have shared is the set of states for which
their bases agree, i.e.
\begin{equation}
R = 1010\, \cdots \,.
\end{equation}

How can Alice Bob know if Eve is interrupting the photons? Consider the ``good"
photons, those where Alice and Bob used the same basis. If Eve is not
interrupting them, then Alice and Bob agree on the state with 100\% probability. 
However, if Eve measures the photons and sends them on to Bob, then Alice and Bob
will have different states some of the time,
as we now show. 

Like Alice and Bob, Eve will have to choose a random basis for each
photon. There is probability 1/2 that she will choose the same basis as the
common basis of Alice and Bob, and
probability 1/2 that she will choose a different basis.
If she chooses the same basis, then the state of the qubit which she measures
and sends on to Bob will be the same as the one Alice sent. Hence, for these
photons, Eve's interception can not be detected. However, from
\begin{align}
H|0\rangle &= {1 \over \sqrt{2}}\left(|0\rangle + |1\rangle \right), \qquad\quad
H|1\rangle = {1 \over \sqrt{2}}\left(|0\rangle - |1\rangle \right), \nonumber \\
|0\rangle &= {1 \over \sqrt{2}}\left(H|0\rangle + H|1\rangle \right),
\qquad
|1\rangle = {1 \over \sqrt{2}}\left(H|0\rangle - H|1\rangle \right).
\end{align}
we see that, out of
the times when Eve chooses a different basis from the common basis of Alice
and Bob,
there is a
probability 1/2 that Eve's intervention will result in her sending on to 
Bob a photon in the opposite state from the one which Alice sent.
Hence, for the photons where Alice and Bob used the same
basis, Eve's intervention results in Alice and Bob having different states
about 1/4 of the time\footnote{There is a probability $1/2$ that Eve measures
in a different basis and for \textit{those} qubits there is a probability $1/2$ that her
measurement changes the state.}.

To see if this is happening, Alice and Bob sacrifice some fraction of
the good photons by sending their values for the state down an insecure
channel. If about 1/4 of the states disagree, then they know that the photons
are being intercepted.  If only a small fraction disagree, Alice and Bob would
have needed to decide beforehand up to what fraction of disagreements they would consider 
an acceptable risk in order to still send the message.

In summary, a quantum key distribution protocol is able to detect an
eavesdropper because measurements in quantum mechanics in general change the
state.

\subsection{BB92 protocol}
\index{quantum key distribution!BB84 protocol}
\index{quantum key distribution!BB92 protocol}
There is a later version, also due to Bennet and Brassard, from 1992 (BB92), in
which only two polarizations are used: $\leftrightarrow$ and
\index{photon!polarization}
\makebox[0pt][l]{$\nearrow$}$\swarrow$. Note that these states are not
orthogonal. Lack of orthogonality is essential for the method to work. If only
orthogonal states are used then there is only one basis, so if Eve knows what
this is
she can measure the states of the photons
in this basis and send then on to Bob without being detected.

The BB92 protocol works as follows. To decide in which state to send the $k$-th
photon, Alice
generates a random bit, $k_i$, which is 0 or 1. If she gets 0 she sends
$|\leftrightarrow\rangle \equiv |0\rangle$ a
$Z$-type photon, whereas if she gets 1 she sends
$|\rotatebox[origin=c]{45}{\Large$\leftrightarrow$}\rangle \equiv H|0\rangle$,
an $X$-type photon.

If Bob were to always measure in the same basis as the one Alice used, i.e.~the $Z$ basis
for $Z$-type photons, and the $X$ basis for $X$-type photons, he would always
get $|0\rangle$ (in whatever basis is being used, $Z$ or $X$).
However, he doesn't know which basis Alice used, so, for each photon,
he chooses a random basis by generating a random bit $l_i$.
As Alice also did, Bob chooses the $Z$-basis if 
$l_i$ is 0 and the $X$ basis if the $l_i=1$. He notes for
which photon he \textit{measures} $|1\rangle$ and sends this information to Alice on a
public channel. This
only happens when they use different bases, i.e.~they generate complementary
random bits, $k_i = 1-l_i$, since if they use the same basis Bob must get
$|0\rangle$.
The shared key is then the set $\{l_i\}$ for
which Bob measures $|1\rangle$. Alice just has to take the complement of her bits for
the same photons to get the same key as Bob. Note that Bob measures $|1\rangle$ either
if Alice chooses a $Z$-basis and Bob an $X$-basis, or vice versa, but
information as to which one is chosen is not transmitted down the public channel. 

If an eavesdropper intercepts the qubits to try to determine this information, the
result is similar to that for the BB84 protocol.
Alice and Bob could check, via a public channel, some of the bits of the key.
If the qubits are being intercepted, Alice and Bob would find that
for about $1/4$ of them, they actually used the same basis. 

\section{Teleportation}
\label{sec:teleport}
\index{teleportation}
Another quantum protocol which has been successfully implemented using photons
is teleportation, in which the state of a qubit, though not the physical qubit
itself, is transported (teleported) from one location to another. The best
reference for this is Mermin~\cite{mermin:07}.

Suppose that Alice has a qubit in a state
\begin{equation}
|\psi\rangle = \alpha |0 \rangle + \beta |1\rangle.
\end{equation}
The values of $\alpha$ and $\beta$ are unknown to her and can not be
determined by measurement because this destroys the superposition.
\index{superposition}
\index{no-cloning theorem}
The no-cloning theorem means that we can't do
repeated measurements on copies of this state to estimate the amplitudes
$\alpha$ and $\beta$.
Alice's qubit may be the result of a (possibly complicated) quantum computation
which she would like to send on to Bob to continue the computation.  Bob is
far away and Alice can not physically transport the qubit to Bob but wants to
send the state.

We assume that and Bob are able to
\begin{itemize}
\item
share an entangled qubit
\begin{equation}
|\beta_{00}\rangle = {1 \over \sqrt{2}} \left(\, |0\rangle_a  |0\rangle_b +
|1\rangle_a  |1\rangle_b\, \right) ,
\end{equation}
where $a$ stands for Alice's qubit and $b$ stands for Bob's, 
\item
and communicate over a classical channel (e.g.~by phone).
\end{itemize}
Hence, together they have a 3-qubit state,
\begin{align}
|\phi_0\rangle &= {1 \over \sqrt{2}} \left(\, \alpha |0 \rangle_a + \beta
|1\rangle_a\, \right)\otimes \left(\, |0\rangle_a  |0\rangle_b +
|1\rangle_a  |1\rangle_b\, \right) \\
&= {1 \over \sqrt{2}} \left(\, \alpha |000 \rangle + \alpha |011 \rangle + 
\beta |100\rangle + \beta |111\rangle \, \right) ,
\end{align}
where the leftmost two qubits refer to Alice and the rightmost qubit to Bob.

\begin{figure}[tbh!]
\begin{center}
\includegraphics[width=8.5cm]{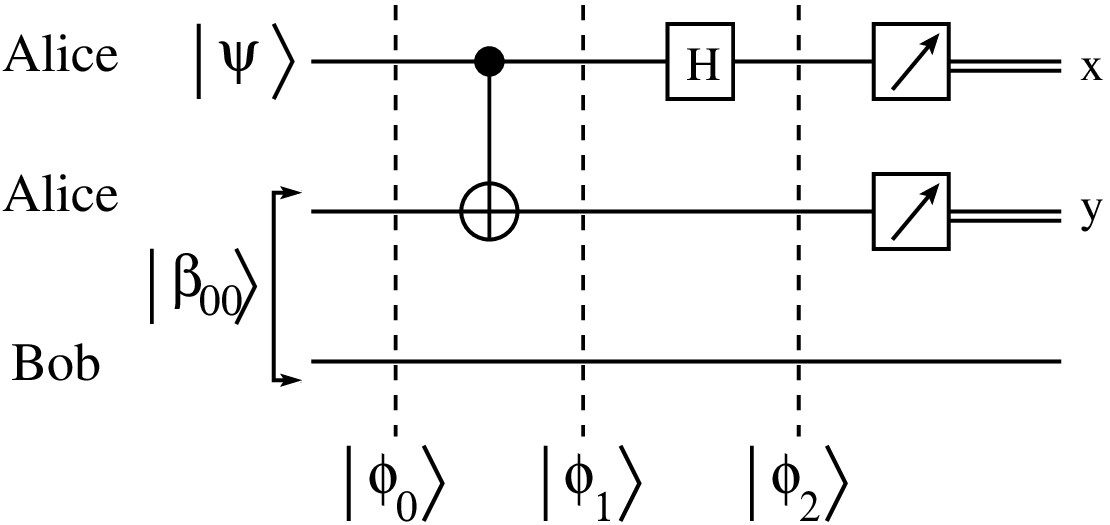}
\caption{
The circuit used in teleportation. The states $|\phi_0\rangle$,
$|\phi_1\rangle$ and $|\phi_2\rangle$ 
are described in the text.
\label{teleport}}
\end{center}
\end{figure}

Alice now applies a Bell measurement\index{Bell measurement}
(discussed in Chapter \ref{ch:bell}) to the two qubits in
her possession, see Fig.~\ref{teleport}.

We obtain $|\phi_1\rangle$ from $|\phi_0\rangle$ by acting with the CNOT gate, so
\begin{equation}
|\phi_1\rangle =
{1 \over \sqrt{2}} \left(\, \alpha |000 \rangle + \alpha |011 \rangle +
\beta |110\rangle + \beta |101\rangle \, \right) ,
\end{equation}
where the leftmost qubit is Alice's control qubit, the middle one is Alice's
target qbit, and the rightmost one is Bob's qubit.

$|\phi_2\rangle$ is obtained from $|\phi_1\rangle$ by applying a Hadamard to
Alice's control qubit, so
\begin{align}
|\phi_2\rangle &=
{\alpha \over \sqrt{2}}\left[{1 \over \sqrt{2}}
\left( |0\rangle_a +|1\rangle_a \right)
\left( |00\rangle_{ab} + |11\rangle_{ab} \right)\right]
+ {\beta \over \sqrt{2}} \left[\left(|0\rangle_a -|1\rangle_a \right)
\left(|10\rangle_{ab} + |01\rangle_{ab}\right) \right] \nonumber \\
&= {1 \over 2} \left[
\alpha \left(|000\rangle + |100\rangle + |011\rangle + |111\rangle
\right) +
\beta \left(|010\rangle - |110\rangle + |001\rangle - |101\rangle
\right)
\right].
\end{align}
Separating out each of the four states of Alice's qubits, $|\phi_2\rangle$ can be
written as
\begin{align}
|\phi_2\rangle =
&{1 \over 2} |00\rangle_a \otimes \left( \alpha|0\rangle_b + \beta |1\rangle_b\right) +
\nonumber \\
&{1 \over 2} |01\rangle_a \otimes \left( \alpha|1\rangle_b + \beta |0\rangle_b\right) +
\nonumber \\
&{1 \over 2} |10\rangle_a \otimes \left( \alpha|0\rangle_b - \beta |1\rangle_b\right) +
\nonumber \\
&{1 \over 2} |11\rangle_a \otimes \left( \alpha|1\rangle_b - \beta |0\rangle_b\right) .
\end{align}

Alice then measures the two qubits in her possession, obtaining results $0$
or $1$ for $x$
and $y$ with equal probability. We see that if she gets $x=y=0$ Bob's
qubit is already in the desired state
\begin{equation}
|\psi\rangle_b = \alpha|0\rangle_b + \beta |1\rangle_b .
\end{equation}
In all cases, after Alice has
communicated her values for $x$ and $y$
to Bob over a classical channel, Bob can act with an appropriate gate on his
qubit to transform it to $|\psi\rangle_b$. The different possibilities are
shown in the table below.

\begin{center}
\begin{tabular}{|c|c|c|}
\hline
$x\quad y$ & State of Bob's qubit & Gate Bob needs to apply \\
\hline\hline
$0 \quad  0$ & $\alpha|0\rangle_b + \beta |1\rangle_b$ & $\mathbbm{1}$ \\
$0 \quad  1$ & $\alpha|1\rangle_b + \beta |0\rangle_b$ & $X$ \\
$1 \quad  0$ & $\alpha|0\rangle_b - \beta |1\rangle_b$ & $Z$ \\
$1 \quad  1$ & $\alpha|1\rangle_b - \beta |0\rangle_b$ & $ZX$ \\
\hline
\end{tabular}
\end{center}
After Bob has acted with the appropriate gate, he will have in his possession
the qubit that has been teleported from Alice.  Note that Alice no longer has
this qubit, only the two classical bits $x$ and $y$, so there is no violation
of the no-cloning theorem. There is also no violation of special relativity
since Alice needs to communicate her measurement results to Bob via a
classical channel.

%
 
\noindent \textit{Final Comment:}\\
There are claims that teleportation has been verified experimentally which I
will now discuss briefly. One would like to show the following:
\begin{itemize}
\item
Alice stores state $|\psi\rangle$.
\item
The state $|\psi\rangle$ is transported to Bob who is far away.
\item
Bob stores state $|\psi\rangle$.
\end{itemize}

To transport qubits over a long distance one needs photons. One can teleport
photons over a large distance while retaining their polarization, but at
\index{photon!polarization}
present one can not conveniently store them. 
One can store other types of qubits, e.g.~trapped ions, but can't entangle
them over large distances, so they can be teleported only locally.  Hence, in
my view, a complete demonstration of teleportation, incorporating all three
bullet points above, has not yet been achieved.


\hrulefill
\section*{Problems}
\input{hw_ch21.tex}

%% file: hw_ch21.tex
\begin{problems}

\item {\bf BB84 Quantum Key Distribution}\\
\index{quantum key distribution!BB84 protocol}
Consider the BB84 Quantum Key Distribution (QKD) protocol discussed in this
chapter.
\textbf{Assume that Eve intercepts every qubit (photon)} that Alice sends,
and then
transmits it to Bob. Like Alice and Bob, Eve chooses one the bases (the
$\mathbbm{1}$ or the $H$ basis) at random.
Alice and Bob compare, over a public channel which can be
intercepted by Eve, which qubits they used the same basis for (Alice for sending
and Bob for measuring.) The values of these qubits (0 or 1) (which Alice and
Bob agree on if Eve did not eavesdrop) form the shared key.
\begin{enumerate}[label=(\roman*)]
\item
For what fraction of the shared key qubits would Alice and Bob get different
results for the qubit due to Eve's interception. (If Eve had not intercepted
the qubits, then Alice and Bob would agree for all qubits in the shared key.)
\item
Supposing that the shared key has $10$ qubits, what is the probability that all
of Alice's and Bob's qubits would agree (in which case Eve's eavesdropping would not be
detected?
\item
What is the probability that all qubits would agree if the shared key has
$100$ qubits?
\end{enumerate}

\end{problems}

%% file: qu_sim8.tex
We are currently in the middle of what is
called the
\index{second quantum revolution}
``second quantum revolution". The first quantum revolution was the
development of quantum mechanics in the 1920's and subsequent
applications to devices like integrated circuits, which use quantum mechanics
in the design of the hardware, but these applications treat
the information, i.e.~the bits, classically.  However, in the second quantum revolution, 
the information itself is treated according to the rules of quantum mechanics.

In this book we have discussed what is called the circuit model (or gate model)
of a quantum computer. The qubits are
initialized, and then acted on by a series of discrete unitary
transformations to solve the problem at hand. This sort of quantum device is
what people normally refer to when they talk about a ``quantum computer".
The circuit model quantum computer was proposed initially by David Deutsch
\cite{deutsch:85}\index{Deutsch, David}.

However, other types of quantum device are being developed as part of the
\index{quantum simulator}
second quantum revolution, which can be termed ``quantum simulators".
It is anticipated that we will have interesting new results from quantum simulators, 
i.e.~results which could not be obtained
by a classical computer, in the next few years. By contrast, the ability to
get interesting new results from a circuit model quantum computer, for example by
decoding information sent down the internet using Shor's algorithm (which
requires factoring a huge integer) will be very far in the future, if
ever.

The idea of a quantum simulator is to use an artificial quantum device to
simulate the quantum system which we want to understand. It was first proposed
by Feynman\cite{feynman:82}\index{Feynman, Richard}. For example a quantum chemist might want to
understand the properties of a certain molecule, or a condensed matter
physicist might want to understand a material with unusual
magnetic or superconducting behavior. Properties of these materials are,
of course, determined by quantum mechanics.  Many problems in nature are not amenable
to analytic (i.e.~pencil and paper) calculations and need to be simulated.
Although many problems can be simulated efficiently on a
classical computer, there remain problems of interest where the
quantum aspects cause serious difficulties for classical simulations. As an
example, we learn in quantum mechanics classes that particles of a
certain type (e.g.~electrons or protons or $\pi$ mesons) are (i) all identical and (ii) are
in one of two classes, bosons or fermions.  For bosons, the state of the
system (wave function) does not change if the two particles are interchanged,
whereas for fermions the state does change sign under particle interchange. This
sign change for fermions can create great difficulty when trying to simulate
\index{sign change under particle  interchange}
fermions on a classical computer. 

\index{analog device}
By and large quantum simulators are \textit{analog} devices. The reason
is that, in order for the system of qubits to model the problem of interest,
there must be interactions between the qubits.
In a classical
(digital) computer they would be represented by floating point numbers with
typically 16 digits or precision\footnote{For many purpose this can be
considered exact but, in any case, the interactions are represented by a
precisely \textit{known} string of bits.}. In a quantum computer, however, interactions are
induced by turning some ``knob" on the experimental apparatus, the nature of
the knob depending on the hardware used for the qubits. For example, in the
case of superconducting qubits, interactions would be determined by the
value of a magnetic field threading superconducting loops. The magnetic field
takes a continuous range of values (i.e.~is analog) and can only be set within
a certain level of precision. 

Above I stated that we will probably have interesting new results from a quantum simulator
before we have new results from a (circuit model) quantum computer. Why is this? A
quantum computer uses quantum parallelism to get its quantum speedup.  This
depends on accurately preserving phase relations between the different pieces
of the state. These phase relations are destroyed by noise, an effect called
decoherence. Present-day qubits are quite noisy. In principle one can include error
\index{decoherence}
correction, but this requires a huge number of physical
qubits for each logical qubit. Thus, in the near future, we will have to
live with noisy qubits. However, as stated, noise is a disaster for circuit
model quantum computers.

Is a modest amount of noise as big a disaster for a quantum simulator? The
answer is ``probably not". For example, suppose we want to simulate the
temperature dependence of the behavior of a material which goes
superconducting. A non-zero temperature means that there is noise due to
thermal fluctuations.  One might hope that a bit of extra (even non-thermal) noise
from the qubits would not change the results all that much, so the results
would, nonetheless, be useful. The next
paragraph discuses another example for which there
is also reason to believe that some noise is not disastrous. 

\index{optimization problems}
A particular type of quantum simulator is one used to solve ``optimization"
problems, where we need to find the maximum (or minimum) of some
\index{objective function}``objective function'' with
constraints. Let's assume
for concreteness that we want
the minimum. Optimization problems are very important in science and
engineering, two widely used applications being speech recognition and image
recognition. Optimization problems are hard when there is ``frustration",
\index{frustration}
i.e.~competition, between different pieces of the function that one has to
minimize. In these cases, if one locally minimizes individual pieces, one will end
up in a ``local minimum"\index{local minimum} rather than the
\index{global minimum}global minimum. It has been proposed
\index{quantum annealing}
to use ``quantum annealing"\footnote{It was earlier proposed to add thermal
fluctuations to solve optimization problems. This approach is called ``thermal
annealing" or ``simulated annealing". Whether quantum annealing is more efficient in finding
ground states than classical algorithms such as simulated annealing is hotly
debated at present.} to try to find the global minimum. We recall from
\index{thermal annealing}\index{simulated annealing|see {thermal annealing}}
Chapter \ref{ch:qu_intro} that if we have two operators which don't commute
then one or both of them must have an uncertainty in any quantum state. Thus
non-commuting operators generate fluctuations.  By making the (classical)
objective function become quantum by adding a non-commuting ``driver" piece to it, one induces
fluctuations, which can help get one out of a local minimum.  In such a
``quantum annealer" the qubits simulate the ``objective function plus driver
function". By letting the
driver piece tend to zero at the end of the simulation, the
model simulated at the end is
just the objective function, and we anticipate that the set of qubits will
then be close to the ground state. Quantum annealing has been pioneered by
a company called \index {D-Wave}D-Wave, which has manufactured machines with around 5000
qubits. These 5000 qubits do not maintain coherence during the time of the
simulation, but it is anticipated that, despite some noise, the induced quantum
fluctuations will help to find the ground state.

To summarize, I anticipate that in the near future qubits will be noisy and we won't be capable
of assembling a huge number of them together. Hence, in the short and intermediate term, 
we will only have ``Noisy Intermediate-Scale
\index{NISQ devices}
Quantum'' (NISQ) devices. I expect that in the next few years we will
be able to get interesting, new\footnote{By ``new" I mean
results that would be impossible
to obtain on a classical computer. By ``interesting" I mean results that scientists would like to
know for their own sake, not just as an illustration of the capabilities of a 
quantum computer.} results from NISQ \textit{simulators}, but probably not
from NISQ circuit model \textit{quantum computers}.

\noindent\textit{Additional note in November 2026.}\\
Perhaps the above remark is too pessimistic.
Recently (March 2026) Google 
warned that, due to advances in quantum hardware,
quantum computers might be able to break RSA encryption by as
early 2029.
Hence, much work is being done to 
develop quantum-safe encryption schemes that can not be broken by some version of Shor's
algorithm, and Google has set an internal deadline to change its own systems
to be quantum-safe by 2029.

I await 2029 with some concern to see if Google's forecast is correct.


\begin{center}
{\Large\bf Appendix}
\end{center}

\begin{subappendices}
\section{The 2025 Physics Nobel Prize}
\label{sec:nobel_2025}
The D-Wave machine, and many other current implementations of quantum computers use
superconducting qubits. Each qubit is a macroscopic circuit, which is big
enough that it can be seen with the naked eye.
This is a surprise, since normally one imagines that quantum effects only
occur in objects that are atomic size or smaller. However, starting with
the work of John Clarke\index{Clarke, John}, Michel Devoret\index{Devoret, Michel},
and John Martinis\index{Martinis, John} in Berkeley around
1984, it is understood that quantum effects \textit{can} occur in superconducting
circuits. Strictly speaking, it is not the size of the system which matters, but rather the
number of degrees of freedom. In a superconductor there is an energy gap
between the ground state and excited states, so when the temperature is much
less than the gap, all excitations are frozen out and the only degree of
freedom is the direction of flow of the persistent current round the superconducting loop. 
Clarke, Devoret and Martinis received the 2025 Physics Nobel Prize\index{Nobel
Prize, 2025} for this work. 

\end{subappendices}